\documentclass[12pt,]{article}
 \usepackage{graphicx}
 \usepackage[cp1251]{inputenc}  
 \usepackage[russian]{babel}    %
\begin{document}
\centerline{\small\it
 Publications of the Pulkovo Observatory (ISSN 0367--7966), Issue 228, DOI:10.31725/0367-7966-2023-228-1}
 \bigskip
 \bigskip

\centerline {\bf GLOBULAR CLUSTERS IN THE CENTRAL REGION }
\centerline {\bf OF THE MILKY WAY GALAXY I. BAR INFLUENCE}
\centerline {\bf ON THE ORBIT PARAMETERS ACCORDING TO GAIA EDR3}

  \bigskip

 \centerline{\bf
            A. T. Bajkova\footnote [1]{bajkova@gaoran.ru} (1),
            A. A. Smirnov (1),
            V. V. Bobylev (1)
            }
  \bigskip
 \centerline{\small \it $^1$ Main (Pulkovo) Astronomical Observatory of the RAS, St. Petersburg, Russia}
  \bigskip

 {\noindent The work is devoted to the analysis of the influence of the galactic bar on the orbital motion of globular clusters in the central region of the Galaxy. For this task, 45 globular clusters were selected, 34 of which belong to the bulge/bar and 11 to the disk. The most accurate astrometric data from the Gaia satellite (Vasiliev and Baumgardt, 2021), as well as new refined average distances (Baumgardt and Vasiliev, 2021), were used to form the 6D-phase space required for orbit integration. The orbits of globular clusters are obtained both in an axisymmetric potential and in a potential including a bar. In this case, the mass, rotation velocity, shape and scale length of the bar were varied. A comparison is made of such orbital parameters as apocentric and pericentric distances, eccentricity and maximum distance from the galactic plane. It is shown that the mass of the bar exerts the greatest influence on the orbital motion, which is expressed mainly in an increase in both the apocentric and pericentric distances in the vast majority of globular clusters. The eccentricities of the orbits in the overwhelming majority also change significantly, and there is a change both upward and downward, especially in the range of values from 0.2 to 0.8. The greatest changes in parameters are observed in globular clusters with high radial velocities and small pericentric distances. The change in orbital parameters depending on the bar rotation velocity is less pronounced. The influence of the geometric parameters of the bar is insignificant in the accepted range of their changes. Several examples show that globular clusters in the bulge are more affected by the bar than those belonging to the disk.
}

\section*{Introduction}Currently, about 170 galactic globular clusters (GCs) are known. According to theoretical estimates, their number in the Galaxy can reach 200 [1]. More than 150 GCs have complete astrometric (positional) and kinematic measurements (proper motions and radial velocities) to form the 6D-phase space needed to build orbits. The study of the orbital motion of GCs is of great importance for studying the evolution of the Galaxy, since they are the oldest objects, whose age reaches 13 billion years.

In most studies, the motion of galactic globular clusters is considered in an axisymmetric stationary potential. However, the real potential of our galaxy is neither axisymmetric nor stationary. And this is primarily due to the fact that a rotating elongated bar is located in the center of the Galaxy.

The purpose of this work is to study the orbital motion of the GCs both in an axisymmetric potential and in a potential including a bar. Since the parameters of the bar are known with great uncertainty, we analyze the influence of the bar at different values of its parameters. Obviously, the orbits that occupy the central part of the Galaxy, mainly the bulge region, with a diameter of about 4 kpc are subject to the greatest influence [2].

To form the 6D-phase space required for orbit integration, we use the latest, most accurate version of the GC proper motion catalog [3] based on Gaia EDR3 measurement data, as well as new average GC distances [4].

Note that this work represents only the first stage of research, which is devoted to the comparison of such orbital parameters as apocentric and pericentric distances, eccentricity, and the maximum distance from the galactic plane, obtained by integration in different potentials. The second stage of the work will be devoted to studying the influence of the bar on the frequency characteristics of the orbits in order to identify globular clusters captured by the bar.

The work is structured as follows. The first section gives a brief description of the accepted models of the potential - the axisymmetric potential and the potential with a bar. The second section provides links to the used astrometric data. The third section is devoted to the selection of globular clusters. In the fourth section, a comparative analysis of the results of orbit integration is given. The Conclusion formulates the main results.

\section{Galactic Potential Model}

\subsection{Axisymmetric potential}
The axisymmetric gravitational potential of the Galaxy is represented as
as the sum of three components~--- central spherical
bulge $\Phi_b(r(R,Z))$, disk $\Phi_d(r(R,Z))$ and massive
spherical dark matter halo $\Phi_h(r(R,Z))$:

 \begin{equation}
 \begin{array}{lll}
  \Phi(R,Z)=\Phi_b(r(R,Z))+\Phi_d(r(R,Z))+\Phi_h(r(R,Z)).
 \label{pot}
 \end{array}
 \end{equation}
Here we use a cylindrical coordinate system ($R,\psi,Z$) with
the origin of coordinates at the center of the Galaxy. In a rectangular system
coordinates $(X,Y,Z)$ with the origin at the center of the Galaxy,
the distance to the star (spherical radius) will be
$r^2=X^2+Y^2+Z^2=R^2+Z^2$, while the $X$ axis is directed from the Sun to the galactic center, the $Y$ axis is perpendicular to the $X$ axis  towards the rotation of the Galaxy, the axis $Z$ is perpendicular to the galactic plane $XY$ towards the north galactic pole.
The gravitational potential is expressed in
units of 100 km$^2$/s$^2$, distances~--- in kpc, masses~--- in
units of galactic mass $M_G=2.325\times 10^7 M_\odot$ ($M_\odot$ is the mass of the Sun),
corresponding gravitational constant $G=1$.

Potentials of the bulge $\Phi_b(r(R,Z))$ and the disk $\Phi_d(r(R,Z))$
are presented in the form proposed by Miyamoto, Nagai~[5]:
 \begin{equation}
  \Phi_b(r)=-\frac{M_b}{(r^2+b_b^2)^{1/2}},
  \label{bulge}
 \end{equation}
 \begin{equation}
 \Phi_d(R,Z)=-\frac{M_d}{\Biggl[R^2+\Bigl(a_d+\sqrt{Z^2+b_d^2}\Bigr)^2\Biggr]^{1/2}},
 \label{disk}
\end{equation}
where $M_b, M_d$~are masses, $b_b, a_d, b_d$~are scale
lengthes of the components in kpc. The halo component is represented according to
Navarro, Frank and White [6]:
 \begin{equation}
  \Phi_h(r)=-\frac{M_h}{r} \ln {\Biggl(1+\frac{r}{a_h}\Biggr)}.
 \label{halo-III}
 \end{equation}
Table~1 contains the values of the model
galactic potential parameters(\ref{bulge})--(\ref{halo-III}), which
were found by Bajkova, Bobylev~[7] using the Galactic rotation curve
[8], built on the basis of objects located on
distances $R$ upto $\sim200$~kpc. Note that when constructing this galactic
rotation curve, there were used the following values of
local parameters: $R_\odot=8.3$~kpc and $V_\odot=244$~km/s. In work
[7], the model (\ref{bulge})--(\ref{halo-III}) is denoted as
model~III. The adopted potential model is the best among the six models considered in [9], since it provides the smallest discrepancy between the data and the model rotation curve.

\subsection{Model of the bar}

As the potential of the central bar was chosen the model of a triaxial ellipsoid~[10]:
\begin{equation}
  \Phi_{bar} = -\frac{M_{bar}}{(q_b^2+X^2+[Ya/b]^2+[Za/c]^2)^{1/2}},
\label{bar}
\end{equation}
where $X=R\cos\vartheta, Y=R\sin\vartheta$, $a, b, c$~are three semiaxes of the bar, $q_b$~ scale length parameter of the bar; $\vartheta=\theta-\Omega_{b}t-\theta_{b}$, $tg(\theta)=Y/X$, $\Omega_{b}$~ is a rotation velocity of the bar, $t $~ is integration time, $\theta_{b}$~is the bar orientation angle relative to the galactic axes $X,Y$, counted from the line connecting the Sun and the center of the Galaxy ($X$ axis) to the major axis of the bar
in the direction of rotation of the galaxy.

The simulation of the galactic potential with a bar was carried out by varying all the parameters of the bar model. Three values of the bar mass were considered: $M_{bar}=430,280,130 M_G$ (in this case, the bar mass was subtracted from the mass of the bulge), three values of the bar rotation velocity:
$\Omega_{b}=30,40,50$~km/s/kpc, three bar lengths: $q_b=4.5,5.0,5.5$ kpc and 5 options (V0,V1,V2,V3,V4) for ratios of the ellipsoid semi-axes that define the shape of the bar (table~1).
Based on the information in numerous literature, the following parameters were used as the basic bar parameters: $M_{bar}=430 M_G$, $\Omega_{b}=40$~km/s/kpc, $q_b=5$ kpc, $ \theta_{b}=25^o$, ratio of the axes of the ellipsoid V0. The last three parameters are adopted, in particular, in [10].

 {\begin{table}[t]                                    
 {\small\baselineskip=1.0ex
{\bf Table~1:} The values of the parameters of the galactic potential model, $M_G=2.325\times 10^7 M_\odot$
  }
 \label{t:model-III}
 \begin{center}\begin{tabular}{|c|r|}\hline
 $M_b$ &   443 M$_G$ \\
 $M_d$ &  2798 M$_G$ \\
 $M_h$ & 12474 M$_G$ \\
 $b_b$ & 0.2672 kpc \\
 $a_d$ &   4.40 kpc  \\
 $b_d$ & 0.3084 kpc  \\
 $a_h$ &    7.7 kpc  \\
\hline\hline
 $M_{bar}$ & 430, 280, 130 M$_G$ \\
 $\Omega_b$ & 30, 40, 50 km/s/kpc \\
 $q_b$     &  4.5, 5.0, 5.5 kpc  \\
 $\theta_{b}$ &  $25^o$   \\\hline
 $a/b$ & 2.38, 3.33, 2.50, 2.50, 3.33  \\
 $a/c$ & 3.03, 2.50, 3.33, 2.50, 3.33  \\
  Options & ~~V0, ~~V1, ~~V2, ~~V3, ~~V4 \\
  \hline
 \end{tabular}\end{center}\end{table}}

To integrate the equations of motion, we used
fourth-order Runge-Kutta algorithm.

The value of the peculiar velocity of the Sun relative to the Local
Standard of Rest was taken equal to
$(u_\odot,v_\odot,w_\odot)=(11.1,12.2,7.3)\pm(0.7,0.5,0.4)$~km/s
according to the work~[11]. The elevation of the Sun above the plane of the Galaxy is assumed to be 17 pc in accordance with [12].

\section{Data}
Data on the proper motions of globular clusters are taken from the new catalog by Vasiliev and Baumgardt (2021) [3], based on Gaia EDR3 observations. Radial velocities
are taken from Vasiliev (2019) [13], which, in turn, were taken mainly from Baumgardt et al. (2019) [14]. The new average distances to globular clusters are taken from Baumgardt and Vasiliev (2021) [4]. A comparative analysis of new data on proper motions and distances with previous versions of the catalogs is given, for example, in [15]. Here, we only note that the accuracy of measuring the new proper motions has increased on average by a factor of two compared to the Gaia DR2 measurements.
As an analysis of the radial velocities presented in the Vasiliev catalog [13] shows, their errors are not significant and for the overwhelming majority of GCs in our sample are less than 1\% of the values of the velocities themselves. More significant errors are still burdened by proper motions and distances. All uncertainties in the data (on positions, distances, proper motions, line-of-sight velocities) are taken into account by the Monte-Carlo method (1000 realizations) when calculating the uncertainties of the components of the 6D-space $(x_o,y_o,z_o,u_o,v_o,w_o)$ (see notation in [15]), which was later used to calculate the orbits. The uncertainties of the orbital parameters: apocentric distance, pericentric distance, eccentricity, and maximum distance from the galactic plane ($apo, peri, Zmax$, and $ecc$, respectively) were also calculated using the Monte-Carlo method (1000 realizations), taking into account the uncertainties of the 6D-phase space components, as well as the uncertainties of the peculiar velocity vector of the Sun according to Sch\"onrich et al. [11]. The equations of motion of a test particle in the gravitational potential of the Galaxy are presented in detail in [15]. The results of calculations of the parameters of the orbits and their uncertainties, made both in the axisymmetric potential and in the potential with a bar with different parameters, are given in tables 2-6.

\section{Selection of globular clusters}

The GC catalog [15] that we have at our disposal includes 152 objects.
The selection of globular clusters from this set belonging to the bulge/bar region was carried out in accordance with a purely geometric criterion considered in [16] and also used by us in [17]. It is very simple and consists in selecting GCs whose orbital apocentric distance does not exceed 3.5 kpc. The orbits are calculated in the axisymmetric potential.

Note that the use of the classification results obtained in [16, 17] for proper motions from the Gaia DR2 catalog and Harris distances [18] according to the version of the catalog data [19] is not entirely correct in this work, since we are dealing with new data. Changes in data (especially in distances) can greatly affect the orbital characteristics of the GCs. Therefore, we carried out a new classification. The application of the restriction on the GC apocentric distance ($apo\leq 3.5$ kpc) made it possible to identify 39 objects, most of which coincided with the results of the previous classification [17, 20], but there are also differences.

Next, from the resulting set of GCs, we singled out the GCs that belong to the disk component of the Galaxy. To do this, we applied the probabilistic method proposed by us in [17] for dividing GCs into subsystems of the Galaxy. It is based on the bimodal distribution of the $L_Z/ecc$ parameter, where $L_Z$ is the vertical ($Z$) component of the angular momentum, $ecc$ is the orbital eccentricity. An illustration of the method is shown in the upper left panel of fig.~1. As a result of applying this method, it was possible to identify 9 GCs belonging to the disk. The remaining 30 belong directly to the bulge/bar. To the obtained sample of 39 objects, we added 6 more GCs belonging to the bulge/bar region according to the classification of Massari et al. (2019) [16], but not included in our sample due to a slight excess of the apocentric distance of the orbits.

Thus, we have formed a sample of 45 GCs in the central region of the Galaxy, of which 34 belong to the bulge/bar, 9 belong to the disk (ESO 456-SC78, Terzan 3, Pismis 26, NGC 6256, 6304, 6569. 6540.6539, 6553), and 2 more objects (NGC 6325, Djorg 2), which have significant negative rotational velocities (retrograde orbits), presumably also belong to the disk.
The complete list of 45 objects is given below in the first column of tables 2--6 with calculated orbital parameters. In the upper part of the tables, 34 bulge/bar objects are presented, in the lower part - 11 disk objects listed above. The first 30 bulge/bar objects are GCs identified as a result of applying the probabilistic separation method, the remaining 4 are GCs according to the classification of Massari et al. (2019). The notation of the orbital parameters and the formulas for their calculation can be found, in particular, in [15], which also contains a table with the full set of parameters calculated in the axisymmetric potential we adopted. Figure~1 also represents the diagrams $"$$L_Z/ecc$ -- total energy $E$$"$, $"$circular velocity -- eccentricity$"$ and $"$radial velocity -- circular velocity$"$, which clearly show the division of the GCs into bulge/bar and disk subsystems.

Figure~2 shows the distribution of the selected 45 GCs in the $X-Y$ and $X-Z$ projections of the galactic coordinate system. The figure also shows sections of the triaxial ellipsoid that describes the bar, with parameters $q_b=5$ kpc, $\theta_{b}= 25^o$, and the ratio of the axes V0.

\begin{figure*}
{\begin{center}
 \includegraphics[width=0.3\textwidth,angle=-90]{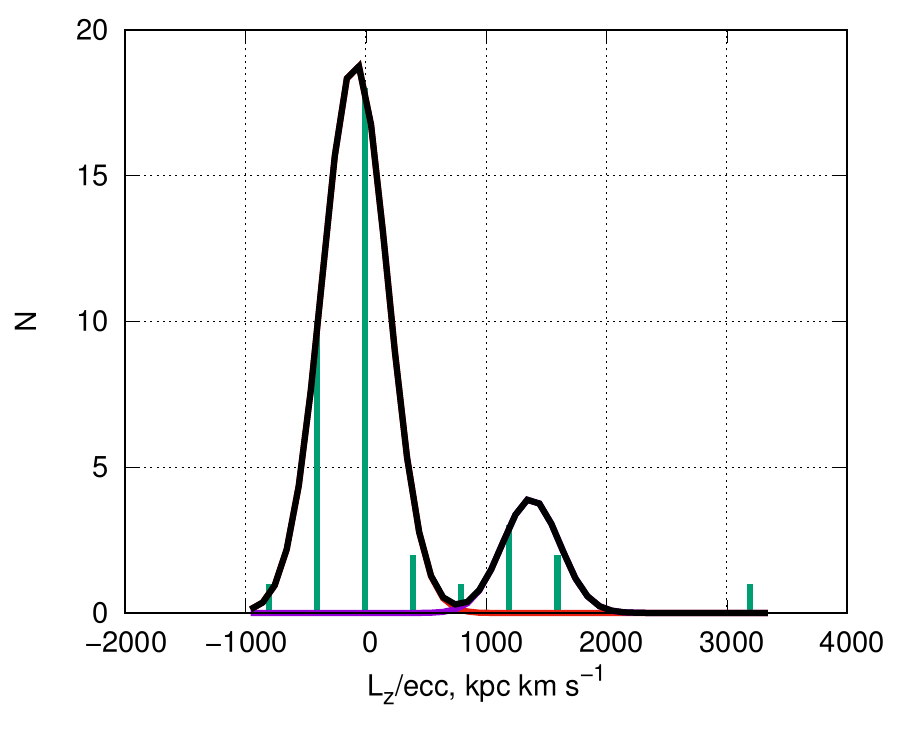}
\includegraphics[width=0.3\textwidth,angle=-90]{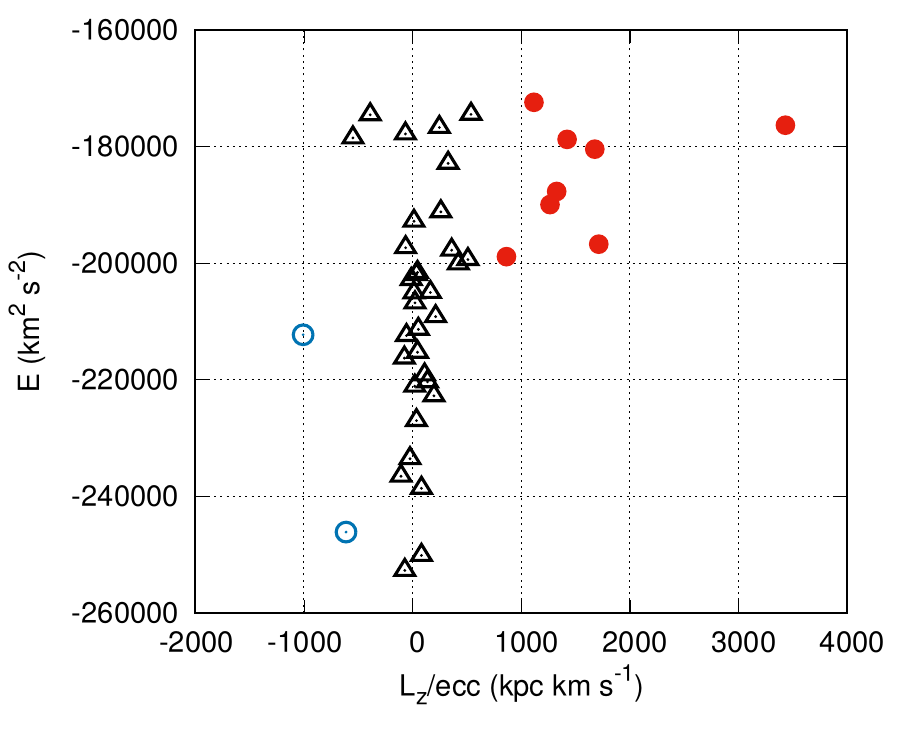}\

\medskip

\includegraphics[width=0.3\textwidth,angle=-90]{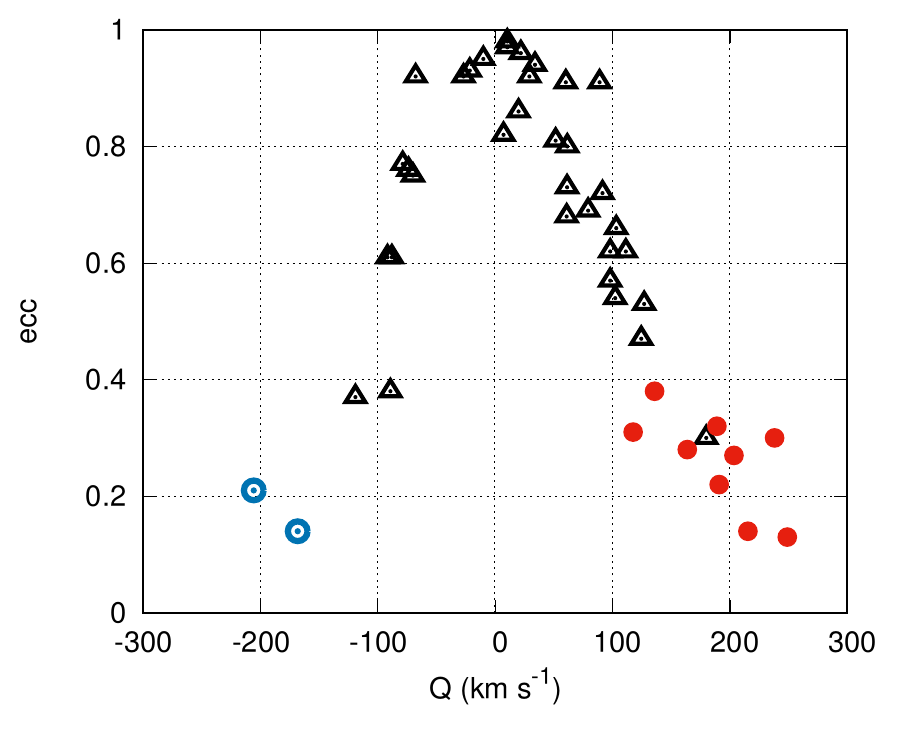}
\includegraphics[width=0.3\textwidth,angle=-90]{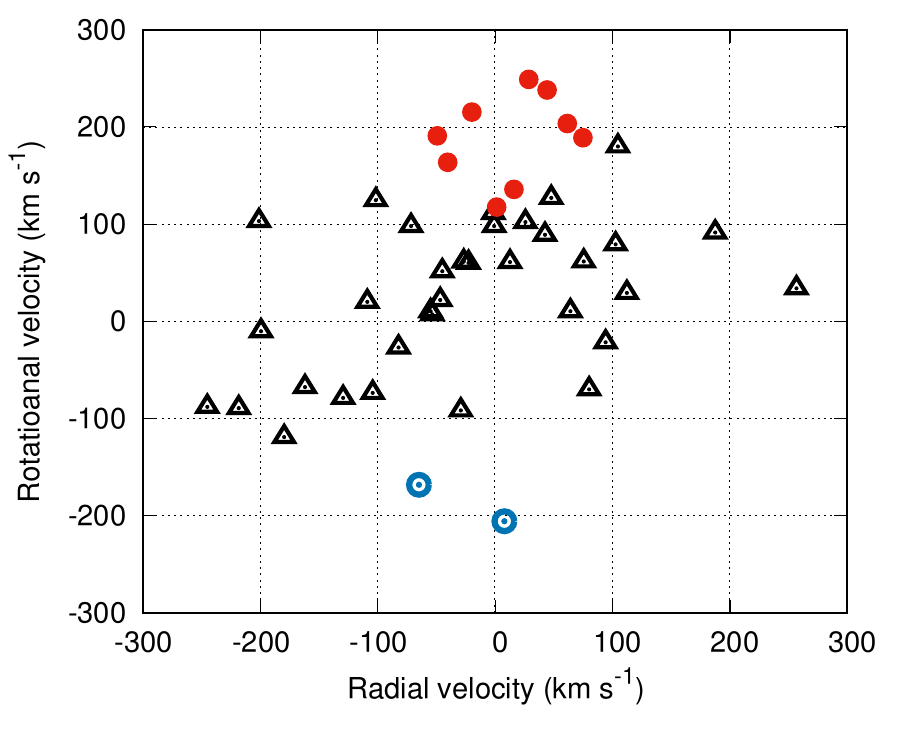}\
             \bigskip
\caption{\small Identification of GCs belonging to a bulge/bar using the probabilistic criterion proposed in [17]. The upper left panel shows the histogram of the GC distribution by the $L_Z/ecc$ parameter. The distribution is clearly bimodal. The left Gaussian determines the probability that the GC belongs to the bulge, the right Gaussian to the disk. The upper right panel shows the $"$$L_Z/ecc$ -- total energy $E$$"$ diagram for the resulting sample of 45 GCs. The lower left panel shows the $"$circular velocity -- eccentricity$"$ diagram. The bottom right panel shows the $"$radial velocity -- circular velocity$"$ diagram. Black triangles denote GCs belonging to the bulge/bar, red closed circles denote GCs belonging to the disk, blue open circles denote GCs with retrograde orbits, also belonging to the disk.}
\label{fL}
\end{center}}
\end{figure*}

\begin{figure*}
{\begin{center}
     \includegraphics[width=0.3\textwidth,angle=-90]{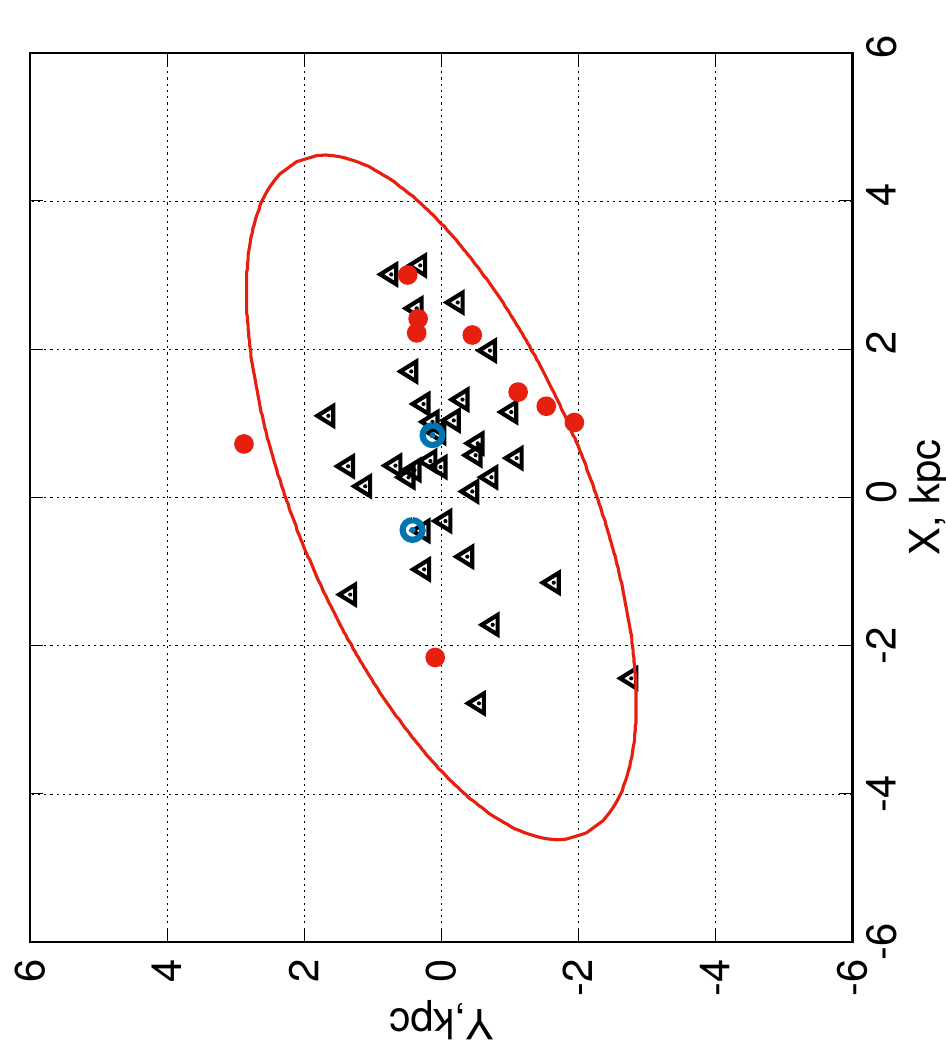}
     \includegraphics[width=0.3\textwidth,angle=-90]{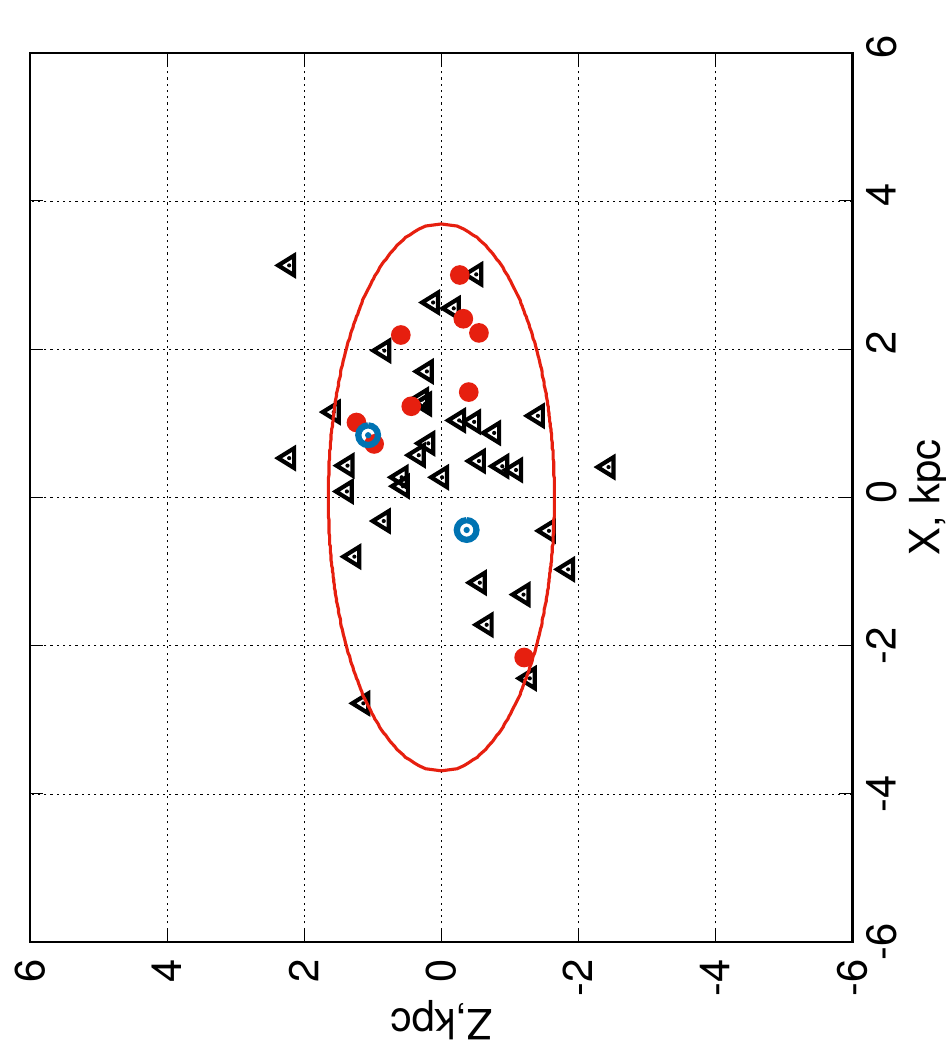}\
        \bigskip
  \caption{\small Distribution of selected 45 GCs in $X-Y$ (left panel) and $X-Z$ (right panel) projections of the galactic coordinate system. The designations of the objects are the same as in the previous figure. The red line shows the cross section of a bar (three-axis ellipsoid) with parameters: $q_b=5$ kpc, $\theta_{b}= 25^o$, axis ratio V0.}
\label{fB}
\end{center}}
\end{figure*}

\section{Results of integration of orbits in different potentials}

This section contains all tabular and graphical material reflecting the simulation of the orbital motion of 45 GCs in various potentials.

The APPENDIX contains images of the orbits in the $X-Y$ and $X-Z$ projections of the galactic coordinate system. The left panels show the orbits obtained in the axisymmetric potential, the right panels show the orbits obtained in the potential including the bar with basic parameters: $M_b=430 M_G$, $\Omega_b=40$ km/s/kpc, $q_b=5$ kpc, $\theta_b=25^o$, ratio of axes V0. In the second case, the orbits are constructed in the system of a rotating bar. The red line shows the projection of the triaxial ellipsoid that describes the bar. The orbits were integrated 2.5 Gyr backward. These images allow visual assessment of the degree of change in the orbital motion. Table~2 lists the parameters of the orbits ($apo, peri, ecc, Z_{max}$) and their uncertainties calculated both in the axisymmetric potential and in the potential with a bar. Figure~3 shows the corresponding comparison diagrams. As can be seen from the figure, as a result of the impact of the bar, the parameters of the orbits have undergone significant changes. Moreover, the vast majority of GCs have significantly increased apocentric distances. The pericentric distances increased mainly for objects with a small value in the axisymmetric potential. The maximum distances from the galactic plane $Z_{max}$ have changed less significantly. The eccentricities of the orbits of the vast majority have changed significantly, and there is a change both in the direction of increase and decrease, especially in the range of values from 0.2 to 0.8. For highly eccentric loops, the change is less significant. As follows from the catalog of orbits given in the APPENDIX, the greatest changes in parameters are observed in GCs with high radial velocities and small pericentric distances.

Further, the influence of individual bar parameters on the orbital characteristics of the GCs was studied within the variations given in table~1.
For each variable bar parameter, tables of orbital parameters and their uncertainties are given, as well as comparison diagrams.

So, table~3 shows the parameters of orbits built in the potential with a bar with different masses, table~4 -- with different bar rotation velocities, table~5 -- with different bar lengths, table 6 -- with different bar axes ratio. Figures~4, 5, 6, 7 show the corresponding comparison diagrams, which give an idea of the degree of influence of the bar parameter on the orbital properties of the GCs.

An analysis of the obtained results suggests that the influence on the parameters of the orbits is determined mainly by the mass of the bar, which is clearly seen from the comparison of fig.~4 and fig.~3, in which there is a similarity in the nature of the change in orbital parameters. At the same time, the larger the mass of the bar, the stronger the change in parameters, which was a completely expected result. This is most pronounced for bulge objects, less pronounced for disk objects.
A change in the rotational velocity of the bar by $\pm10$ km/s/kpc has a moderate effect on the change in the orbital parameters. It is most noticeable for disk objects. And a change in the size and shape of the bar within the given limits has a very small effect on the considered orbital parameters. So, the main factor of the large scatter of the orbit parameters obtained in the axisymmetric and non-axisymmetric potentials (fig.~3) relative to the comparison line is the large mass of the bar, which is actually equal to the mass of the entire bulge.

Figures 8 and 9 illustrate the conclusions drawn on the example of 6 objects. Thus, fig.~8 shows the orbits for the GCs NGC6266, NGC 6522, Terzan 4 belonging to the bulge/bar, and fig.~9 -- for GCs NGC 6540, NGC 6553, NGC 6569 belonging to the disk.
In these figures, orbits are given from left to right 1) for various bar masses: $M_b= 430, 280, 130$ (in units of $M_G$), shown in purple, green and mustard colors, respectively; 2) for different angular velocities of bar rotation: $\Omega_b=30, 40, 50$ km/s/kpc (violet, green and mustard colors); 3) for different bar lengths: $q=4.5, 5.0, 5.5$ kpc (violet, green and mustard colors); 4)For different bar axis ratios: V0 V1, V2, V3, V4 (purple, green, mustard, yellow and blue).

As can be seen from fig.~8, indeed, the strongest effect on the orbital motion of the bulge/bar GCs is the influence of the bar mass (especially noticeable for Terzan 4), then comes the influence of the bar rotation velocity (noticeable for NGC 6522), the influence of other bar parameters is insignificant. In the case of disk objects (fig.~9), the influence of the bar mass is weaker than on bulge/bar objects, but still it is noticeable, insignificantly only for NGC 6553. The influence of the bar rotation velocity is more pronounced (especially for NGC 6569). The other parameters of the bar have little effect. In the case of retrograde disk objects (NGC 6325, Djorg2), as follows from tables 3 and 4, the influence of the bar mass is significant, and the influence of changes in the bar rotation velocity is insignificant.

\section{Conclusion}

In this paper, we consider the influence of a bar on the orbital motion of globular clusters in the central region of our galaxy based on the latest data on proper motions from the Gaia EDR3 catalog [3] and new average distances [4].
A selection of 45 globular clusters was made, 34 of which belong to the bulge/bar and 11 to the disk of the Galaxy.
The orbital characteristics of the GCs obtained in an axisymmetric potential and in a potential including a bar in the form of a triaxial ellipsoid are compared. Orbits are integrated 2.5 Gyr backward. The paper presents all the numerical and graphical material on the simulation of the motion of the selected GCs in the potential of the Galaxy with different bar parameters.
The degree of influence of the bar on the orbital parameters of the GCs as a whole, as well as each individual parameter (mass, rotation velocity, shape and size of the bar) is studied by varying the parameters within reasonable limits based on information known from the literature.

The simulation showed that the bar mass has the greatest influence on the orbital motion, which is expressed mainly in an increase in both apocentric and pericentric distances for the vast majority of GCs. The eccentricities of the orbits in the overwhelming majority also change significantly, and there is a change both upward and downward, especially in the range of values from 0.2 to 0.8. As follows from the catalog of orbits given in the APPENDIX, the greatest changes in parameters are observed in GCs with high radial velocities and small pericentric distances. The change in orbital parameters depending on the bar rotation velocity is less pronounced. The influence of the geometric parameters of the bar is insignificant in the accepted range of their changes. Several examples show that bulge/bar GCs are more affected by the bar than disk GCs.

The authors are grateful to the referee for a number of useful remarks, which made it possible to improve the article.

\bigskip

\medskip{REFERENCES}\medskip {\small

1. K.F. Ogorodnikov, {\it Dynamics of stellar systems} (M.:Fizmatgiz, 1965).

2. G.A. Gontcharov, A.T. Bajkova, Astron. Lett. {\bf 39}, 689 (2013).

3. E. Vasiliev and H. Baumgardt, MNRAS 505, 5978 (2021).

4. H. Baumgardt and E. Vasiliev, MNRAS 505, 5957 (2021).

5. M. Miyamoto and R. Nagai, PASJ {\bf 27}, 533 (1975).

6. J.F. Navarro, C.S. Frenk and S.D.M. White, Astrophys. J. {\bf 490}, 493 (1997).

7. A.T. Bajkova and V.V. Bobylev, Astron. Lett. {\bf 42}, 567 (2016).

8. P. Bhattacharjee, S. Chaudhury S. and S. Kundu, Astrophys. J. {\bf 785}, 63 (2014).

9. A.T. Bajkova and V.V. Bobylev V.V., Open Ast. {\bf 26}, 72 (2017).

10. J. Palou$\breve{s}$, B. Jungwiert and J.~Kopeck\'y, Astron. Astrophys. {\bf 274}, 189 (1993).

11. R. Sch\"onrich, J. Binney and W. Dehnen, MNRAS  {\bf 403}, 1829 (2010).

12. V.V. Bobylev and A.T. Bajkova, Astron. Lett. {\bf 42}, 1 (2016).

13. E. Vasiliev, MNRAS {\bf 484}, 2832 (2019).

14. H. Baumgardt, M. Hilker, A. Solima and A. Bellini, MNRAS, {\bf 482}, 5138 (2019).

15. A.T. Bajkova and V.V. Bobylev, arXiv: 2212.00739 (2022) (IzvPulkovo, Issue 227 (2022); doi:10.31725/0367-7966-2022-227-2).

16. D. Massari, H.H. Koppelman and A. Helmi, Astron. Astrophys. {\bf 630}, L4 (2019).

17. A.T. Bajkova, G. Carraro, V.I. Korchagin, N.O. Budanova and V.V. Bobylev, Astrophys. J. {\bf 895}, 69 (2020).

18. W. Harris, AJ, {\bf 112}, 1487 (1996).

19. W. Harris, arXiv: 1012.3224 (2010).

20. A.T. Bajkova and V.V. Bobylev, Res. Astron. Astrophys. {\bf 21}, 173 (2021).


\begin{figure*}
{\begin{center}

\includegraphics[width=0.3\textwidth,angle=-90]{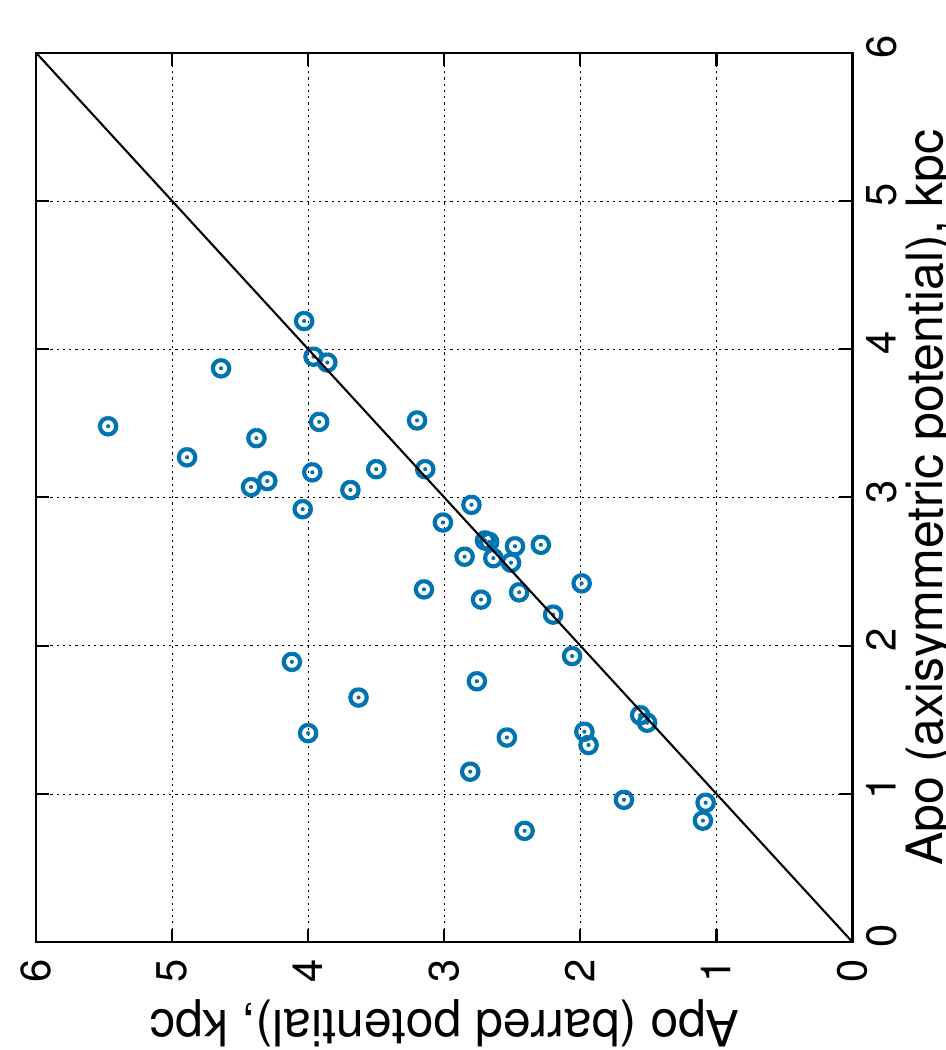}
\includegraphics[width=0.3\textwidth,angle=-90]{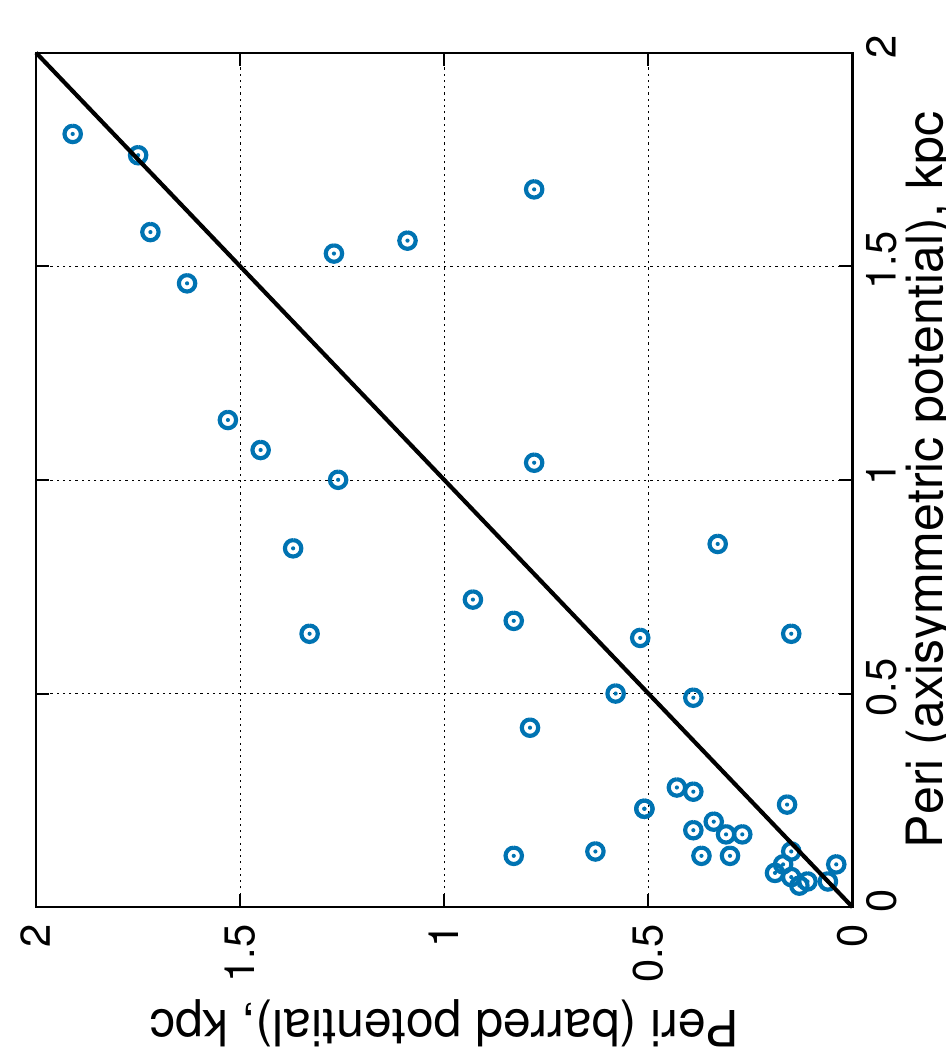}\

\medskip

\includegraphics[width=0.3\textwidth,angle=-90]{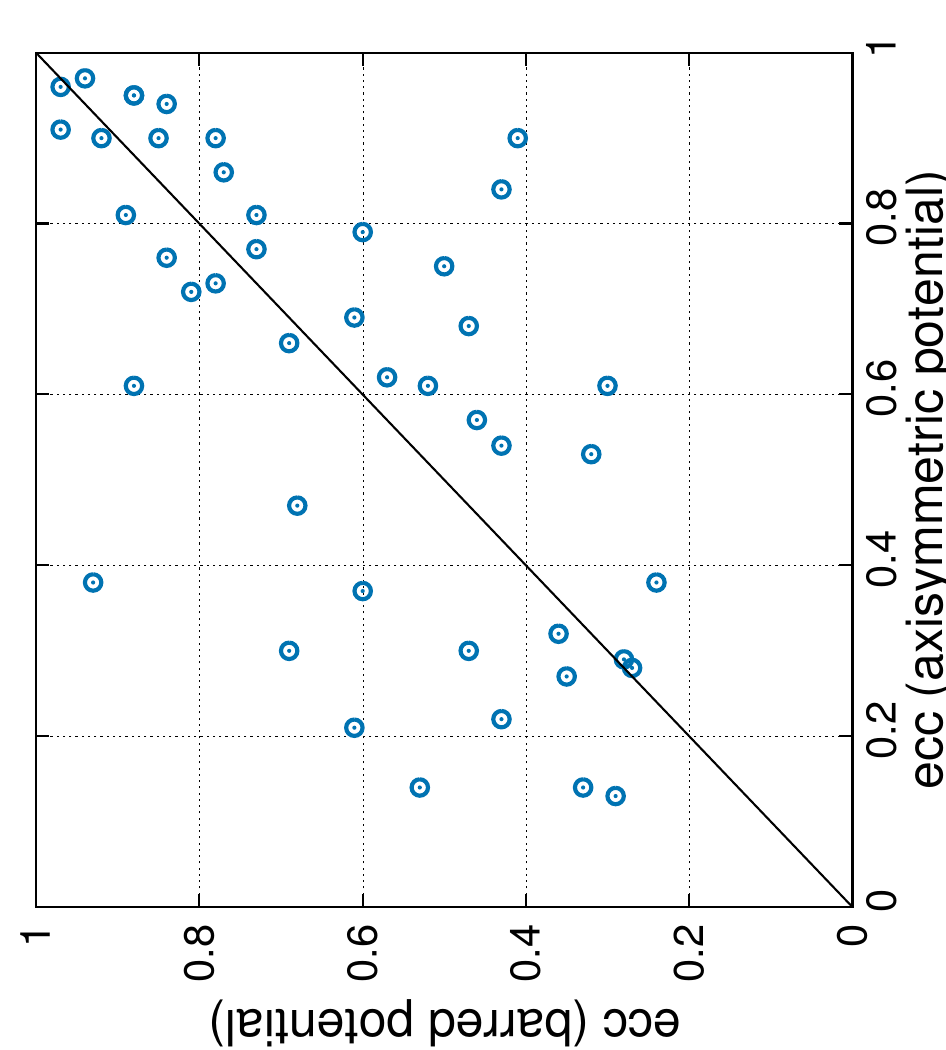}
\includegraphics[width=0.3\textwidth,angle=-90]{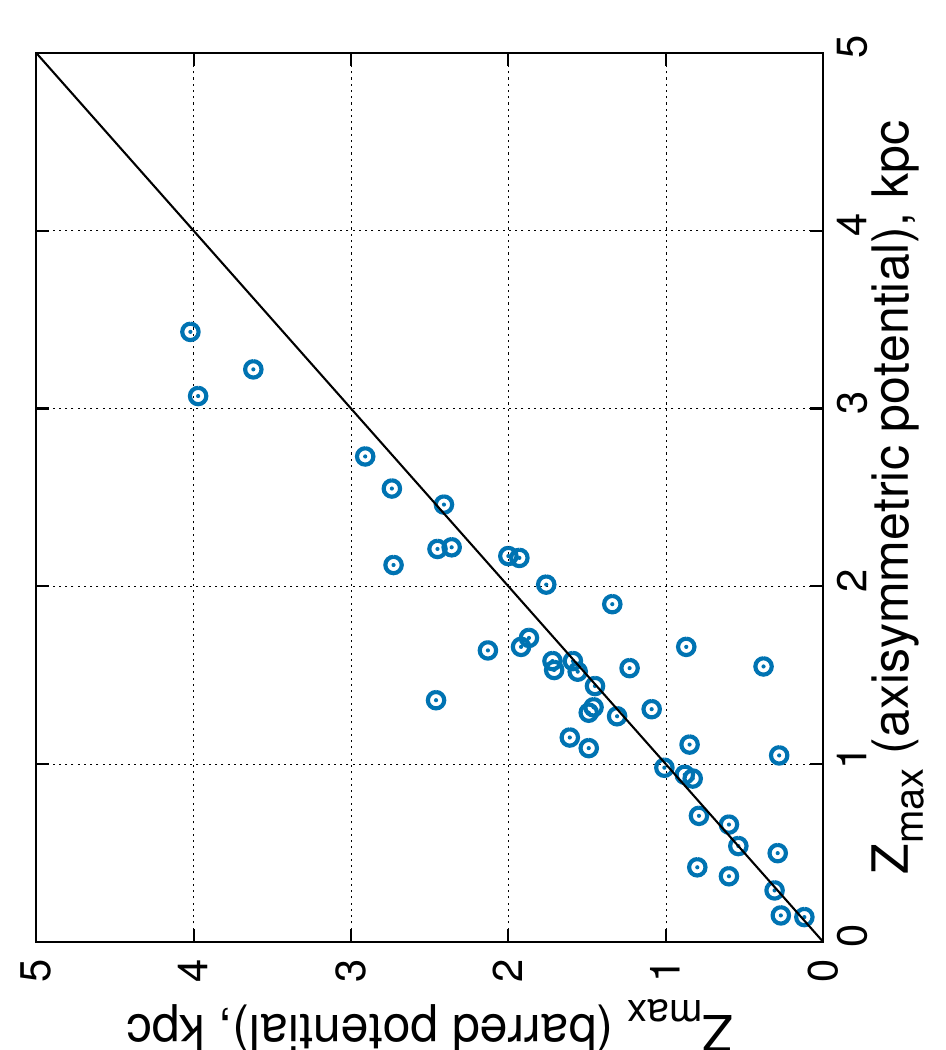}\

\caption{\small Comparison of the GC orbital parameters in the axisymmetric potential and the potential with a bar. Bar parameters: $M_b=430 M_G$, $\Omega_b=40$ km/s/kpc, $q_b=5$ kpc, $\theta_b=25^o$, axis ratio V0. Each panel has a line of coincise.}
\end{center}}
\end{figure*}

\begin{table*}
	\label{t:prop}
		\begin{scriptsize} \begin{center}
{\scriptsize {\bf Table 2:} Parameters of the orbits of 45 GCs constructed in the axisymmetric potential (index $a$) and in the potential with a bar (index $b$). Bar parameters: $M_b=430 M_G$,~$\Omega_b=40$ km/s/kpc, $q_b=5$~kpc,~$\theta_b=25^o$, ratio of ellipsoid axes V0.}

\bigskip

\begin{tabular}{|l|c|c||c|c||c|c||c|c|}\hline
 ID   & $apo^a$ &$apo^b$&$peri^a$ &$peri^b$ &$ecc^a$&$ecc^b$ &$Z_{max}^a$&$Z_{max}^b$ \\
      &[kpc]&[kpc]&[kpc]&[kpc]& & &[kpc]&[kpc] \\\hline
NGC 6144  & $ 3.40^{+0.09}_{-0.09}$& $ 4.38^{+0.07}_{-0.07}$& $ 1.56^{+0.07}_{-0.06}$& $ 1.09^{+0.07}_{-0.10}$& $ 0.37^{+0.03}_{-0.03}$& $ 0.60^{+0.03}_{-0.02}$& $ 3.22^{+0.11}_{-0.11}$& $ 3.62^{+0.14}_{-0.12}$\\
E452-11  & $ 2.94^{+0.11}_{-0.21}$& $ 2.80^{+0.21}_{-0.30}$& $ 0.10^{+0.04}_{-0.04}$& $ 0.19^{+0.06}_{-0.14}$& $ 0.94^{+0.02}_{-0.03}$& $ 0.88^{+0.08}_{-0.04}$& $ 2.15^{+0.07}_{-0.06}$& $ 1.93^{+0.07}_{-0.01}$\\
NGC 6266  & $ 2.70^{+0.07}_{-0.07}$& $ 2.67^{+0.07}_{-0.08}$& $ 0.84^{+0.04}_{-0.05}$& $ 1.37^{+0.05}_{-0.03}$& $ 0.53^{+0.01}_{-0.01}$& $ 0.32^{+0.01}_{-0.02}$& $ 0.98^{+0.02}_{-0.01}$& $ 1.01^{+0.02}_{-0.01}$\\
NGC 6273  & $ 3.48^{+0.12}_{-0.06}$& $ 5.47^{+0.19}_{-0.14}$& $ 0.85^{+0.04}_{-0.01}$& $ 0.33^{+0.09}_{-0.17}$& $ 0.61^{+0.01}_{-0.02}$& $ 0.88^{+0.06}_{-0.02}$& $ 3.43^{+0.12}_{-0.10}$& $ 4.02^{+0.28}_{-0.15}$\\
NGC 6293  & $ 3.19^{+0.21}_{-0.34}$& $ 3.50^{+0.25}_{-0.14}$& $ 0.11^{+0.12}_{-0.01}$& $ 0.27^{+0.10}_{-0.14}$& $ 0.93^{+0.01}_{-0.07}$& $ 0.85^{+0.08}_{-0.03}$& $ 2.20^{+0.17}_{-0.16}$& $ 2.45^{+0.30}_{-0.06}$\\
NGC 6342  & $ 1.76^{+0.18}_{-0.12}$& $ 2.76^{+0.03}_{-0.23}$& $ 0.63^{+0.07}_{-0.09}$& $ 0.52^{+0.15}_{-0.01}$& $ 0.47^{+0.09}_{-0.06}$& $ 0.68^{+0.00}_{-0.09}$& $ 1.53^{+0.13}_{-0.09}$& $ 1.71^{+0.08}_{-0.06}$\\
NGC 6355  & $ 1.41^{+0.32}_{-0.08}$& $ 4.00^{+0.00}_{-0.40}$& $ 0.64^{+0.05}_{-0.10}$& $ 0.15^{+0.10}_{-0.10}$& $ 0.38^{+0.12}_{-0.03}$& $ 0.93^{+0.04}_{-0.05}$& $ 1.36^{+0.22}_{-0.07}$& $ 2.46^{+0.09}_{-0.12}$\\
Terzan 2  & $ 0.96^{+0.29}_{-0.24}$& $ 1.68^{+0.14}_{-0.49}$& $ 0.13^{+0.03}_{-0.03}$& $ 0.15^{+0.21}_{-0.00}$& $ 0.76^{+0.09}_{-0.12}$& $ 0.84^{+0.00}_{-0.29}$& $ 0.37^{+0.26}_{-0.10}$& $ 0.60^{+0.07}_{-0.11}$\\
Terzan 4  & $ 0.94^{+0.27}_{-0.20}$& $ 1.08^{+0.29}_{-0.19}$& $ 0.18^{+0.01}_{-0.03}$& $ 0.39^{+0.13}_{-0.09}$& $ 0.68^{+0.07}_{-0.04}$& $ 0.47^{+0.08}_{-0.09}$& $ 0.66^{+0.17}_{-0.12}$& $ 0.60^{+0.08}_{-0.04}$\\
BH 229   & $ 2.31^{+0.55}_{-0.00}$& $ 2.73^{+0.15}_{-0.12}$& $ 0.24^{+0.05}_{-0.24}$& $ 0.16^{+0.12}_{-0.05}$& $ 0.81^{+0.19}_{-0.01}$& $ 0.89^{+0.03}_{-0.07}$& $ 2.12^{+0.16}_{-0.19}$& $ 2.73^{+0.12}_{-0.16}$\\
Liller 1  & $ 0.82^{+0.21}_{-0.05}$& $ 1.10^{+0.21}_{-0.10}$& $ 0.12^{+0.09}_{-0.07}$& $ 0.37^{+0.22}_{-0.18}$& $ 0.75^{+0.13}_{-0.09}$& $ 0.50^{+0.21}_{-0.18}$& $ 0.15^{+0.06}_{-0.02}$& $ 0.27^{+0.04}_{-0.17}$\\
NGC 6380  & $ 2.37^{+0.20}_{-0.16}$& $ 2.45^{+0.15}_{-0.08}$& $ 0.10^{+0.07}_{-0.03}$& $ 0.30^{+0.07}_{-0.05}$& $ 0.92^{+0.02}_{-0.04}$& $ 0.78^{+0.03}_{-0.03}$& $ 1.66^{+0.05}_{-0.09}$& $ 0.87^{+0.06}_{-0.02}$\\
Terzan 1  & $ 2.83^{+0.19}_{-0.17}$& $ 3.01^{+0.20}_{-0.20}$& $ 0.67^{+0.08}_{-0.06}$& $ 0.83^{+0.06}_{-0.06}$& $ 0.62^{+0.02}_{-0.03}$& $ 0.57^{+0.02}_{-0.02}$& $ 0.14^{+0.01}_{-0.01}$& $ 0.12^{+0.00}_{-0.01}$\\
NGC 6401  & $ 1.95^{+0.21}_{-0.23}$& $ 4.12^{+0.11}_{-0.20}$& $ 0.05^{+0.14}_{-0.02}$& $ 0.17^{+0.09}_{-0.12}$& $ 0.95^{+0.01}_{-0.13}$& $ 0.92^{+0.05}_{-0.04}$& $ 1.52^{+0.03}_{-0.17}$& $ 1.56^{+0.62}_{-0.08}$\\
Pal 6    & $ 2.88^{+0.68}_{-1.35}$& $ 4.04^{+0.34}_{-0.76}$& $ 0.06^{+0.18}_{-0.04}$& $ 0.06^{+0.19}_{-0.01}$& $ 0.96^{+0.04}_{-0.31}$& $ 0.97^{+0.00}_{-0.08}$& $ 2.23^{+0.39}_{-0.98}$& $ 2.36^{+0.27}_{-0.57}$\\
Terzan 5  & $ 1.90^{+0.19}_{-0.15}$& $ 2.06^{+0.16}_{-0.13}$& $ 0.23^{+0.02}_{-0.05}$& $ 0.51^{+0.05}_{-0.08}$& $ 0.79^{+0.03}_{-0.01}$& $ 0.60^{+0.04}_{-0.01}$& $ 1.04^{+0.07}_{-0.12}$& $ 0.28^{+0.03}_{-0.01}$\\
NGC 6440  & $ 1.45^{+0.08}_{-0.11}$& $ 1.51^{+0.13}_{-0.04}$& $ 0.10^{+0.04}_{-0.06}$& $ 0.13^{+0.10}_{-0.07}$& $ 0.87^{+0.08}_{-0.06}$& $ 0.84^{+0.09}_{-0.10}$& $ 1.16^{+0.08}_{-0.13}$& $ 0.85^{+0.07}_{-0.06}$\\
Terzan 6  & $ 1.33^{+0.39}_{-0.50}$& $ 1.94^{+0.28}_{-0.32}$& $ 0.17^{+0.07}_{-0.05}$& $ 0.31^{+0.07}_{-0.11}$& $ 0.77^{+0.05}_{-0.11}$& $ 0.73^{+0.07}_{-0.05}$& $ 0.50^{+0.33}_{-0.19}$& $ 0.29^{+0.19}_{-0.04}$\\
NGC 6453  & $ 2.59^{+0.35}_{-0.07}$& $ 2.64^{+0.39}_{-0.17}$& $ 0.20^{+0.16}_{-0.13}$& $ 0.34^{+0.05}_{-0.18}$& $ 0.86^{+0.09}_{-0.10}$& $ 0.77^{+0.12}_{-0.01}$& $ 2.18^{+0.24}_{-0.26}$& $ 2.00^{+0.10}_{-0.17}$\\
Terzan 9  & $ 2.71^{+0.38}_{-0.30}$& $ 2.70^{+0.31}_{-0.38}$& $ 0.28^{+0.08}_{-0.08}$& $ 0.43^{+0.13}_{-0.16}$& $ 0.82^{+0.04}_{-0.05}$& $ 0.73^{+0.09}_{-0.09}$& $ 1.55^{+0.25}_{-0.32}$& $ 0.38^{+0.03}_{-0.04}$\\
NGC 6522  & $ 1.42^{+0.18}_{-0.17}$& $ 1.97^{+0.10}_{-0.12}$& $ 0.42^{+0.13}_{-0.14}$& $ 0.79^{+0.13}_{-0.17}$& $ 0.54^{+0.09}_{-0.05}$& $ 0.43^{+0.08}_{-0.06}$& $ 1.09^{+0.14}_{-0.15}$& $ 1.49^{+0.07}_{-0.08}$\\
NGC 6528  & $ 1.15^{+0.48}_{-0.39}$& $ 2.81^{+0.32}_{-0.17}$& $ 0.23^{+0.15}_{-0.09}$& $ 0.51^{+0.07}_{-0.07}$& $ 0.66^{+0.19}_{-0.32}$& $ 0.69^{+0.03}_{-0.01}$& $ 0.92^{+0.23}_{-0.23}$& $ 0.83^{+0.08}_{-0.02}$\\
NGC 6558  & $ 1.65^{+0.27}_{-0.21}$& $ 3.63^{+0.00}_{-0.52}$& $ 0.27^{+0.08}_{-0.06}$& $ 0.39^{+0.02}_{-0.14}$& $ 0.72^{+0.08}_{-0.10}$& $ 0.81^{+0.04}_{-0.02}$& $ 1.31^{+0.12}_{-0.13}$& $ 1.09^{+0.05}_{-0.02}$\\
NGC 6624  & $ 1.66^{+0.02}_{-0.18}$& $ 1.56^{+0.04}_{-0.01}$& $ 0.07^{+0.12}_{-0.00}$& $ 0.63^{+0.04}_{-0.12}$& $ 0.92^{+0.00}_{-0.14}$& $ 0.43^{+0.08}_{-0.03}$& $ 1.29^{+0.03}_{-0.03}$& $ 1.49^{+0.03}_{-0.01}$\\
NGC 6626  & $ 3.19^{+0.24}_{-0.07}$& $ 3.14^{+0.19}_{-0.06}$& $ 0.49^{+0.04}_{-0.07}$& $ 0.39^{+0.23}_{-0.18}$& $ 0.73^{+0.05}_{-0.02}$& $ 0.78^{+0.10}_{-0.10}$& $ 1.90^{+0.07}_{-0.11}$& $ 1.34^{+0.00}_{-0.14}$\\
NGC 6638  & $ 2.55^{+0.34}_{-0.14}$& $ 2.29^{+0.46}_{-0.10}$& $ 0.09^{+0.01}_{-0.05}$& $ 0.15^{+0.04}_{-0.09}$& $ 0.93^{+0.04}_{-0.01}$& $ 0.88^{+0.08}_{-0.03}$& $ 2.02^{+0.20}_{-0.26}$& $ 1.76^{+0.03}_{-0.29}$\\
NGC 6637  & $ 2.43^{+0.07}_{-0.27}$& $ 1.99^{+0.04}_{-0.02}$& $ 0.10^{+0.08}_{-0.01}$& $ 0.83^{+0.05}_{-0.08}$& $ 0.92^{+0.01}_{-0.08}$& $ 0.41^{+0.05}_{-0.03}$& $ 1.72^{+0.03}_{-0.02}$& $ 1.87^{+0.03}_{-0.02}$\\
NGC 6642  & $ 2.24^{+0.08}_{-0.09}$& $ 2.20^{+0.08}_{-0.04}$& $ 0.07^{+0.06}_{-0.01}$& $ 0.04^{+0.14}_{-0.02}$& $ 0.94^{+0.01}_{-0.05}$& $ 0.97^{+0.00}_{-0.12}$& $ 1.54^{+0.11}_{-0.01}$& $ 1.23^{+0.16}_{-0.09}$\\
NGC 6717  & $ 2.67^{+0.13}_{-0.11}$& $ 2.48^{+0.06}_{-0.03}$& $ 0.64^{+0.05}_{-0.05}$& $ 1.33^{+0.12}_{-0.03}$& $ 0.61^{+0.04}_{-0.03}$& $ 0.30^{+0.02}_{-0.04}$& $ 1.44^{+0.02}_{-0.03}$& $ 1.45^{+0.03}_{-0.03}$\\
NGC 6723  & $ 3.11^{+0.06}_{-0.05}$& $ 4.30^{+0.05}_{-0.03}$& $ 1.68^{+0.04}_{-0.02}$& $ 0.78^{+0.07}_{-0.13}$& $ 0.30^{+0.00}_{-0.01}$& $ 0.69^{+0.05}_{-0.02}$& $ 3.07^{+0.04}_{-0.05}$& $ 3.97^{+0.06}_{-0.08}$\\
NGC 6171  & $ 3.95^{+0.05}_{-0.06}$& $ 3.96^{+0.06}_{-0.05}$& $ 1.07^{+0.04}_{-0.04}$& $ 1.45^{+0.07}_{-0.03}$& $ 0.57^{+0.02}_{-0.01}$& $ 0.46^{+0.01}_{-0.02}$& $ 2.46^{+0.03}_{-0.04}$& $ 2.41^{+0.03}_{-0.04}$\\
NGC 6316  & $ 3.91^{+0.54}_{-0.29}$& $ 3.86^{+0.35}_{-0.38}$& $ 0.72^{+0.19}_{-0.13}$& $ 0.93^{+0.11}_{-0.30}$& $ 0.69^{+0.03}_{-0.03}$& $ 0.61^{+0.10}_{-0.02}$& $ 1.58^{+0.14}_{-0.05}$& $ 1.72^{+0.06}_{-0.11}$\\
NGC 6388  & $ 4.19^{+0.07}_{-0.09}$& $ 4.03^{+0.17}_{-0.08}$& $ 1.00^{+0.07}_{-0.06}$& $ 1.26^{+0.05}_{-0.11}$& $ 0.61^{+0.02}_{-0.01}$& $ 0.52^{+0.04}_{-0.01}$& $ 1.58^{+0.02}_{-0.05}$& $ 1.59^{+0.08}_{-0.03}$\\
NGC 6539  & $ 3.51^{+0.18}_{-0.16}$& $ 3.92^{+0.22}_{-0.10}$& $ 1.92^{+0.23}_{-0.20}$& $ 2.18^{+0.20}_{-0.20}$& $ 0.29^{+0.05}_{-0.04}$& $ 0.28^{+0.06}_{-0.03}$& $ 2.55^{+0.22}_{-0.07}$& $ 2.74^{+0.20}_{-0.13}$\\
NGC 6553  & $ 3.87^{+0.11}_{-0.11}$& $ 4.64^{+0.09}_{-0.11}$& $ 2.96^{+0.12}_{-0.13}$& $ 2.54^{+0.08}_{-0.11}$& $ 0.13^{+0.01}_{-0.00}$& $ 0.29^{+0.01}_{-0.00}$& $ 0.29^{+0.01}_{-0.01}$& $ 0.31^{+0.01}_{-0.01}$\\
NGC 6652  & $ 3.52^{+0.12}_{-0.12}$& $ 3.20^{+0.11}_{-0.10}$& $ 0.06^{+0.02}_{-0.03}$& $ 0.11^{+0.03}_{-0.05}$& $ 0.96^{+0.02}_{-0.01}$& $ 0.94^{+0.02}_{-0.02}$& $ 2.75^{+0.05}_{-0.09}$& $ 2.91^{+0.04}_{-0.08}$\\
Terzan 3  & $ 3.07^{+0.20}_{-0.11}$& $ 4.42^{+0.19}_{-0.21}$& $ 2.33^{+0.07}_{-0.22}$& $ 2.23^{+0.17}_{-0.23}$& $ 0.14^{+0.05}_{-0.01}$& $ 0.33^{+0.04}_{-0.03}$& $ 1.66^{+0.13}_{-0.06}$& $ 1.92^{+0.11}_{-0.10}$\\
NGC 6256  & $ 2.38^{+0.09}_{-0.07}$& $ 3.15^{+0.15}_{-0.10}$& $ 1.53^{+0.32}_{-0.35}$& $ 1.27^{+0.32}_{-0.27}$& $ 0.22^{+0.12}_{-0.09}$& $ 0.43^{+0.08}_{-0.08}$& $ 0.71^{+0.06}_{-0.07}$& $ 0.79^{+0.14}_{-0.05}$\\
NGC 6304  & $ 3.05^{+0.16}_{-0.12}$& $ 3.69^{+0.14}_{-0.16}$& $ 1.58^{+0.11}_{-0.11}$& $ 1.72^{+0.29}_{-0.29}$& $ 0.32^{+0.01}_{-0.01}$& $ 0.36^{+0.08}_{-0.07}$& $ 0.94^{+0.04}_{-0.02}$& $ 0.88^{+0.10}_{-0.02}$\\
Pismis 26  & $ 3.27^{+0.48}_{-0.35}$& $ 4.89^{+0.51}_{-0.24}$& $ 1.76^{+0.18}_{-0.18}$& $ 1.75^{+0.18}_{-0.23}$& $ 0.30^{+0.03}_{-0.01}$& $ 0.47^{+0.06}_{-0.02}$& $ 1.64^{+0.23}_{-0.16}$& $ 2.13^{+0.13}_{-0.07}$\\
NGC 6540  & $ 2.56^{+0.18}_{-0.28}$& $ 2.51^{+0.25}_{-0.22}$& $ 1.14^{+0.13}_{-0.19}$& $ 1.53^{+0.06}_{-0.11}$& $ 0.38^{+0.04}_{-0.02}$& $ 0.24^{+0.06}_{-0.04}$& $ 0.54^{+0.20}_{-0.04}$& $ 0.54^{+0.04}_{-0.02}$\\
NGC 6569  & $ 2.60^{+0.22}_{-0.20}$& $ 2.85^{+0.29}_{-0.14}$& $ 1.46^{+0.22}_{-0.19}$& $ 1.63^{+0.06}_{-0.47}$& $ 0.28^{+0.05}_{-0.05}$& $ 0.27^{+0.18}_{-0.02}$& $ 1.27^{+0.03}_{-0.03}$& $ 1.31^{+0.07}_{-0.03}$\\
E456-78  & $ 3.17^{+0.27}_{-0.23}$& $ 3.97^{+0.14}_{-0.21}$& $ 1.81^{+0.24}_{-0.23}$& $ 1.91^{+0.20}_{-0.17}$& $ 0.27^{+0.03}_{-0.02}$& $ 0.35^{+0.02}_{-0.03}$& $ 1.32^{+0.10}_{-0.10}$& $ 1.46^{+0.12}_{-0.08}$\\
NGC 6325  & $ 1.38^{+0.27}_{-0.11}$& $ 2.54^{+0.29}_{-0.29}$& $ 1.04^{+0.20}_{-0.29}$& $ 0.78^{+0.34}_{-0.30}$& $ 0.14^{+0.18}_{-0.06}$& $ 0.53^{+0.15}_{-0.14}$& $ 1.15^{+0.12}_{-0.04}$& $ 1.61^{+0.14}_{-0.08}$\\
Djorg 2   & $ 0.75^{+0.10}_{-0.04}$& $ 2.41^{+0.07}_{-0.04}$& $ 0.50^{+0.14}_{-0.13}$& $ 0.58^{+0.12}_{-0.15}$& $ 0.21^{+0.12}_{-0.09}$& $ 0.61^{+0.08}_{-0.05}$& $ 0.42^{+0.03}_{-0.01}$& $ 0.80^{+0.03}_{-0.01}$\\\hline
 \end{tabular}
   \end{center}
   \end{scriptsize}
  \end{table*}

 \begin{table*}
  \rotatebox{90}{
		\begin{minipage}{1.5\linewidth}
		\label{t:prop}
		\begin{tiny} \begin{center}
{\scriptsize {\bf Table 3:} Parameters of the orbits of 45 GCs in the potential with a bar versus the mass of the bar: \\

$M_b=430 M_G$ (index "1"), $M_b=280 M_G$ (index "2"), $M_b=130 M_G$ (index "3").}

\bigskip

\begin{tabular}{|l|c|c|c||c|c|c||c|c|c||c|c|c|}\hline
 ID  & $apo^1$ &$apo^2$&$apo^3$&$peri^1$ &$peri^2$&$peri^3$ &$ecc^1$&
  $ecc^2$ &$ecc^3$ &$Z_{max}^1$&$Z_{max}^2$&$Z_{max}^3$ \\
    &[kpc]&[kpc]&[kpc]&[kpc]&[kpc]&[kpc]& & & &[kpc]&[kpc]&[kpc] \\\hline
NGC 6144  & $ 4.38^{+0.07}_{-0.07}$& $ 4.00^{+0.06}_{-0.07}$& $ 3.66^{+0.07}_{-0.08}$& $ 1.09^{+0.07}_{-0.10}$& $ 1.24^{+0.09}_{-0.07}$& $ 1.44^{+0.05}_{-0.09}$& $ 0.60^{+0.03}_{-0.02}$& $ 0.53^{+0.02}_{-0.03}$& $ 0.43^{+0.03}_{-0.01}$& $ 3.62^{+0.14}_{-0.12}$& $ 3.59^{+0.10}_{-0.12}$& $ 3.40^{+0.10}_{-0.10}$\\
E452-11  & $ 2.80^{+0.21}_{-0.30}$& $ 3.24^{+0.15}_{-0.19}$& $ 3.03^{+0.14}_{-0.17}$& $ 0.19^{+0.06}_{-0.14}$& $ 0.07^{+0.06}_{-0.02}$& $ 0.07^{+0.04}_{-0.02}$& $ 0.88^{+0.08}_{-0.04}$& $ 0.96^{+0.01}_{-0.04}$& $ 0.95^{+0.02}_{-0.02}$& $ 1.93^{+0.07}_{-0.01}$& $ 1.87^{+0.18}_{-0.02}$& $ 2.09^{+0.15}_{-0.03}$\\
NGC 6266  & $ 2.67^{+0.07}_{-0.08}$& $ 2.61^{+0.21}_{-0.11}$& $ 2.82^{+0.08}_{-0.05}$& $ 1.37^{+0.05}_{-0.03}$& $ 1.07^{+0.05}_{-0.07}$& $ 0.80^{+0.07}_{-0.06}$& $ 0.32^{+0.01}_{-0.02}$& $ 0.42^{+0.05}_{-0.03}$& $ 0.56^{+0.02}_{-0.02}$& $ 1.01^{+0.02}_{-0.01}$& $ 1.02^{+0.00}_{-0.04}$& $ 1.20^{+0.00}_{-0.20}$\\
NGC 6273  & $ 5.47^{+0.19}_{-0.14}$& $ 4.68^{+0.15}_{-0.06}$& $ 4.02^{+0.17}_{-0.07}$& $ 0.33^{+0.09}_{-0.17}$& $ 0.32^{+0.13}_{-0.11}$& $ 0.62^{+0.08}_{-0.13}$& $ 0.88^{+0.06}_{-0.02}$& $ 0.87^{+0.05}_{-0.05}$& $ 0.73^{+0.06}_{-0.03}$& $ 4.02^{+0.28}_{-0.15}$& $ 4.56^{+0.05}_{-0.56}$& $ 3.89^{+0.19}_{-0.17}$\\
NGC 6293  & $ 3.50^{+0.25}_{-0.14}$& $ 3.13^{+0.36}_{-0.16}$& $ 3.61^{+0.00}_{-0.70}$& $ 0.27^{+0.10}_{-0.14}$& $ 0.17^{+0.18}_{-0.07}$& $ 0.11^{+0.18}_{-0.01}$& $ 0.85^{+0.08}_{-0.03}$& $ 0.90^{+0.04}_{-0.09}$& $ 0.94^{+0.00}_{-0.11}$& $ 2.45^{+0.30}_{-0.06}$& $ 2.85^{+0.02}_{-0.39}$& $ 2.49^{+0.03}_{-0.30}$\\
NGC 6342  & $ 2.76^{+0.03}_{-0.23}$& $ 2.01^{+0.16}_{-0.11}$& $ 1.79^{+0.65}_{-0.00}$& $ 0.52^{+0.15}_{-0.01}$& $ 0.77^{+0.12}_{-0.11}$& $ 0.75^{+0.00}_{-0.59}$& $ 0.68^{+0.00}_{-0.09}$& $ 0.45^{+0.08}_{-0.09}$& $ 0.41^{+0.47}_{-0.00}$& $ 1.71^{+0.08}_{-0.06}$& $ 1.71^{+0.11}_{-0.09}$& $ 1.61^{+0.18}_{-0.08}$\\
NGC 6355  & $ 4.00^{+0.00}_{-0.40}$& $ 2.75^{+0.43}_{-0.17}$& $ 2.01^{+0.52}_{-0.23}$& $ 0.15^{+0.10}_{-0.10}$& $ 0.08^{+0.07}_{-0.03}$& $ 0.45^{+0.08}_{-0.17}$& $ 0.93^{+0.04}_{-0.05}$& $ 0.94^{+0.03}_{-0.04}$& $ 0.64^{+0.15}_{-0.08}$& $ 2.46^{+0.09}_{-0.12}$& $ 2.50^{+0.03}_{-0.23}$& $ 1.86^{+0.12}_{-0.11}$\\
Terzan 2  & $ 1.68^{+0.14}_{-0.49}$& $ 1.11^{+0.51}_{-0.30}$& $ 1.00^{+0.36}_{-0.27}$& $ 0.15^{+0.21}_{-0.00}$& $ 0.21^{+0.06}_{-0.07}$& $ 0.15^{+0.05}_{-0.03}$& $ 0.84^{+0.00}_{-0.29}$& $ 0.69^{+0.16}_{-0.19}$& $ 0.73^{+0.11}_{-0.15}$& $ 0.60^{+0.07}_{-0.11}$& $ 0.48^{+0.40}_{-0.11}$& $ 0.60^{+0.20}_{-0.30}$\\
Terzan 4  & $ 1.08^{+0.29}_{-0.19}$& $ 1.00^{+0.32}_{-0.24}$& $ 0.98^{+0.35}_{-0.26}$& $ 0.39^{+0.13}_{-0.09}$& $ 0.23^{+0.04}_{-0.12}$& $ 0.19^{+0.03}_{-0.05}$& $ 0.47^{+0.08}_{-0.09}$& $ 0.63^{+0.19}_{-0.11}$& $ 0.68^{+0.09}_{-0.09}$& $ 0.60^{+0.08}_{-0.04}$& $ 0.66^{+0.20}_{-0.14}$& $ 0.64^{+0.25}_{-0.12}$\\
BH 229   & $ 2.73^{+0.15}_{-0.12}$& $ 2.37^{+0.14}_{-0.17}$& $ 2.08^{+0.18}_{-0.10}$& $ 0.16^{+0.12}_{-0.05}$& $ 0.57^{+0.08}_{-0.09}$& $ 0.63^{+0.02}_{-0.10}$& $ 0.89^{+0.03}_{-0.07}$& $ 0.61^{+0.04}_{-0.03}$& $ 0.54^{+0.06}_{-0.01}$& $ 2.73^{+0.12}_{-0.16}$& $ 2.35^{+0.13}_{-0.18}$& $ 2.07^{+0.15}_{-0.13}$\\
Liller 1  & $ 1.10^{+0.21}_{-0.10}$& $ 0.90^{+0.20}_{-0.05}$& $ 0.84^{+0.20}_{-0.06}$& $ 0.37^{+0.22}_{-0.18}$& $ 0.19^{+0.16}_{-0.14}$& $ 0.14^{+0.10}_{-0.07}$& $ 0.50^{+0.21}_{-0.18}$& $ 0.65^{+0.20}_{-0.12}$& $ 0.72^{+0.12}_{-0.10}$& $ 0.27^{+0.04}_{-0.17}$& $ 0.16^{+0.07}_{-0.02}$& $ 0.15^{+0.08}_{-0.03}$\\
NGC 6380  & $ 2.45^{+0.15}_{-0.08}$& $ 2.51^{+0.18}_{-0.13}$& $ 2.28^{+0.31}_{-0.04}$& $ 0.30^{+0.07}_{-0.05}$& $ 0.05^{+0.09}_{-0.00}$& $ 0.08^{+0.09}_{-0.02}$& $ 0.78^{+0.03}_{-0.03}$& $ 0.96^{+0.00}_{-0.06}$& $ 0.93^{+0.02}_{-0.06}$& $ 0.87^{+0.06}_{-0.02}$& $ 1.72^{+0.06}_{-0.28}$& $ 1.65^{+0.07}_{-0.11}$\\
Terzan 1  & $ 3.01^{+0.20}_{-0.20}$& $ 2.91^{+0.19}_{-0.15}$& $ 2.86^{+0.18}_{-0.19}$& $ 0.83^{+0.06}_{-0.06}$& $ 0.77^{+0.08}_{-0.07}$& $ 0.69^{+0.10}_{-0.11}$& $ 0.57^{+0.02}_{-0.02}$& $ 0.58^{+0.02}_{-0.02}$& $ 0.61^{+0.04}_{-0.03}$& $ 0.12^{+0.00}_{-0.01}$& $ 0.13^{+0.02}_{-0.02}$& $ 0.13^{+0.01}_{-0.01}$\\
NGC 6401  & $ 4.12^{+0.11}_{-0.20}$& $ 3.29^{+0.16}_{-0.38}$& $ 2.45^{+0.19}_{-0.20}$& $ 0.17^{+0.09}_{-0.12}$& $ 0.03^{+0.21}_{-0.00}$& $ 0.05^{+0.13}_{-0.01}$& $ 0.92^{+0.05}_{-0.04}$& $ 0.98^{+0.00}_{-0.12}$& $ 0.96^{+0.00}_{-0.09}$& $ 1.56^{+0.62}_{-0.08}$& $ 1.96^{+0.18}_{-0.14}$& $ 1.78^{+0.02}_{-0.18}$\\
Pal 6    & $ 4.04^{+0.34}_{-0.76}$& $ 3.55^{+0.82}_{-0.70}$& $ 3.22^{+0.82}_{-0.87}$& $ 0.06^{+0.19}_{-0.01}$& $ 0.03^{+0.13}_{-0.00}$& $ 0.05^{+0.10}_{-0.03}$& $ 0.97^{+0.00}_{-0.08}$& $ 0.98^{+0.00}_{-0.07}$& $ 0.97^{+0.03}_{-0.11}$& $ 2.36^{+0.27}_{-0.57}$& $ 2.51^{+0.20}_{-0.96}$& $ 2.41^{+0.47}_{-0.67}$\\
Terzan 5  & $ 2.06^{+0.16}_{-0.13}$& $ 2.01^{+0.16}_{-0.13}$& $ 1.91^{+0.21}_{-0.16}$& $ 0.51^{+0.05}_{-0.08}$& $ 0.34^{+0.04}_{-0.05}$& $ 0.13^{+0.07}_{-0.04}$& $ 0.60^{+0.04}_{-0.01}$& $ 0.71^{+0.03}_{-0.02}$& $ 0.88^{+0.02}_{-0.05}$& $ 0.28^{+0.03}_{-0.01}$& $ 0.27^{+0.31}_{-0.10}$& $ 1.25^{+0.03}_{-0.22}$\\
NGC 6440  & $ 1.51^{+0.13}_{-0.04}$& $ 1.67^{+0.11}_{-0.14}$& $ 1.58^{+0.08}_{-0.13}$& $ 0.13^{+0.10}_{-0.07}$& $ 0.10^{+0.05}_{-0.08}$& $ 0.05^{+0.08}_{-0.03}$& $ 0.84^{+0.09}_{-0.10}$& $ 0.89^{+0.09}_{-0.06}$& $ 0.94^{+0.03}_{-0.10}$& $ 0.85^{+0.07}_{-0.06}$& $ 1.18^{+0.09}_{-0.10}$& $ 1.18^{+0.10}_{-0.11}$\\
Terzan 6  & $ 1.94^{+0.28}_{-0.32}$& $ 1.62^{+0.36}_{-0.43}$& $ 1.44^{+0.47}_{-0.45}$& $ 0.31^{+0.07}_{-0.11}$& $ 0.24^{+0.09}_{-0.05}$& $ 0.20^{+0.06}_{-0.05}$& $ 0.73^{+0.07}_{-0.05}$& $ 0.74^{+0.05}_{-0.11}$& $ 0.75^{+0.07}_{-0.10}$& $ 0.29^{+0.19}_{-0.04}$& $ 0.66^{+0.26}_{-0.25}$& $ 0.56^{+0.44}_{-0.16}$\\
NGC 6453  & $ 2.64^{+0.39}_{-0.17}$& $ 2.47^{+0.27}_{-0.32}$& $ 2.45^{+0.21}_{-0.32}$& $ 0.34^{+0.05}_{-0.18}$& $ 0.31^{+0.09}_{-0.08}$& $ 0.53^{+0.03}_{-0.22}$& $ 0.77^{+0.12}_{-0.01}$& $ 0.78^{+0.06}_{-0.09}$& $ 0.64^{+0.14}_{-0.04}$& $ 2.00^{+0.10}_{-0.17}$& $ 2.19^{+0.19}_{-0.25}$& $ 2.14^{+0.20}_{-0.20}$\\
Terzan 9  & $ 2.70^{+0.31}_{-0.38}$& $ 2.72^{+0.30}_{-0.31}$& $ 2.75^{+0.29}_{-0.33}$& $ 0.43^{+0.13}_{-0.16}$& $ 0.38^{+0.13}_{-0.12}$& $ 0.30^{+0.11}_{-0.09}$& $ 0.73^{+0.09}_{-0.09}$& $ 0.75^{+0.08}_{-0.07}$& $ 0.81^{+0.04}_{-0.07}$& $ 0.38^{+0.03}_{-0.04}$& $ 1.02^{+0.12}_{-0.72}$& $ 0.34^{+1.24}_{-0.00}$\\
NGC 6522  & $ 1.97^{+0.10}_{-0.12}$& $ 1.65^{+0.20}_{-0.08}$& $ 1.53^{+0.24}_{-0.18}$& $ 0.79^{+0.13}_{-0.17}$& $ 0.66^{+0.20}_{-0.17}$& $ 0.44^{+0.10}_{-0.16}$& $ 0.43^{+0.08}_{-0.06}$& $ 0.43^{+0.11}_{-0.08}$& $ 0.55^{+0.12}_{-0.04}$& $ 1.49^{+0.07}_{-0.08}$& $ 1.27^{+0.13}_{-0.08}$& $ 1.17^{+0.17}_{-0.15}$\\
NGC 6528  & $ 2.81^{+0.32}_{-0.17}$& $ 2.12^{+0.26}_{-0.50}$& $ 1.56^{+0.34}_{-0.54}$& $ 0.51^{+0.07}_{-0.07}$& $ 0.13^{+0.05}_{-0.04}$& $ 0.10^{+0.21}_{-0.03}$& $ 0.69^{+0.03}_{-0.01}$& $ 0.88^{+0.03}_{-0.05}$& $ 0.88^{+0.07}_{-0.36}$& $ 0.83^{+0.08}_{-0.02}$& $ 1.21^{+0.15}_{-0.13}$& $ 1.02^{+0.25}_{-0.18}$\\
NGC 6558  & $ 3.63^{+0.00}_{-0.52}$& $ 2.28^{+0.51}_{-0.22}$& $ 2.19^{+0.13}_{-0.27}$& $ 0.39^{+0.02}_{-0.14}$& $ 0.30^{+0.04}_{-0.05}$& $ 0.10^{+0.13}_{-0.02}$& $ 0.81^{+0.04}_{-0.02}$& $ 0.76^{+0.05}_{-0.01}$& $ 0.91^{+0.01}_{-0.11}$& $ 1.09^{+0.05}_{-0.02}$& $ 1.44^{+0.04}_{-0.05}$& $ 1.49^{+0.07}_{-0.17}$\\
NGC 6624  & $ 1.56^{+0.04}_{-0.01}$& $ 1.55^{+0.53}_{-0.00}$& $ 1.53^{+0.19}_{-0.08}$& $ 0.63^{+0.04}_{-0.12}$& $ 0.19^{+0.00}_{-0.16}$& $ 0.28^{+0.02}_{-0.11}$& $ 0.43^{+0.08}_{-0.03}$& $ 0.78^{+0.20}_{-0.00}$& $ 0.69^{+0.13}_{-0.02}$& $ 1.49^{+0.03}_{-0.01}$& $ 1.44^{+0.06}_{-0.01}$& $ 1.33^{+0.05}_{-0.03}$\\
NGC 6626  & $ 3.14^{+0.19}_{-0.06}$& $ 3.14^{+0.43}_{-0.00}$& $ 3.18^{+0.46}_{-0.00}$& $ 0.39^{+0.23}_{-0.18}$& $ 0.64^{+0.02}_{-0.11}$& $ 0.55^{+0.05}_{-0.05}$& $ 0.78^{+0.10}_{-0.10}$& $ 0.66^{+0.08}_{-0.00}$& $ 0.70^{+0.05}_{-0.00}$& $ 1.34^{+0.00}_{-0.14}$& $ 1.69^{+0.00}_{-0.28}$& $ 1.78^{+0.01}_{-0.13}$\\
NGC 6638  & $ 2.29^{+0.46}_{-0.10}$& $ 2.86^{+0.25}_{-0.35}$& $ 2.74^{+0.26}_{-0.21}$& $ 0.15^{+0.04}_{-0.09}$& $ 0.11^{+0.01}_{-0.07}$& $ 0.06^{+0.04}_{-0.02}$& $ 0.88^{+0.08}_{-0.03}$& $ 0.93^{+0.04}_{-0.02}$& $ 0.96^{+0.01}_{-0.03}$& $ 1.76^{+0.03}_{-0.29}$& $ 1.58^{+0.39}_{-0.00}$& $ 2.05^{+0.15}_{-0.26}$\\
NGC 6637  & $ 1.99^{+0.04}_{-0.02}$& $ 1.83^{+0.04}_{-0.02}$& $ 1.80^{+0.69}_{-0.00}$& $ 0.83^{+0.05}_{-0.08}$& $ 0.63^{+0.07}_{-0.06}$& $ 0.64^{+0.00}_{-0.51}$& $ 0.41^{+0.05}_{-0.03}$& $ 0.49^{+0.03}_{-0.04}$& $ 0.47^{+0.43}_{-0.00}$& $ 1.87^{+0.03}_{-0.02}$& $ 1.77^{+0.02}_{-0.02}$& $ 1.70^{+0.09}_{-0.00}$\\
NGC 6642  & $ 2.20^{+0.08}_{-0.04}$& $ 2.46^{+0.12}_{-0.12}$& $ 2.26^{+0.16}_{-0.04}$& $ 0.04^{+0.14}_{-0.02}$& $ 0.08^{+0.07}_{-0.02}$& $ 0.06^{+0.10}_{-0.00}$& $ 0.97^{+0.00}_{-0.12}$& $ 0.93^{+0.02}_{-0.04}$& $ 0.95^{+0.00}_{-0.07}$& $ 1.23^{+0.16}_{-0.09}$& $ 1.59^{+0.04}_{-0.21}$& $ 1.71^{+0.00}_{-0.23}$\\
NGC 6717  & $ 2.48^{+0.06}_{-0.03}$& $ 2.72^{+0.00}_{-0.27}$& $ 2.56^{+0.12}_{-0.07}$& $ 1.33^{+0.12}_{-0.03}$& $ 1.01^{+0.11}_{-0.05}$& $ 0.79^{+0.05}_{-0.03}$& $ 0.30^{+0.02}_{-0.04}$& $ 0.46^{+0.00}_{-0.06}$& $ 0.53^{+0.02}_{-0.03}$& $ 1.45^{+0.03}_{-0.03}$& $ 1.43^{+0.03}_{-0.02}$& $ 1.50^{+0.00}_{-0.08}$\\
NGC 6723  & $ 4.30^{+0.05}_{-0.03}$& $ 3.81^{+0.06}_{-0.05}$& $ 3.44^{+0.03}_{-0.04}$& $ 0.78^{+0.07}_{-0.13}$& $ 1.21^{+0.02}_{-0.03}$& $ 1.46^{+0.11}_{-0.11}$& $ 0.69^{+0.05}_{-0.02}$& $ 0.52^{+0.01}_{-0.01}$& $ 0.40^{+0.03}_{-0.03}$& $ 3.97^{+0.06}_{-0.08}$& $ 3.69^{+0.05}_{-0.05}$& $ 3.38^{+0.02}_{-0.05}$\\
NGC 6171  & $ 3.96^{+0.06}_{-0.05}$& $ 3.93^{+0.08}_{-0.08}$& $ 3.99^{+0.06}_{-0.04}$& $ 1.45^{+0.07}_{-0.03}$& $ 0.94^{+0.48}_{-0.00}$& $ 1.27^{+0.04}_{-0.05}$& $ 0.46^{+0.01}_{-0.02}$& $ 0.61^{+0.00}_{-0.14}$& $ 0.52^{+0.01}_{-0.01}$& $ 2.41^{+0.03}_{-0.04}$& $ 2.61^{+0.00}_{-0.17}$& $ 2.51^{+0.05}_{-0.03}$\\
NGC 6316  & $ 3.86^{+0.35}_{-0.38}$& $ 4.37^{+0.17}_{-0.52}$& $ 4.08^{+0.37}_{-0.41}$& $ 0.93^{+0.11}_{-0.30}$& $ 0.63^{+0.17}_{-0.16}$& $ 0.72^{+0.14}_{-0.23}$& $ 0.61^{+0.10}_{-0.02}$& $ 0.75^{+0.04}_{-0.06}$& $ 0.70^{+0.07}_{-0.03}$& $ 1.72^{+0.06}_{-0.11}$& $ 1.59^{+0.18}_{-0.03}$& $ 1.60^{+0.19}_{-0.03}$\\
NGC 6388  & $ 4.03^{+0.17}_{-0.08}$& $ 4.28^{+0.07}_{-0.11}$& $ 4.20^{+0.07}_{-0.08}$& $ 1.26^{+0.05}_{-0.11}$& $ 1.04^{+0.18}_{-0.08}$& $ 1.09^{+0.08}_{-0.05}$& $ 0.52^{+0.04}_{-0.01}$& $ 0.61^{+0.03}_{-0.06}$& $ 0.59^{+0.01}_{-0.02}$& $ 1.59^{+0.08}_{-0.03}$& $ 1.64^{+0.05}_{-0.14}$& $ 1.50^{+0.03}_{-0.04}$\\
NGC 6539  & $ 3.92^{+0.22}_{-0.10}$& $ 3.88^{+0.16}_{-0.25}$& $ 3.58^{+0.26}_{-0.11}$& $ 2.18^{+0.20}_{-0.20}$& $ 2.01^{+0.23}_{-0.36}$& $ 1.99^{+0.16}_{-0.23}$& $ 0.28^{+0.06}_{-0.03}$& $ 0.32^{+0.08}_{-0.06}$& $ 0.29^{+0.06}_{-0.03}$& $ 2.74^{+0.20}_{-0.13}$& $ 2.70^{+0.20}_{-0.12}$& $ 2.66^{+0.23}_{-0.10}$\\
NGC 6553  & $ 4.64^{+0.09}_{-0.11}$& $ 4.37^{+0.10}_{-0.11}$& $ 4.12^{+0.09}_{-0.12}$& $ 2.54^{+0.08}_{-0.11}$& $ 2.71^{+0.16}_{-0.20}$& $ 2.78^{+0.18}_{-0.47}$& $ 0.29^{+0.01}_{-0.00}$& $ 0.23^{+0.04}_{-0.02}$& $ 0.19^{+0.08}_{-0.02}$& $ 0.31^{+0.01}_{-0.01}$& $ 0.30^{+0.01}_{-0.00}$& $ 0.30^{+0.01}_{-0.01}$\\
NGC 6652  & $ 3.20^{+0.11}_{-0.10}$& $ 3.77^{+0.33}_{-0.23}$& $ 3.55^{+0.29}_{-0.10}$& $ 0.11^{+0.03}_{-0.05}$& $ 0.02^{+0.08}_{-0.00}$& $ 0.05^{+0.06}_{-0.02}$& $ 0.94^{+0.02}_{-0.02}$& $ 0.99^{+0.00}_{-0.04}$& $ 0.97^{+0.01}_{-0.03}$& $ 2.91^{+0.04}_{-0.08}$& $ 2.74^{+0.12}_{-0.04}$& $ 2.83^{+0.04}_{-0.11}$\\
Terzan 3  & $ 4.42^{+0.19}_{-0.21}$& $ 3.81^{+0.21}_{-0.17}$& $ 3.37^{+0.16}_{-0.11}$& $ 2.23^{+0.17}_{-0.23}$& $ 2.22^{+0.16}_{-0.54}$& $ 2.33^{+0.10}_{-0.17}$& $ 0.33^{+0.04}_{-0.03}$& $ 0.26^{+0.12}_{-0.02}$& $ 0.18^{+0.04}_{-0.02}$& $ 1.92^{+0.11}_{-0.10}$& $ 1.82^{+0.12}_{-0.05}$& $ 1.73^{+0.12}_{-0.08}$\\
NGC 6256  & $ 3.15^{+0.15}_{-0.10}$& $ 2.77^{+0.10}_{-0.10}$& $ 2.50^{+0.13}_{-0.05}$& $ 1.27^{+0.32}_{-0.27}$& $ 1.69^{+0.24}_{-0.30}$& $ 1.62^{+0.25}_{-0.38}$& $ 0.43^{+0.08}_{-0.08}$& $ 0.24^{+0.09}_{-0.06}$& $ 0.21^{+0.13}_{-0.06}$& $ 0.79^{+0.14}_{-0.05}$& $ 0.71^{+0.09}_{-0.02}$& $ 0.74^{+0.07}_{-0.08}$\\
NGC 6304  & $ 3.69^{+0.14}_{-0.16}$& $ 3.40^{+0.13}_{-0.13}$& $ 3.18^{+0.16}_{-0.12}$& $ 1.72^{+0.29}_{-0.29}$& $ 1.74^{+0.06}_{-0.11}$& $ 1.66^{+0.12}_{-0.11}$& $ 0.36^{+0.08}_{-0.07}$& $ 0.32^{+0.02}_{-0.00}$& $ 0.31^{+0.02}_{-0.01}$& $ 0.88^{+0.10}_{-0.02}$& $ 0.98^{+0.02}_{-0.11}$& $ 0.96^{+0.04}_{-0.03}$\\
Pismis 26  & $ 4.89^{+0.51}_{-0.24}$& $ 4.20^{+0.40}_{-0.34}$& $ 3.70^{+0.43}_{-0.32}$& $ 1.75^{+0.18}_{-0.23}$& $ 1.73^{+0.25}_{-0.40}$& $ 1.77^{+0.10}_{-0.23}$& $ 0.47^{+0.06}_{-0.02}$& $ 0.42^{+0.08}_{-0.04}$& $ 0.35^{+0.06}_{-0.01}$& $ 2.13^{+0.13}_{-0.07}$& $ 1.92^{+0.21}_{-0.16}$& $ 1.75^{+0.23}_{-0.11}$\\
NGC 6540  & $ 2.51^{+0.25}_{-0.22}$& $ 2.53^{+0.26}_{-0.26}$& $ 2.53^{+0.23}_{-0.23}$& $ 1.53^{+0.06}_{-0.11}$& $ 1.42^{+0.13}_{-0.13}$& $ 1.26^{+0.17}_{-0.25}$& $ 0.24^{+0.06}_{-0.04}$& $ 0.28^{+0.03}_{-0.03}$& $ 0.34^{+0.09}_{-0.06}$& $ 0.54^{+0.04}_{-0.02}$& $ 0.53^{+0.03}_{-0.03}$& $ 0.52^{+0.10}_{-0.05}$\\
NGC 6569  & $ 2.85^{+0.29}_{-0.14}$& $ 2.78^{+0.24}_{-0.17}$& $ 2.65^{+0.27}_{-0.20}$& $ 1.63^{+0.06}_{-0.47}$& $ 1.66^{+0.18}_{-0.16}$& $ 1.68^{+0.09}_{-0.36}$& $ 0.27^{+0.18}_{-0.02}$& $ 0.25^{+0.03}_{-0.02}$& $ 0.22^{+0.10}_{-0.00}$& $ 1.31^{+0.07}_{-0.03}$& $ 1.30^{+0.05}_{-0.03}$& $ 1.30^{+0.03}_{-0.05}$\\
E456-78  & $ 3.97^{+0.14}_{-0.21}$& $ 3.62^{+0.24}_{-0.26}$& $ 3.36^{+0.24}_{-0.24}$& $ 1.91^{+0.20}_{-0.17}$& $ 1.53^{+0.58}_{-0.13}$& $ 1.87^{+0.14}_{-0.24}$& $ 0.35^{+0.02}_{-0.03}$& $ 0.40^{+0.01}_{-0.11}$& $ 0.28^{+0.04}_{-0.00}$& $ 1.46^{+0.12}_{-0.08}$& $ 1.45^{+0.07}_{-0.11}$& $ 1.36^{+0.10}_{-0.07}$\\
NGC 6325  & $ 2.54^{+0.29}_{-0.29}$& $ 1.96^{+0.40}_{-0.32}$& $ 1.53^{+0.44}_{-0.19}$& $ 0.78^{+0.34}_{-0.30}$& $ 0.90^{+0.25}_{-0.24}$& $ 1.10^{+0.15}_{-0.23}$& $ 0.53^{+0.15}_{-0.14}$& $ 0.37^{+0.11}_{-0.09}$& $ 0.16^{+0.14}_{-0.03}$& $ 1.61^{+0.14}_{-0.08}$& $ 1.50^{+0.14}_{-0.11}$& $ 1.27^{+0.17}_{-0.09}$\\
Djorg 2   & $ 2.41^{+0.07}_{-0.04}$& $ 1.36^{+0.33}_{-0.17}$& $ 0.83^{+0.21}_{-0.06}$& $ 0.58^{+0.12}_{-0.15}$& $ 0.57^{+0.08}_{-0.15}$& $ 0.62^{+0.16}_{-0.11}$& $ 0.61^{+0.08}_{-0.05}$& $ 0.41^{+0.13}_{-0.02}$& $ 0.14^{+0.11}_{-0.04}$& $ 0.80^{+0.03}_{-0.01}$& $ 0.60^{+0.07}_{-0.03}$& $ 0.47^{+0.06}_{-0.01}$\\\hline
 \end{tabular}
   \end{center}
   \end{tiny}
  \end{minipage}
  }
  \end{table*}

  \begin{figure*}
{\begin{center}
   \includegraphics[width=0.3\textwidth,angle=-90]{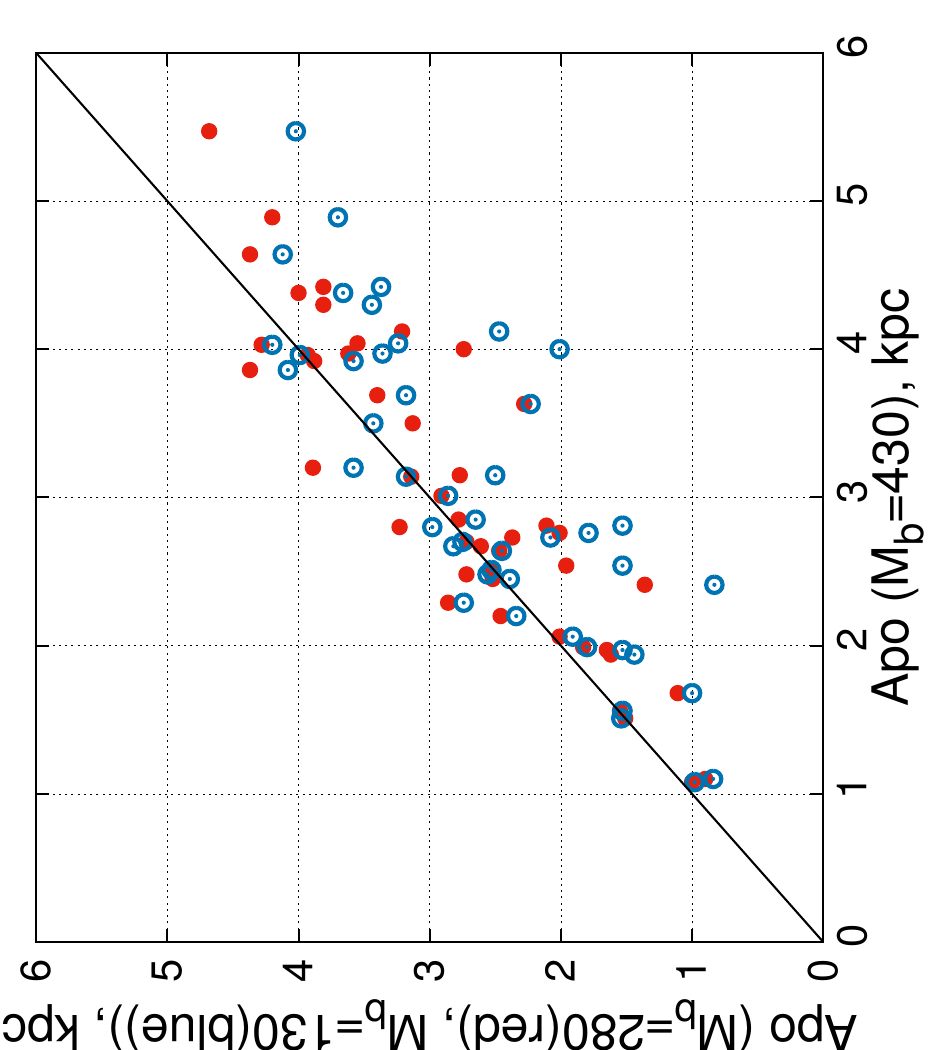}  \includegraphics[width=0.3\textwidth,angle=-90]{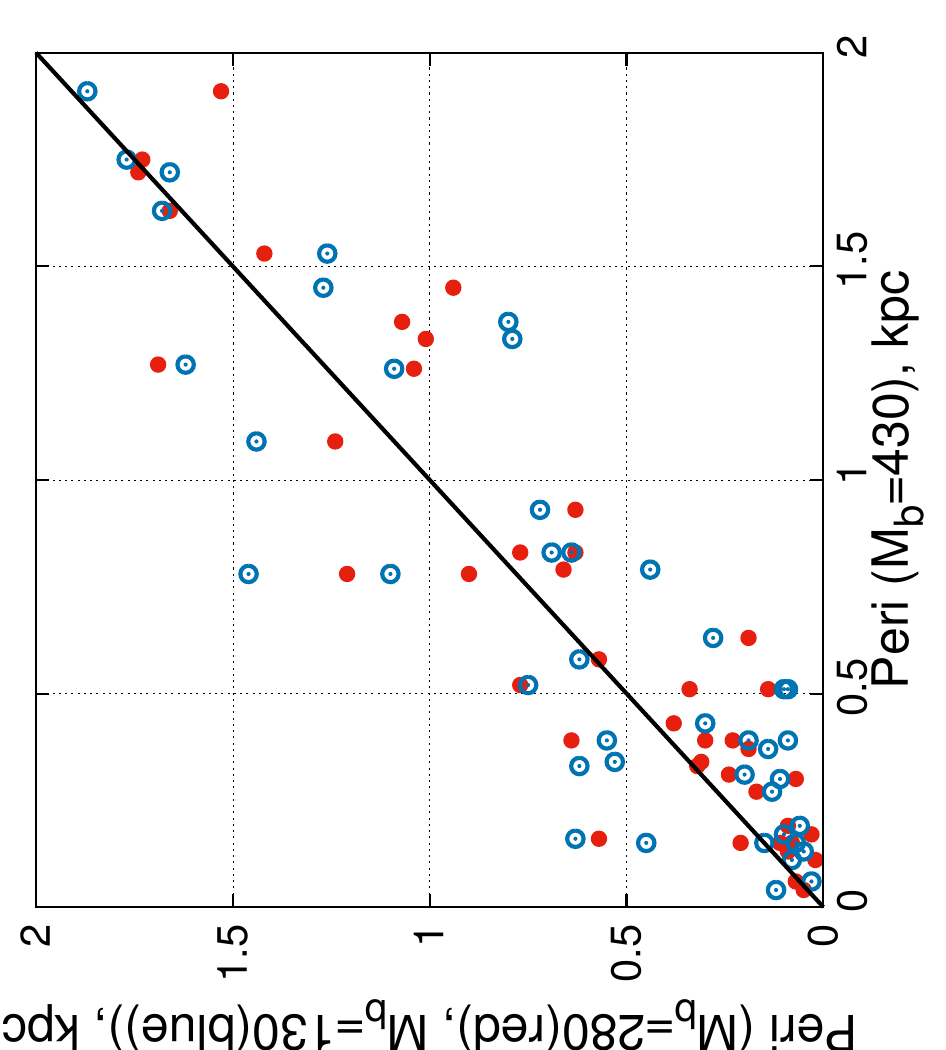}\

\medskip

\includegraphics[width=0.3\textwidth,angle=-90]{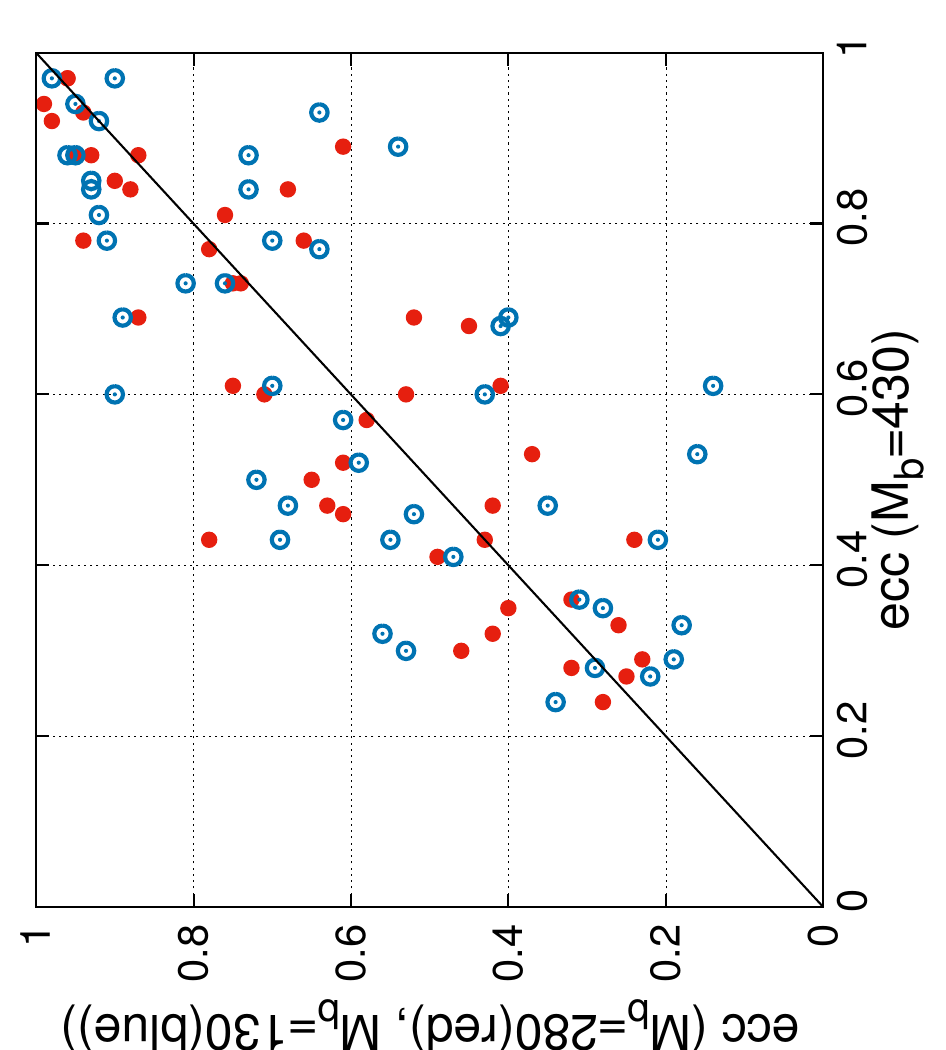}   \includegraphics[width=0.3\textwidth,angle=-90]{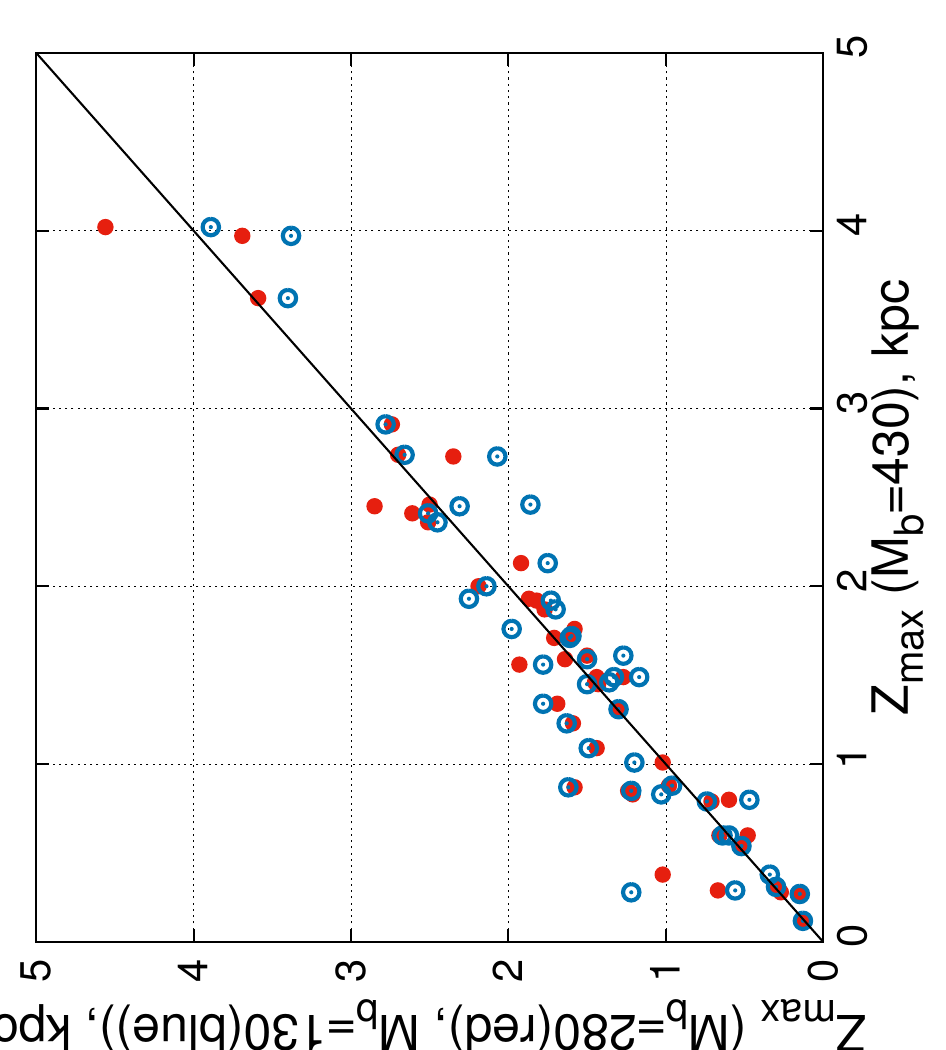}\

\caption{\small Comparison of the orbital parameters of the GCs in the potential with a bar with different mass. Other bar parameters: $\Omega_b=40$ km/s/kpc, $q_b=5$ kpc, $\theta_b=25^o$, axis ratio V0. Each panel has a line of coincise.}
\end{center}}
\end{figure*}

  \begin{table*}
  \rotatebox{90}
{
	\label{t:prop}
		\begin{minipage}{1.5\linewidth}
		\label{t:prop}
		\begin{tiny} \begin{center}
{\scriptsize {\bf Table 4:} Parameters of the orbits of 45 GCs in the potential with a bar versus the bar rotation velocity: \\

$\Omega_b=30$ km/s/kpc (index "1"), $\Omega_b=40$ km/s/kpc (index "2"), $\Omega_b=50$ km/s/kpc (index " 3").}

\bigskip

\begin{tabular}{|l|c|c|c||c|c|c||c|c|c||c|c|c|}\hline
ID   &$apo^1$ & $apo^2$ & $apo^3$ & $peri^1$ & $peri^2$ &$peri^3$ & $ecc^1$ & $ecc^2$ & $ecc^3$ & $Z_{max}^1$ & $Z_{max}^2$ & $Z_{max}^3$ \\
    &[kpc]&[kpc]&[kpc]&[kpc]&[kpc]&[kpc]& & & &[kpc]&[kpc]&[kpc] \\\hline
NGC 6144  & $ 4.43^{+0.09}_{-0.07}$& $ 4.38^{+0.07}_{-0.07}$& $ 4.47^{+0.07}_{-0.09}$& $ 1.00^{+0.09}_{-0.10}$& $ 1.09^{+0.07}_{-0.10}$& $ 1.06^{+0.06}_{-0.15}$& $ 0.63^{+0.04}_{-0.03}$& $ 0.60^{+0.03}_{-0.02}$& $ 0.62^{+0.04}_{-0.03}$& $ 3.63^{+0.08}_{-0.10}$& $ 3.62^{+0.14}_{-0.12}$& $ 3.66^{+0.05}_{-0.13}$\\
E452-11  & $ 2.73^{+0.21}_{-0.12}$& $ 2.80^{+0.21}_{-0.30}$& $ 2.77^{+0.15}_{-0.12}$& $ 0.15^{+0.03}_{-0.07}$& $ 0.19^{+0.06}_{-0.14}$& $ 0.12^{+0.12}_{-0.04}$& $ 0.90^{+0.04}_{-0.02}$& $ 0.88^{+0.08}_{-0.04}$& $ 0.92^{+0.02}_{-0.08}$& $ 1.96^{+0.05}_{-0.09}$& $ 1.93^{+0.07}_{-0.01}$& $ 1.90^{+0.10}_{-0.05}$\\
NGC 6266  & $ 2.67^{+0.07}_{-0.07}$& $ 2.67^{+0.07}_{-0.08}$& $ 2.73^{+0.07}_{-0.09}$& $ 1.37^{+0.03}_{-0.06}$& $ 1.37^{+0.05}_{-0.03}$& $ 1.36^{+0.03}_{-0.03}$& $ 0.32^{+0.02}_{-0.01}$& $ 0.32^{+0.01}_{-0.02}$& $ 0.33^{+0.02}_{-0.02}$& $ 1.01^{+0.01}_{-0.01}$& $ 1.01^{+0.02}_{-0.01}$& $ 1.02^{+0.02}_{-0.01}$\\
NGC 6273  & $ 5.49^{+0.16}_{-0.06}$& $ 5.47^{+0.19}_{-0.14}$& $ 5.72^{+0.09}_{-0.10}$& $ 0.32^{+0.11}_{-0.19}$& $ 0.33^{+0.09}_{-0.17}$& $ 0.19^{+0.11}_{-0.09}$& $ 0.89^{+0.06}_{-0.03}$& $ 0.88^{+0.06}_{-0.02}$& $ 0.94^{+0.03}_{-0.04}$& $ 4.13^{+0.56}_{-0.14}$& $ 4.02^{+0.28}_{-0.15}$& $ 4.19^{+0.40}_{-0.15}$\\
NGC 6293  & $ 3.42^{+0.29}_{-0.05}$& $ 3.50^{+0.25}_{-0.14}$& $ 3.57^{+0.29}_{-0.14}$& $ 0.38^{+0.08}_{-0.22}$& $ 0.27^{+0.10}_{-0.14}$& $ 0.20^{+0.09}_{-0.09}$& $ 0.80^{+0.11}_{-0.03}$& $ 0.85^{+0.08}_{-0.03}$& $ 0.89^{+0.05}_{-0.03}$& $ 2.46^{+0.15}_{-0.09}$& $ 2.45^{+0.30}_{-0.06}$& $ 2.57^{+0.14}_{-0.11}$\\
NGC 6342  & $ 2.48^{+0.13}_{-0.07}$& $ 2.76^{+0.03}_{-0.23}$& $ 2.55^{+0.16}_{-0.08}$& $ 0.62^{+0.07}_{-0.05}$& $ 0.52^{+0.15}_{-0.01}$& $ 0.67^{+0.14}_{-0.07}$& $ 0.60^{+0.04}_{-0.04}$& $ 0.68^{+0.00}_{-0.09}$& $ 0.58^{+0.05}_{-0.07}$& $ 1.73^{+0.05}_{-0.05}$& $ 1.71^{+0.08}_{-0.06}$& $ 1.75^{+0.09}_{-0.03}$\\
NGC 6355  & $ 3.59^{+0.26}_{-0.04}$& $ 4.00^{+0.00}_{-0.40}$& $ 3.77^{+0.14}_{-0.11}$& $ 0.20^{+0.03}_{-0.10}$& $ 0.15^{+0.10}_{-0.10}$& $ 0.20^{+0.04}_{-0.10}$& $ 0.90^{+0.04}_{-0.01}$& $ 0.93^{+0.04}_{-0.05}$& $ 0.90^{+0.05}_{-0.02}$& $ 2.45^{+0.31}_{-0.01}$& $ 2.46^{+0.09}_{-0.12}$& $ 2.49^{+0.08}_{-0.10}$\\
Terzan 2  & $ 1.37^{+0.40}_{-0.21}$& $ 1.68^{+0.14}_{-0.49}$& $ 1.37^{+0.39}_{-0.26}$& $ 0.32^{+0.08}_{-0.06}$& $ 0.15^{+0.21}_{-0.00}$& $ 0.32^{+0.10}_{-0.08}$& $ 0.62^{+0.12}_{-0.12}$& $ 0.84^{+0.00}_{-0.29}$& $ 0.62^{+0.13}_{-0.16}$& $ 0.54^{+0.09}_{-0.06}$& $ 0.60^{+0.07}_{-0.11}$& $ 0.55^{+0.10}_{-0.09}$\\
Terzan 4  & $ 1.09^{+0.34}_{-0.22}$& $ 1.08^{+0.29}_{-0.19}$& $ 1.15^{+0.19}_{-0.24}$& $ 0.34^{+0.09}_{-0.07}$& $ 0.39^{+0.13}_{-0.09}$& $ 0.47^{+0.07}_{-0.12}$& $ 0.52^{+0.09}_{-0.08}$& $ 0.47^{+0.08}_{-0.09}$& $ 0.42^{+0.09}_{-0.08}$& $ 0.61^{+0.07}_{-0.04}$& $ 0.60^{+0.08}_{-0.04}$& $ 0.60^{+0.09}_{-0.04}$\\
BH 229   & $ 2.78^{+0.09}_{-0.18}$& $ 2.73^{+0.15}_{-0.12}$& $ 2.73^{+0.18}_{-0.14}$& $ 0.10^{+0.13}_{-0.02}$& $ 0.16^{+0.12}_{-0.05}$& $ 0.25^{+0.09}_{-0.12}$& $ 0.93^{+0.01}_{-0.08}$& $ 0.89^{+0.03}_{-0.07}$& $ 0.83^{+0.07}_{-0.04}$& $ 2.74^{+0.10}_{-0.17}$& $ 2.73^{+0.12}_{-0.16}$& $ 2.71^{+0.17}_{-0.15}$\\
Liller 1  & $ 1.10^{+0.22}_{-0.06}$& $ 1.10^{+0.21}_{-0.10}$& $ 1.10^{+0.22}_{-0.08}$& $ 0.37^{+0.19}_{-0.26}$& $ 0.37^{+0.22}_{-0.18}$& $ 0.37^{+0.28}_{-0.21}$& $ 0.50^{+0.31}_{-0.14}$& $ 0.50^{+0.21}_{-0.18}$& $ 0.50^{+0.24}_{-0.21}$& $ 0.28^{+0.10}_{-0.17}$& $ 0.27^{+0.04}_{-0.17}$& $ 0.27^{+0.06}_{-0.17}$\\
NGC 6380  & $ 2.53^{+0.13}_{-0.14}$& $ 2.45^{+0.15}_{-0.08}$& $ 2.50^{+0.16}_{-0.10}$& $ 0.20^{+0.15}_{-0.00}$& $ 0.30^{+0.07}_{-0.05}$& $ 0.12^{+0.16}_{-0.02}$& $ 0.85^{+0.00}_{-0.09}$& $ 0.78^{+0.03}_{-0.03}$& $ 0.91^{+0.01}_{-0.10}$& $ 0.92^{+0.03}_{-0.07}$& $ 0.87^{+0.06}_{-0.02}$& $ 0.95^{+0.22}_{-0.11}$\\
Terzan 1  & $ 3.00^{+0.16}_{-0.18}$& $ 3.01^{+0.20}_{-0.20}$& $ 2.93^{+0.17}_{-0.19}$& $ 0.82^{+0.07}_{-0.06}$& $ 0.83^{+0.06}_{-0.06}$& $ 0.90^{+0.05}_{-0.07}$& $ 0.57^{+0.02}_{-0.02}$& $ 0.57^{+0.02}_{-0.02}$& $ 0.53^{+0.02}_{-0.02}$& $ 0.13^{+0.00}_{-0.02}$& $ 0.12^{+0.00}_{-0.01}$& $ 0.12^{+0.00}_{-0.01}$\\
NGC 6401  & $ 3.76^{+0.34}_{-0.02}$& $ 4.12^{+0.11}_{-0.20}$& $ 3.78^{+0.36}_{-0.04}$& $ 0.06^{+0.13}_{-0.03}$& $ 0.17^{+0.09}_{-0.12}$& $ 0.15^{+0.10}_{-0.05}$& $ 0.97^{+0.01}_{-0.06}$& $ 0.92^{+0.05}_{-0.04}$& $ 0.92^{+0.03}_{-0.04}$& $ 1.65^{+0.12}_{-0.03}$& $ 1.56^{+0.62}_{-0.08}$& $ 2.43^{+0.00}_{-0.79}$\\
Pal 6    & $ 3.59^{+0.73}_{-0.36}$& $ 4.04^{+0.34}_{-0.76}$& $ 3.70^{+0.94}_{-0.39}$& $ 0.06^{+0.22}_{-0.00}$& $ 0.06^{+0.19}_{-0.01}$& $ 0.10^{+0.11}_{-0.06}$& $ 0.97^{+0.00}_{-0.11}$& $ 0.97^{+0.00}_{-0.08}$& $ 0.95^{+0.03}_{-0.04}$& $ 2.55^{+0.26}_{-0.68}$& $ 2.36^{+0.27}_{-0.57}$& $ 1.90^{+0.74}_{-0.24}$\\
Terzan 5  & $ 2.05^{+0.11}_{-0.15}$& $ 2.06^{+0.16}_{-0.13}$& $ 2.08^{+0.17}_{-0.15}$& $ 0.51^{+0.04}_{-0.07}$& $ 0.51^{+0.05}_{-0.08}$& $ 0.48^{+0.01}_{-0.07}$& $ 0.60^{+0.03}_{-0.01}$& $ 0.60^{+0.04}_{-0.01}$& $ 0.62^{+0.06}_{-0.01}$& $ 0.28^{+0.00}_{-0.02}$& $ 0.28^{+0.03}_{-0.01}$& $ 0.28^{+0.01}_{-0.01}$\\
NGC 6440  & $ 1.55^{+0.14}_{-0.11}$& $ 1.51^{+0.13}_{-0.04}$& $ 1.51^{+0.13}_{-0.05}$& $ 0.15^{+0.11}_{-0.10}$& $ 0.13^{+0.10}_{-0.07}$& $ 0.13^{+0.10}_{-0.10}$& $ 0.83^{+0.10}_{-0.12}$& $ 0.84^{+0.09}_{-0.10}$& $ 0.85^{+0.11}_{-0.12}$& $ 0.83^{+0.07}_{-0.04}$& $ 0.85^{+0.07}_{-0.06}$& $ 0.84^{+0.11}_{-0.05}$\\
Terzan 6  & $ 1.94^{+0.31}_{-0.38}$& $ 1.94^{+0.28}_{-0.32}$& $ 1.94^{+0.34}_{-0.25}$& $ 0.33^{+0.04}_{-0.10}$& $ 0.31^{+0.07}_{-0.11}$& $ 0.33^{+0.08}_{-0.07}$& $ 0.71^{+0.06}_{-0.03}$& $ 0.73^{+0.07}_{-0.05}$& $ 0.71^{+0.05}_{-0.04}$& $ 0.31^{+0.19}_{-0.05}$& $ 0.29^{+0.19}_{-0.04}$& $ 0.29^{+0.14}_{-0.05}$\\
NGC 6453  & $ 2.77^{+0.24}_{-0.21}$& $ 2.64^{+0.39}_{-0.17}$& $ 2.87^{+0.13}_{-0.24}$& $ 0.18^{+0.11}_{-0.03}$& $ 0.34^{+0.05}_{-0.18}$& $ 0.22^{+0.08}_{-0.09}$& $ 0.88^{+0.02}_{-0.08}$& $ 0.77^{+0.12}_{-0.01}$& $ 0.86^{+0.05}_{-0.06}$& $ 1.90^{+0.13}_{-0.12}$& $ 2.00^{+0.10}_{-0.17}$& $ 1.93^{+0.19}_{-0.10}$\\
Terzan 9  & $ 2.73^{+0.31}_{-0.44}$& $ 2.70^{+0.31}_{-0.38}$& $ 2.72^{+0.24}_{-0.42}$& $ 0.45^{+0.15}_{-0.14}$& $ 0.43^{+0.13}_{-0.16}$& $ 0.52^{+0.10}_{-0.19}$& $ 0.72^{+0.07}_{-0.10}$& $ 0.73^{+0.09}_{-0.09}$& $ 0.68^{+0.09}_{-0.05}$& $ 0.36^{+0.06}_{-0.03}$& $ 0.38^{+0.03}_{-0.04}$& $ 0.36^{+0.09}_{-0.06}$\\
NGC 6522  & $ 1.97^{+0.16}_{-0.09}$& $ 1.97^{+0.10}_{-0.12}$& $ 2.24^{+0.01}_{-0.31}$& $ 0.84^{+0.18}_{-0.14}$& $ 0.79^{+0.13}_{-0.17}$& $ 0.61^{+0.39}_{-0.13}$& $ 0.40^{+0.06}_{-0.06}$& $ 0.43^{+0.08}_{-0.06}$& $ 0.57^{+0.04}_{-0.20}$& $ 1.49^{+0.10}_{-0.07}$& $ 1.49^{+0.07}_{-0.08}$& $ 1.48^{+0.09}_{-0.07}$\\
NGC 6528  & $ 2.82^{+0.14}_{-0.08}$& $ 2.81^{+0.32}_{-0.17}$& $ 2.81^{+0.20}_{-0.09}$& $ 0.53^{+0.05}_{-0.09}$& $ 0.51^{+0.07}_{-0.07}$& $ 0.57^{+0.04}_{-0.08}$& $ 0.68^{+0.05}_{-0.01}$& $ 0.69^{+0.03}_{-0.01}$& $ 0.66^{+0.04}_{-0.01}$& $ 0.85^{+0.05}_{-0.03}$& $ 0.83^{+0.08}_{-0.02}$& $ 0.82^{+0.08}_{-0.00}$\\
NGC 6558  & $ 2.91^{+0.32}_{-0.00}$& $ 3.63^{+0.00}_{-0.52}$& $ 3.09^{+0.22}_{-0.03}$& $ 0.29^{+0.06}_{-0.09}$& $ 0.39^{+0.02}_{-0.14}$& $ 0.33^{+0.10}_{-0.05}$& $ 0.82^{+0.05}_{-0.02}$& $ 0.81^{+0.04}_{-0.02}$& $ 0.81^{+0.02}_{-0.04}$& $ 1.11^{+0.07}_{-0.03}$& $ 1.09^{+0.05}_{-0.02}$& $ 1.11^{+0.04}_{-0.04}$\\
NGC 6624  & $ 1.60^{+0.03}_{-0.02}$& $ 1.56^{+0.04}_{-0.01}$& $ 1.58^{+0.03}_{-0.02}$& $ 0.49^{+0.11}_{-0.09}$& $ 0.63^{+0.04}_{-0.12}$& $ 0.58^{+0.06}_{-0.08}$& $ 0.53^{+0.07}_{-0.08}$& $ 0.43^{+0.08}_{-0.03}$& $ 0.46^{+0.06}_{-0.04}$& $ 1.51^{+0.03}_{-0.01}$& $ 1.49^{+0.03}_{-0.01}$& $ 1.50^{+0.02}_{-0.03}$\\
NGC 6626  & $ 3.20^{+0.09}_{-0.12}$& $ 3.14^{+0.19}_{-0.06}$& $ 3.19^{+0.16}_{-0.14}$& $ 0.72^{+0.05}_{-0.15}$& $ 0.39^{+0.23}_{-0.18}$& $ 0.72^{+0.07}_{-0.08}$& $ 0.63^{+0.06}_{-0.02}$& $ 0.78^{+0.10}_{-0.10}$& $ 0.63^{+0.05}_{-0.03}$& $ 1.20^{+0.05}_{-0.03}$& $ 1.34^{+0.00}_{-0.14}$& $ 1.19^{+0.04}_{-0.03}$\\
NGC 6638  & $ 2.41^{+0.29}_{-0.15}$& $ 2.29^{+0.46}_{-0.10}$& $ 2.43^{+0.40}_{-0.18}$& $ 0.10^{+0.03}_{-0.05}$& $ 0.15^{+0.04}_{-0.09}$& $ 0.09^{+0.05}_{-0.05}$& $ 0.92^{+0.04}_{-0.03}$& $ 0.88^{+0.08}_{-0.03}$& $ 0.93^{+0.04}_{-0.04}$& $ 1.67^{+0.12}_{-0.24}$& $ 1.76^{+0.03}_{-0.29}$& $ 1.69^{+0.05}_{-0.21}$\\
NGC 6637  & $ 1.98^{+0.04}_{-0.02}$& $ 1.99^{+0.04}_{-0.02}$& $ 1.99^{+0.04}_{-0.03}$& $ 0.79^{+0.10}_{-0.06}$& $ 0.83^{+0.05}_{-0.08}$& $ 0.78^{+0.06}_{-0.09}$& $ 0.43^{+0.03}_{-0.05}$& $ 0.41^{+0.05}_{-0.03}$& $ 0.44^{+0.05}_{-0.03}$& $ 1.86^{+0.03}_{-0.03}$& $ 1.87^{+0.03}_{-0.02}$& $ 1.87^{+0.04}_{-0.02}$\\
NGC 6642  & $ 2.31^{+0.06}_{-0.09}$& $ 2.20^{+0.08}_{-0.04}$& $ 2.20^{+0.15}_{-0.05}$& $ 0.13^{+0.22}_{-0.00}$& $ 0.04^{+0.14}_{-0.02}$& $ 0.10^{+0.12}_{-0.05}$& $ 0.90^{+0.00}_{-0.17}$& $ 0.97^{+0.00}_{-0.12}$& $ 0.92^{+0.03}_{-0.09}$& $ 1.15^{+0.02}_{-0.05}$& $ 1.23^{+0.16}_{-0.09}$& $ 1.15^{+0.11}_{-0.07}$\\
NGC 6717  & $ 2.46^{+0.08}_{-0.02}$& $ 2.48^{+0.06}_{-0.03}$& $ 2.57^{+0.07}_{-0.05}$& $ 1.33^{+0.00}_{-0.11}$& $ 1.33^{+0.12}_{-0.03}$& $ 1.13^{+0.09}_{-0.07}$& $ 0.30^{+0.04}_{-0.00}$& $ 0.30^{+0.02}_{-0.04}$& $ 0.39^{+0.03}_{-0.03}$& $ 1.44^{+0.02}_{-0.03}$& $ 1.45^{+0.03}_{-0.03}$& $ 1.44^{+0.02}_{-0.02}$\\
NGC 6723  & $ 4.29^{+0.11}_{-0.04}$& $ 4.30^{+0.05}_{-0.03}$& $ 4.45^{+0.11}_{-0.09}$& $ 0.72^{+0.16}_{-0.25}$& $ 0.78^{+0.07}_{-0.13}$& $ 0.85^{+0.09}_{-0.13}$& $ 0.71^{+0.10}_{-0.05}$& $ 0.69^{+0.05}_{-0.02}$& $ 0.68^{+0.04}_{-0.03}$& $ 3.93^{+0.09}_{-0.07}$& $ 3.97^{+0.06}_{-0.08}$& $ 3.99^{+0.09}_{-0.07}$\\
NGC 6171  & $ 3.98^{+0.06}_{-0.06}$& $ 3.96^{+0.06}_{-0.05}$& $ 3.97^{+0.05}_{-0.10}$& $ 1.56^{+0.02}_{-0.14}$& $ 1.45^{+0.07}_{-0.03}$& $ 1.34^{+0.05}_{-0.11}$& $ 0.44^{+0.03}_{-0.01}$& $ 0.46^{+0.01}_{-0.02}$& $ 0.49^{+0.03}_{-0.01}$& $ 2.44^{+0.02}_{-0.04}$& $ 2.41^{+0.03}_{-0.04}$& $ 2.41^{+0.04}_{-0.03}$\\
NGC 6316  & $ 3.89^{+0.39}_{-0.35}$& $ 3.86^{+0.35}_{-0.38}$& $ 3.96^{+0.92}_{-0.61}$& $ 0.86^{+0.16}_{-0.20}$& $ 0.93^{+0.11}_{-0.30}$& $ 0.71^{+0.17}_{-0.20}$& $ 0.64^{+0.06}_{-0.05}$& $ 0.61^{+0.10}_{-0.02}$& $ 0.70^{+0.07}_{-0.04}$& $ 1.66^{+0.09}_{-0.04}$& $ 1.72^{+0.06}_{-0.11}$& $ 1.63^{+0.13}_{-0.06}$\\
NGC 6388  & $ 3.96^{+0.11}_{-0.08}$& $ 4.03^{+0.17}_{-0.08}$& $ 4.31^{+0.04}_{-0.11}$& $ 1.14^{+0.04}_{-0.06}$& $ 1.26^{+0.05}_{-0.11}$& $ 1.19^{+0.11}_{-0.07}$& $ 0.55^{+0.02}_{-0.01}$& $ 0.52^{+0.04}_{-0.01}$& $ 0.57^{+0.02}_{-0.04}$& $ 2.20^{+0.08}_{-0.02}$& $ 1.59^{+0.08}_{-0.03}$& $ 1.30^{+0.05}_{-0.01}$\\
NGC 6539  & $ 4.01^{+0.31}_{-0.18}$& $ 3.92^{+0.22}_{-0.10}$& $ 4.31^{+1.20}_{-0.33}$& $ 1.89^{+0.29}_{-0.40}$& $ 2.18^{+0.20}_{-0.20}$& $ 2.17^{+0.16}_{-0.21}$& $ 0.36^{+0.11}_{-0.07}$& $ 0.28^{+0.06}_{-0.03}$& $ 0.33^{+0.10}_{-0.02}$& $ 2.76^{+0.19}_{-0.12}$& $ 2.74^{+0.20}_{-0.13}$& $ 2.83^{+0.48}_{-0.20}$\\
NGC 6553  & $ 4.66^{+0.07}_{-0.07}$& $ 4.64^{+0.09}_{-0.11}$& $ 6.82^{+0.38}_{-1.39}$& $ 2.31^{+0.08}_{-0.08}$& $ 2.54^{+0.08}_{-0.11}$& $ 2.29^{+0.05}_{-0.08}$& $ 0.34^{+0.01}_{-0.01}$& $ 0.29^{+0.01}_{-0.00}$& $ 0.50^{+0.02}_{-0.09}$& $ 0.31^{+0.01}_{-0.01}$& $ 0.31^{+0.01}_{-0.01}$& $ 0.38^{+0.01}_{-0.05}$\\
NGC 6652  & $ 3.22^{+0.12}_{-0.10}$& $ 3.20^{+0.11}_{-0.10}$& $ 3.33^{+0.03}_{-0.12}$& $ 0.05^{+0.06}_{-0.01}$& $ 0.11^{+0.03}_{-0.05}$& $ 0.11^{+0.06}_{-0.03}$& $ 0.97^{+0.00}_{-0.03}$& $ 0.94^{+0.02}_{-0.02}$& $ 0.94^{+0.01}_{-0.04}$& $ 2.88^{+0.06}_{-0.07}$& $ 2.91^{+0.04}_{-0.08}$& $ 2.87^{+0.09}_{-0.05}$\\
Terzan 3  & $ 4.15^{+0.19}_{-0.10}$& $ 4.42^{+0.19}_{-0.21}$& $ 5.91^{+0.15}_{-0.15}$& $ 2.30^{+0.15}_{-0.22}$& $ 2.23^{+0.17}_{-0.23}$& $ 2.32^{+0.14}_{-0.13}$& $ 0.29^{+0.05}_{-0.03}$& $ 0.33^{+0.04}_{-0.03}$& $ 0.44^{+0.03}_{-0.04}$& $ 1.87^{+0.12}_{-0.07}$& $ 1.92^{+0.11}_{-0.10}$& $ 2.25^{+0.10}_{-0.11}$\\
NGC 6256  & $ 3.14^{+0.10}_{-0.08}$& $ 3.15^{+0.15}_{-0.10}$& $ 3.21^{+0.22}_{-0.07}$& $ 1.55^{+0.30}_{-0.19}$& $ 1.27^{+0.32}_{-0.27}$& $ 1.78^{+0.25}_{-0.22}$& $ 0.34^{+0.05}_{-0.08}$& $ 0.43^{+0.08}_{-0.08}$& $ 0.29^{+0.06}_{-0.04}$& $ 0.73^{+0.08}_{-0.04}$& $ 0.79^{+0.14}_{-0.05}$& $ 0.73^{+0.07}_{-0.03}$\\
NGC 6304  & $ 3.63^{+0.12}_{-0.18}$& $ 3.69^{+0.14}_{-0.16}$& $ 3.83^{+0.21}_{-0.17}$& $ 1.83^{+0.07}_{-0.11}$& $ 1.72^{+0.29}_{-0.29}$& $ 1.84^{+0.07}_{-0.10}$& $ 0.33^{+0.01}_{-0.01}$& $ 0.36^{+0.08}_{-0.07}$& $ 0.35^{+0.01}_{-0.00}$& $ 0.89^{+0.05}_{-0.03}$& $ 0.88^{+0.10}_{-0.02}$& $ 0.90^{+0.05}_{-0.03}$\\
Pismis 26  & $ 4.88^{+0.27}_{-0.31}$& $ 4.89^{+0.51}_{-0.24}$& $ 6.23^{+0.29}_{-0.43}$& $ 1.70^{+0.05}_{-0.36}$& $ 1.75^{+0.18}_{-0.23}$& $ 1.75^{+0.16}_{-0.20}$& $ 0.48^{+0.09}_{-0.01}$& $ 0.47^{+0.06}_{-0.02}$& $ 0.56^{+0.05}_{-0.04}$& $ 2.04^{+0.26}_{-0.17}$& $ 2.13^{+0.13}_{-0.07}$& $ 2.32^{+0.16}_{-0.12}$\\
NGC 6540  & $ 2.53^{+0.28}_{-0.21}$& $ 2.51^{+0.25}_{-0.22}$& $ 2.56^{+0.27}_{-0.26}$& $ 1.59^{+0.12}_{-0.10}$& $ 1.53^{+0.06}_{-0.11}$& $ 1.61^{+0.15}_{-0.20}$& $ 0.23^{+0.04}_{-0.04}$& $ 0.24^{+0.06}_{-0.04}$& $ 0.23^{+0.06}_{-0.04}$& $ 0.54^{+0.03}_{-0.02}$& $ 0.54^{+0.04}_{-0.02}$& $ 0.54^{+0.04}_{-0.02}$\\
NGC 6569  & $ 3.02^{+0.16}_{-0.18}$& $ 2.85^{+0.29}_{-0.14}$& $ 2.98^{+0.20}_{-0.21}$& $ 1.71^{+0.07}_{-0.26}$& $ 1.63^{+0.06}_{-0.47}$& $ 1.60^{+0.11}_{-0.17}$& $ 0.28^{+0.06}_{-0.01}$& $ 0.27^{+0.18}_{-0.02}$& $ 0.30^{+0.03}_{-0.00}$& $ 1.32^{+0.08}_{-0.04}$& $ 1.31^{+0.07}_{-0.03}$& $ 1.30^{+0.05}_{-0.03}$\\
E456-78  & $ 3.99^{+0.17}_{-0.24}$& $ 3.97^{+0.14}_{-0.21}$& $ 4.05^{+0.81}_{-0.45}$& $ 1.65^{+0.26}_{-0.19}$& $ 1.91^{+0.20}_{-0.17}$& $ 1.74^{+0.12}_{-0.22}$& $ 0.41^{+0.03}_{-0.04}$& $ 0.35^{+0.02}_{-0.03}$& $ 0.40^{+0.06}_{-0.01}$& $ 1.48^{+0.11}_{-0.07}$& $ 1.46^{+0.12}_{-0.08}$& $ 1.55^{+0.14}_{-0.14}$\\
NGC 6325  & $ 2.56^{+0.32}_{-0.35}$& $ 2.54^{+0.29}_{-0.29}$& $ 2.54^{+0.50}_{-0.29}$& $ 0.65^{+0.27}_{-0.38}$& $ 0.78^{+0.34}_{-0.30}$& $ 0.78^{+0.23}_{-0.30}$& $ 0.59^{+0.22}_{-0.12}$& $ 0.53^{+0.15}_{-0.14}$& $ 0.53^{+0.17}_{-0.10}$& $ 1.94^{+0.16}_{-0.18}$& $ 1.61^{+0.14}_{-0.08}$& $ 1.59^{+0.20}_{-0.10}$\\
Djorg 2   & $ 2.43^{+0.12}_{-0.04}$& $ 2.41^{+0.07}_{-0.04}$& $ 2.48^{+0.08}_{-0.07}$& $ 0.55^{+0.18}_{-0.19}$& $ 0.58^{+0.12}_{-0.15}$& $ 0.47^{+0.15}_{-0.15}$& $ 0.63^{+0.11}_{-0.08}$& $ 0.61^{+0.08}_{-0.05}$& $ 0.68^{+0.09}_{-0.08}$& $ 0.80^{+0.08}_{-0.02}$& $ 0.80^{+0.03}_{-0.01}$& $ 1.24^{+0.00}_{-0.40}$\\
\hline
 \end{tabular}
   \end{center}
   \end{tiny}
  \end{minipage}
  }
  \end{table*}

\begin{figure*}
{\begin{center}
   \includegraphics[width=0.3\textwidth,angle=-90]{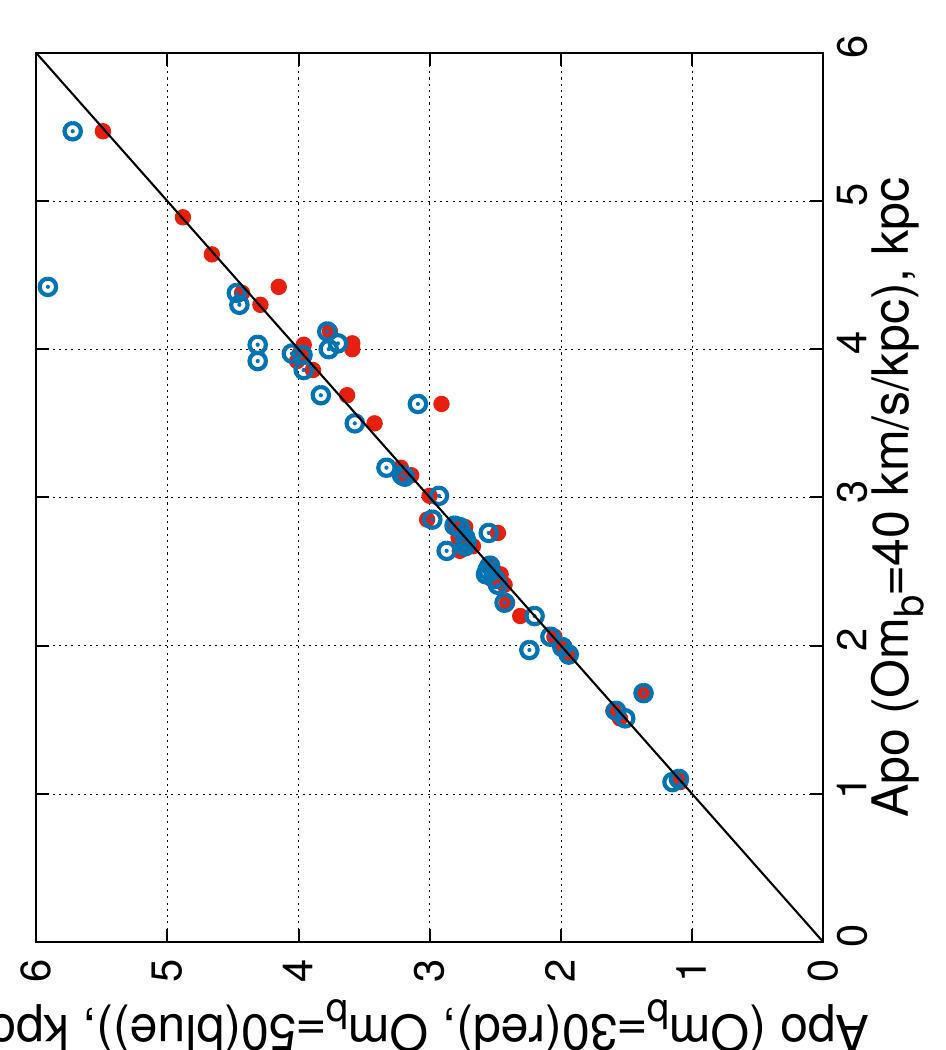}
\includegraphics[width=0.3\textwidth,angle=-90]{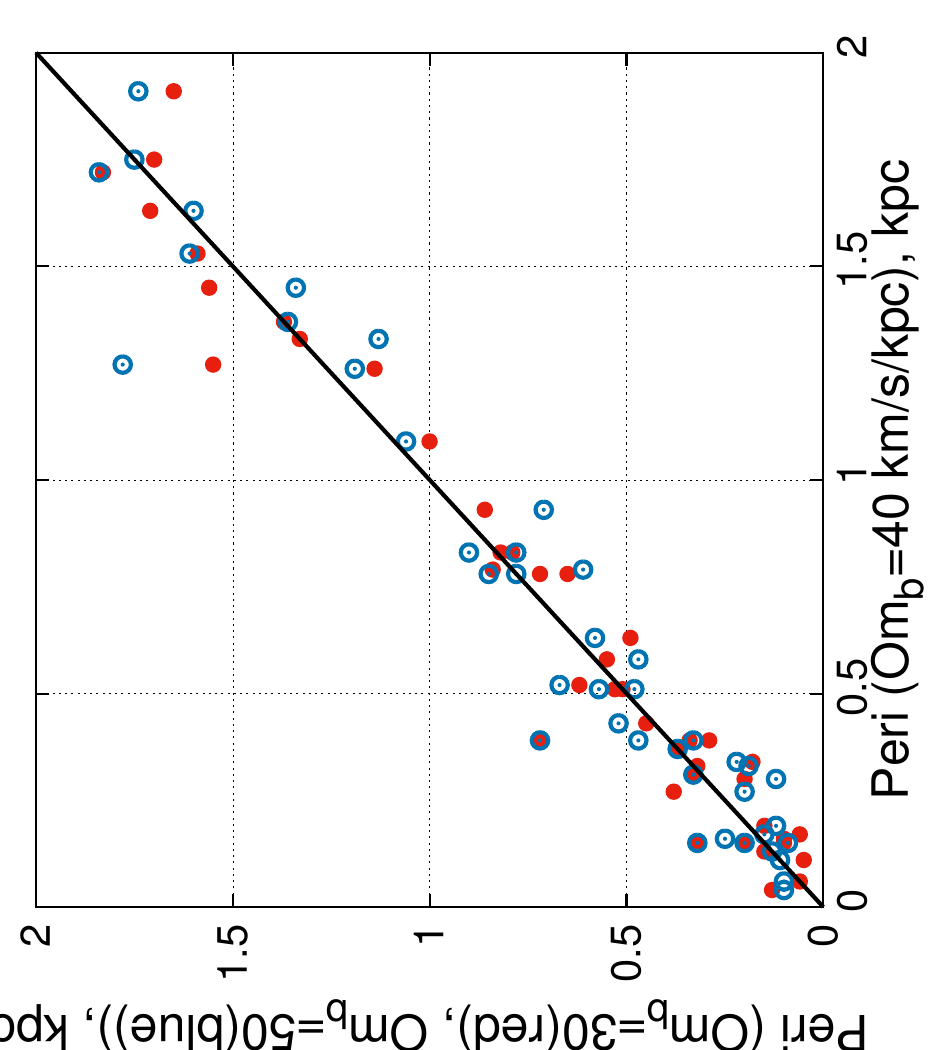}\

 \medskip
   \includegraphics[width=0.3\textwidth,angle=-90]{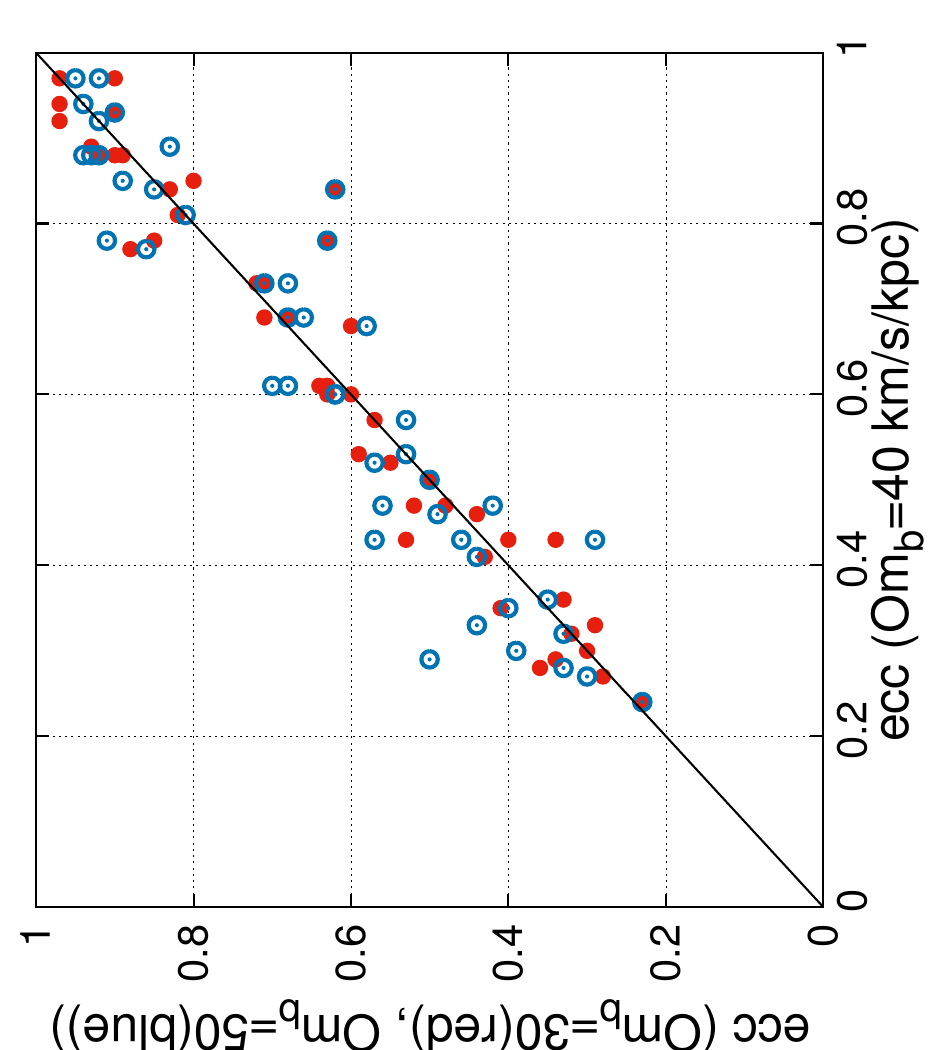}   \includegraphics[width=0.3\textwidth,angle=-90]{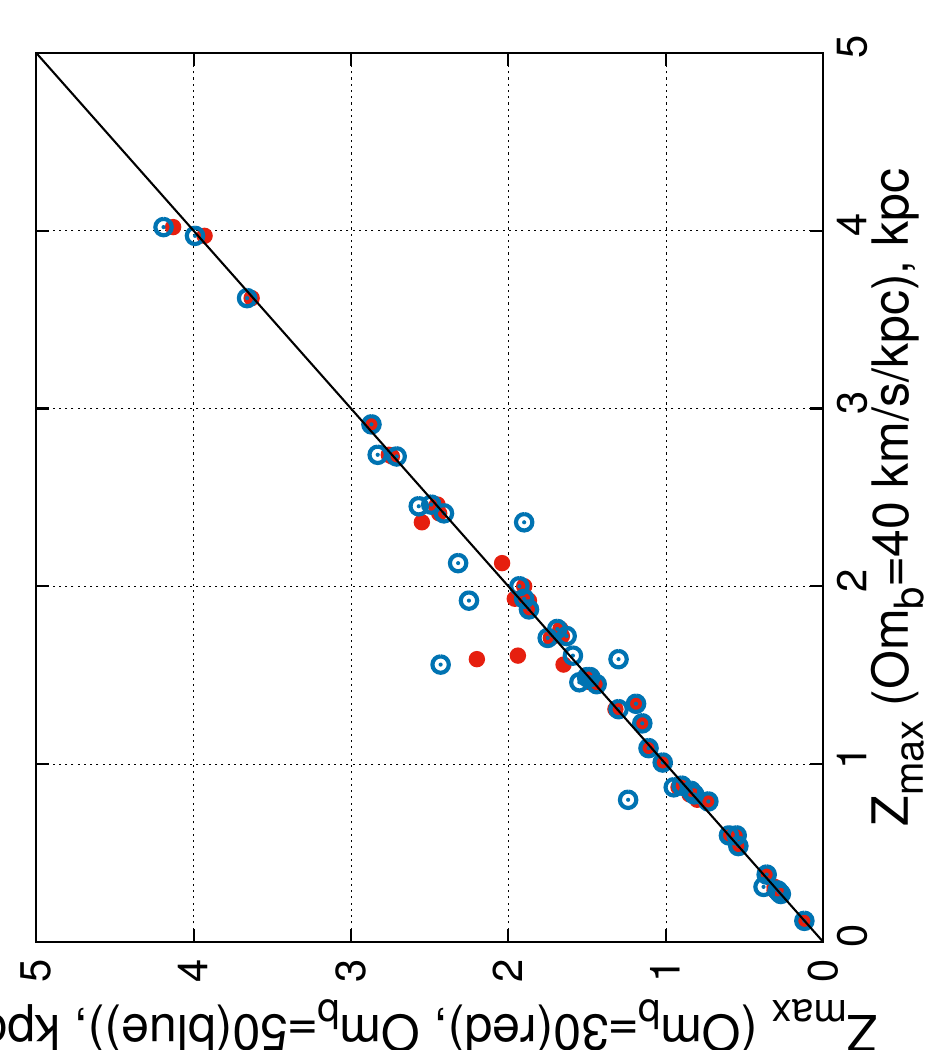}\

\caption{\small Comparison of the orbital parameters of the GCs in the potential with a bar with different rotation velocities. Other bar parameters: $M_b=430 M_G$, $q_b=5$ kpc, $\theta_b=25^o$, axis ratio V0. Each panel has a line of coincise.}
\end{center}}
\end{figure*}

  \begin{table*}
\rotatebox{90}{
	\label{t:prop}
		\begin{minipage}{1.5\linewidth}
		\label{t:prop}
		\begin{tiny} \begin{center}
{\scriptsize {\bf Table 5:} Parameters of the orbits of 45 GCs in the potential with a bar the length of the bar: \\

$q_b=4.5$ kpc (index "1"), $q_b=5$ kpc (index "2"), $q_b=5.5$ kpc (index "3").}

\bigskip

\begin{tabular}{|l|c|c|c||c|c|c||c|c|c||c|c|c|}\hline
 ID  & $apo^1$ &$apo^2$&$apo^3$&$peri^1$ &$peri^2$&$peri^3$ &$ecc^1$&
  $ecc^2$ &$ecc^3$ &$Z_{max}^1$&$Z_{max}^2$&$Z_{max}^3$ \\
    &[kpc]&[kpc]&[kpc]&[kpc]&[kpc]&[kpc]& & & &[kpc]&[kpc]&[kpc] \\\hline
NGC 6144  & $ 4.35^{+0.06}_{-0.06}$& $ 4.38^{+0.07}_{-0.07}$& $ 4.40^{+0.08}_{-0.06}$& $ 1.06^{+0.09}_{-0.07}$& $ 1.09^{+0.07}_{-0.10}$& $ 1.18^{+0.00}_{-0.22}$& $ 0.61^{+0.02}_{-0.03}$& $ 0.60^{+0.03}_{-0.02}$& $ 0.58^{+0.06}_{-0.00}$& $ 3.60^{+0.09}_{-0.11}$& $ 3.62^{+0.14}_{-0.12}$& $ 3.63^{+0.12}_{-0.13}$\\
E452-11  & $ 2.95^{+0.11}_{-0.42}$& $ 2.80^{+0.21}_{-0.30}$& $ 3.15^{+0.00}_{-0.61}$& $ 0.15^{+0.10}_{-0.10}$& $ 0.19^{+0.06}_{-0.14}$& $ 0.21^{+0.04}_{-0.13}$& $ 0.91^{+0.05}_{-0.07}$& $ 0.88^{+0.08}_{-0.04}$& $ 0.88^{+0.06}_{-0.04}$& $ 1.95^{+0.07}_{-0.01}$& $ 1.93^{+0.07}_{-0.01}$& $ 1.91^{+0.10}_{-0.00}$\\
NGC 6266  & $ 2.69^{+0.07}_{-0.09}$& $ 2.67^{+0.07}_{-0.08}$& $ 2.66^{+0.07}_{-0.09}$& $ 1.33^{+0.06}_{-0.02}$& $ 1.37^{+0.05}_{-0.03}$& $ 1.42^{+0.01}_{-0.06}$& $ 0.34^{+0.01}_{-0.03}$& $ 0.32^{+0.01}_{-0.02}$& $ 0.31^{+0.01}_{-0.02}$& $ 1.01^{+0.02}_{-0.01}$& $ 1.01^{+0.02}_{-0.01}$& $ 1.02^{+0.01}_{-0.01}$\\
NGC 6273  & $ 5.40^{+0.10}_{-0.06}$& $ 5.47^{+0.19}_{-0.14}$& $ 5.51^{+0.22}_{-0.17}$& $ 0.32^{+0.10}_{-0.19}$& $ 0.33^{+0.09}_{-0.17}$& $ 0.25^{+0.16}_{-0.09}$& $ 0.89^{+0.06}_{-0.03}$& $ 0.88^{+0.06}_{-0.02}$& $ 0.91^{+0.03}_{-0.05}$& $ 4.04^{+0.32}_{-0.11}$& $ 4.02^{+0.28}_{-0.15}$& $ 4.01^{+0.20}_{-0.14}$\\
NGC 6293  & $ 3.45^{+0.33}_{-0.09}$& $ 3.50^{+0.25}_{-0.14}$& $ 3.50^{+0.24}_{-0.10}$& $ 0.27^{+0.13}_{-0.12}$& $ 0.27^{+0.10}_{-0.14}$& $ 0.31^{+0.05}_{-0.16}$& $ 0.85^{+0.06}_{-0.05}$& $ 0.85^{+0.08}_{-0.03}$& $ 0.84^{+0.08}_{-0.02}$& $ 2.50^{+0.20}_{-0.26}$& $ 2.45^{+0.30}_{-0.06}$& $ 2.51^{+0.28}_{-0.07}$\\
NGC 6342  & $ 2.73^{+0.11}_{-0.20}$& $ 2.76^{+0.03}_{-0.23}$& $ 2.47^{+0.32}_{-0.00}$& $ 0.64^{+0.02}_{-0.12}$& $ 0.52^{+0.15}_{-0.01}$& $ 0.57^{+0.14}_{-0.04}$& $ 0.62^{+0.06}_{-0.02}$& $ 0.68^{+0.00}_{-0.09}$& $ 0.63^{+0.04}_{-0.06}$& $ 1.71^{+0.09}_{-0.07}$& $ 1.71^{+0.08}_{-0.06}$& $ 1.79^{+0.05}_{-0.12}$\\
NGC 6355  & $ 3.61^{+0.33}_{-0.00}$& $ 4.00^{+0.00}_{-0.40}$& $ 4.10^{+0.00}_{-0.51}$& $ 0.10^{+0.11}_{-0.05}$& $ 0.15^{+0.10}_{-0.10}$& $ 0.06^{+0.23}_{-0.02}$& $ 0.95^{+0.02}_{-0.05}$& $ 0.93^{+0.04}_{-0.05}$& $ 0.97^{+0.01}_{-0.11}$& $ 2.46^{+0.11}_{-0.19}$& $ 2.46^{+0.09}_{-0.12}$& $ 2.42^{+0.13}_{-0.05}$\\
Terzan 2  & $ 1.48^{+0.39}_{-0.26}$& $ 1.68^{+0.14}_{-0.49}$& $ 1.39^{+0.43}_{-0.22}$& $ 0.24^{+0.12}_{-0.08}$& $ 0.15^{+0.21}_{-0.00}$& $ 0.29^{+0.11}_{-0.07}$& $ 0.72^{+0.13}_{-0.17}$& $ 0.84^{+0.00}_{-0.29}$& $ 0.65^{+0.13}_{-0.14}$& $ 0.56^{+0.13}_{-0.07}$& $ 0.60^{+0.07}_{-0.11}$& $ 0.55^{+0.12}_{-0.07}$\\
Terzan 4  & $ 1.10^{+0.33}_{-0.20}$& $ 1.08^{+0.29}_{-0.19}$& $ 1.07^{+0.36}_{-0.18}$& $ 0.36^{+0.16}_{-0.06}$& $ 0.39^{+0.13}_{-0.09}$& $ 0.41^{+0.15}_{-0.07}$& $ 0.50^{+0.06}_{-0.09}$& $ 0.47^{+0.08}_{-0.09}$& $ 0.45^{+0.07}_{-0.09}$& $ 0.60^{+0.10}_{-0.05}$& $ 0.60^{+0.08}_{-0.04}$& $ 0.60^{+0.11}_{-0.05}$\\
BH 229   & $ 2.69^{+0.15}_{-0.12}$& $ 2.73^{+0.15}_{-0.12}$& $ 2.78^{+0.14}_{-0.15}$& $ 0.17^{+0.08}_{-0.09}$& $ 0.16^{+0.12}_{-0.05}$& $ 0.21^{+0.13}_{-0.06}$& $ 0.88^{+0.06}_{-0.04}$& $ 0.89^{+0.03}_{-0.07}$& $ 0.86^{+0.03}_{-0.07}$& $ 2.69^{+0.12}_{-0.15}$& $ 2.73^{+0.12}_{-0.16}$& $ 2.75^{+0.14}_{-0.15}$\\
Liller 1  & $ 1.14^{+0.14}_{-0.10}$& $ 1.10^{+0.21}_{-0.10}$& $ 1.10^{+0.20}_{-0.05}$& $ 0.33^{+0.28}_{-0.19}$& $ 0.37^{+0.22}_{-0.18}$& $ 0.37^{+0.20}_{-0.23}$& $ 0.55^{+0.21}_{-0.23}$& $ 0.50^{+0.21}_{-0.18}$& $ 0.50^{+0.26}_{-0.15}$& $ 0.30^{+0.05}_{-0.19}$& $ 0.27^{+0.04}_{-0.17}$& $ 0.28^{+0.07}_{-0.17}$\\
NGC 6380  & $ 2.44^{+0.17}_{-0.08}$& $ 2.45^{+0.15}_{-0.08}$& $ 2.46^{+0.15}_{-0.08}$& $ 0.30^{+0.07}_{-0.08}$& $ 0.30^{+0.07}_{-0.05}$& $ 0.32^{+0.06}_{-0.06}$& $ 0.78^{+0.05}_{-0.03}$& $ 0.78^{+0.03}_{-0.03}$& $ 0.77^{+0.04}_{-0.03}$& $ 0.87^{+0.07}_{-0.02}$& $ 0.87^{+0.06}_{-0.02}$& $ 0.88^{+0.04}_{-0.03}$\\
Terzan 1  & $ 3.02^{+0.15}_{-0.18}$& $ 3.01^{+0.20}_{-0.20}$& $ 3.00^{+0.23}_{-0.17}$& $ 0.77^{+0.04}_{-0.06}$& $ 0.83^{+0.06}_{-0.06}$& $ 0.88^{+0.06}_{-0.06}$& $ 0.59^{+0.03}_{-0.02}$& $ 0.57^{+0.02}_{-0.02}$& $ 0.55^{+0.03}_{-0.03}$& $ 0.12^{+0.00}_{-0.01}$& $ 0.12^{+0.00}_{-0.01}$& $ 0.12^{+0.00}_{-0.01}$\\
NGC 6401  & $ 4.10^{+0.10}_{-0.16}$& $ 4.12^{+0.11}_{-0.20}$& $ 4.15^{+0.13}_{-0.25}$& $ 0.14^{+0.10}_{-0.08}$& $ 0.17^{+0.09}_{-0.12}$& $ 0.05^{+0.20}_{-0.01}$& $ 0.93^{+0.04}_{-0.04}$& $ 0.92^{+0.05}_{-0.04}$& $ 0.98^{+0.00}_{-0.09}$& $ 1.74^{+0.30}_{-0.27}$& $ 1.56^{+0.62}_{-0.08}$& $ 1.65^{+0.50}_{-0.15}$\\
Pal 6    & $ 3.90^{+0.33}_{-0.61}$& $ 4.04^{+0.34}_{-0.76}$& $ 3.67^{+0.70}_{-0.43}$& $ 0.09^{+0.09}_{-0.04}$& $ 0.06^{+0.19}_{-0.01}$& $ 0.08^{+0.14}_{-0.03}$& $ 0.96^{+0.01}_{-0.05}$& $ 0.97^{+0.00}_{-0.08}$& $ 0.96^{+0.01}_{-0.07}$& $ 2.38^{+0.17}_{-0.62}$& $ 2.36^{+0.27}_{-0.57}$& $ 2.40^{+0.24}_{-0.57}$\\
Terzan 5  & $ 2.04^{+0.18}_{-0.11}$& $ 2.06^{+0.16}_{-0.13}$& $ 2.06^{+0.14}_{-0.15}$& $ 0.33^{+0.20}_{-0.00}$& $ 0.51^{+0.05}_{-0.08}$& $ 0.51^{+0.05}_{-0.06}$& $ 0.72^{+0.00}_{-0.12}$& $ 0.60^{+0.04}_{-0.01}$& $ 0.60^{+0.03}_{-0.02}$& $ 0.29^{+0.02}_{-0.02}$& $ 0.28^{+0.03}_{-0.01}$& $ 0.28^{+0.02}_{-0.01}$\\
NGC 6440  & $ 1.50^{+0.13}_{-0.04}$& $ 1.51^{+0.13}_{-0.04}$& $ 1.51^{+0.11}_{-0.04}$& $ 0.13^{+0.13}_{-0.07}$& $ 0.13^{+0.10}_{-0.07}$& $ 0.13^{+0.15}_{-0.09}$& $ 0.84^{+0.09}_{-0.13}$& $ 0.84^{+0.09}_{-0.10}$& $ 0.84^{+0.10}_{-0.15}$& $ 0.85^{+0.06}_{-0.07}$& $ 0.85^{+0.07}_{-0.06}$& $ 0.86^{+0.08}_{-0.08}$\\
Terzan 6  & $ 1.94^{+0.33}_{-0.28}$& $ 1.94^{+0.28}_{-0.32}$& $ 1.95^{+0.30}_{-0.31}$& $ 0.28^{+0.08}_{-0.11}$& $ 0.31^{+0.07}_{-0.11}$& $ 0.33^{+0.08}_{-0.08}$& $ 0.75^{+0.08}_{-0.05}$& $ 0.73^{+0.07}_{-0.05}$& $ 0.71^{+0.05}_{-0.05}$& $ 0.29^{+0.19}_{-0.04}$& $ 0.29^{+0.19}_{-0.04}$& $ 0.29^{+0.15}_{-0.05}$\\
NGC 6453  & $ 2.76^{+0.34}_{-0.29}$& $ 2.64^{+0.39}_{-0.17}$& $ 2.66^{+0.34}_{-0.19}$& $ 0.03^{+0.31}_{-0.00}$& $ 0.34^{+0.05}_{-0.18}$& $ 0.28^{+0.09}_{-0.11}$& $ 0.98^{+0.00}_{-0.19}$& $ 0.77^{+0.12}_{-0.01}$& $ 0.81^{+0.07}_{-0.05}$& $ 1.96^{+0.15}_{-0.21}$& $ 2.00^{+0.10}_{-0.17}$& $ 2.01^{+0.12}_{-0.16}$\\
Terzan 9  & $ 2.70^{+0.38}_{-0.37}$& $ 2.70^{+0.31}_{-0.38}$& $ 2.71^{+0.38}_{-0.34}$& $ 0.39^{+0.11}_{-0.12}$& $ 0.43^{+0.13}_{-0.16}$& $ 0.45^{+0.13}_{-0.13}$& $ 0.75^{+0.08}_{-0.08}$& $ 0.73^{+0.09}_{-0.09}$& $ 0.72^{+0.08}_{-0.09}$& $ 0.36^{+0.05}_{-0.01}$& $ 0.38^{+0.03}_{-0.04}$& $ 0.37^{+0.03}_{-0.03}$\\
NGC 6522  & $ 1.94^{+0.13}_{-0.09}$& $ 1.97^{+0.10}_{-0.12}$& $ 1.99^{+0.16}_{-0.12}$& $ 0.77^{+0.17}_{-0.15}$& $ 0.79^{+0.13}_{-0.17}$& $ 0.81^{+0.16}_{-0.14}$& $ 0.43^{+0.08}_{-0.07}$& $ 0.43^{+0.08}_{-0.06}$& $ 0.42^{+0.07}_{-0.06}$& $ 1.46^{+0.08}_{-0.06}$& $ 1.49^{+0.07}_{-0.08}$& $ 1.51^{+0.10}_{-0.08}$\\
NGC 6528  & $ 2.77^{+0.26}_{-0.11}$& $ 2.81^{+0.32}_{-0.17}$& $ 2.84^{+0.34}_{-0.15}$& $ 0.49^{+0.06}_{-0.04}$& $ 0.51^{+0.07}_{-0.07}$& $ 0.52^{+0.09}_{-0.09}$& $ 0.70^{+0.01}_{-0.01}$& $ 0.69^{+0.03}_{-0.01}$& $ 0.69^{+0.04}_{-0.02}$& $ 0.82^{+0.08}_{-0.00}$& $ 0.83^{+0.08}_{-0.02}$& $ 0.83^{+0.09}_{-0.01}$\\
NGC 6558  & $ 3.26^{+0.29}_{-0.24}$& $ 3.63^{+0.00}_{-0.52}$& $ 3.57^{+0.00}_{-0.42}$& $ 0.37^{+0.04}_{-0.11}$& $ 0.39^{+0.02}_{-0.14}$& $ 0.35^{+0.04}_{-0.09}$& $ 0.79^{+0.05}_{-0.00}$& $ 0.81^{+0.04}_{-0.02}$& $ 0.82^{+0.03}_{-0.03}$& $ 1.09^{+0.04}_{-0.03}$& $ 1.09^{+0.05}_{-0.02}$& $ 1.09^{+0.04}_{-0.02}$\\
NGC 6624  & $ 1.59^{+0.03}_{-0.02}$& $ 1.56^{+0.04}_{-0.01}$& $ 1.57^{+0.03}_{-0.01}$& $ 0.48^{+0.13}_{-0.11}$& $ 0.63^{+0.04}_{-0.12}$& $ 0.62^{+0.04}_{-0.07}$& $ 0.54^{+0.09}_{-0.09}$& $ 0.43^{+0.08}_{-0.03}$& $ 0.43^{+0.05}_{-0.02}$& $ 1.51^{+0.01}_{-0.02}$& $ 1.49^{+0.03}_{-0.01}$& $ 1.50^{+0.02}_{-0.02}$\\
NGC 6626  & $ 3.15^{+0.18}_{-0.07}$& $ 3.14^{+0.19}_{-0.06}$& $ 3.20^{+0.13}_{-0.09}$& $ 0.44^{+0.10}_{-0.23}$& $ 0.39^{+0.23}_{-0.18}$& $ 0.32^{+0.41}_{-0.07}$& $ 0.76^{+0.12}_{-0.05}$& $ 0.78^{+0.10}_{-0.10}$& $ 0.82^{+0.03}_{-0.19}$& $ 1.30^{+0.03}_{-0.10}$& $ 1.34^{+0.00}_{-0.14}$& $ 1.26^{+0.05}_{-0.04}$\\
NGC 6638  & $ 2.29^{+0.49}_{-0.10}$& $ 2.29^{+0.46}_{-0.10}$& $ 2.42^{+0.29}_{-0.20}$& $ 0.12^{+0.04}_{-0.06}$& $ 0.15^{+0.04}_{-0.09}$& $ 0.06^{+0.13}_{-0.00}$& $ 0.90^{+0.06}_{-0.03}$& $ 0.88^{+0.08}_{-0.03}$& $ 0.95^{+0.01}_{-0.10}$& $ 1.78^{+0.02}_{-0.27}$& $ 1.76^{+0.03}_{-0.29}$& $ 1.74^{+0.04}_{-0.28}$\\
NGC 6637  & $ 1.99^{+0.04}_{-0.02}$& $ 1.99^{+0.04}_{-0.02}$& $ 1.99^{+0.04}_{-0.02}$& $ 0.84^{+0.06}_{-0.11}$& $ 0.83^{+0.05}_{-0.08}$& $ 0.83^{+0.04}_{-0.11}$& $ 0.41^{+0.06}_{-0.04}$& $ 0.41^{+0.05}_{-0.03}$& $ 0.41^{+0.06}_{-0.02}$& $ 1.86^{+0.04}_{-0.03}$& $ 1.87^{+0.03}_{-0.02}$& $ 1.87^{+0.04}_{-0.02}$\\
NGC 6642  & $ 2.20^{+0.15}_{-0.04}$& $ 2.20^{+0.08}_{-0.04}$& $ 2.22^{+0.07}_{-0.04}$& $ 0.04^{+0.14}_{-0.00}$& $ 0.04^{+0.14}_{-0.02}$& $ 0.09^{+0.08}_{-0.05}$& $ 0.96^{+0.00}_{-0.11}$& $ 0.97^{+0.00}_{-0.12}$& $ 0.92^{+0.04}_{-0.06}$& $ 1.24^{+0.24}_{-0.06}$& $ 1.23^{+0.16}_{-0.09}$& $ 1.19^{+0.19}_{-0.06}$\\
NGC 6717  & $ 2.51^{+0.06}_{-0.06}$& $ 2.48^{+0.06}_{-0.03}$& $ 2.49^{+0.05}_{-0.06}$& $ 1.31^{+0.09}_{-0.06}$& $ 1.33^{+0.12}_{-0.03}$& $ 1.37^{+0.08}_{-0.04}$& $ 0.32^{+0.02}_{-0.05}$& $ 0.30^{+0.02}_{-0.04}$& $ 0.29^{+0.02}_{-0.03}$& $ 1.44^{+0.04}_{-0.02}$& $ 1.45^{+0.03}_{-0.03}$& $ 1.45^{+0.03}_{-0.03}$\\
NGC 6723  & $ 4.32^{+0.02}_{-0.09}$& $ 4.30^{+0.05}_{-0.03}$& $ 4.30^{+0.04}_{-0.04}$& $ 0.78^{+0.00}_{-0.26}$& $ 0.78^{+0.07}_{-0.13}$& $ 0.77^{+0.10}_{-0.08}$& $ 0.69^{+0.09}_{-0.00}$& $ 0.69^{+0.05}_{-0.02}$& $ 0.70^{+0.02}_{-0.04}$& $ 3.96^{+0.06}_{-0.09}$& $ 3.97^{+0.06}_{-0.08}$& $ 3.98^{+0.05}_{-0.08}$\\
NGC 6171  & $ 3.95^{+0.05}_{-0.07}$& $ 3.96^{+0.06}_{-0.05}$& $ 3.96^{+0.07}_{-0.05}$& $ 1.40^{+0.07}_{-0.04}$& $ 1.45^{+0.07}_{-0.03}$& $ 1.53^{+0.04}_{-0.07}$& $ 0.48^{+0.01}_{-0.02}$& $ 0.46^{+0.01}_{-0.02}$& $ 0.44^{+0.02}_{-0.01}$& $ 2.39^{+0.04}_{-0.02}$& $ 2.41^{+0.03}_{-0.04}$& $ 2.41^{+0.03}_{-0.03}$\\
NGC 6316  & $ 3.85^{+0.38}_{-0.35}$& $ 3.86^{+0.35}_{-0.38}$& $ 3.86^{+0.36}_{-0.30}$& $ 0.83^{+0.20}_{-0.29}$& $ 0.93^{+0.11}_{-0.30}$& $ 0.92^{+0.13}_{-0.23}$& $ 0.65^{+0.10}_{-0.06}$& $ 0.61^{+0.10}_{-0.02}$& $ 0.62^{+0.07}_{-0.03}$& $ 1.65^{+0.10}_{-0.05}$& $ 1.72^{+0.06}_{-0.11}$& $ 1.70^{+0.04}_{-0.10}$\\
NGC 6388  & $ 4.06^{+0.26}_{-0.16}$& $ 4.03^{+0.17}_{-0.08}$& $ 4.19^{+0.05}_{-0.14}$& $ 1.19^{+0.08}_{-0.12}$& $ 1.26^{+0.05}_{-0.11}$& $ 1.17^{+0.11}_{-0.05}$& $ 0.55^{+0.05}_{-0.04}$& $ 0.52^{+0.04}_{-0.01}$& $ 0.56^{+0.02}_{-0.03}$& $ 1.70^{+0.13}_{-0.14}$& $ 1.59^{+0.08}_{-0.03}$& $ 1.67^{+0.00}_{-0.13}$\\
NGC 6539  & $ 3.91^{+0.24}_{-0.14}$& $ 3.92^{+0.22}_{-0.10}$& $ 3.93^{+0.21}_{-0.14}$& $ 2.15^{+0.16}_{-0.19}$& $ 2.18^{+0.20}_{-0.20}$& $ 2.22^{+0.18}_{-0.22}$& $ 0.29^{+0.05}_{-0.03}$& $ 0.28^{+0.06}_{-0.03}$& $ 0.28^{+0.05}_{-0.04}$& $ 2.74^{+0.19}_{-0.19}$& $ 2.74^{+0.20}_{-0.13}$& $ 2.76^{+0.19}_{-0.16}$\\
NGC 6553  & $ 4.61^{+0.11}_{-0.09}$& $ 4.64^{+0.09}_{-0.11}$& $ 4.67^{+0.10}_{-0.11}$& $ 2.51^{+0.11}_{-0.09}$& $ 2.54^{+0.08}_{-0.11}$& $ 2.56^{+0.11}_{-0.10}$& $ 0.29^{+0.01}_{-0.00}$& $ 0.29^{+0.01}_{-0.00}$& $ 0.29^{+0.01}_{-0.01}$& $ 0.31^{+0.01}_{-0.01}$& $ 0.31^{+0.01}_{-0.01}$& $ 0.31^{+0.01}_{-0.01}$\\
NGC 6652  & $ 3.23^{+0.16}_{-0.15}$& $ 3.20^{+0.11}_{-0.10}$& $ 3.19^{+0.13}_{-0.08}$& $ 0.15^{+0.00}_{-0.09}$& $ 0.11^{+0.03}_{-0.05}$& $ 0.15^{+0.00}_{-0.07}$& $ 0.91^{+0.05}_{-0.00}$& $ 0.94^{+0.02}_{-0.02}$& $ 0.91^{+0.04}_{-0.00}$& $ 2.84^{+0.11}_{-0.03}$& $ 2.91^{+0.04}_{-0.08}$& $ 2.96^{+0.01}_{-0.11}$\\
Terzan 3  & $ 4.42^{+0.20}_{-0.16}$& $ 4.42^{+0.19}_{-0.21}$& $ 4.27^{+0.33}_{-0.02}$& $ 2.23^{+0.15}_{-0.23}$& $ 2.23^{+0.17}_{-0.23}$& $ 2.23^{+0.13}_{-0.25}$& $ 0.33^{+0.04}_{-0.02}$& $ 0.33^{+0.04}_{-0.03}$& $ 0.31^{+0.07}_{-0.00}$& $ 1.92^{+0.08}_{-0.10}$& $ 1.92^{+0.11}_{-0.10}$& $ 1.92^{+0.11}_{-0.06}$\\
NGC 6256  & $ 3.19^{+0.16}_{-0.12}$& $ 3.15^{+0.15}_{-0.10}$& $ 3.13^{+0.14}_{-0.07}$& $ 1.23^{+0.32}_{-0.27}$& $ 1.27^{+0.32}_{-0.27}$& $ 1.26^{+0.42}_{-0.22}$& $ 0.45^{+0.07}_{-0.09}$& $ 0.43^{+0.08}_{-0.08}$& $ 0.43^{+0.06}_{-0.11}$& $ 0.74^{+0.20}_{-0.00}$& $ 0.79^{+0.14}_{-0.05}$& $ 0.90^{+0.02}_{-0.16}$\\
NGC 6304  & $ 3.64^{+0.14}_{-0.16}$& $ 3.69^{+0.14}_{-0.16}$& $ 3.65^{+0.15}_{-0.14}$& $ 1.72^{+0.22}_{-0.73}$& $ 1.72^{+0.29}_{-0.29}$& $ 1.82^{+0.20}_{-0.27}$& $ 0.36^{+0.22}_{-0.06}$& $ 0.36^{+0.08}_{-0.07}$& $ 0.34^{+0.06}_{-0.05}$& $ 0.89^{+0.10}_{-0.01}$& $ 0.88^{+0.10}_{-0.02}$& $ 0.89^{+0.09}_{-0.04}$\\
Pismis 26  & $ 4.85^{+0.71}_{-0.38}$& $ 4.89^{+0.51}_{-0.24}$& $ 4.93^{+0.36}_{-0.31}$& $ 1.57^{+0.32}_{-0.04}$& $ 1.75^{+0.18}_{-0.23}$& $ 1.72^{+0.16}_{-0.23}$& $ 0.51^{+0.02}_{-0.06}$& $ 0.47^{+0.06}_{-0.02}$& $ 0.48^{+0.05}_{-0.03}$& $ 2.21^{+0.12}_{-0.22}$& $ 2.13^{+0.13}_{-0.07}$& $ 2.08^{+0.14}_{-0.07}$\\
NGC 6540  & $ 2.52^{+0.26}_{-0.21}$& $ 2.51^{+0.25}_{-0.22}$& $ 2.52^{+0.26}_{-0.20}$& $ 1.48^{+0.08}_{-0.07}$& $ 1.53^{+0.06}_{-0.11}$& $ 1.57^{+0.07}_{-0.09}$& $ 0.26^{+0.04}_{-0.04}$& $ 0.24^{+0.06}_{-0.04}$& $ 0.23^{+0.06}_{-0.04}$& $ 0.54^{+0.03}_{-0.02}$& $ 0.54^{+0.04}_{-0.02}$& $ 0.55^{+0.03}_{-0.03}$\\
NGC 6569  & $ 2.84^{+0.28}_{-0.11}$& $ 2.85^{+0.29}_{-0.14}$& $ 2.90^{+0.28}_{-0.20}$& $ 1.60^{+0.01}_{-0.45}$& $ 1.63^{+0.06}_{-0.47}$& $ 1.45^{+0.23}_{-0.27}$& $ 0.28^{+0.17}_{-0.01}$& $ 0.27^{+0.18}_{-0.02}$& $ 0.33^{+0.11}_{-0.07}$& $ 1.31^{+0.07}_{-0.03}$& $ 1.31^{+0.07}_{-0.03}$& $ 1.33^{+0.05}_{-0.05}$\\
E456-78  & $ 3.96^{+0.14}_{-0.24}$& $ 3.97^{+0.14}_{-0.21}$& $ 3.97^{+0.17}_{-0.21}$& $ 1.90^{+0.28}_{-0.19}$& $ 1.91^{+0.20}_{-0.17}$& $ 1.91^{+0.20}_{-0.20}$& $ 0.35^{+0.02}_{-0.05}$& $ 0.35^{+0.02}_{-0.03}$& $ 0.35^{+0.03}_{-0.03}$& $ 1.44^{+0.10}_{-0.09}$& $ 1.46^{+0.12}_{-0.08}$& $ 1.48^{+0.11}_{-0.09}$\\
NGC 6325  & $ 2.51^{+0.33}_{-0.30}$& $ 2.54^{+0.29}_{-0.29}$& $ 2.54^{+0.46}_{-0.27}$& $ 0.79^{+0.28}_{-0.28}$& $ 0.78^{+0.34}_{-0.30}$& $ 0.80^{+0.29}_{-0.35}$& $ 0.52^{+0.15}_{-0.13}$& $ 0.53^{+0.15}_{-0.14}$& $ 0.52^{+0.18}_{-0.10}$& $ 1.66^{+0.13}_{-0.10}$& $ 1.61^{+0.14}_{-0.08}$& $ 1.61^{+0.16}_{-0.07}$\\
Djorg 2   & $ 2.39^{+0.07}_{-0.05}$& $ 2.41^{+0.07}_{-0.04}$& $ 2.43^{+0.08}_{-0.05}$& $ 0.56^{+0.12}_{-0.17}$& $ 0.58^{+0.12}_{-0.15}$& $ 0.59^{+0.14}_{-0.19}$& $ 0.62^{+0.10}_{-0.06}$& $ 0.61^{+0.08}_{-0.05}$& $ 0.61^{+0.11}_{-0.06}$& $ 0.80^{+0.05}_{-0.02}$& $ 0.80^{+0.03}_{-0.01}$& $ 0.80^{+0.05}_{-0.01}$\\
\hline
 \end{tabular}
   \end{center}
   \end{tiny}
  \end{minipage}
  }
  \end{table*}

 \begin{figure*}
{\begin{center}
   \includegraphics[width=0.3\textwidth,angle=-90]{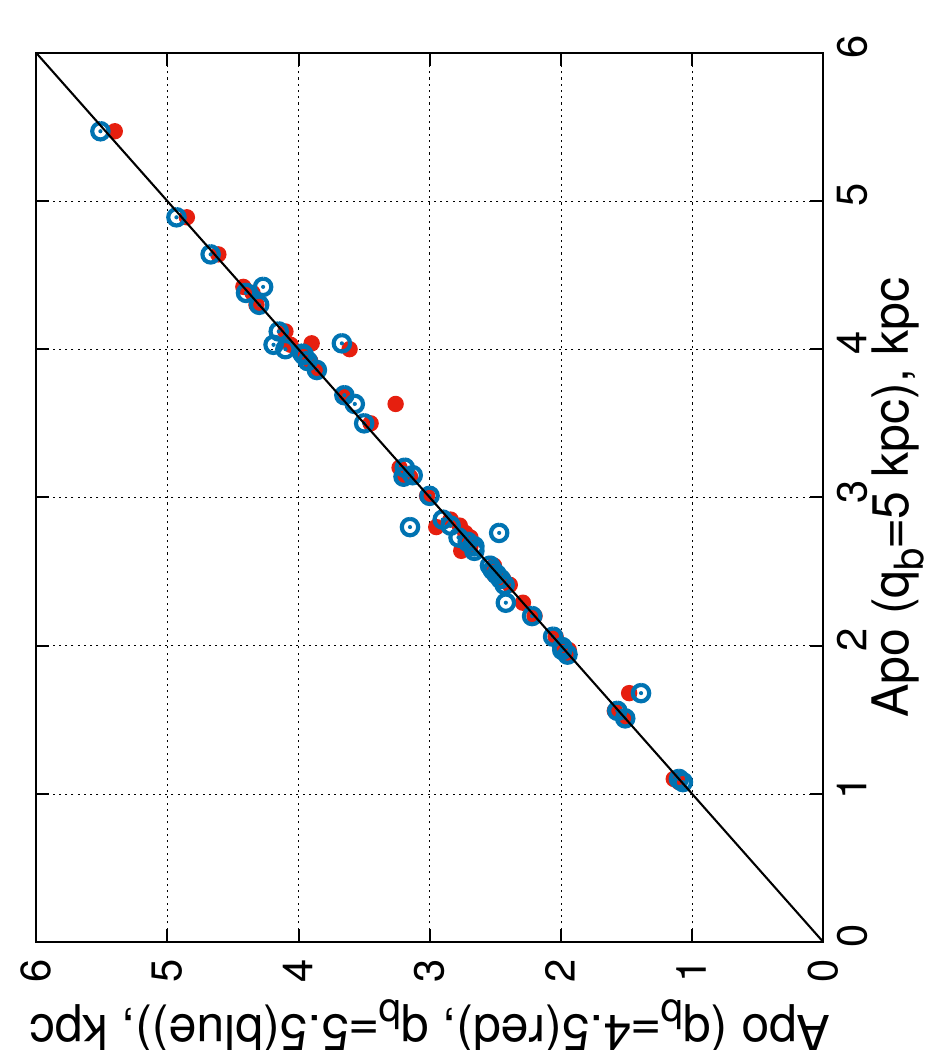}  \includegraphics[width=0.3\textwidth,angle=-90]{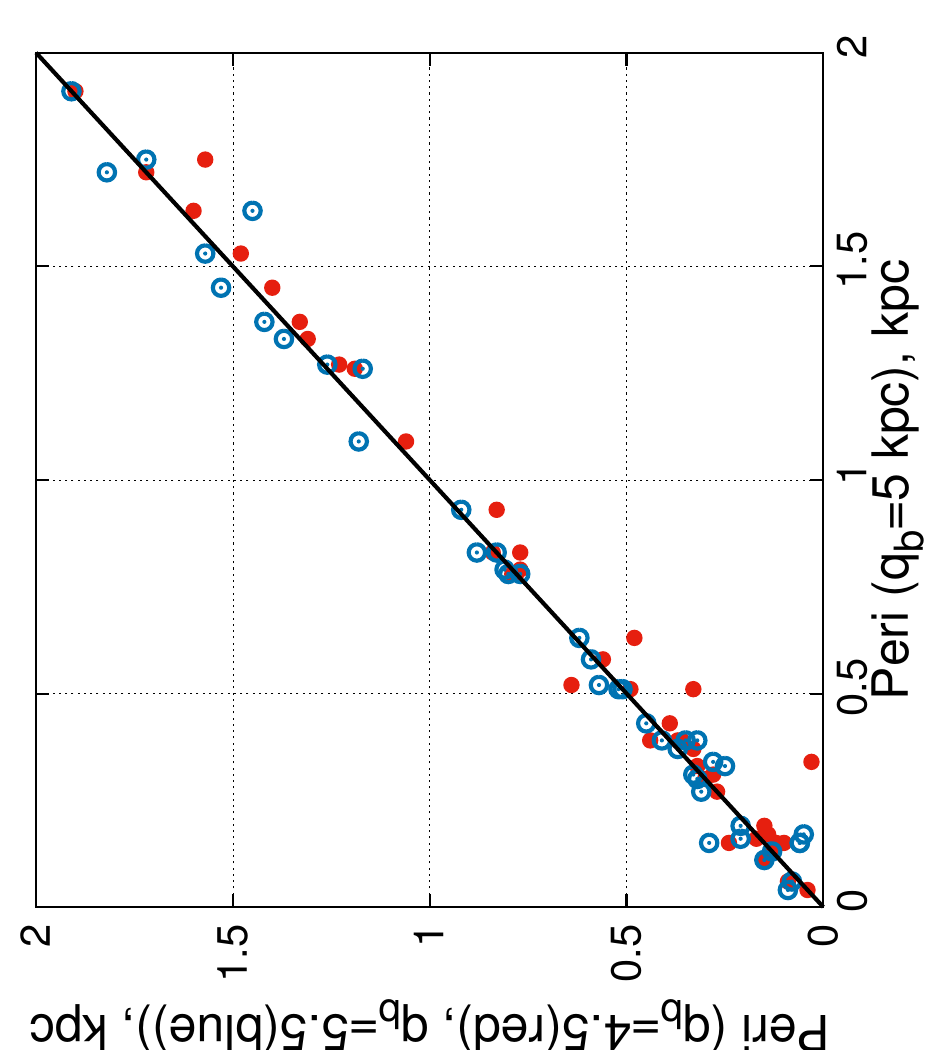}\

   \medskip
   \includegraphics[width=0.3\textwidth,angle=-90]{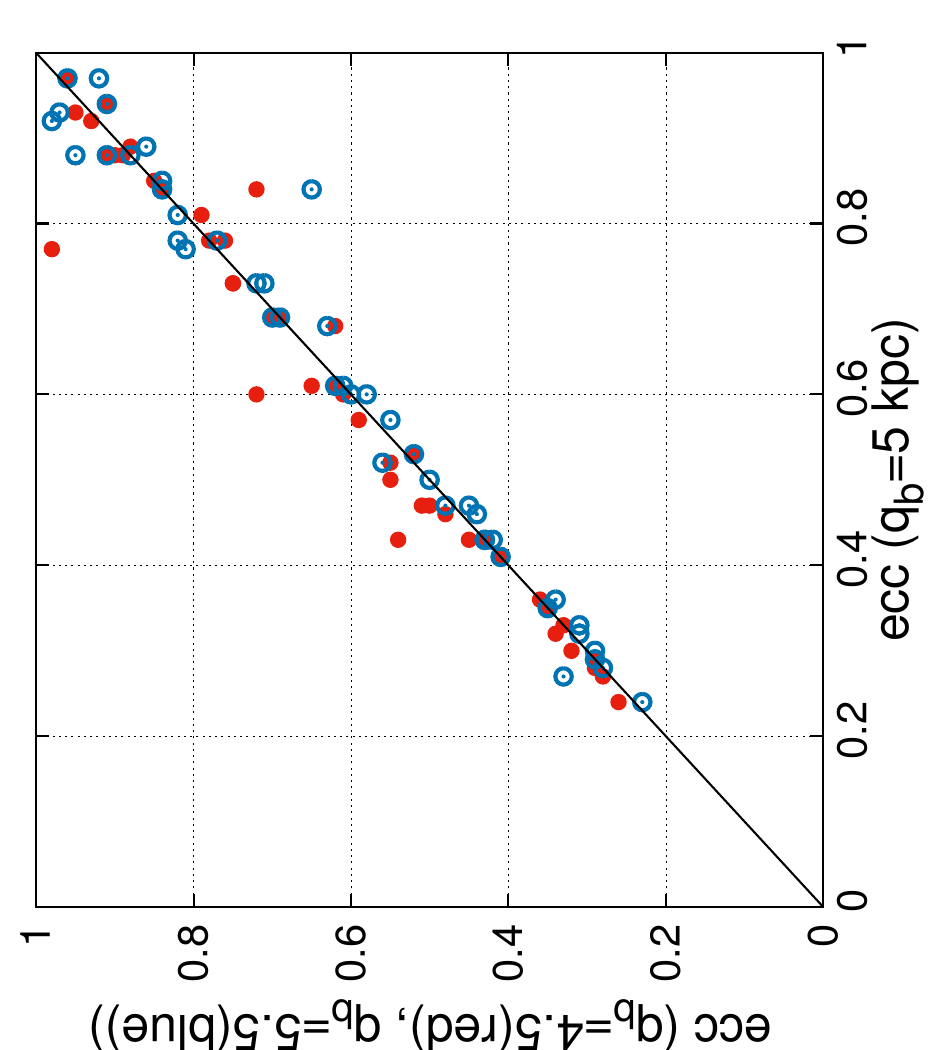}   \includegraphics[width=0.3\textwidth,angle=-90]{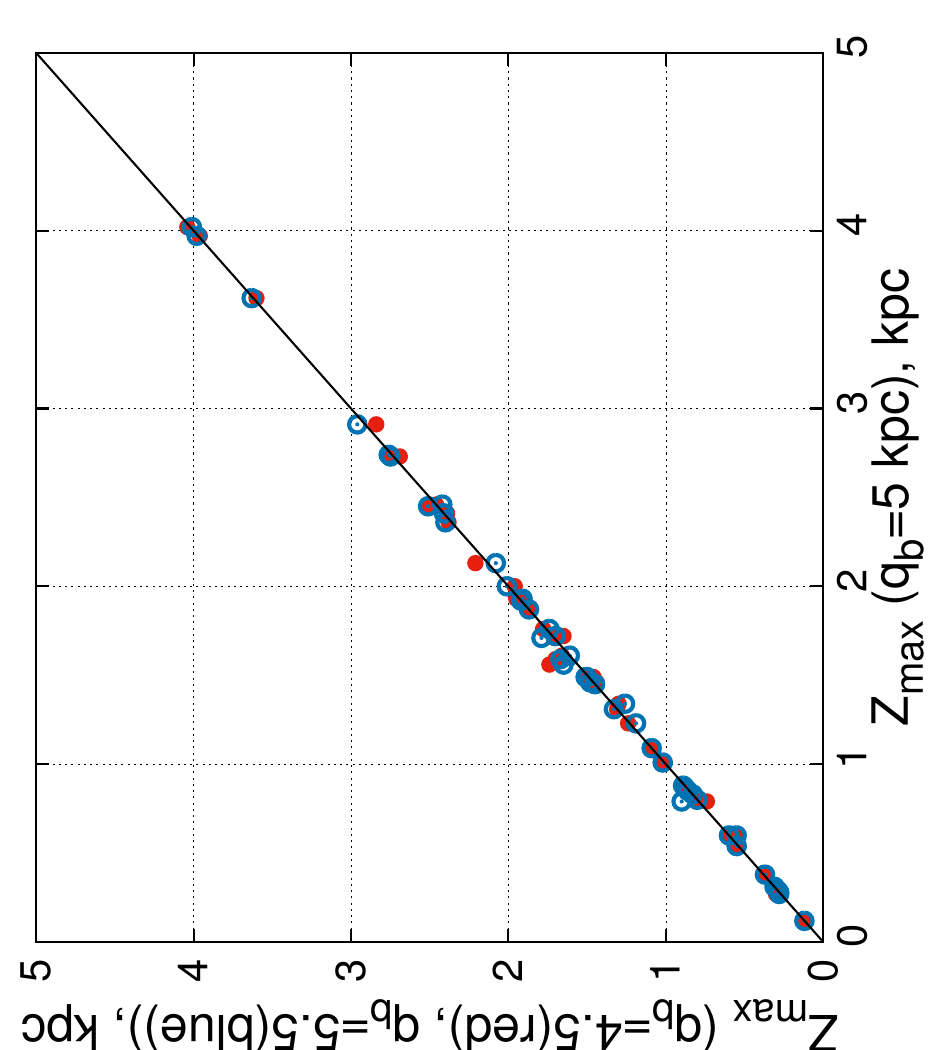}\

\caption{\small Comparison of the orbital parameters of the GCs in the potential with a bar of different lengths. Other bar parameters: $M_b=430 M_G$, $\Omega_b=40$ km/s/kpc, $\theta_b=25^o$, axis ratio V0. Each panel has a line of coincise.}
\end{center}}
\end{figure*}

 \begin{table*}
  \rotatebox{90}{
		\begin{minipage}{1.5\linewidth}
		\label{t:prop}
		\begin{tiny} \begin{center}
{\scriptsize {\bf Table 6:} Orbital parameters of 45 GCs in potential with a bar with different axis ratios: V1 ("1"), V2 ("2"), V3 ("3"), V4 ("4").}

\bigskip

\begin{tabular}{|l c c c c|c c c c|c c c c|c c c c|}\hline
 ID  & $apo^1$ & $apo^2$ & $apo^3$ & $apo^4$ & $peri^1$ & $peri^2$ & $peri^3$ & $peri^4$ & $ecc^1$ &
  $ecc^2$ & $ecc^3$ & $ecc^4$ & $Z_{max}^1$ & $Z_{max}^2$ & $Z_{max}^3$ & $Z_{max}^4$ \\
    &[kpc]&[kpc]&[kpc]&[kpc]&[kpc]&[kpc]&[kpc]&[kpc]& & & & &[kpc]&[kpc]&[kpc]&[kpc] \\\hline
NGC 6144  & $ 4.36^{+0.04}_{-0.12}$& $ 4.38^{+0.07}_{-0.06}$& $ 4.34^{+0.09}_{-0.06}$& $ 4.36^{+0.13}_{-0.06}$& $ 1.13^{+0.10}_{-0.08}$& $ 1.04^{+0.13}_{-0.03}$& $ 1.06^{+0.10}_{-0.07}$& $ 1.11^{+0.08}_{-0.11}$& $ 0.59^{+0.02}_{-0.04}$& $ 0.62^{+0.01}_{-0.04}$& $ 0.61^{+0.02}_{-0.04}$& $ 0.59^{+0.04}_{-0.02}$& $ 3.87^{+0.05}_{-0.30}$& $ 3.61^{+0.13}_{-0.11}$& $ 3.61^{+0.28}_{-0.10}$& $ 3.66^{+0.38}_{-0.12}$\\
E452-11  & $ 2.71^{+0.51}_{-0.16}$& $ 2.94^{+0.07}_{-0.37}$& $ 2.64^{+0.35}_{-0.19}$& $ 2.82^{+0.52}_{-0.10}$& $ 0.09^{+0.13}_{-0.01}$& $ 0.17^{+0.08}_{-0.10}$& $ 0.09^{+0.12}_{-0.01}$& $ 0.24^{+0.02}_{-0.18}$& $ 0.94^{+0.01}_{-0.08}$& $ 0.89^{+0.06}_{-0.05}$& $ 0.94^{+0.00}_{-0.08}$& $ 0.85^{+0.10}_{-0.00}$& $ 1.96^{+0.10}_{-0.03}$& $ 1.96^{+0.06}_{-0.03}$& $ 1.96^{+0.07}_{-0.05}$& $ 1.99^{+0.03}_{-0.04}$\\
NGC 6266  & $ 2.72^{+0.05}_{-0.08}$& $ 2.67^{+0.08}_{-0.07}$& $ 2.69^{+0.09}_{-0.08}$& $ 2.71^{+0.06}_{-0.08}$& $ 1.31^{+0.06}_{-0.03}$& $ 1.37^{+0.04}_{-0.03}$& $ 1.37^{+0.05}_{-0.03}$& $ 1.32^{+0.05}_{-0.03}$& $ 0.35^{+0.01}_{-0.03}$& $ 0.32^{+0.02}_{-0.01}$& $ 0.32^{+0.02}_{-0.02}$& $ 0.34^{+0.02}_{-0.01}$& $ 1.02^{+0.02}_{-0.01}$& $ 1.01^{+0.02}_{-0.01}$& $ 1.02^{+0.01}_{-0.01}$& $ 1.02^{+0.01}_{-0.02}$\\
NGC 6273  & $ 5.34^{+0.17}_{-0.10}$& $ 5.47^{+0.13}_{-0.07}$& $ 5.41^{+0.10}_{-0.06}$& $ 5.40^{+0.12}_{-0.06}$& $ 0.27^{+0.13}_{-0.11}$& $ 0.28^{+0.12}_{-0.13}$& $ 0.27^{+0.17}_{-0.09}$& $ 0.30^{+0.12}_{-0.14}$& $ 0.90^{+0.04}_{-0.04}$& $ 0.90^{+0.05}_{-0.04}$& $ 0.91^{+0.03}_{-0.06}$& $ 0.89^{+0.05}_{-0.03}$& $ 4.00^{+0.47}_{-0.12}$& $ 4.05^{+0.20}_{-0.09}$& $ 3.94^{+0.18}_{-0.08}$& $ 4.08^{+0.12}_{-0.15}$\\
NGC 6293  & $ 3.49^{+0.24}_{-0.17}$& $ 3.48^{+0.35}_{-0.13}$& $ 3.51^{+0.23}_{-0.16}$& $ 3.53^{+0.33}_{-0.20}$& $ 0.21^{+0.11}_{-0.12}$& $ 0.29^{+0.11}_{-0.15}$& $ 0.19^{+0.13}_{-0.08}$& $ 0.28^{+0.08}_{-0.15}$& $ 0.89^{+0.05}_{-0.05}$& $ 0.85^{+0.07}_{-0.04}$& $ 0.90^{+0.04}_{-0.06}$& $ 0.85^{+0.08}_{-0.03}$& $ 2.52^{+0.27}_{-0.14}$& $ 2.48^{+0.28}_{-0.19}$& $ 2.56^{+0.21}_{-0.15}$& $ 2.39^{+0.41}_{-0.06}$\\
NGC 6342  & $ 2.36^{+0.29}_{-0.00}$& $ 2.73^{+0.08}_{-0.19}$& $ 2.50^{+0.23}_{-0.10}$& $ 2.52^{+0.37}_{-0.00}$& $ 1.02^{+0.00}_{-0.54}$& $ 0.68^{+0.00}_{-0.16}$& $ 0.46^{+0.30}_{-0.04}$& $ 0.52^{+0.12}_{-0.09}$& $ 0.40^{+0.29}_{-0.00}$& $ 0.60^{+0.08}_{-0.01}$& $ 0.69^{+0.03}_{-0.16}$& $ 0.66^{+0.07}_{-0.05}$& $ 1.74^{+0.07}_{-0.07}$& $ 1.72^{+0.07}_{-0.06}$& $ 1.77^{+0.06}_{-0.12}$& $ 1.75^{+0.05}_{-0.08}$\\
NGC 6355  & $ 3.66^{+0.24}_{-0.12}$& $ 3.65^{+0.38}_{-0.00}$& $ 3.71^{+0.16}_{-0.11}$& $ 3.72^{+0.28}_{-0.11}$& $ 0.48^{+0.00}_{-0.34}$& $ 0.14^{+0.09}_{-0.08}$& $ 0.11^{+0.30}_{-0.01}$& $ 0.14^{+0.15}_{-0.10}$& $ 0.77^{+0.16}_{-0.00}$& $ 0.93^{+0.04}_{-0.04}$& $ 0.94^{+0.01}_{-0.14}$& $ 0.93^{+0.05}_{-0.07}$& $ 2.24^{+0.18}_{-0.06}$& $ 2.48^{+0.14}_{-0.19}$& $ 2.42^{+0.03}_{-0.12}$& $ 2.39^{+0.17}_{-0.09}$\\
Terzan 2  & $ 1.47^{+0.41}_{-0.31}$& $ 1.67^{+0.11}_{-0.52}$& $ 1.51^{+0.32}_{-0.32}$& $ 1.51^{+0.34}_{-0.32}$& $ 0.23^{+0.13}_{-0.12}$& $ 0.17^{+0.21}_{-0.00}$& $ 0.21^{+0.16}_{-0.03}$& $ 0.22^{+0.14}_{-0.08}$& $ 0.73^{+0.16}_{-0.20}$& $ 0.82^{+0.00}_{-0.30}$& $ 0.76^{+0.06}_{-0.22}$& $ 0.75^{+0.11}_{-0.20}$& $ 0.56^{+0.13}_{-0.09}$& $ 0.58^{+0.08}_{-0.11}$& $ 0.59^{+0.06}_{-0.10}$& $ 0.57^{+0.11}_{-0.08}$\\
Terzan 4  & $ 1.14^{+0.40}_{-0.20}$& $ 1.09^{+0.28}_{-0.19}$& $ 1.09^{+0.24}_{-0.21}$& $ 1.16^{+0.36}_{-0.19}$& $ 0.29^{+0.16}_{-0.07}$& $ 0.38^{+0.12}_{-0.08}$& $ 0.38^{+0.12}_{-0.08}$& $ 0.31^{+0.17}_{-0.08}$& $ 0.59^{+0.07}_{-0.10}$& $ 0.48^{+0.08}_{-0.09}$& $ 0.49^{+0.06}_{-0.11}$& $ 0.58^{+0.06}_{-0.11}$& $ 0.72^{+0.03}_{-0.13}$& $ 0.60^{+0.08}_{-0.05}$& $ 0.61^{+0.06}_{-0.05}$& $ 0.60^{+0.11}_{-0.03}$\\
BH 229   & $ 2.82^{+0.11}_{-0.10}$& $ 2.72^{+0.15}_{-0.11}$& $ 2.79^{+0.13}_{-0.15}$& $ 2.76^{+0.15}_{-0.12}$& $ 0.41^{+0.09}_{-0.09}$& $ 0.03^{+0.24}_{-0.00}$& $ 0.32^{+0.07}_{-0.10}$& $ 0.20^{+0.17}_{-0.07}$& $ 0.75^{+0.04}_{-0.04}$& $ 0.98^{+0.00}_{-0.15}$& $ 0.79^{+0.06}_{-0.03}$& $ 0.87^{+0.03}_{-0.09}$& $ 2.77^{+0.12}_{-0.11}$& $ 2.70^{+0.15}_{-0.13}$& $ 2.76^{+0.13}_{-0.15}$& $ 2.74^{+0.13}_{-0.14}$\\
Liller 1  & $ 1.15^{+0.15}_{-0.12}$& $ 1.09^{+0.24}_{-0.05}$& $ 1.09^{+0.18}_{-0.06}$& $ 1.14^{+0.12}_{-0.09}$& $ 0.32^{+0.24}_{-0.22}$& $ 0.37^{+0.23}_{-0.24}$& $ 0.37^{+0.25}_{-0.18}$& $ 0.32^{+0.28}_{-0.16}$& $ 0.57^{+0.26}_{-0.23}$& $ 0.50^{+0.29}_{-0.18}$& $ 0.50^{+0.21}_{-0.20}$& $ 0.56^{+0.19}_{-0.24}$& $ 0.24^{+0.13}_{-0.13}$& $ 0.28^{+0.10}_{-0.19}$& $ 0.25^{+0.04}_{-0.14}$& $ 0.28^{+0.05}_{-0.17}$\\
NGC 6380  & $ 2.44^{+0.16}_{-0.04}$& $ 2.44^{+0.17}_{-0.09}$& $ 2.44^{+0.18}_{-0.08}$& $ 2.44^{+0.16}_{-0.03}$& $ 0.28^{+0.07}_{-0.13}$& $ 0.29^{+0.09}_{-0.07}$& $ 0.30^{+0.08}_{-0.08}$& $ 0.29^{+0.08}_{-0.11}$& $ 0.80^{+0.09}_{-0.04}$& $ 0.79^{+0.05}_{-0.05}$& $ 0.78^{+0.05}_{-0.04}$& $ 0.78^{+0.09}_{-0.04}$& $ 0.81^{+0.08}_{-0.01}$& $ 0.87^{+0.07}_{-0.01}$& $ 0.84^{+0.06}_{-0.02}$& $ 0.84^{+0.09}_{-0.00}$\\
Terzan 1  & $ 3.11^{+0.25}_{-0.20}$& $ 3.02^{+0.25}_{-0.17}$& $ 3.02^{+0.21}_{-0.21}$& $ 3.10^{+0.26}_{-0.19}$& $ 0.67^{+0.05}_{-0.06}$& $ 0.81^{+0.04}_{-0.06}$& $ 0.81^{+0.05}_{-0.06}$& $ 0.67^{+0.04}_{-0.05}$& $ 0.65^{+0.02}_{-0.02}$& $ 0.58^{+0.03}_{-0.02}$& $ 0.58^{+0.03}_{-0.03}$& $ 0.64^{+0.03}_{-0.01}$& $ 0.12^{+0.01}_{-0.01}$& $ 0.12^{+0.00}_{-0.01}$& $ 0.12^{+0.00}_{-0.01}$& $ 0.12^{+0.01}_{-0.01}$\\
NGC 6401  & $ 4.07^{+0.00}_{-0.28}$& $ 4.13^{+0.12}_{-0.12}$& $ 3.77^{+0.42}_{-0.00}$& $ 4.09^{+0.13}_{-0.16}$& $ 0.12^{+0.09}_{-0.07}$& $ 0.13^{+0.12}_{-0.07}$& $ 0.13^{+0.10}_{-0.08}$& $ 0.11^{+0.14}_{-0.05}$& $ 0.94^{+0.03}_{-0.04}$& $ 0.94^{+0.03}_{-0.05}$& $ 0.93^{+0.04}_{-0.04}$& $ 0.95^{+0.02}_{-0.06}$& $ 1.66^{+0.54}_{-0.07}$& $ 1.66^{+0.56}_{-0.13}$& $ 2.34^{+0.00}_{-0.77}$& $ 1.53^{+0.64}_{-0.03}$\\
Pal 6    & $ 3.88^{+0.41}_{-0.61}$& $ 3.61^{+0.84}_{-0.37}$& $ 3.96^{+0.39}_{-0.63}$& $ 4.02^{+0.24}_{-0.88}$& $ 0.10^{+0.17}_{-0.05}$& $ 0.06^{+0.16}_{-0.01}$& $ 0.06^{+0.18}_{-0.02}$& $ 0.11^{+0.11}_{-0.07}$& $ 0.95^{+0.02}_{-0.08}$& $ 0.96^{+0.01}_{-0.07}$& $ 0.97^{+0.00}_{-0.08}$& $ 0.95^{+0.03}_{-0.07}$& $ 2.20^{+0.50}_{-0.30}$& $ 1.96^{+0.66}_{-0.20}$& $ 1.86^{+0.84}_{-0.00}$& $ 1.79^{+0.77}_{-0.05}$\\
Terzan 5  & $ 2.06^{+0.15}_{-0.15}$& $ 2.06^{+0.16}_{-0.12}$& $ 2.06^{+0.13}_{-0.13}$& $ 2.05^{+0.14}_{-0.13}$& $ 0.34^{+0.18}_{-0.00}$& $ 0.51^{+0.04}_{-0.08}$& $ 0.51^{+0.04}_{-0.09}$& $ 0.34^{+0.17}_{-0.03}$& $ 0.72^{+0.00}_{-0.12}$& $ 0.60^{+0.05}_{-0.01}$& $ 0.60^{+0.05}_{-0.01}$& $ 0.72^{+0.02}_{-0.12}$& $ 0.30^{+0.08}_{-0.08}$& $ 0.28^{+0.03}_{-0.01}$& $ 0.28^{+0.02}_{-0.01}$& $ 0.29^{+0.13}_{-0.08}$\\
NGC 6440  & $ 1.53^{+0.11}_{-0.05}$& $ 1.52^{+0.12}_{-0.06}$& $ 1.52^{+0.16}_{-0.07}$& $ 1.53^{+0.14}_{-0.06}$& $ 0.11^{+0.15}_{-0.06}$& $ 0.13^{+0.10}_{-0.09}$& $ 0.15^{+0.09}_{-0.10}$& $ 0.15^{+0.10}_{-0.10}$& $ 0.86^{+0.08}_{-0.14}$& $ 0.84^{+0.10}_{-0.10}$& $ 0.82^{+0.12}_{-0.09}$& $ 0.82^{+0.12}_{-0.10}$& $ 0.86^{+0.11}_{-0.07}$& $ 0.85^{+0.08}_{-0.06}$& $ 0.85^{+0.10}_{-0.07}$& $ 0.86^{+0.10}_{-0.08}$\\
Terzan 6  & $ 1.96^{+0.40}_{-0.27}$& $ 1.94^{+0.34}_{-0.28}$& $ 1.94^{+0.36}_{-0.25}$& $ 1.96^{+0.37}_{-0.25}$& $ 0.21^{+0.10}_{-0.13}$& $ 0.29^{+0.08}_{-0.09}$& $ 0.29^{+0.10}_{-0.09}$& $ 0.21^{+0.12}_{-0.11}$& $ 0.80^{+0.11}_{-0.05}$& $ 0.74^{+0.07}_{-0.05}$& $ 0.74^{+0.06}_{-0.06}$& $ 0.81^{+0.08}_{-0.08}$& $ 0.62^{+0.06}_{-0.32}$& $ 0.29^{+0.18}_{-0.05}$& $ 0.29^{+0.19}_{-0.05}$& $ 0.31^{+0.33}_{-0.05}$\\
NGC 6453  & $ 2.66^{+0.46}_{-0.20}$& $ 2.85^{+0.23}_{-0.35}$& $ 2.66^{+0.32}_{-0.17}$& $ 2.62^{+0.47}_{-0.14}$& $ 0.31^{+0.12}_{-0.12}$& $ 0.21^{+0.14}_{-0.08}$& $ 0.33^{+0.05}_{-0.16}$& $ 0.05^{+0.30}_{-0.00}$& $ 0.79^{+0.08}_{-0.06}$& $ 0.86^{+0.05}_{-0.08}$& $ 0.78^{+0.11}_{-0.03}$& $ 0.96^{+0.00}_{-0.18}$& $ 2.03^{+0.14}_{-0.12}$& $ 2.00^{+0.10}_{-0.22}$& $ 2.01^{+0.16}_{-0.10}$& $ 1.98^{+0.16}_{-0.17}$\\
Terzan 9  & $ 2.71^{+0.44}_{-0.37}$& $ 2.70^{+0.43}_{-0.28}$& $ 2.71^{+0.36}_{-0.37}$& $ 2.71^{+0.42}_{-0.37}$& $ 0.31^{+0.14}_{-0.14}$& $ 0.41^{+0.13}_{-0.15}$& $ 0.41^{+0.12}_{-0.16}$& $ 0.32^{+0.12}_{-0.13}$& $ 0.80^{+0.09}_{-0.10}$& $ 0.74^{+0.10}_{-0.08}$& $ 0.73^{+0.10}_{-0.07}$& $ 0.79^{+0.08}_{-0.09}$& $ 0.80^{+0.00}_{-0.47}$& $ 0.37^{+0.04}_{-0.03}$& $ 0.37^{+0.15}_{-0.07}$& $ 0.38^{+0.26}_{-0.10}$\\
NGC 6522  & $ 2.01^{+0.19}_{-0.12}$& $ 1.96^{+0.12}_{-0.12}$& $ 2.00^{+0.14}_{-0.13}$& $ 1.97^{+0.13}_{-0.14}$& $ 0.62^{+0.15}_{-0.36}$& $ 0.79^{+0.15}_{-0.17}$& $ 0.77^{+0.14}_{-0.20}$& $ 0.72^{+0.14}_{-0.24}$& $ 0.53^{+0.25}_{-0.08}$& $ 0.43^{+0.07}_{-0.06}$& $ 0.44^{+0.11}_{-0.06}$& $ 0.46^{+0.14}_{-0.06}$& $ 1.53^{+0.09}_{-0.06}$& $ 1.47^{+0.07}_{-0.07}$& $ 1.51^{+0.09}_{-0.07}$& $ 1.49^{+0.08}_{-0.08}$\\
NGC 6528  & $ 2.84^{+0.38}_{-0.22}$& $ 2.80^{+0.35}_{-0.19}$& $ 2.83^{+0.31}_{-0.18}$& $ 2.81^{+0.48}_{-0.18}$& $ 0.39^{+0.08}_{-0.09}$& $ 0.52^{+0.06}_{-0.08}$& $ 0.45^{+0.08}_{-0.07}$& $ 0.44^{+0.06}_{-0.06}$& $ 0.76^{+0.05}_{-0.03}$& $ 0.69^{+0.03}_{-0.01}$& $ 0.72^{+0.03}_{-0.02}$& $ 0.73^{+0.04}_{-0.02}$& $ 0.85^{+0.08}_{-0.05}$& $ 0.82^{+0.12}_{-0.00}$& $ 0.82^{+0.08}_{-0.02}$& $ 0.82^{+0.13}_{-0.01}$\\
NGC 6558  & $ 3.07^{+0.34}_{-0.18}$& $ 3.61^{+0.00}_{-0.47}$& $ 3.01^{+0.28}_{-0.15}$& $ 3.68^{+0.00}_{-0.73}$& $ 0.36^{+0.09}_{-0.10}$& $ 0.39^{+0.04}_{-0.10}$& $ 0.33^{+0.07}_{-0.07}$& $ 0.39^{+0.06}_{-0.13}$& $ 0.79^{+0.05}_{-0.03}$& $ 0.80^{+0.03}_{-0.02}$& $ 0.80^{+0.04}_{-0.02}$& $ 0.81^{+0.03}_{-0.04}$& $ 1.11^{+0.11}_{-0.05}$& $ 1.09^{+0.03}_{-0.02}$& $ 1.10^{+0.03}_{-0.03}$& $ 1.10^{+0.04}_{-0.02}$\\
NGC 6624  & $ 1.58^{+0.03}_{-0.02}$& $ 1.57^{+0.05}_{-0.00}$& $ 1.58^{+0.03}_{-0.02}$& $ 1.56^{+0.03}_{-0.02}$& $ 0.53^{+0.06}_{-0.06}$& $ 0.60^{+0.06}_{-0.20}$& $ 0.59^{+0.06}_{-0.06}$& $ 0.60^{+0.03}_{-0.07}$& $ 0.50^{+0.05}_{-0.04}$& $ 0.45^{+0.15}_{-0.04}$& $ 0.45^{+0.05}_{-0.03}$& $ 0.45^{+0.05}_{-0.03}$& $ 1.51^{+0.02}_{-0.02}$& $ 1.50^{+0.03}_{-0.01}$& $ 1.50^{+0.03}_{-0.02}$& $ 1.49^{+0.02}_{-0.02}$\\
NGC 6626  & $ 3.49^{+0.09}_{-0.30}$& $ 3.14^{+0.21}_{-0.08}$& $ 3.40^{+0.00}_{-0.33}$& $ 3.46^{+0.00}_{-0.38}$& $ 0.15^{+0.26}_{-0.07}$& $ 0.33^{+0.30}_{-0.13}$& $ 0.36^{+0.18}_{-0.16}$& $ 0.35^{+0.07}_{-0.24}$& $ 0.92^{+0.03}_{-0.14}$& $ 0.81^{+0.07}_{-0.13}$& $ 0.81^{+0.08}_{-0.10}$& $ 0.81^{+0.12}_{-0.04}$& $ 1.23^{+0.09}_{-0.05}$& $ 1.34^{+0.00}_{-0.13}$& $ 1.25^{+0.04}_{-0.04}$& $ 1.24^{+0.13}_{-0.02}$\\
NGC 6638  & $ 2.39^{+0.52}_{-0.14}$& $ 2.47^{+0.28}_{-0.26}$& $ 2.40^{+0.35}_{-0.20}$& $ 2.42^{+0.54}_{-0.22}$& $ 0.14^{+0.01}_{-0.10}$& $ 0.09^{+0.08}_{-0.04}$& $ 0.07^{+0.10}_{-0.01}$& $ 0.12^{+0.03}_{-0.06}$& $ 0.89^{+0.08}_{-0.00}$& $ 0.93^{+0.03}_{-0.06}$& $ 0.94^{+0.02}_{-0.08}$& $ 0.90^{+0.06}_{-0.02}$& $ 1.68^{+0.12}_{-0.08}$& $ 1.79^{+0.03}_{-0.27}$& $ 1.67^{+0.11}_{-0.14}$& $ 1.74^{+0.10}_{-0.18}$\\
NGC 6637  & $ 2.06^{+0.01}_{-0.04}$& $ 1.99^{+0.04}_{-0.02}$& $ 2.00^{+0.03}_{-0.03}$& $ 2.01^{+0.03}_{-0.03}$& $ 0.88^{+0.11}_{-0.07}$& $ 0.82^{+0.04}_{-0.10}$& $ 0.86^{+0.08}_{-0.09}$& $ 0.88^{+0.07}_{-0.09}$& $ 0.40^{+0.03}_{-0.05}$& $ 0.42^{+0.05}_{-0.03}$& $ 0.40^{+0.05}_{-0.04}$& $ 0.39^{+0.05}_{-0.04}$& $ 1.85^{+0.03}_{-0.03}$& $ 1.87^{+0.03}_{-0.02}$& $ 1.86^{+0.03}_{-0.03}$& $ 1.85^{+0.03}_{-0.03}$\\
NGC 6642  & $ 2.25^{+0.14}_{-0.03}$& $ 2.20^{+0.09}_{-0.03}$& $ 2.22^{+0.07}_{-0.03}$& $ 2.23^{+0.13}_{-0.04}$& $ 0.02^{+0.22}_{-0.00}$& $ 0.04^{+0.13}_{-0.00}$& $ 0.09^{+0.09}_{-0.05}$& $ 0.11^{+0.08}_{-0.07}$& $ 0.98^{+0.00}_{-0.17}$& $ 0.96^{+0.01}_{-0.10}$& $ 0.92^{+0.04}_{-0.06}$& $ 0.91^{+0.05}_{-0.06}$& $ 1.60^{+0.00}_{-0.36}$& $ 1.51^{+0.00}_{-0.35}$& $ 1.59^{+0.00}_{-0.40}$& $ 1.56^{+0.00}_{-0.35}$\\
NGC 6717  & $ 2.65^{+0.05}_{-0.08}$& $ 2.49^{+0.06}_{-0.04}$& $ 2.54^{+0.04}_{-0.03}$& $ 2.50^{+0.08}_{-0.01}$& $ 1.35^{+0.11}_{-0.18}$& $ 1.31^{+0.08}_{-0.03}$& $ 1.53^{+0.01}_{-0.14}$& $ 1.37^{+0.10}_{-0.13}$& $ 0.33^{+0.06}_{-0.04}$& $ 0.31^{+0.02}_{-0.03}$& $ 0.25^{+0.05}_{-0.01}$& $ 0.29^{+0.06}_{-0.03}$& $ 1.43^{+0.03}_{-0.02}$& $ 1.44^{+0.03}_{-0.02}$& $ 1.45^{+0.03}_{-0.03}$& $ 1.46^{+0.01}_{-0.04}$\\
NGC 6723  & $ 4.23^{+0.04}_{-0.05}$& $ 4.32^{+0.04}_{-0.04}$& $ 4.26^{+0.06}_{-0.05}$& $ 4.34^{+0.05}_{-0.09}$& $ 0.15^{+0.62}_{-0.00}$& $ 0.73^{+0.12}_{-0.10}$& $ 0.73^{+0.06}_{-0.16}$& $ 0.59^{+0.13}_{-0.15}$& $ 0.93^{+0.00}_{-0.24}$& $ 0.71^{+0.03}_{-0.04}$& $ 0.71^{+0.05}_{-0.02}$& $ 0.76^{+0.05}_{-0.04}$& $ 3.97^{+0.08}_{-0.09}$& $ 3.97^{+0.09}_{-0.09}$& $ 3.96^{+0.08}_{-0.08}$& $ 3.97^{+0.09}_{-0.07}$\\
NGC 6171  & $ 3.95^{+0.05}_{-0.07}$& $ 3.95^{+0.08}_{-0.05}$& $ 3.96^{+0.07}_{-0.06}$& $ 3.96^{+0.08}_{-0.05}$& $ 1.35^{+0.03}_{-0.08}$& $ 1.47^{+0.02}_{-0.06}$& $ 1.41^{+0.05}_{-0.04}$& $ 1.42^{+0.00}_{-0.10}$& $ 0.49^{+0.02}_{-0.01}$& $ 0.46^{+0.02}_{-0.01}$& $ 0.48^{+0.01}_{-0.02}$& $ 0.47^{+0.03}_{-0.00}$& $ 2.41^{+0.04}_{-0.04}$& $ 2.40^{+0.04}_{-0.03}$& $ 2.41^{+0.03}_{-0.03}$& $ 2.42^{+0.02}_{-0.04}$\\
NGC 6316  & $ 3.86^{+0.49}_{-0.30}$& $ 3.87^{+0.46}_{-0.35}$& $ 3.87^{+0.40}_{-0.32}$& $ 3.87^{+0.39}_{-0.33}$& $ 0.86^{+0.23}_{-0.28}$& $ 0.88^{+0.18}_{-0.26}$& $ 0.86^{+0.22}_{-0.22}$& $ 0.83^{+0.17}_{-0.28}$& $ 0.63^{+0.11}_{-0.05}$& $ 0.63^{+0.09}_{-0.05}$& $ 0.64^{+0.07}_{-0.06}$& $ 0.65^{+0.09}_{-0.05}$& $ 1.69^{+0.13}_{-0.07}$& $ 1.70^{+0.07}_{-0.08}$& $ 1.67^{+0.12}_{-0.05}$& $ 1.72^{+0.11}_{-0.10}$\\
NGC 6388  & $ 4.19^{+0.11}_{-0.15}$& $ 4.05^{+0.22}_{-0.17}$& $ 4.21^{+0.09}_{-0.16}$& $ 4.09^{+0.31}_{-0.14}$& $ 1.16^{+0.02}_{-0.08}$& $ 1.22^{+0.07}_{-0.11}$& $ 1.17^{+0.08}_{-0.04}$& $ 1.14^{+0.07}_{-0.10}$& $ 0.56^{+0.03}_{-0.01}$& $ 0.54^{+0.04}_{-0.03}$& $ 0.57^{+0.01}_{-0.03}$& $ 0.56^{+0.06}_{-0.03}$& $ 1.59^{+0.08}_{-0.10}$& $ 1.67^{+0.10}_{-0.12}$& $ 1.52^{+0.11}_{-0.01}$& $ 1.72^{+0.10}_{-0.14}$\\
NGC 6539  & $ 3.92^{+0.31}_{-0.17}$& $ 3.88^{+0.24}_{-0.10}$& $ 3.99^{+0.21}_{-0.17}$& $ 3.97^{+0.21}_{-0.22}$& $ 2.10^{+0.16}_{-0.22}$& $ 2.19^{+0.15}_{-0.21}$& $ 2.15^{+0.18}_{-0.19}$& $ 2.12^{+0.14}_{-0.23}$& $ 0.30^{+0.06}_{-0.03}$& $ 0.28^{+0.05}_{-0.03}$& $ 0.30^{+0.05}_{-0.04}$& $ 0.30^{+0.06}_{-0.03}$& $ 2.84^{+0.21}_{-0.14}$& $ 2.73^{+0.20}_{-0.13}$& $ 2.80^{+0.19}_{-0.14}$& $ 2.80^{+0.17}_{-0.15}$\\
NGC 6553  & $ 4.68^{+0.13}_{-0.09}$& $ 4.64^{+0.10}_{-0.10}$& $ 4.65^{+0.08}_{-0.13}$& $ 4.67^{+0.10}_{-0.12}$& $ 2.33^{+0.06}_{-0.12}$& $ 2.51^{+0.08}_{-0.09}$& $ 2.51^{+0.07}_{-0.12}$& $ 2.33^{+0.04}_{-0.19}$& $ 0.34^{+0.01}_{-0.01}$& $ 0.30^{+0.01}_{-0.01}$& $ 0.30^{+0.01}_{-0.01}$& $ 0.33^{+0.03}_{-0.00}$& $ 0.31^{+0.01}_{-0.01}$& $ 0.31^{+0.01}_{-0.01}$& $ 0.31^{+0.01}_{-0.01}$& $ 0.31^{+0.01}_{-0.01}$\\
NGC 6652  & $ 3.20^{+0.10}_{-0.06}$& $ 3.21^{+0.13}_{-0.10}$& $ 3.18^{+0.09}_{-0.07}$& $ 3.23^{+0.10}_{-0.09}$& $ 0.06^{+0.11}_{-0.00}$& $ 0.11^{+0.03}_{-0.04}$& $ 0.13^{+0.06}_{-0.03}$& $ 0.18^{+0.00}_{-0.10}$& $ 0.96^{+0.00}_{-0.06}$& $ 0.93^{+0.02}_{-0.01}$& $ 0.92^{+0.02}_{-0.03}$& $ 0.90^{+0.05}_{-0.00}$& $ 2.98^{+0.02}_{-0.09}$& $ 2.88^{+0.08}_{-0.05}$& $ 2.94^{+0.03}_{-0.08}$& $ 2.96^{+0.03}_{-0.11}$\\
Terzan 3  & $ 4.78^{+0.10}_{-0.31}$& $ 4.37^{+0.29}_{-0.14}$& $ 4.49^{+0.18}_{-0.18}$& $ 4.59^{+0.23}_{-0.16}$& $ 2.06^{+0.32}_{-0.14}$& $ 2.21^{+0.16}_{-0.25}$& $ 2.15^{+0.22}_{-0.19}$& $ 2.02^{+0.29}_{-0.13}$& $ 0.40^{+0.01}_{-0.07}$& $ 0.33^{+0.05}_{-0.02}$& $ 0.35^{+0.04}_{-0.04}$& $ 0.39^{+0.03}_{-0.05}$& $ 1.94^{+0.15}_{-0.08}$& $ 1.94^{+0.10}_{-0.09}$& $ 1.91^{+0.12}_{-0.07}$& $ 1.92^{+0.12}_{-0.07}$\\
NGC 6256  & $ 3.38^{+0.12}_{-0.14}$& $ 3.15^{+0.18}_{-0.08}$& $ 3.18^{+0.17}_{-0.08}$& $ 3.27^{+0.21}_{-0.06}$& $ 0.83^{+0.39}_{-0.22}$& $ 1.22^{+0.34}_{-0.30}$& $ 1.20^{+0.36}_{-0.25}$& $ 0.93^{+0.33}_{-0.29}$& $ 0.61^{+0.06}_{-0.12}$& $ 0.44^{+0.10}_{-0.08}$& $ 0.45^{+0.08}_{-0.09}$& $ 0.56^{+0.10}_{-0.09}$& $ 0.87^{+0.05}_{-0.06}$& $ 0.93^{+0.01}_{-0.18}$& $ 0.88^{+0.03}_{-0.10}$& $ 0.92^{+0.05}_{-0.07}$\\
NGC 6304  & $ 3.81^{+0.08}_{-0.30}$& $ 3.68^{+0.10}_{-0.18}$& $ 3.65^{+0.19}_{-0.20}$& $ 3.70^{+0.18}_{-0.17}$& $ 1.18^{+0.43}_{-0.82}$& $ 1.75^{+0.24}_{-0.52}$& $ 1.72^{+0.28}_{-0.67}$& $ 1.46^{+0.20}_{-0.93}$& $ 0.53^{+0.29}_{-0.12}$& $ 0.36^{+0.14}_{-0.07}$& $ 0.36^{+0.20}_{-0.06}$& $ 0.43^{+0.32}_{-0.04}$& $ 0.98^{+0.12}_{-0.03}$& $ 0.88^{+0.11}_{-0.01}$& $ 1.02^{+0.04}_{-0.10}$& $ 1.06^{+0.00}_{-0.14}$\\
Pismis 26  & $ 5.12^{+1.31}_{-0.76}$& $ 4.88^{+0.41}_{-0.28}$& $ 5.00^{+0.41}_{-0.34}$& $ 5.03^{+0.70}_{-0.44}$& $ 1.78^{+0.18}_{-0.33}$& $ 1.78^{+0.11}_{-0.27}$& $ 1.60^{+0.30}_{-0.13}$& $ 1.66^{+0.19}_{-0.28}$& $ 0.48^{+0.08}_{-0.00}$& $ 0.46^{+0.07}_{-0.01}$& $ 0.51^{+0.03}_{-0.05}$& $ 0.50^{+0.07}_{-0.02}$& $ 2.05^{+0.47}_{-0.13}$& $ 2.15^{+0.11}_{-0.12}$& $ 2.09^{+0.18}_{-0.11}$& $ 2.12^{+0.24}_{-0.09}$\\
NGC 6540  & $ 2.56^{+0.20}_{-0.29}$& $ 2.52^{+0.24}_{-0.26}$& $ 2.50^{+0.22}_{-0.30}$& $ 2.51^{+0.25}_{-0.20}$& $ 1.38^{+0.06}_{-0.08}$& $ 1.51^{+0.06}_{-0.10}$& $ 1.54^{+0.01}_{-0.16}$& $ 1.43^{+0.02}_{-0.13}$& $ 0.30^{+0.04}_{-0.06}$& $ 0.25^{+0.05}_{-0.05}$& $ 0.24^{+0.05}_{-0.03}$& $ 0.27^{+0.07}_{-0.02}$& $ 0.55^{+0.03}_{-0.03}$& $ 0.54^{+0.04}_{-0.02}$& $ 0.57^{+0.00}_{-0.06}$& $ 0.57^{+0.04}_{-0.06}$\\
NGC 6569  & $ 2.83^{+0.32}_{-0.06}$& $ 2.84^{+0.34}_{-0.10}$& $ 2.84^{+0.25}_{-0.11}$& $ 2.83^{+0.22}_{-0.11}$& $ 1.54^{+0.00}_{-0.61}$& $ 1.60^{+0.06}_{-0.59}$& $ 1.60^{+0.02}_{-0.51}$& $ 1.51^{+0.05}_{-0.49}$& $ 0.30^{+0.25}_{-0.00}$& $ 0.28^{+0.23}_{-0.02}$& $ 0.28^{+0.20}_{-0.02}$& $ 0.31^{+0.20}_{-0.04}$& $ 1.32^{+0.12}_{-0.05}$& $ 1.32^{+0.07}_{-0.03}$& $ 1.31^{+0.08}_{-0.03}$& $ 1.32^{+0.07}_{-0.03}$\\
E456-78  & $ 4.03^{+0.23}_{-0.22}$& $ 3.91^{+0.19}_{-0.17}$& $ 3.99^{+0.18}_{-0.23}$& $ 4.08^{+0.12}_{-0.25}$& $ 1.81^{+0.27}_{-0.12}$& $ 1.94^{+0.16}_{-0.24}$& $ 1.87^{+0.25}_{-0.20}$& $ 1.88^{+0.23}_{-0.16}$& $ 0.38^{+0.01}_{-0.04}$& $ 0.34^{+0.04}_{-0.02}$& $ 0.36^{+0.03}_{-0.04}$& $ 0.37^{+0.02}_{-0.05}$& $ 1.46^{+0.12}_{-0.08}$& $ 1.46^{+0.10}_{-0.08}$& $ 1.47^{+0.11}_{-0.08}$& $ 1.47^{+0.11}_{-0.09}$\\
NGC 6325  & $ 2.44^{+0.42}_{-0.24}$& $ 2.53^{+0.34}_{-0.28}$& $ 2.52^{+0.30}_{-0.30}$& $ 2.49^{+0.36}_{-0.32}$& $ 0.78^{+0.33}_{-0.27}$& $ 0.79^{+0.32}_{-0.21}$& $ 0.79^{+0.26}_{-0.33}$& $ 0.78^{+0.34}_{-0.30}$& $ 0.52^{+0.14}_{-0.13}$& $ 0.52^{+0.11}_{-0.13}$& $ 0.52^{+0.17}_{-0.11}$& $ 0.52^{+0.16}_{-0.15}$& $ 1.65^{+0.14}_{-0.10}$& $ 1.64^{+0.10}_{-0.08}$& $ 1.61^{+0.14}_{-0.07}$& $ 1.67^{+0.13}_{-0.10}$\\
Djorg 2   & $ 2.43^{+0.08}_{-0.03}$& $ 2.40^{+0.09}_{-0.04}$& $ 2.43^{+0.06}_{-0.06}$& $ 2.41^{+0.09}_{-0.04}$& $ 0.47^{+0.15}_{-0.19}$& $ 0.59^{+0.16}_{-0.12}$& $ 0.54^{+0.11}_{-0.21}$& $ 0.51^{+0.13}_{-0.18}$& $ 0.67^{+0.12}_{-0.07}$& $ 0.61^{+0.06}_{-0.08}$& $ 0.64^{+0.11}_{-0.06}$& $ 0.65^{+0.11}_{-0.06}$& $ 0.78^{+0.08}_{-0.02}$& $ 0.81^{+0.03}_{-0.02}$& $ 0.78^{+0.05}_{-0.02}$& $ 0.81^{+0.06}_{-0.02}$\\\hline
 \end{tabular}
   \end{center}
   \end{tiny}
  \end{minipage}
  }
  \end{table*}

\begin{figure*}
{\begin{center}
   \includegraphics[width=0.3\textwidth,angle=-90]{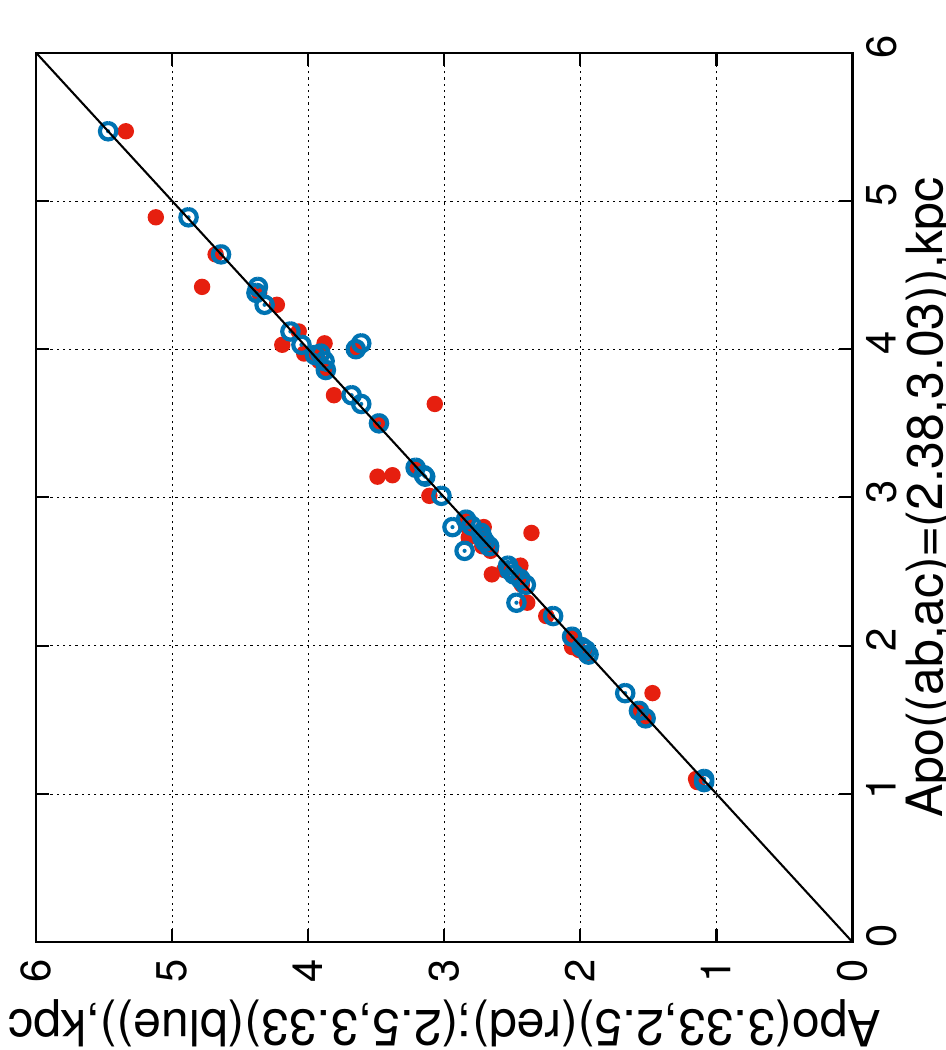}   \includegraphics[width=0.3\textwidth,angle=-90]{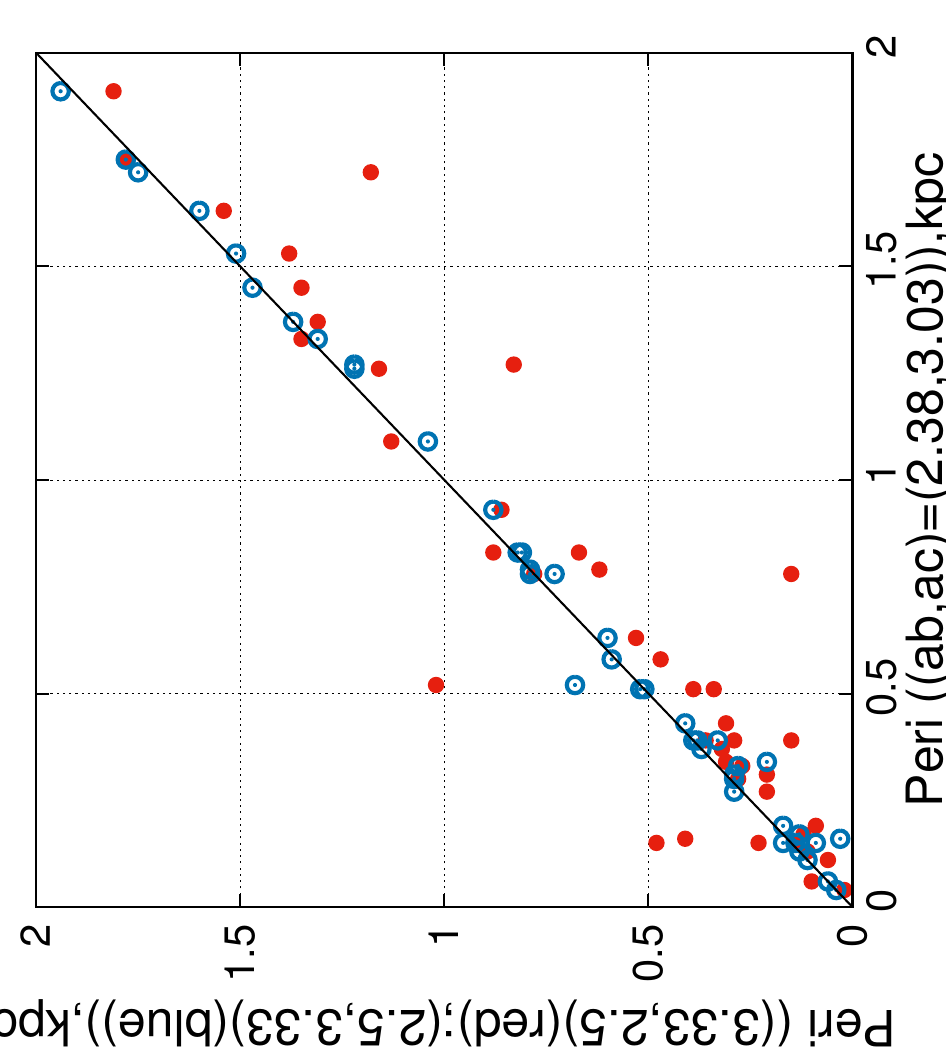}\

\medskip
     \includegraphics[width=0.3\textwidth,angle=-90]{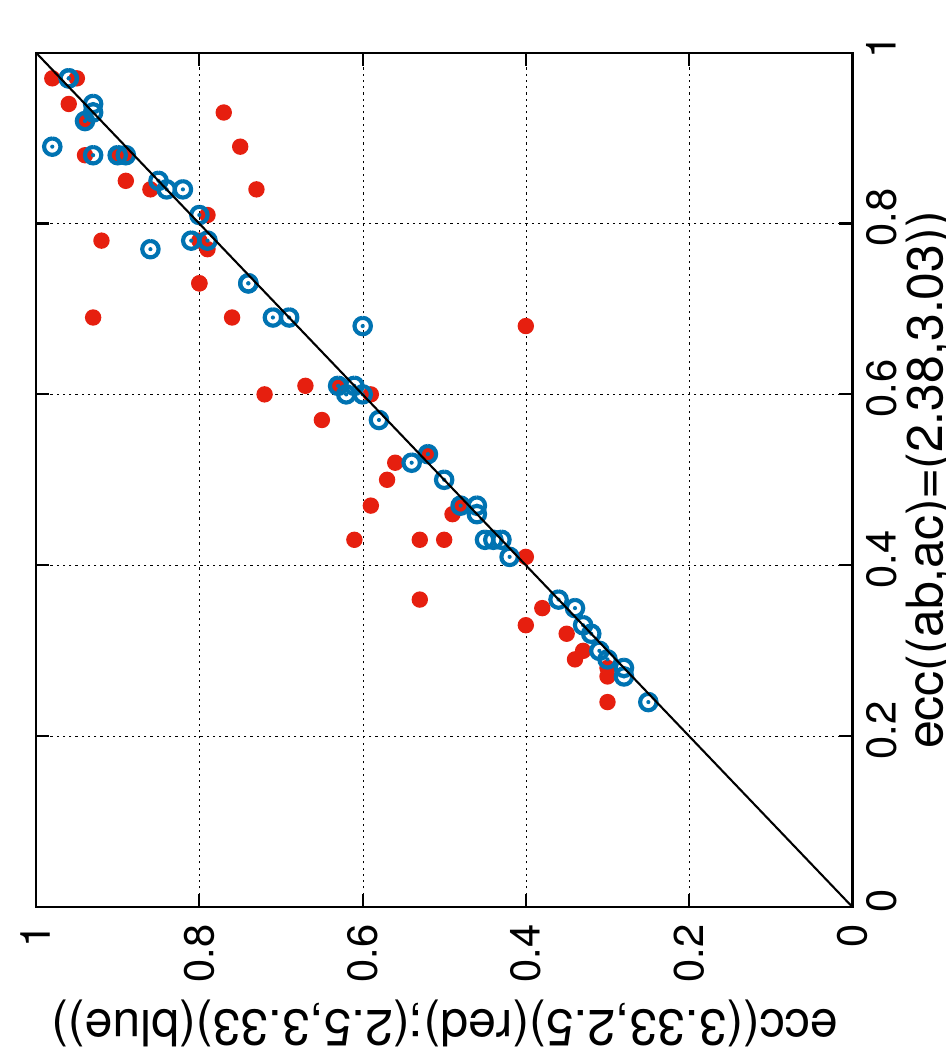}   \includegraphics[width=0.3\textwidth,angle=-90]{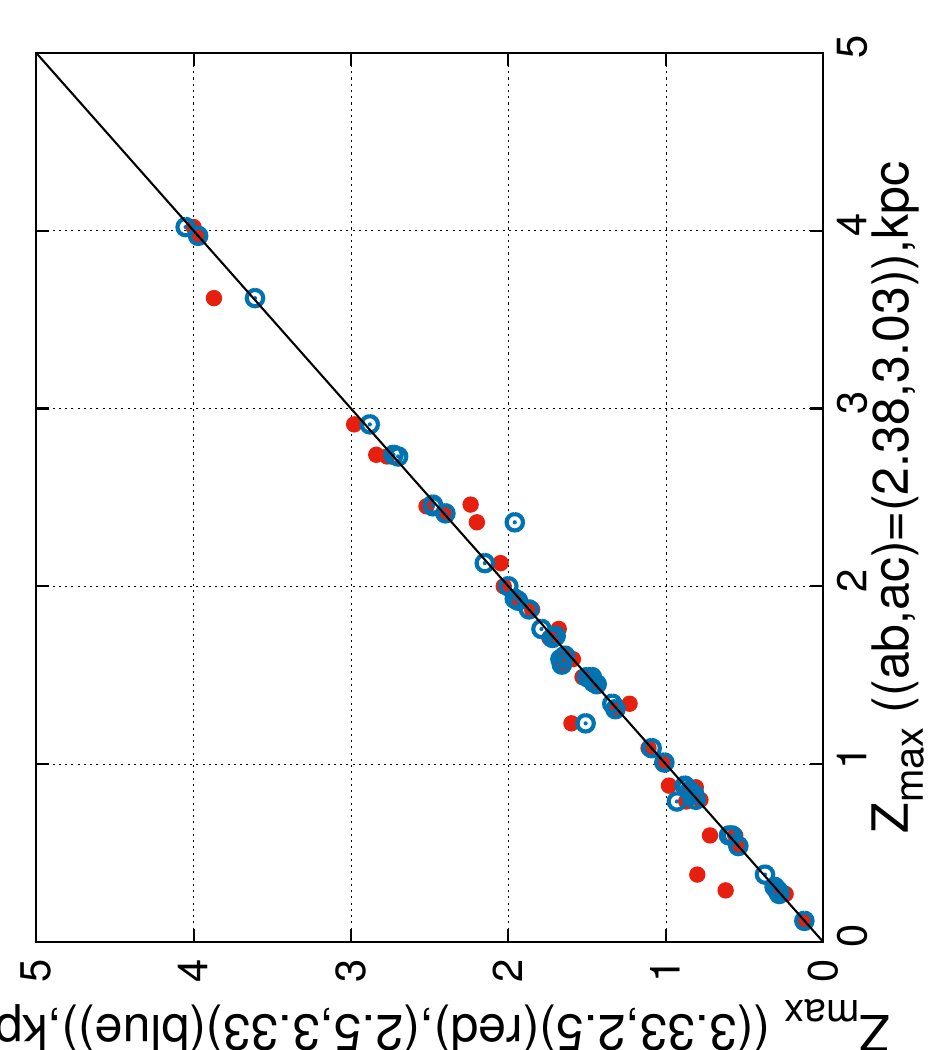}\
  \includegraphics[width=0.3\textwidth,angle=-90]{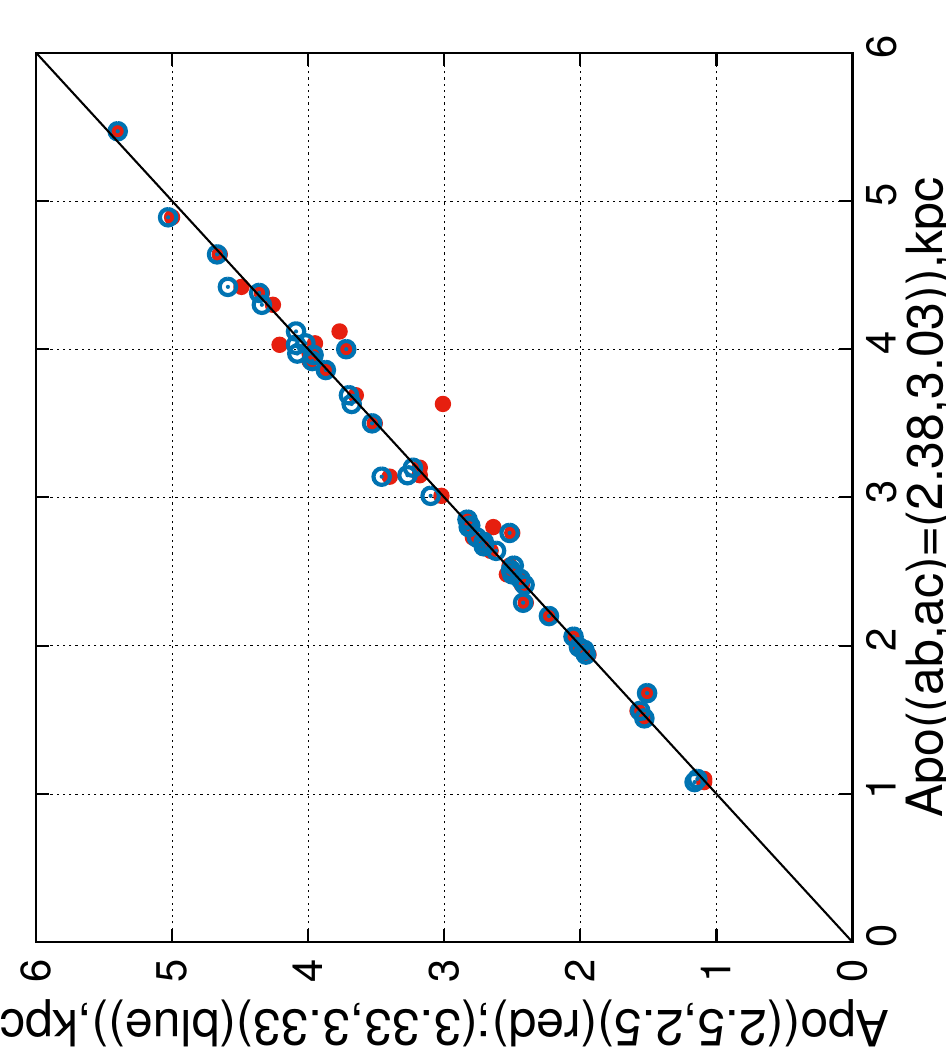}   \includegraphics[width=0.3\textwidth,angle=-90]{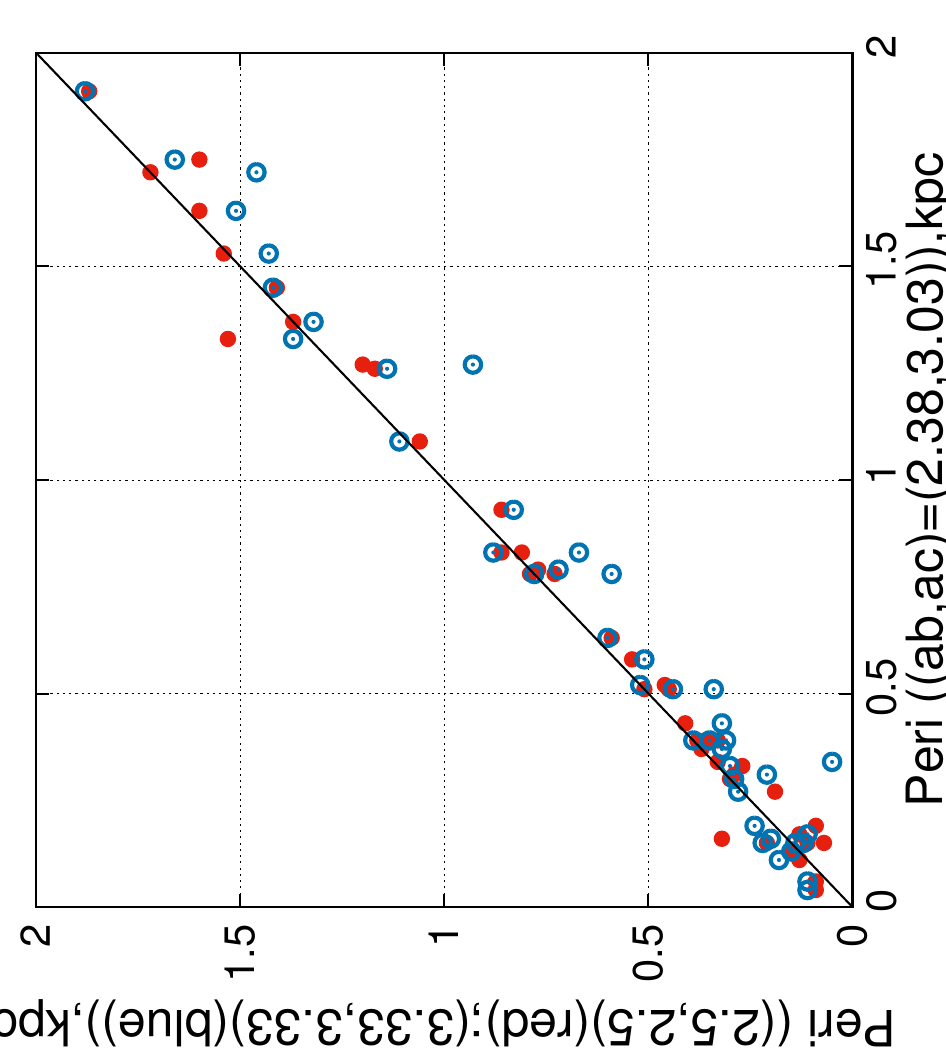}\

  \medskip
  \includegraphics[width=0.3\textwidth,angle=-90]{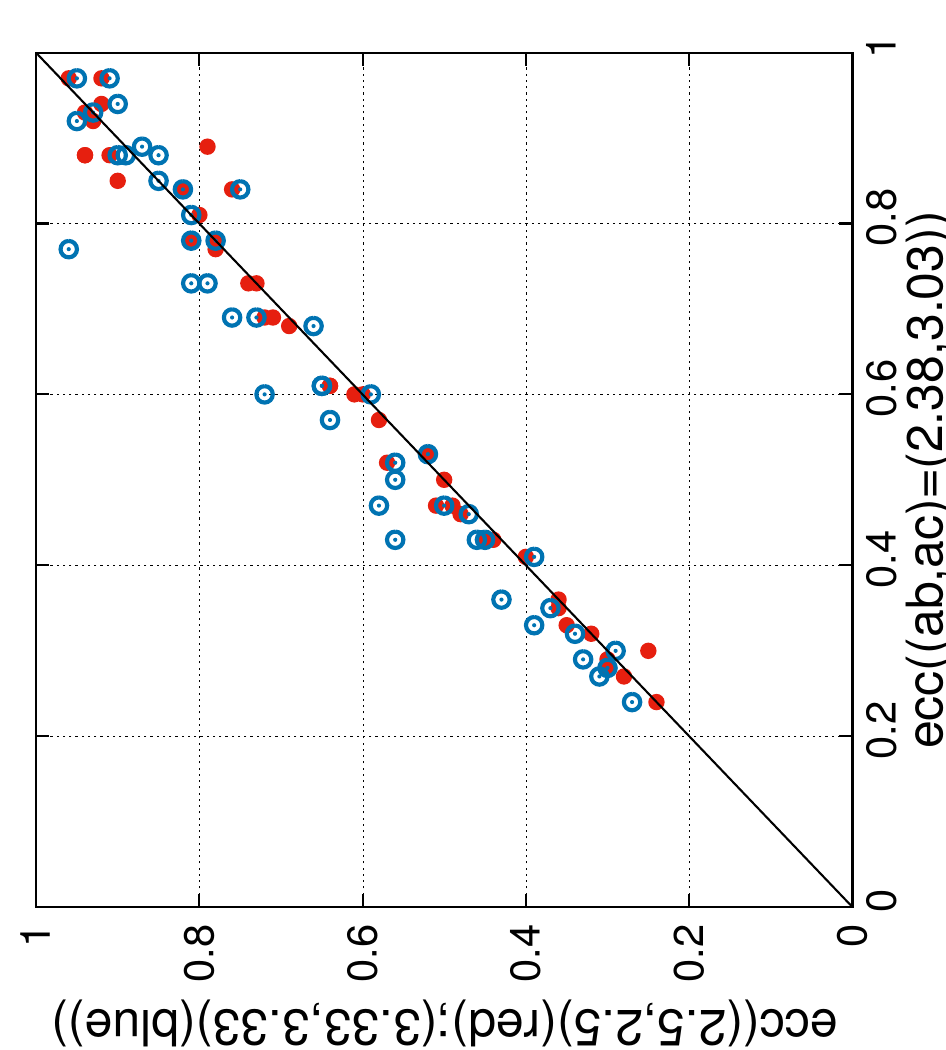}   \includegraphics[width=0.3\textwidth,angle=-90]{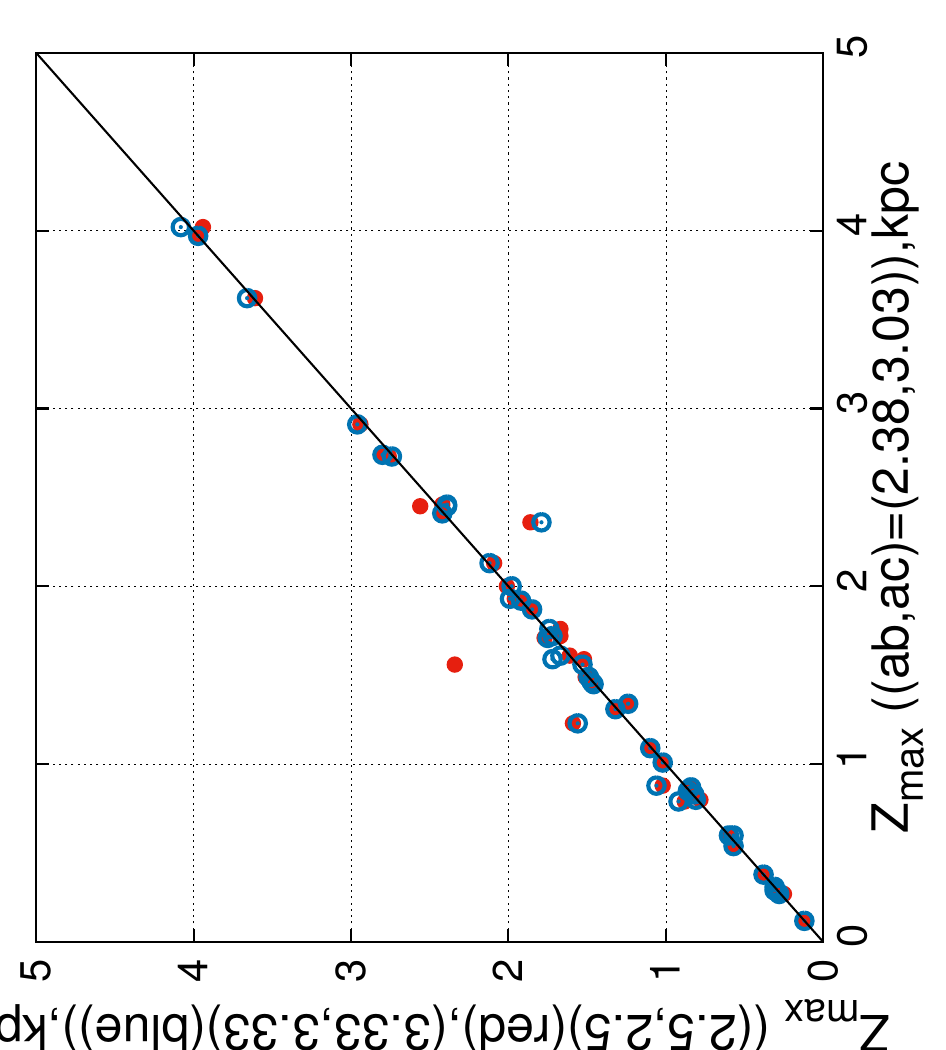}\
\caption{\small Comparison of the orbital parameters of the GCs in the potential with a bar with different axis ratios. Other bar parameters: $M_b=430 M_G$, $\Omega_b=40$ km/s/kpc, $q_b=5$ kpc, $\theta_b=25^o$. Each panel has a line of coincise.}
\label{fD}
\end{center}}
\end{figure*}

\begin{figure*}
{\begin{center}
   \includegraphics[width=0.2\textwidth,angle=-90]{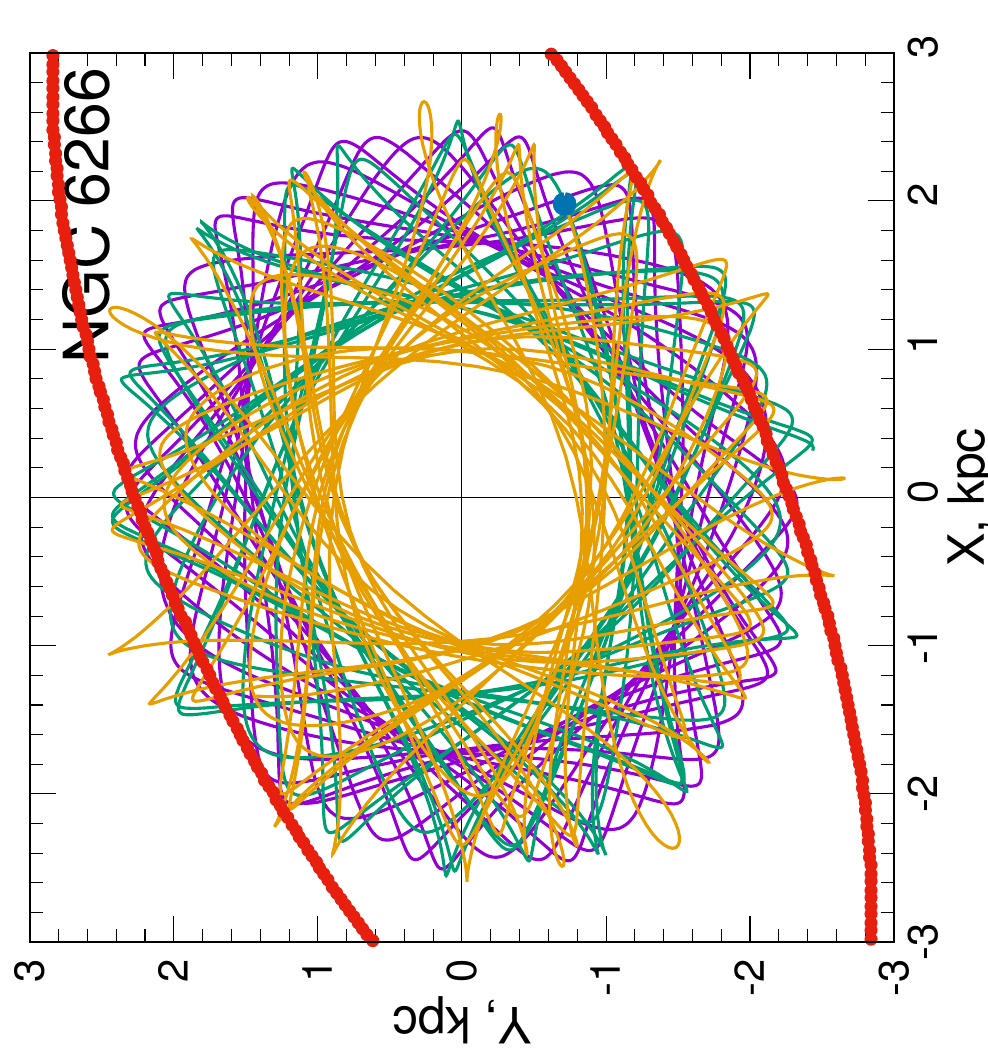}
   \includegraphics[width=0.2\textwidth,angle=-90]{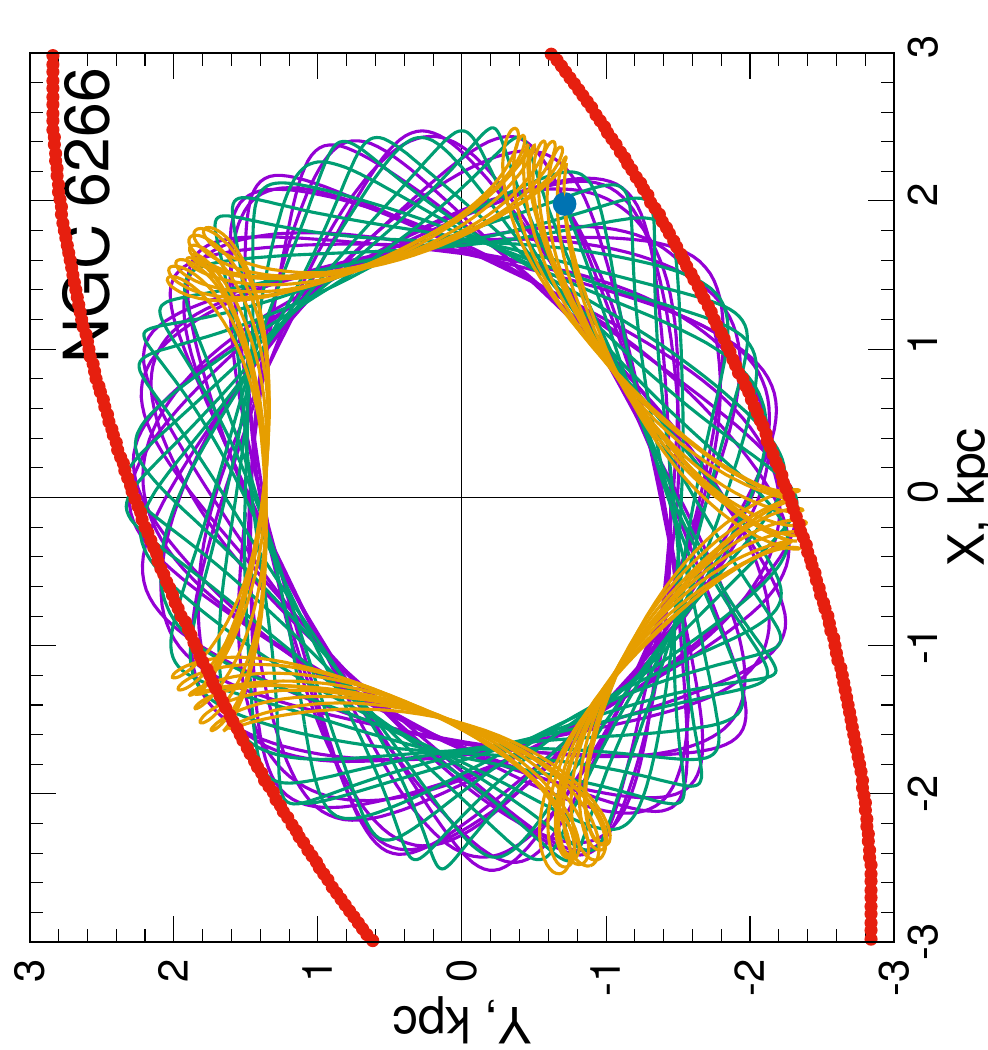}
   \includegraphics[width=0.2\textwidth,angle=-90]{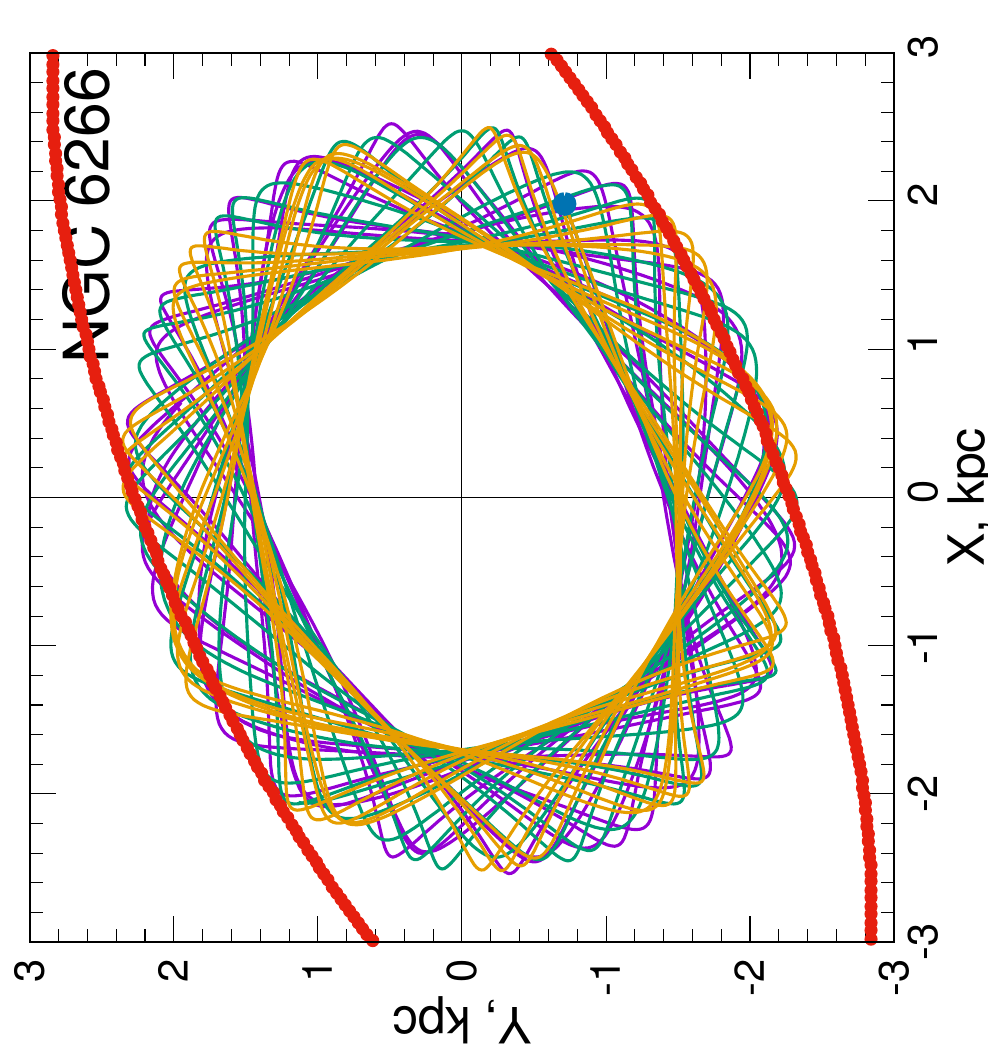}
   \includegraphics[width=0.2\textwidth,angle=-90]{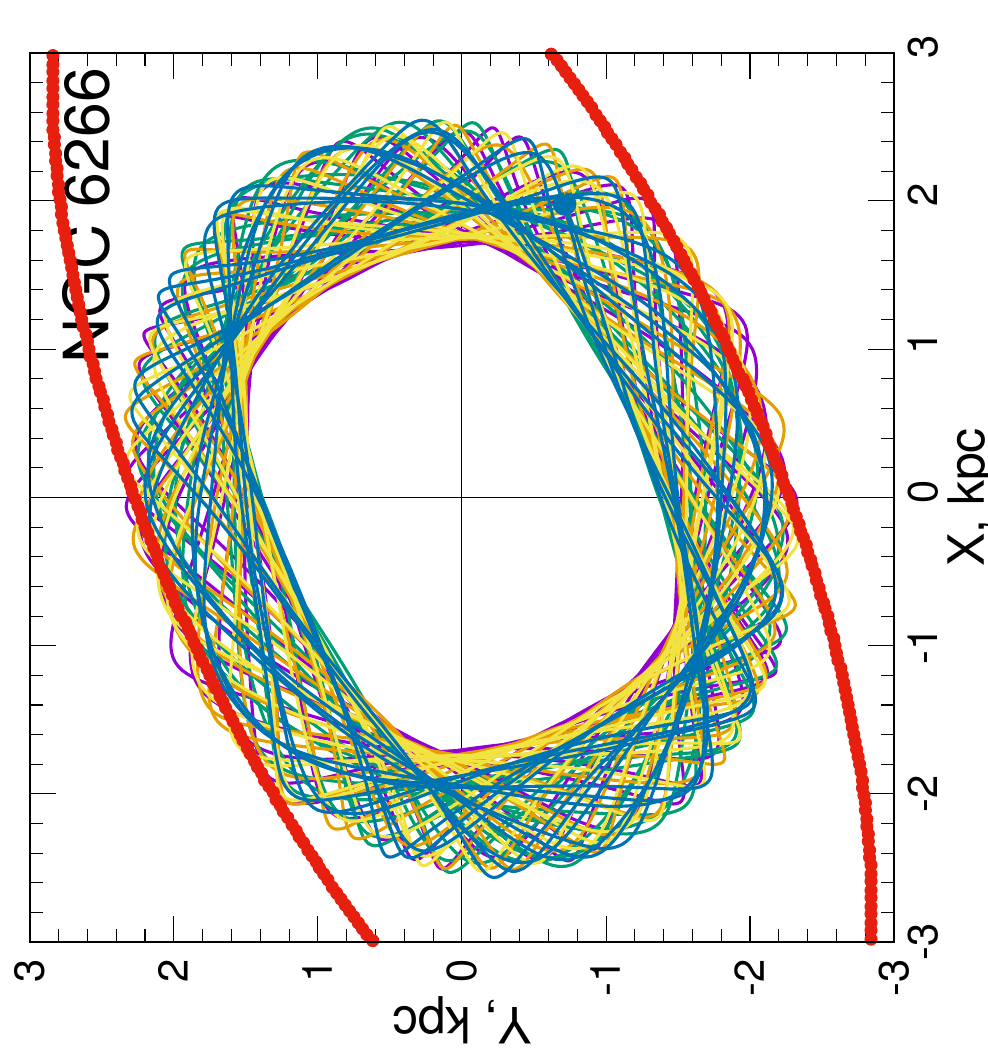}\

   \includegraphics[width=0.2\textwidth,angle=-90]{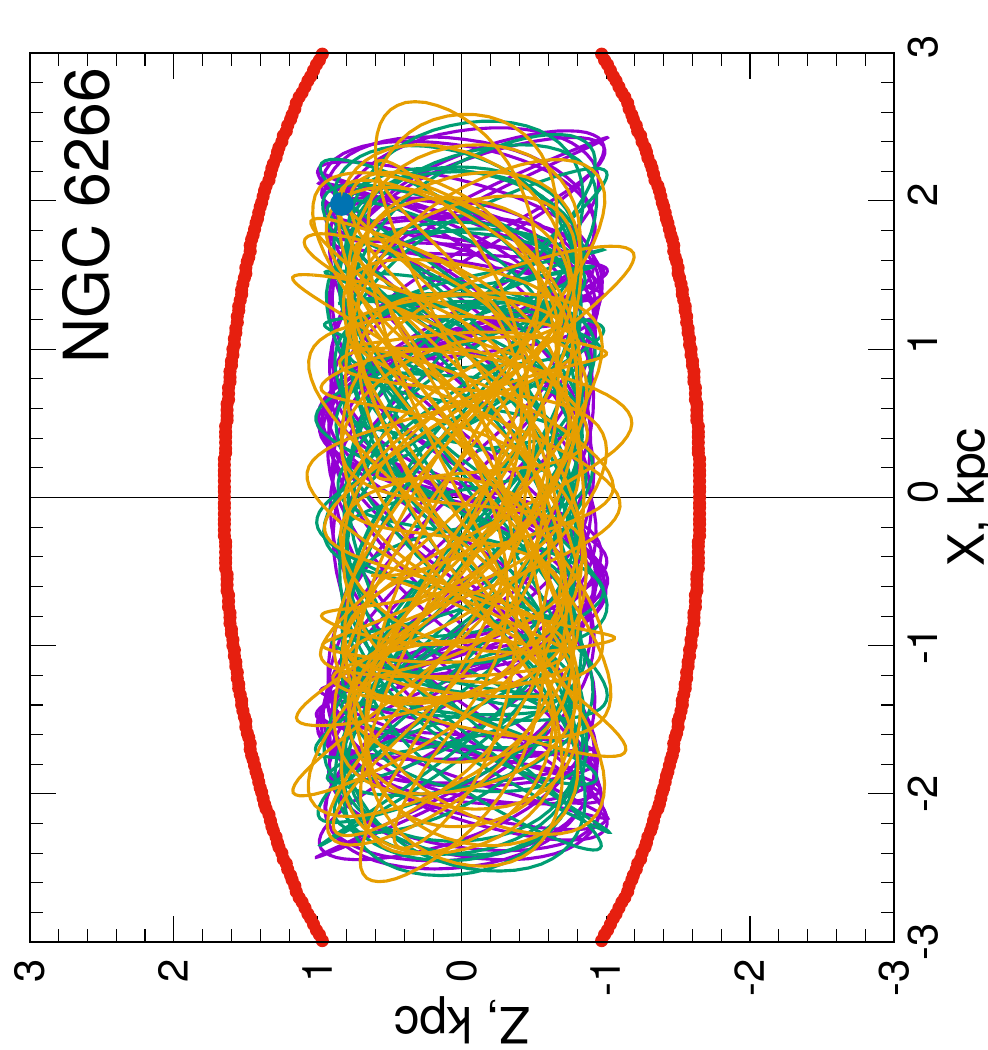}
   \includegraphics[width=0.2\textwidth,angle=-90]{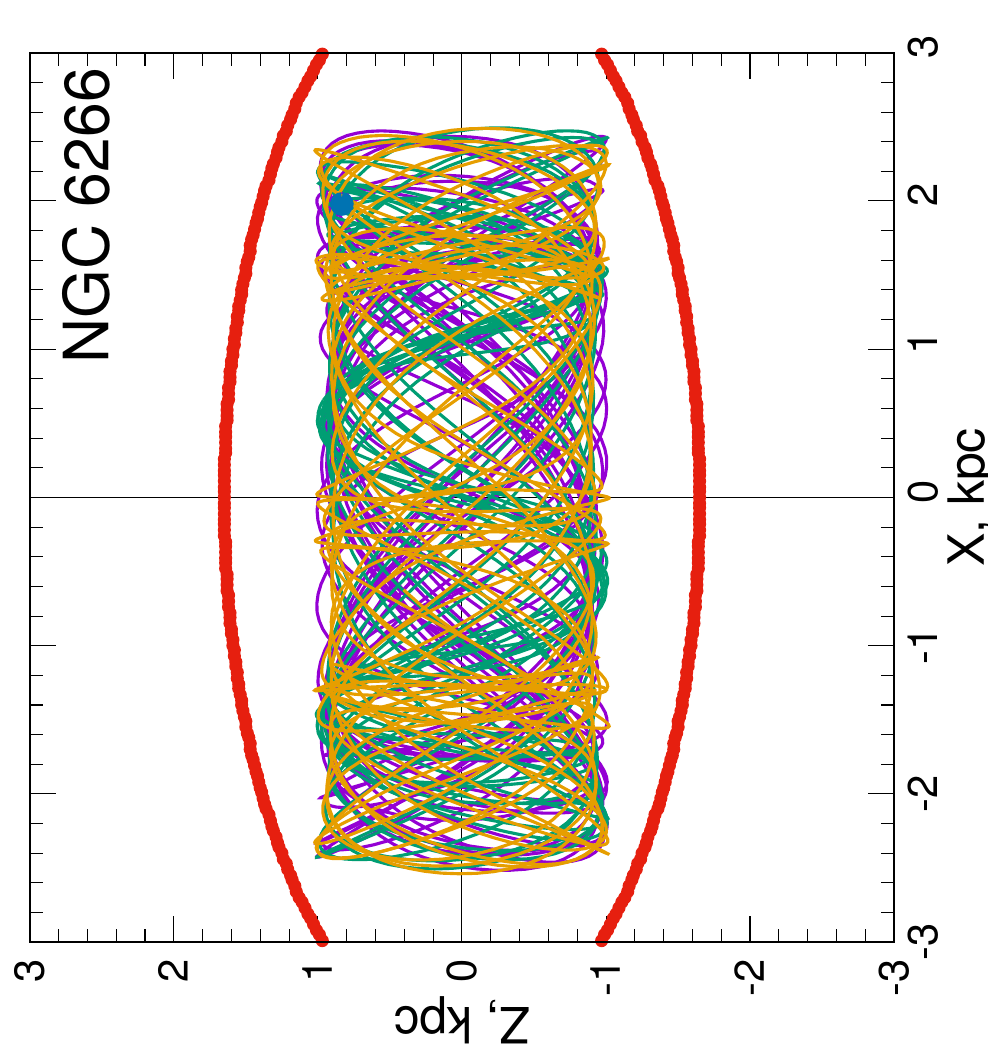}
   \includegraphics[width=0.2\textwidth,angle=-90]{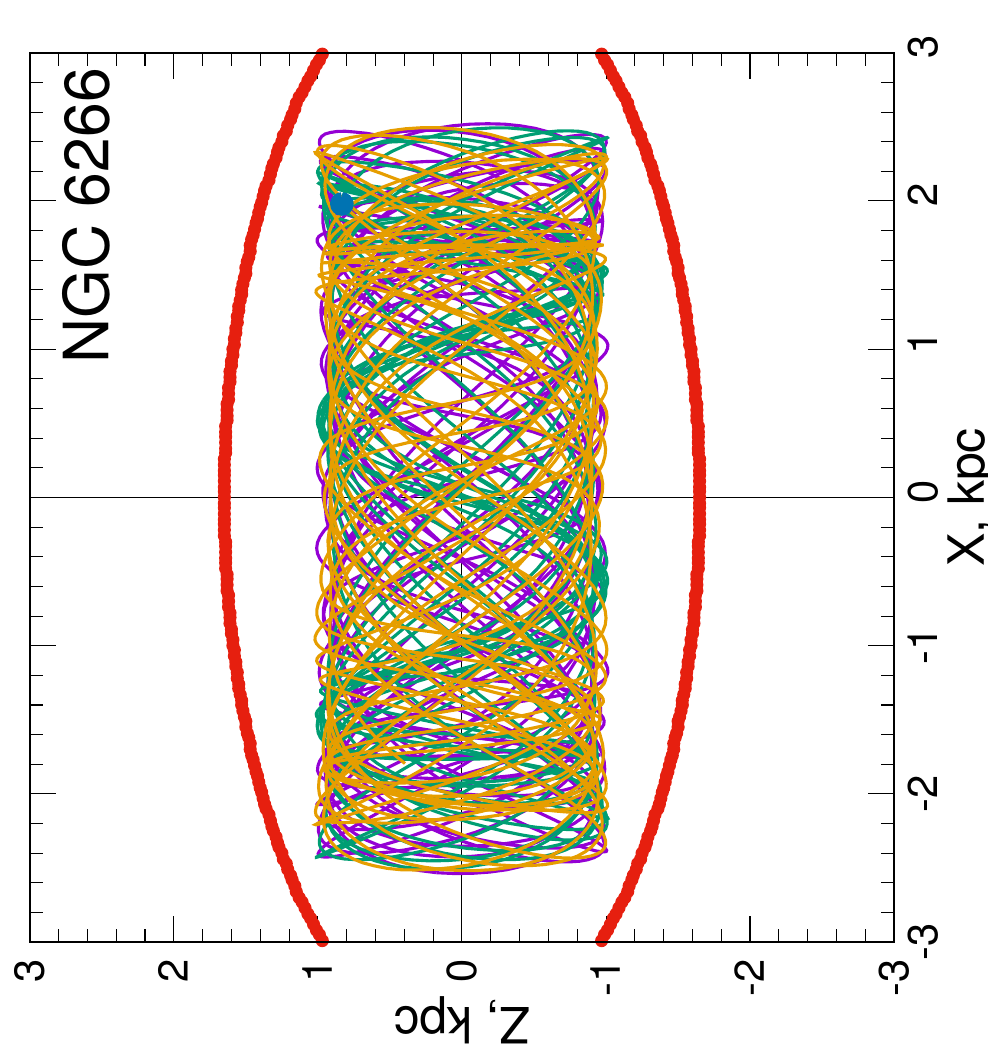}
   \includegraphics[width=0.2\textwidth,angle=-90]{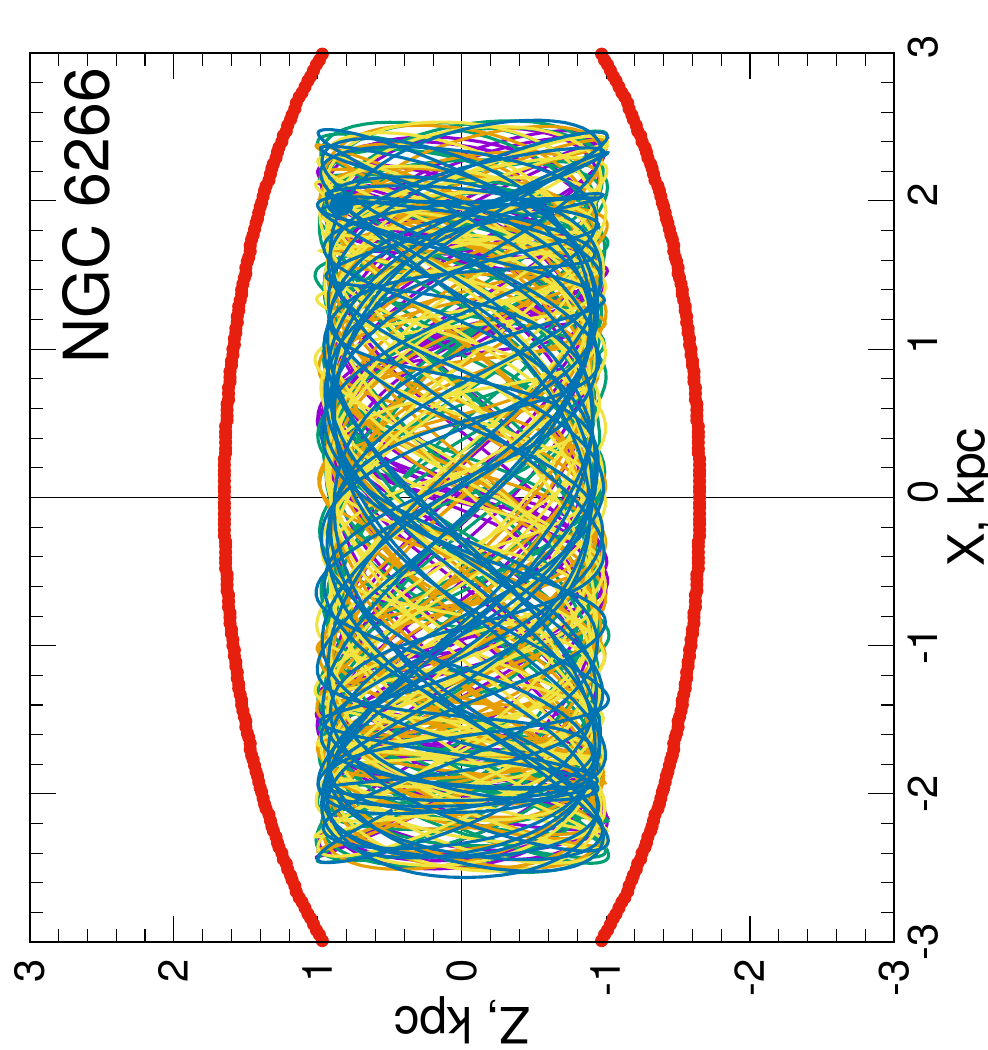}\

   \includegraphics[width=0.2\textwidth,angle=-90]{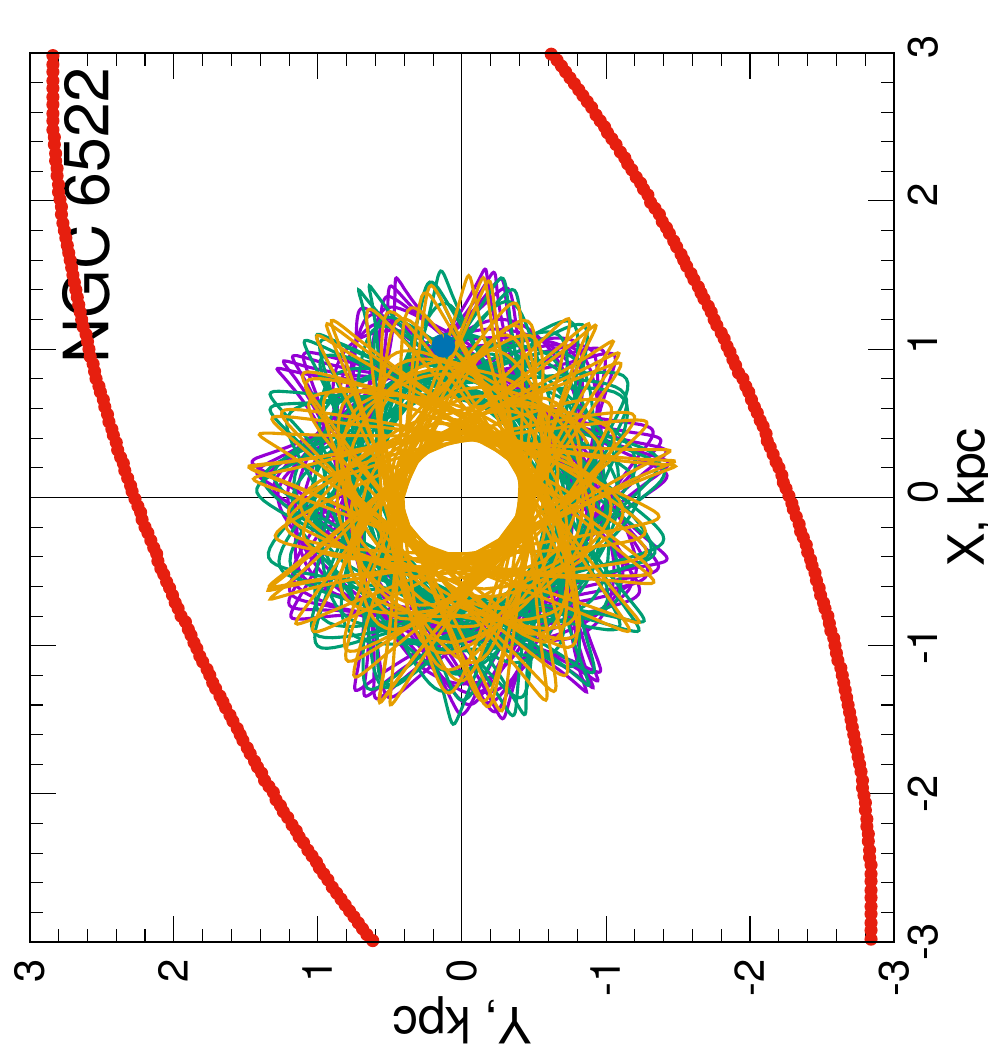}
   \includegraphics[width=0.2\textwidth,angle=-90]{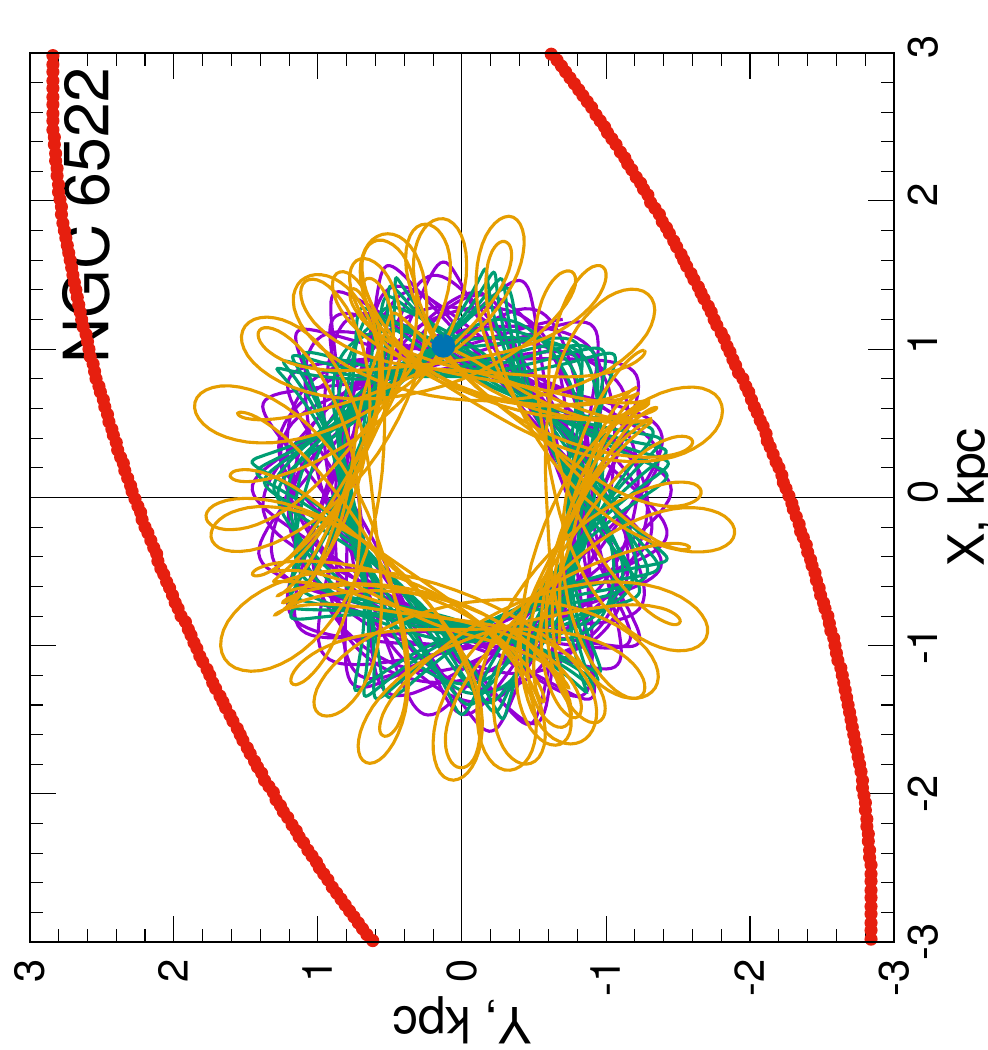}
   \includegraphics[width=0.2\textwidth,angle=-90]{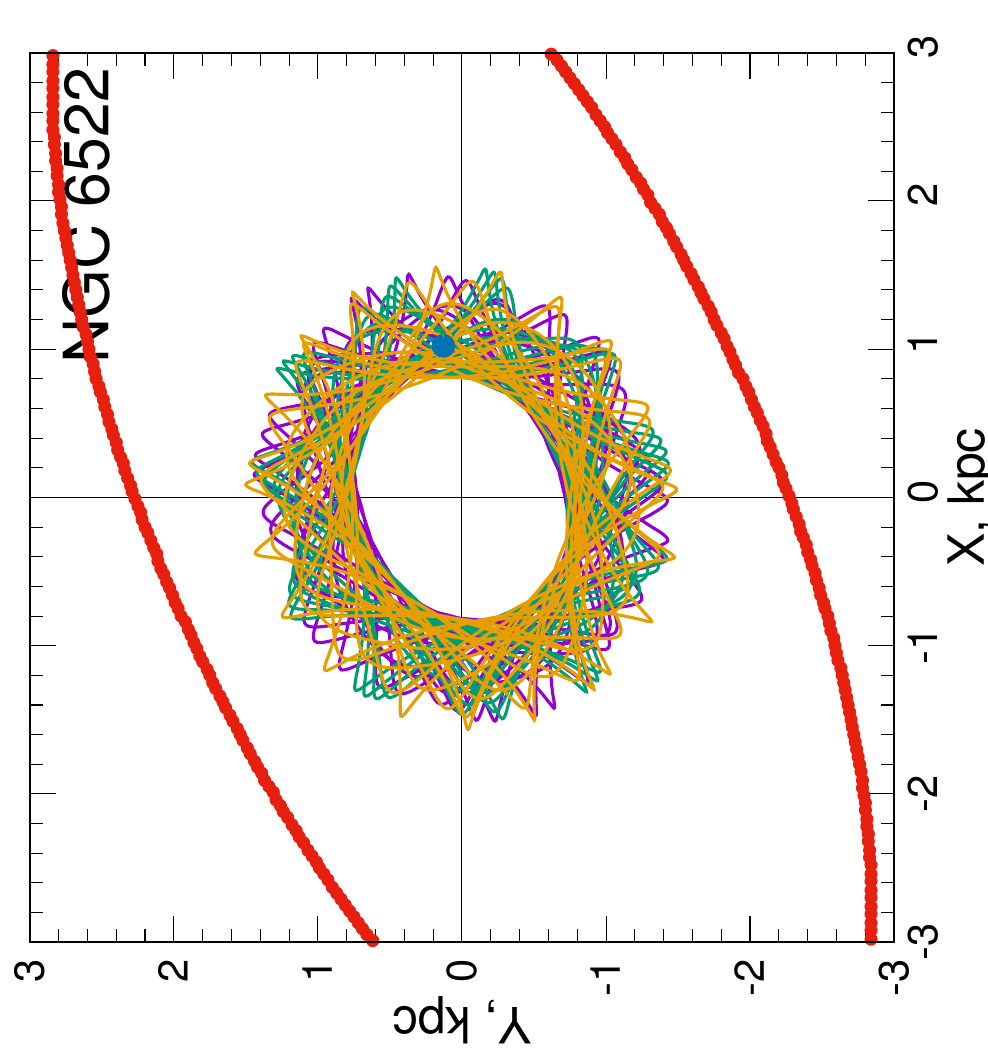}
   \includegraphics[width=0.2\textwidth,angle=-90]{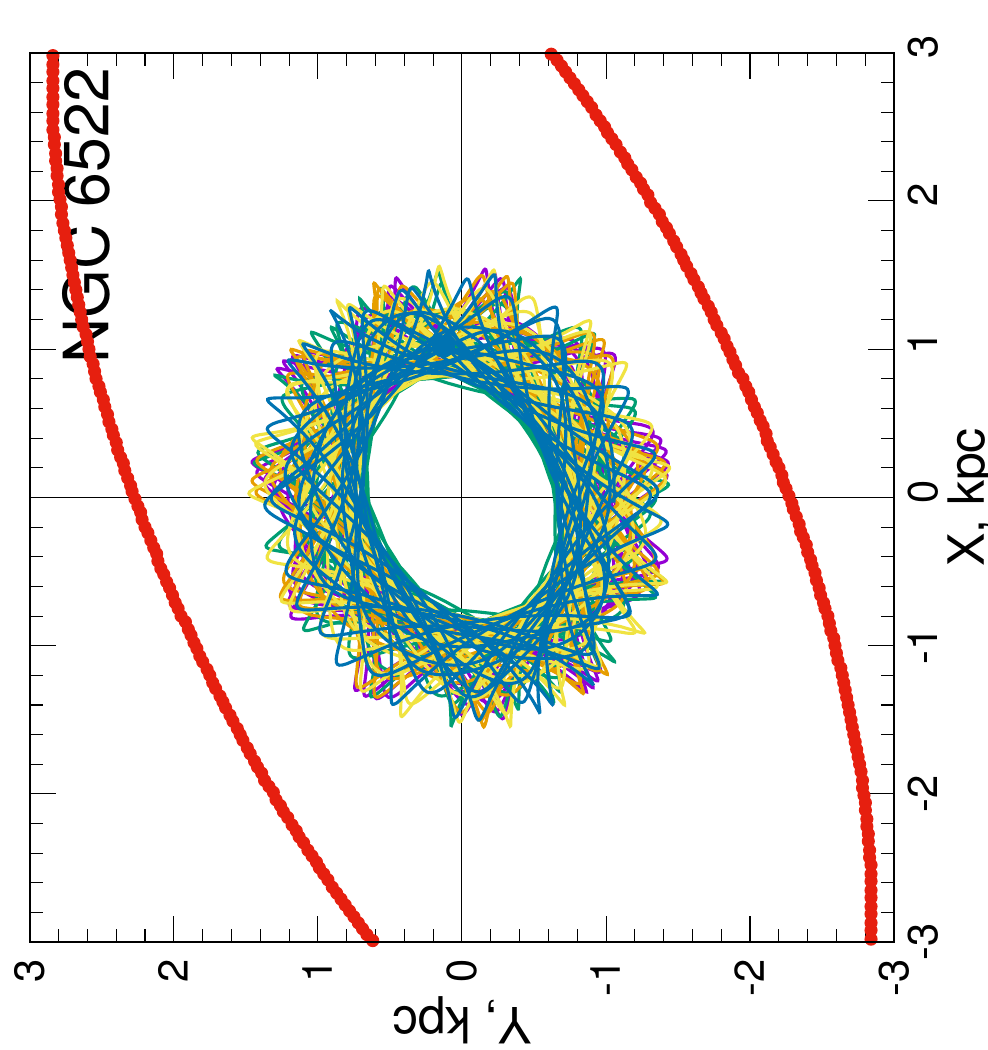}\

   \includegraphics[width=0.2\textwidth,angle=-90]{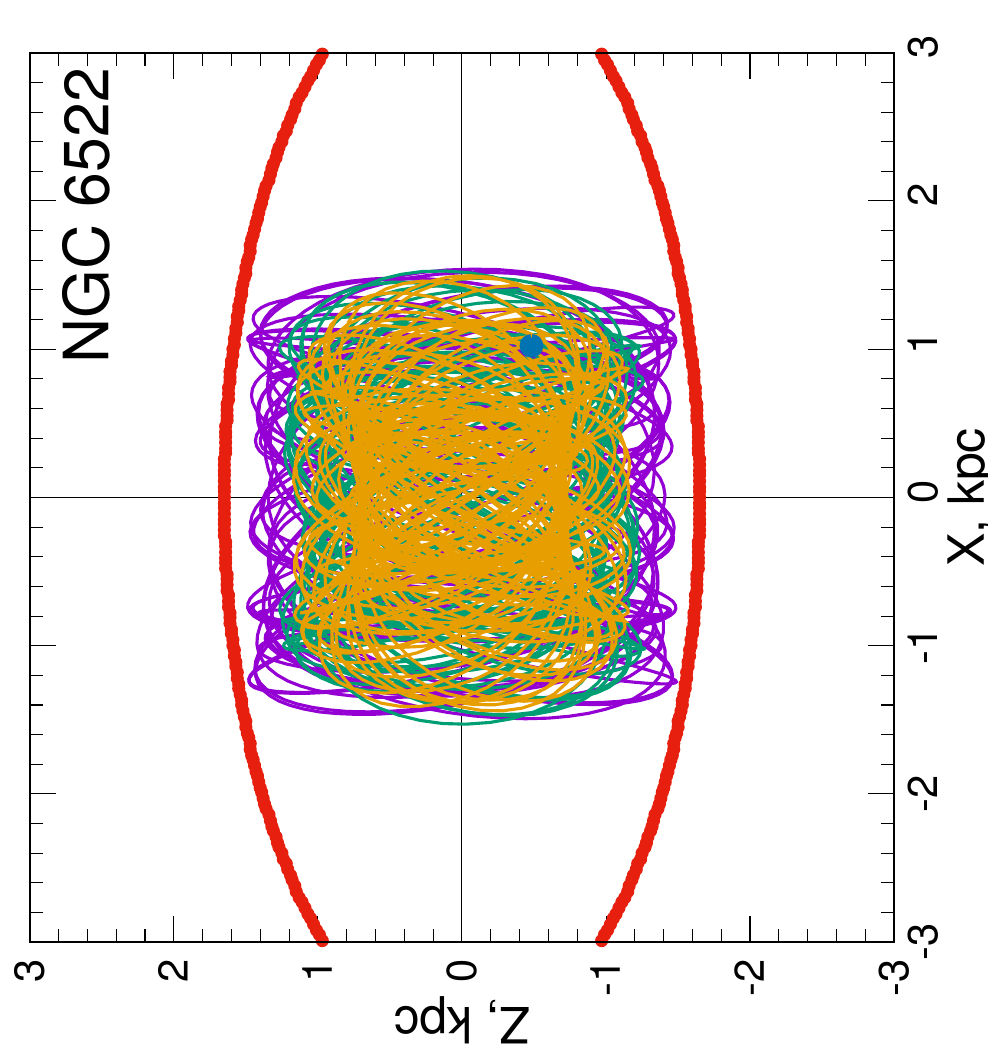}
   \includegraphics[width=0.2\textwidth,angle=-90]{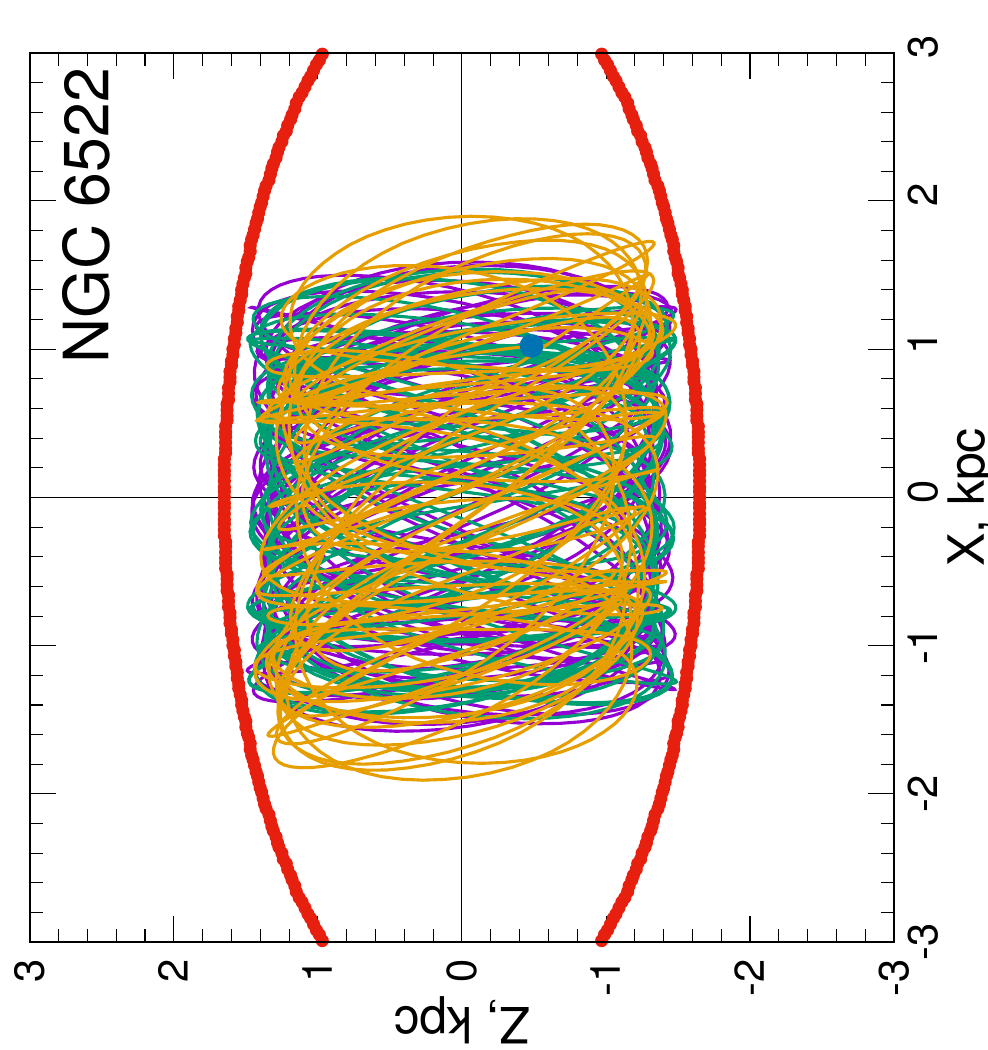}
   \includegraphics[width=0.2\textwidth,angle=-90]{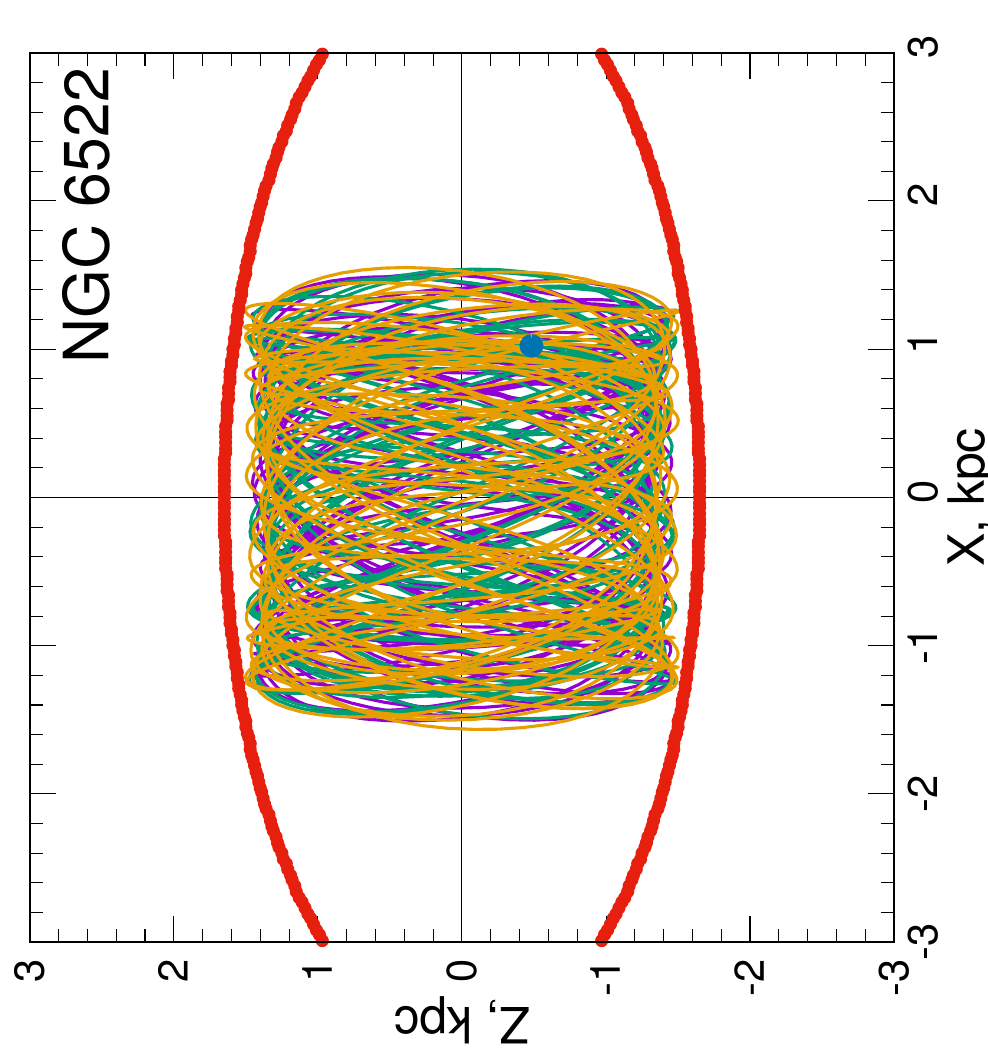}
   \includegraphics[width=0.2\textwidth,angle=-90]{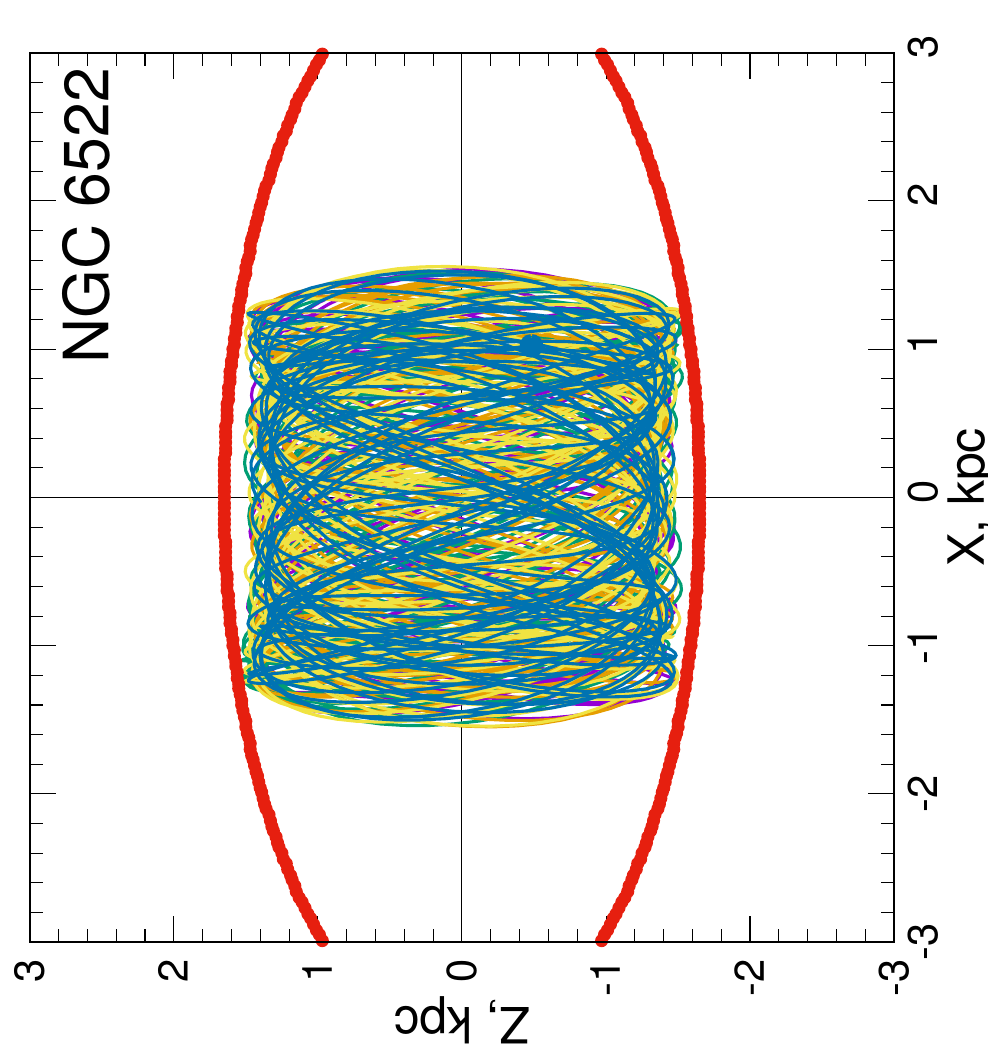}\

   \includegraphics[width=0.2\textwidth,angle=-90]{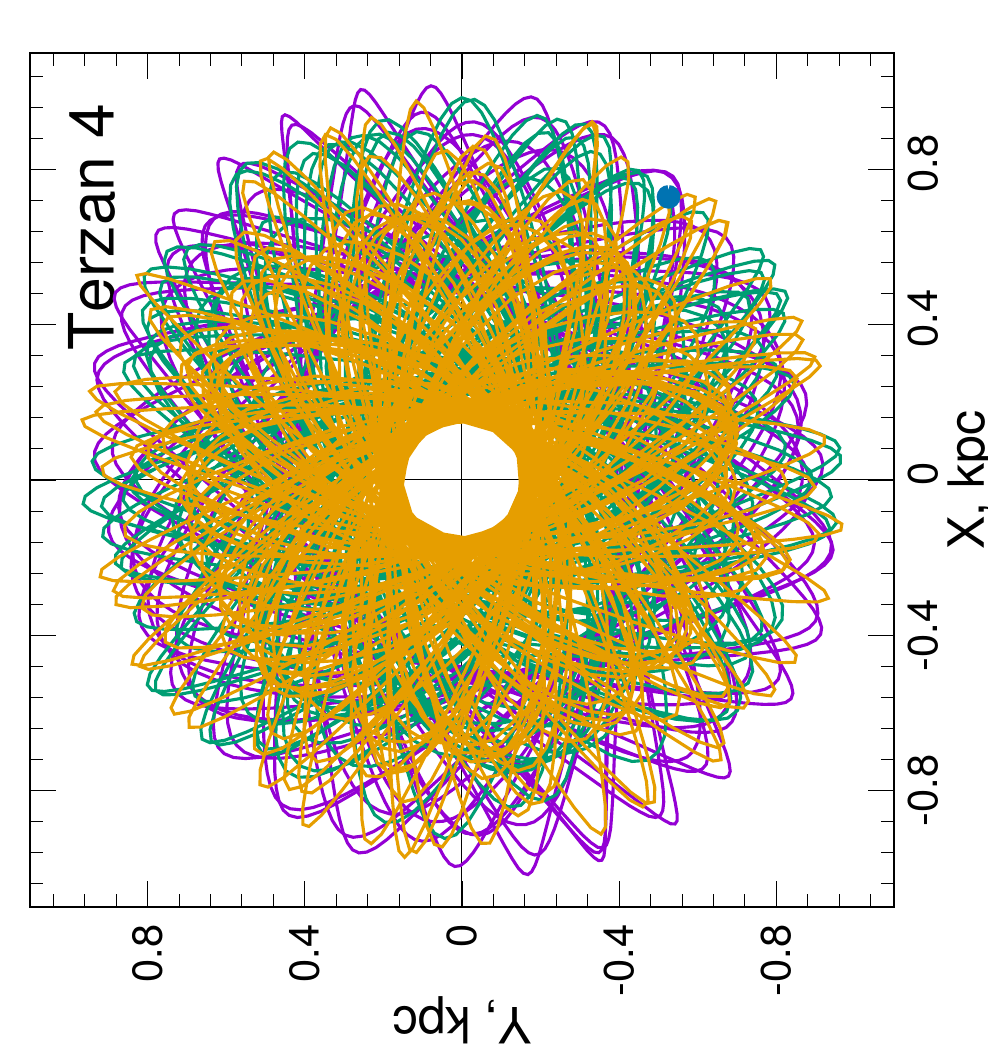}
   \includegraphics[width=0.2\textwidth,angle=-90]{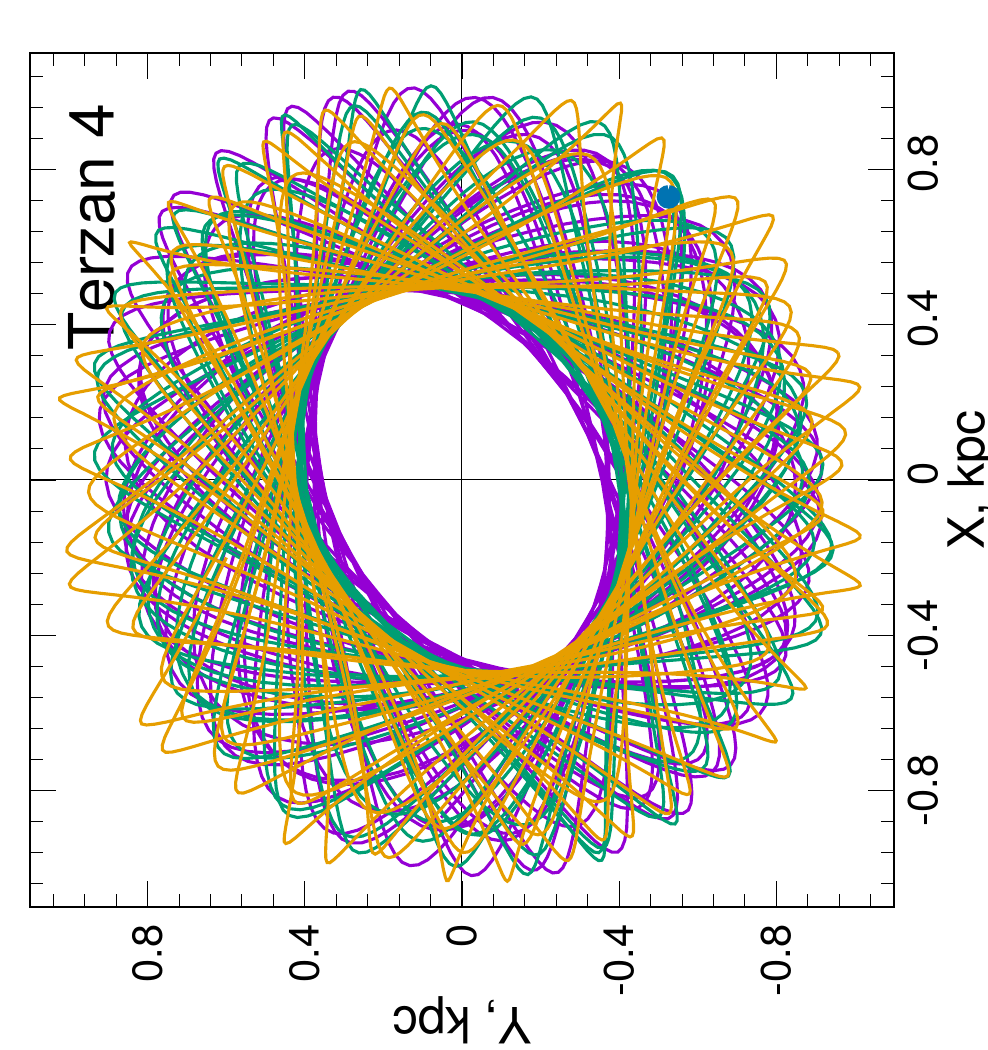}
   \includegraphics[width=0.2\textwidth,angle=-90]{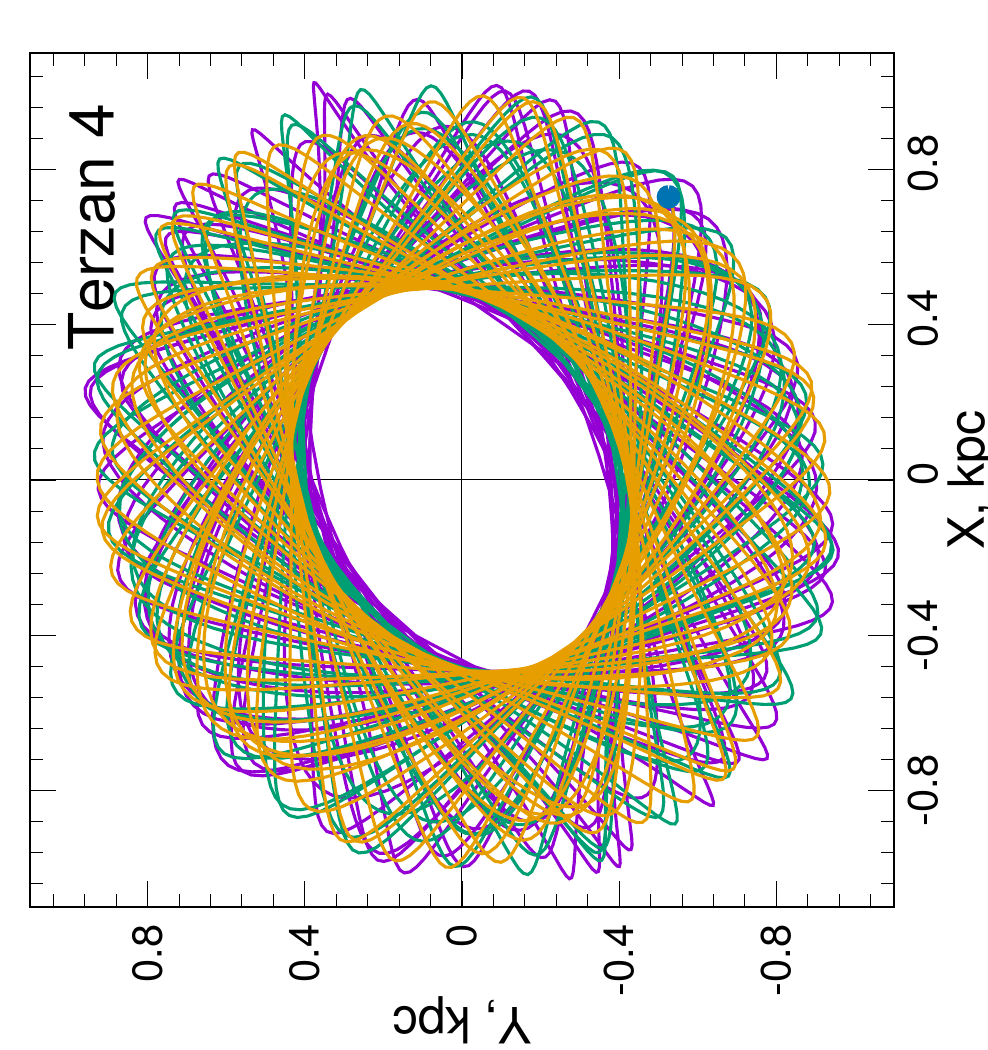}
   \includegraphics[width=0.2\textwidth,angle=-90]{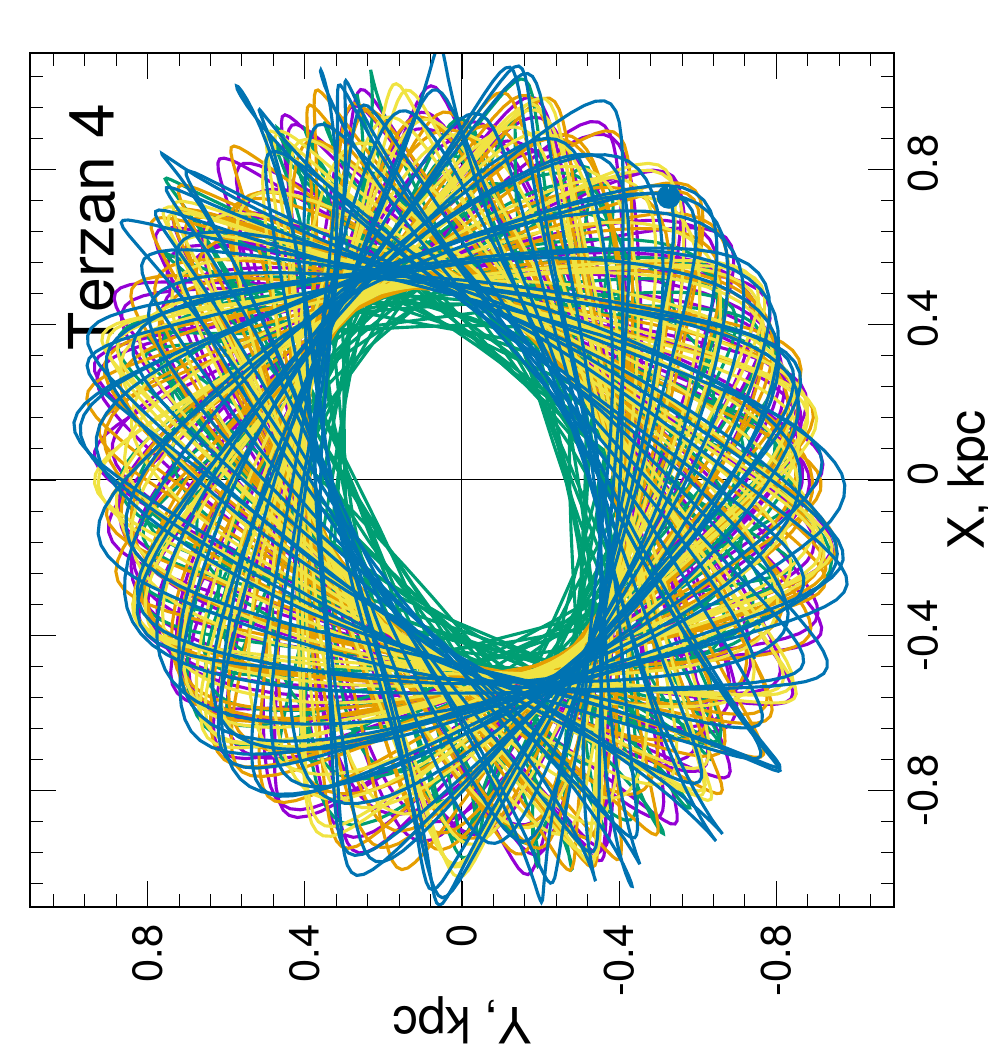}\

   \includegraphics[width=0.2\textwidth,angle=-90]{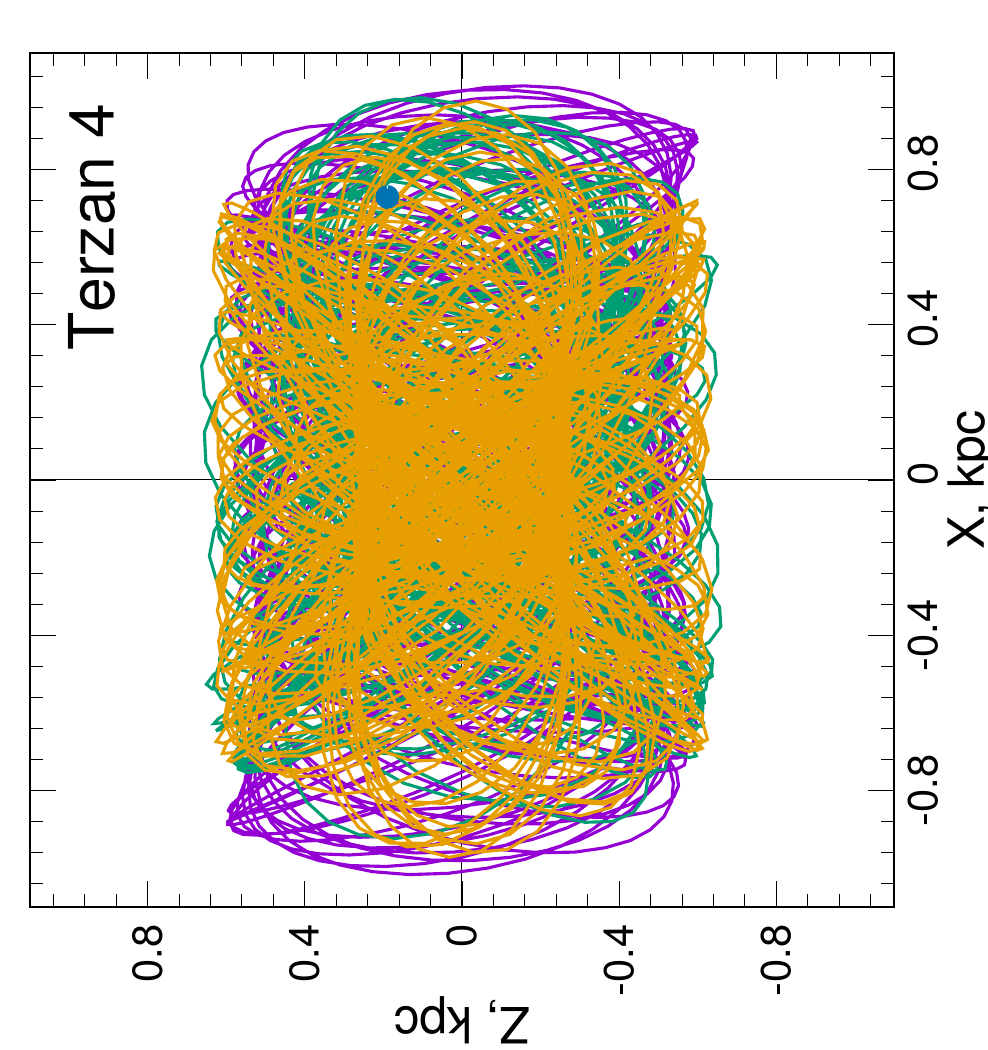}
   \includegraphics[width=0.2\textwidth,angle=-90]{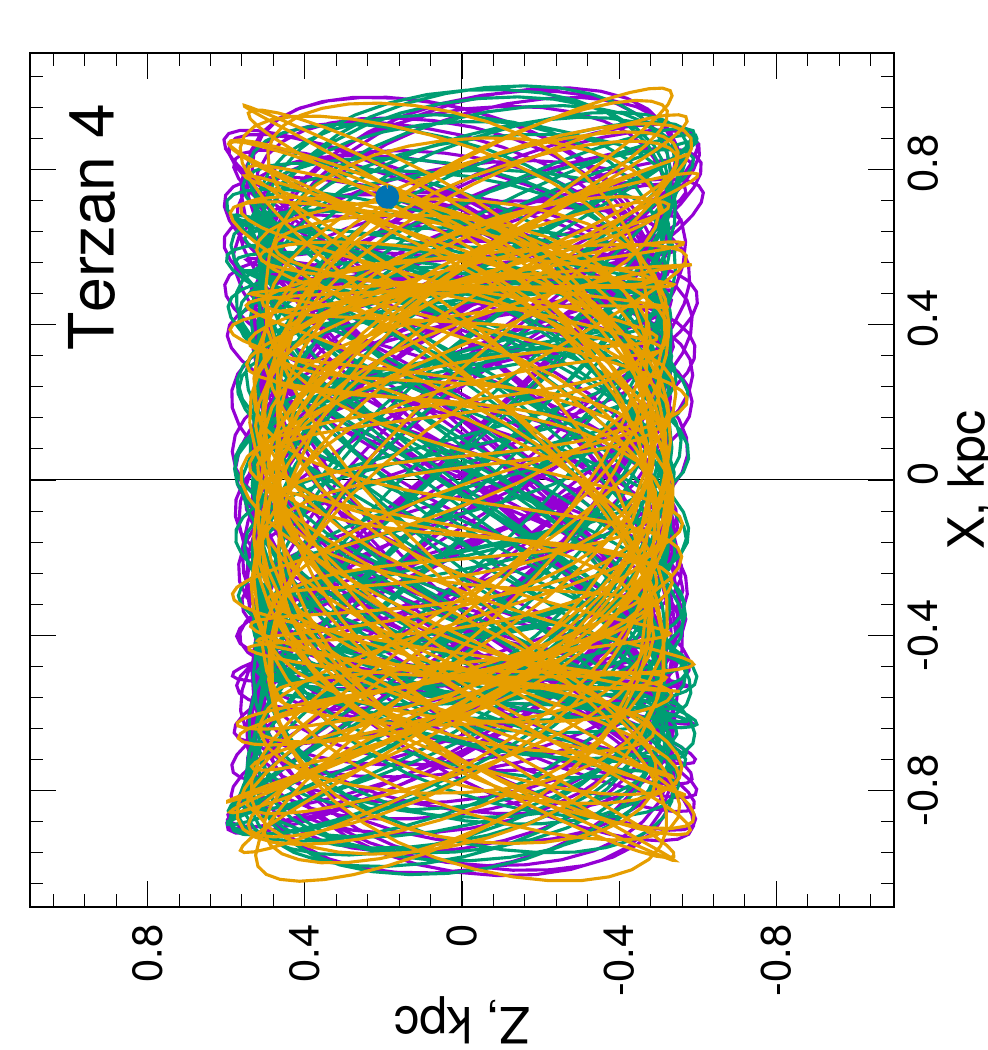}
   \includegraphics[width=0.2\textwidth,angle=-90]{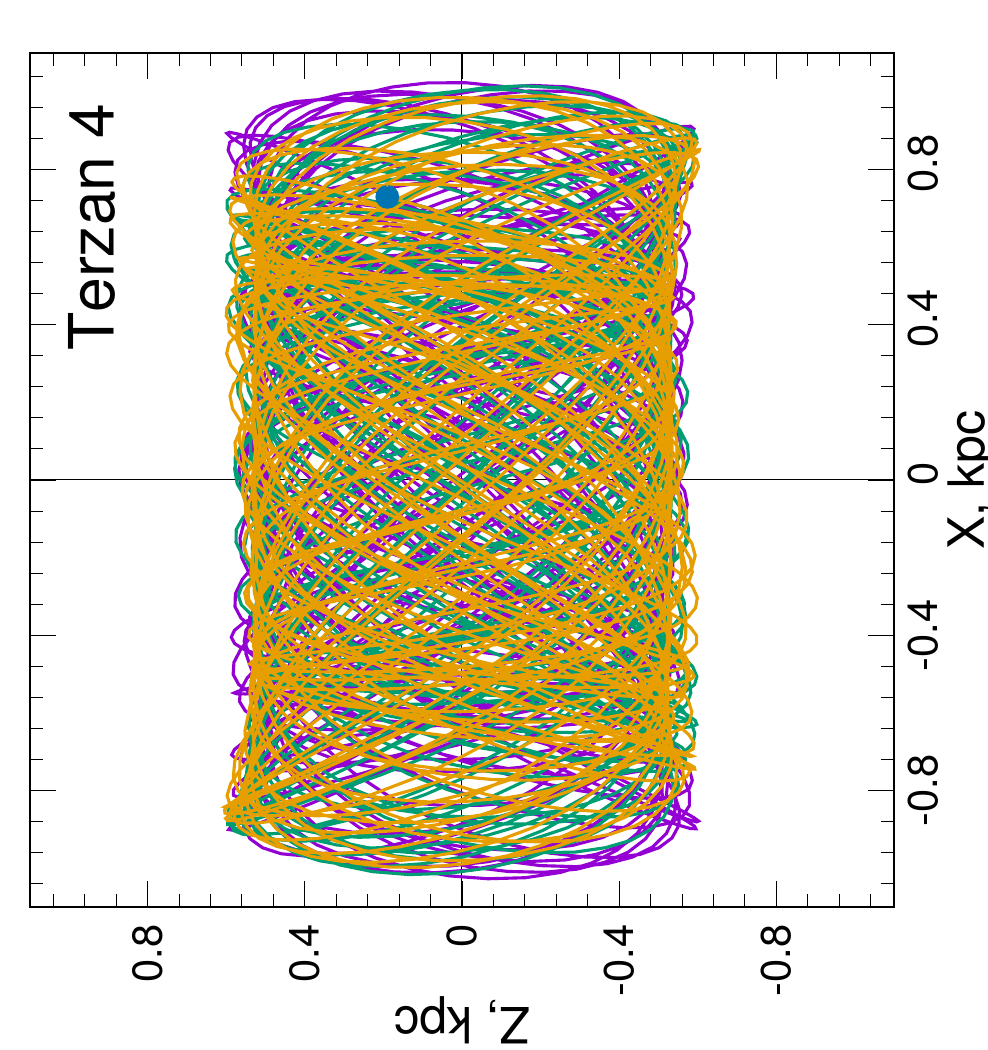}
   \includegraphics[width=0.2\textwidth,angle=-90]{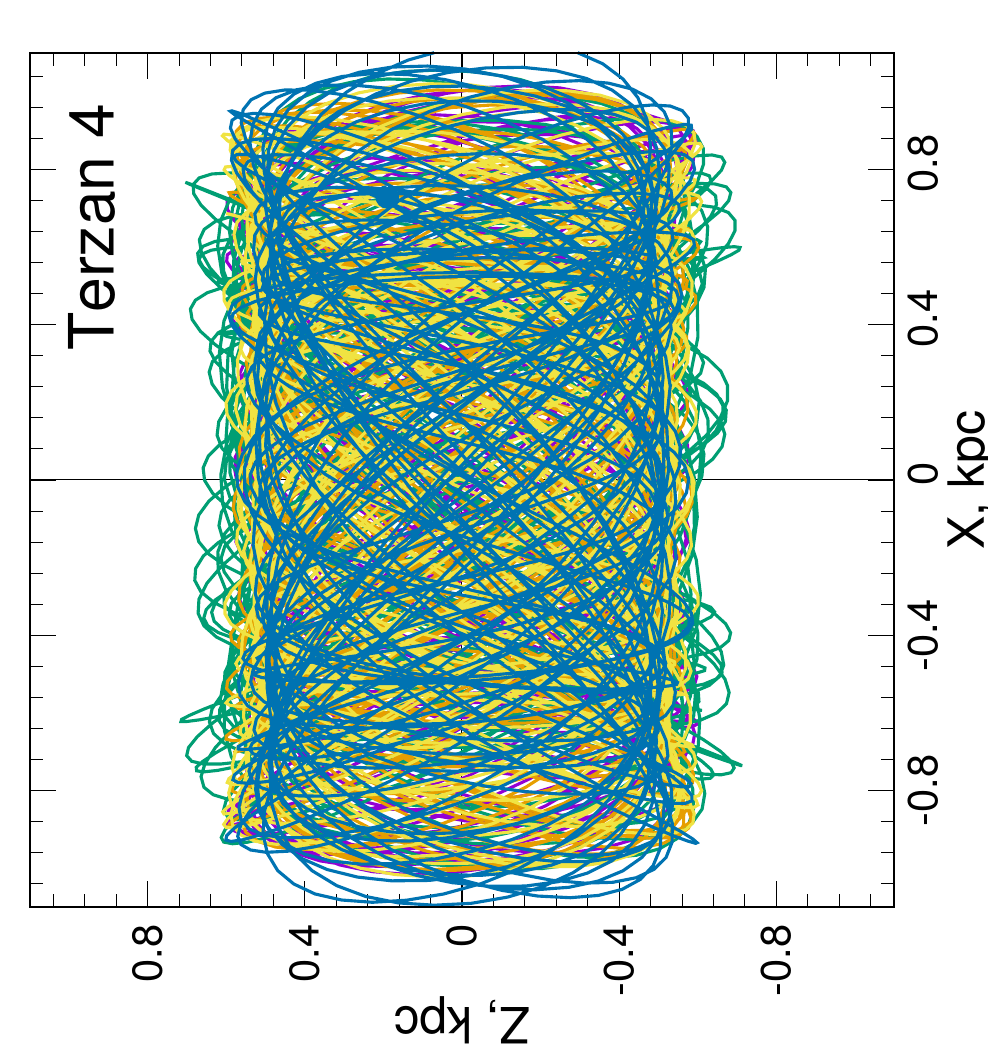}\
        \caption{\small Orbits (in the rotating bar system) of the bulge/bar GCs NGC 6266, NGC 6522, Terzan 4 in projections onto the $X-Y$ and $X-Z$ galactic planes (upper and lower rows for each GC, respectively). The red lines show the sections of the bar. The blue circle indicates the beginning of the orbit. Deciphering the colors of the orbits is given in the text of the article.}
\label{fD}
\end{center}}
\end{figure*}

\begin{figure*}
{\begin{center}

  \includegraphics[width=0.2\textwidth,angle=-90]{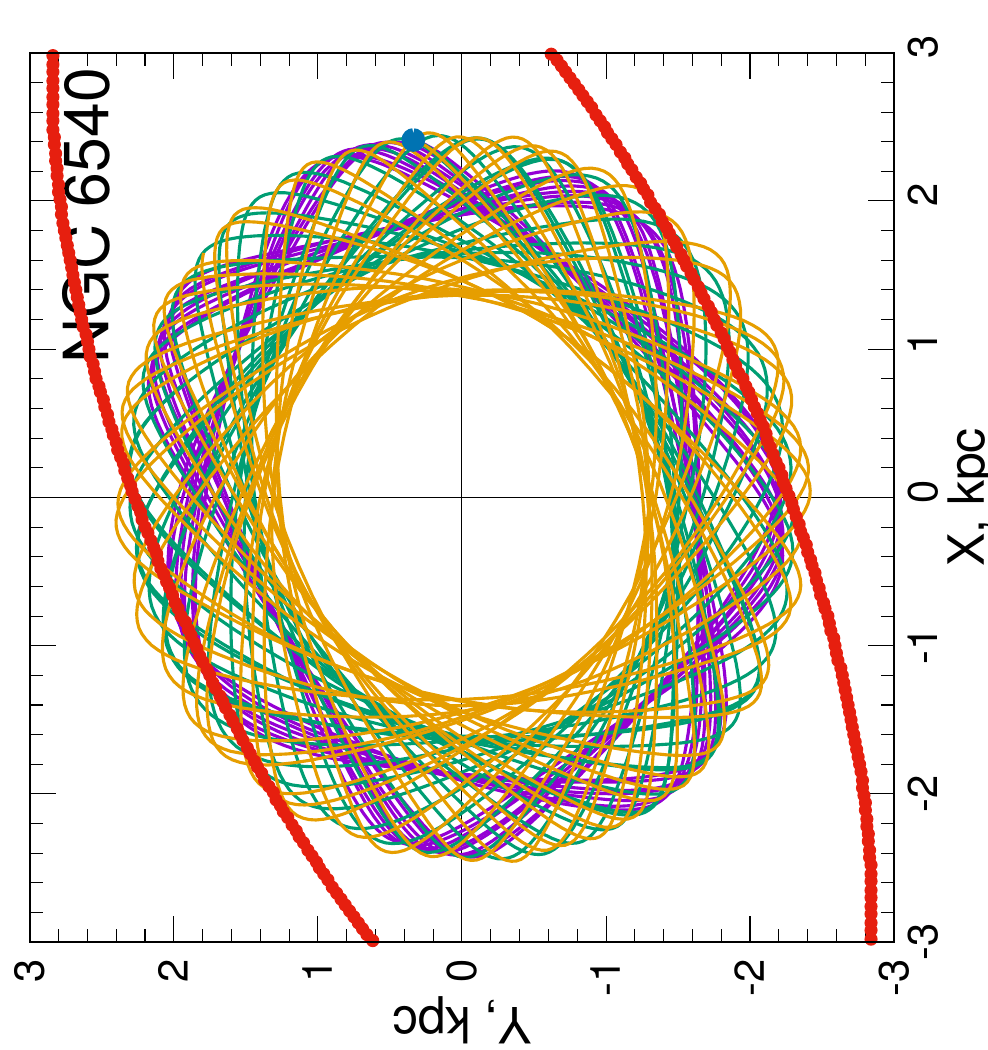}
   \includegraphics[width=0.2\textwidth,angle=-90]{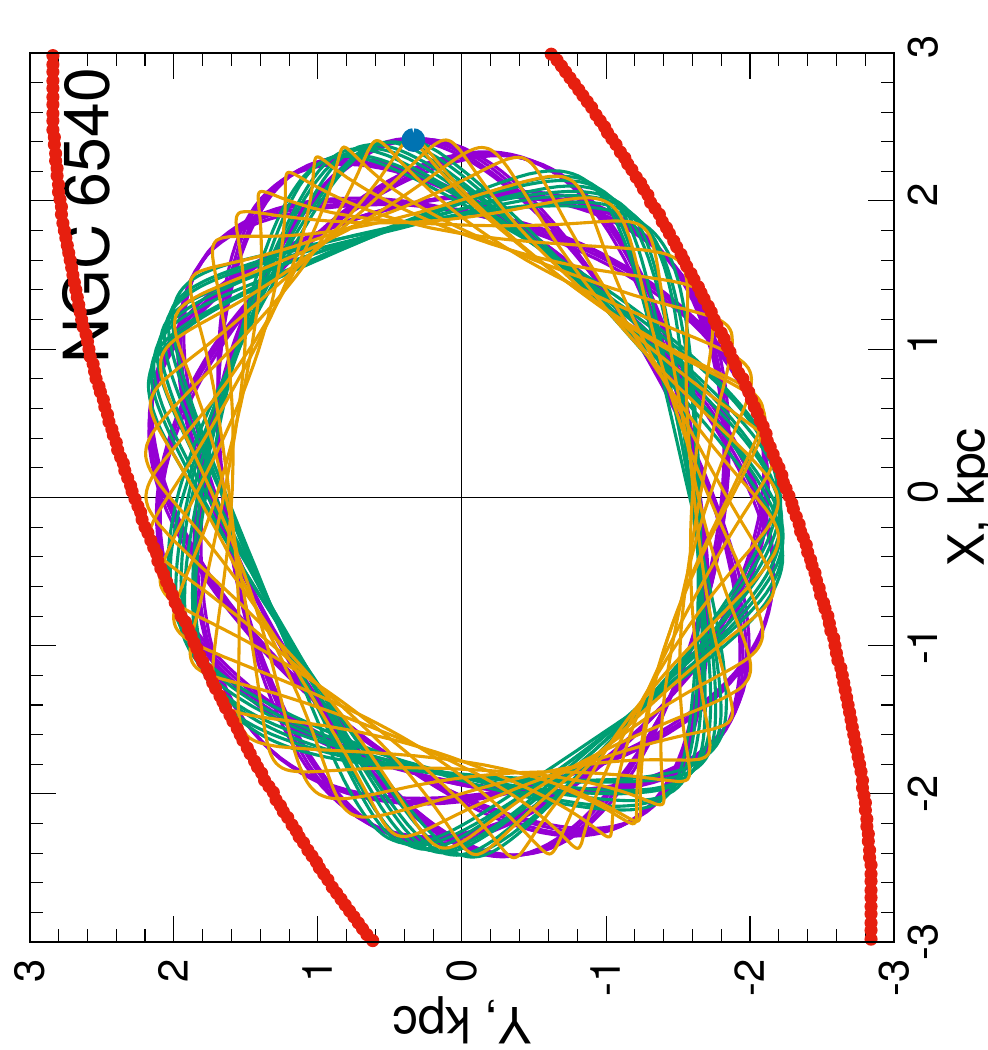}
   \includegraphics[width=0.2\textwidth,angle=-90]{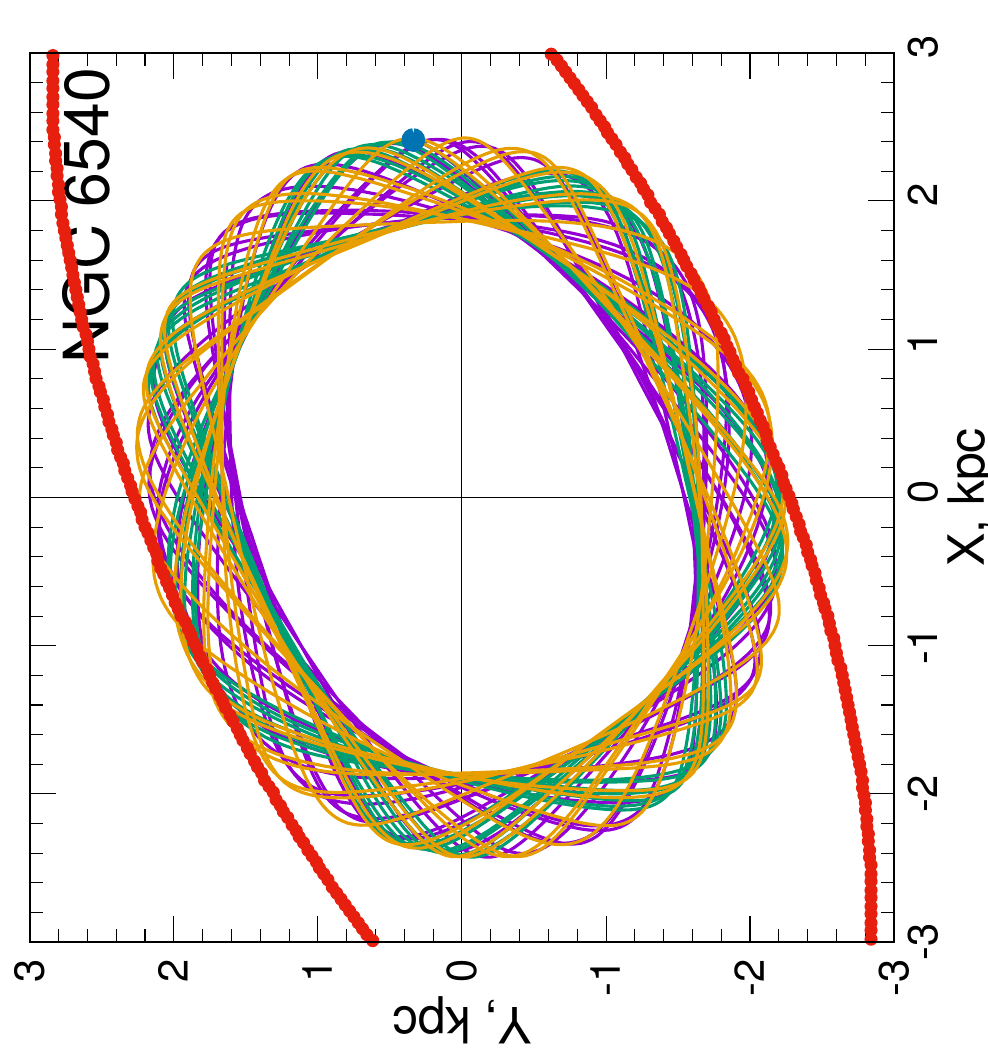}
   \includegraphics[width=0.2\textwidth,angle=-90]{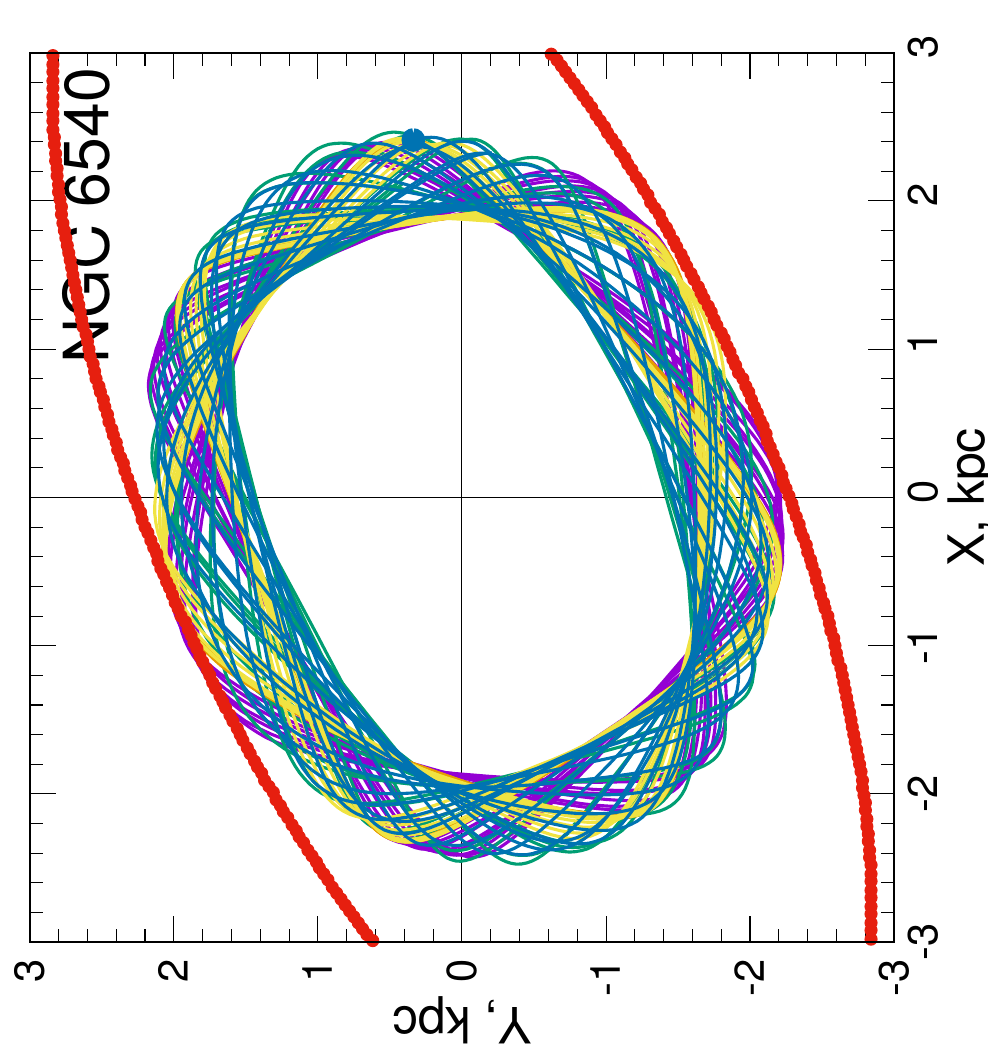}\

   \includegraphics[width=0.2\textwidth,angle=-90]{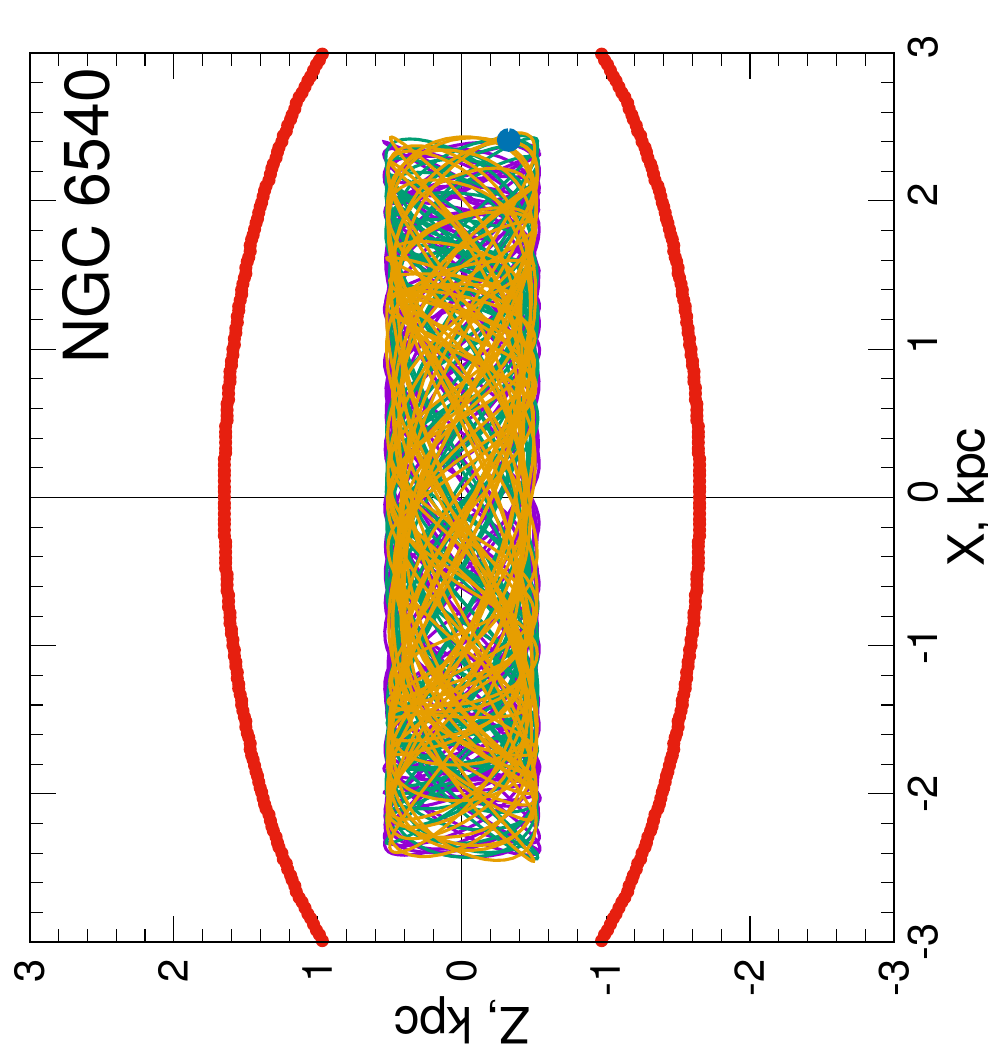}
   \includegraphics[width=0.2\textwidth,angle=-90]{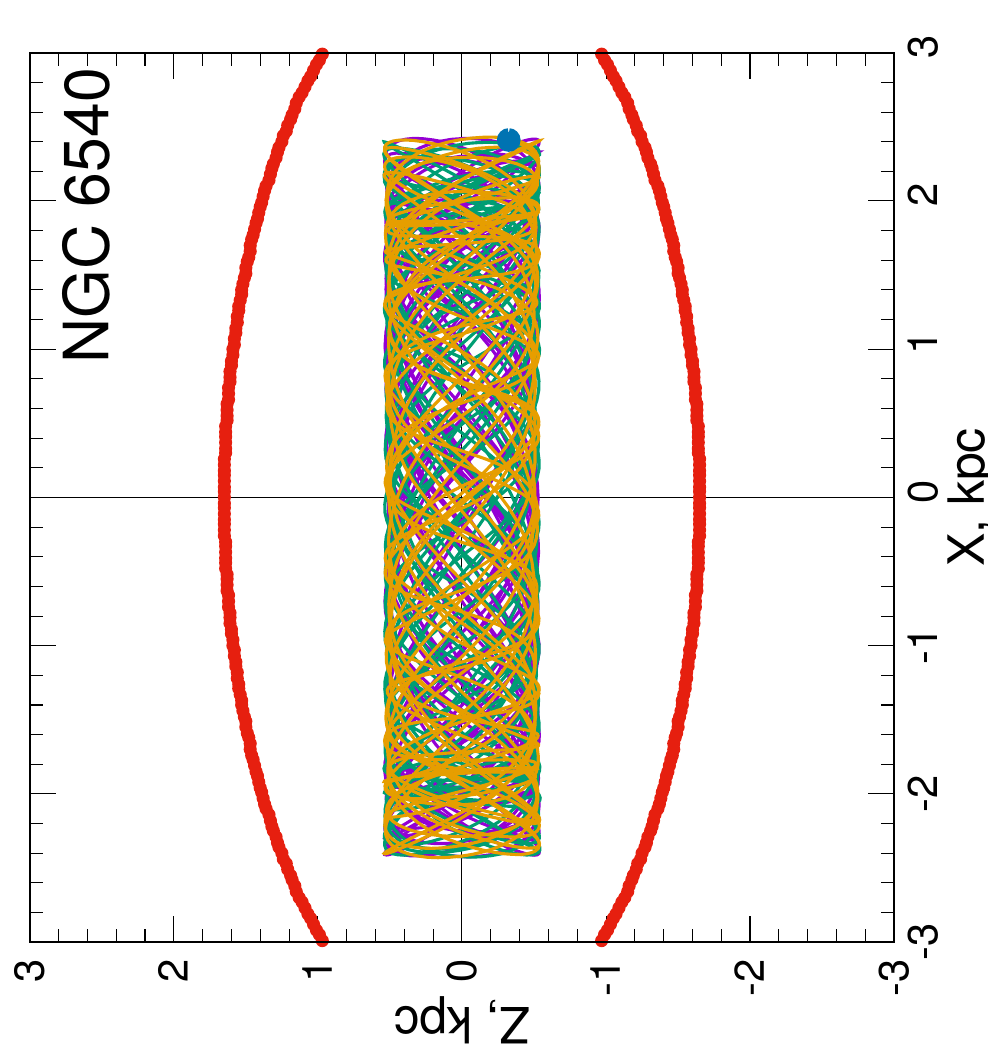}
   \includegraphics[width=0.2\textwidth,angle=-90]{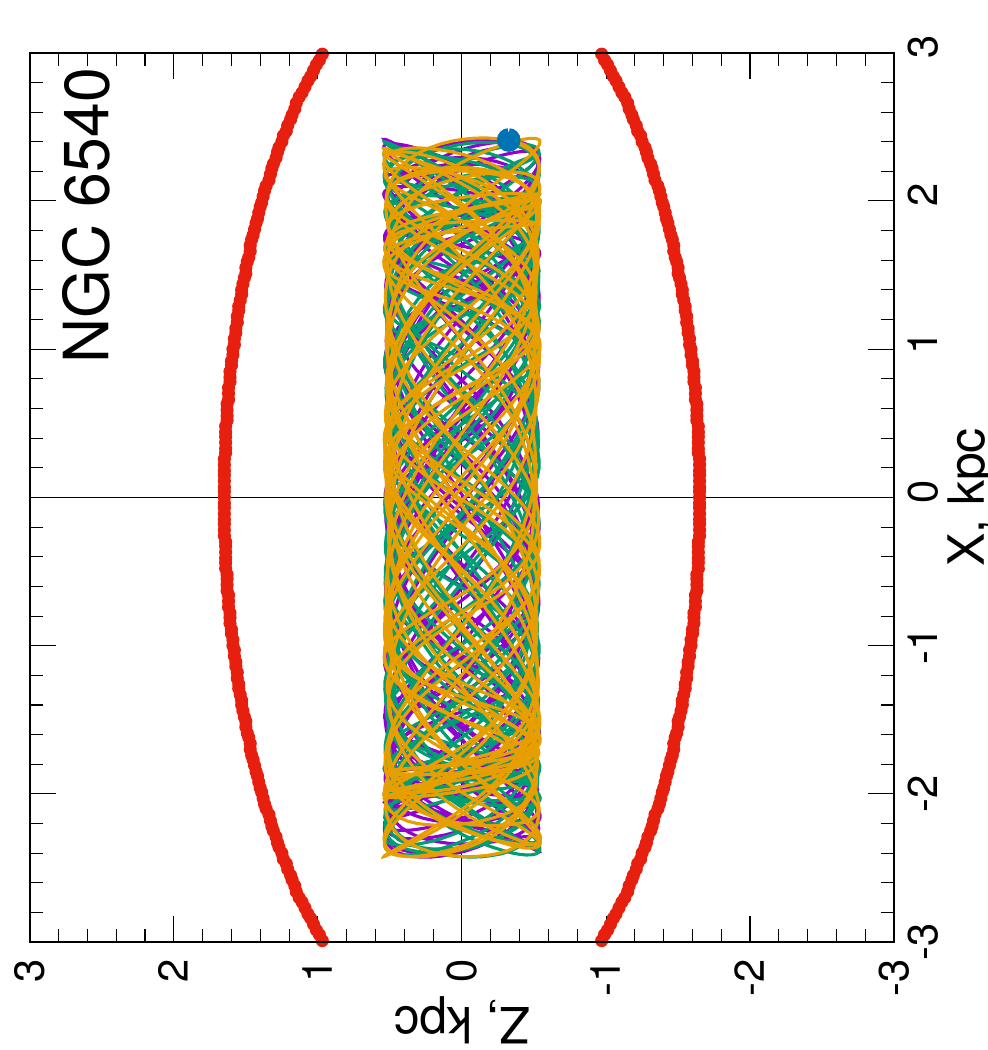}
   \includegraphics[width=0.2\textwidth,angle=-90]{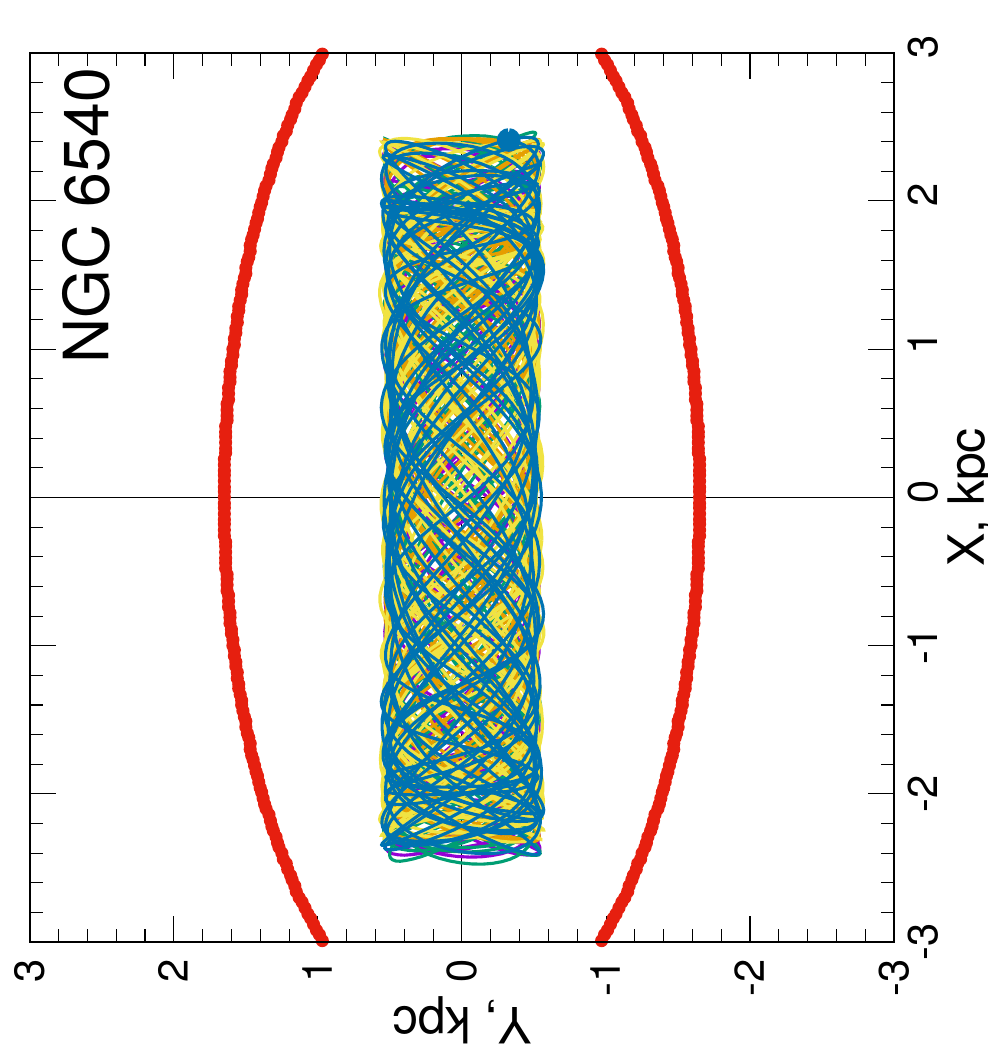}\

   \includegraphics[width=0.2\textwidth,angle=-90]{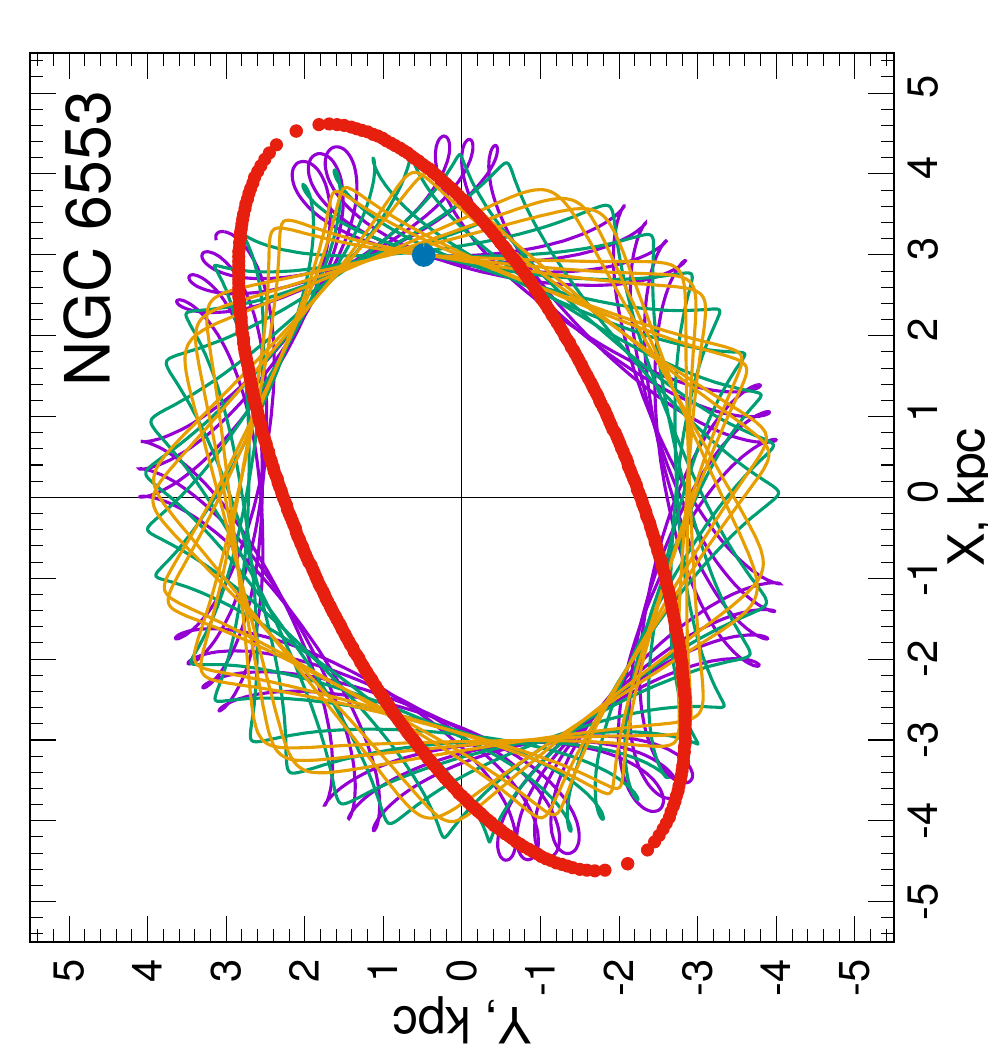}
   \includegraphics[width=0.2\textwidth,angle=-90]{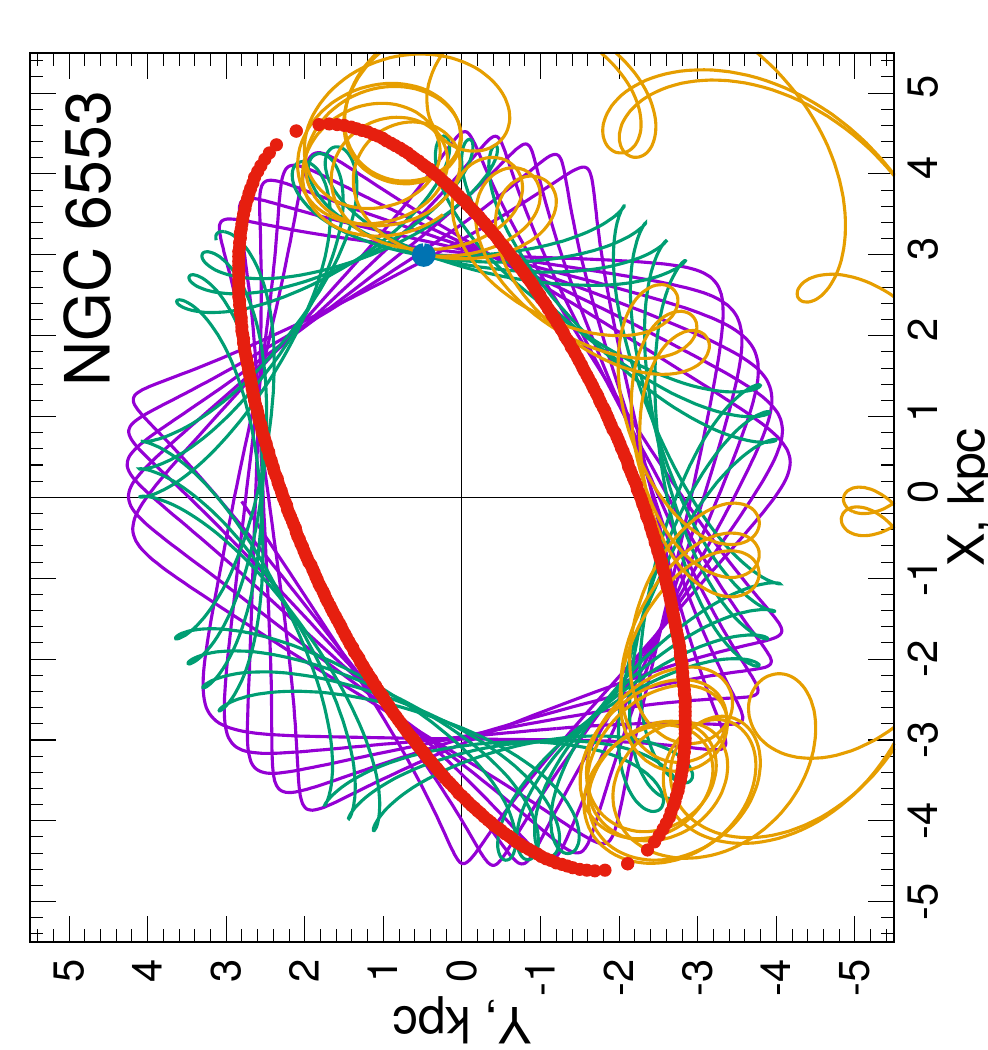}
   \includegraphics[width=0.2\textwidth,angle=-90]{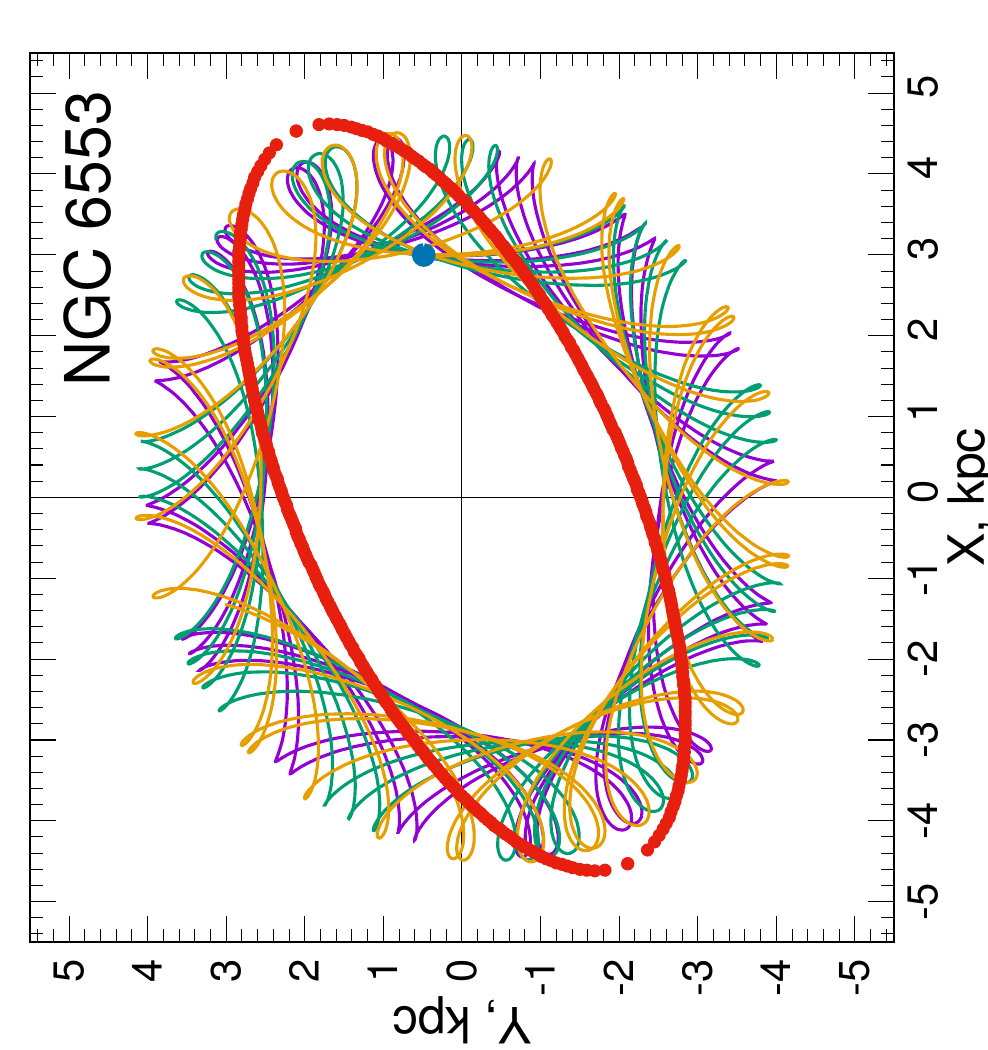}
   \includegraphics[width=0.2\textwidth,angle=-90]{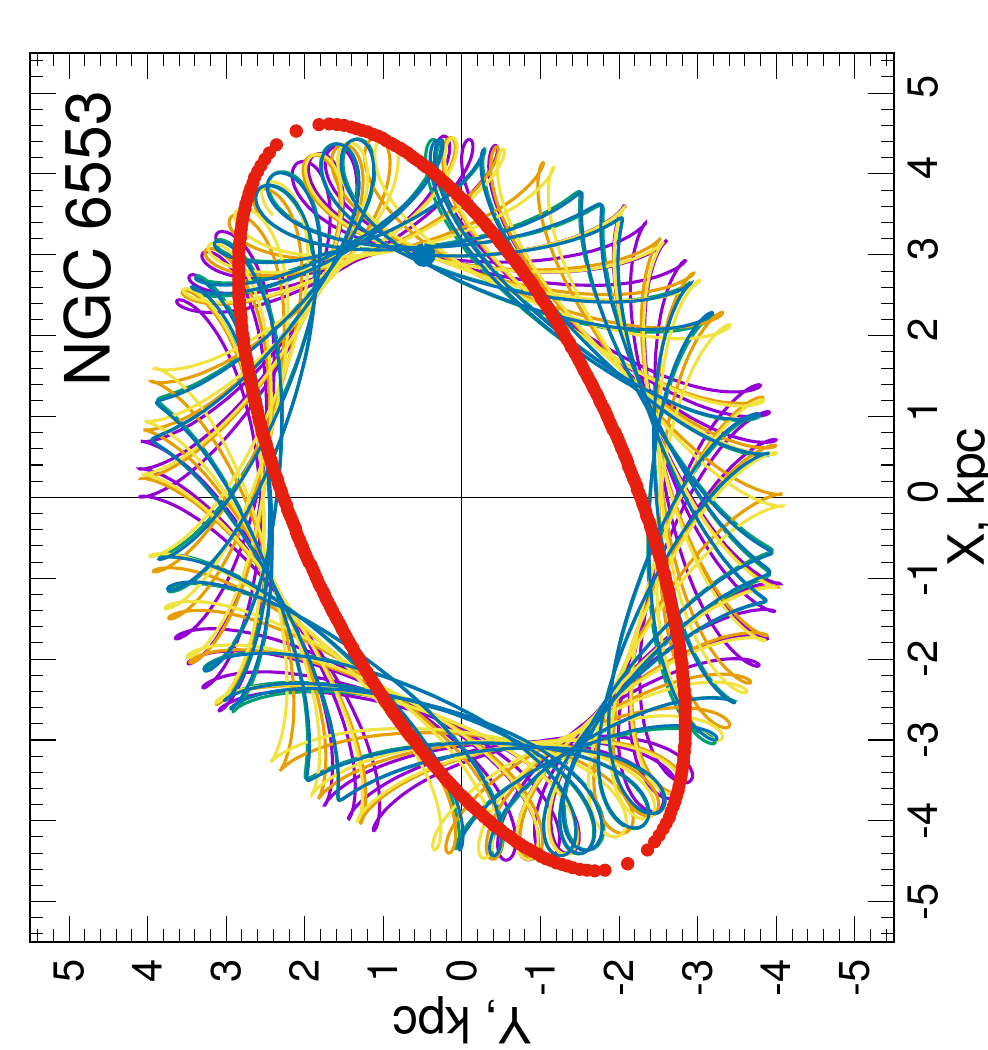}\

   \includegraphics[width=0.2\textwidth,angle=-90]{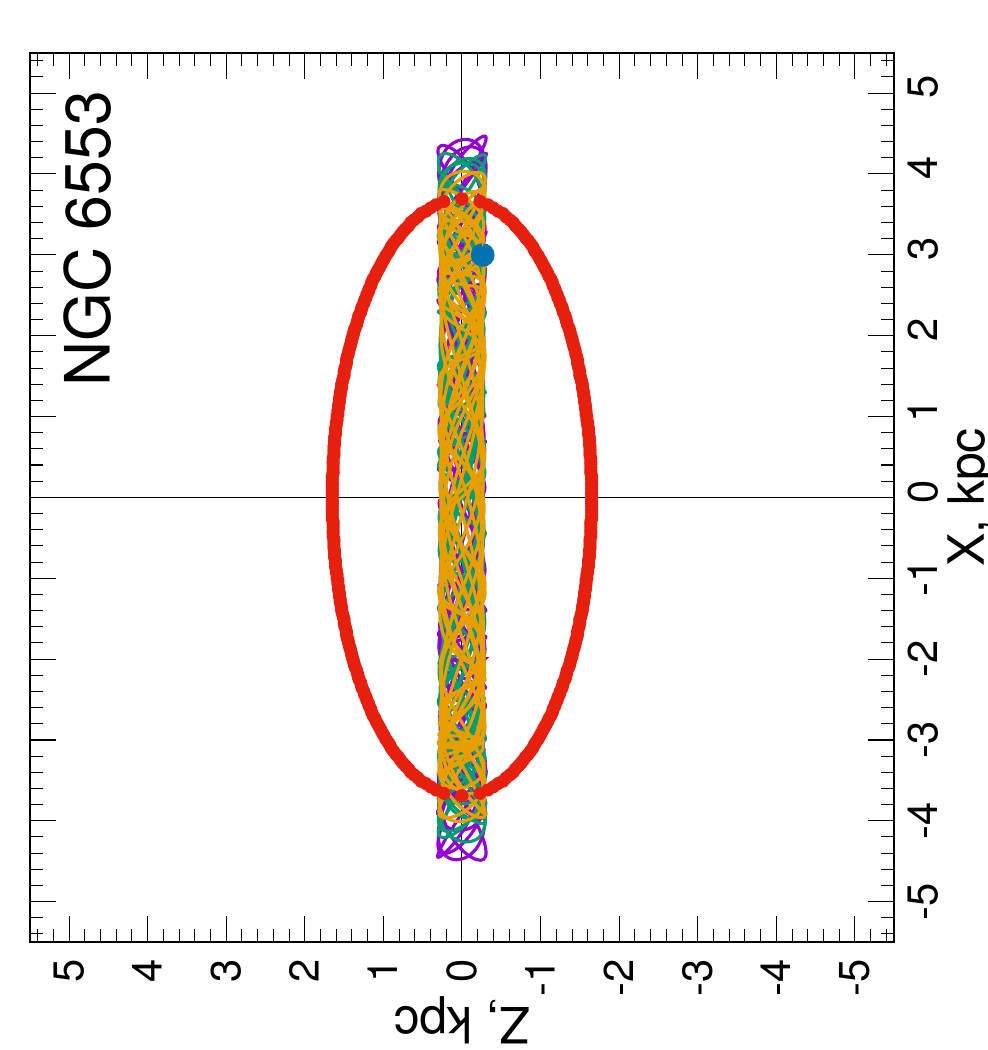}
   \includegraphics[width=0.2\textwidth,angle=-90]{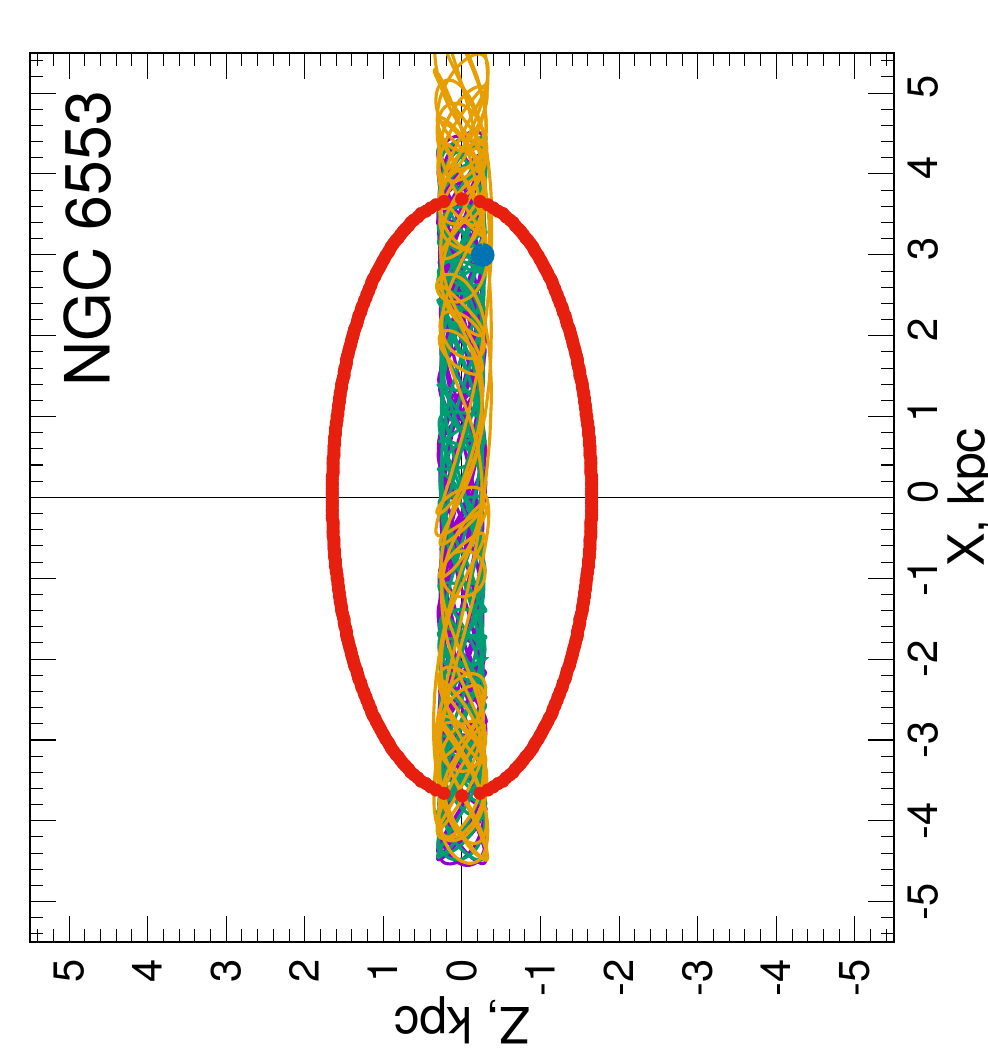}
   \includegraphics[width=0.2\textwidth,angle=-90]{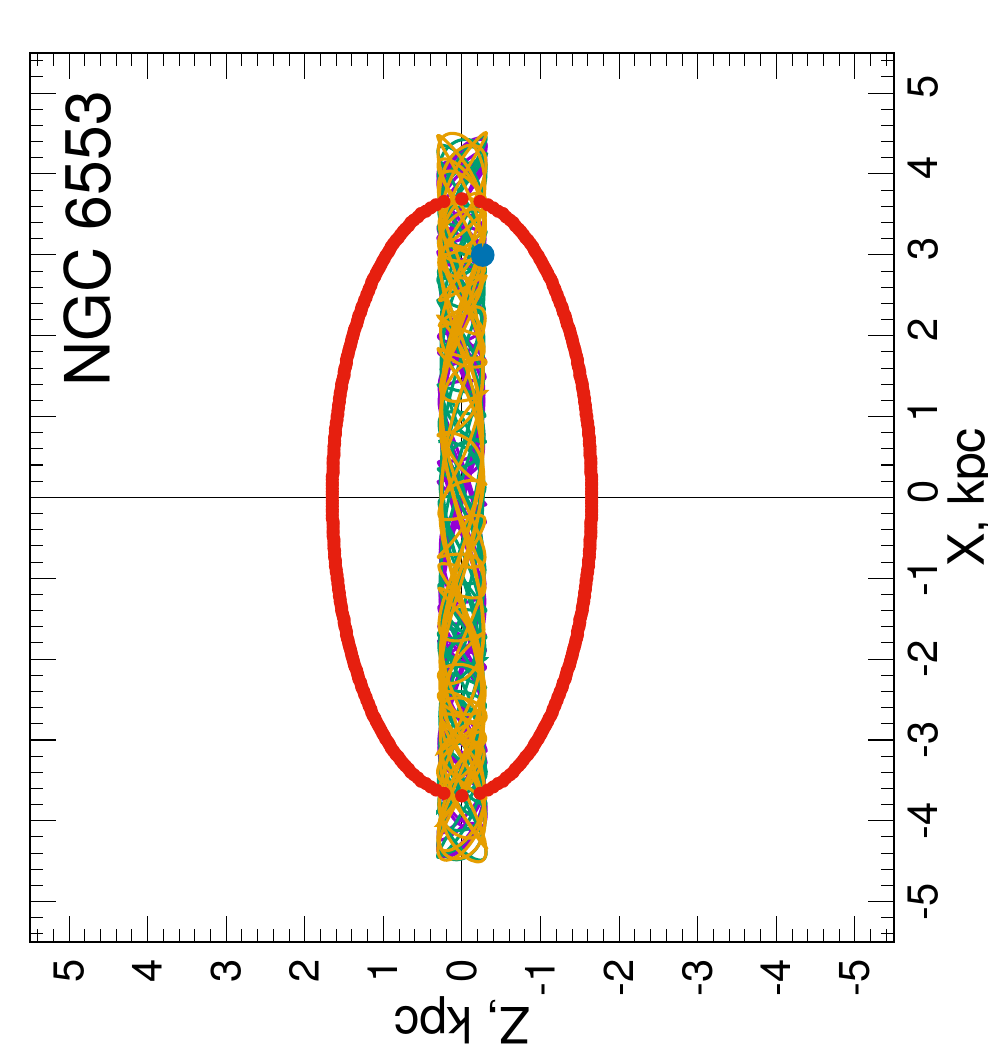}
   \includegraphics[width=0.2\textwidth,angle=-90]{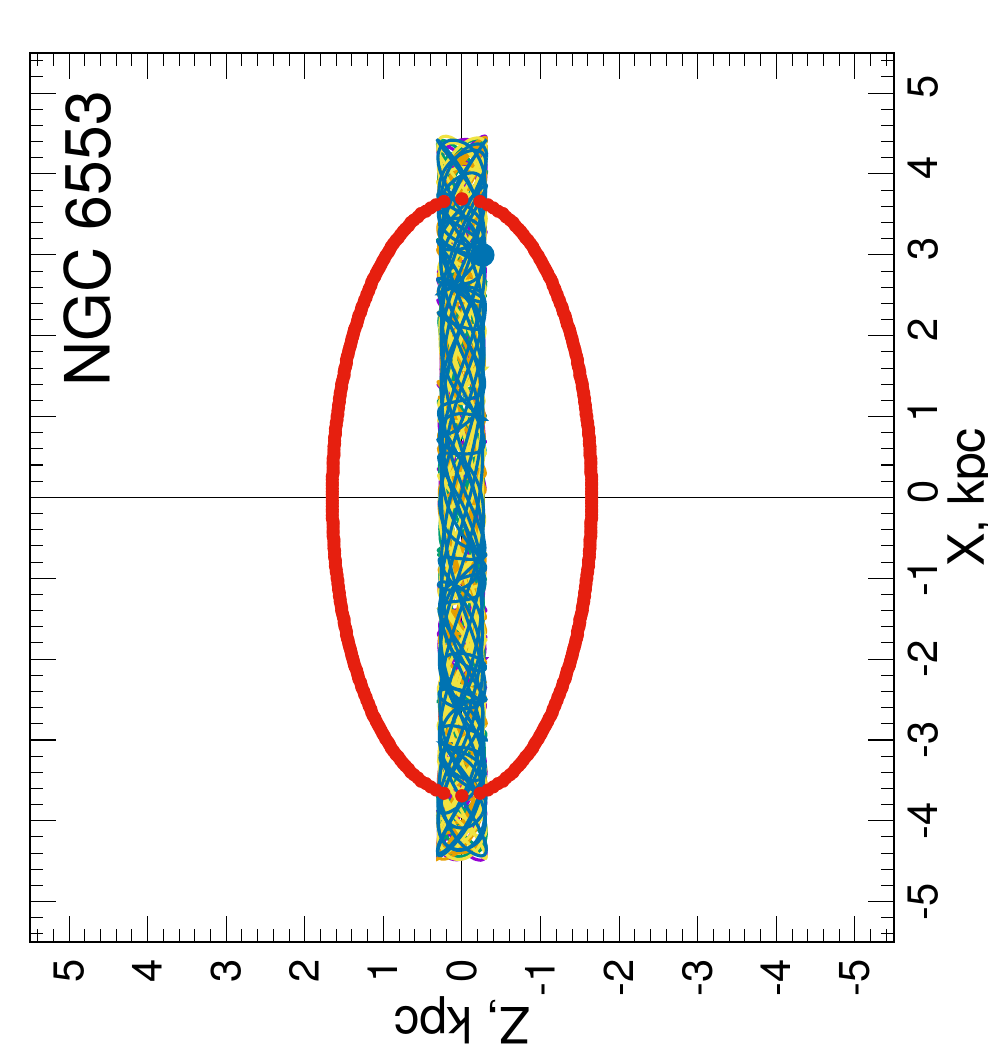}\

   \includegraphics[width=0.2\textwidth,angle=-90]{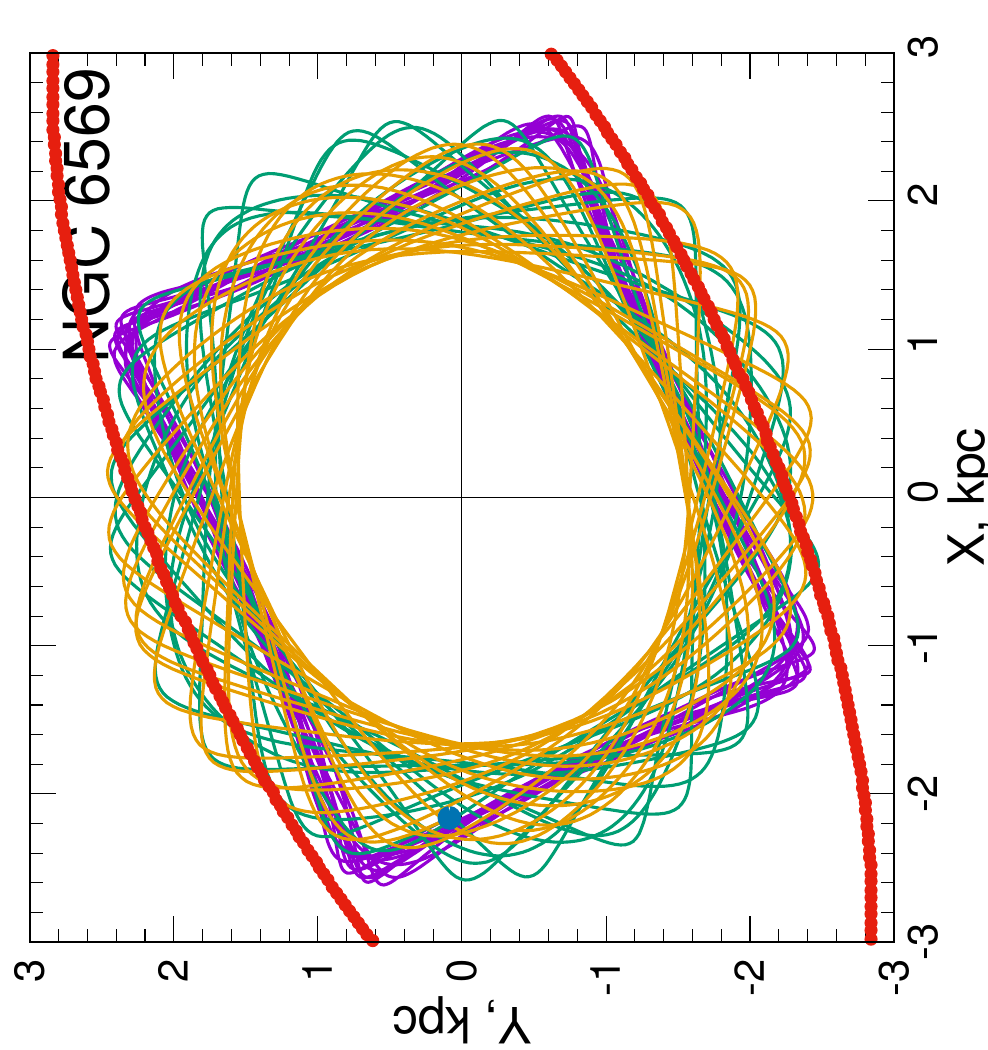}
   \includegraphics[width=0.2\textwidth,angle=-90]{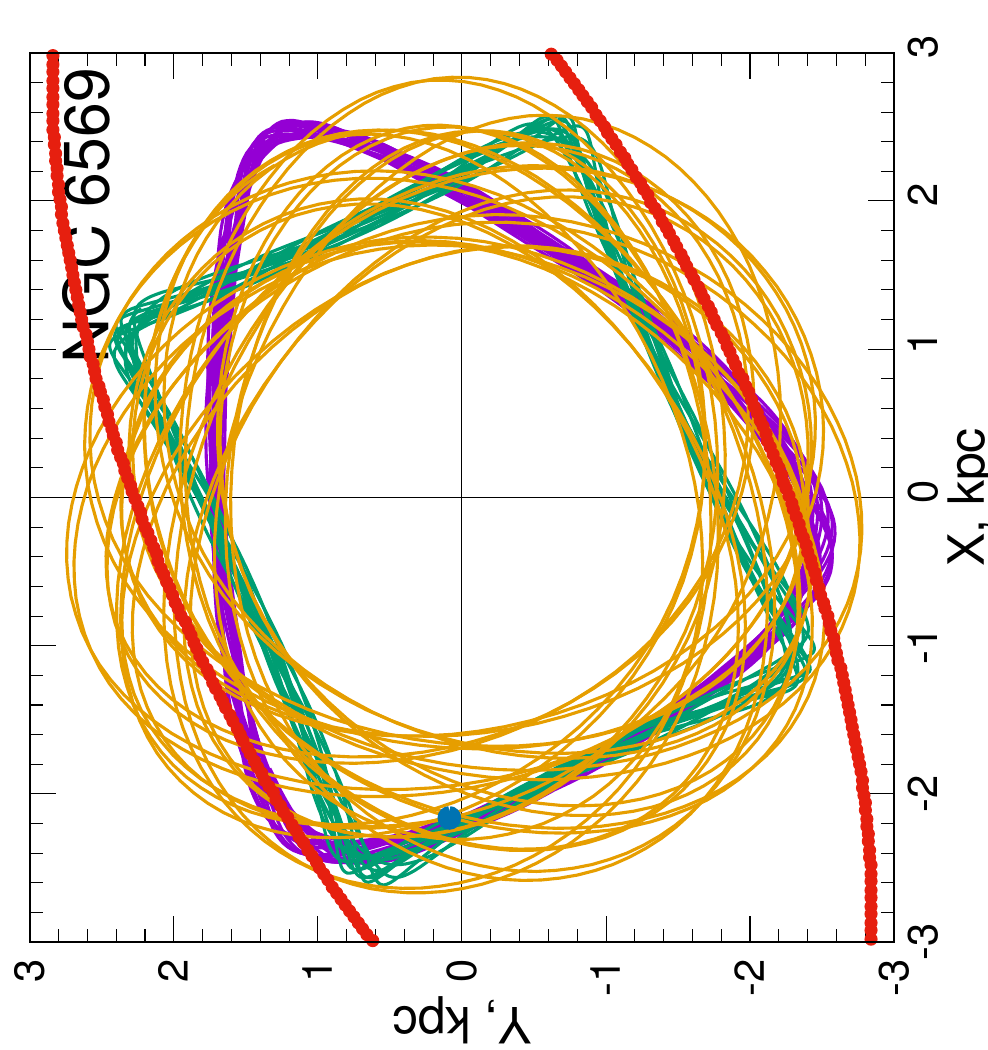}
   \includegraphics[width=0.2\textwidth,angle=-90]{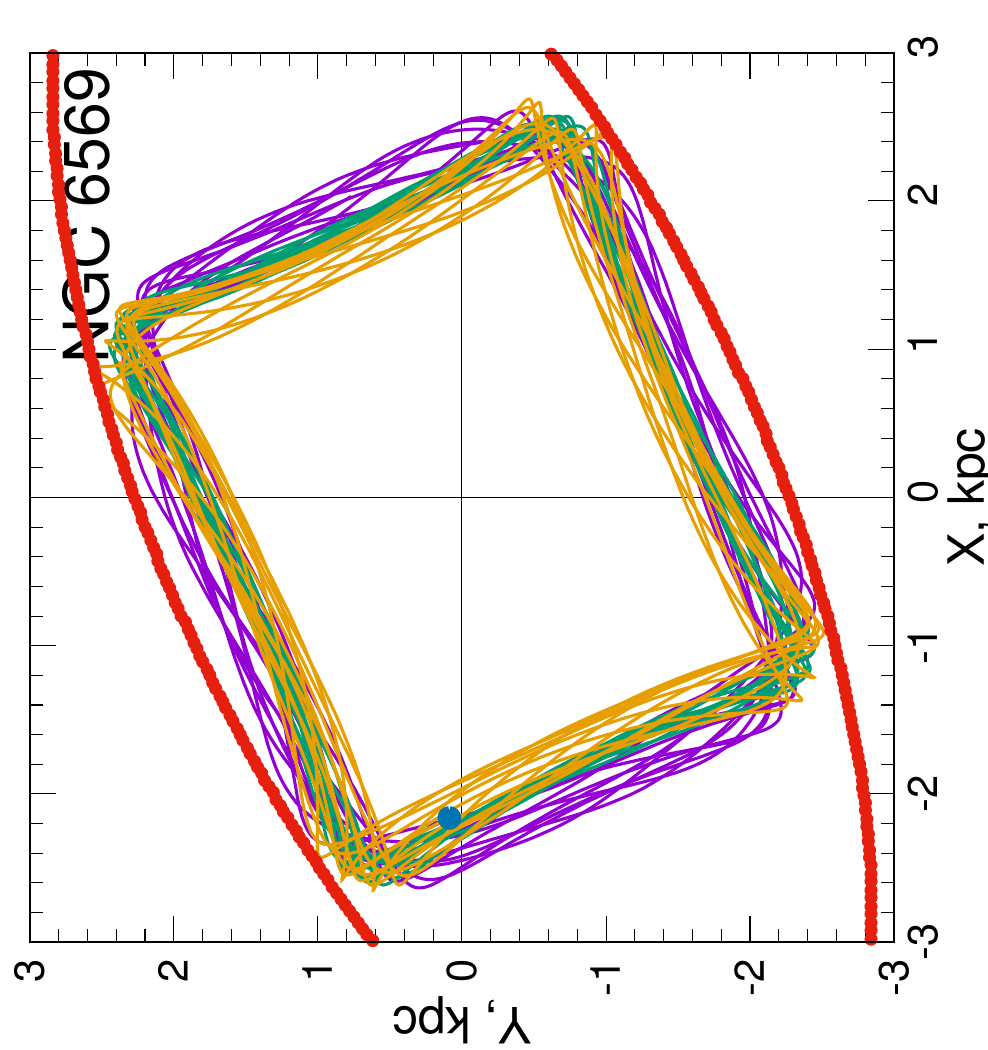}
   \includegraphics[width=0.2\textwidth,angle=-90]{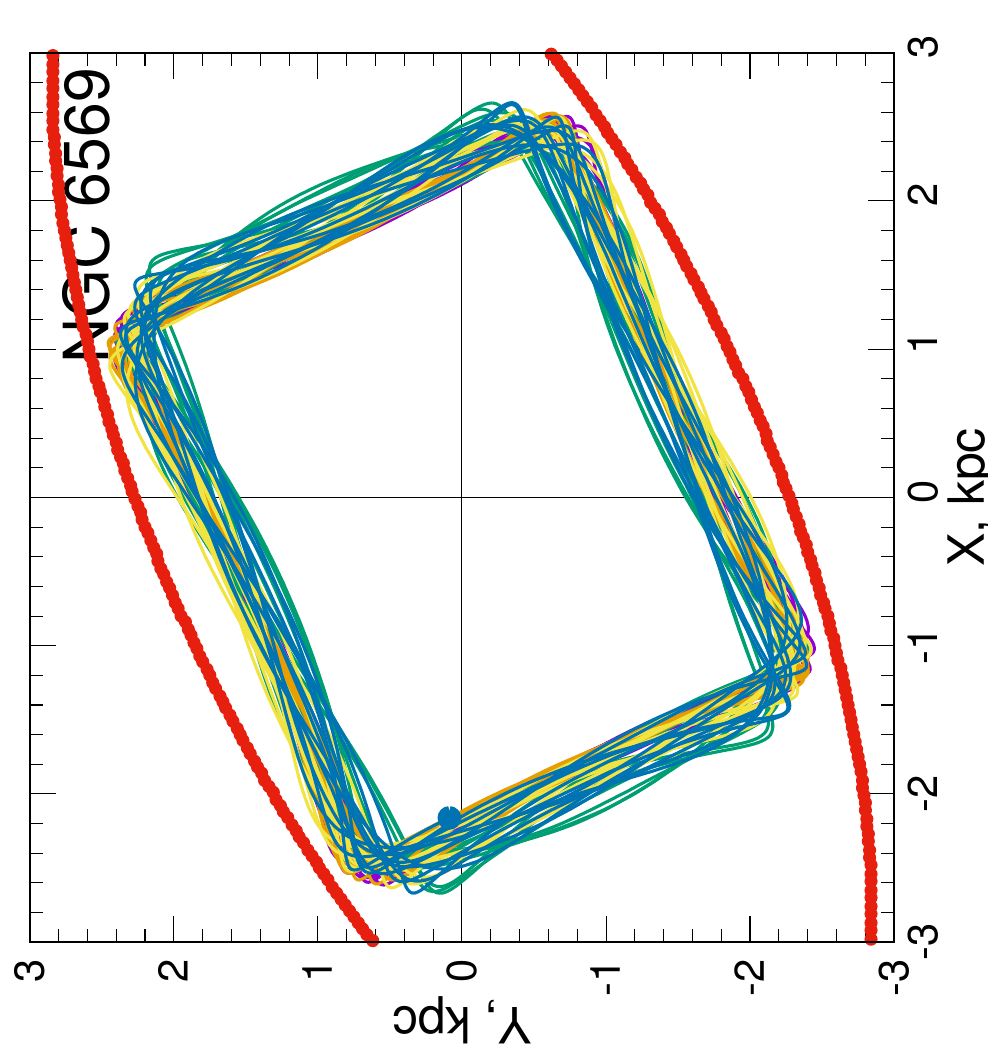}\

   \includegraphics[width=0.2\textwidth,angle=-90]{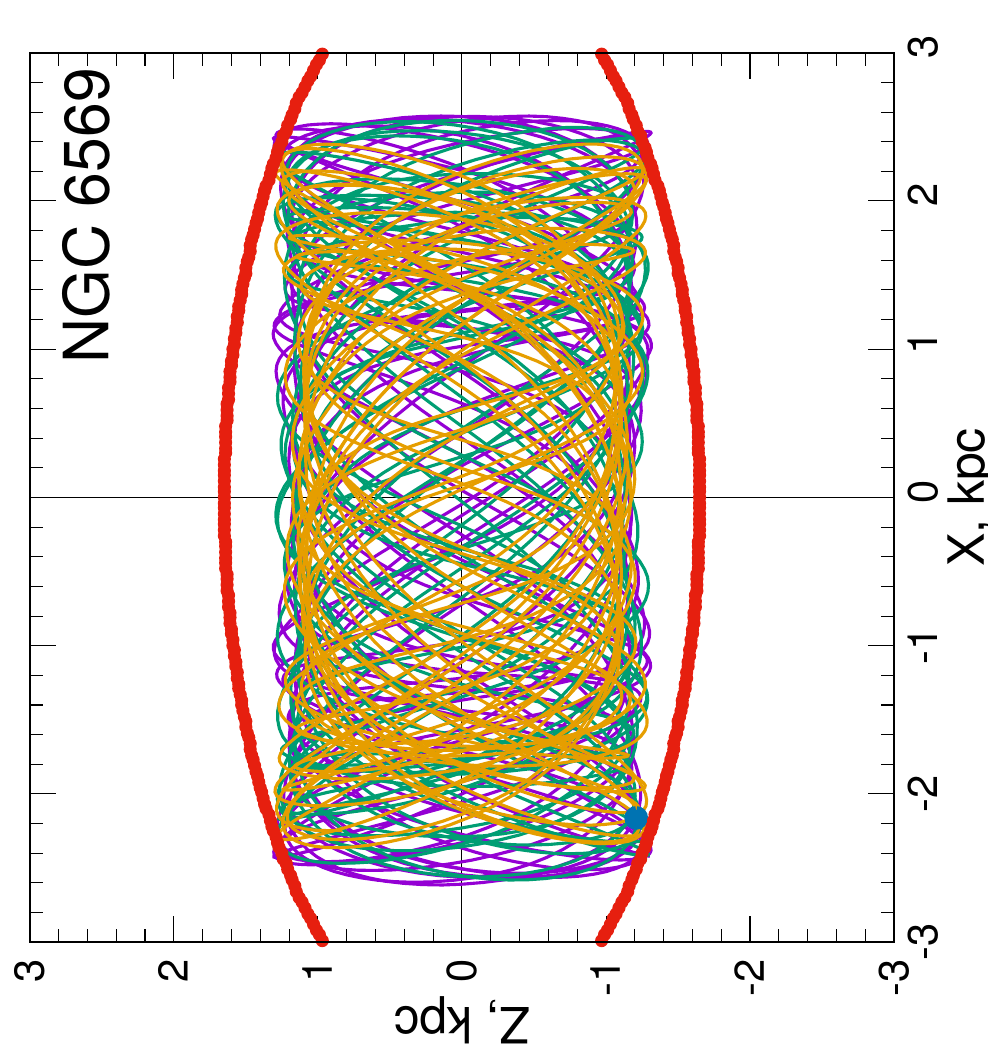}
   \includegraphics[width=0.2\textwidth,angle=-90]{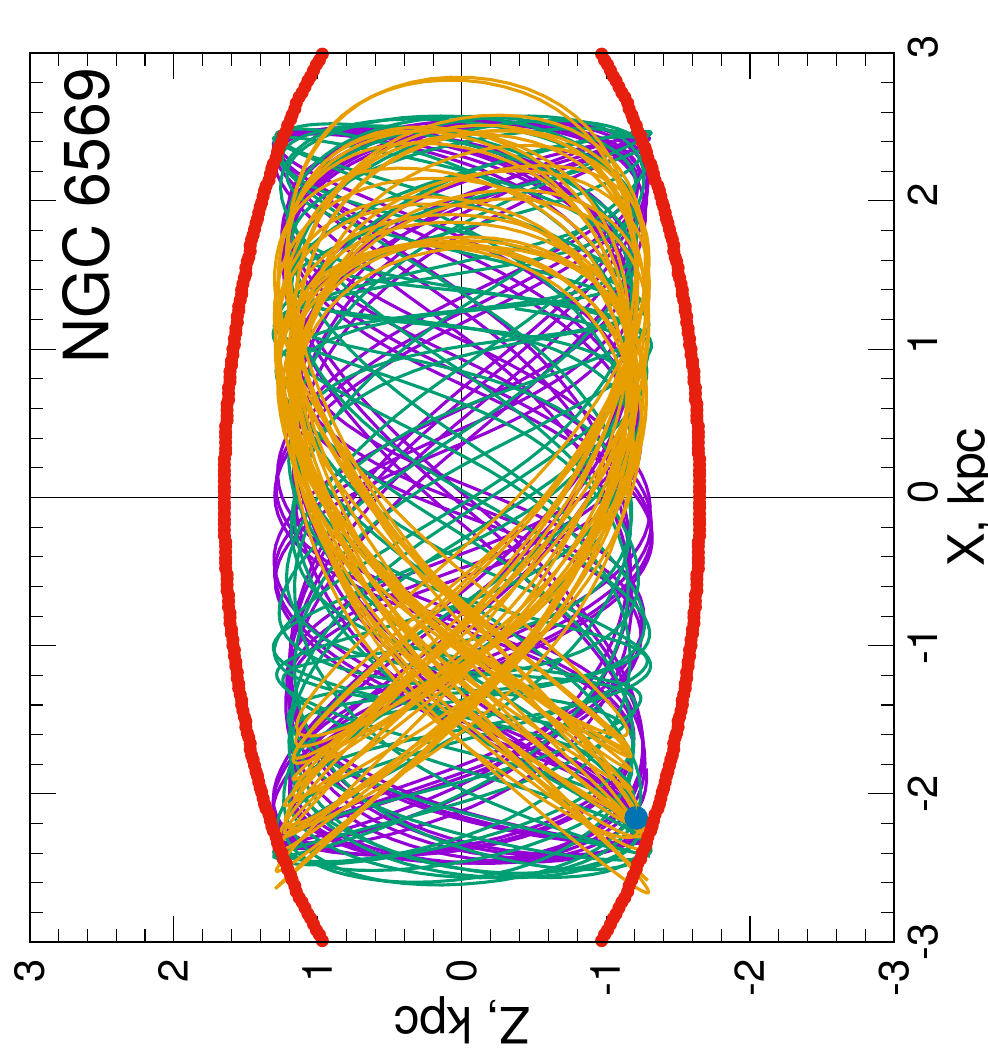}
   \includegraphics[width=0.2\textwidth,angle=-90]{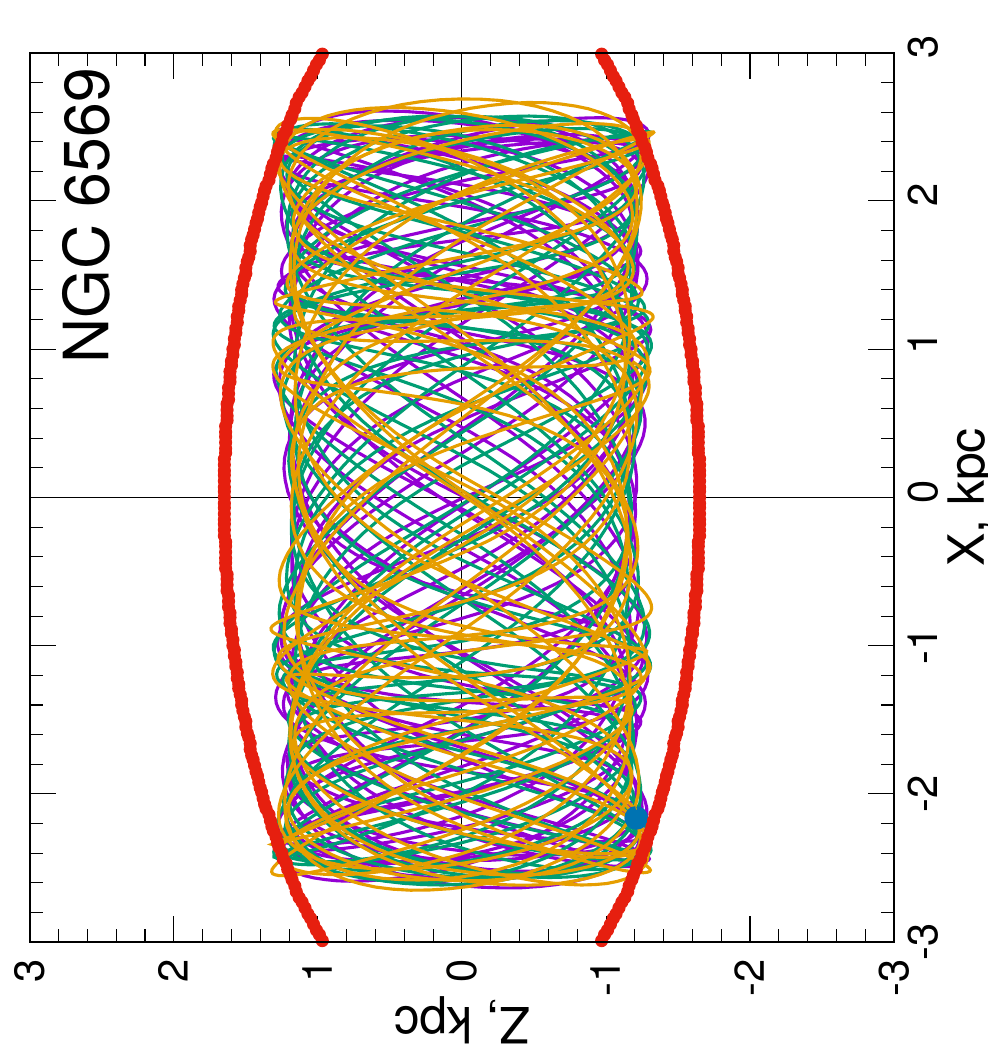}
   \includegraphics[width=0.2\textwidth,angle=-90]{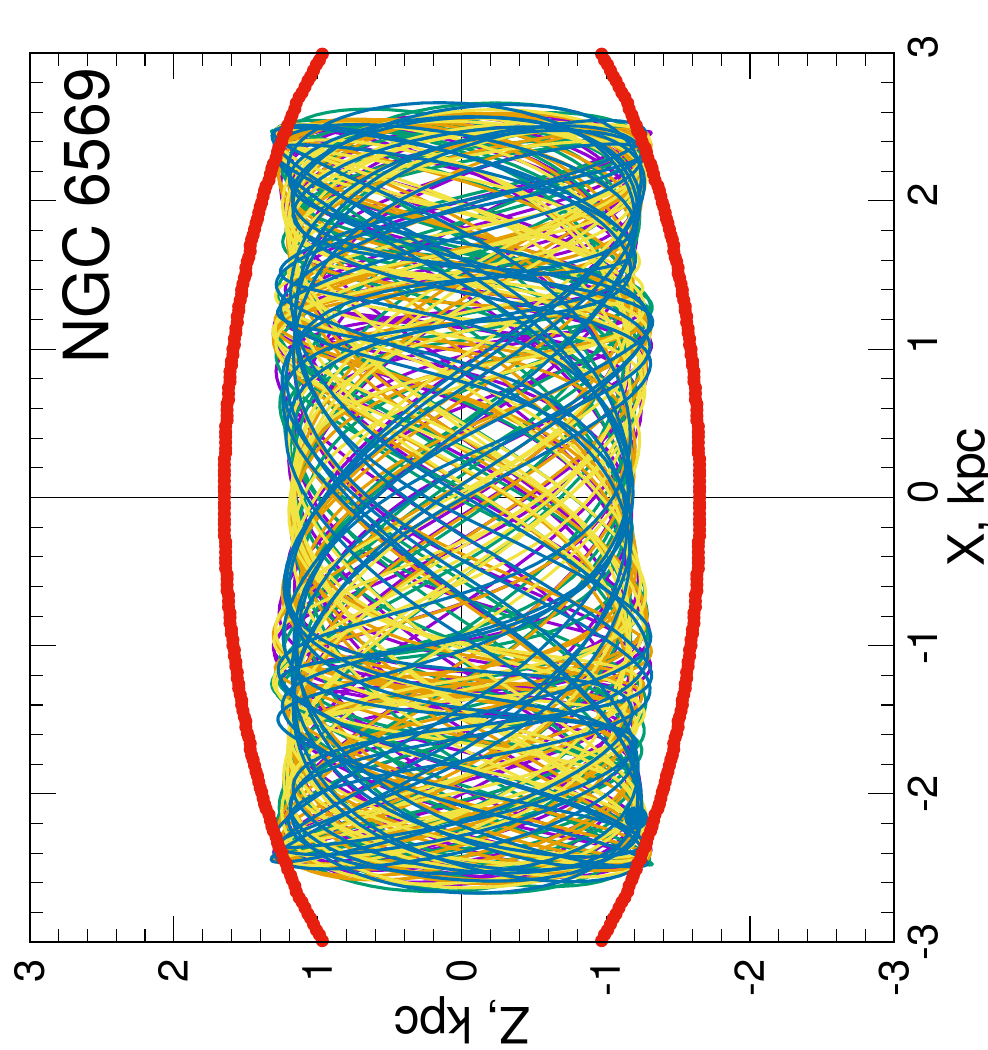}\

   \caption{\small Orbits in the rotating bar system of the GCs NGC 6540, NGC 6553, NGC 6569 belonging to the disk. All designations are as in the previous figure. }
\label{fD}
\end{center}}
\end{figure*}

\bigskip

\centerline{{\bf APPENDIX} (8 last pages):}

\bigskip

\noindent{Orbits of selected 45 GCs in $X-Y$ and $X-Z$ projections of the galactic coordinate system. The left panels show the orbits obtained in the axisymmetric potential, the right panels show the orbits obtained in the potential including the bar with the following parameters: $M_b=430 M_G$, $\Omega_b=40$ km/s/kpc, $q_b=5$ kpc, $\theta_b=25^o$, ratio of axes V0. In the second case, the orbits are constructed in the system of a rotating bar. The red line shows the sections of the bar. The orbits were integrated 2.5 Gyr backward. The beginning of the orbits is marked with a blue circle.}

\begin{figure*}
{\begin{center}
   \includegraphics[width=0.225\textwidth,angle=-90]{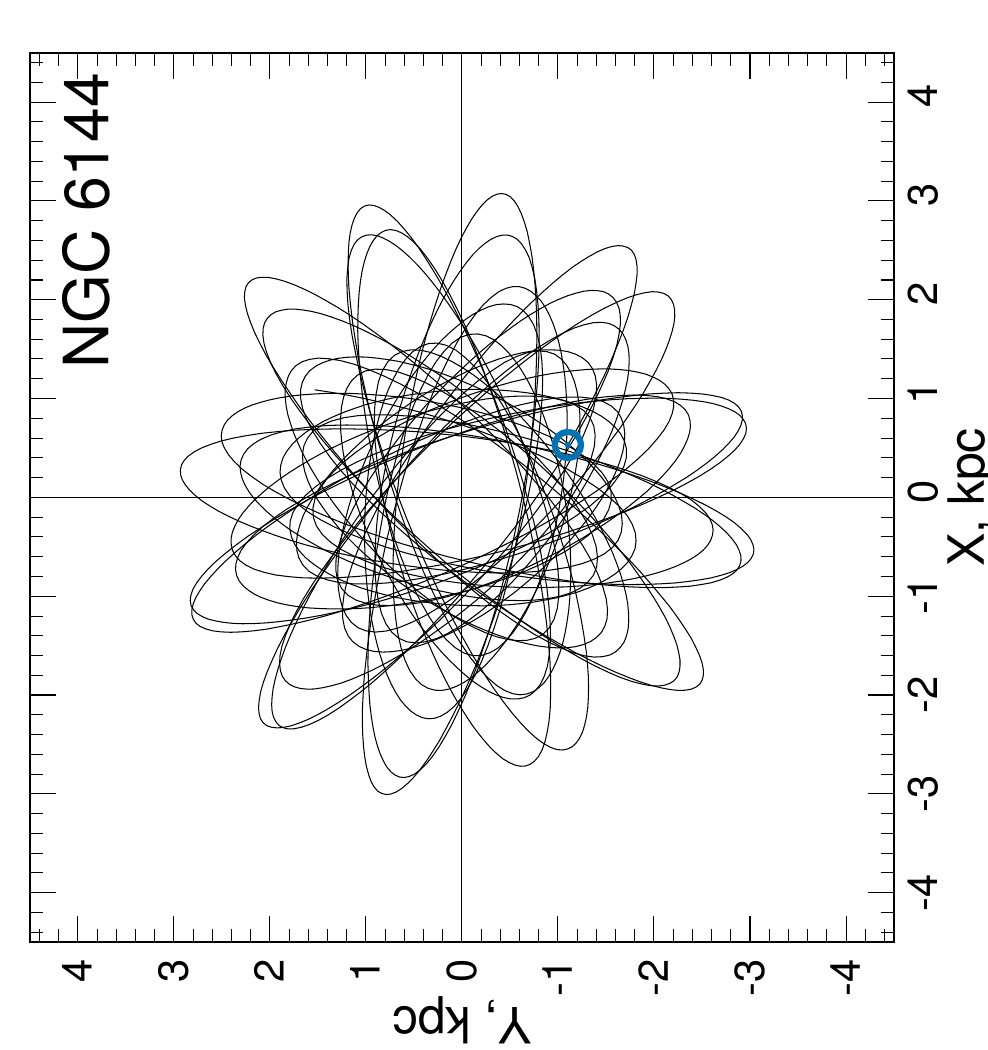}
     \includegraphics[width=0.225\textwidth,angle=-90]{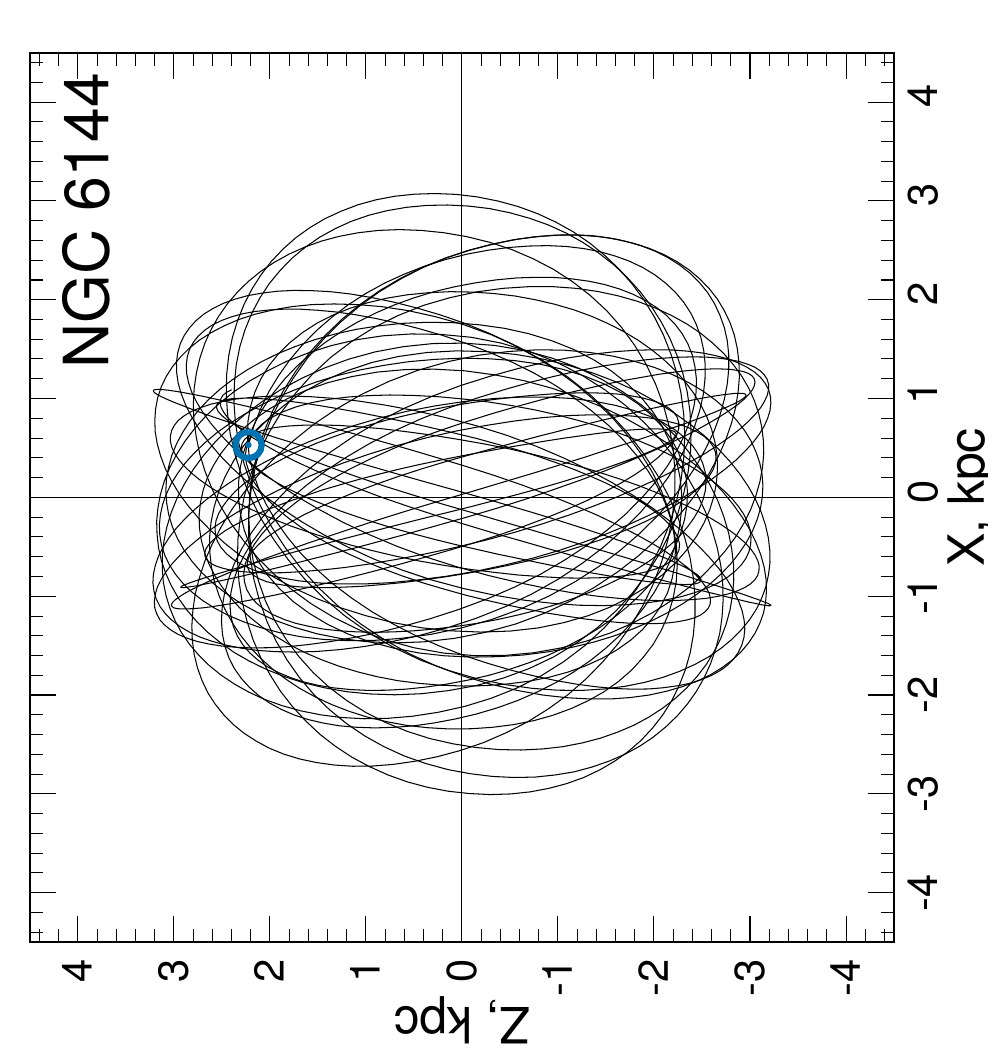}
        \includegraphics[width=0.225\textwidth,angle=-90]{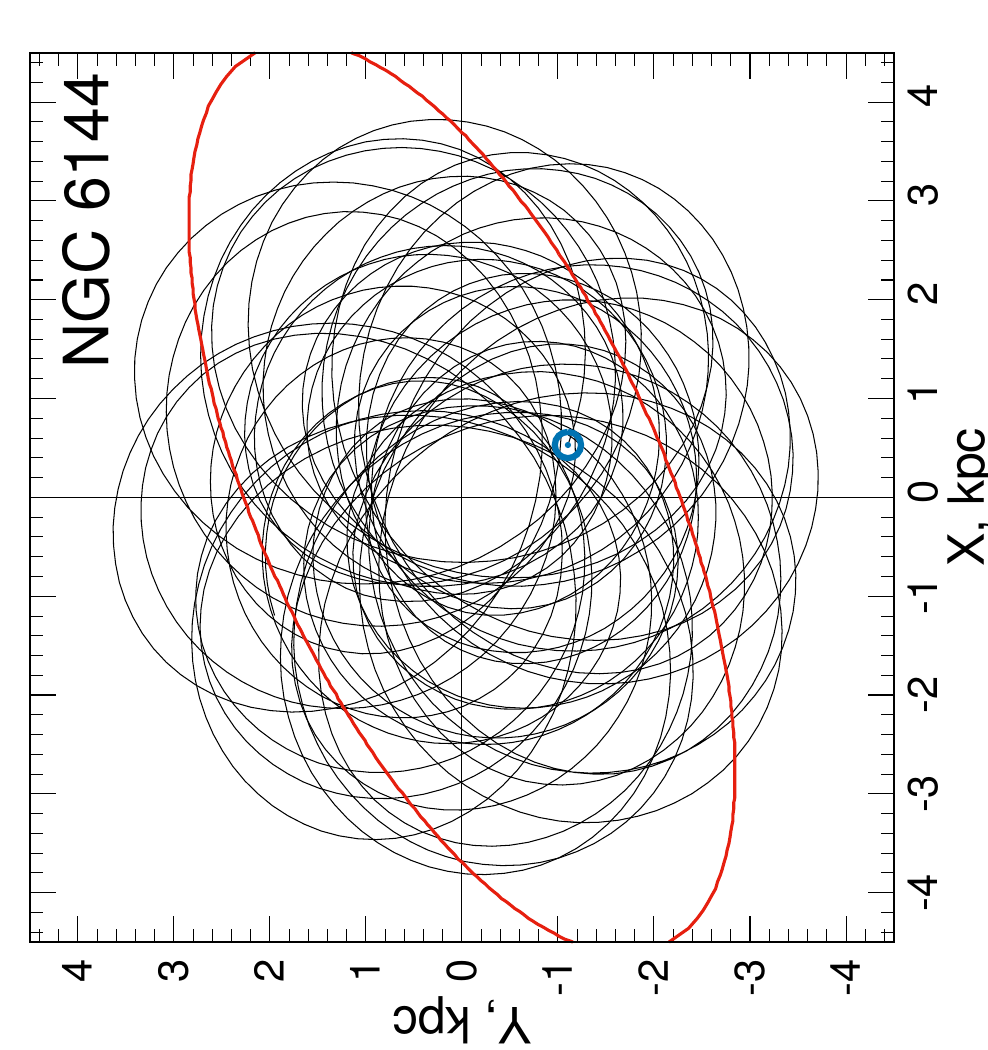}
     \includegraphics[width=0.225\textwidth,angle=-90]{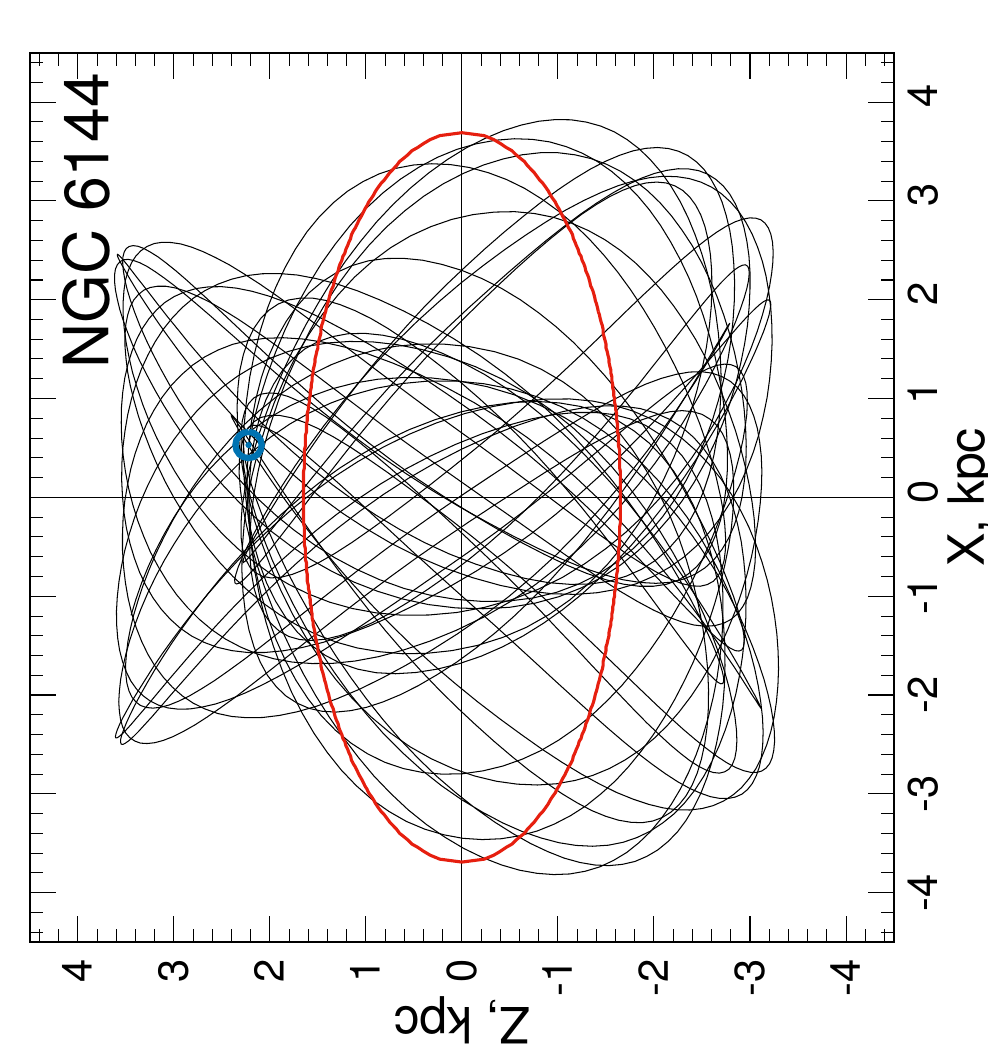}\
   \includegraphics[width=0.225\textwidth,angle=-90]{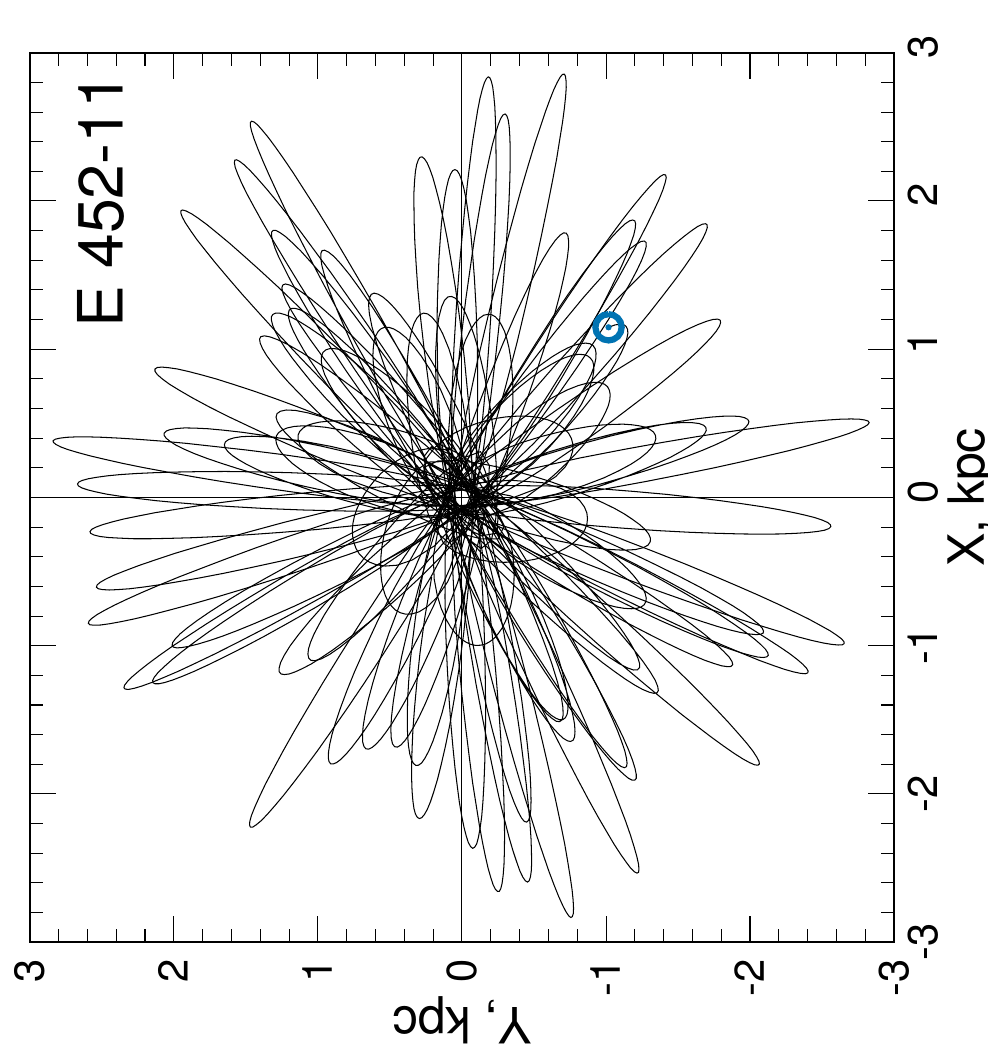}
     \includegraphics[width=0.225\textwidth,angle=-90]{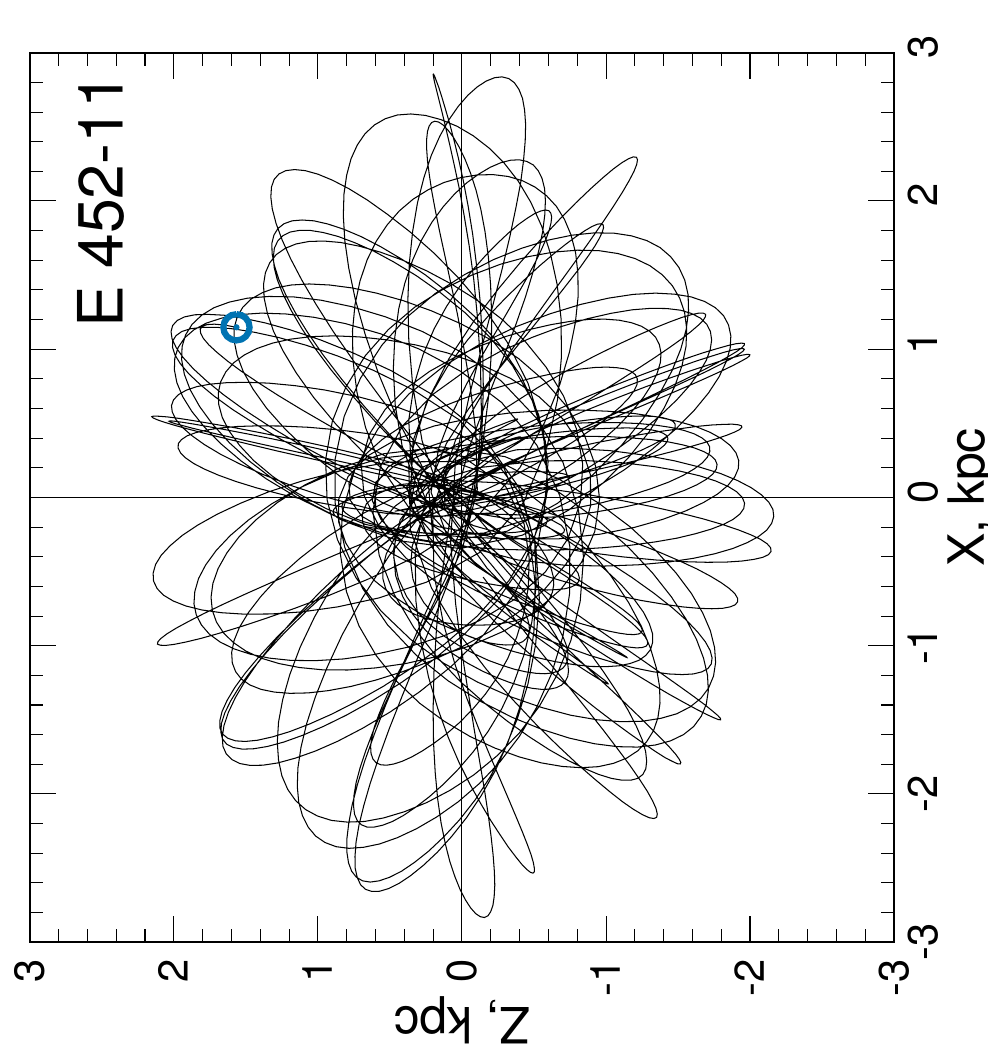}
   \includegraphics[width=0.225\textwidth,angle=-90]{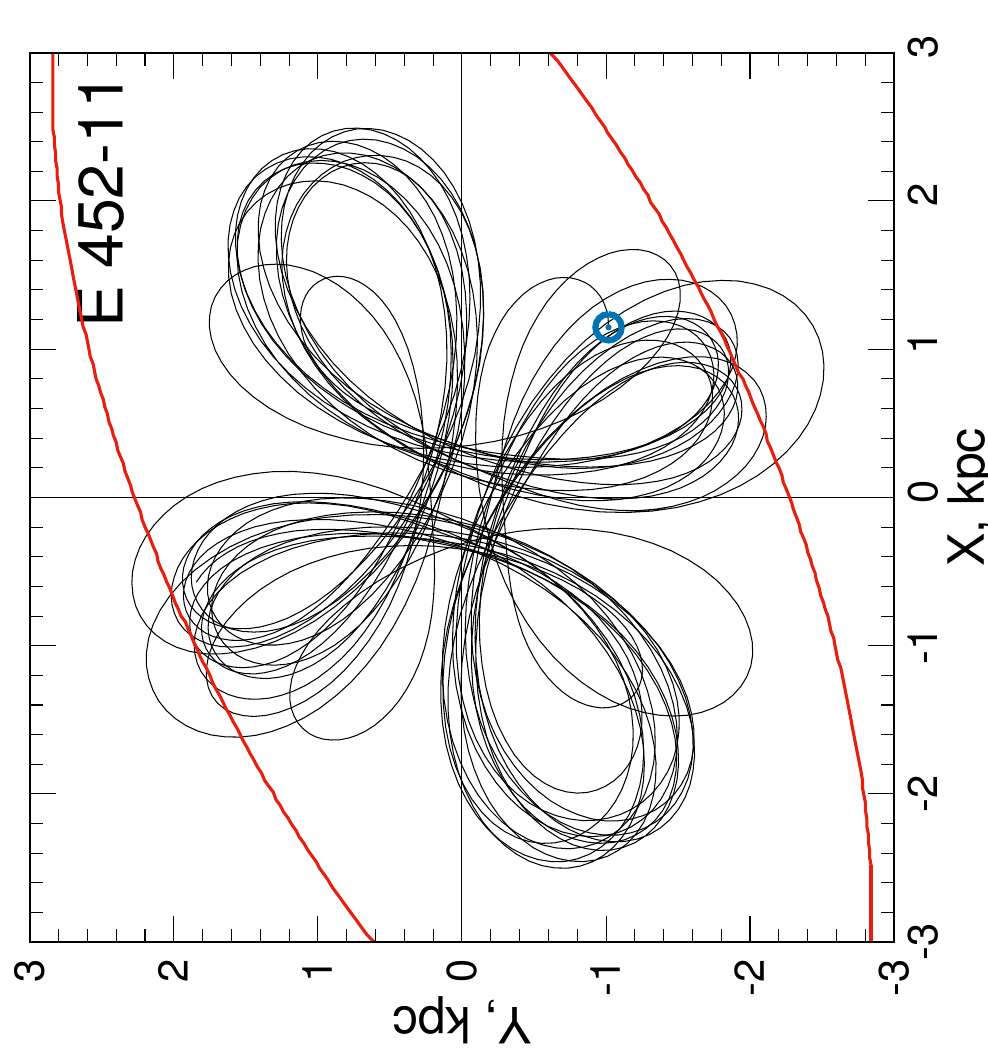}
     \includegraphics[width=0.225\textwidth,angle=-90]{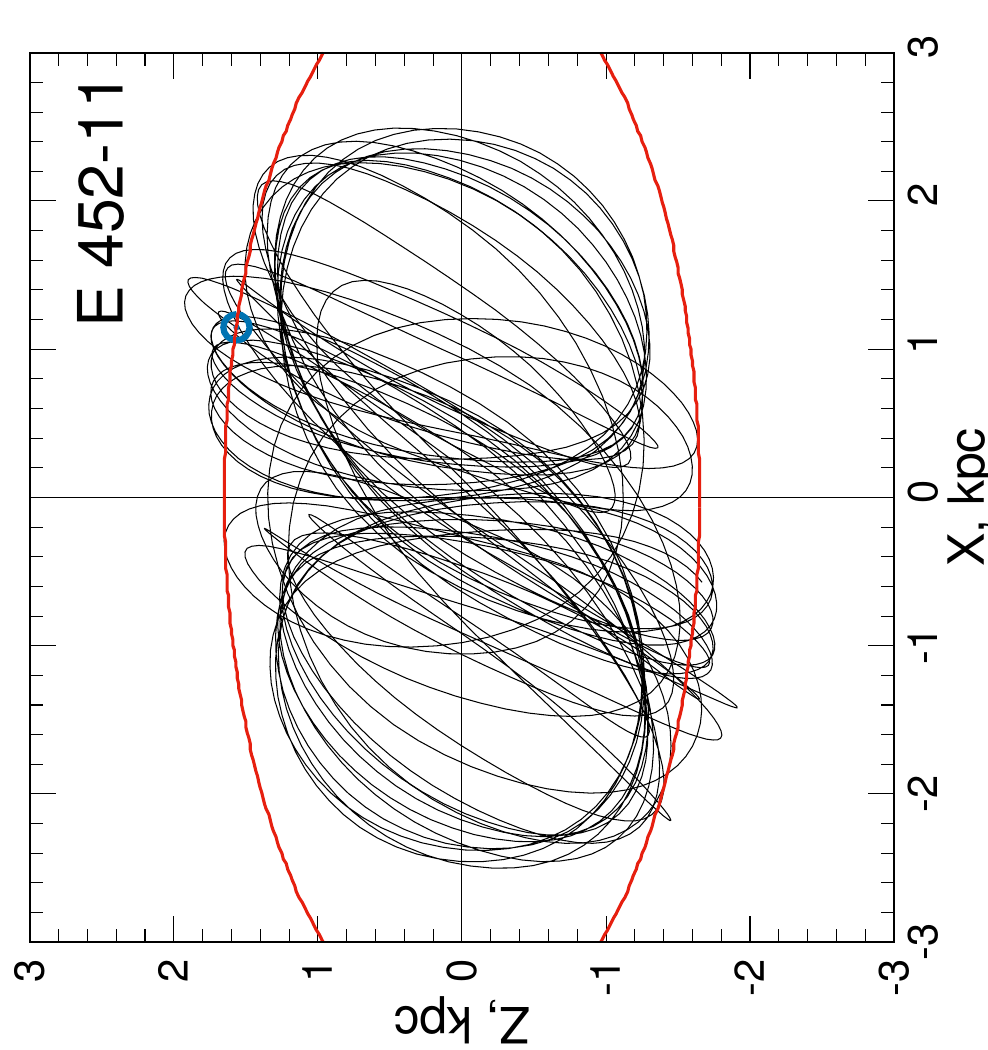}\
   \includegraphics[width=0.225\textwidth,angle=-90]{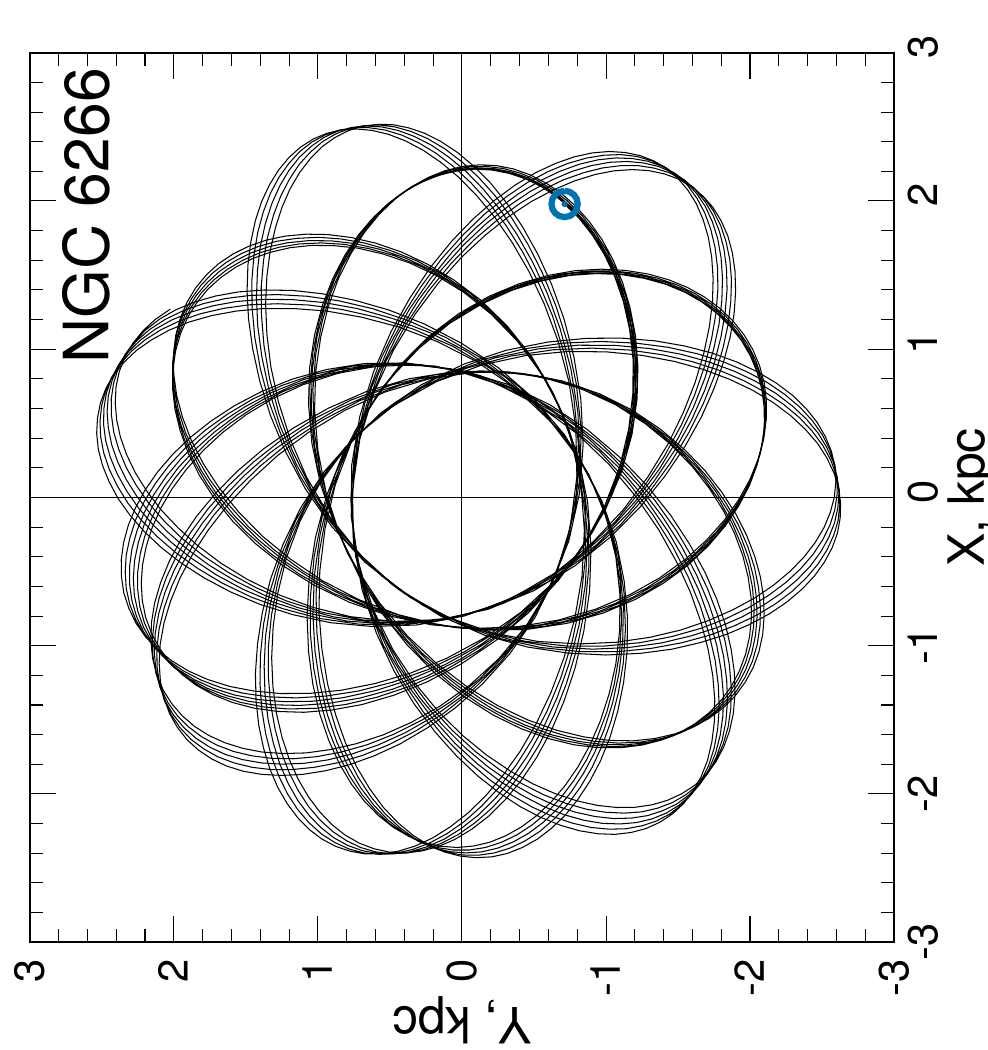}
     \includegraphics[width=0.225\textwidth,angle=-90]{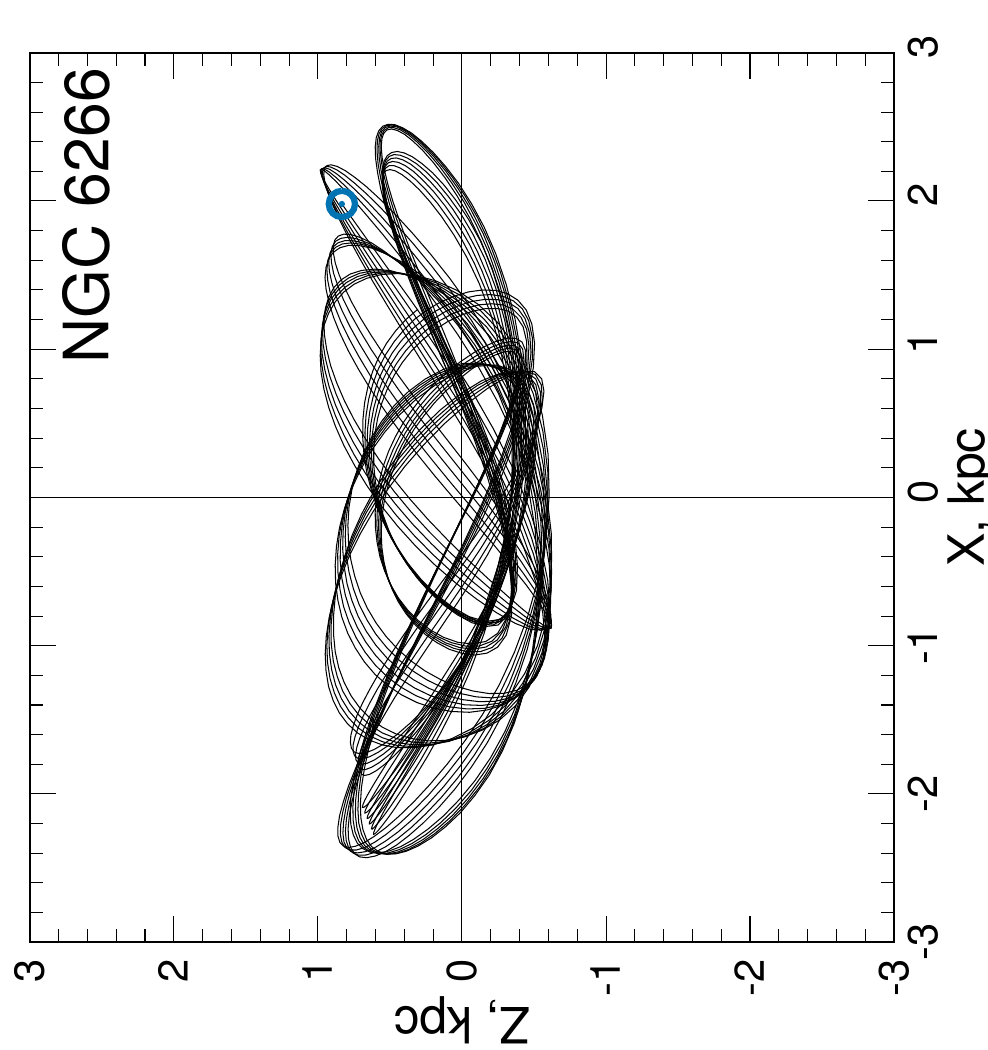}
   \includegraphics[width=0.225\textwidth,angle=-90]{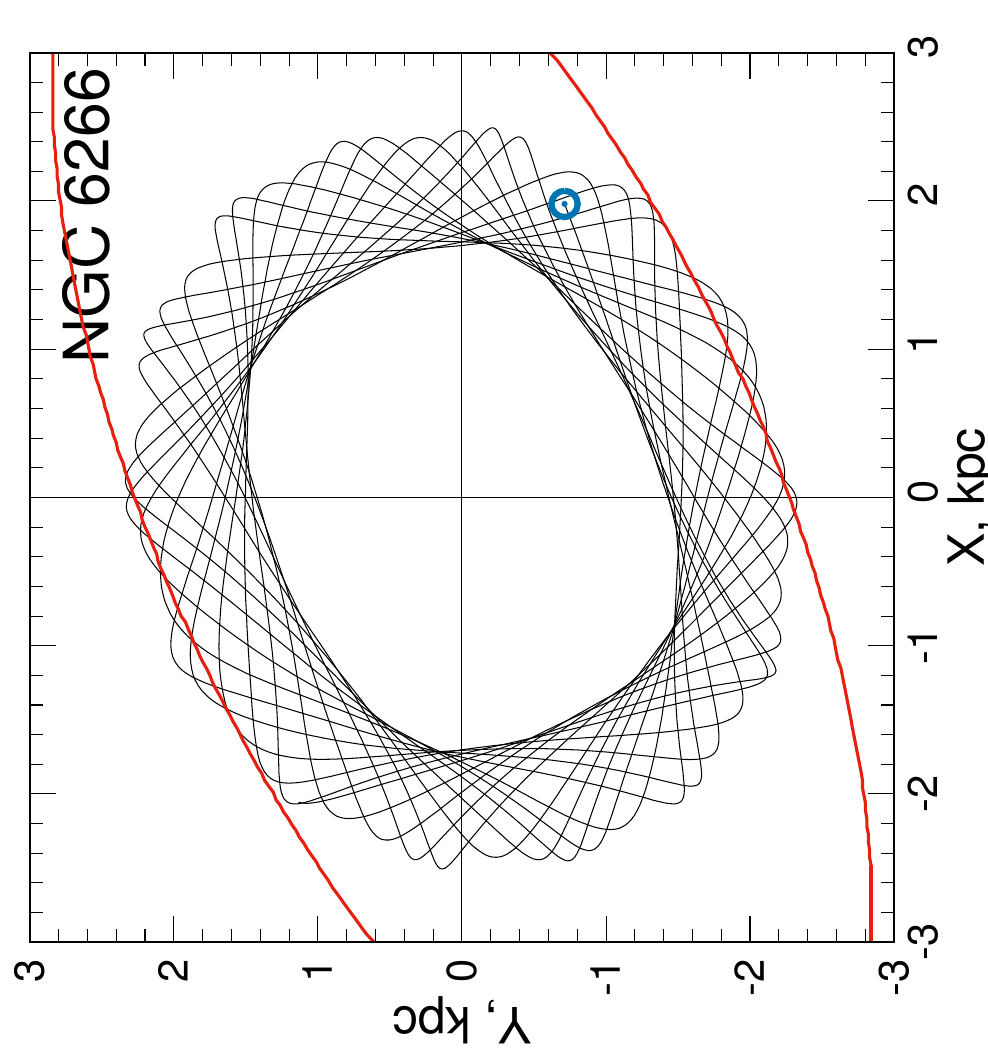}
     \includegraphics[width=0.225\textwidth,angle=-90]{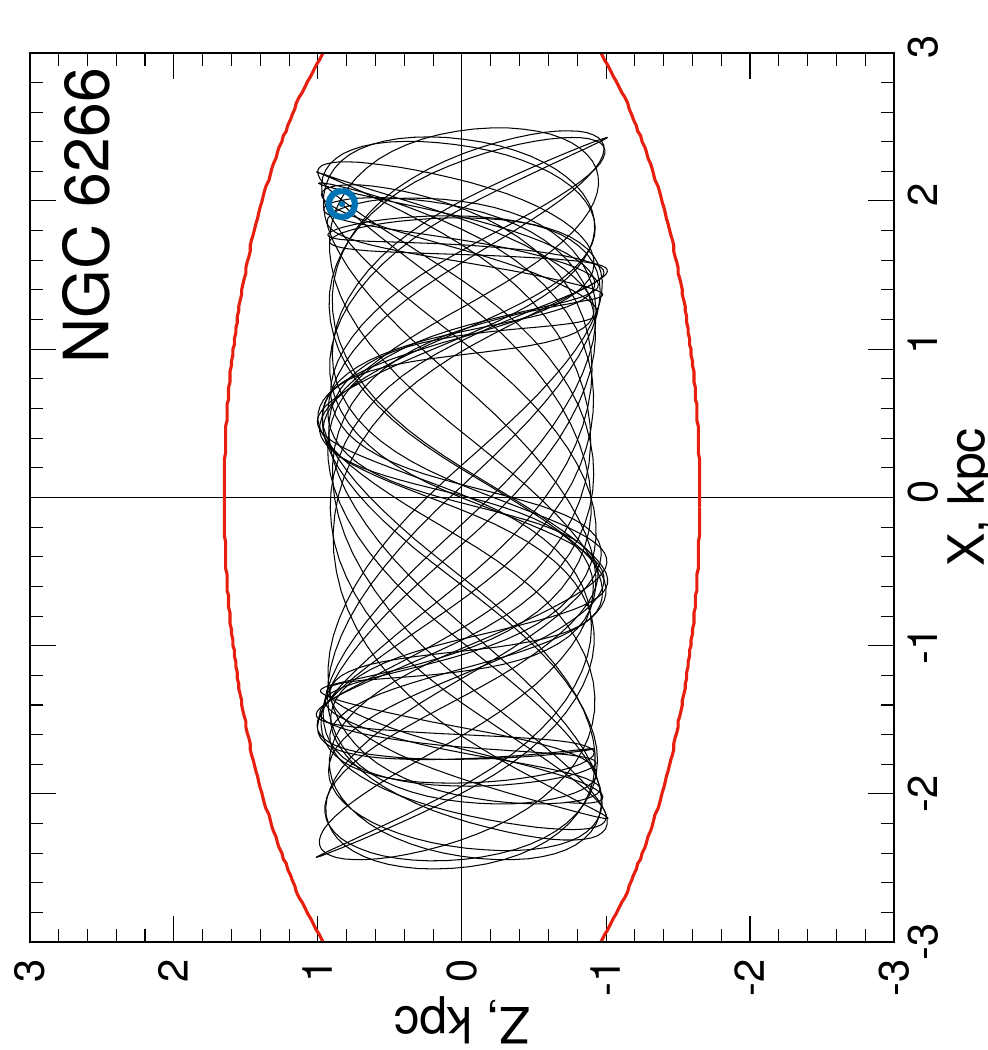}\
   \includegraphics[width=0.225\textwidth,angle=-90]{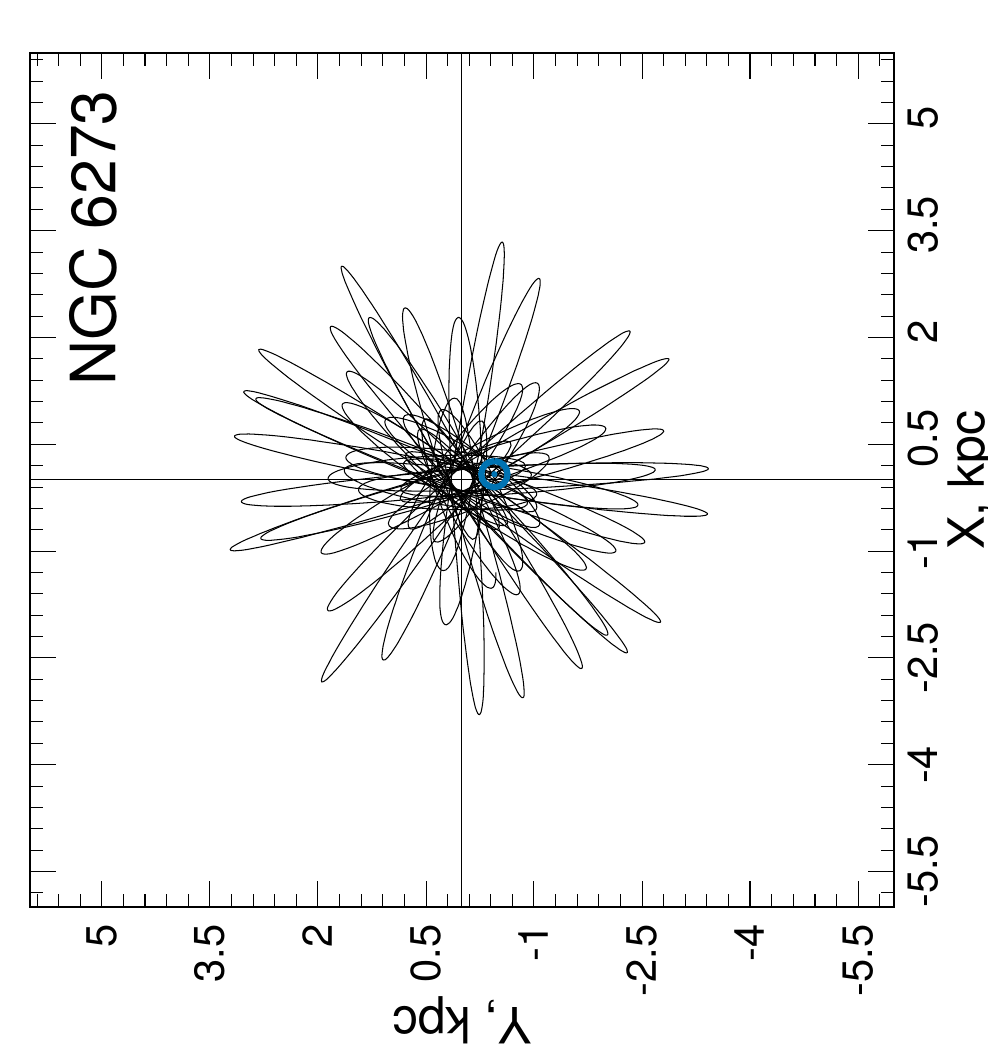}
     \includegraphics[width=0.225\textwidth,angle=-90]{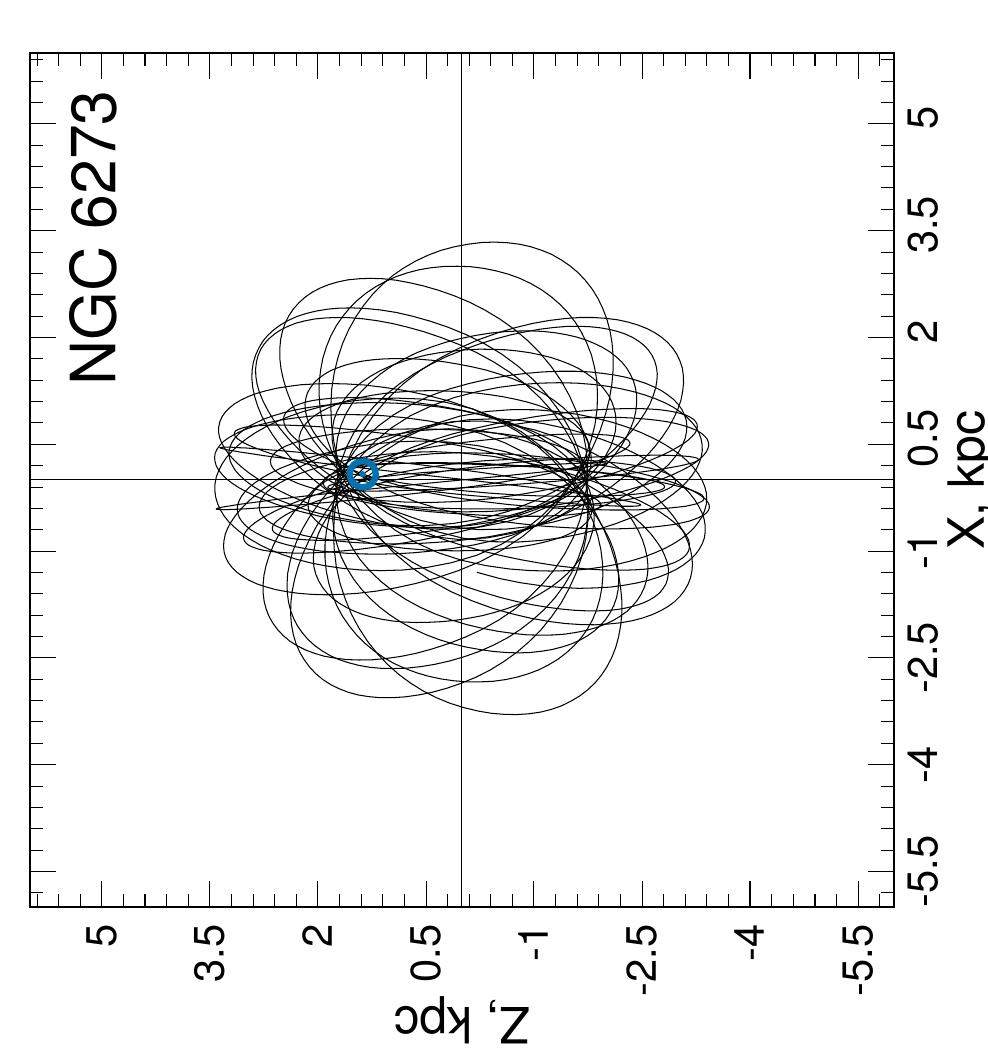}
    \includegraphics[width=0.225\textwidth,angle=-90]{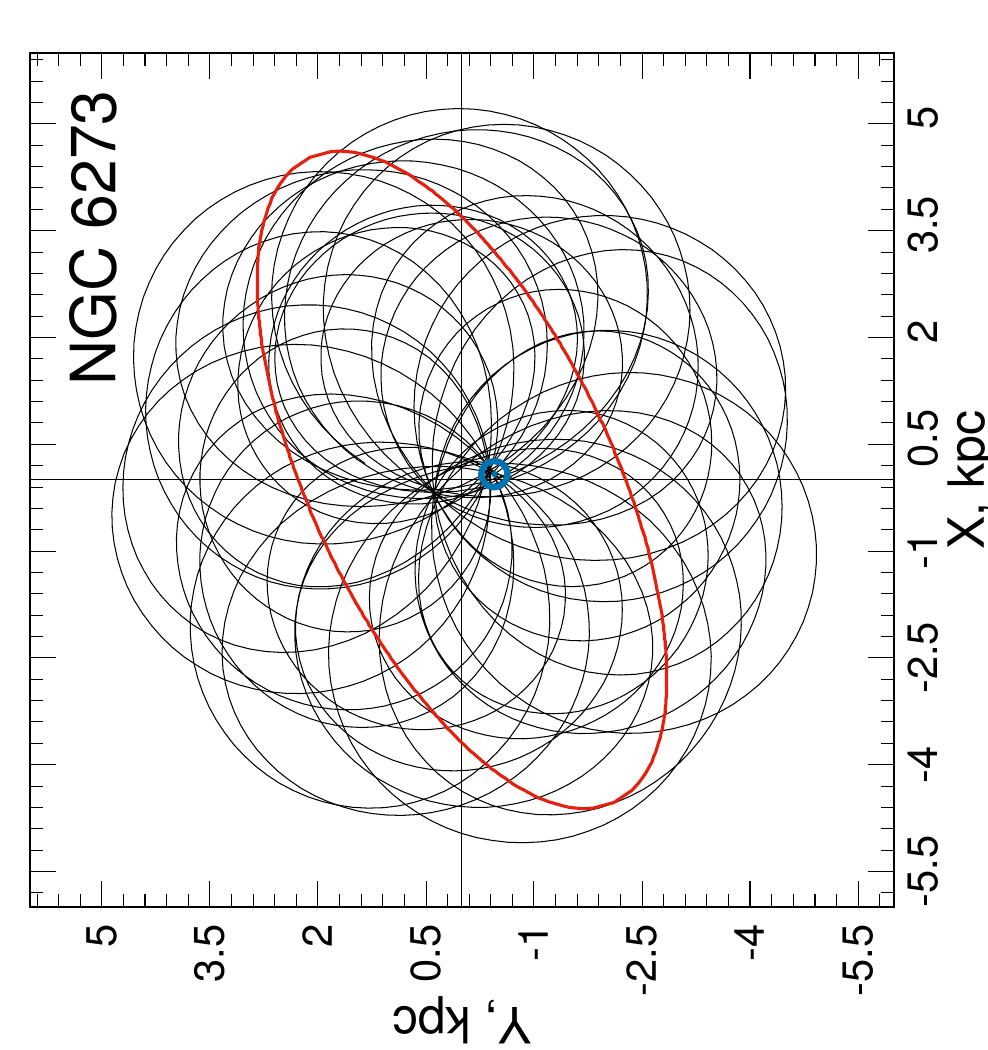}
     \includegraphics[width=0.225\textwidth,angle=-90]{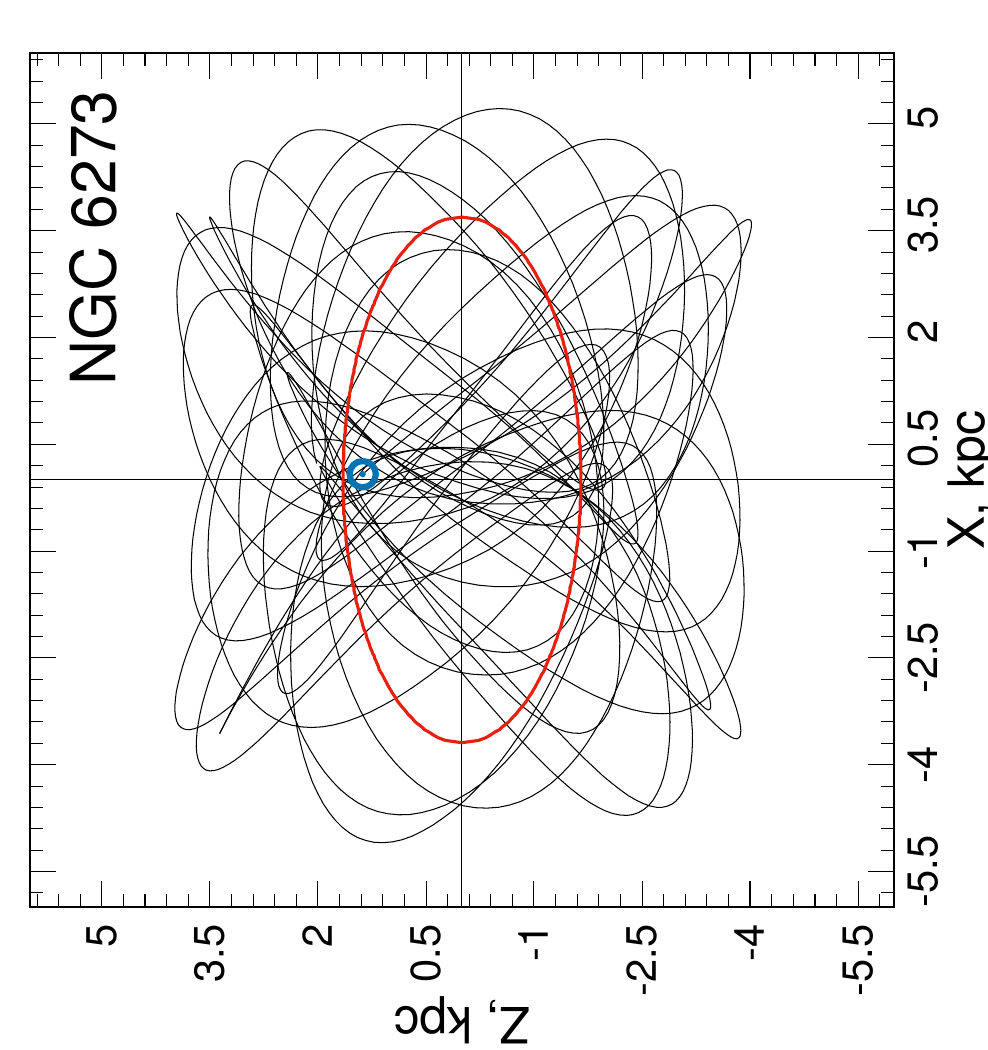}\
 \includegraphics[width=0.225\textwidth,angle=-90]{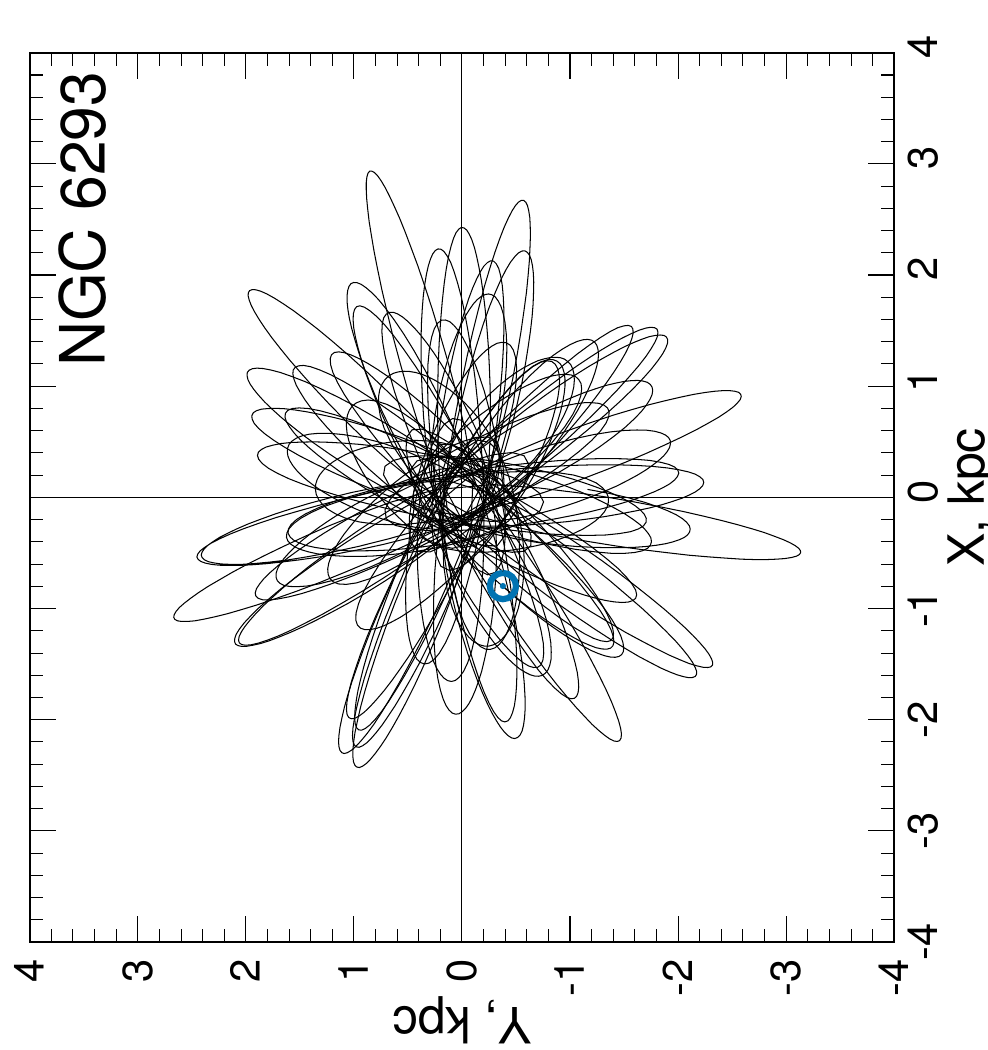}
     \includegraphics[width=0.225\textwidth,angle=-90]{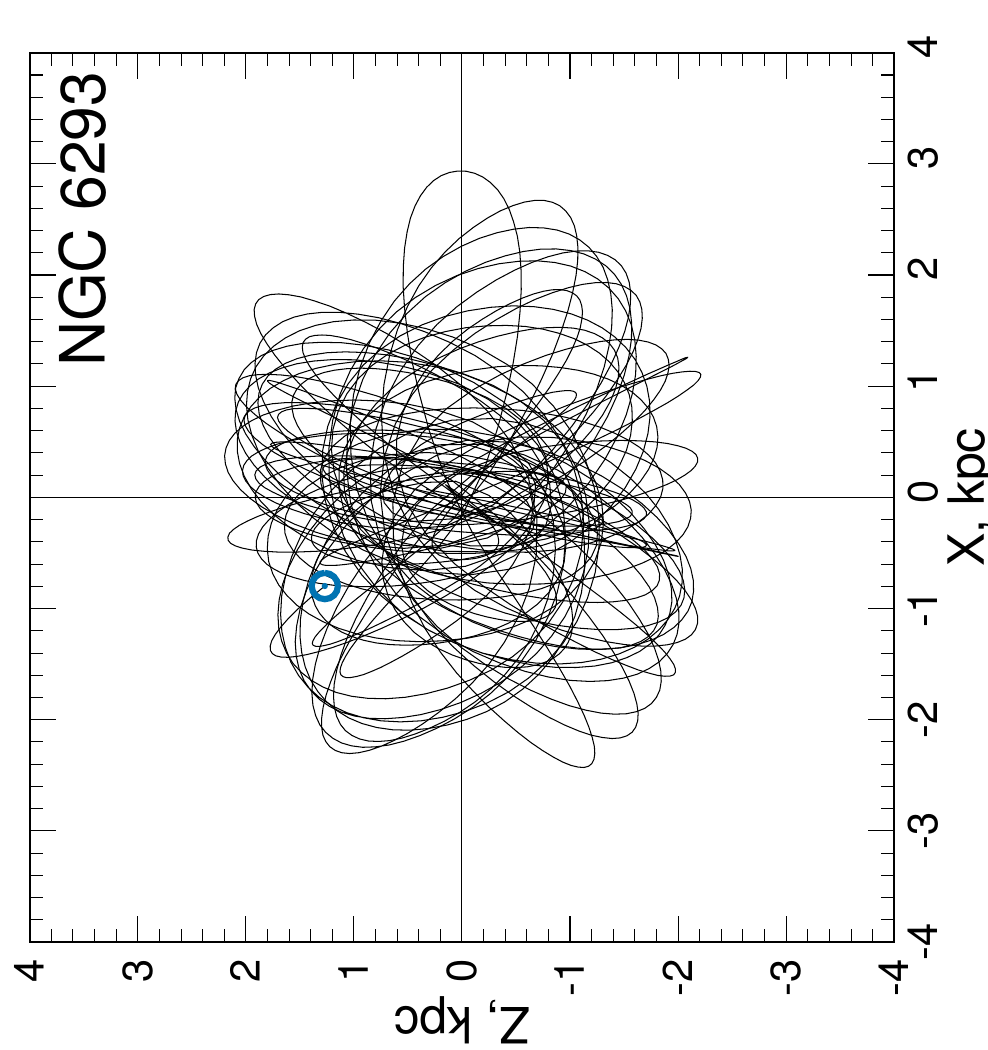}
 \includegraphics[width=0.225\textwidth,angle=-90]{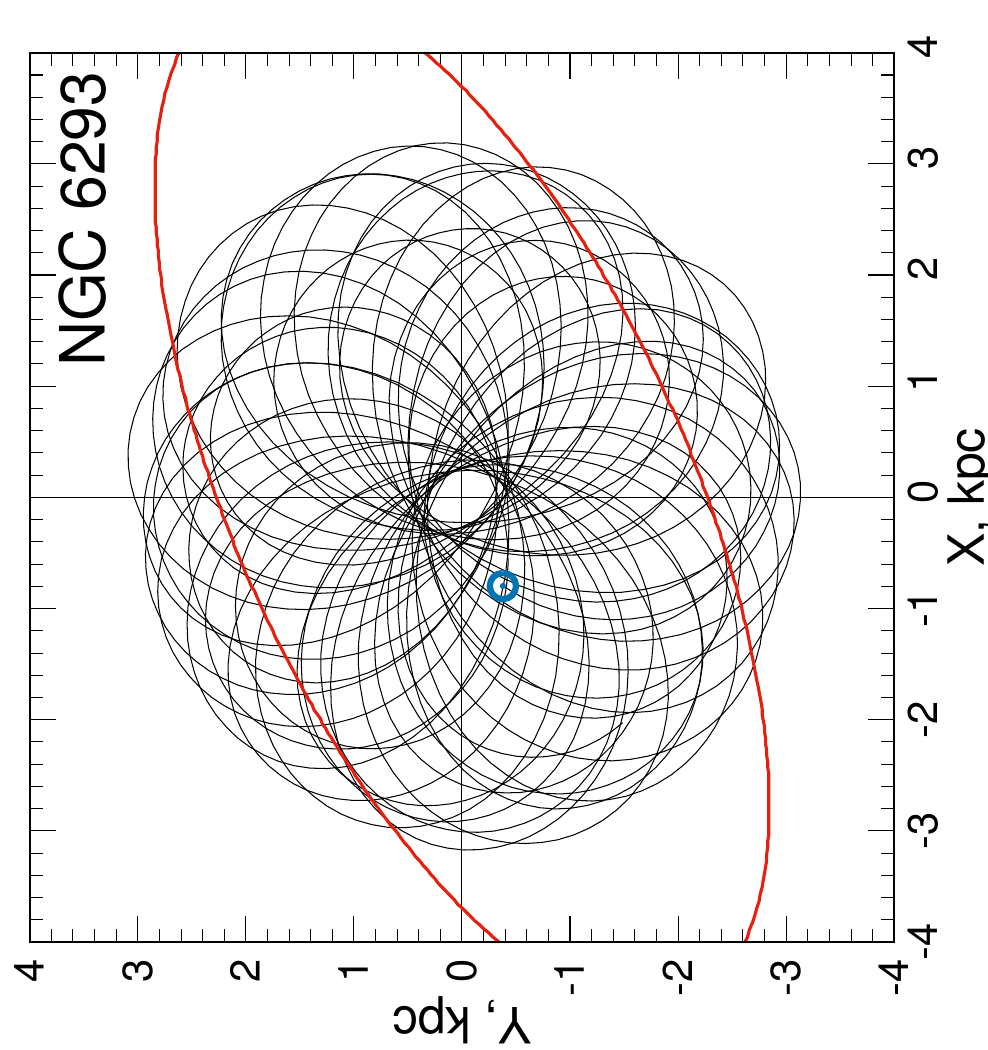}
     \includegraphics[width=0.225\textwidth,angle=-90]{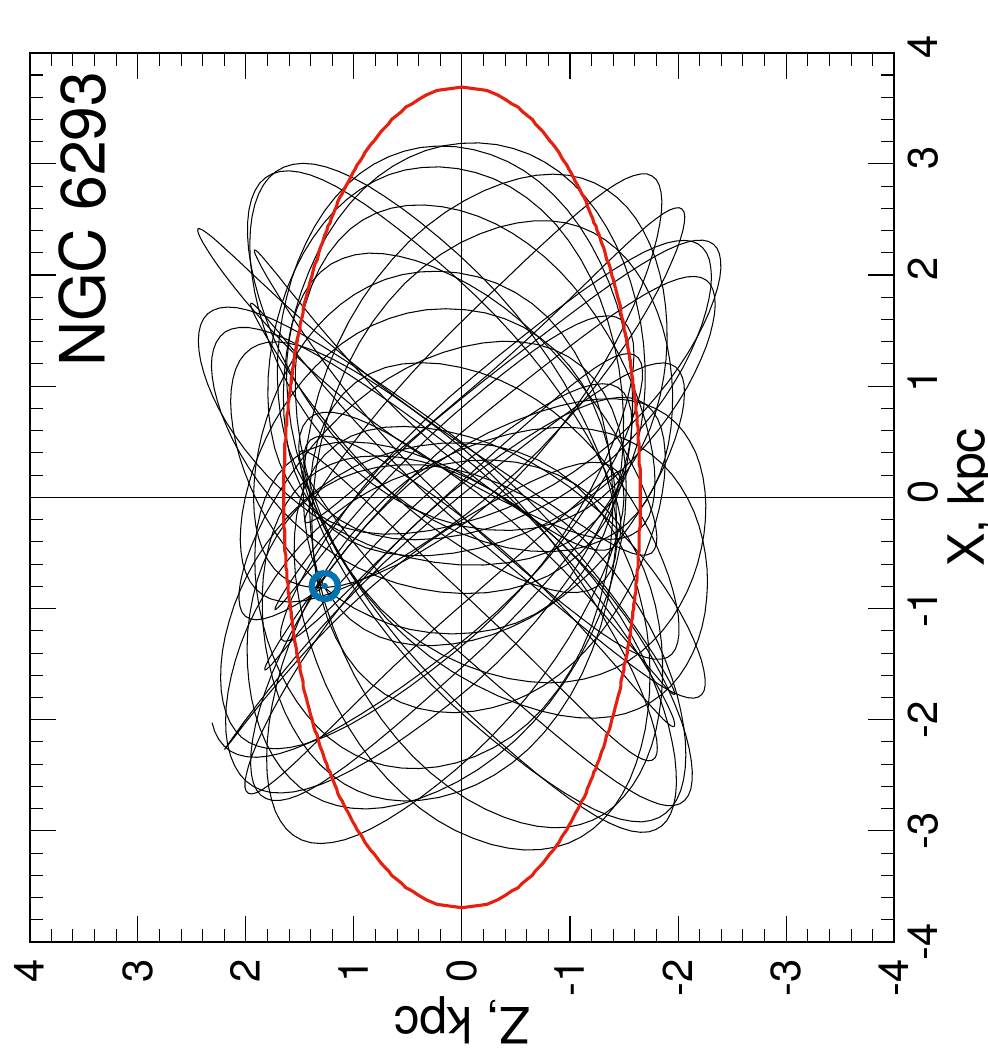}\

\medskip

 \centerline{APPENDIX}
\label{fD}
\end{center}}
\end{figure*}

\begin{figure*}
{\begin{center}
    \includegraphics[width=0.225\textwidth,angle=-90]{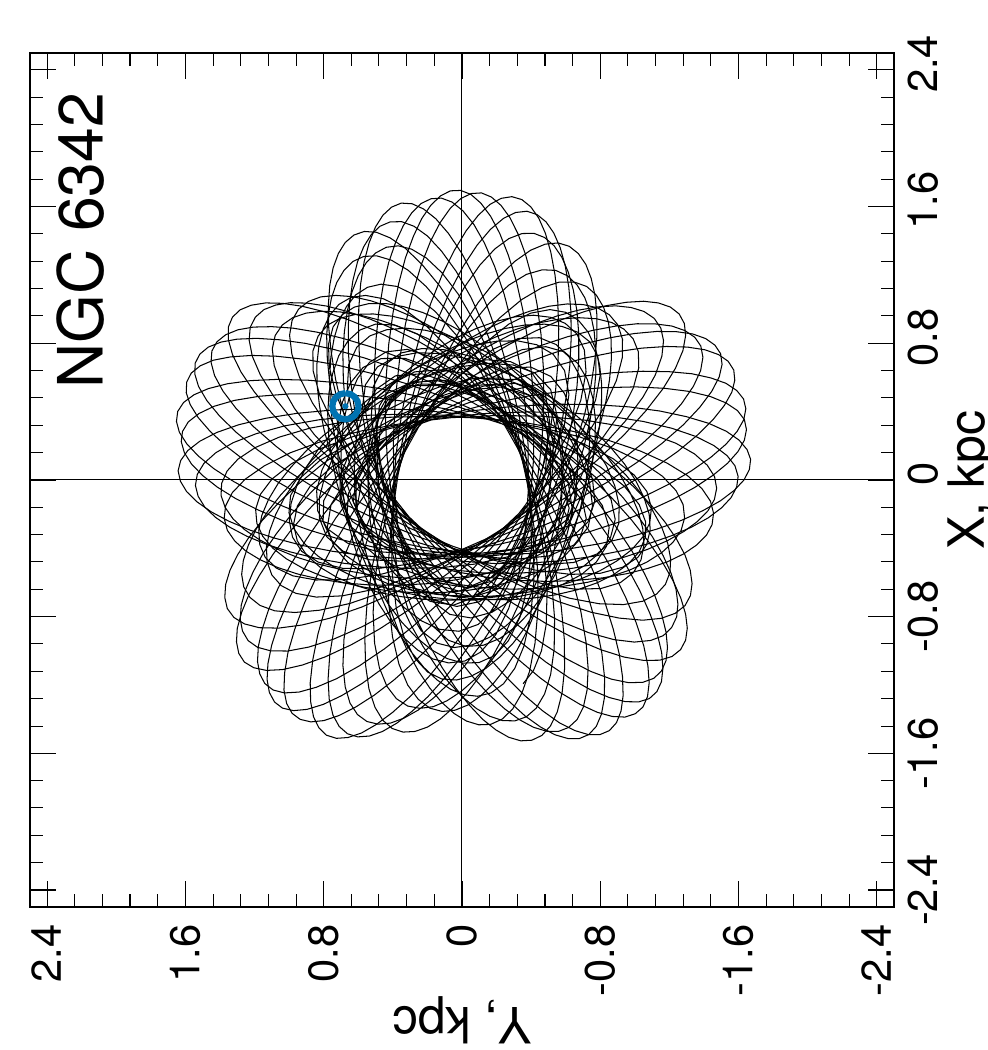}
     \includegraphics[width=0.225\textwidth,angle=-90]{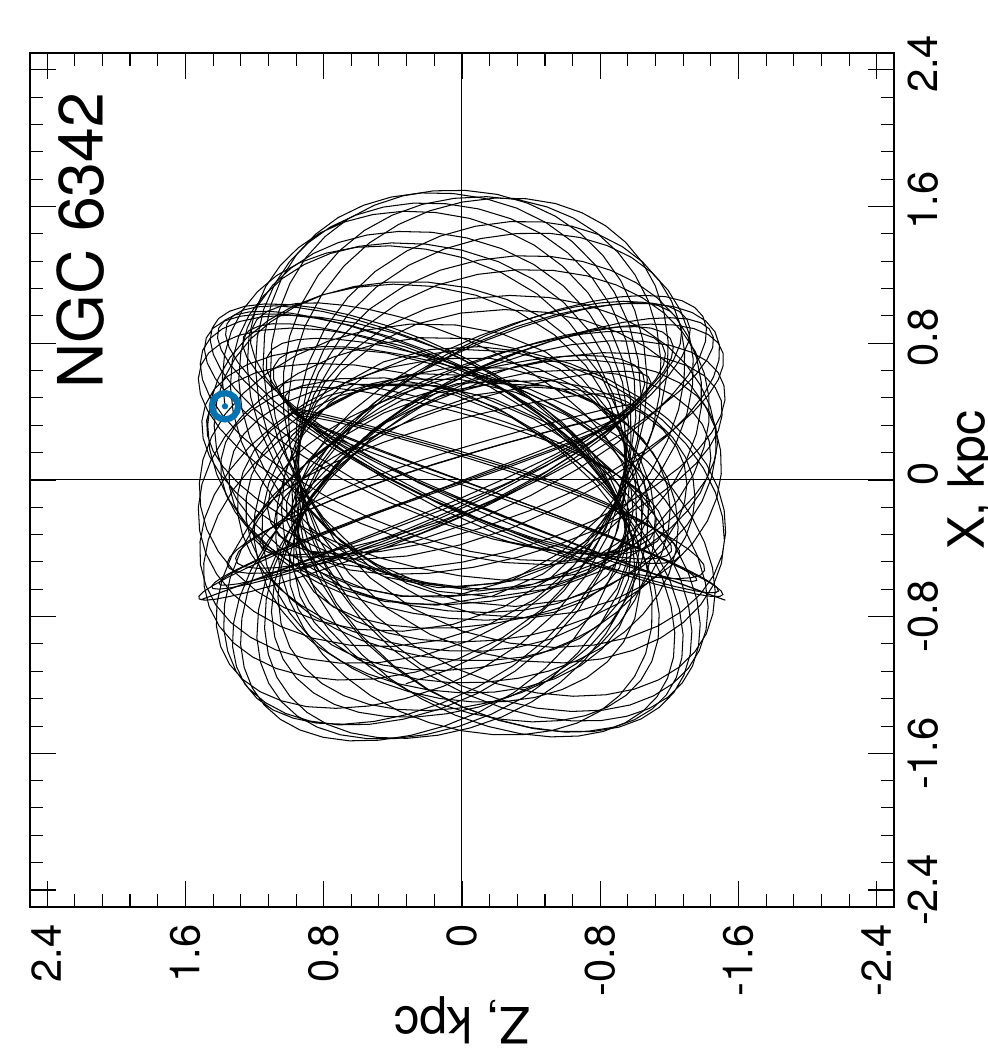}
    \includegraphics[width=0.225\textwidth,angle=-90]{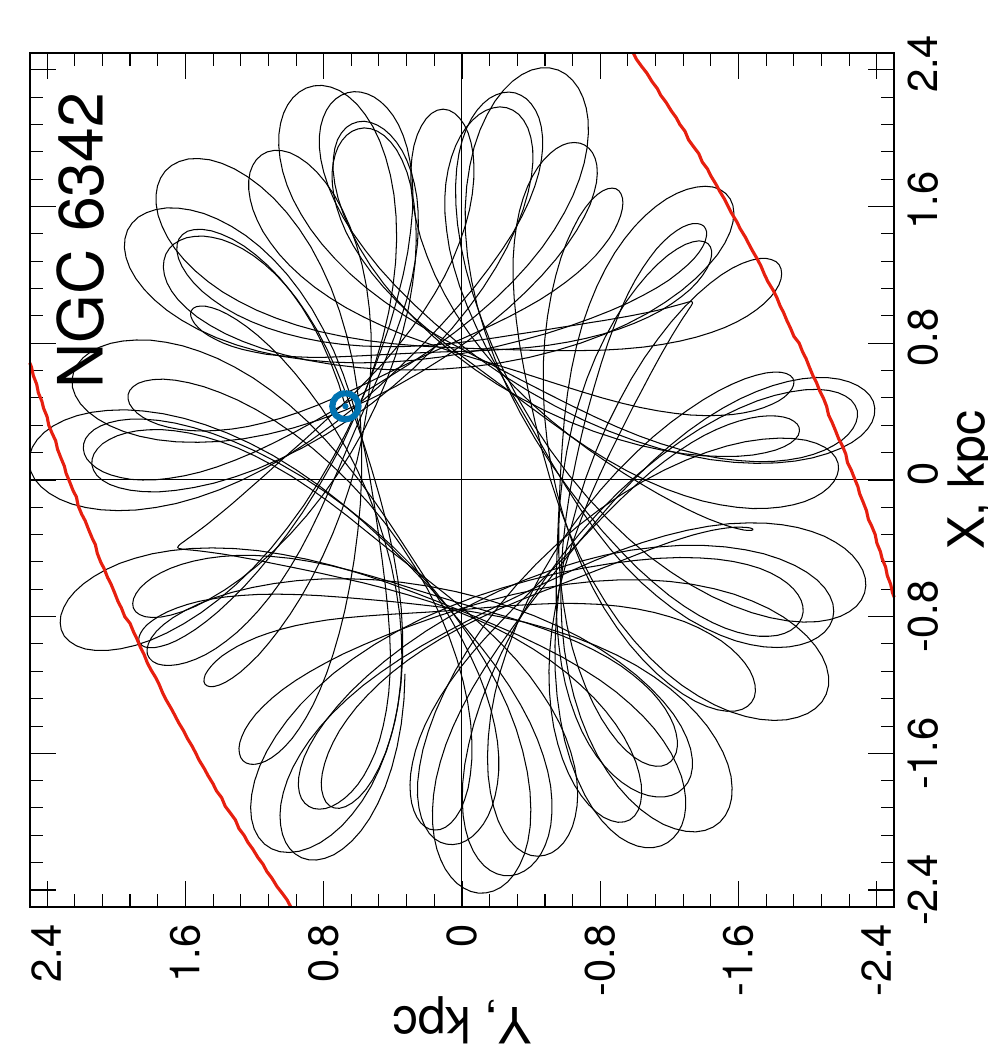}
     \includegraphics[width=0.225\textwidth,angle=-90]{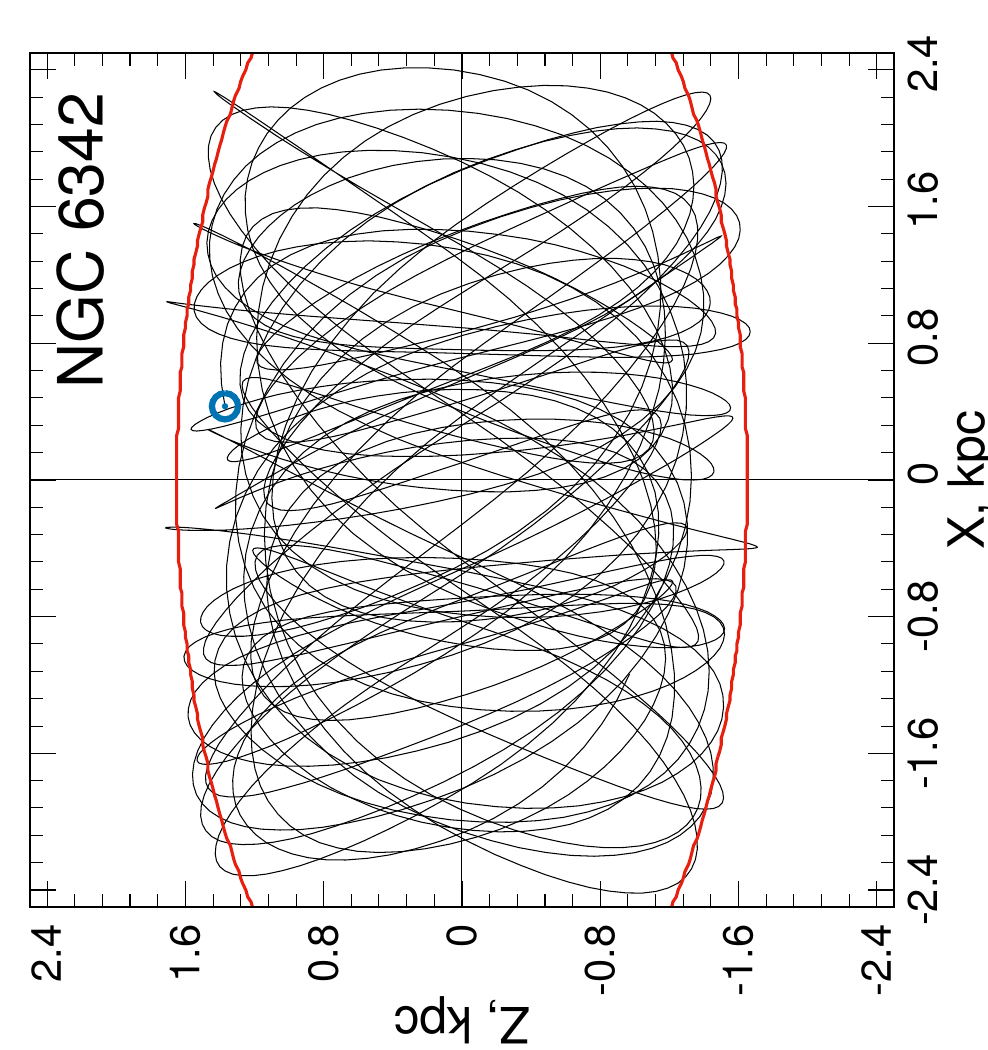}\
   \includegraphics[width=0.225\textwidth,angle=-90]{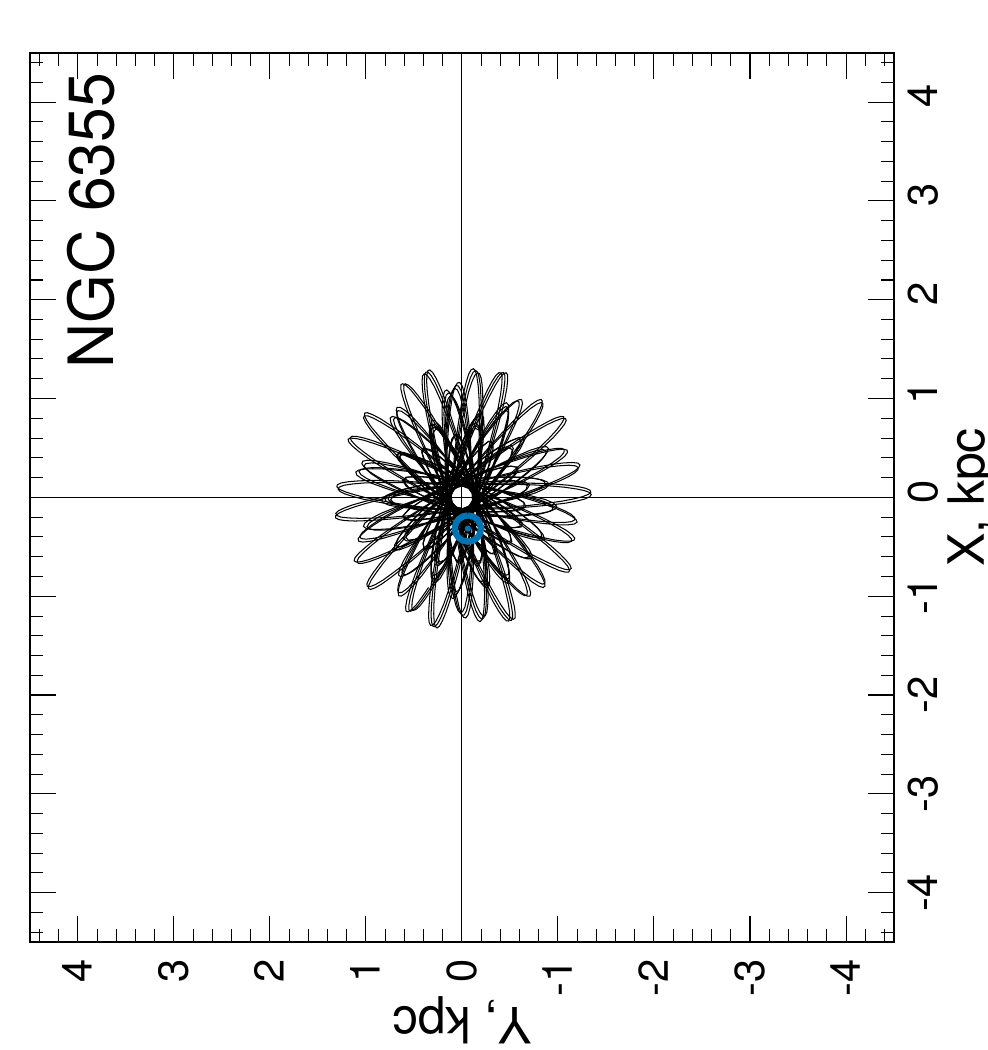}
     \includegraphics[width=0.225\textwidth,angle=-90]{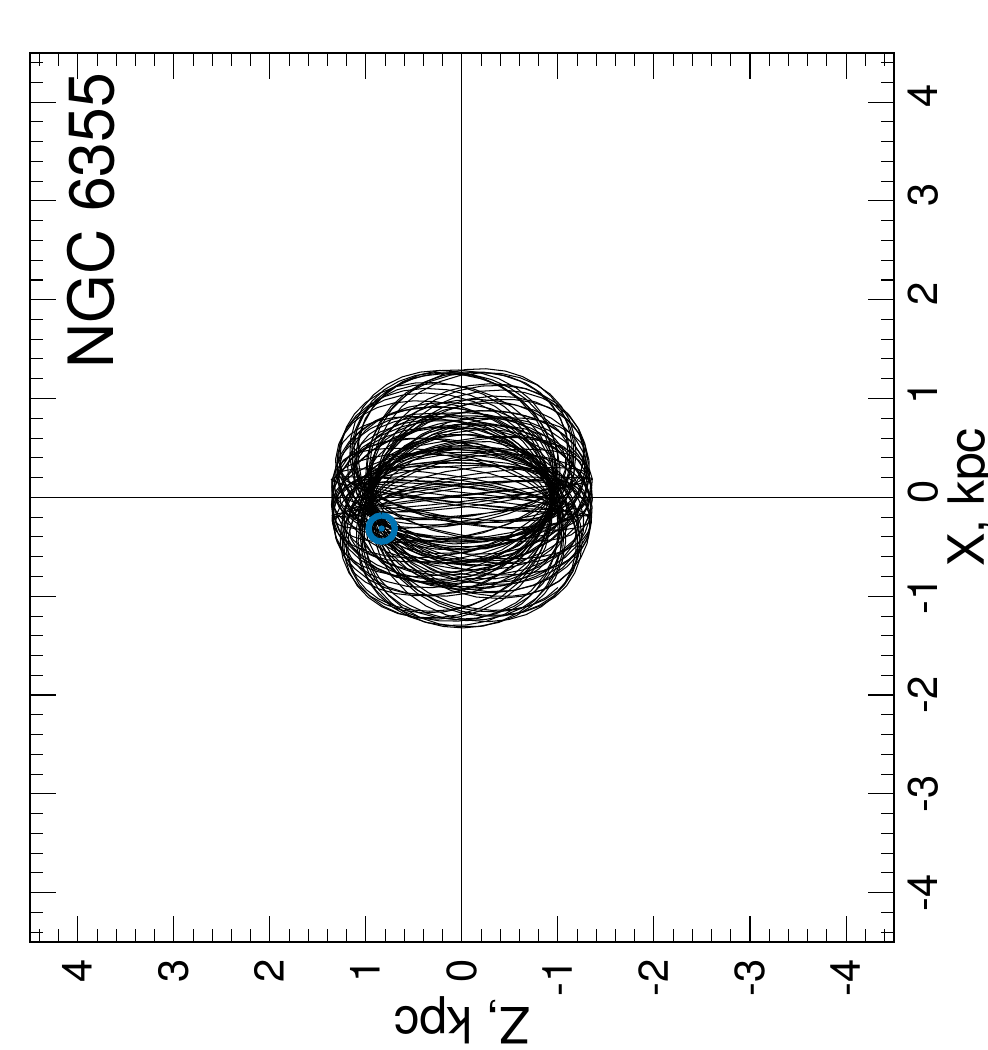}
   \includegraphics[width=0.225\textwidth,angle=-90]{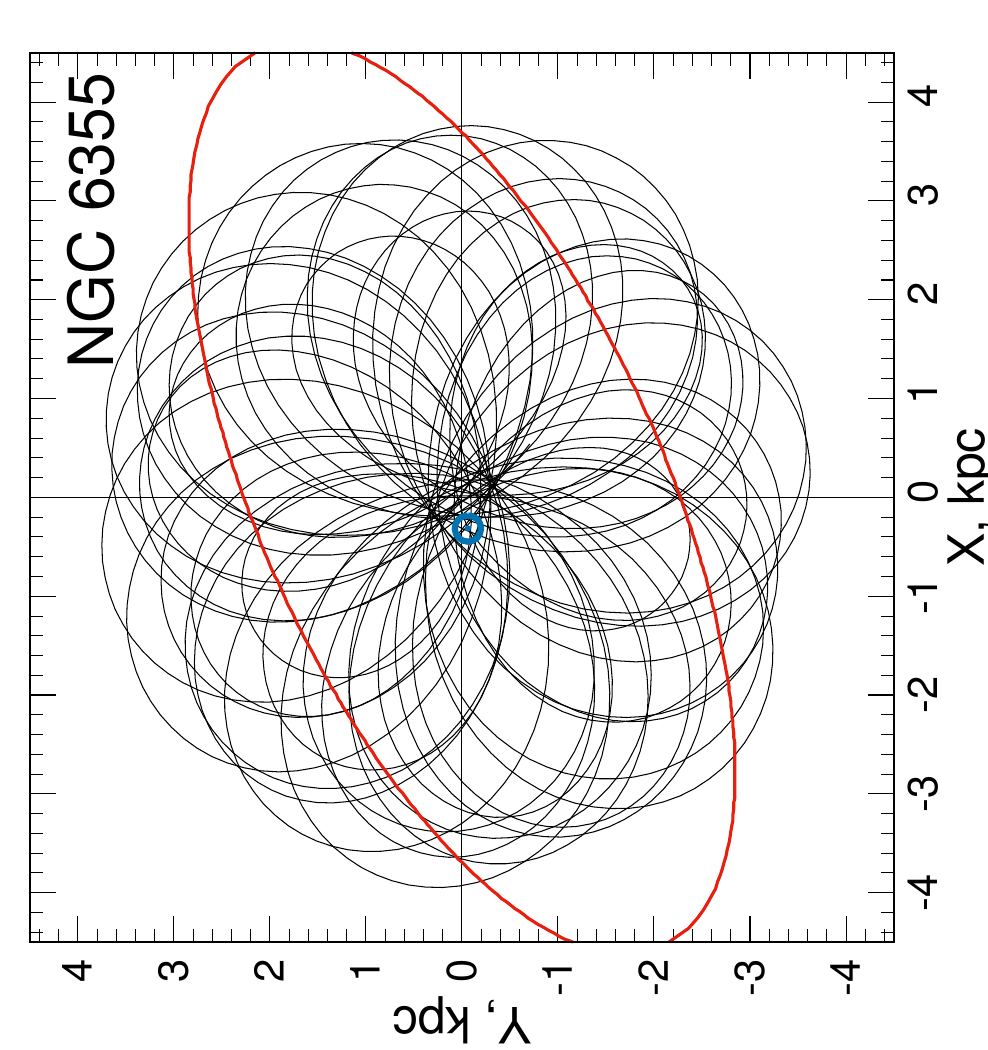}
     \includegraphics[width=0.225\textwidth,angle=-90]{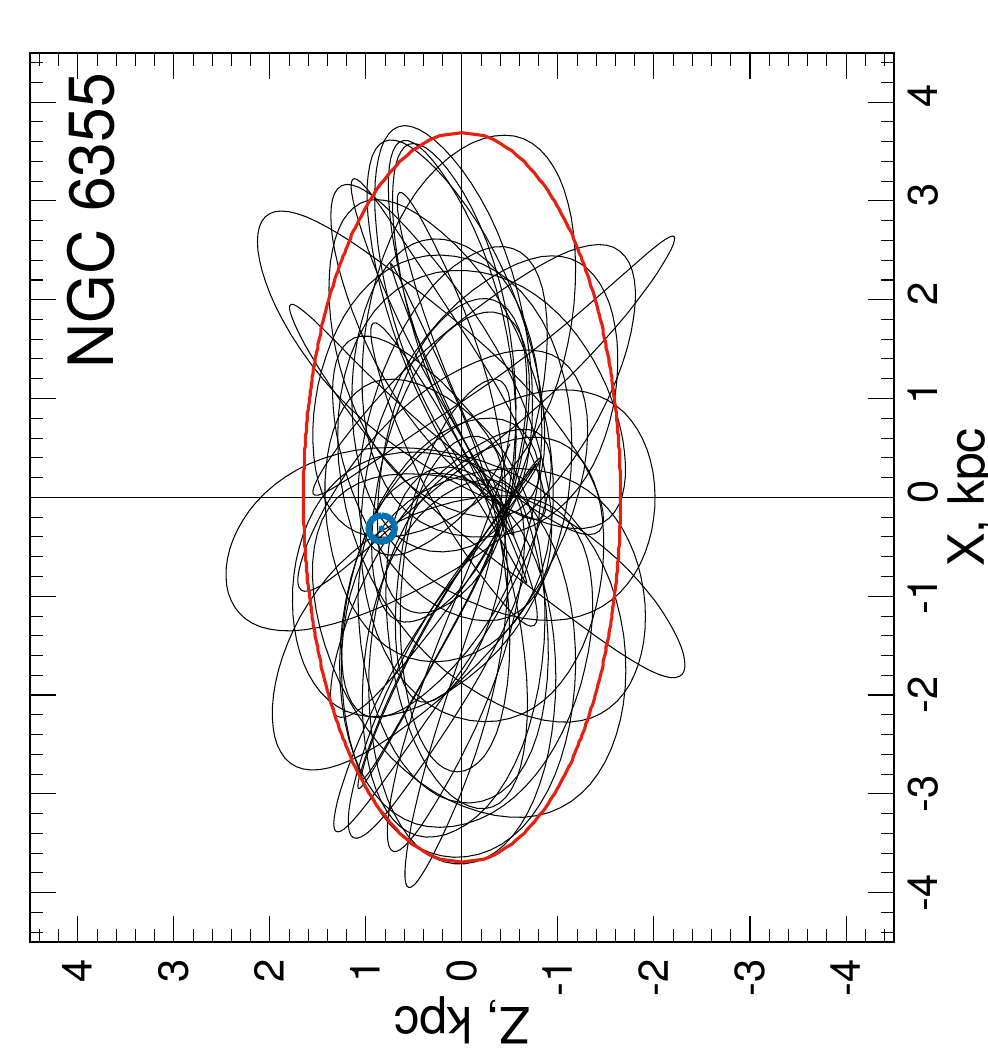}\
 \includegraphics[width=0.225\textwidth,angle=-90]{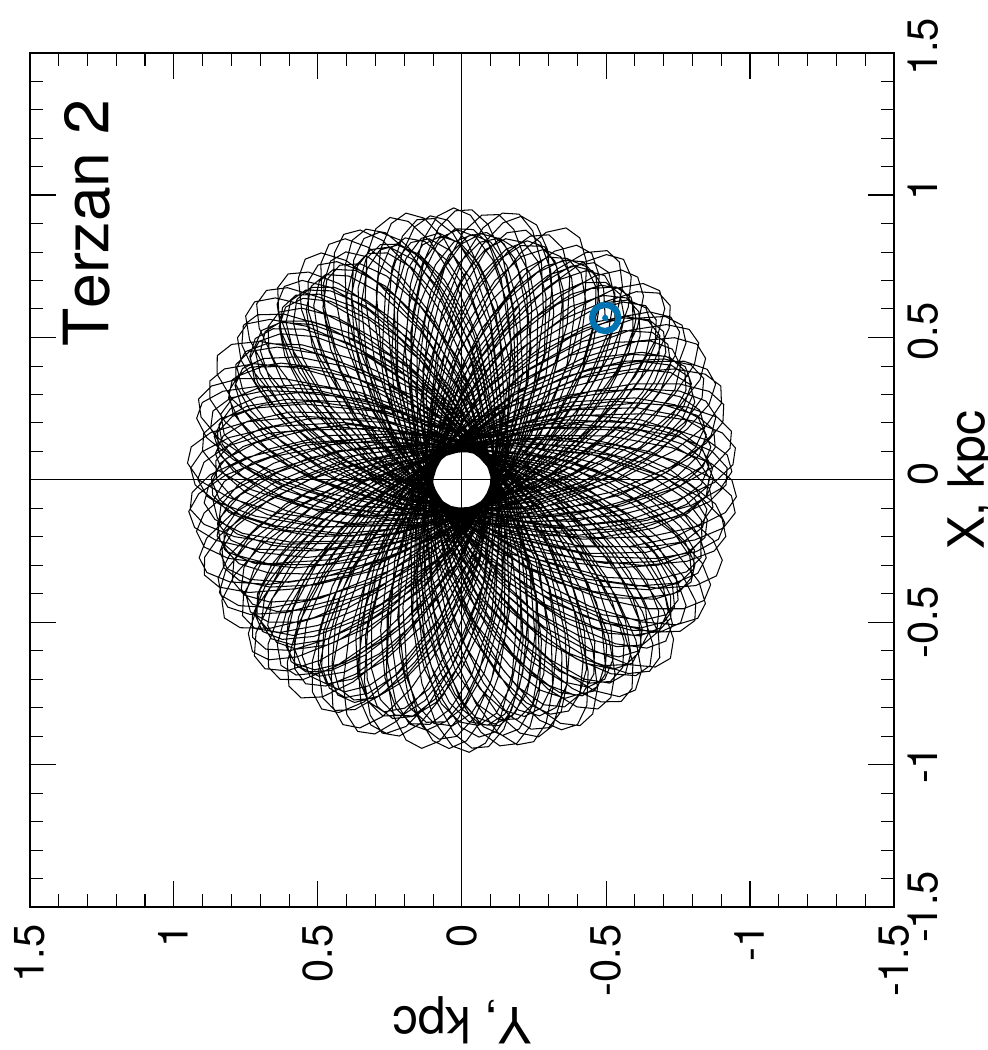}
     \includegraphics[width=0.225\textwidth,angle=-90]{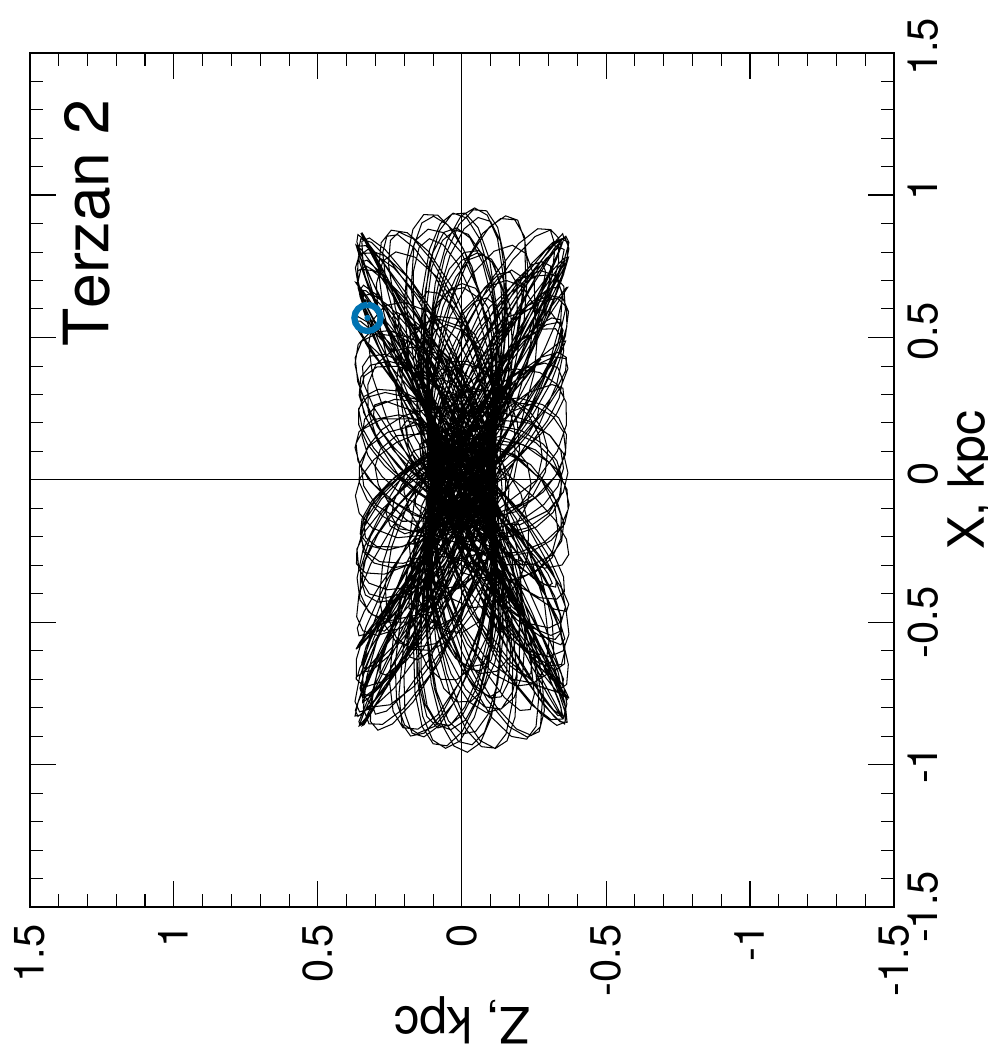}
 \includegraphics[width=0.225\textwidth,angle=-90]{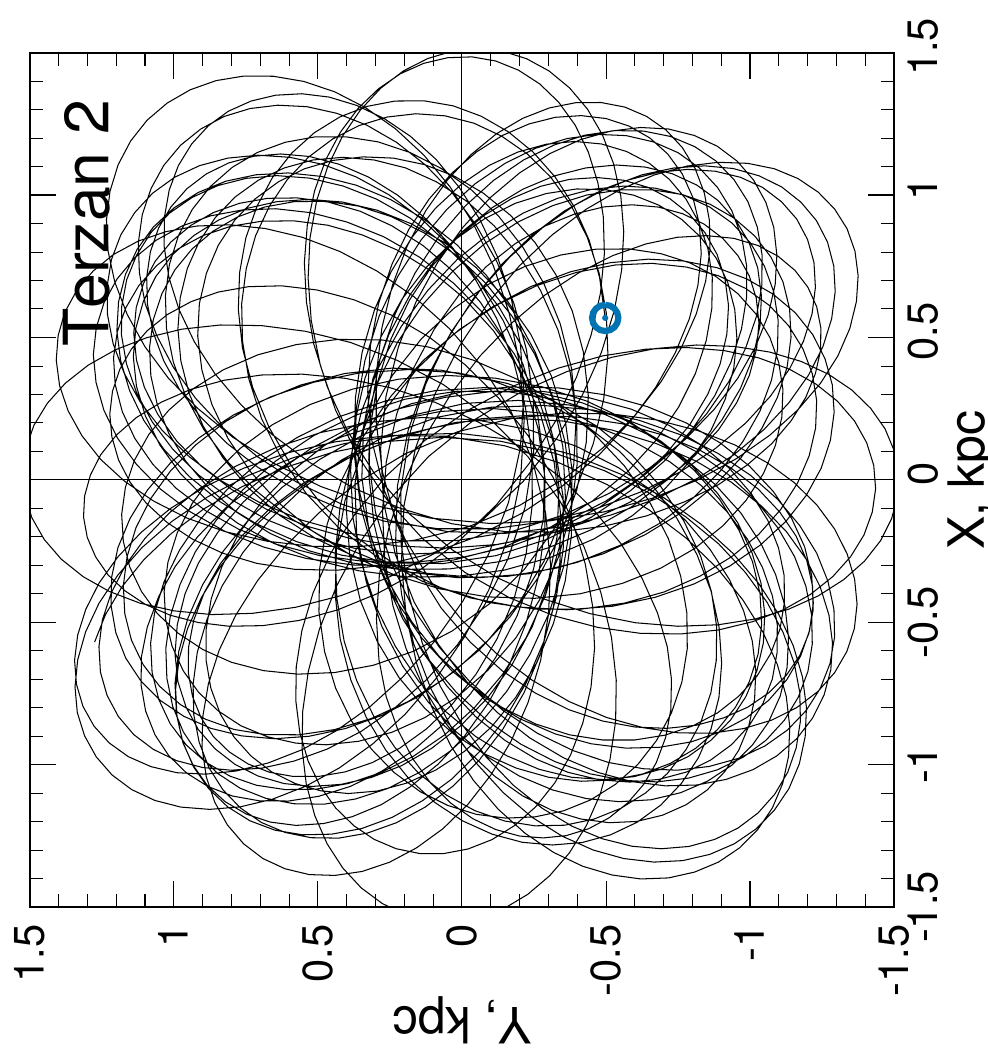}
     \includegraphics[width=0.225\textwidth,angle=-90]{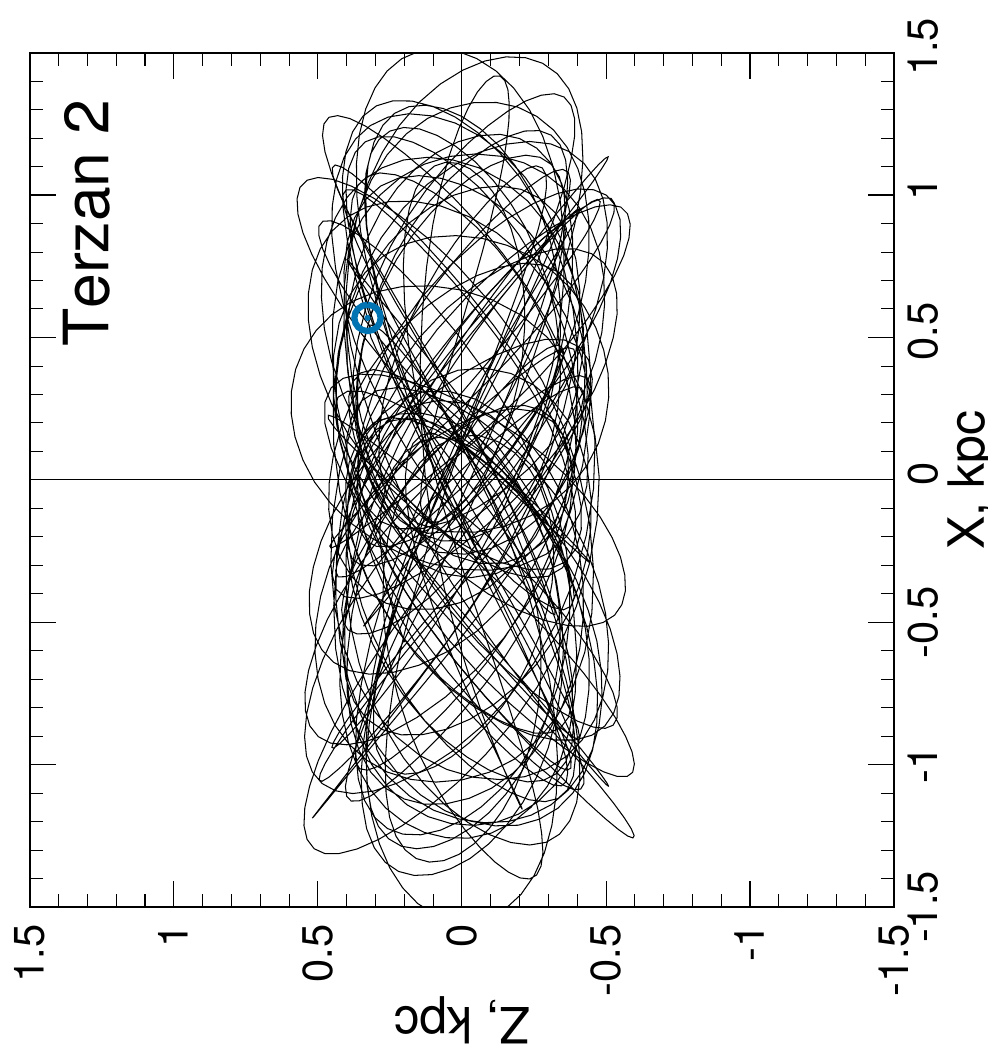}\
         \includegraphics[width=0.225\textwidth,angle=-90]{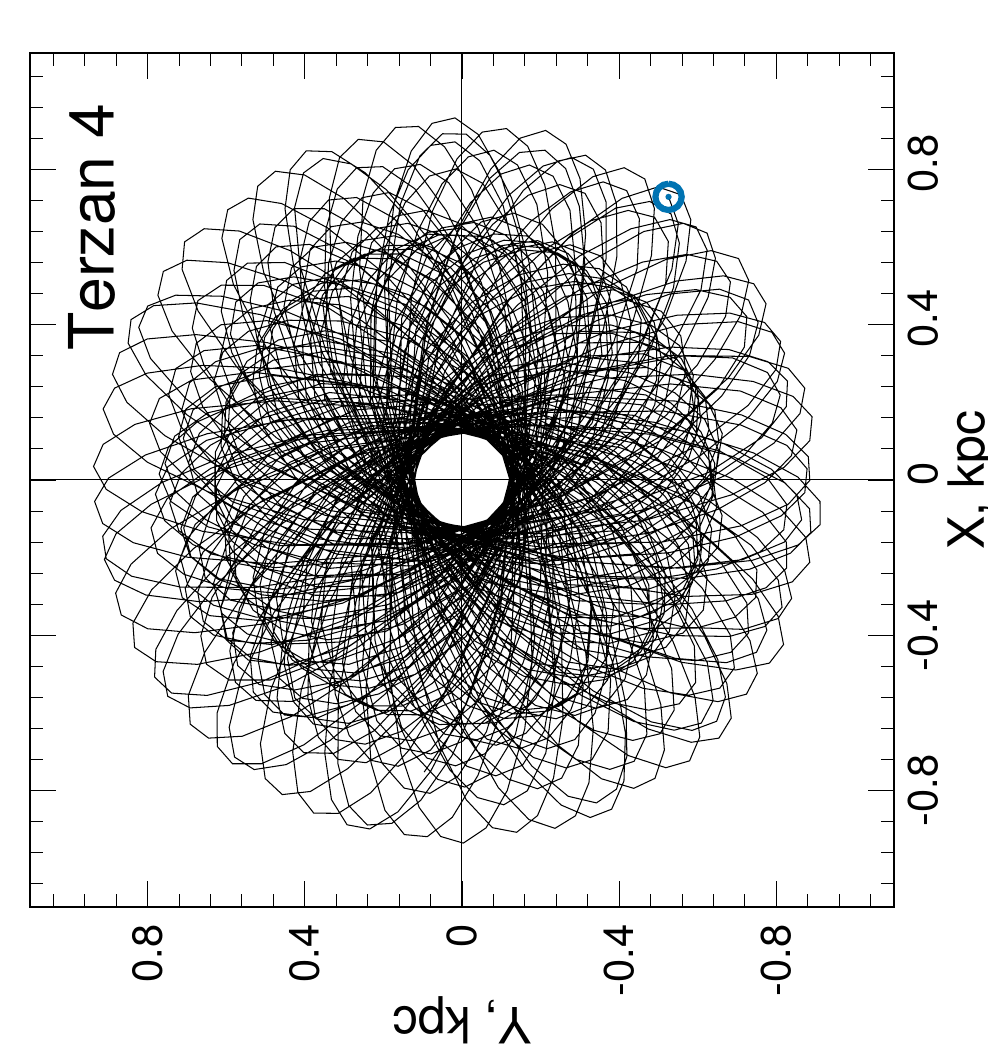}
     \includegraphics[width=0.225\textwidth,angle=-90]{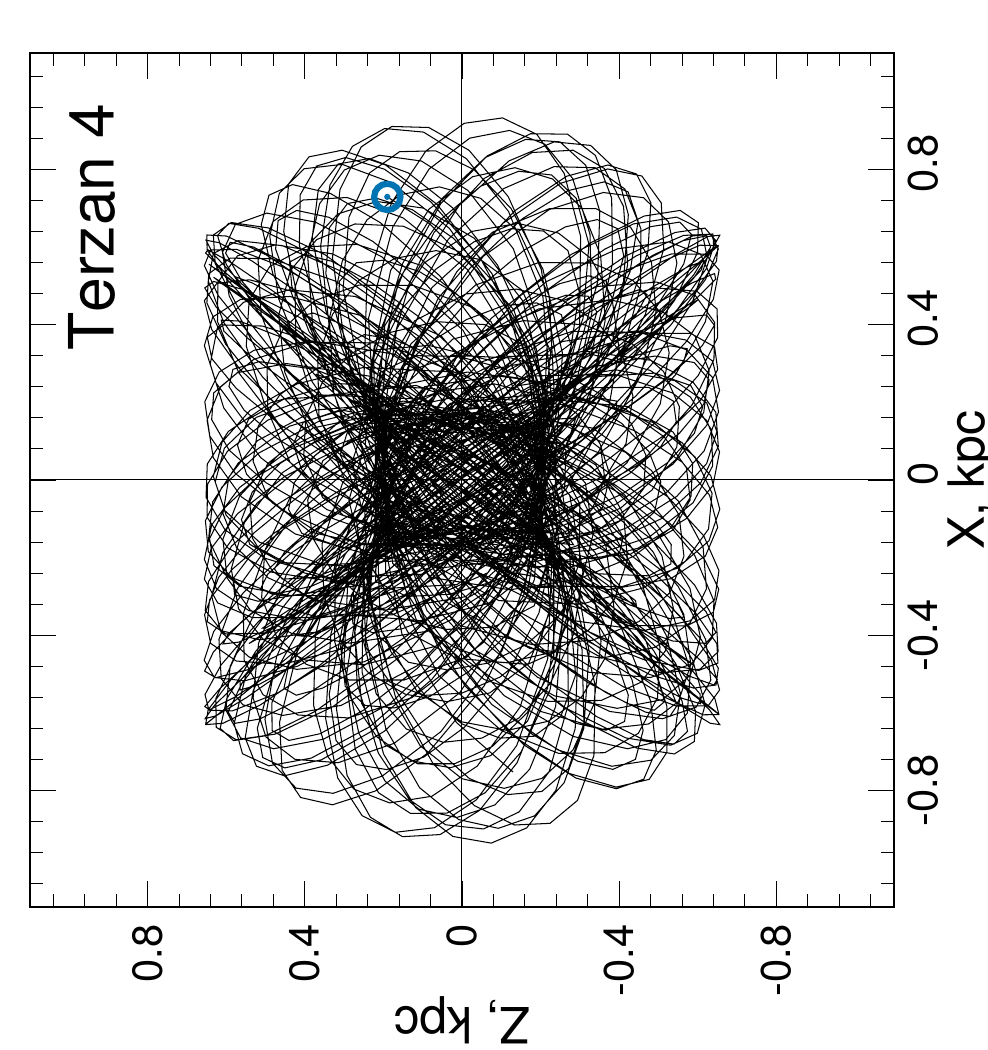}
         \includegraphics[width=0.225\textwidth,angle=-90]{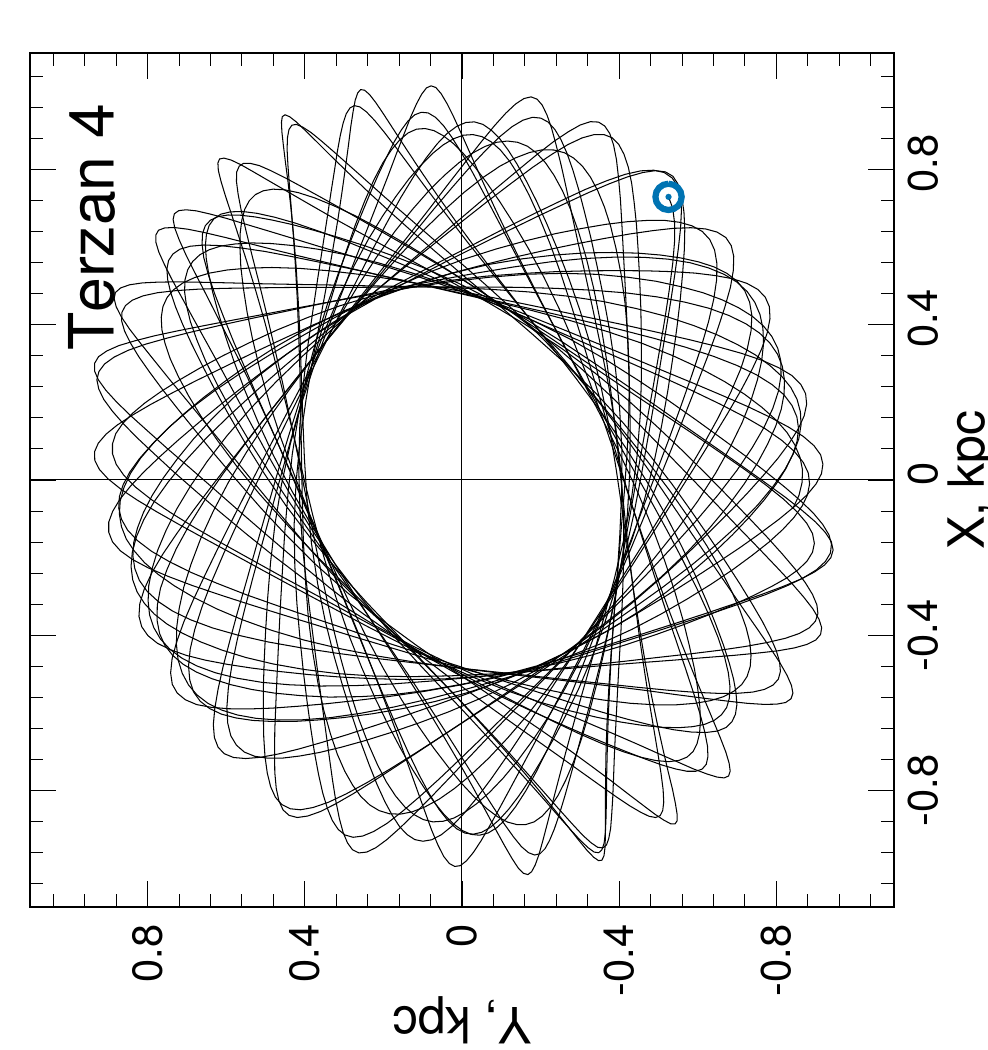}
     \includegraphics[width=0.225\textwidth,angle=-90]{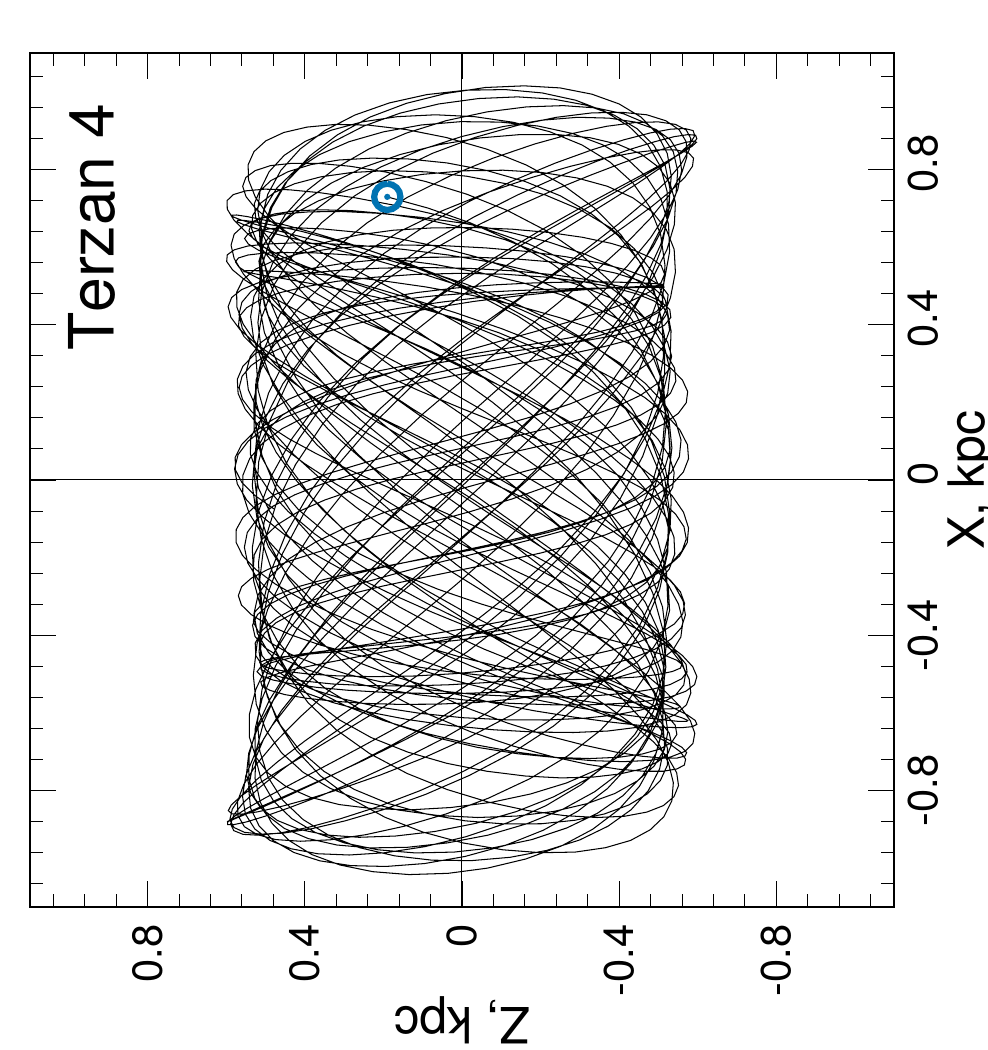}\
   \includegraphics[width=0.225\textwidth,angle=-90]{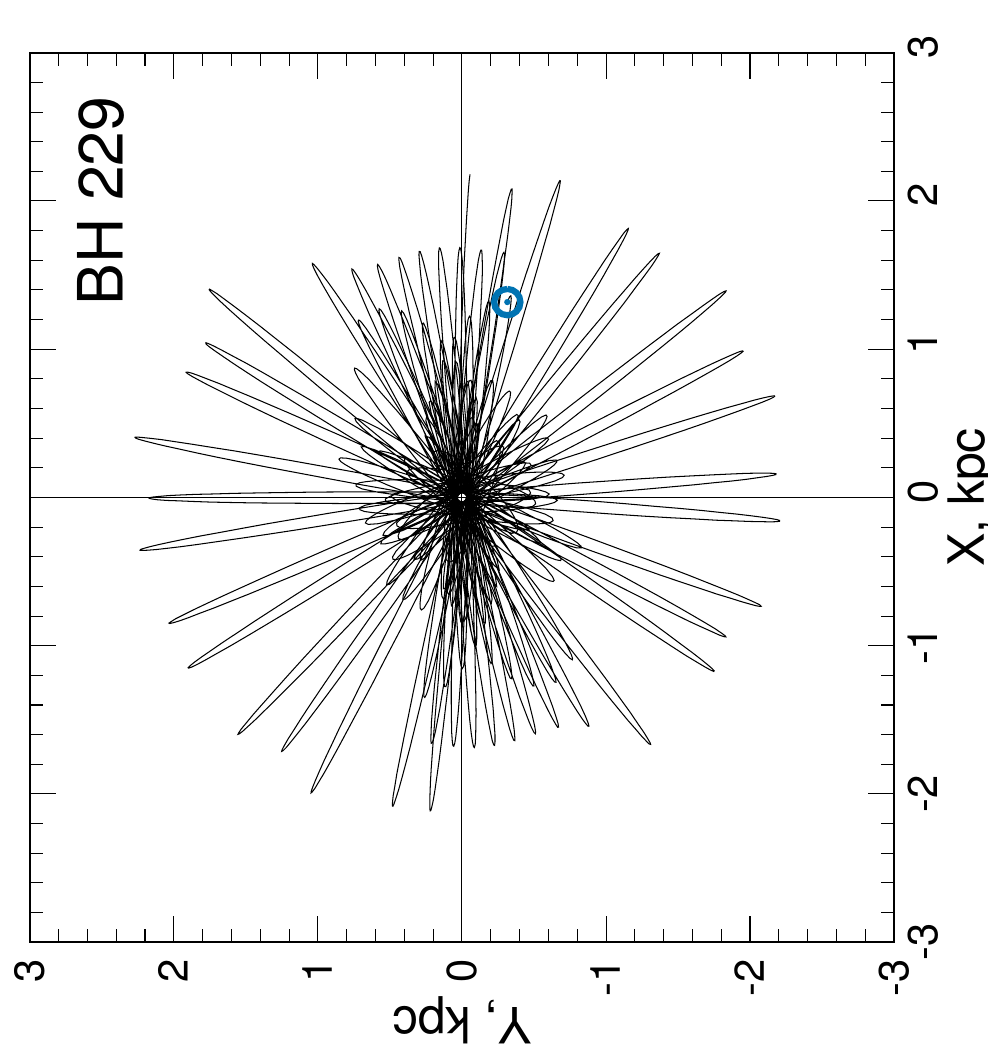}
     \includegraphics[width=0.225\textwidth,angle=-90]{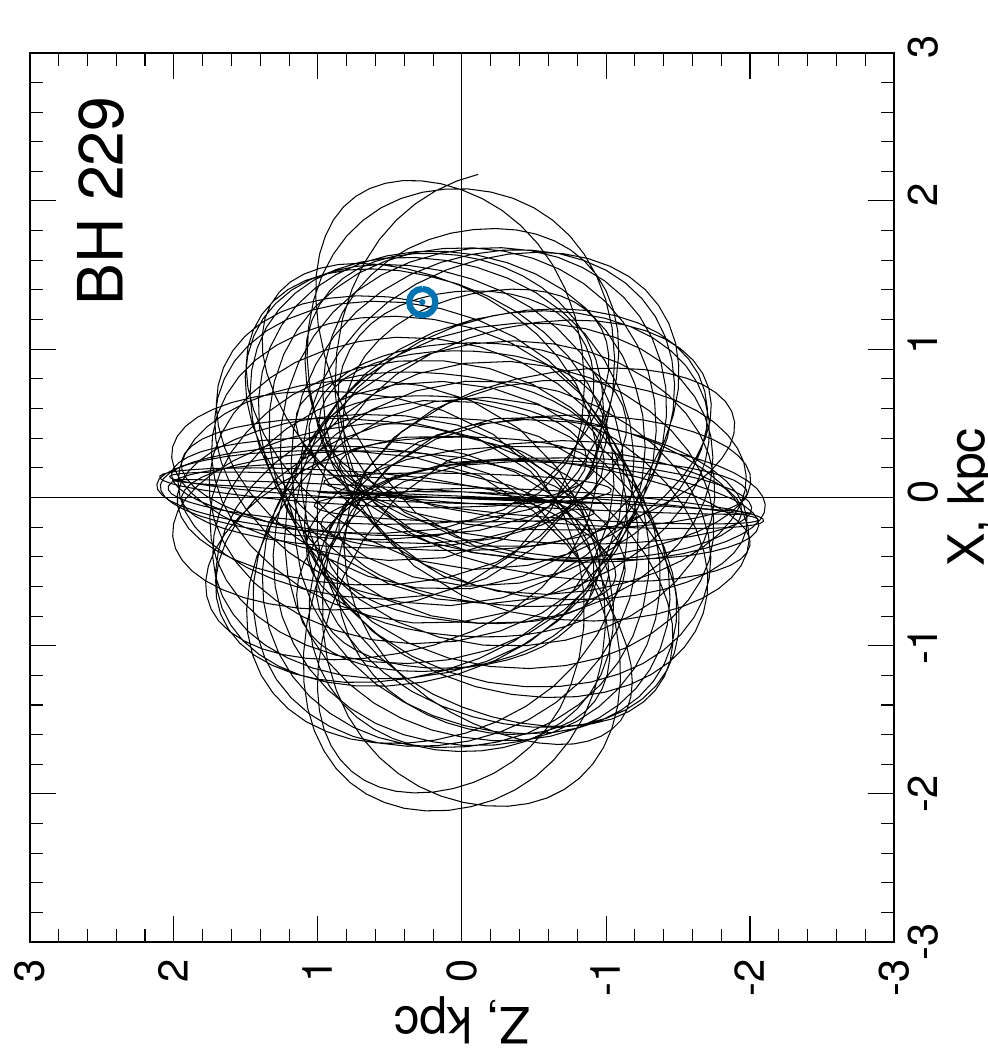}
   \includegraphics[width=0.225\textwidth,angle=-90]{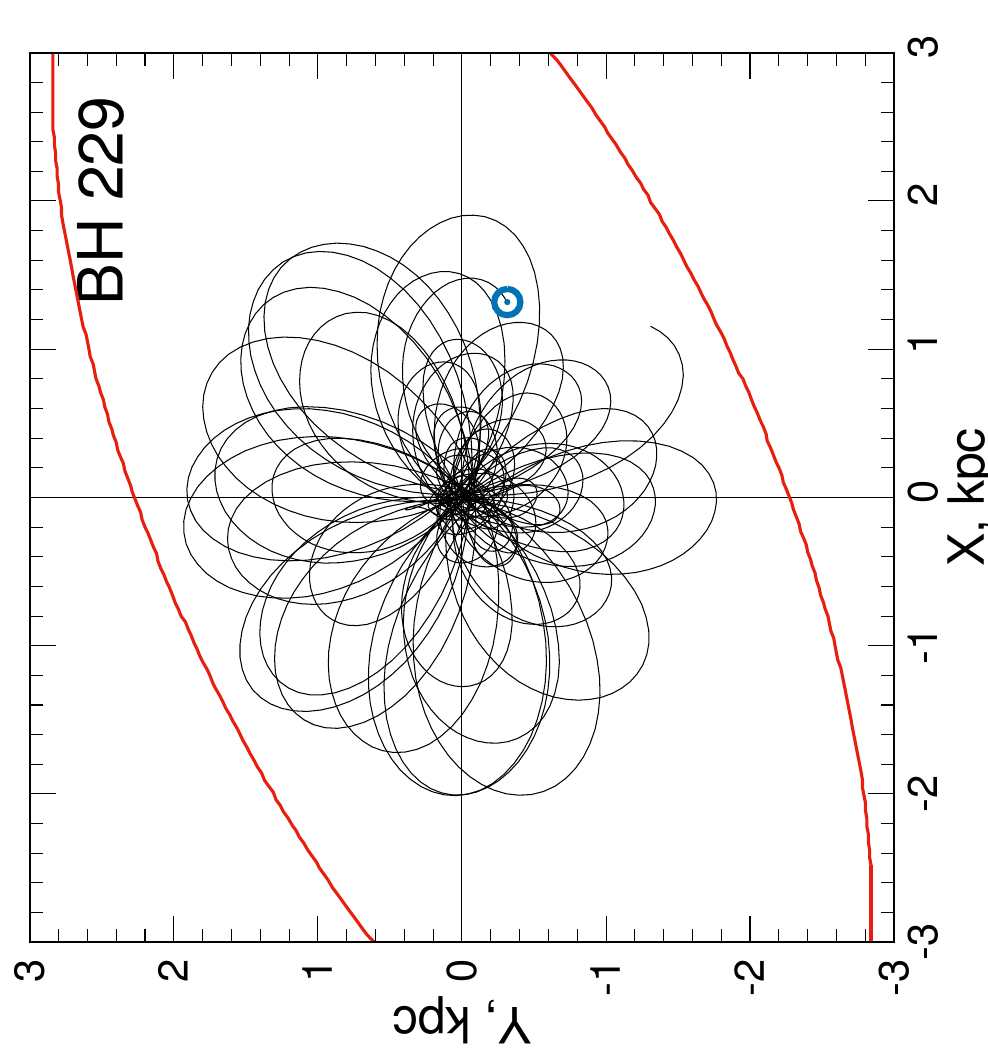}
     \includegraphics[width=0.225\textwidth,angle=-90]{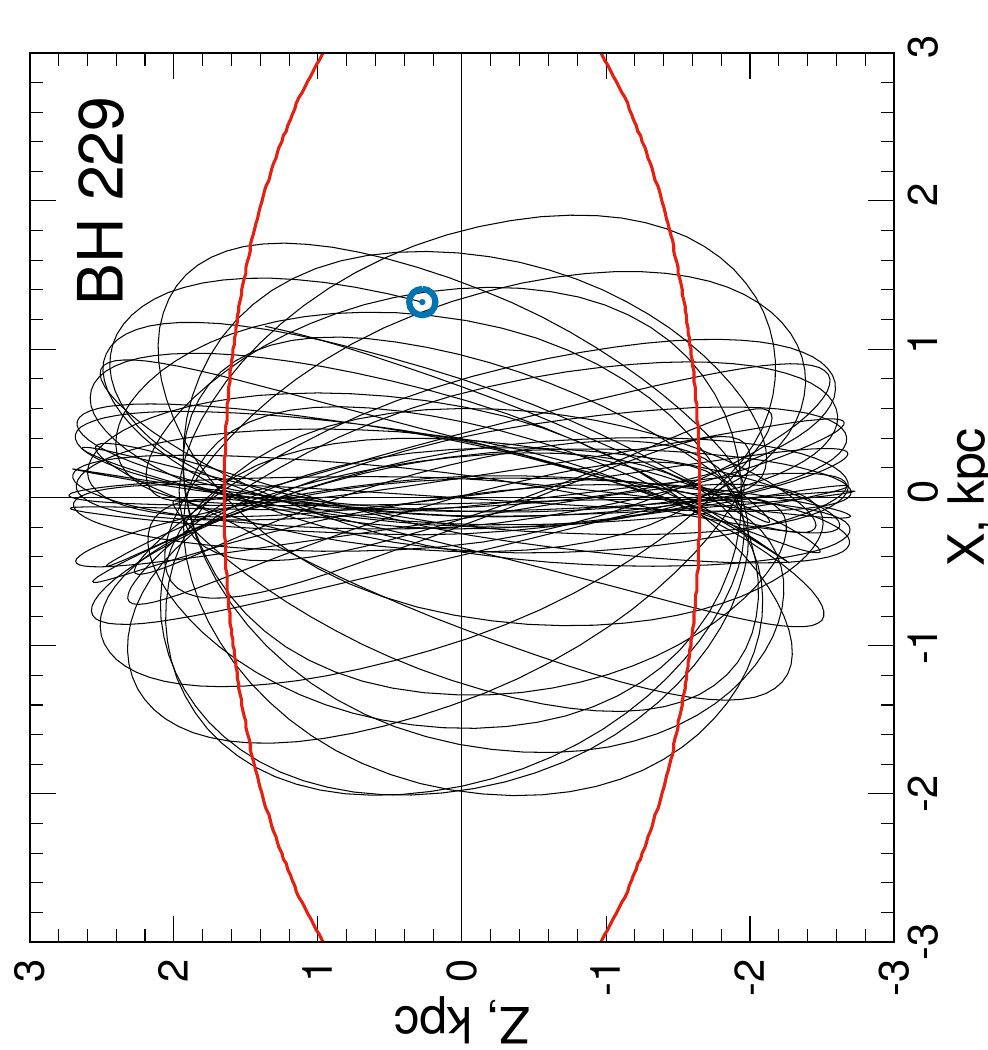}\
        \includegraphics[width=0.225\textwidth,angle=-90]{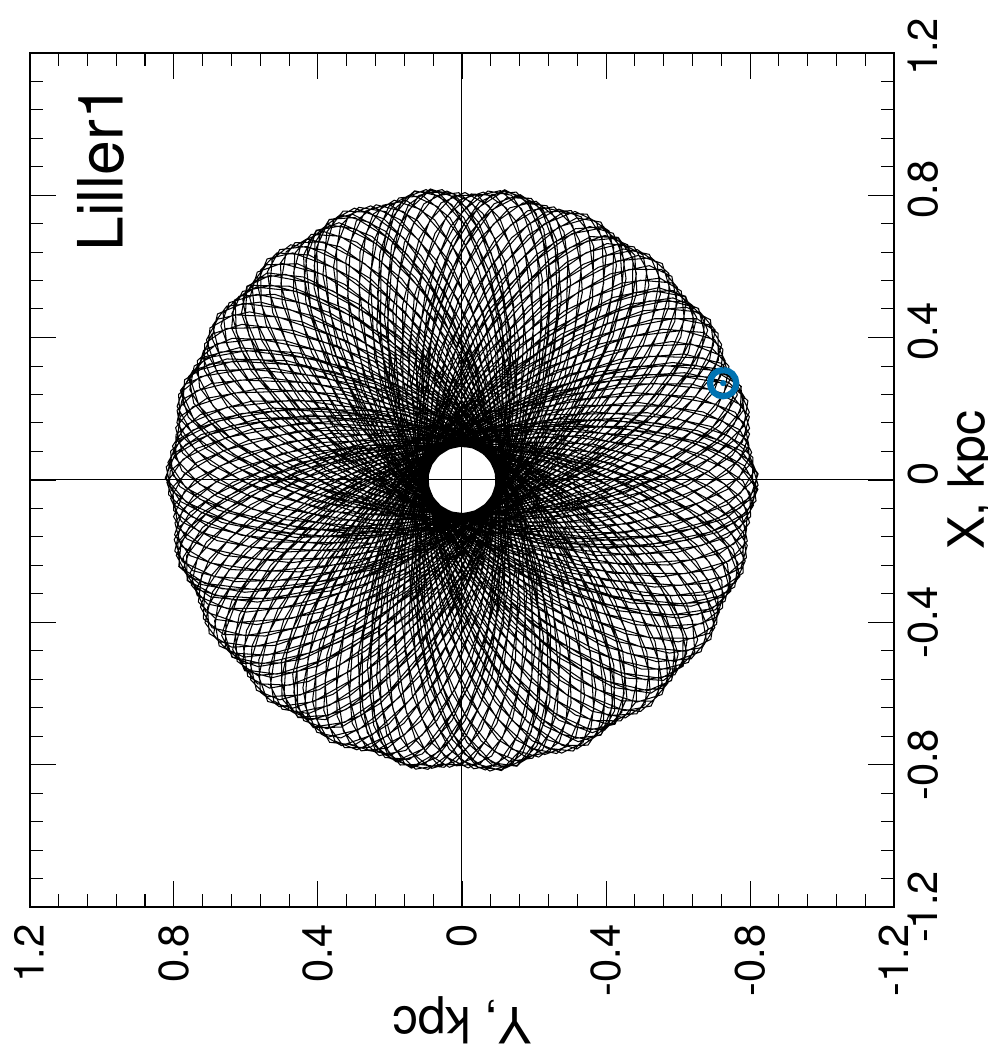}
     \includegraphics[width=0.225\textwidth,angle=-90]{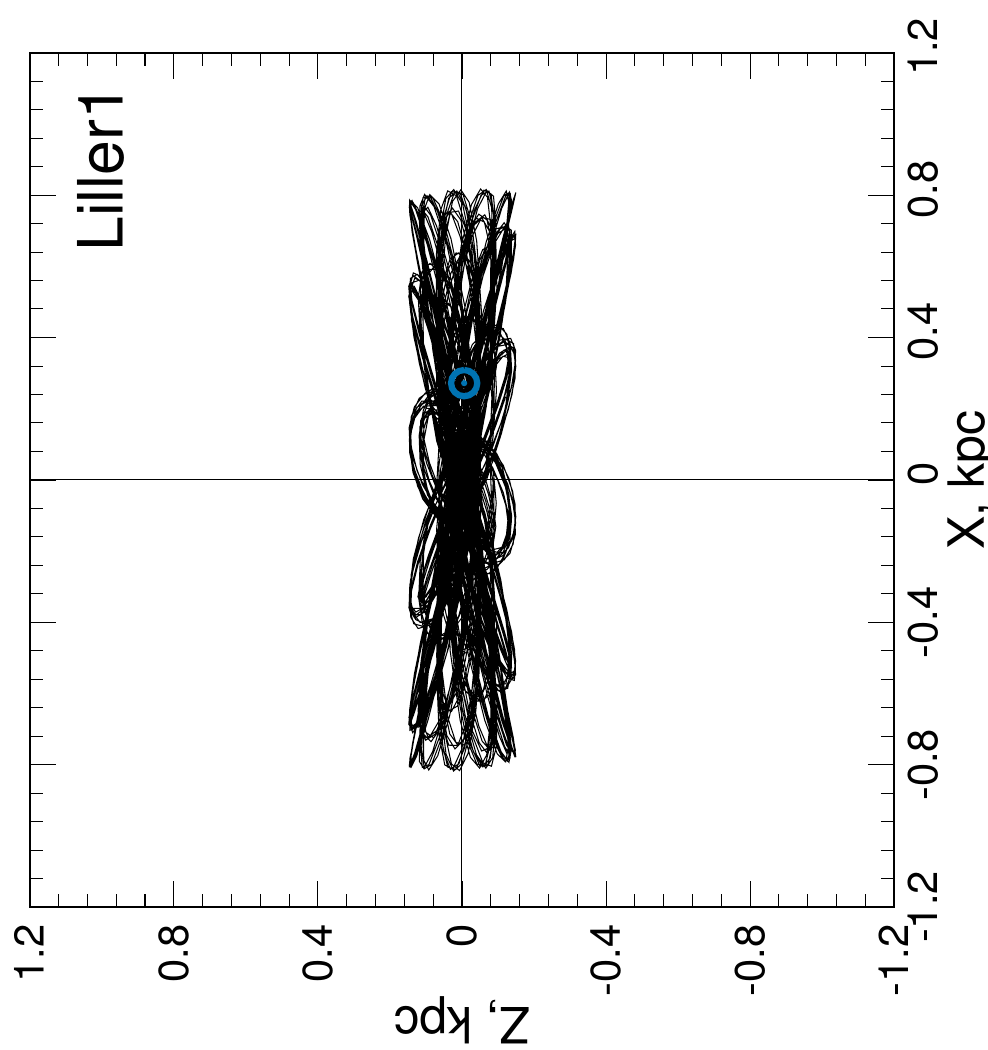}
        \includegraphics[width=0.225\textwidth,angle=-90]{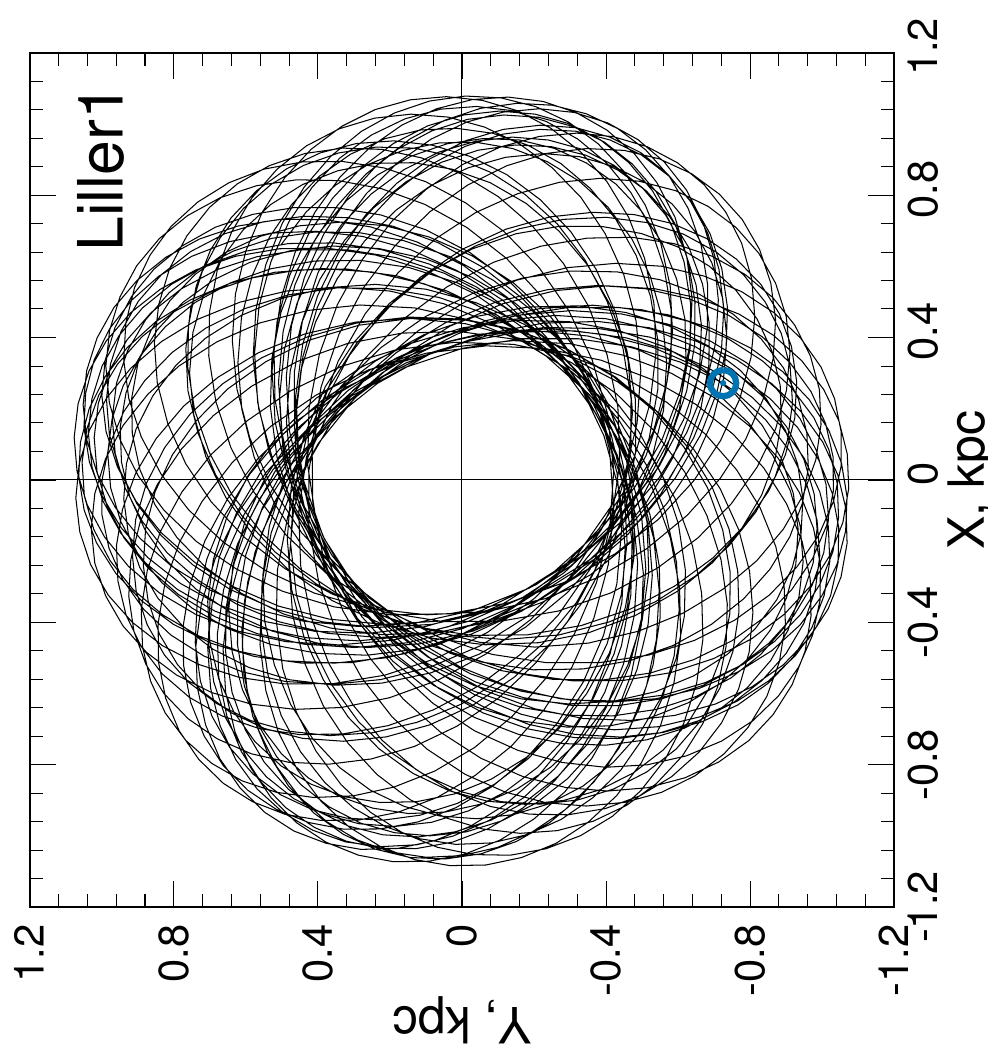}
     \includegraphics[width=0.225\textwidth,angle=-90]{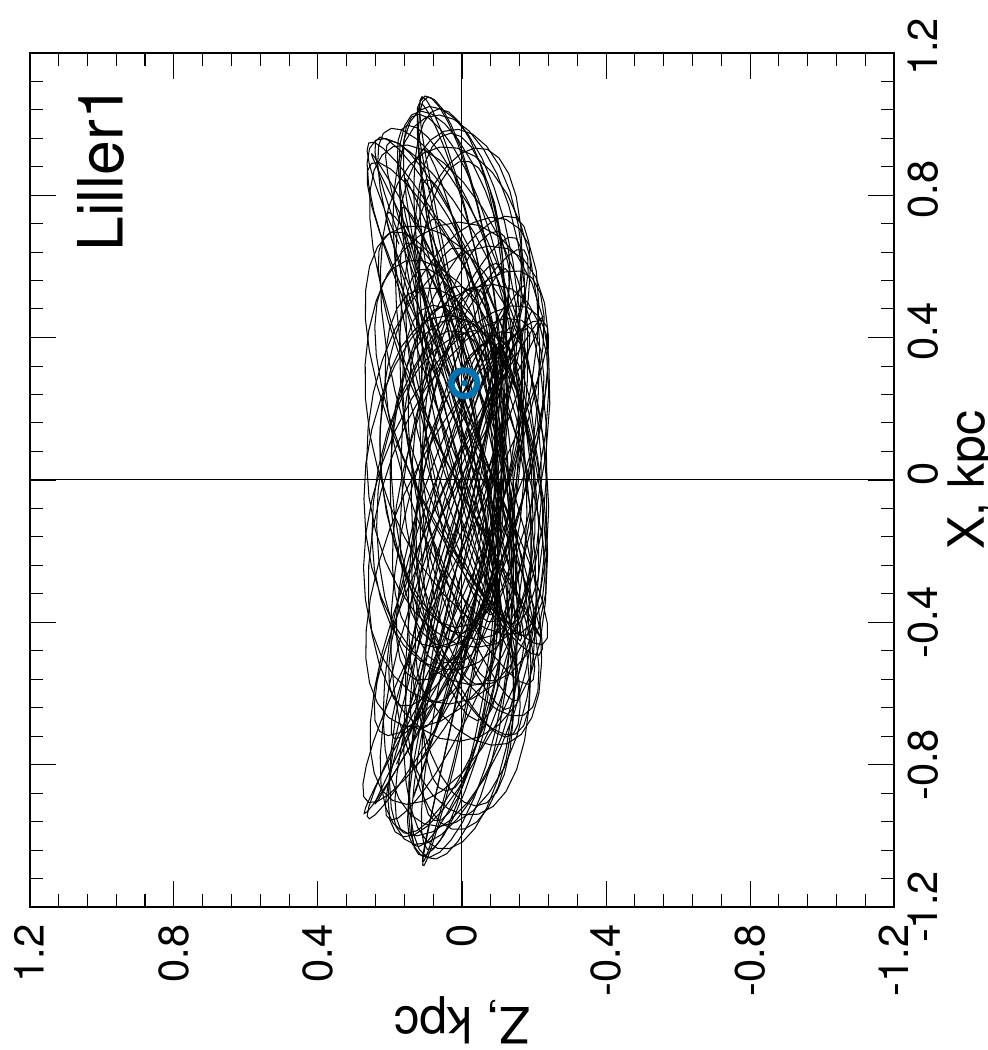}\

\medskip

 \centerline{APPENDIX. Continued}
\label{fD}
\end{center}}
\end{figure*}

\begin{figure*}
{\begin{center}
   \includegraphics[width=0.225\textwidth,angle=-90]{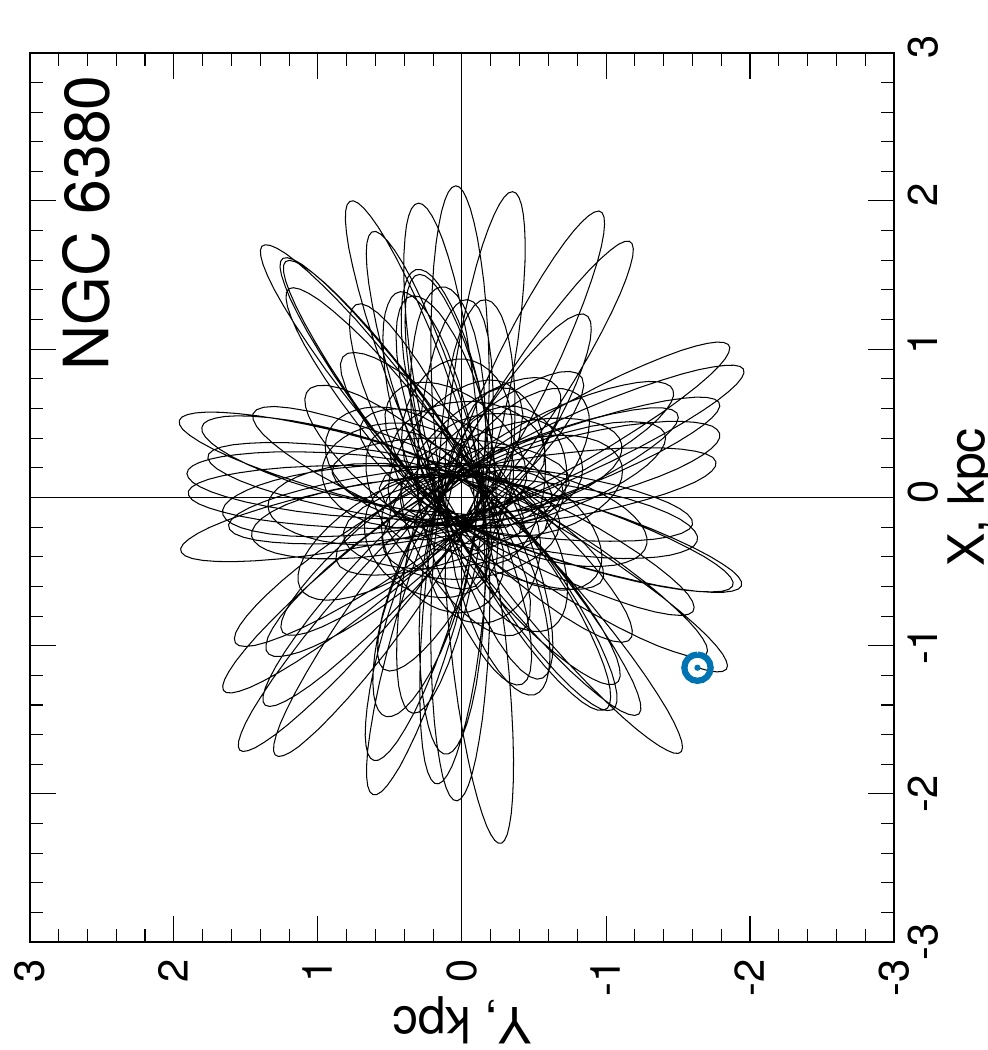}
     \includegraphics[width=0.225\textwidth,angle=-90]{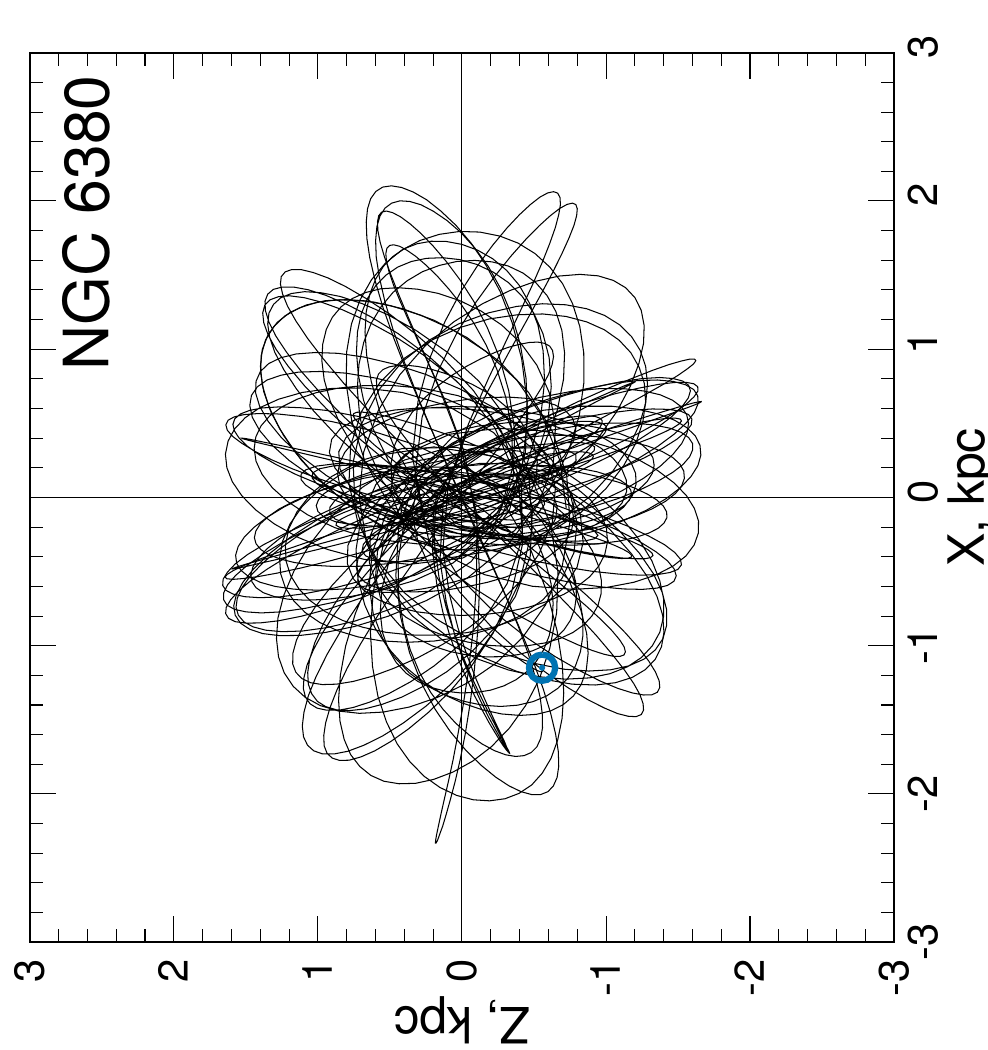}
         \includegraphics[width=0.225\textwidth,angle=-90]{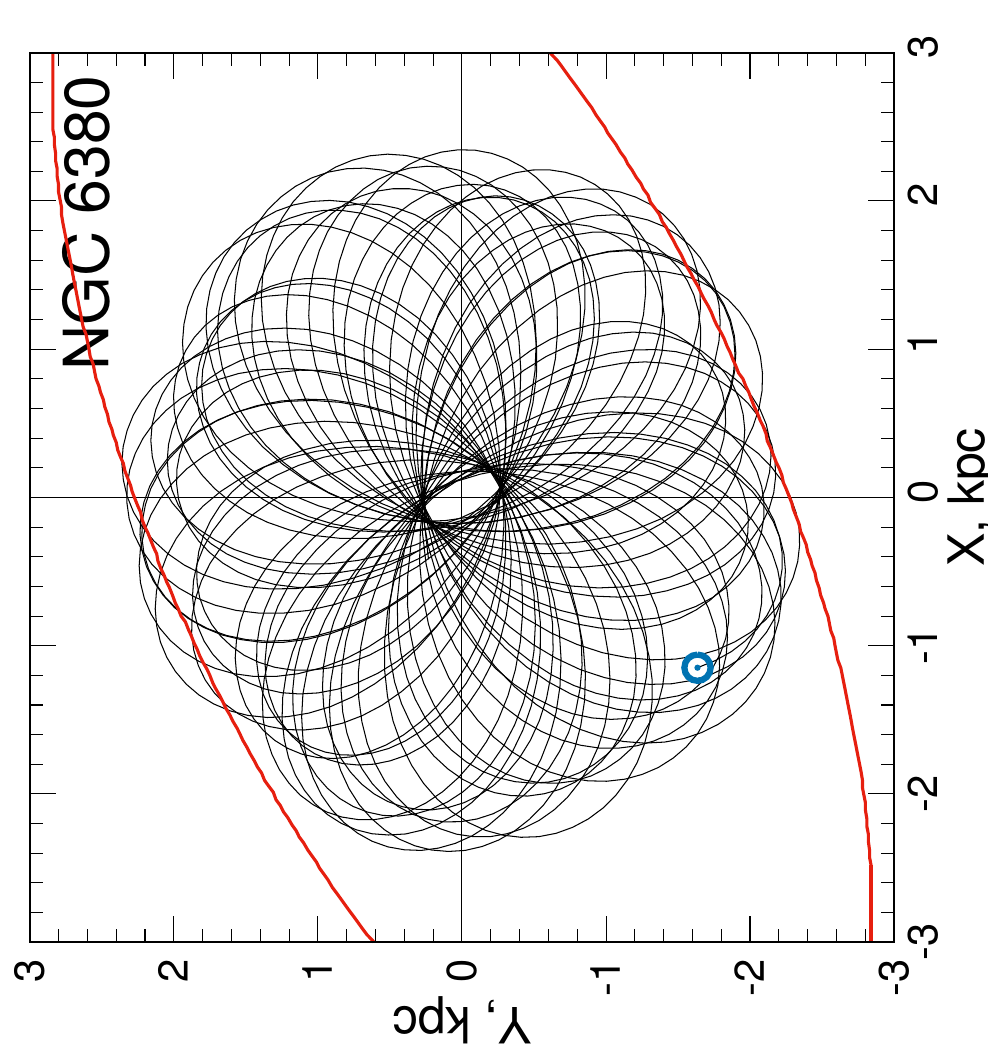}
     \includegraphics[width=0.225\textwidth,angle=-90]{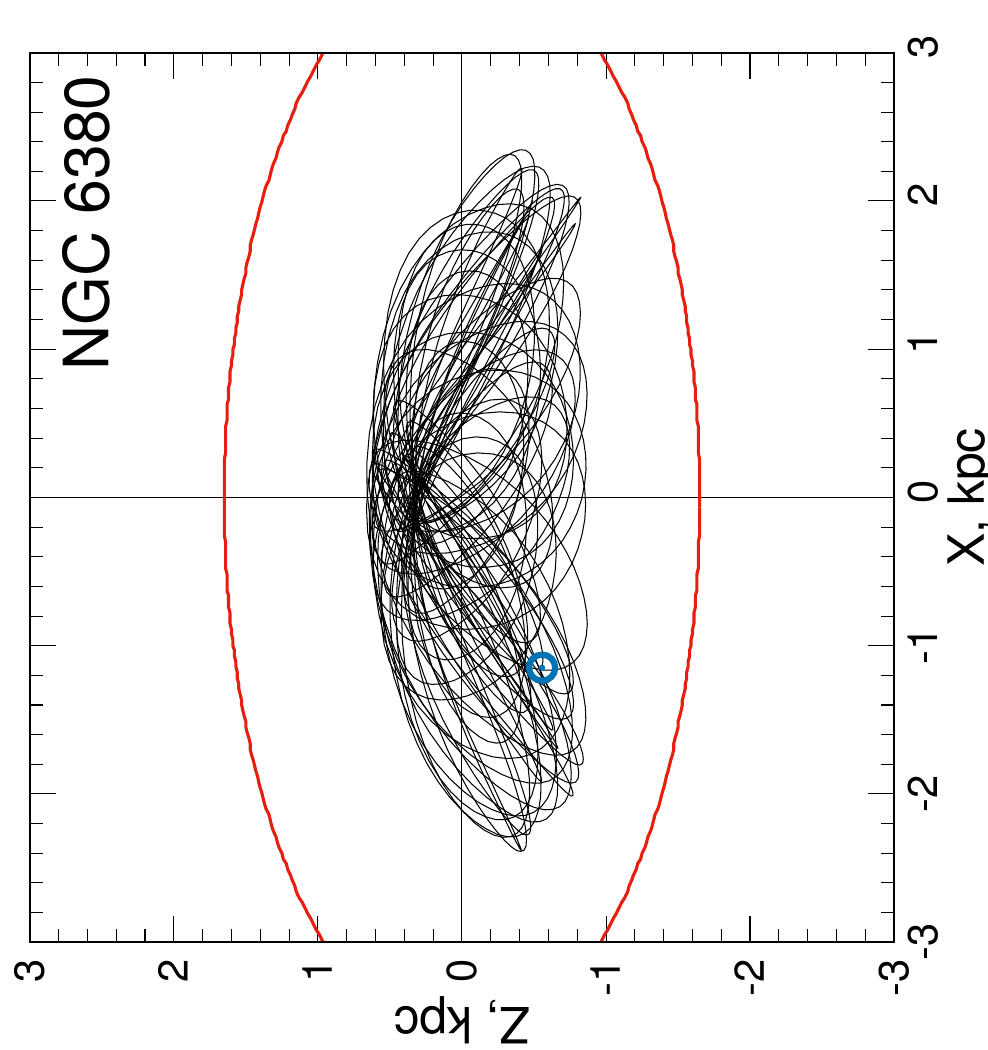}\
   \includegraphics[width=0.225\textwidth,angle=-90]{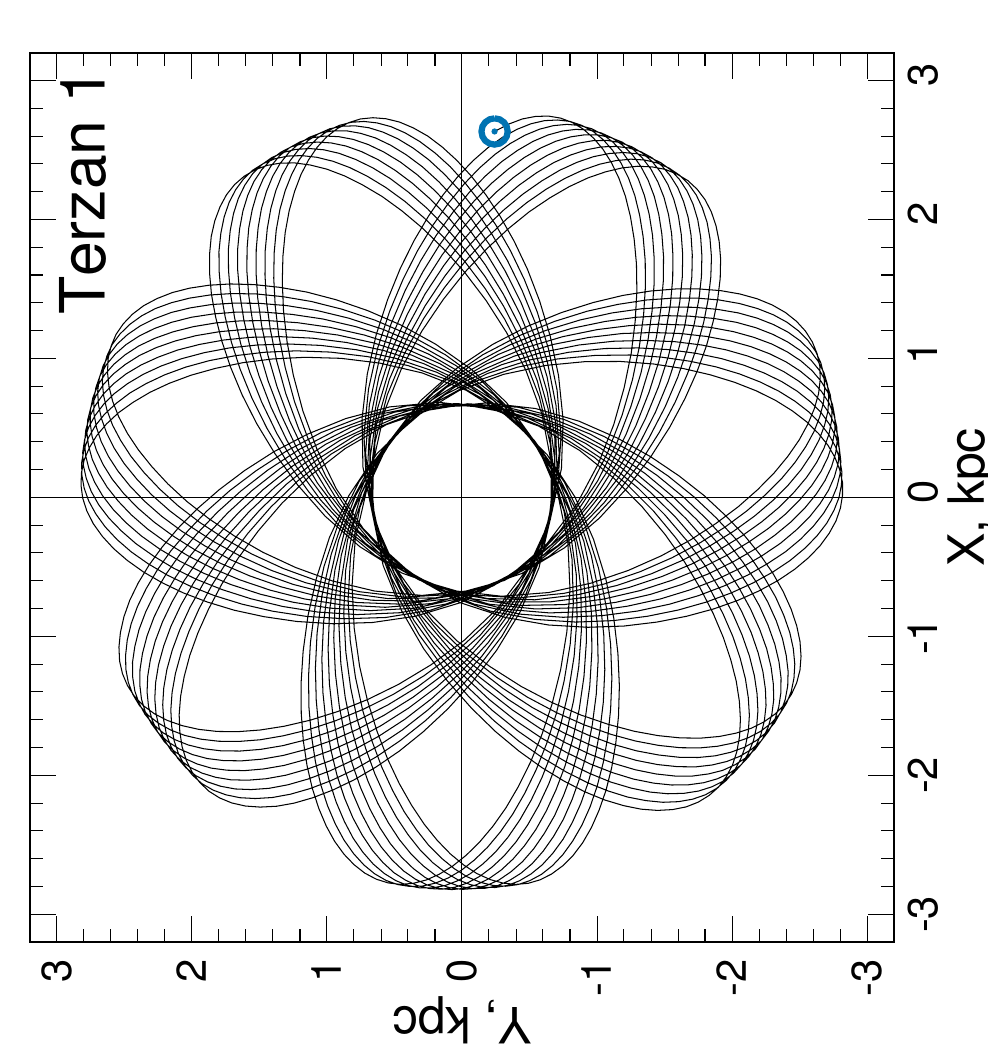}
     \includegraphics[width=0.225\textwidth,angle=-90]{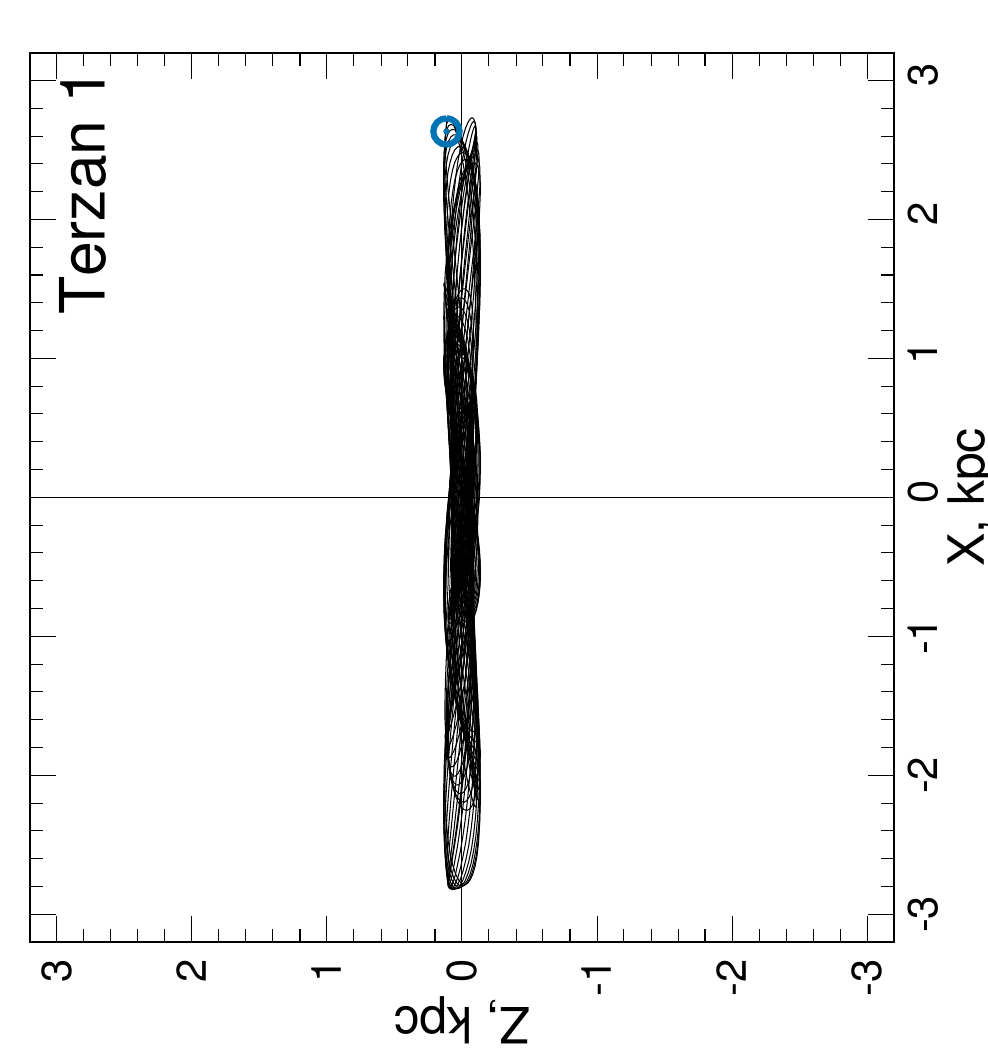}
   \includegraphics[width=0.225\textwidth,angle=-90]{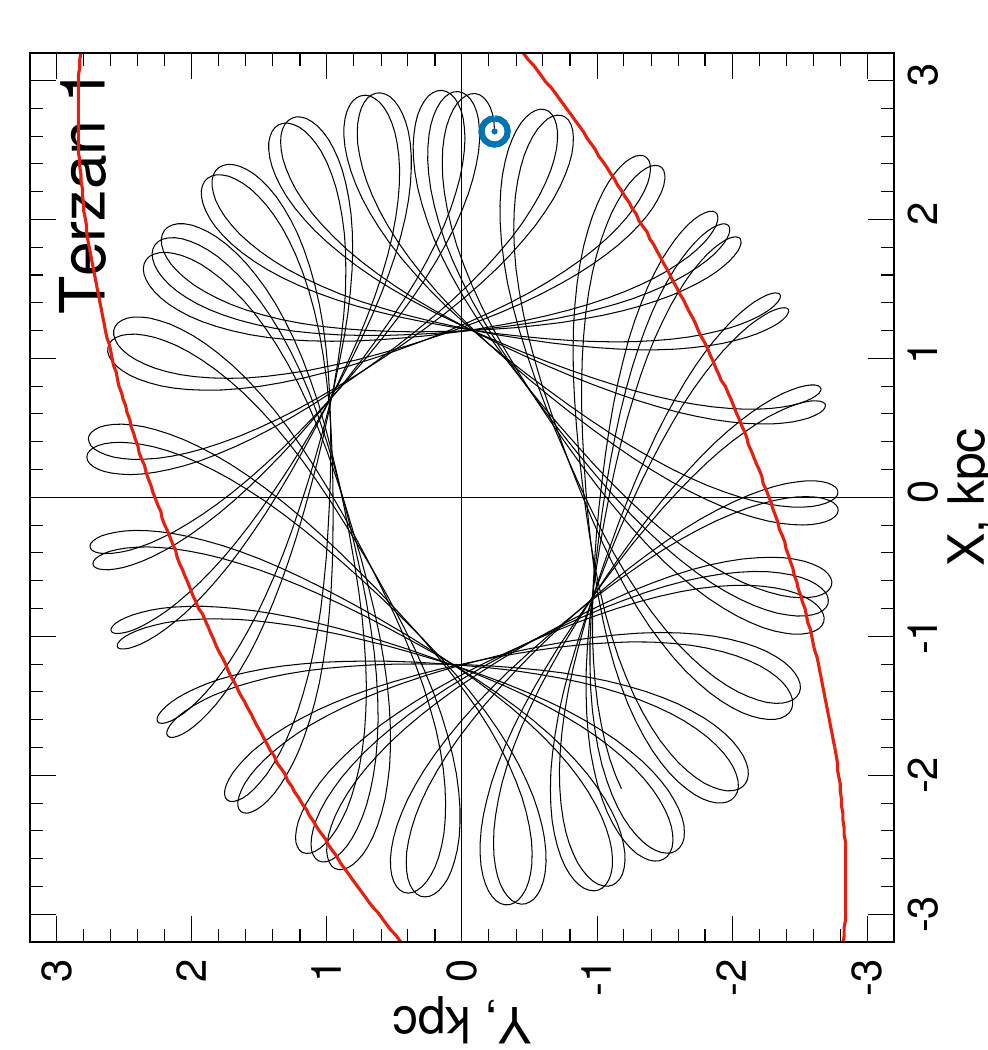}
     \includegraphics[width=0.225\textwidth,angle=-90]{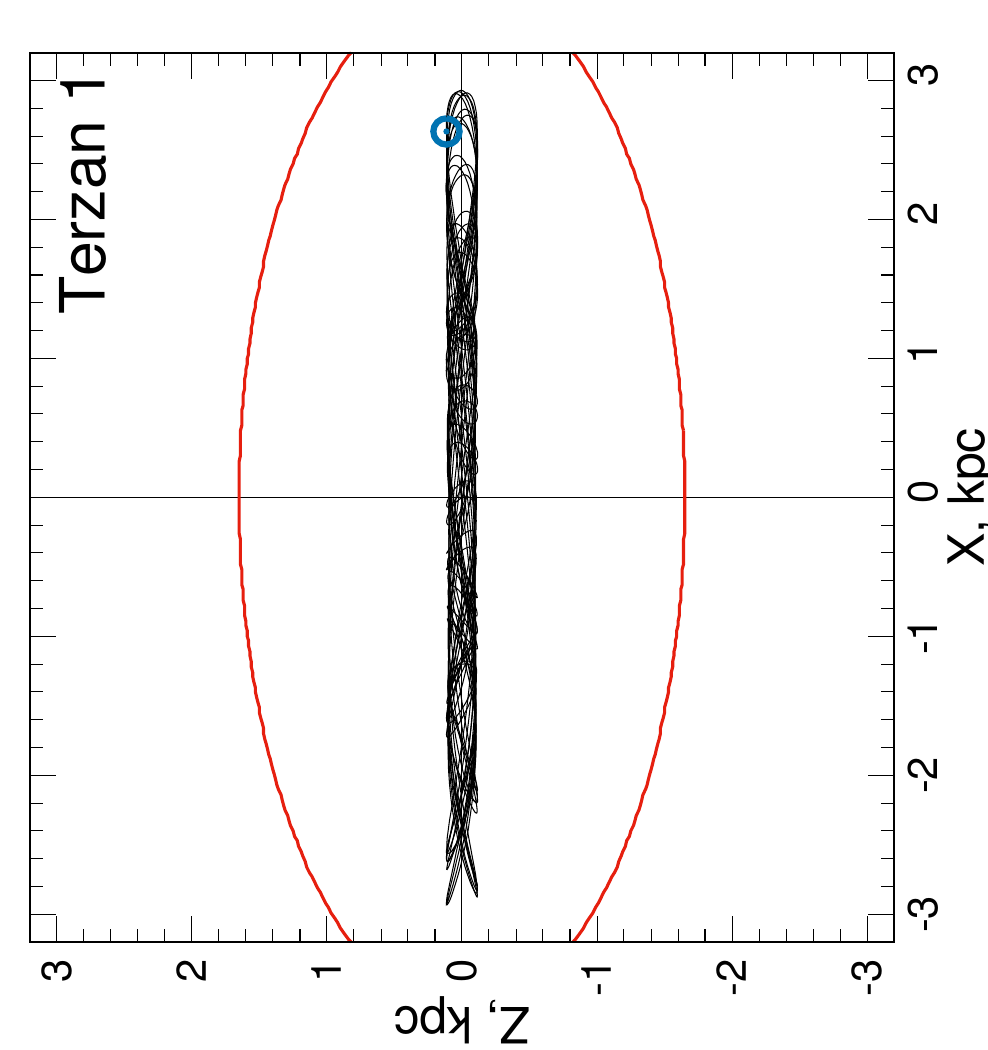}\
 \includegraphics[width=0.225\textwidth,angle=-90]{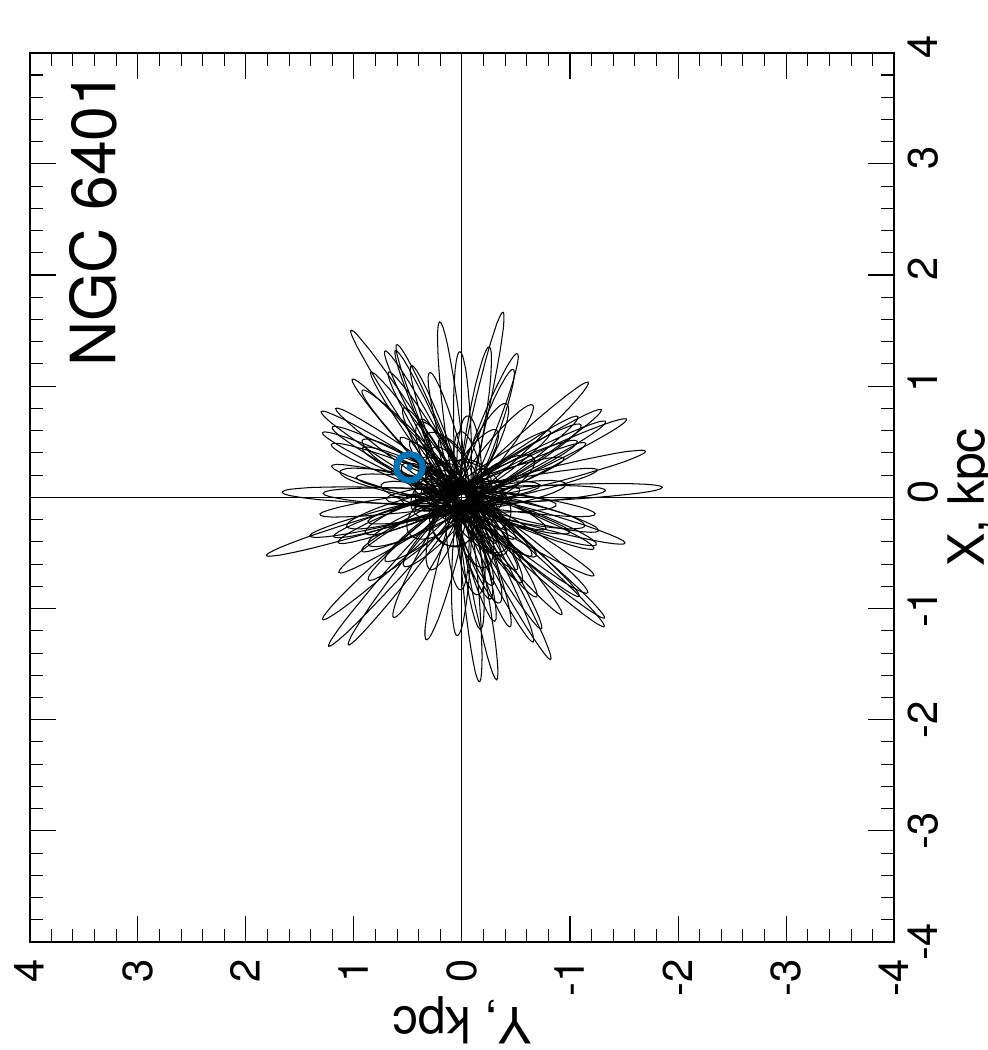}
     \includegraphics[width=0.225\textwidth,angle=-90]{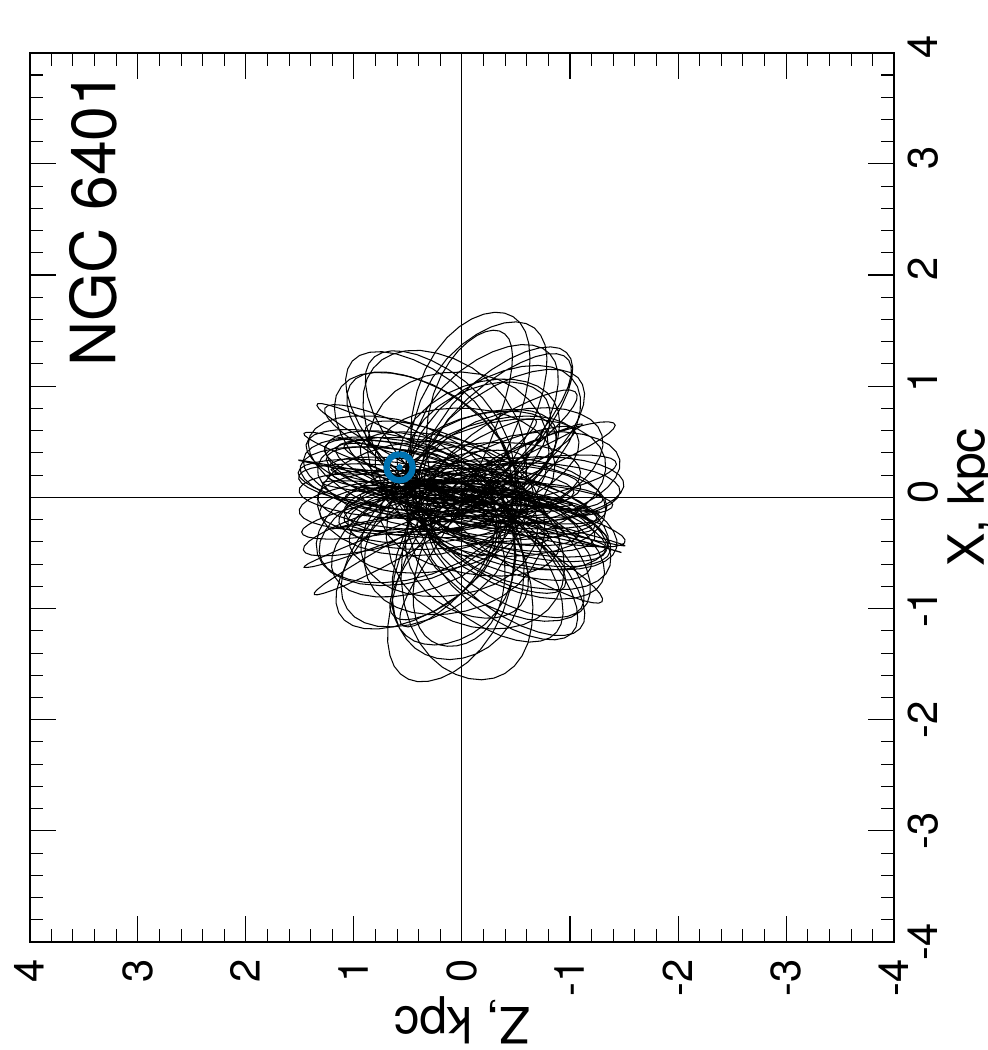}
      \includegraphics[width=0.225\textwidth,angle=-90]{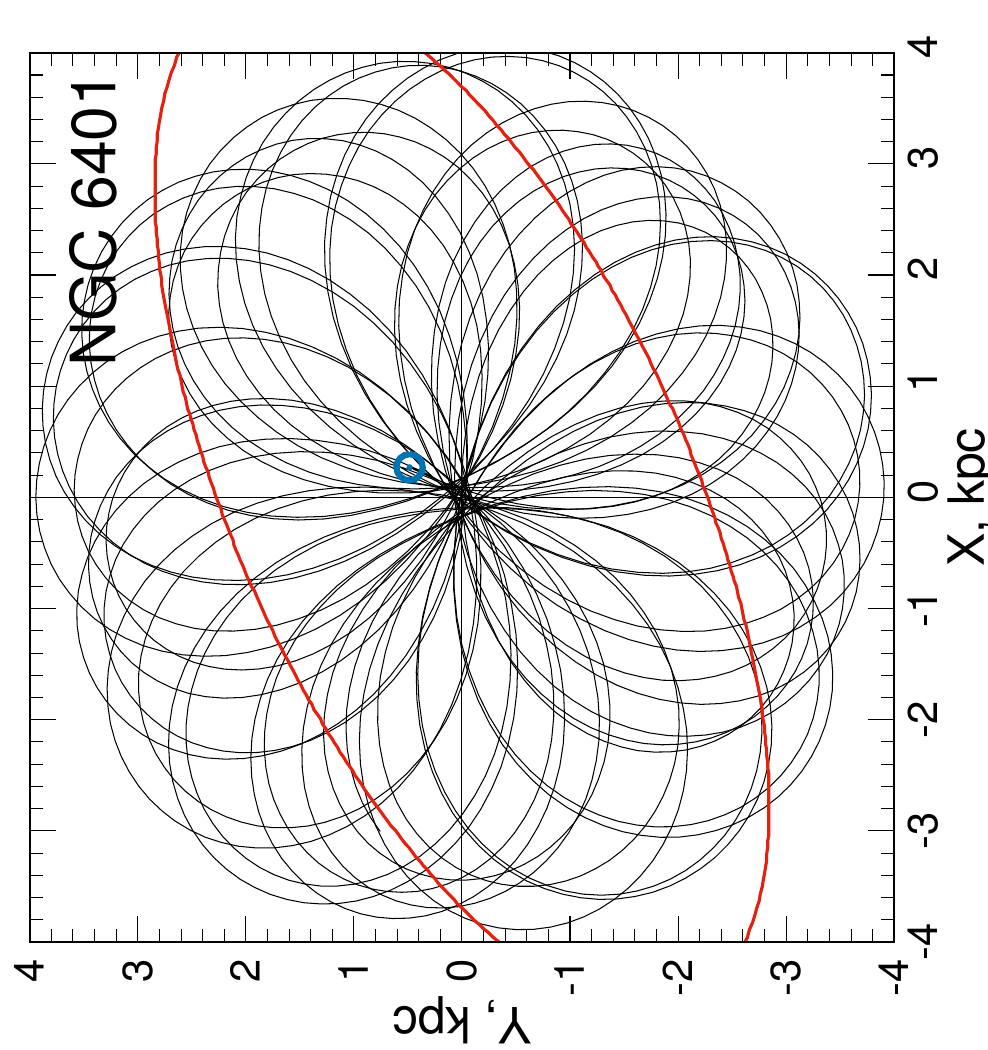}
     \includegraphics[width=0.225\textwidth,angle=-90]{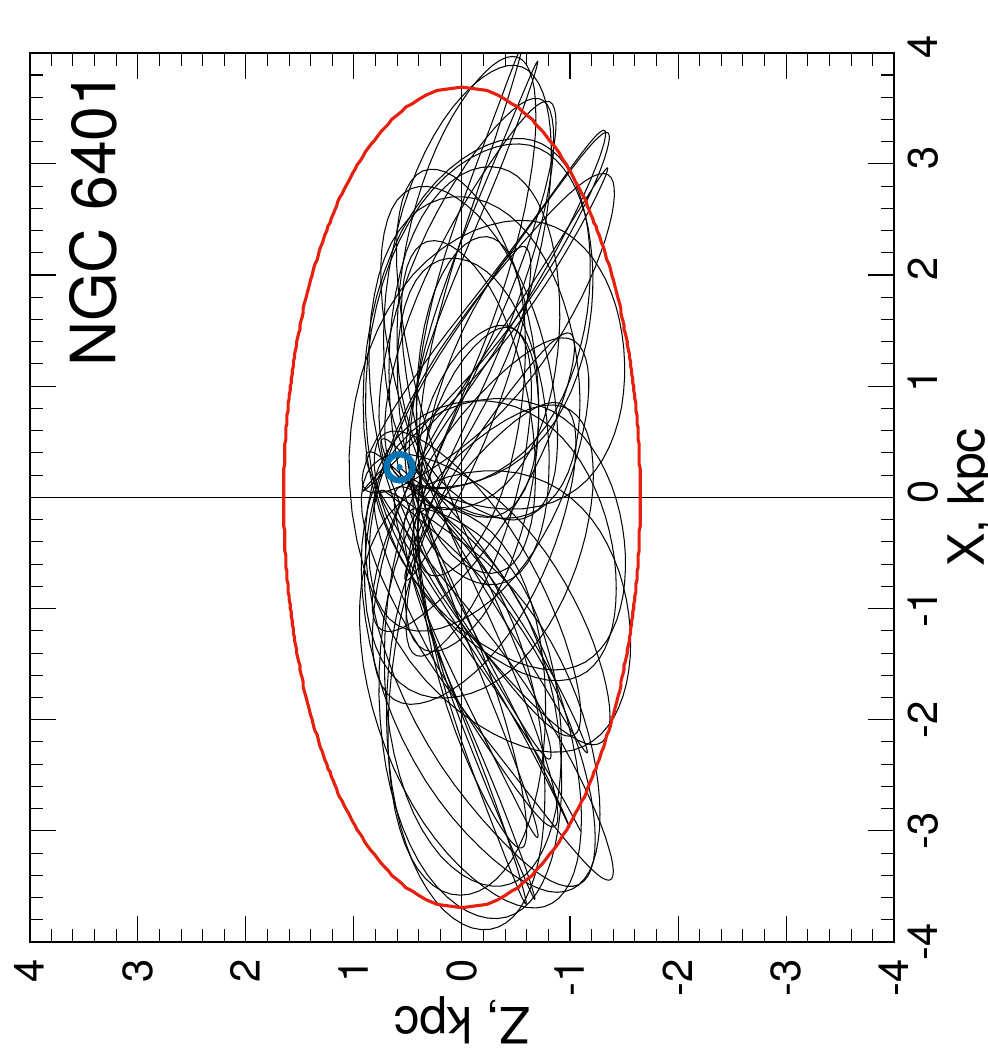}\
    \includegraphics[width=0.225\textwidth,angle=-90]{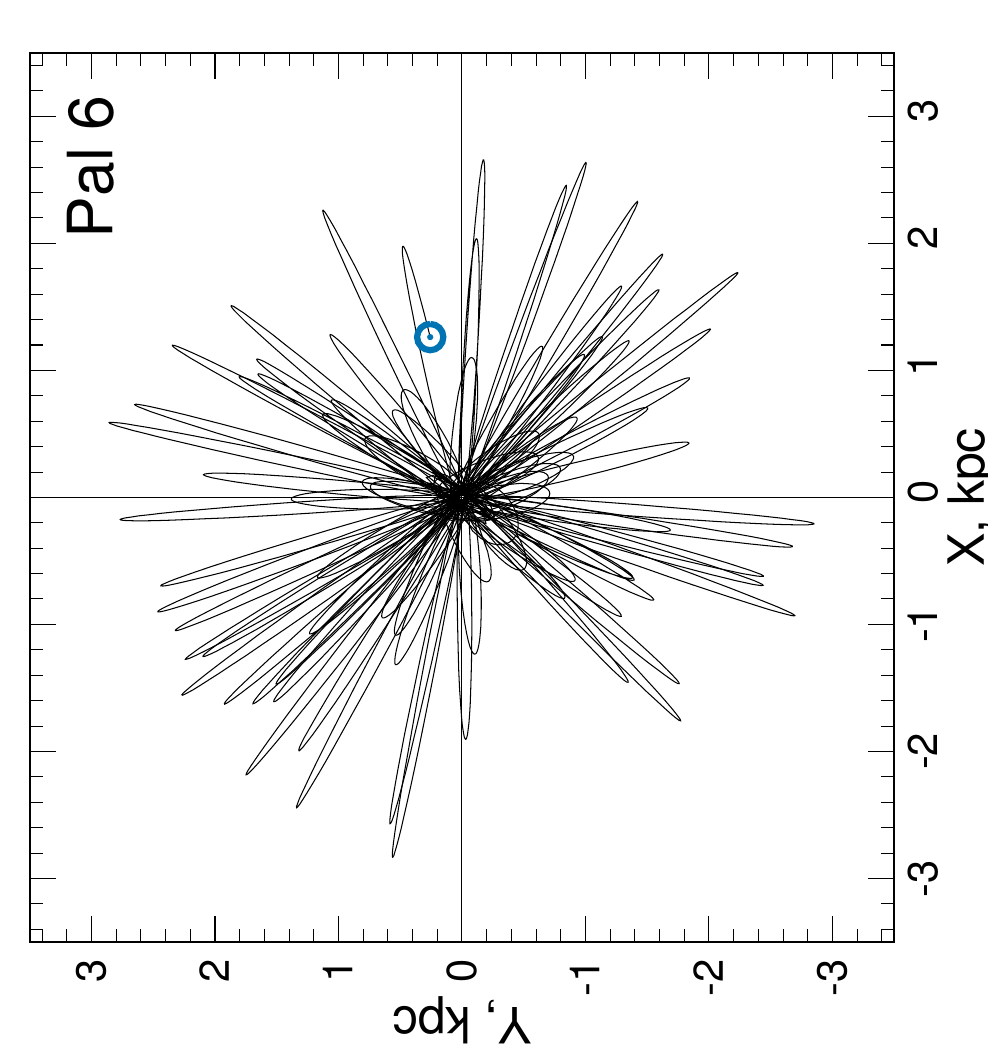}
     \includegraphics[width=0.225\textwidth,angle=-90]{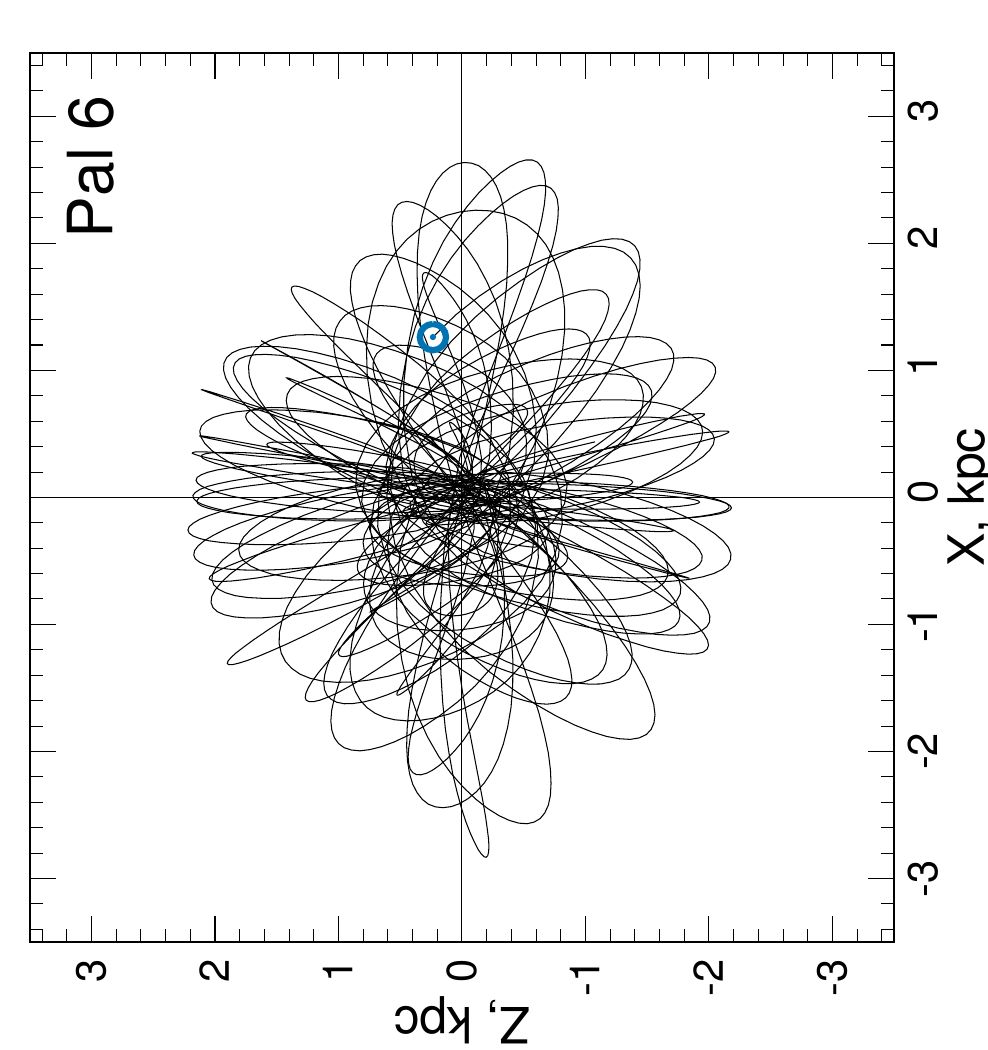}
    \includegraphics[width=0.225\textwidth,angle=-90]{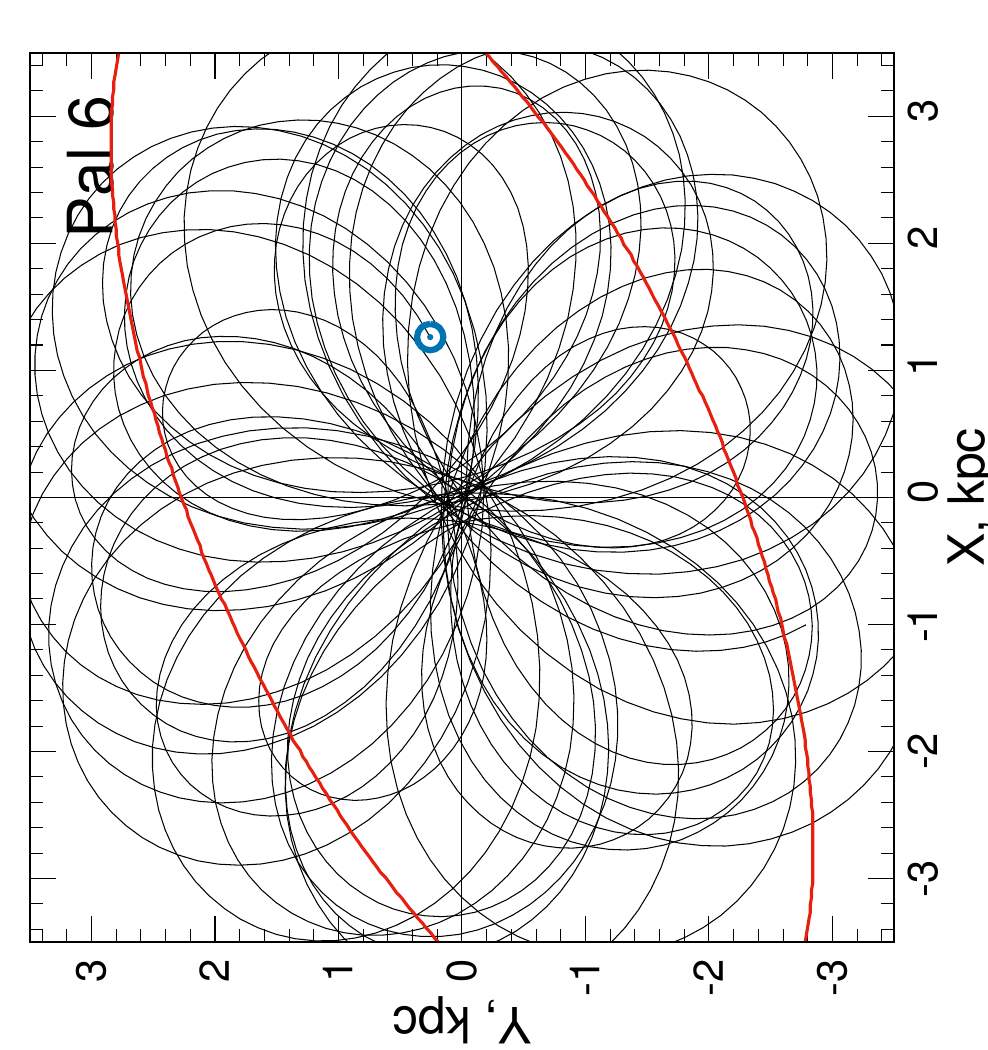}
     \includegraphics[width=0.225\textwidth,angle=-90]{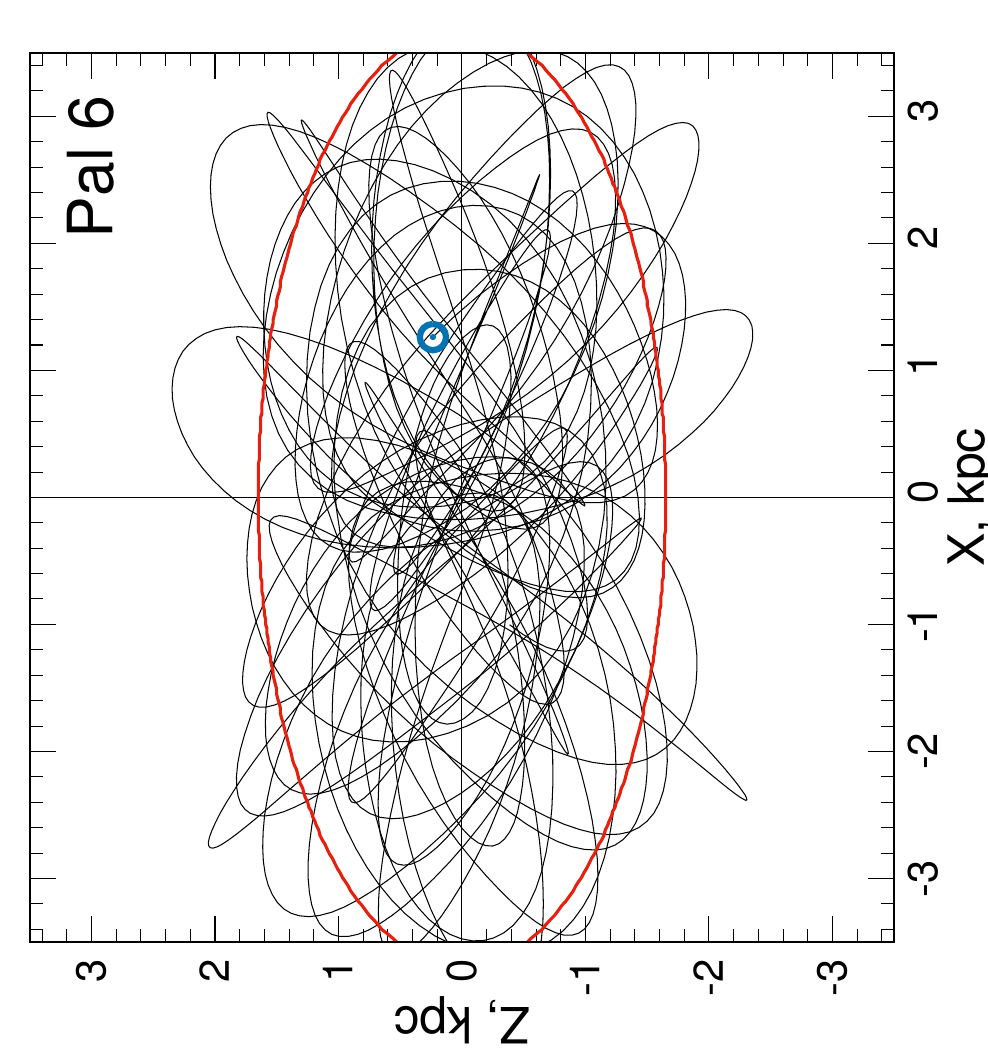}\
         \includegraphics[width=0.225\textwidth,angle=-90]{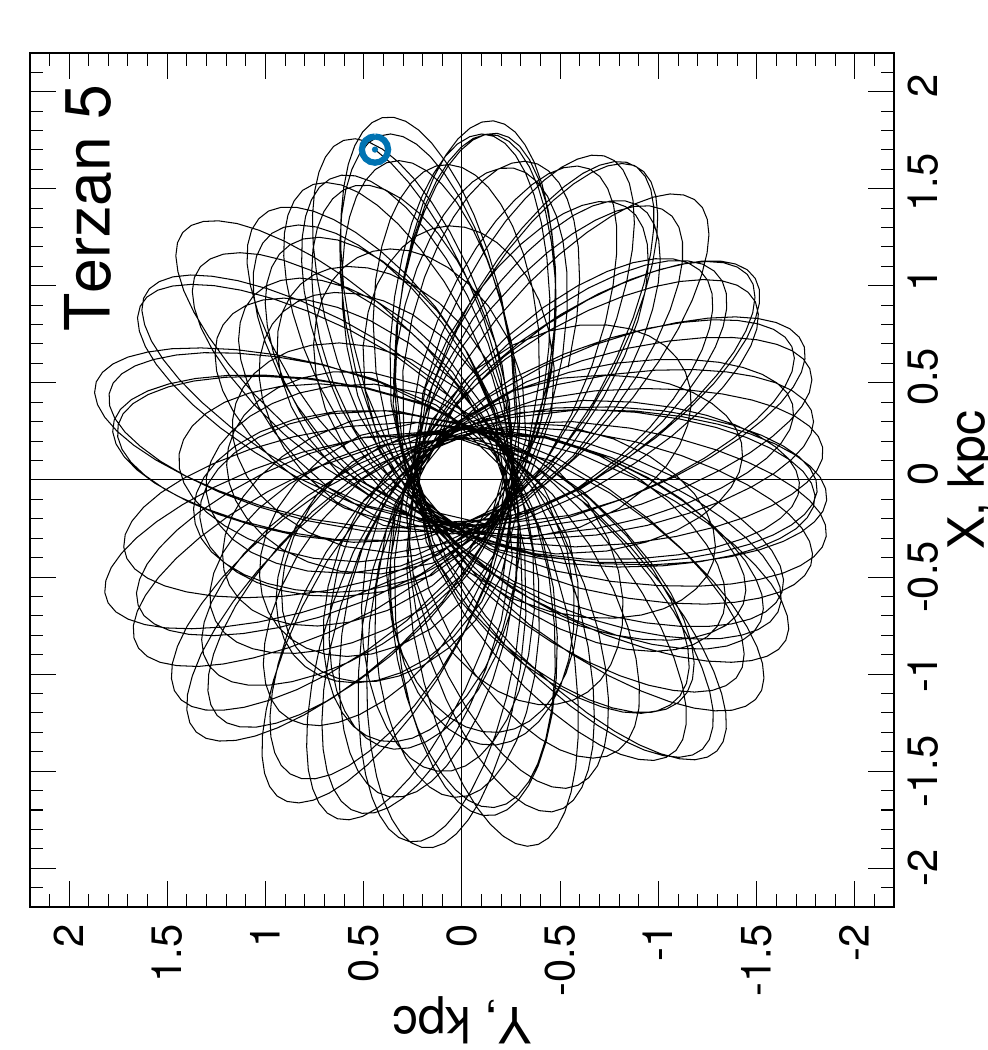}
     \includegraphics[width=0.225\textwidth,angle=-90]{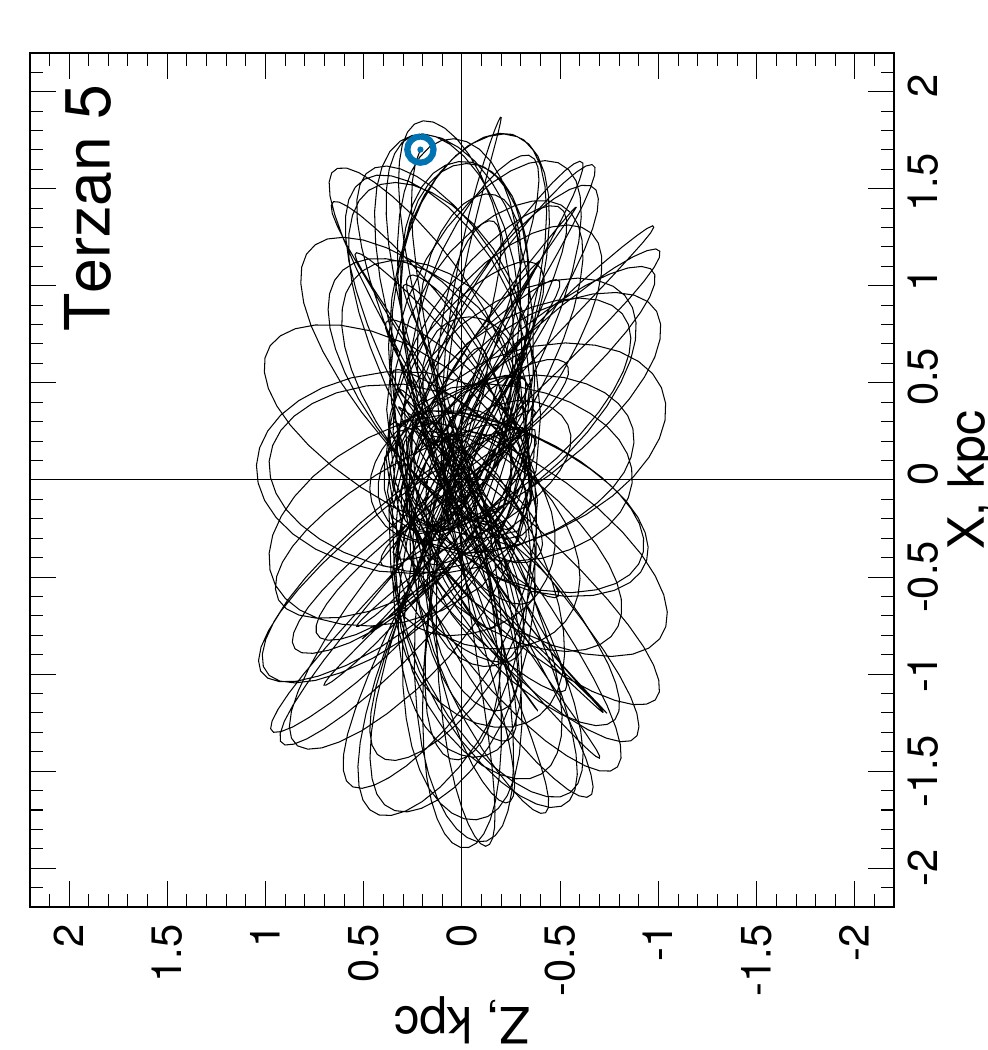}
              \includegraphics[width=0.225\textwidth,angle=-90]{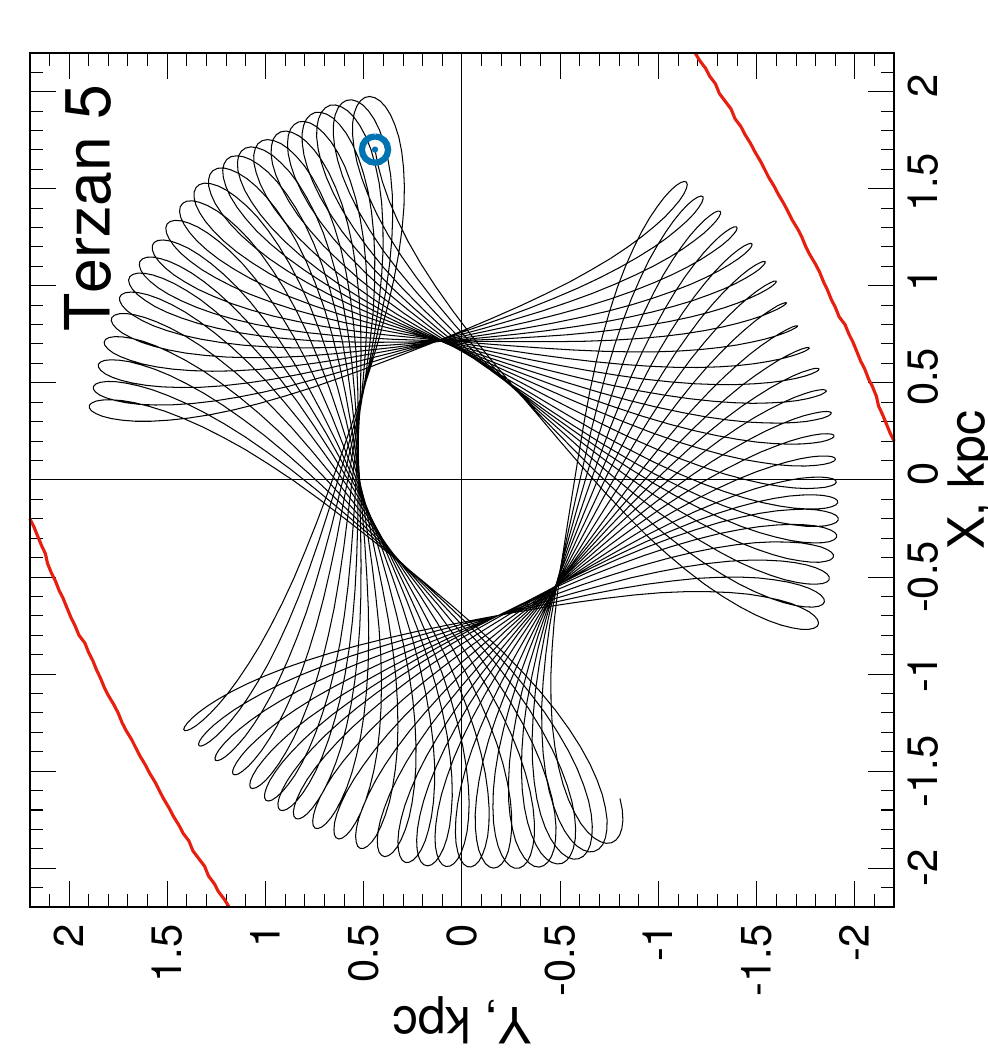}
     \includegraphics[width=0.225\textwidth,angle=-90]{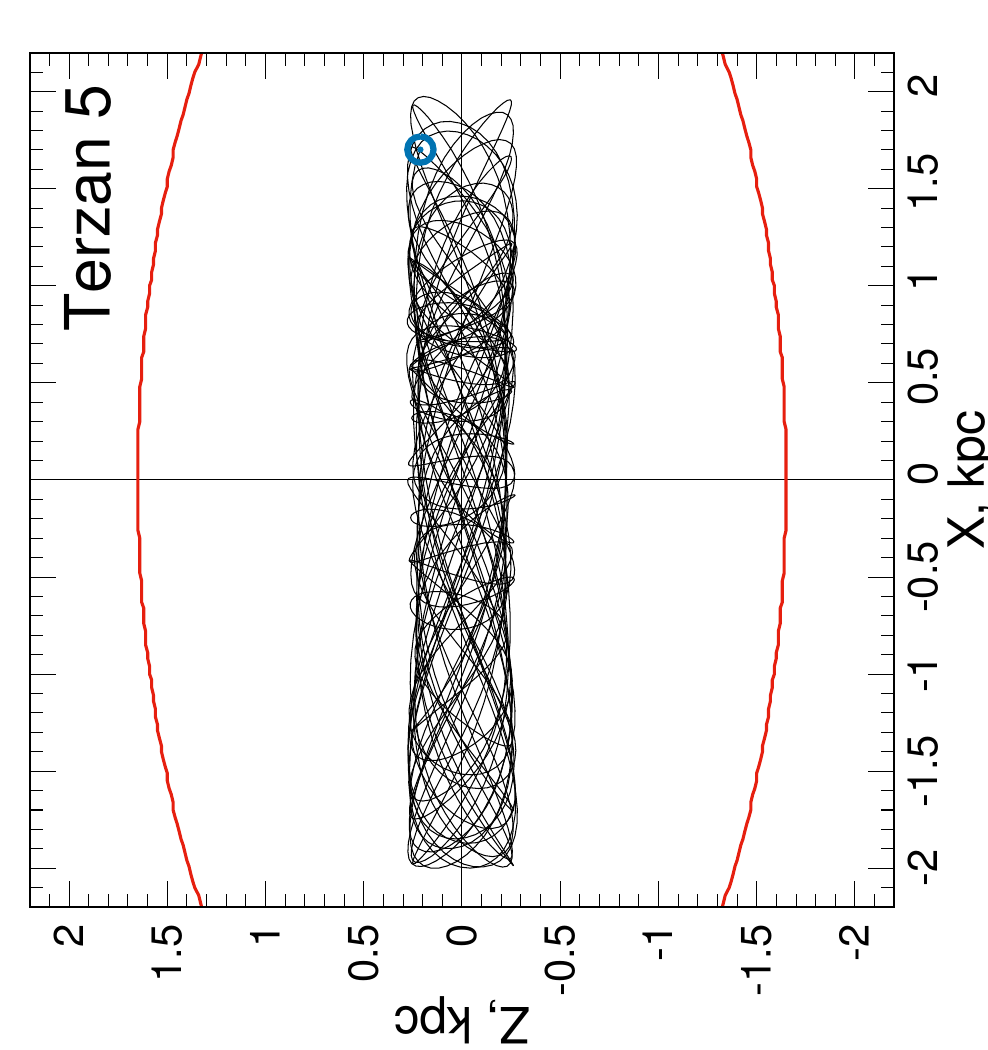}\
   \includegraphics[width=0.225\textwidth,angle=-90]{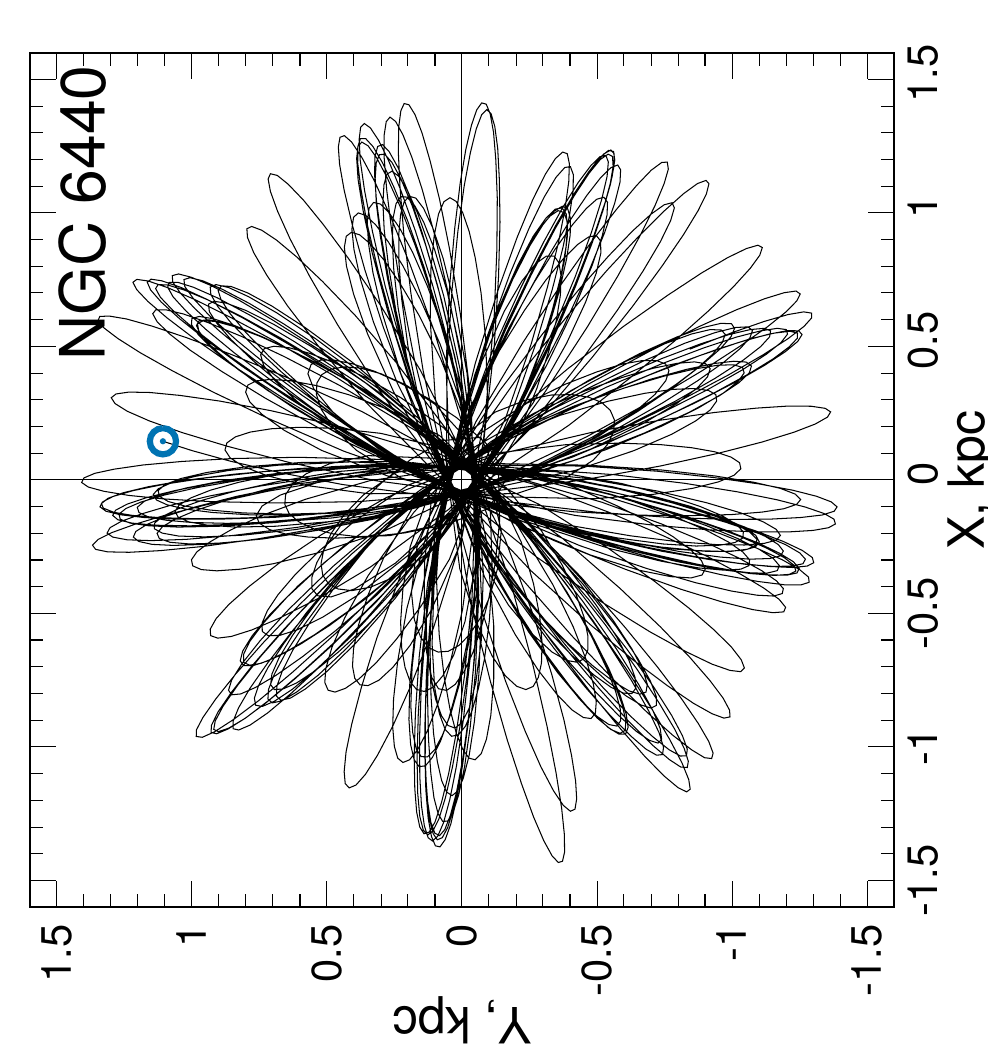}
     \includegraphics[width=0.225\textwidth,angle=-90]{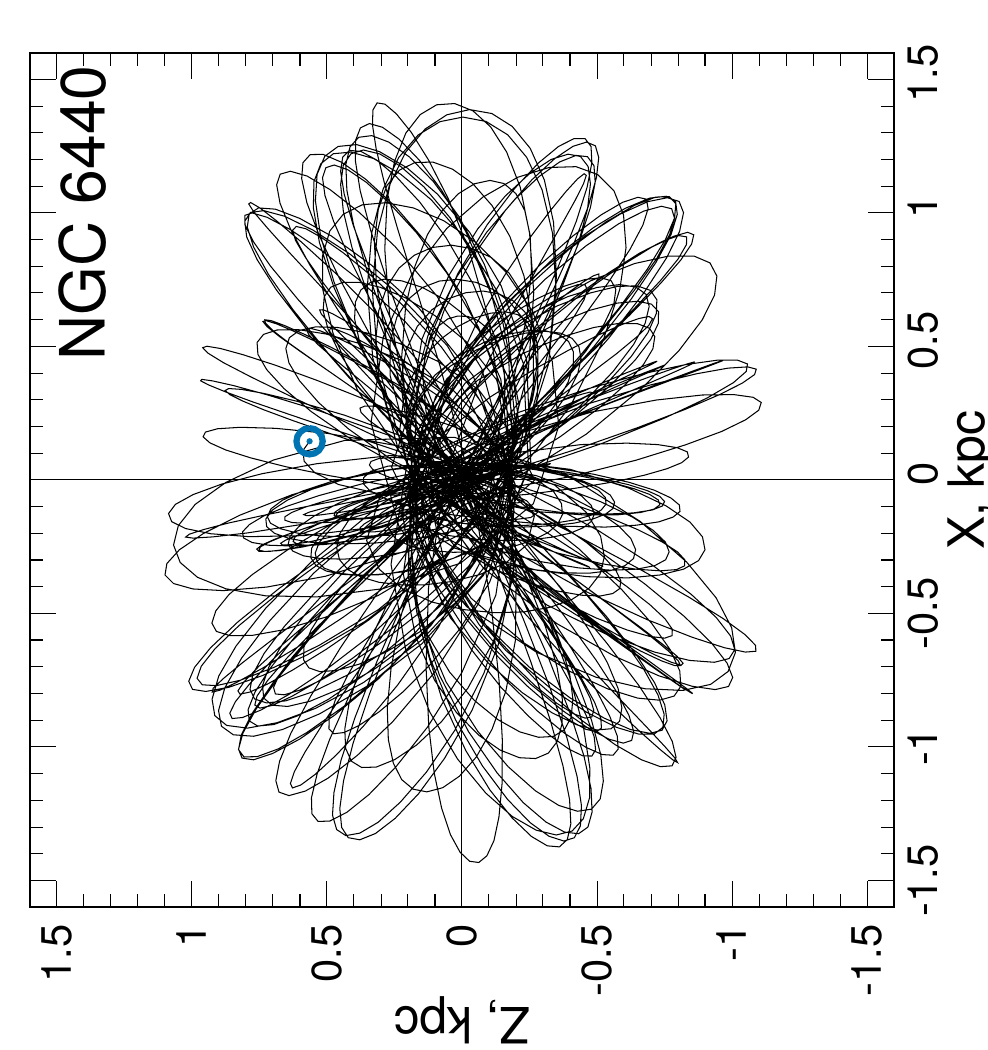}
   \includegraphics[width=0.225\textwidth,angle=-90]{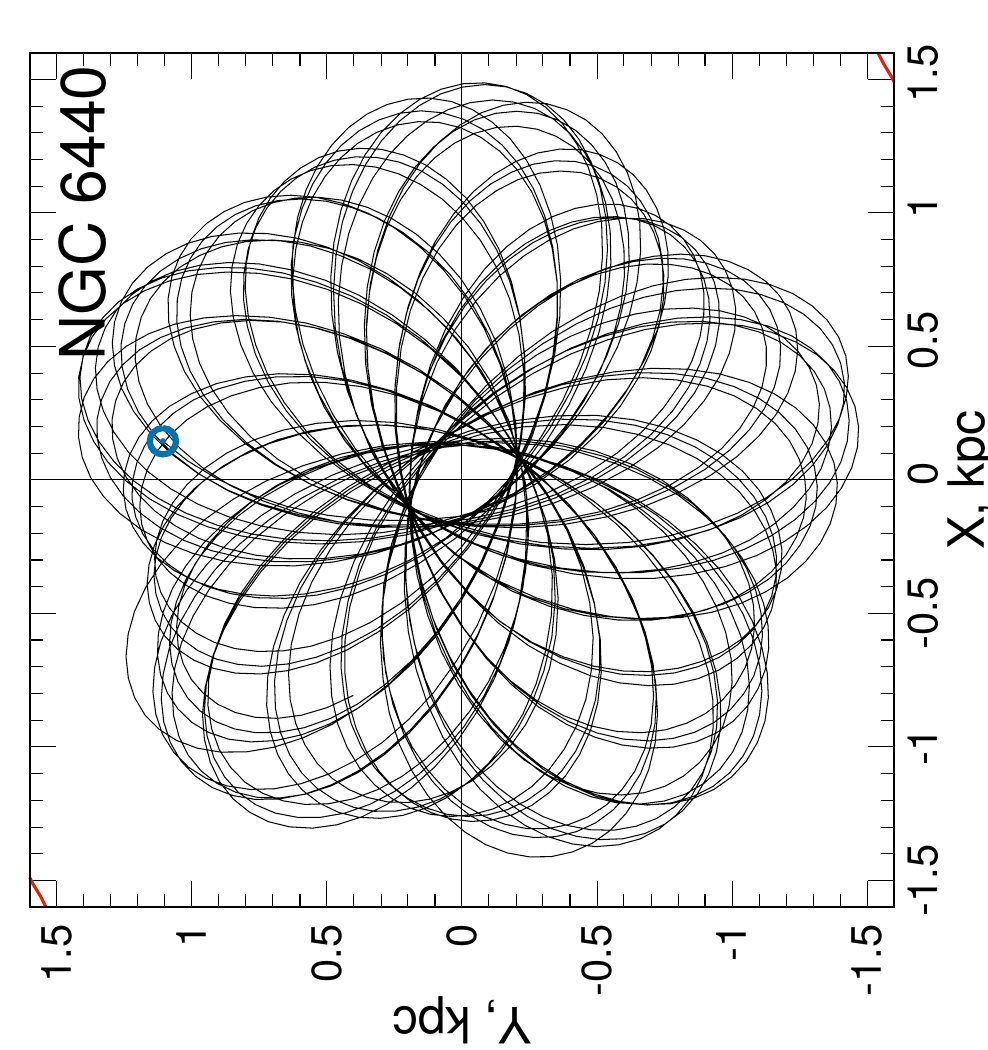}
     \includegraphics[width=0.225\textwidth,angle=-90]{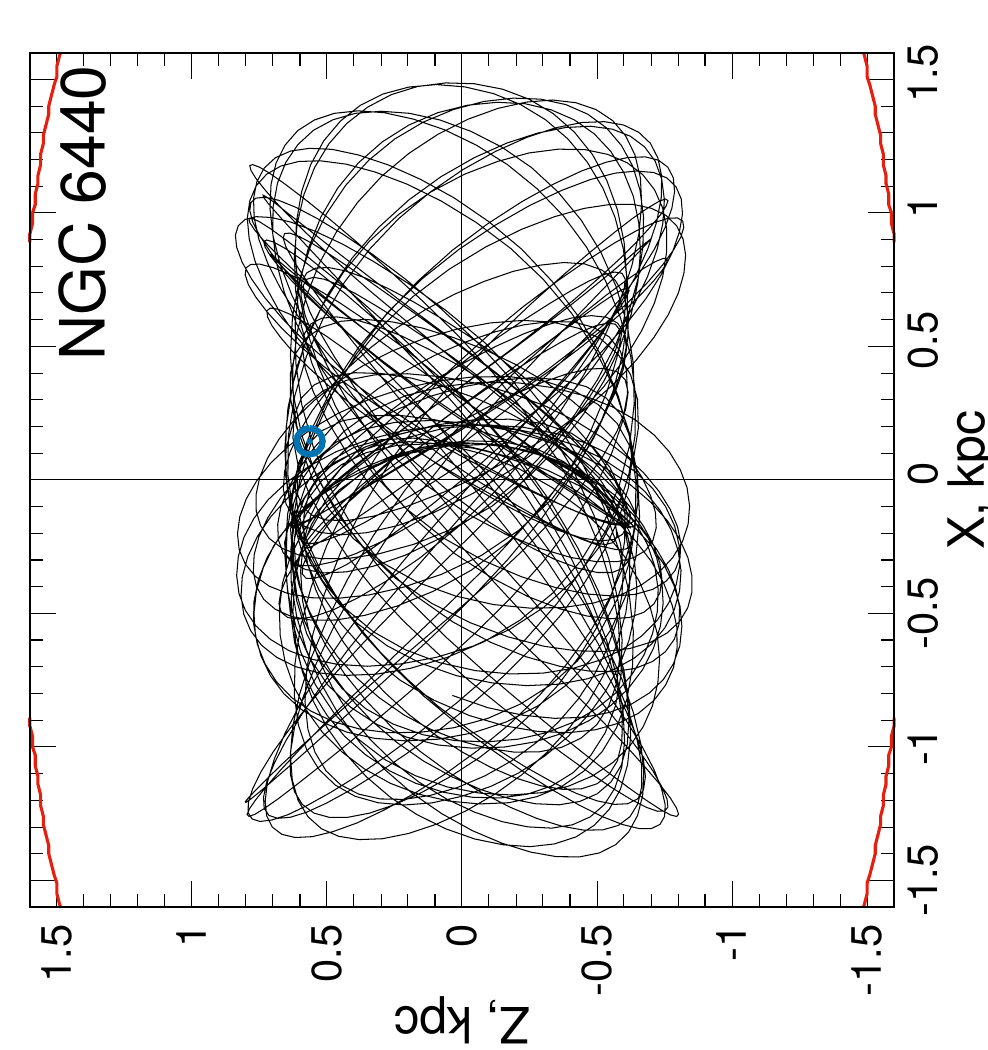}\

\medskip

 \centerline{APPENDIX. Continued}
\label{fD}
\end{center}}
\end{figure*}

\begin{figure*}
{\begin{center}
  \includegraphics[width=0.225\textwidth,angle=-90]{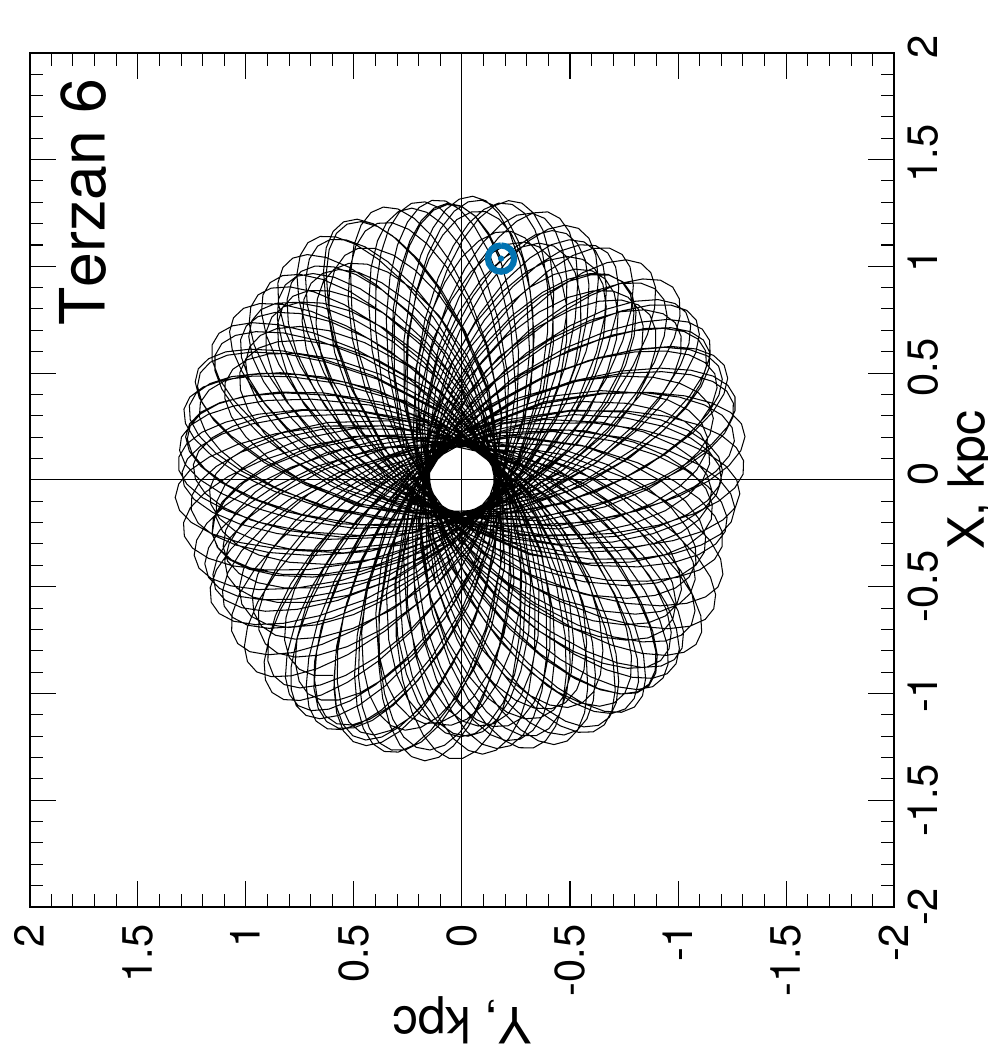}
     \includegraphics[width=0.225\textwidth,angle=-90]{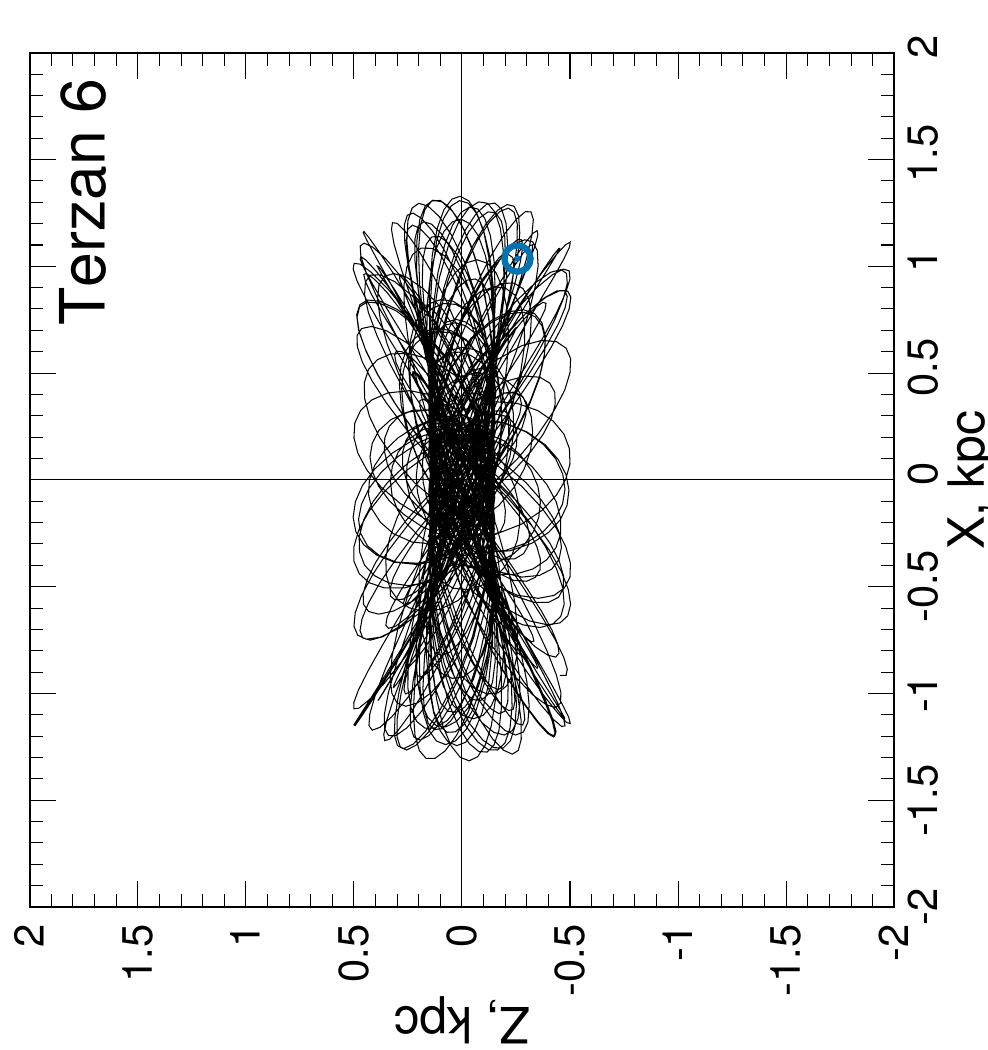}
   \includegraphics[width=0.225\textwidth,angle=-90]{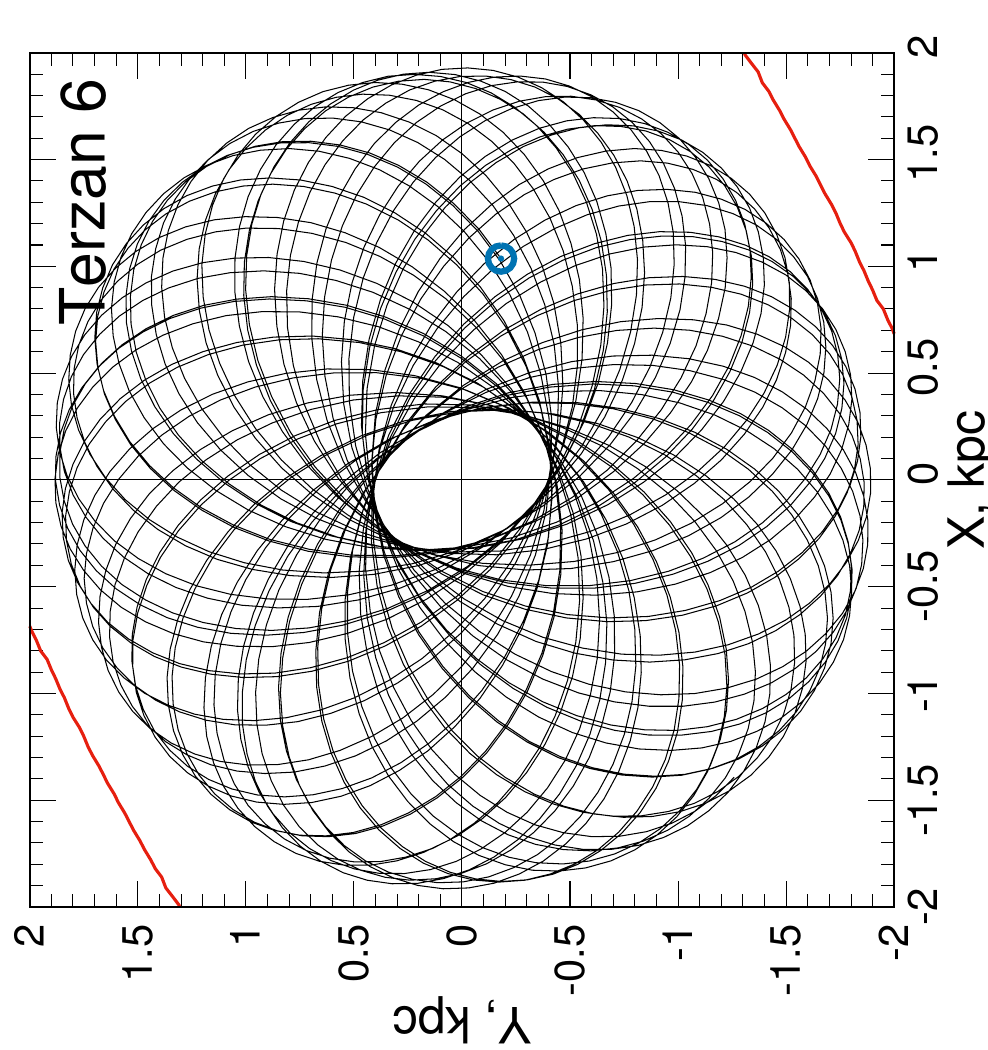}
     \includegraphics[width=0.225\textwidth,angle=-90]{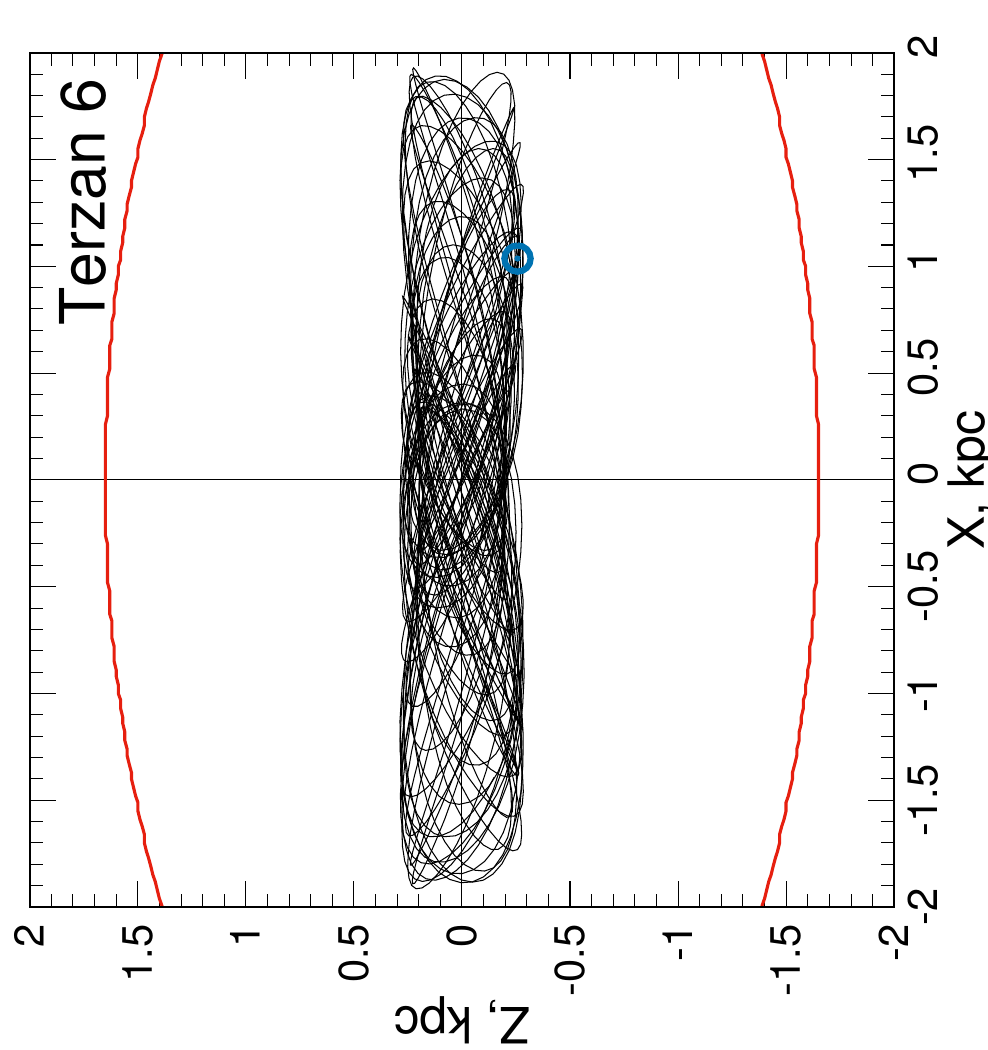}\
    \includegraphics[width=0.225\textwidth,angle=-90]{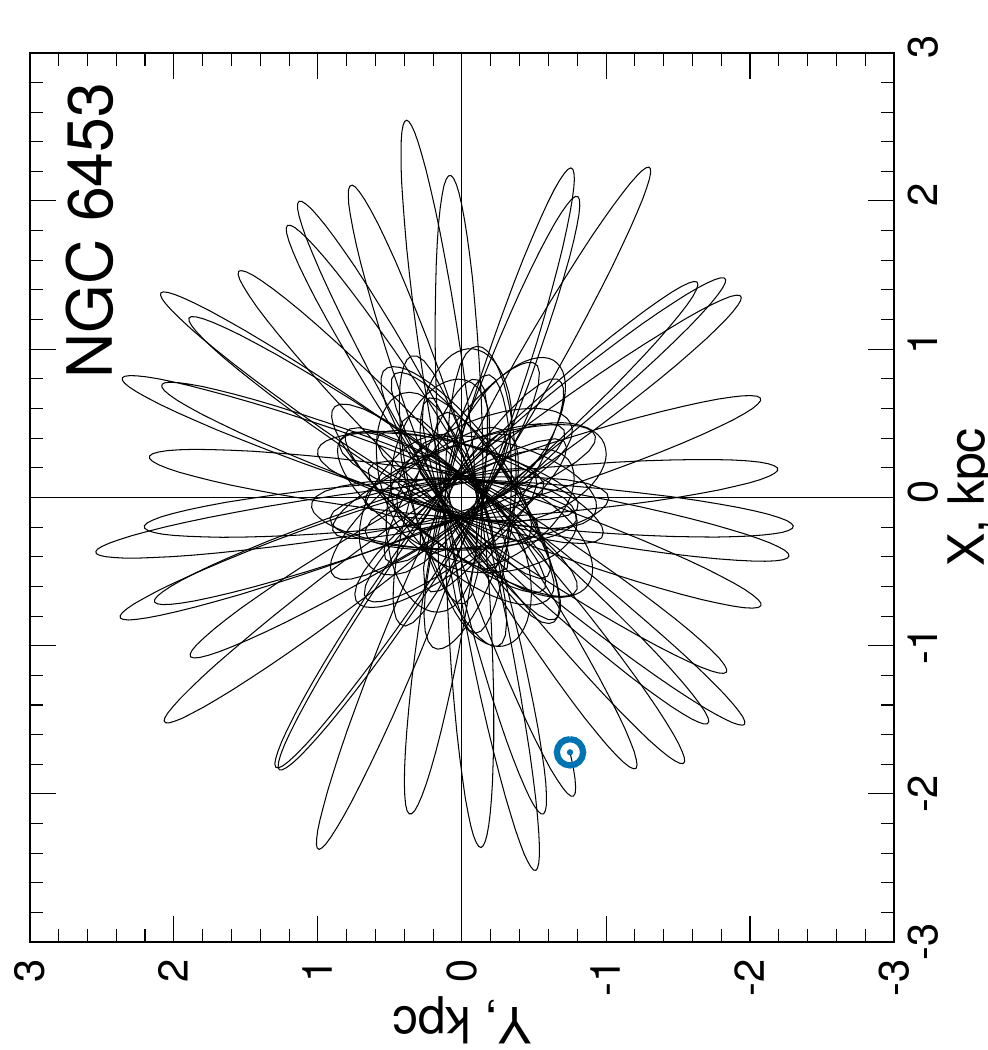}
     \includegraphics[width=0.225\textwidth,angle=-90]{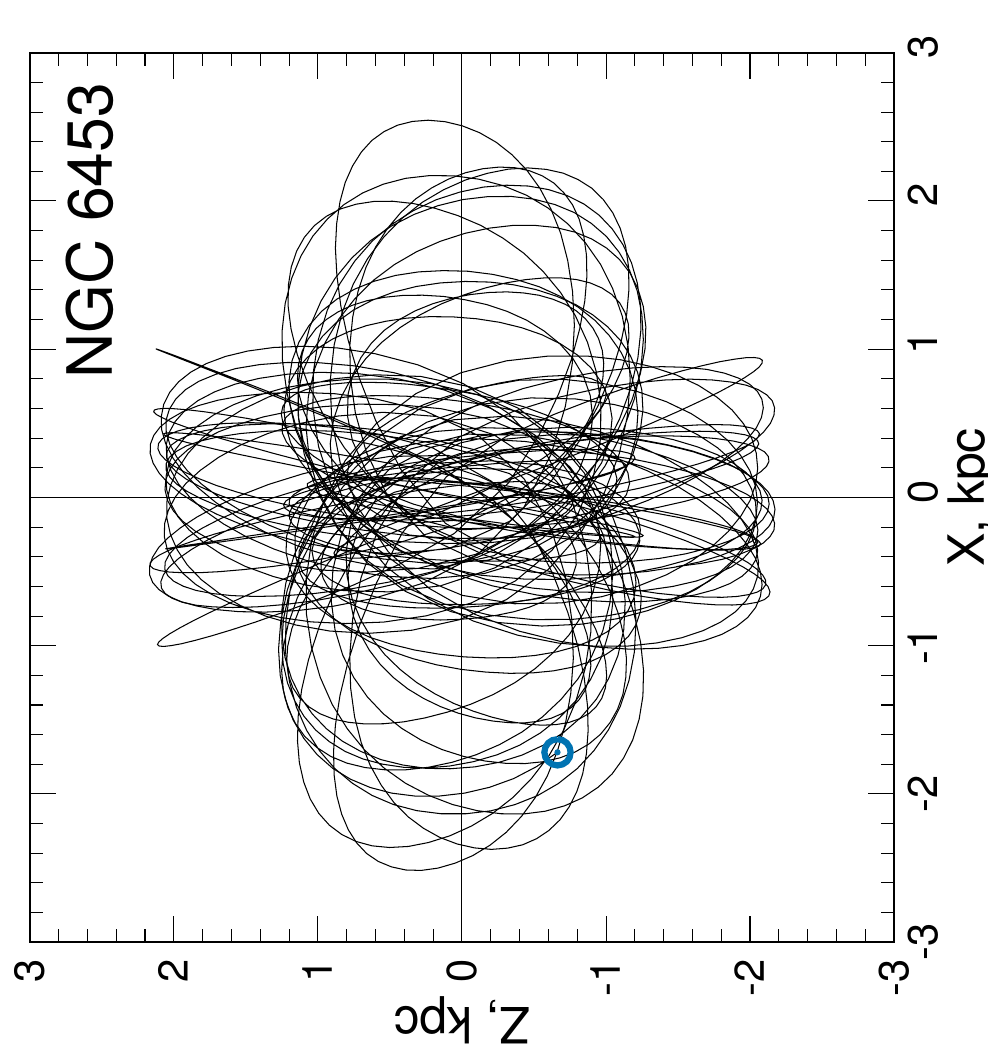}
     \includegraphics[width=0.225\textwidth,angle=-90]{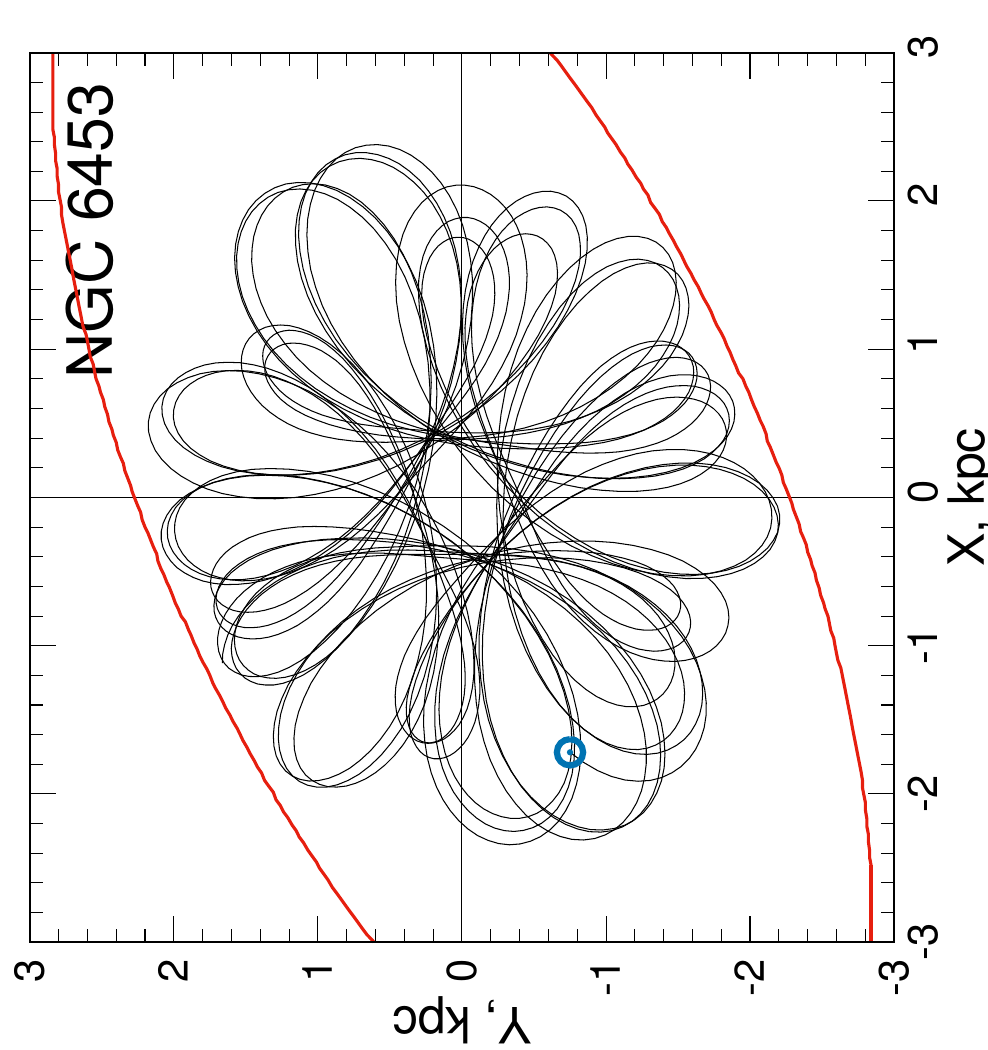}
     \includegraphics[width=0.225\textwidth,angle=-90]{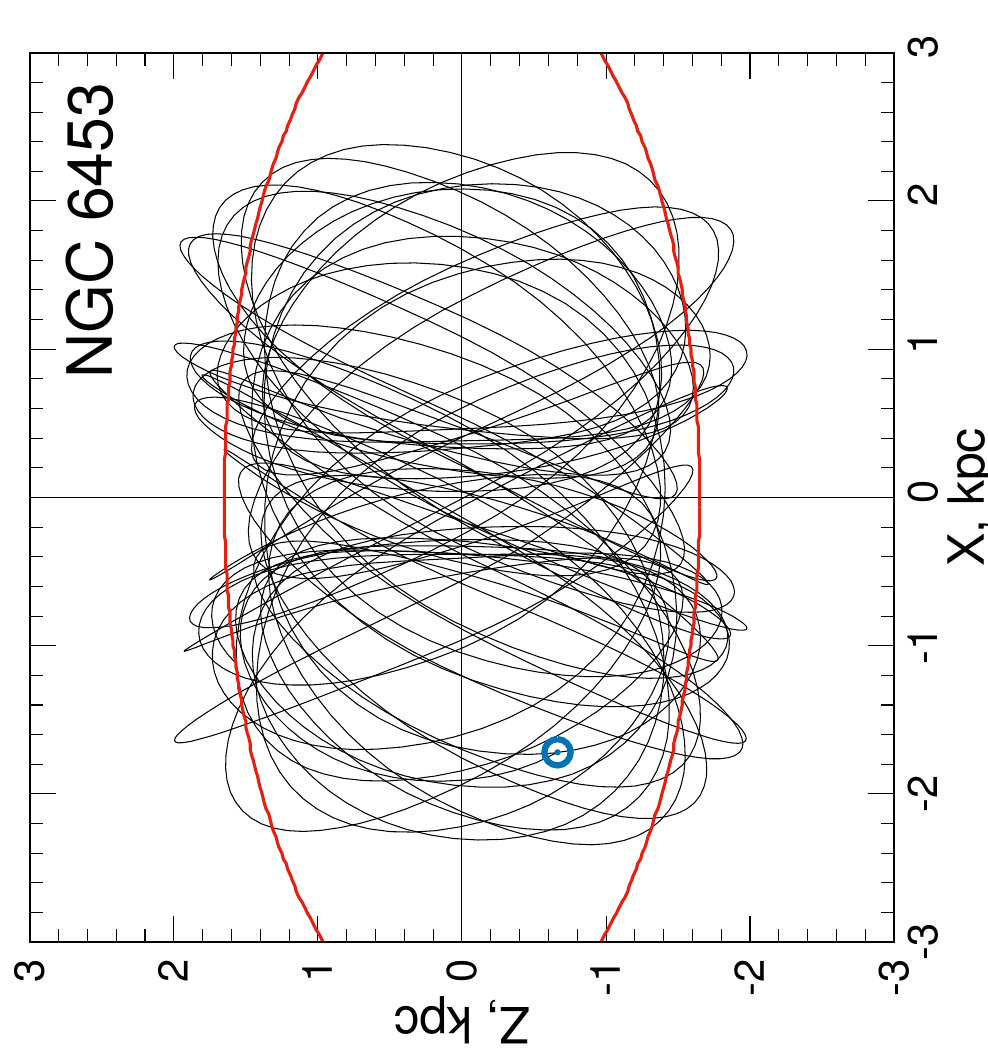}\
   \includegraphics[width=0.225\textwidth,angle=-90]{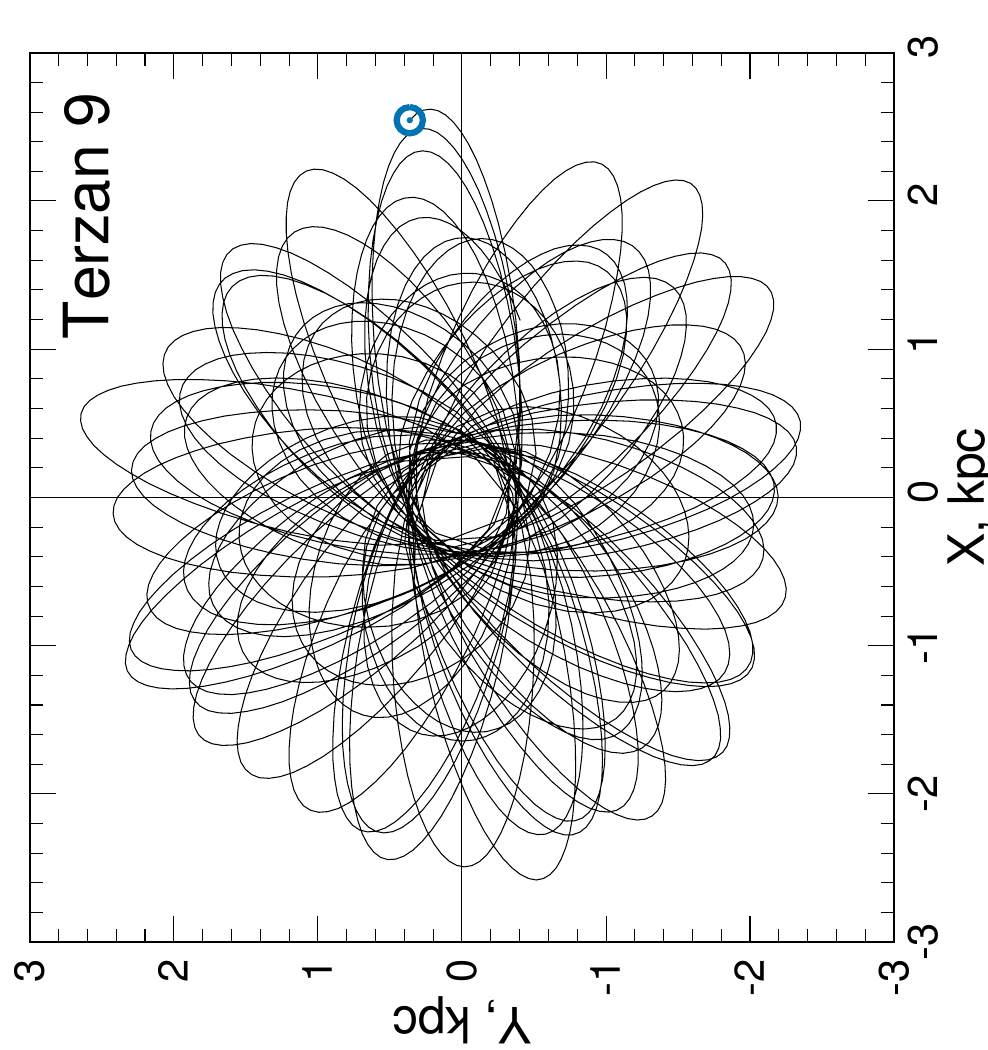}
     \includegraphics[width=0.225\textwidth,angle=-90]{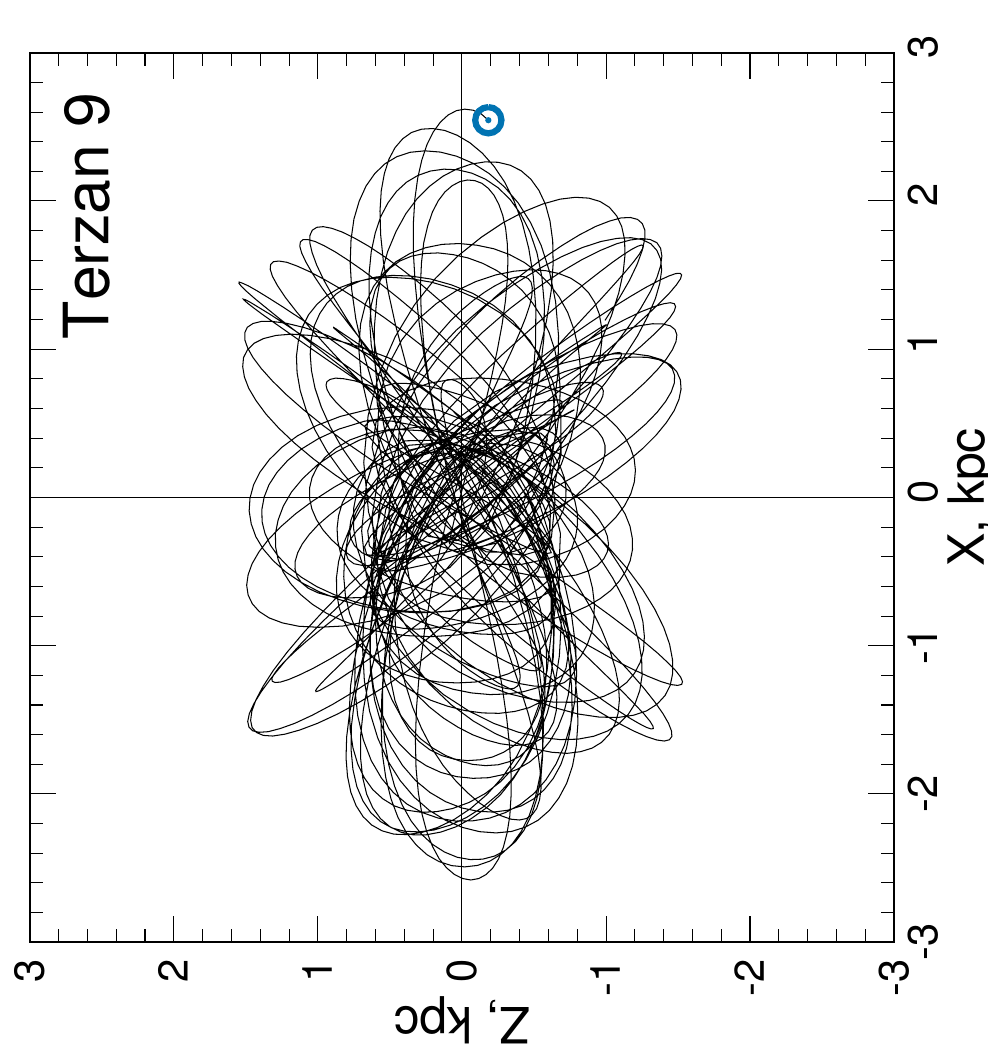}
        \includegraphics[width=0.225\textwidth,angle=-90]{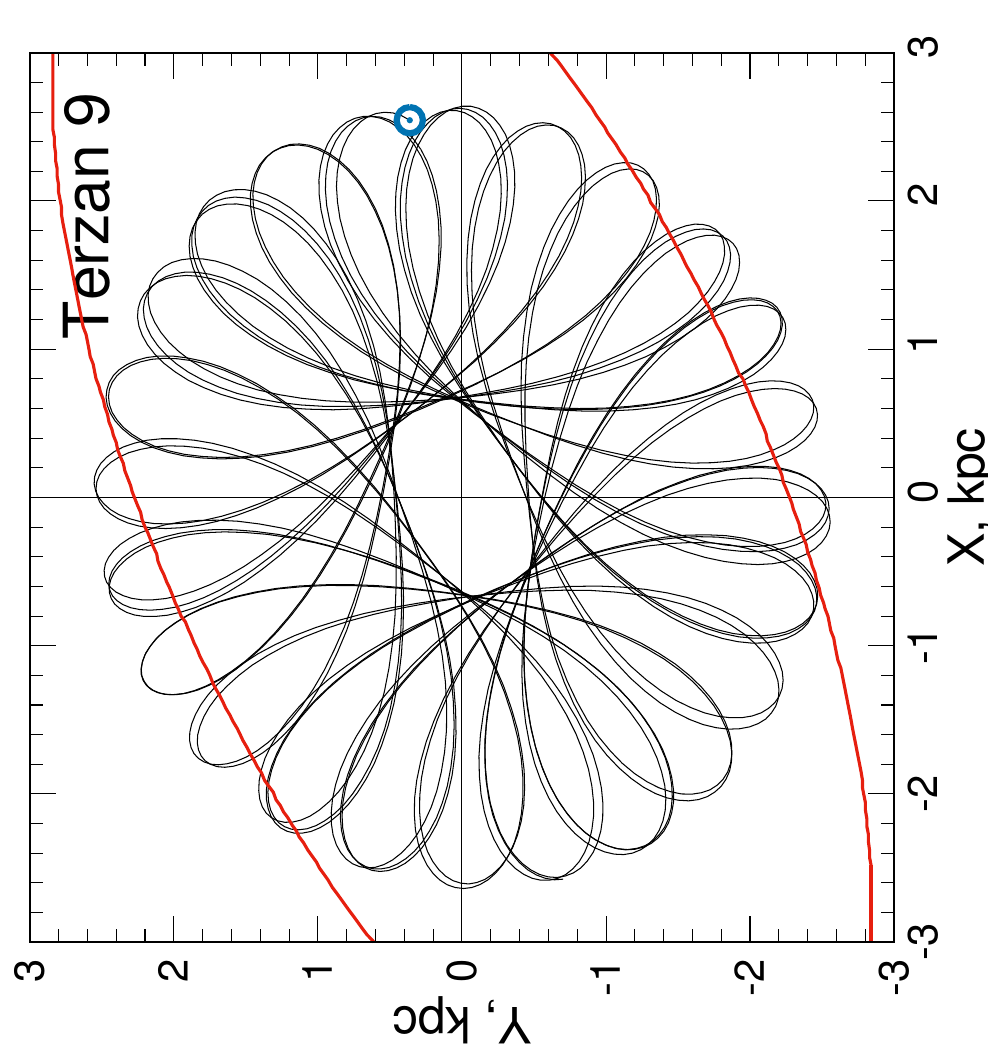}
     \includegraphics[width=0.225\textwidth,angle=-90]{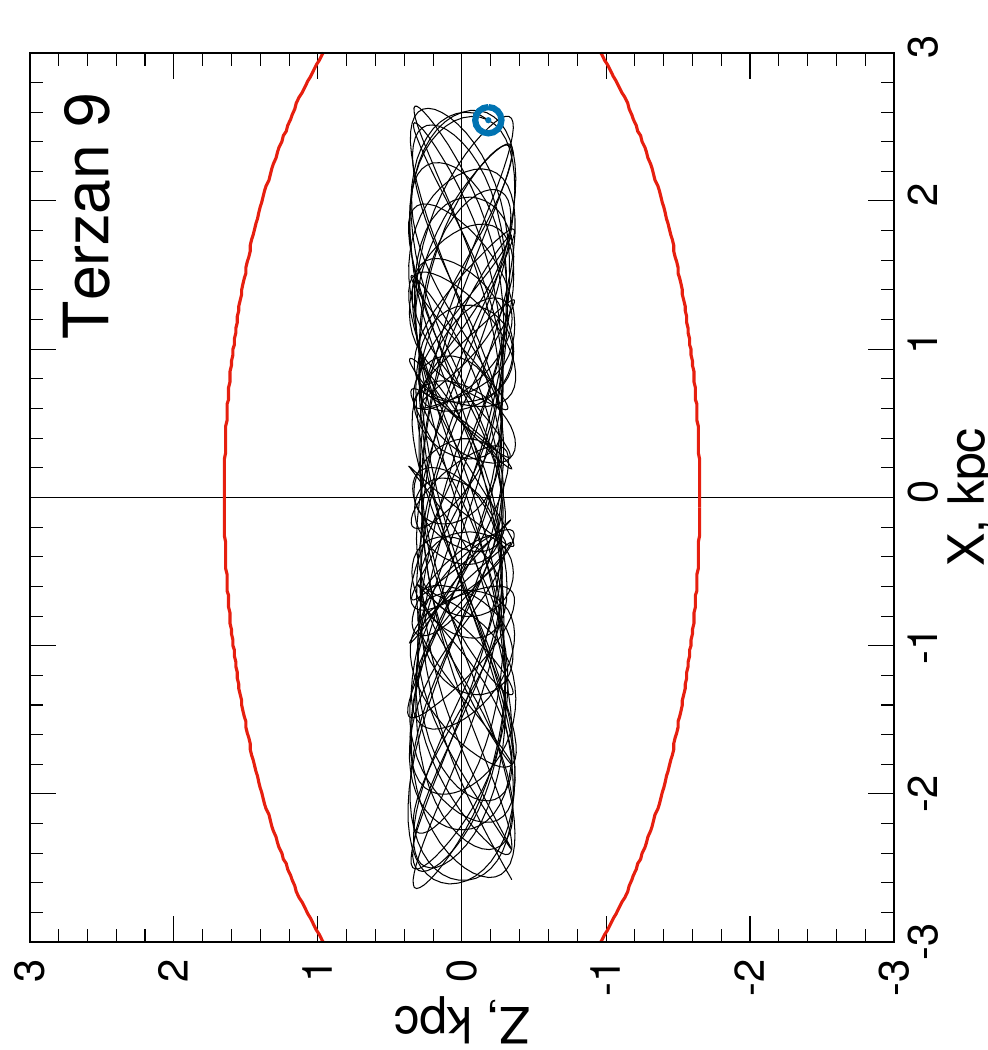}\
        \includegraphics[width=0.225\textwidth,angle=-90]{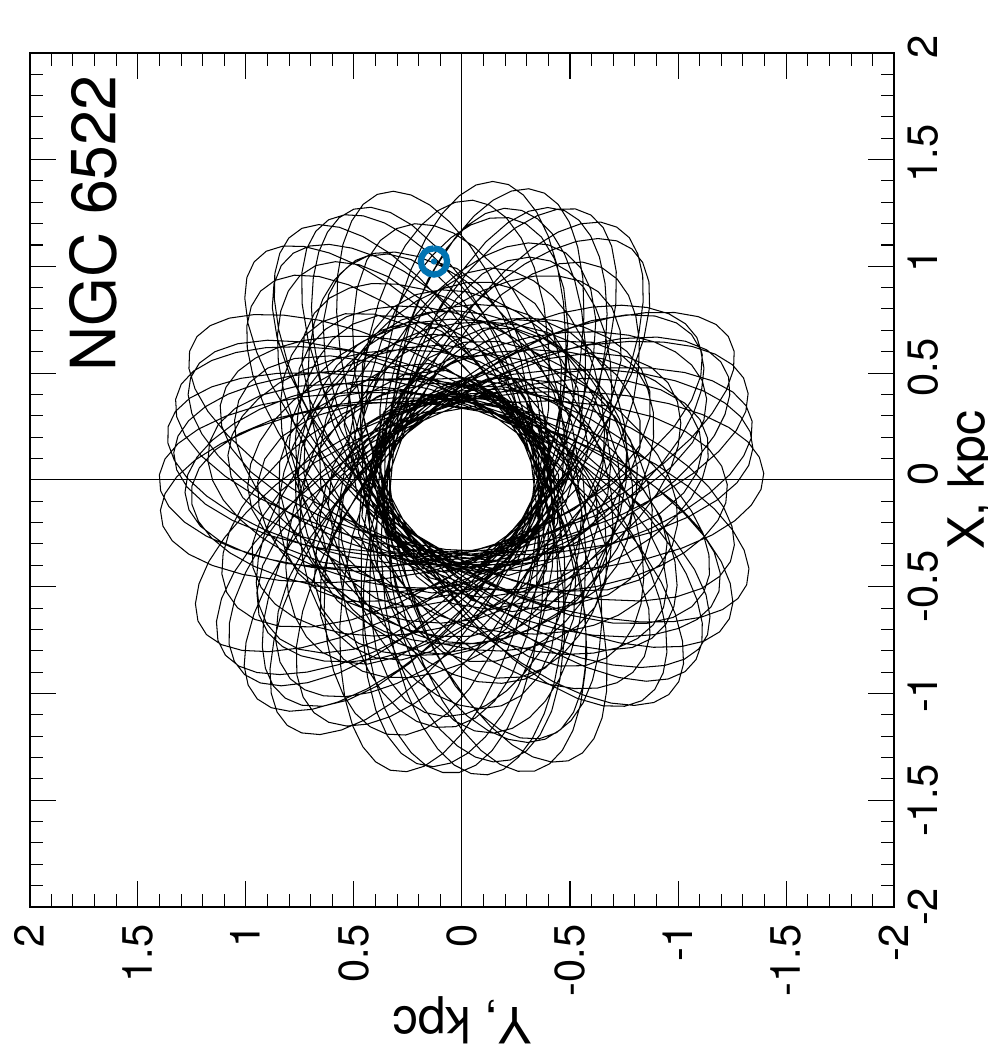}
     \includegraphics[width=0.225\textwidth,angle=-90]{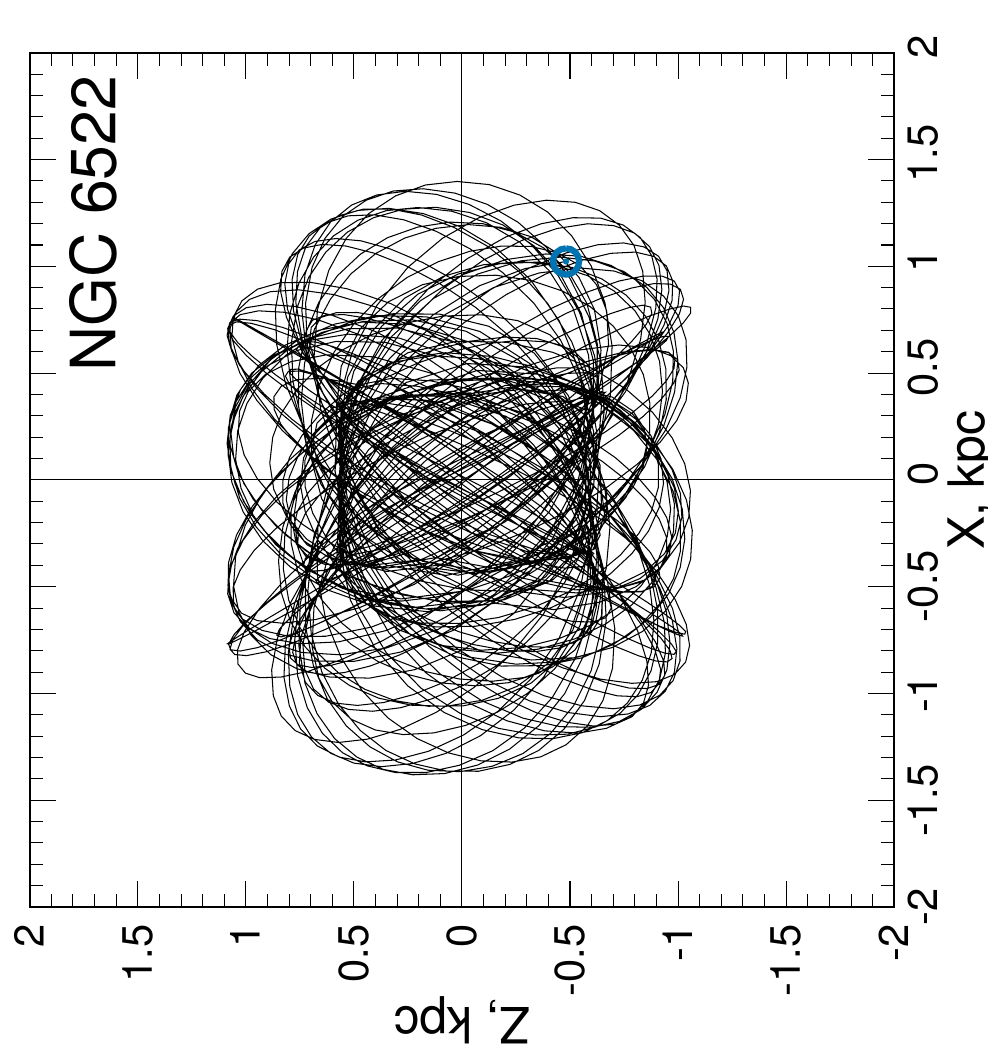}
        \includegraphics[width=0.225\textwidth,angle=-90]{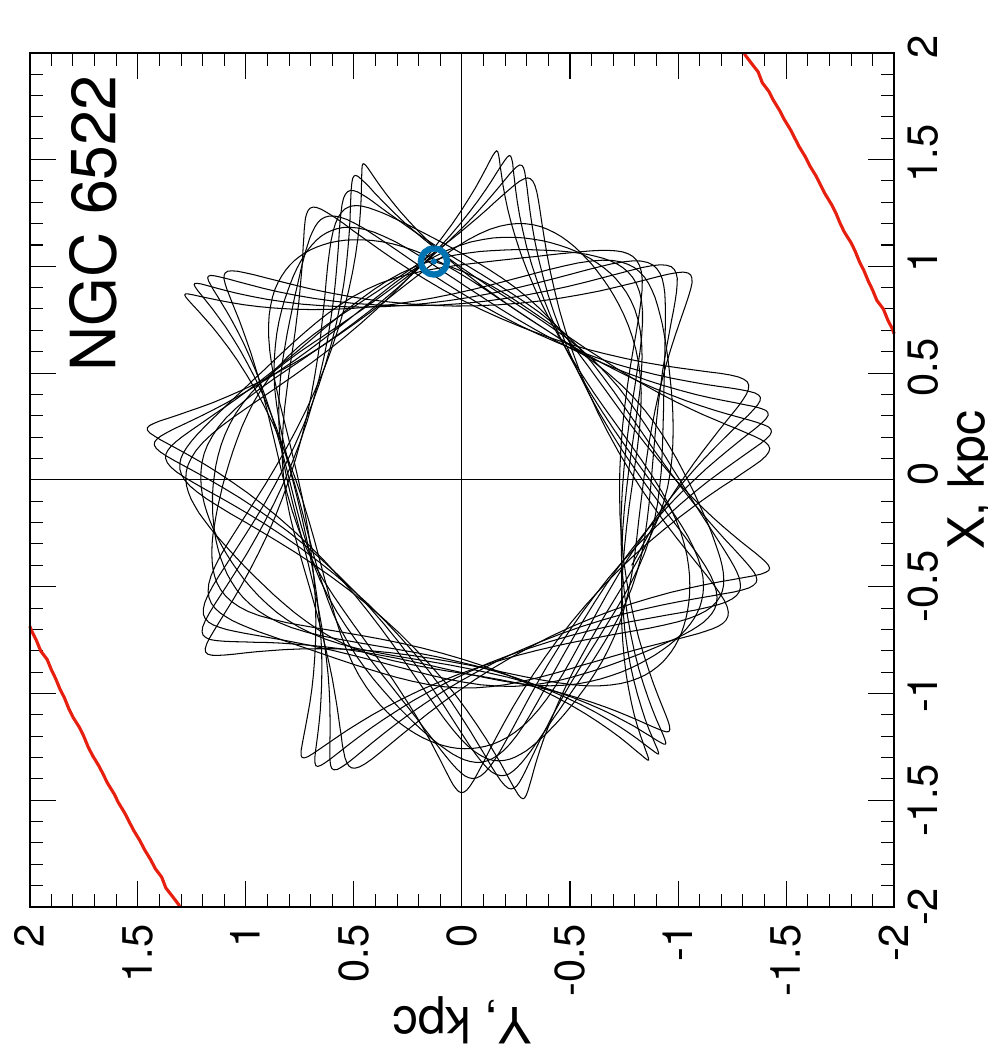}
     \includegraphics[width=0.225\textwidth,angle=-90]{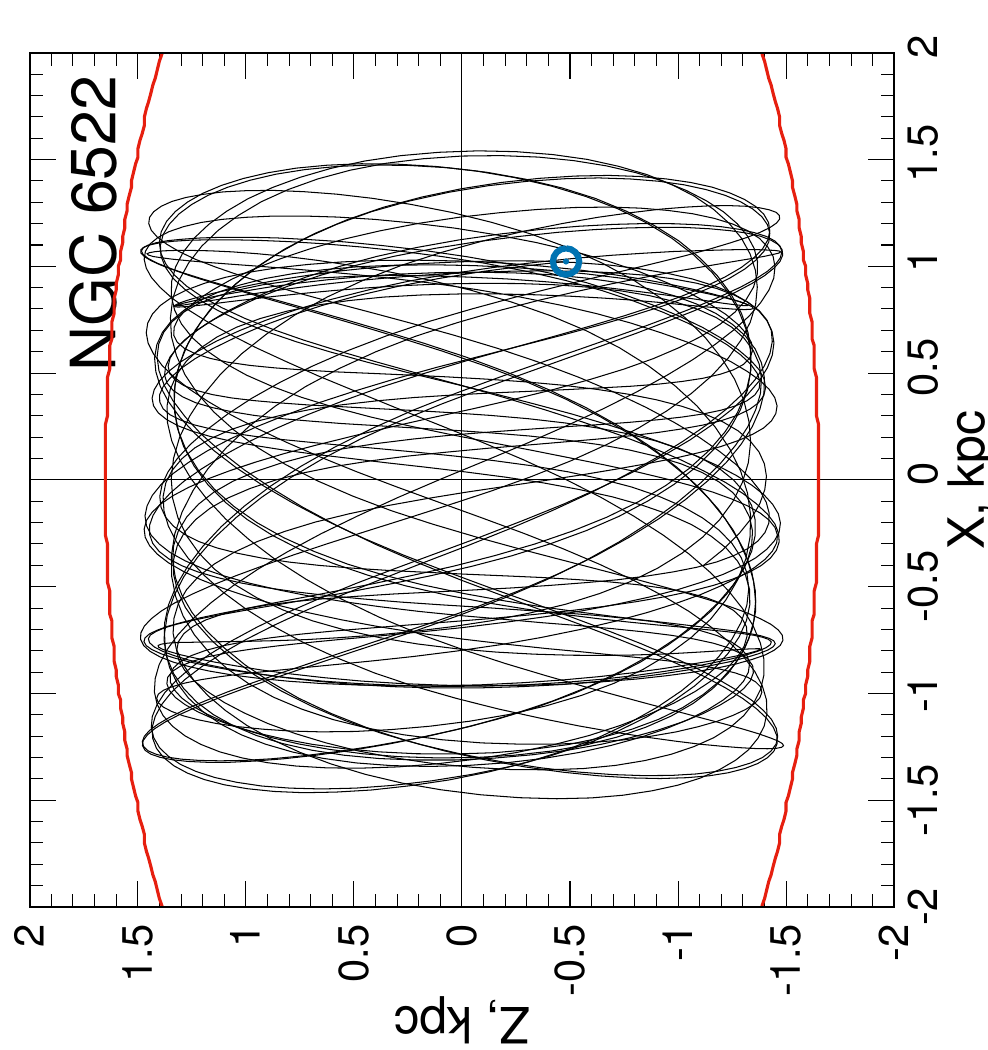}\
          \includegraphics[width=0.225\textwidth,angle=-90]{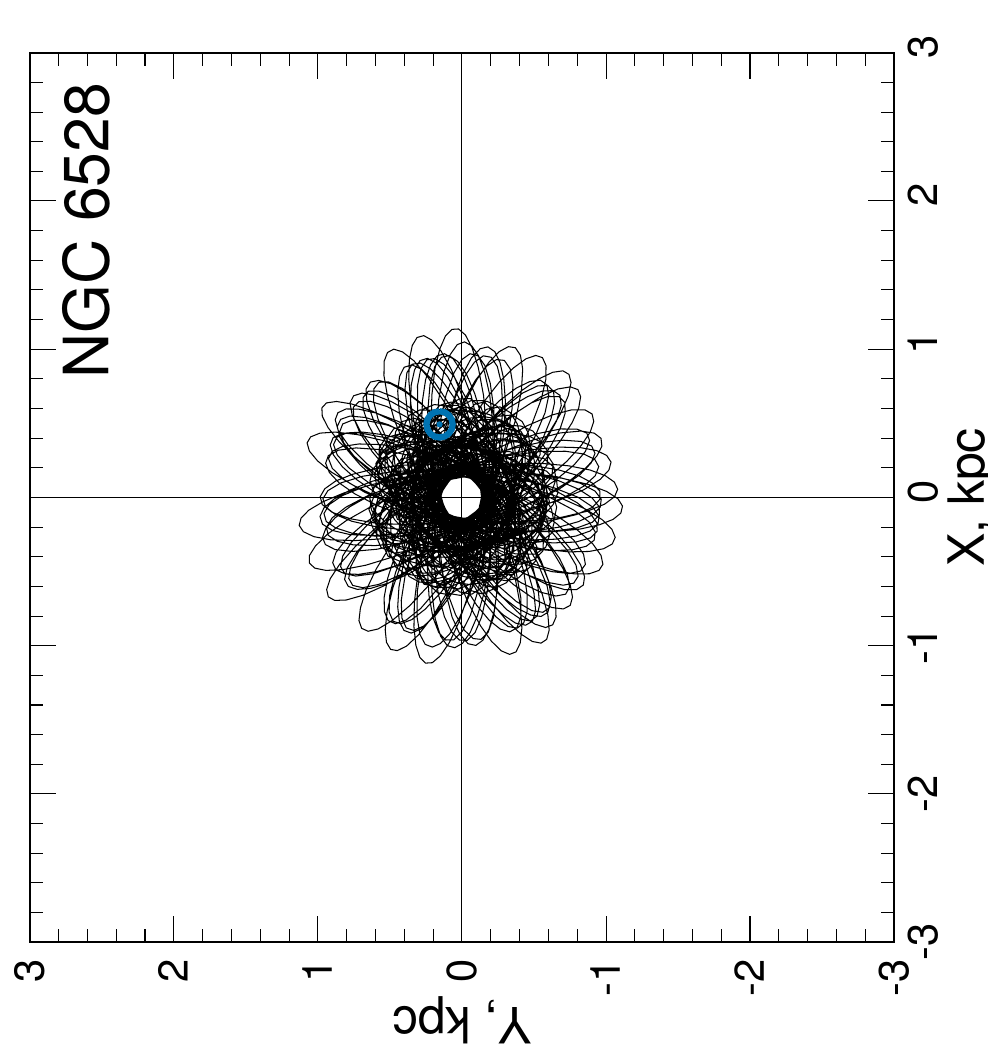}
     \includegraphics[width=0.225\textwidth,angle=-90]{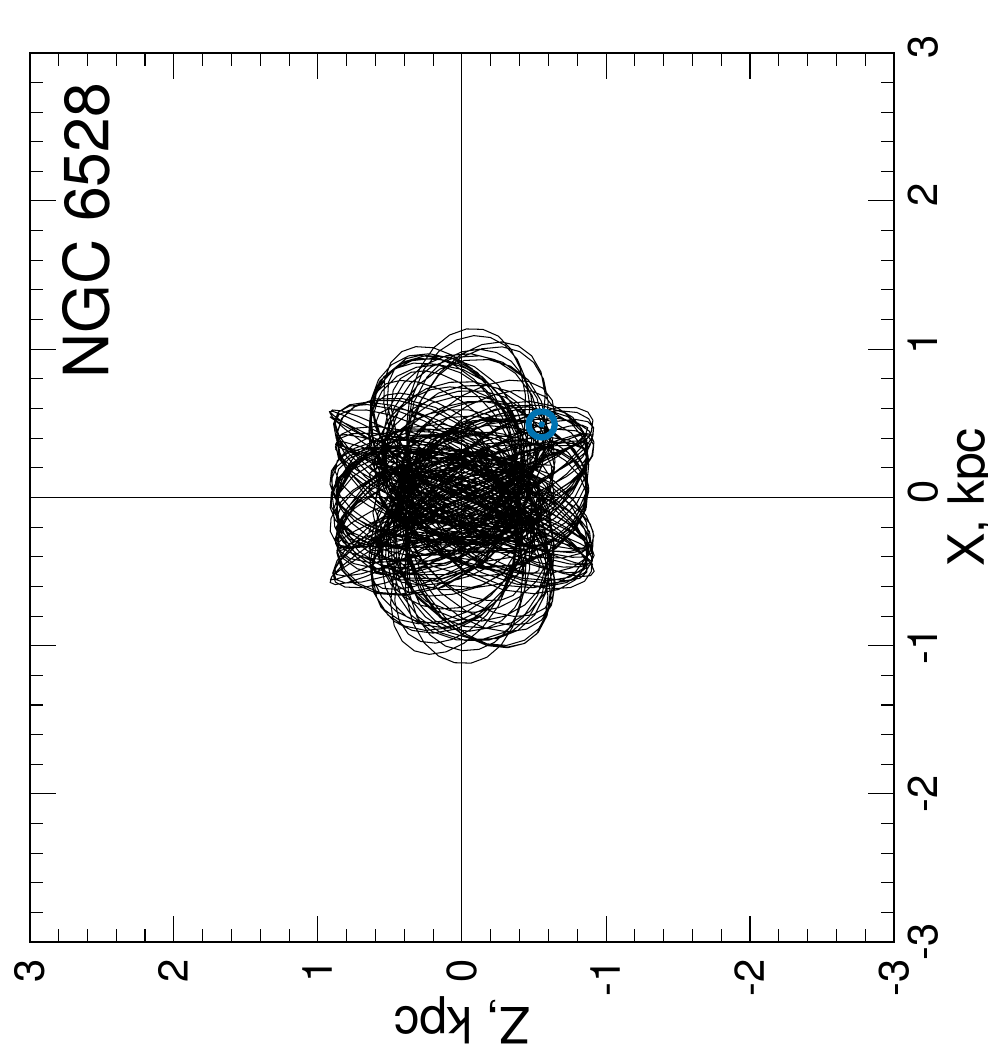}
               \includegraphics[width=0.225\textwidth,angle=-90]{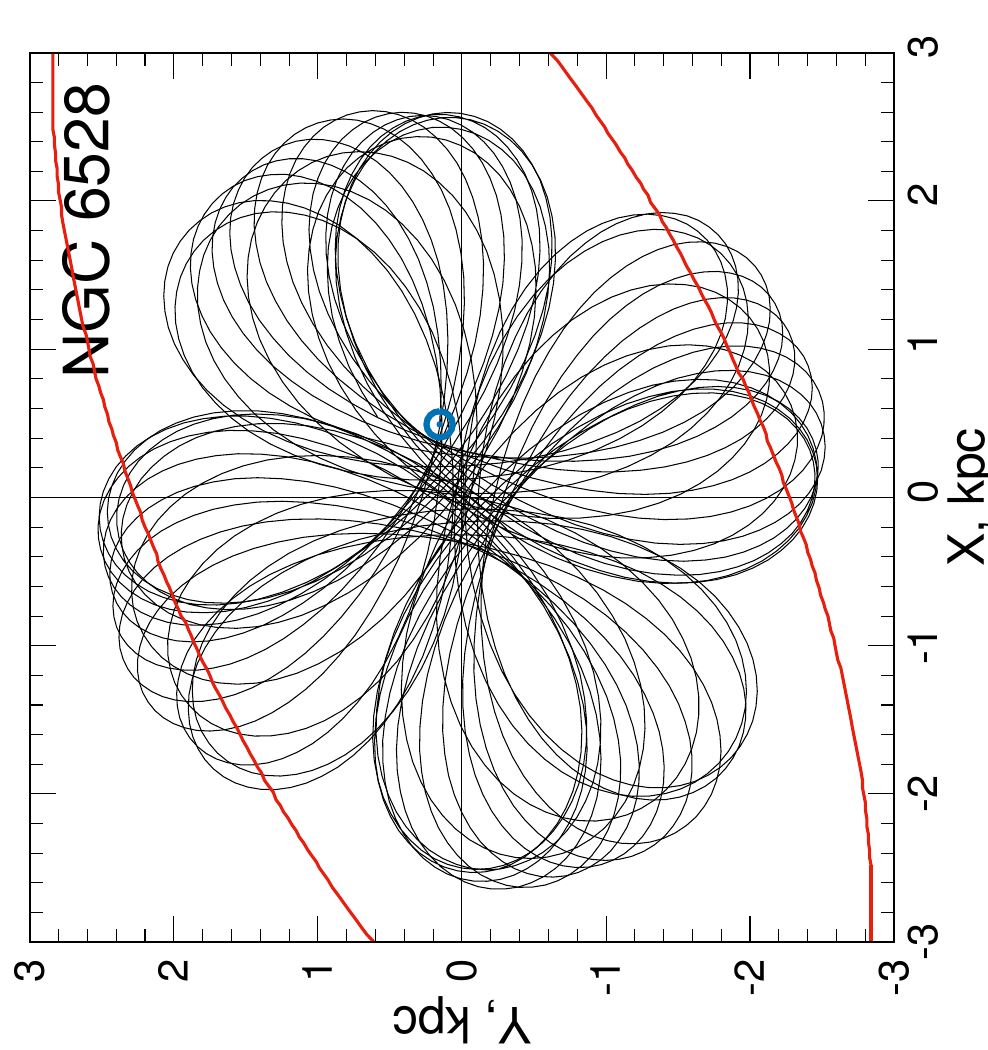}
     \includegraphics[width=0.225\textwidth,angle=-90]{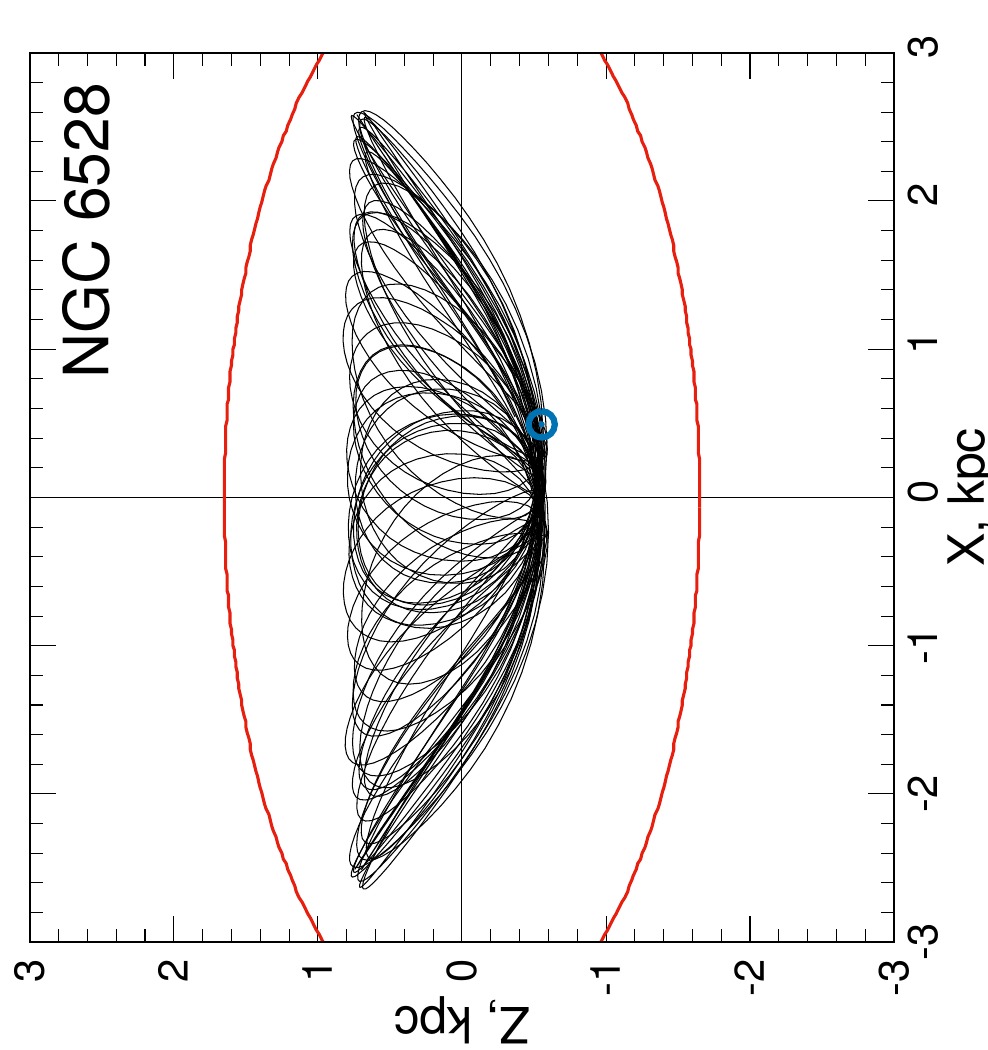}\
        \includegraphics[width=0.225\textwidth,angle=-90]{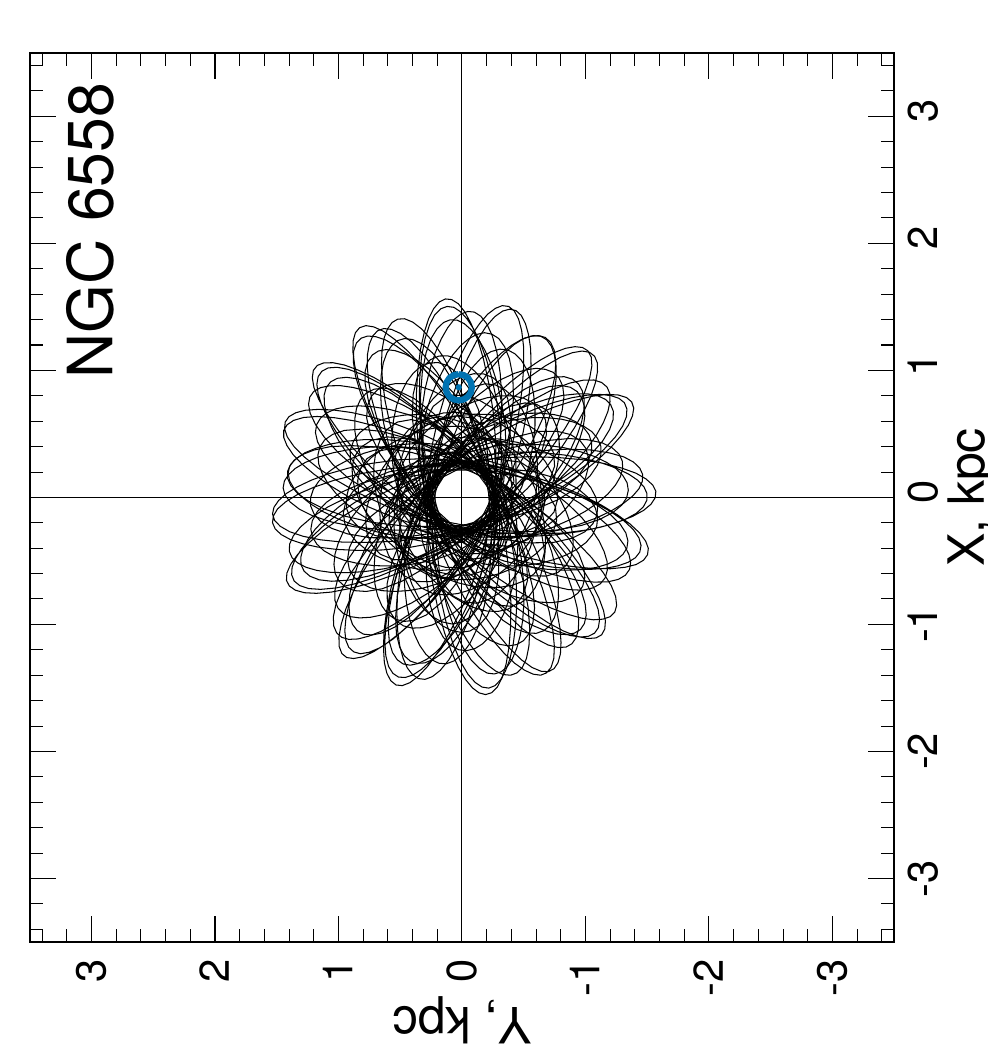}
     \includegraphics[width=0.225\textwidth,angle=-90]{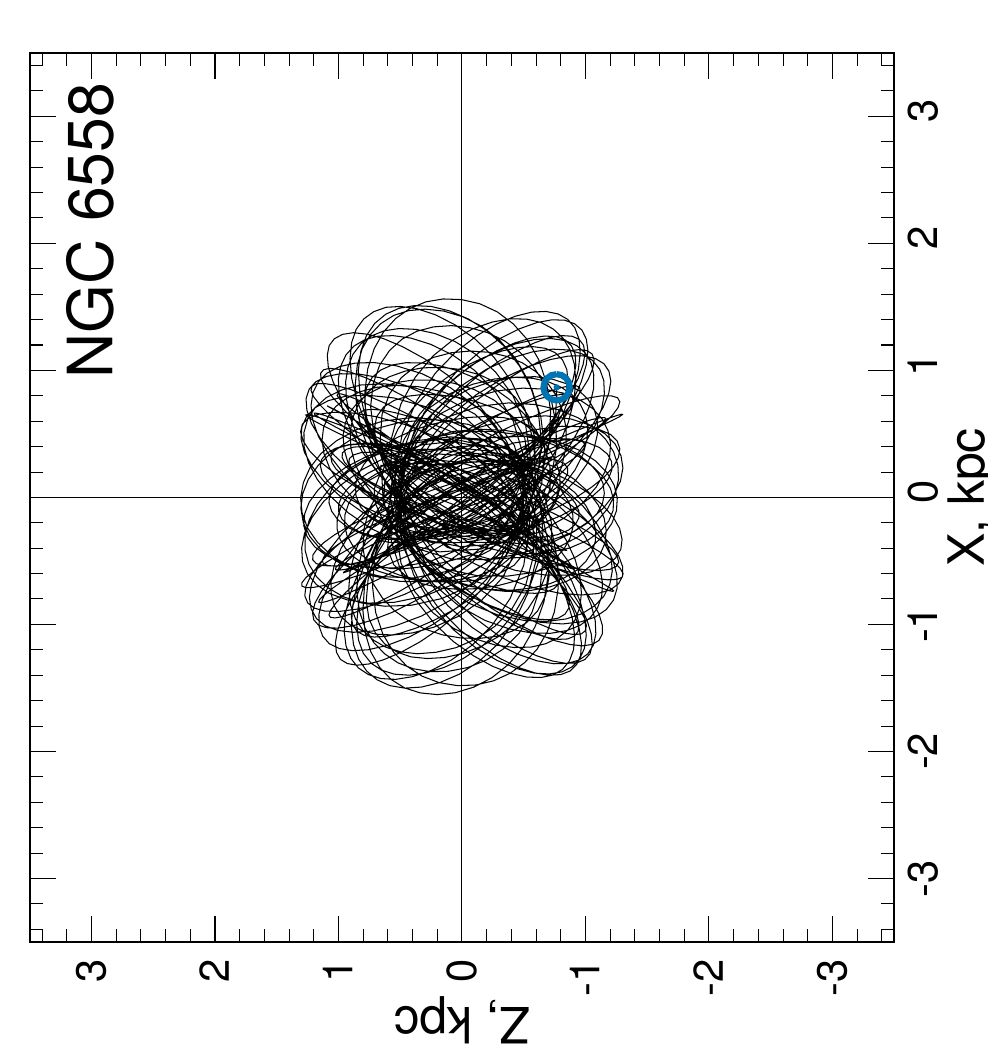}
        \includegraphics[width=0.225\textwidth,angle=-90]{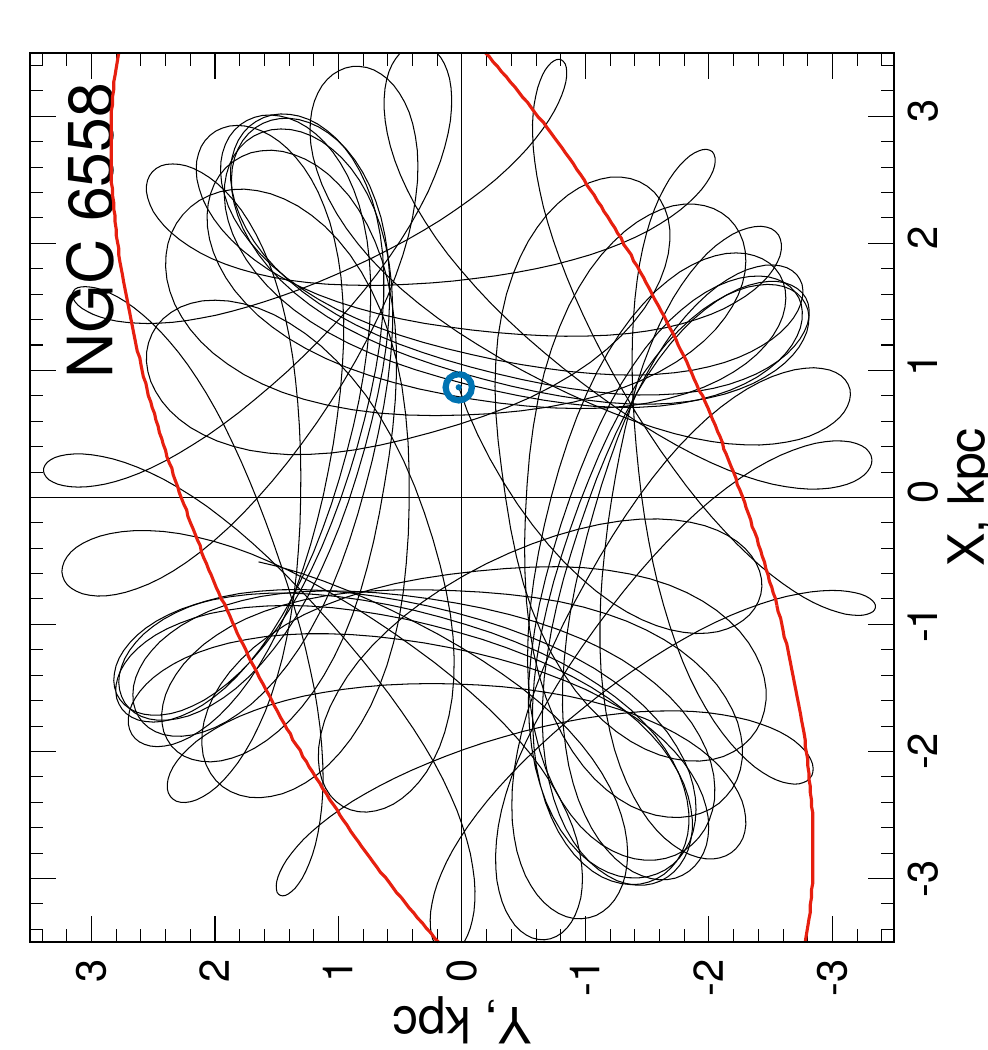}
     \includegraphics[width=0.225\textwidth,angle=-90]{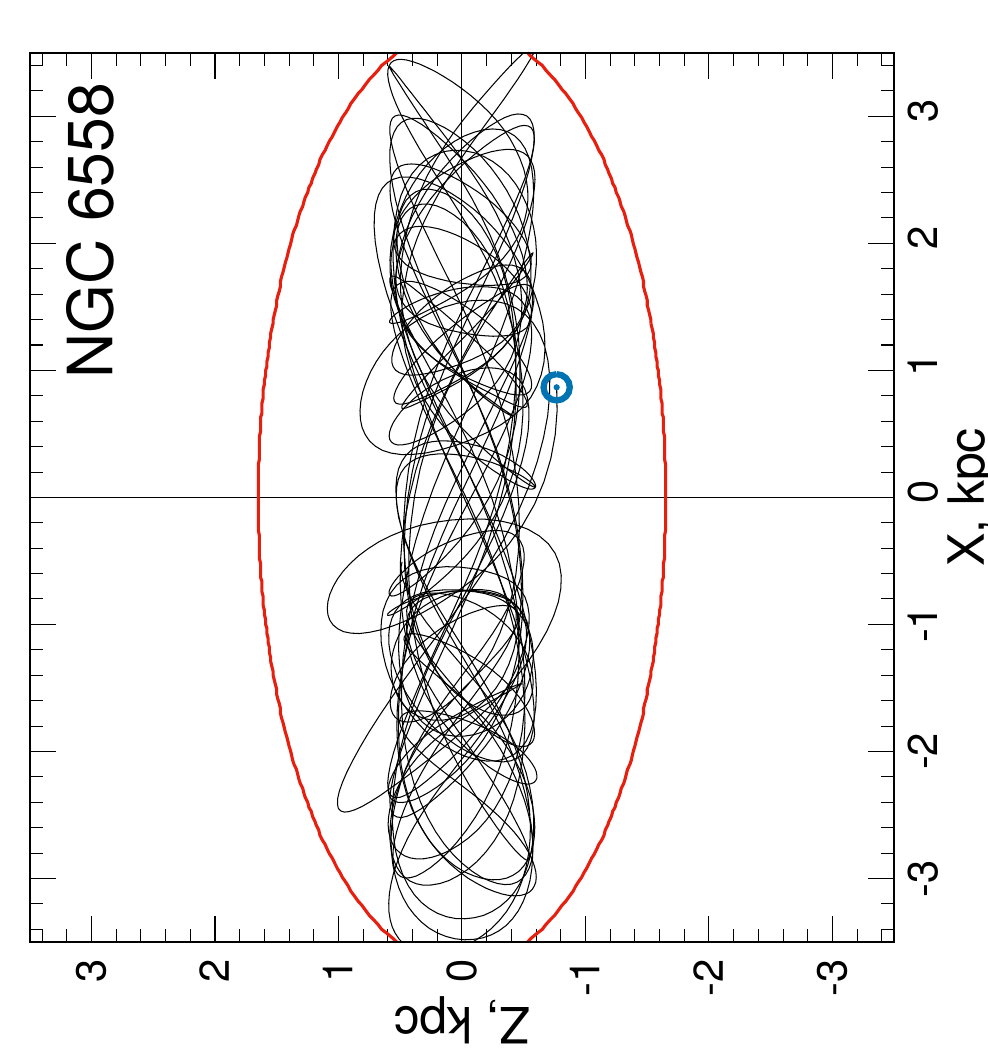}\

\medskip

 \centerline{APPENDIX. Continued}
\label{fD}
\end{center}}
\end{figure*}

\begin{figure*}
{\begin{center}
    \includegraphics[width=0.225\textwidth,angle=-90]{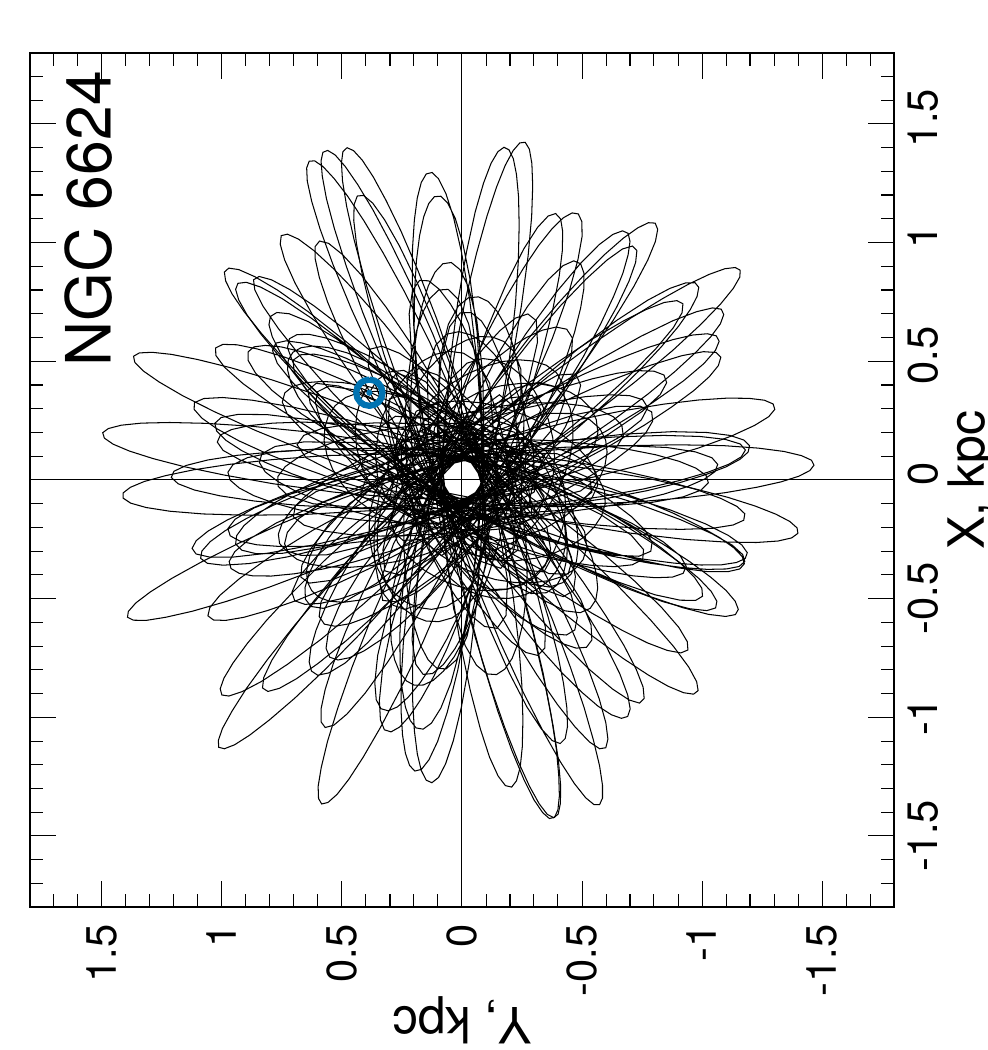}
     \includegraphics[width=0.225\textwidth,angle=-90]{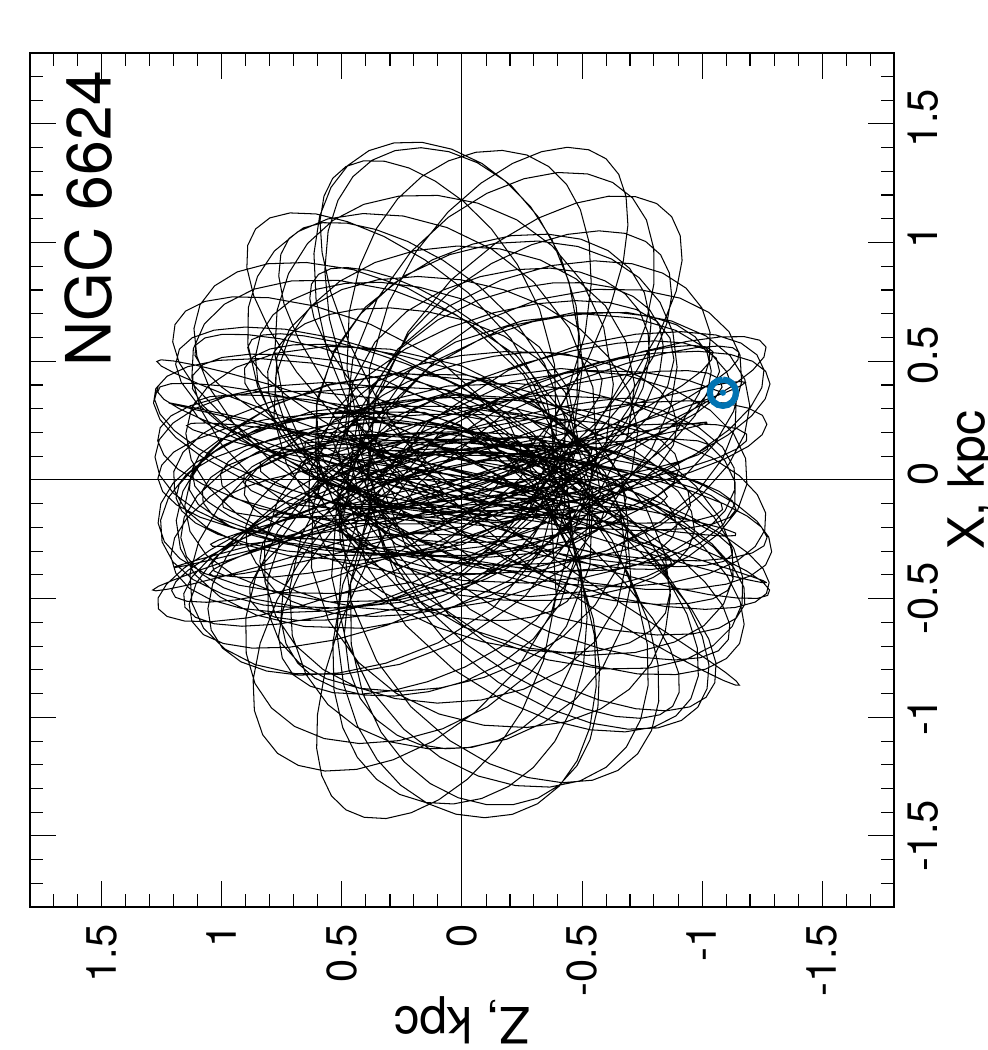}
         \includegraphics[width=0.225\textwidth,angle=-90]{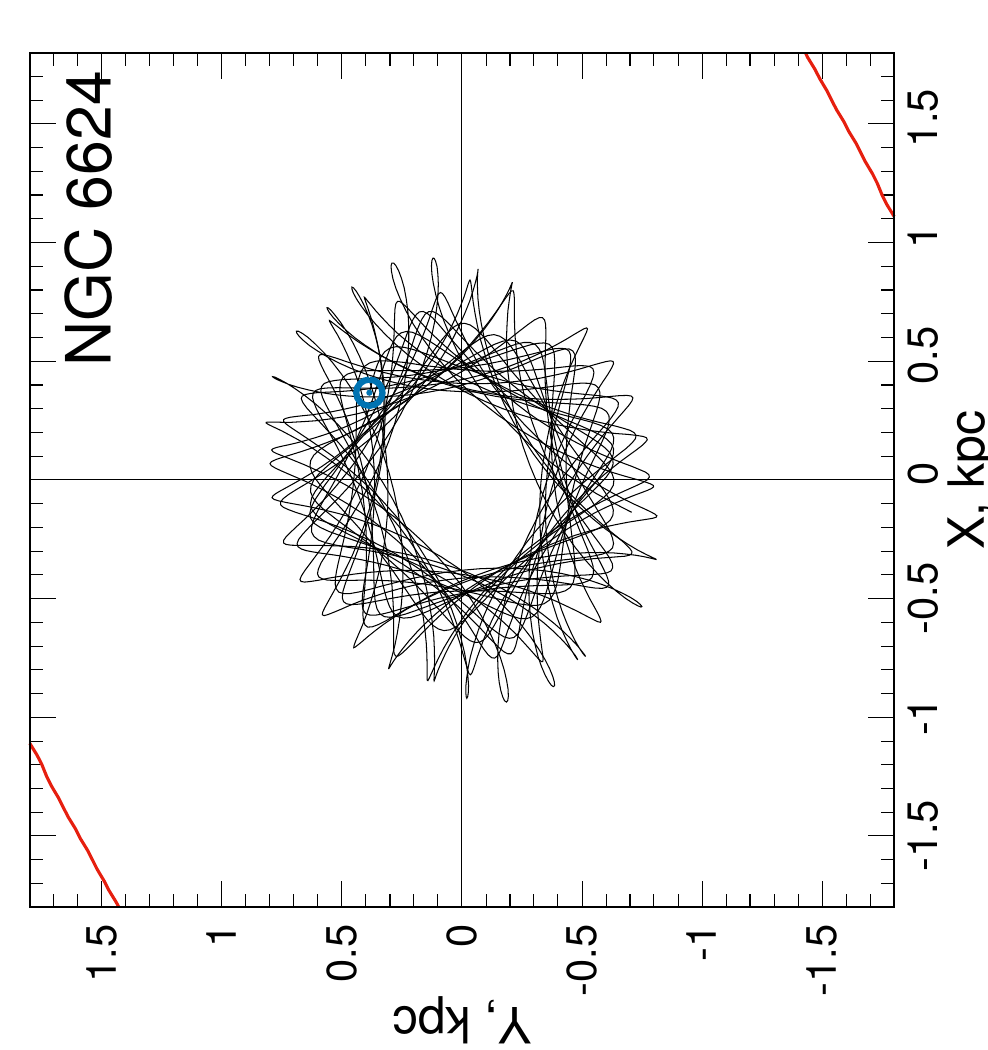}
     \includegraphics[width=0.225\textwidth,angle=-90]{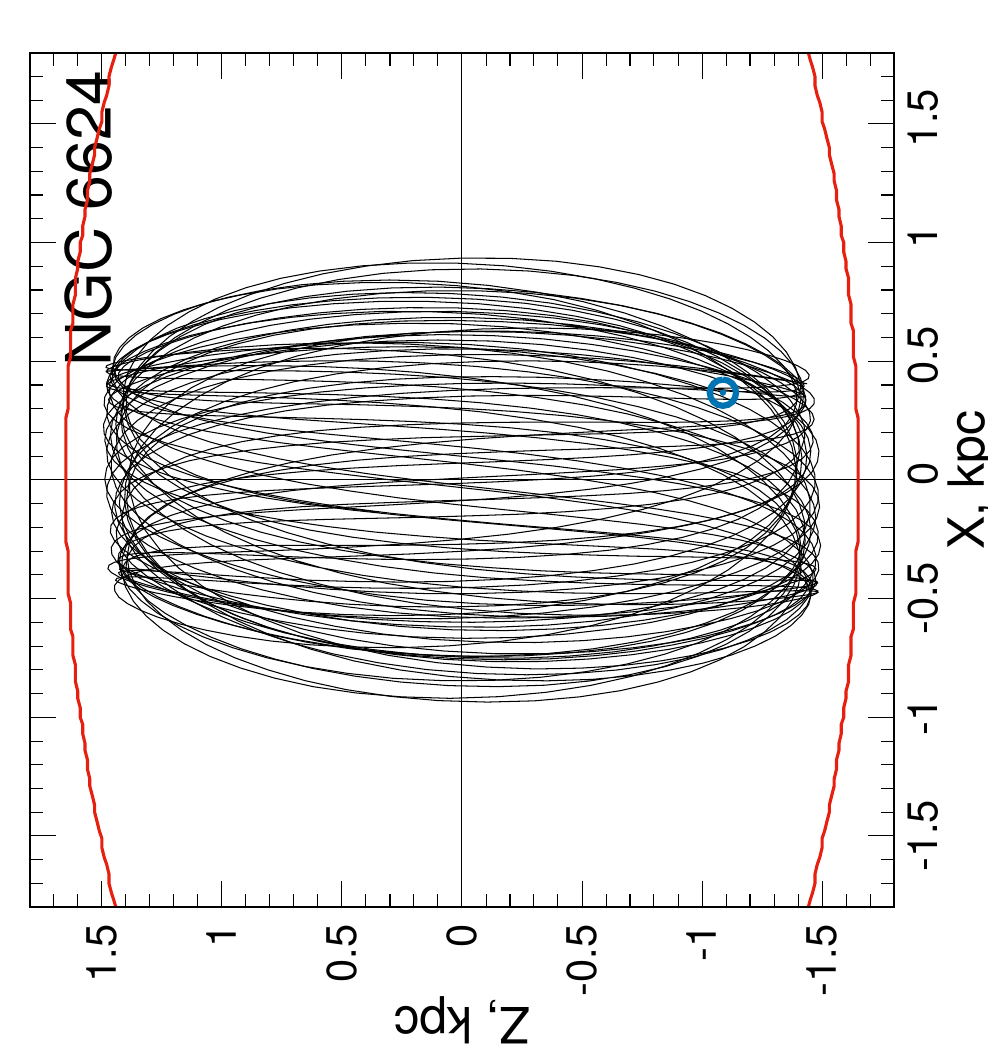}\
      \includegraphics[width=0.225\textwidth,angle=-90]{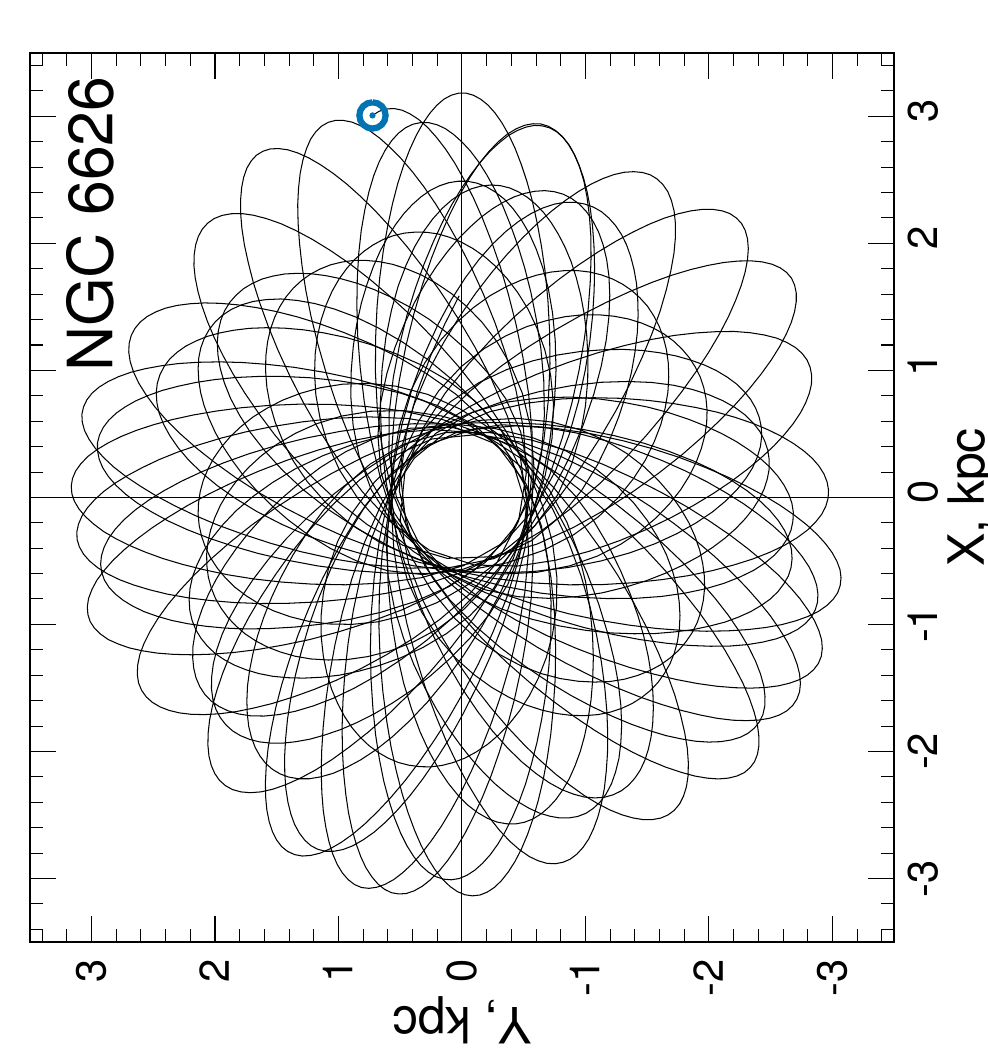}
     \includegraphics[width=0.225\textwidth,angle=-90]{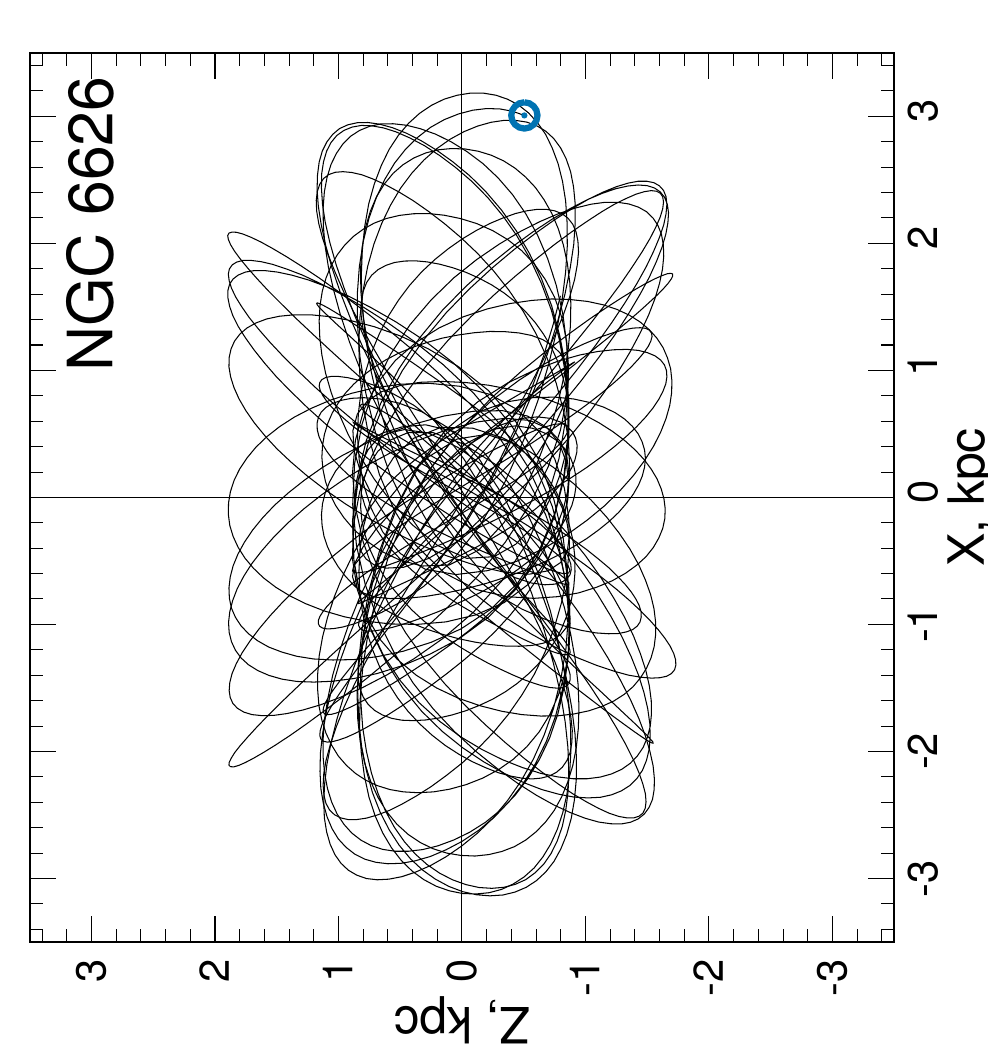}
      \includegraphics[width=0.225\textwidth,angle=-90]{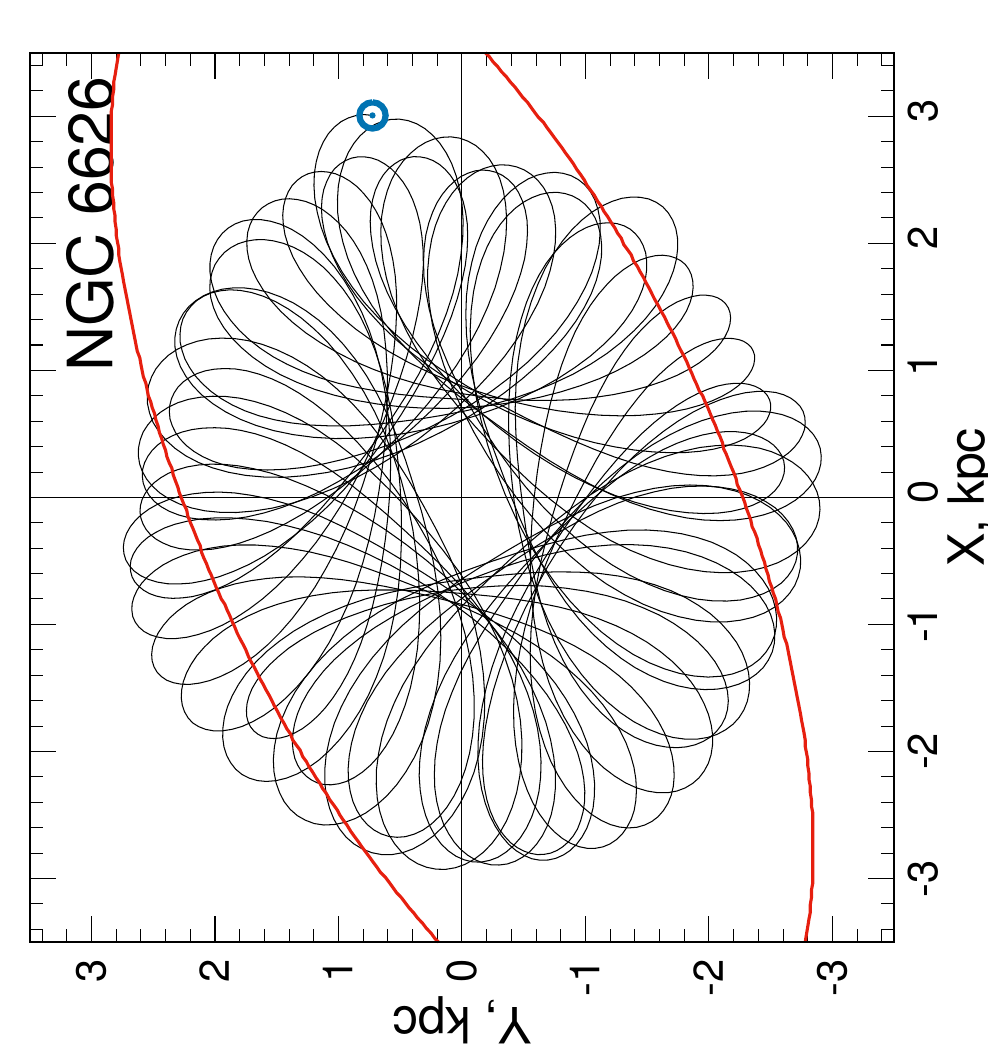}
     \includegraphics[width=0.225\textwidth,angle=-90]{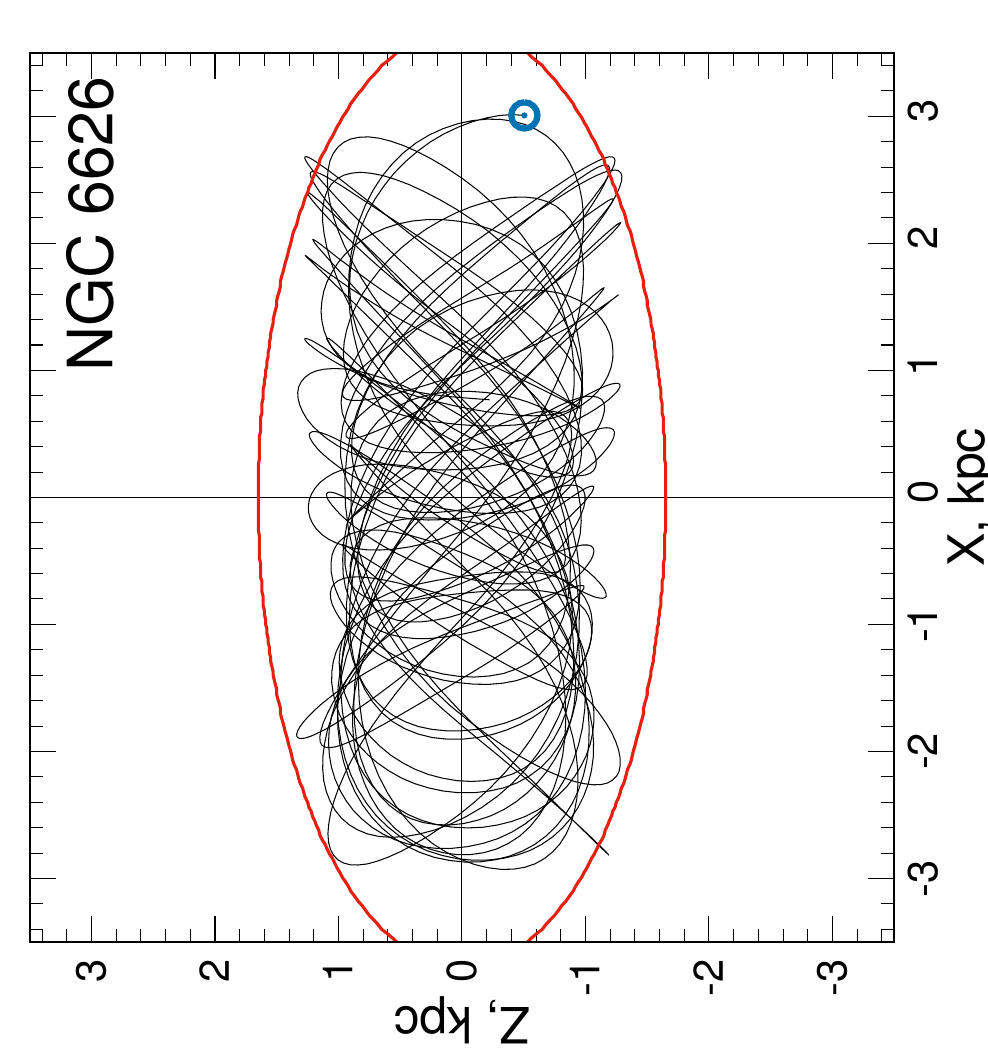}\
   \includegraphics[width=0.225\textwidth,angle=-90]{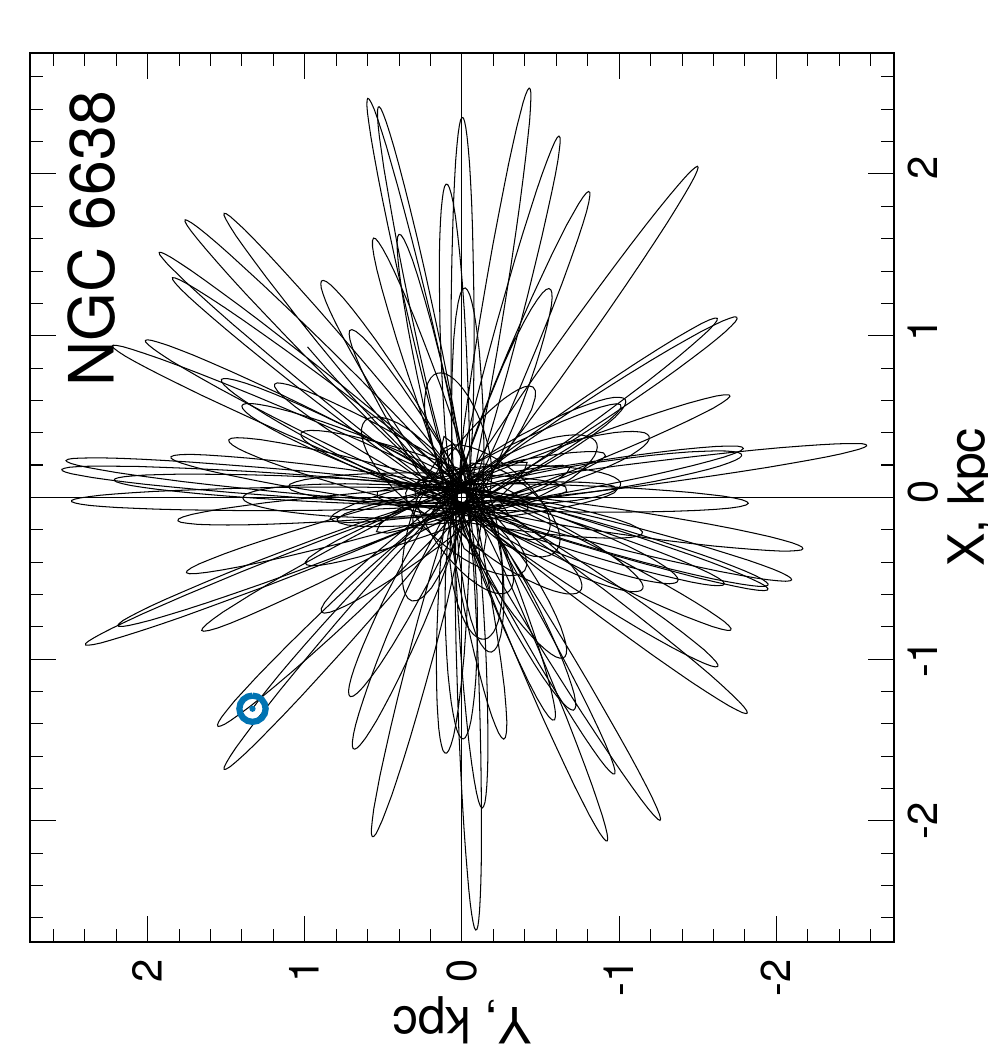}
     \includegraphics[width=0.225\textwidth,angle=-90]{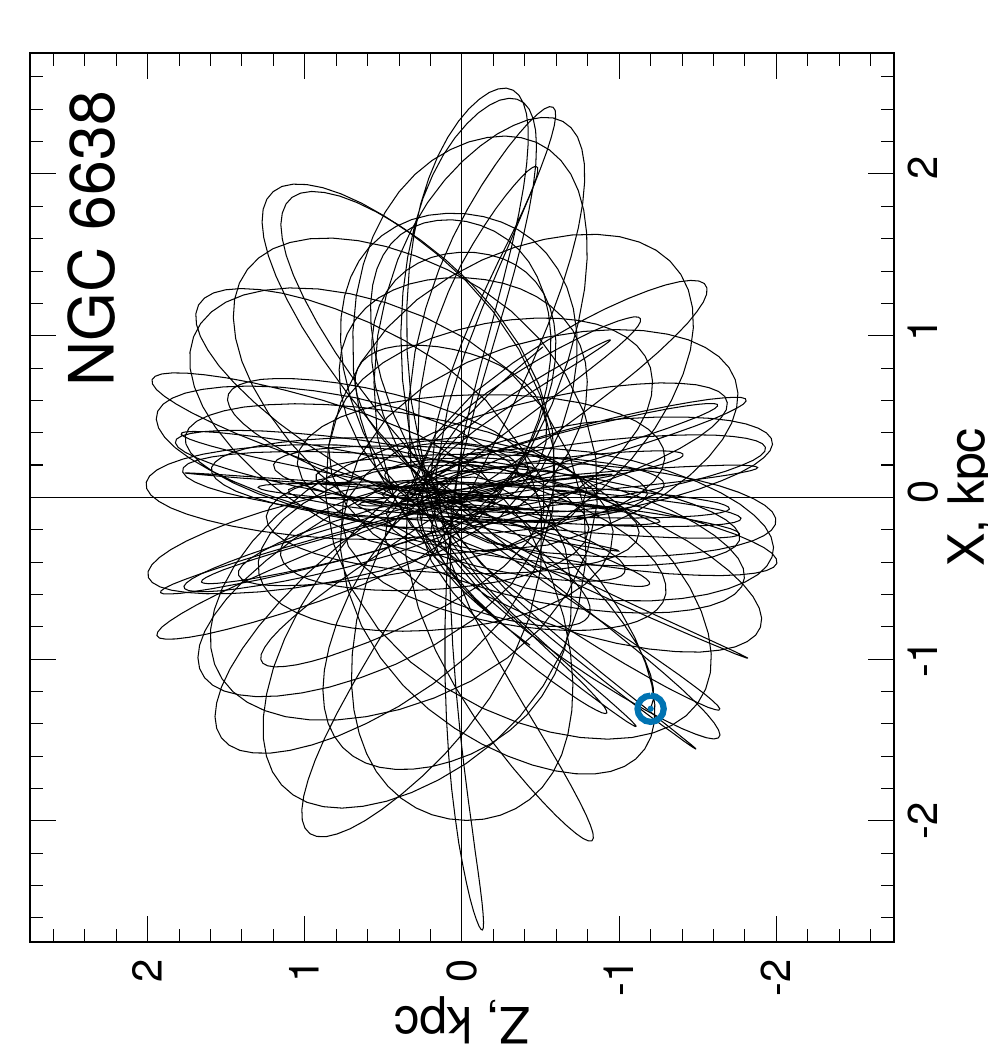}
        \includegraphics[width=0.225\textwidth,angle=-90]{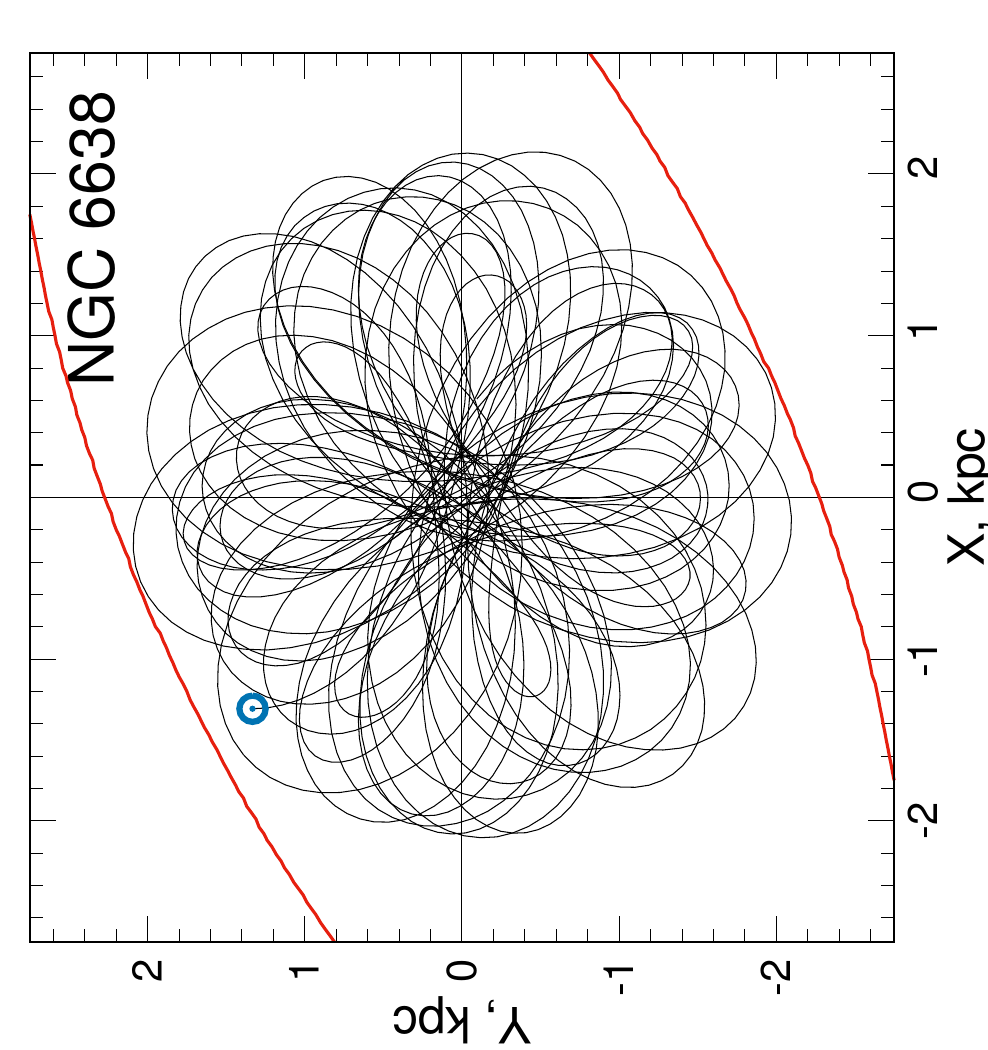}
     \includegraphics[width=0.225\textwidth,angle=-90]{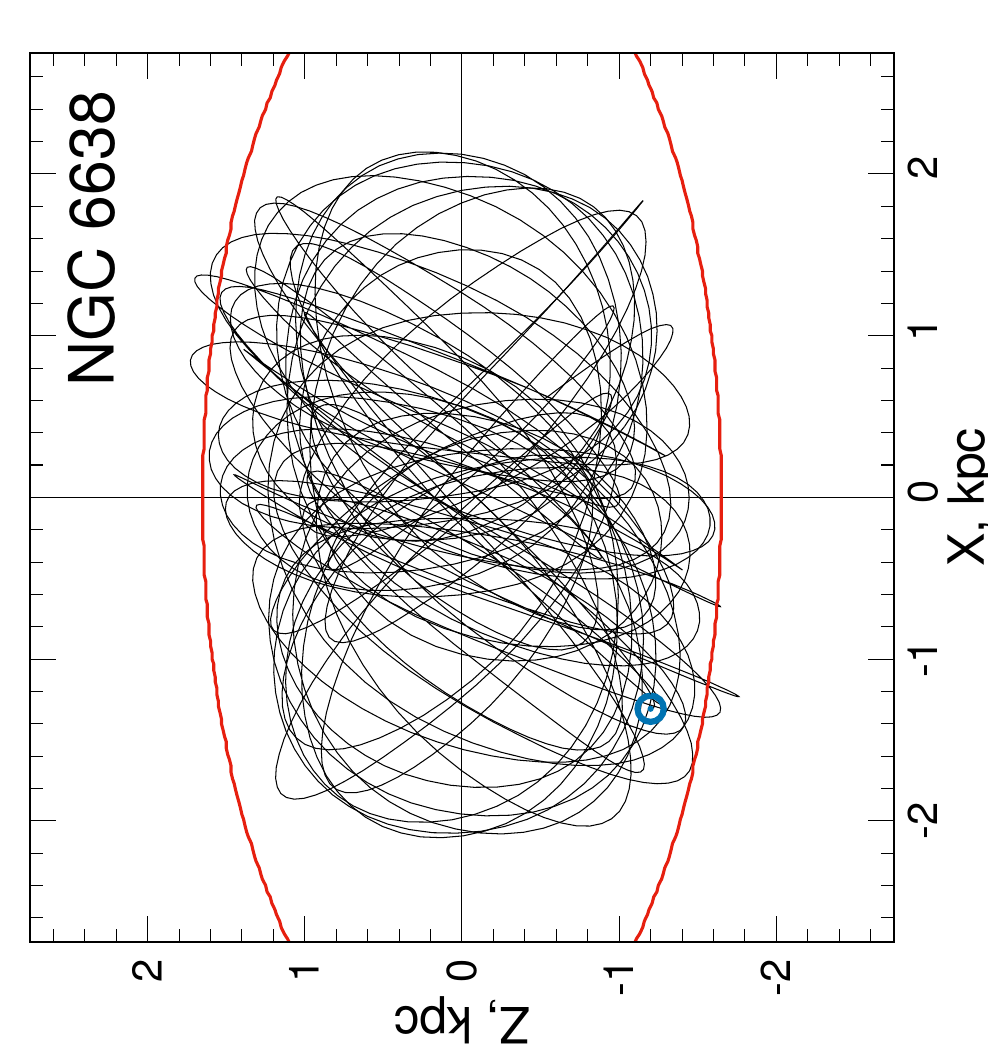}\
    \includegraphics[width=0.225\textwidth,angle=-90]{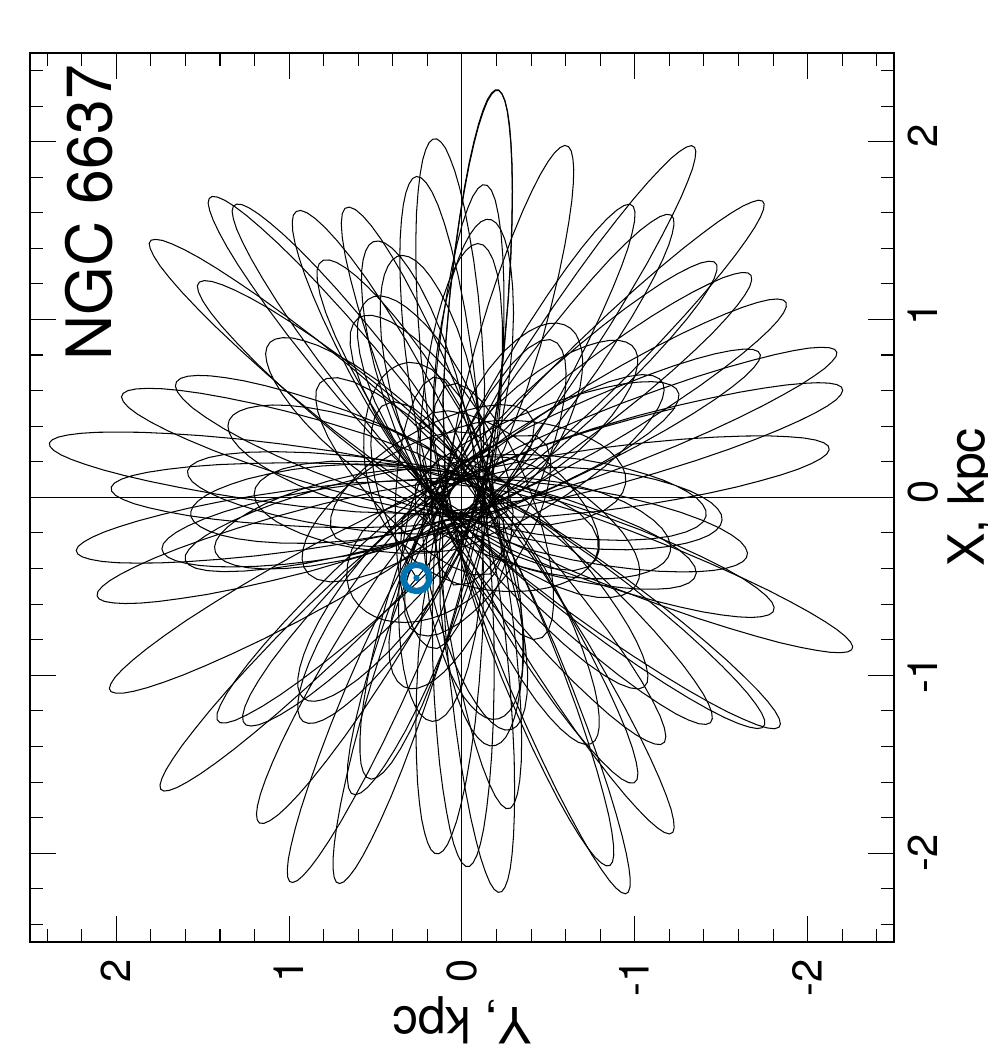}
     \includegraphics[width=0.225\textwidth,angle=-90]{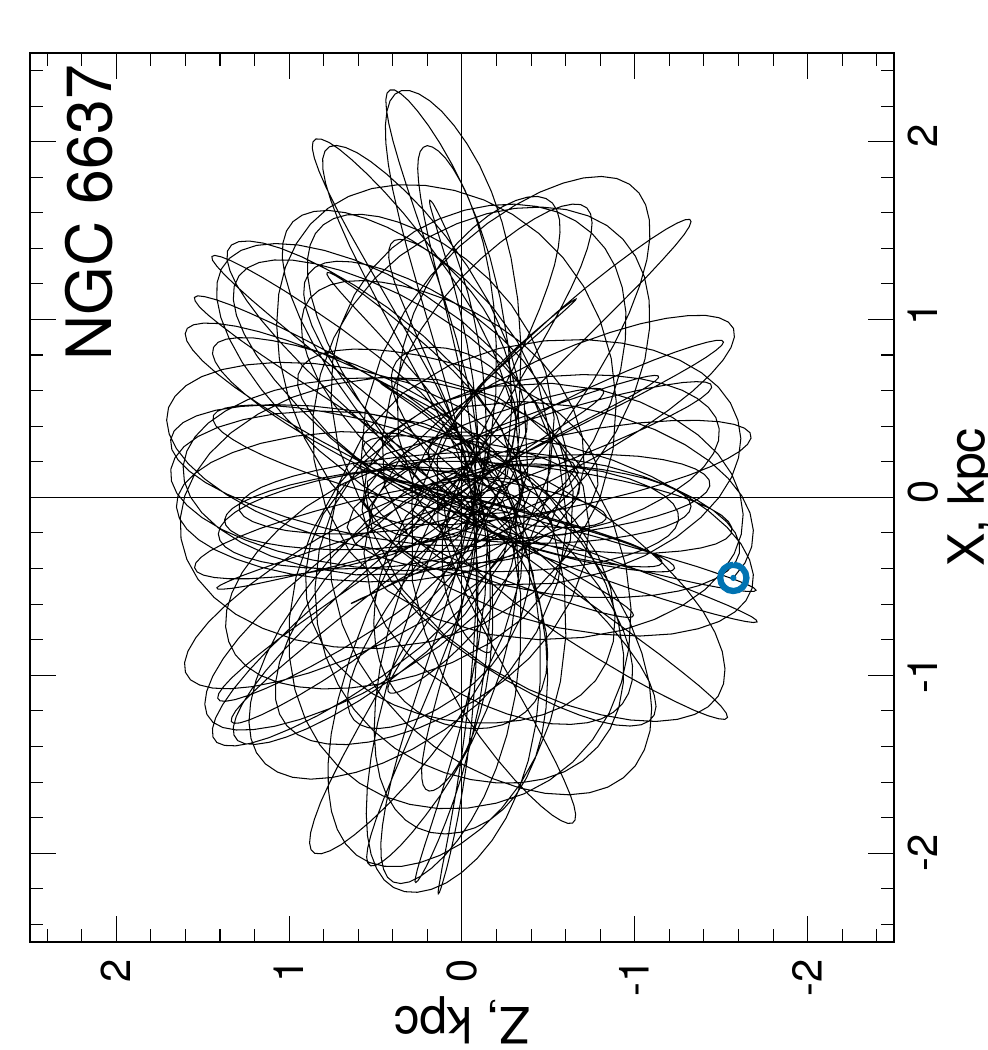}
    \includegraphics[width=0.225\textwidth,angle=-90]{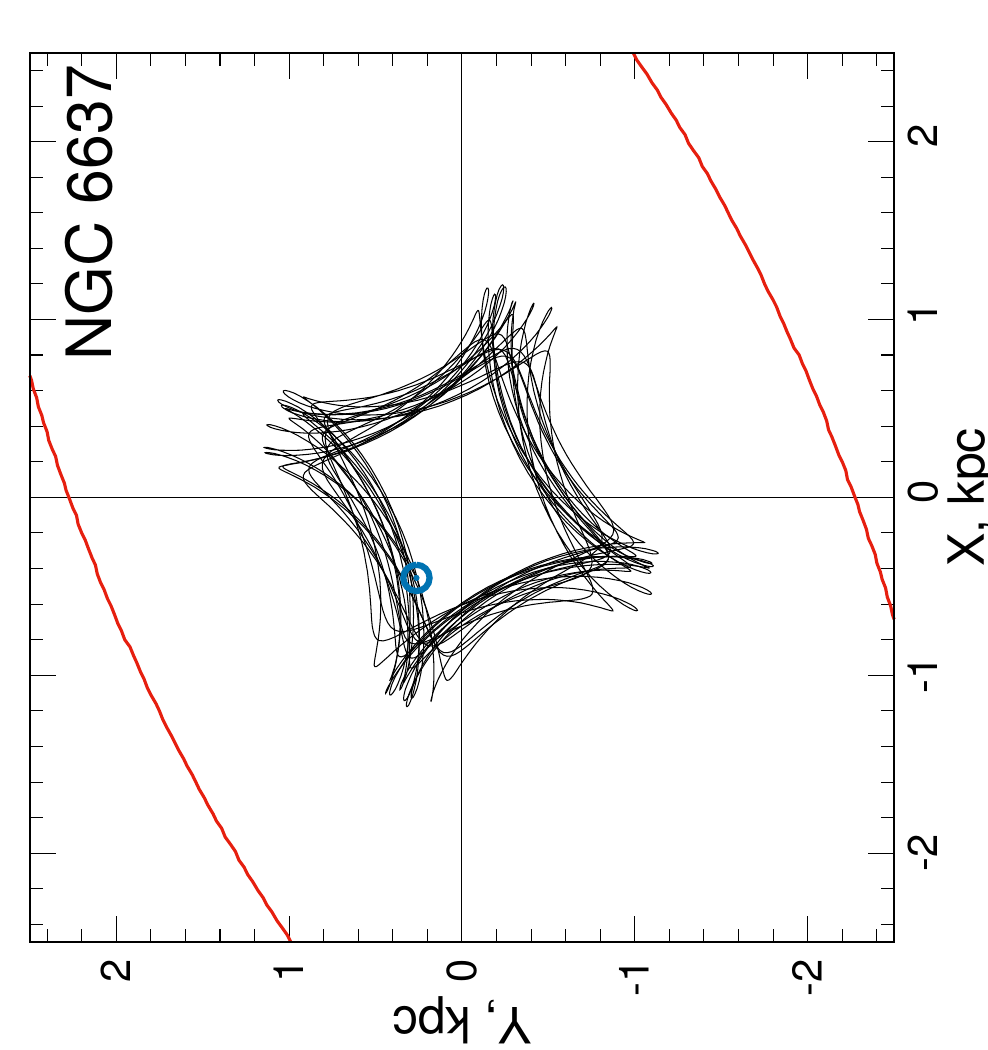}
     \includegraphics[width=0.225\textwidth,angle=-90]{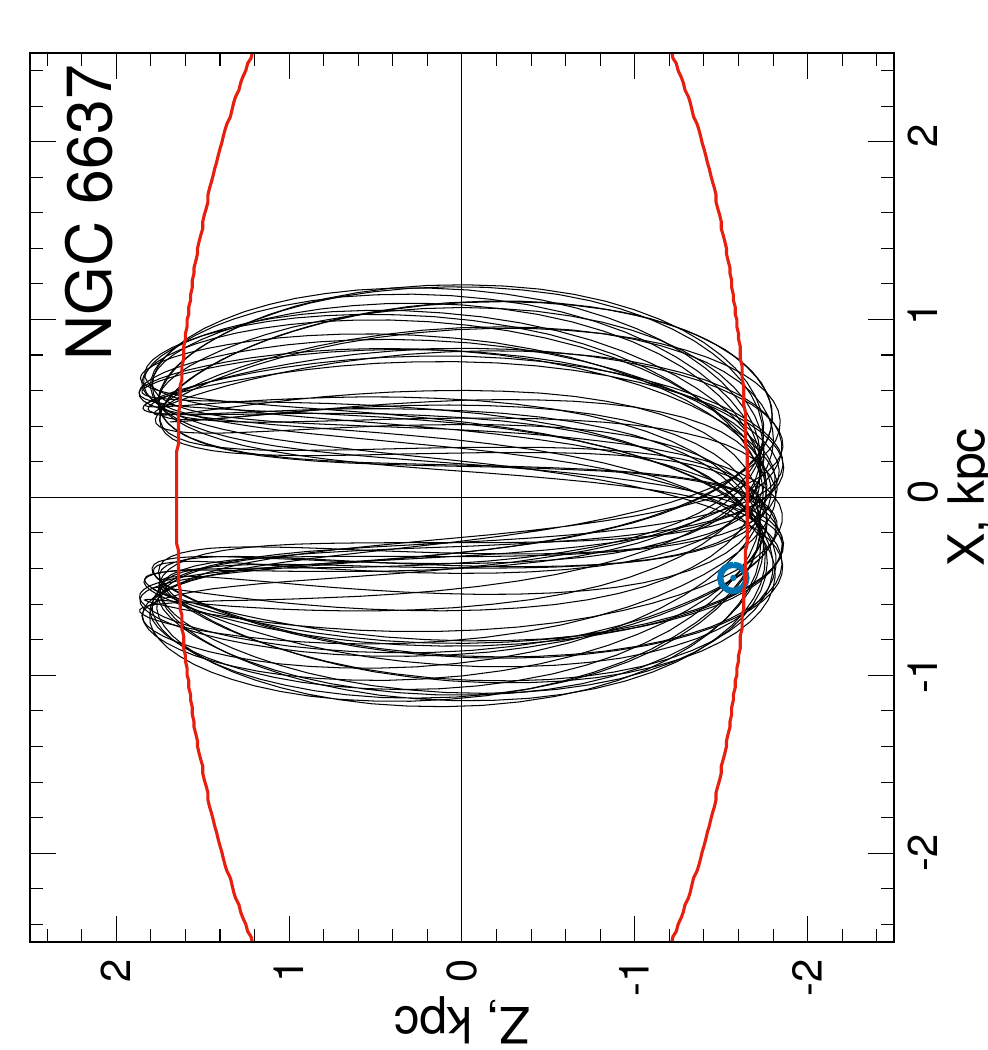}\
    \includegraphics[width=0.225\textwidth,angle=-90]{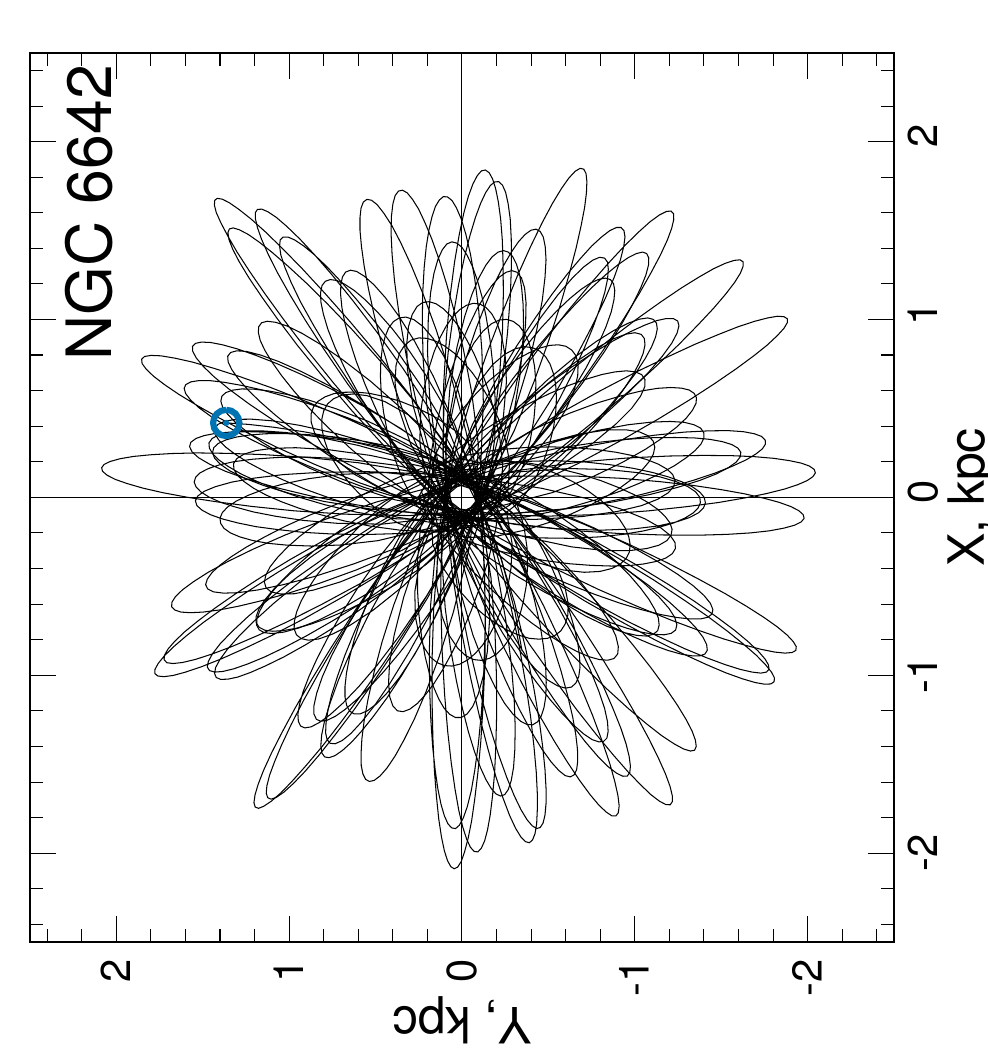}
     \includegraphics[width=0.225\textwidth,angle=-90]{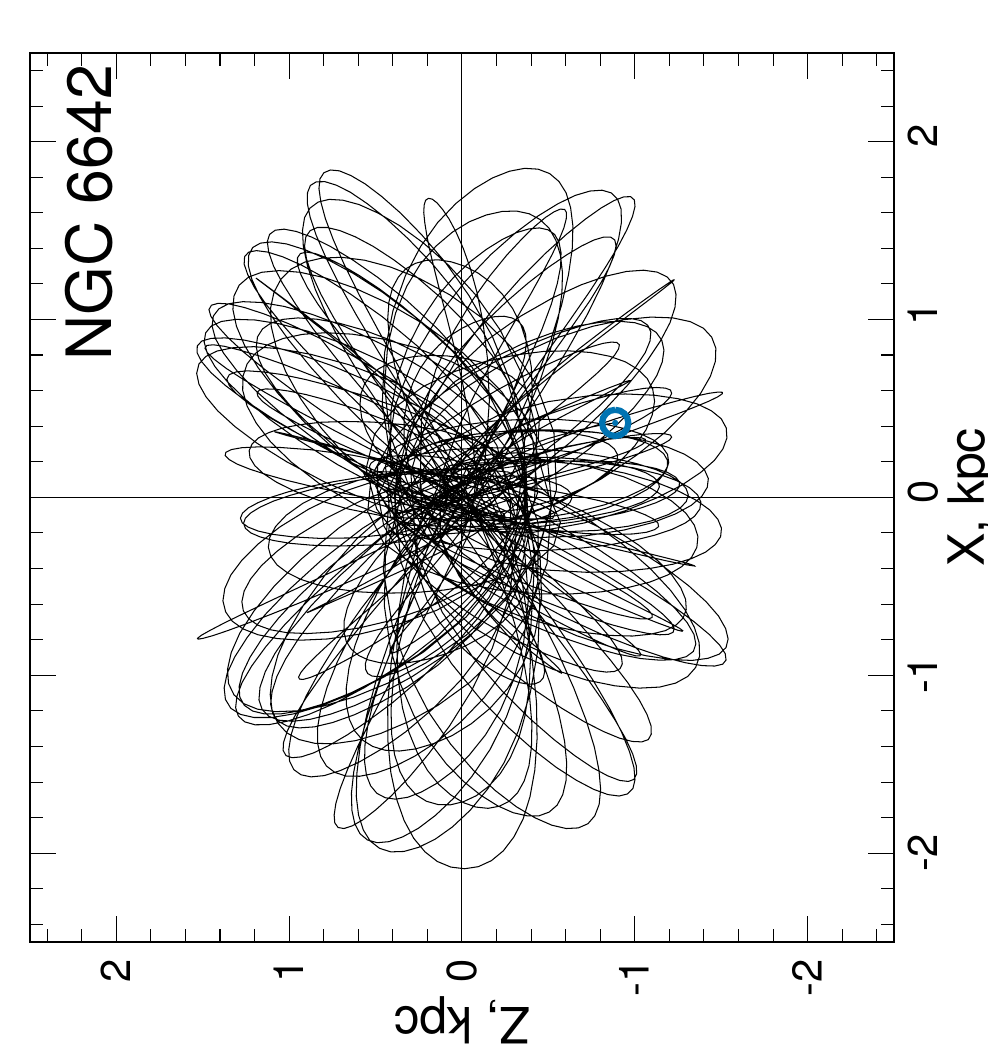}
         \includegraphics[width=0.225\textwidth,angle=-90]{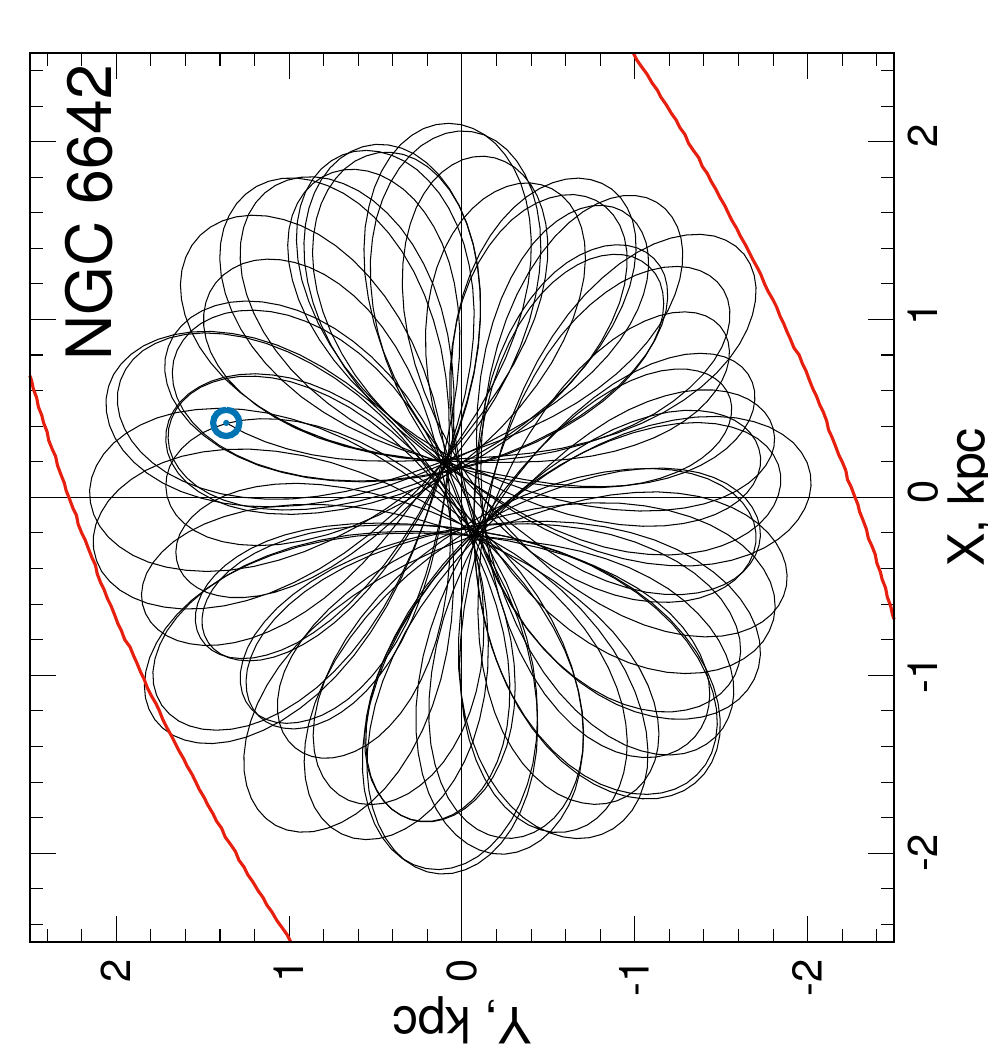}
     \includegraphics[width=0.225\textwidth,angle=-90]{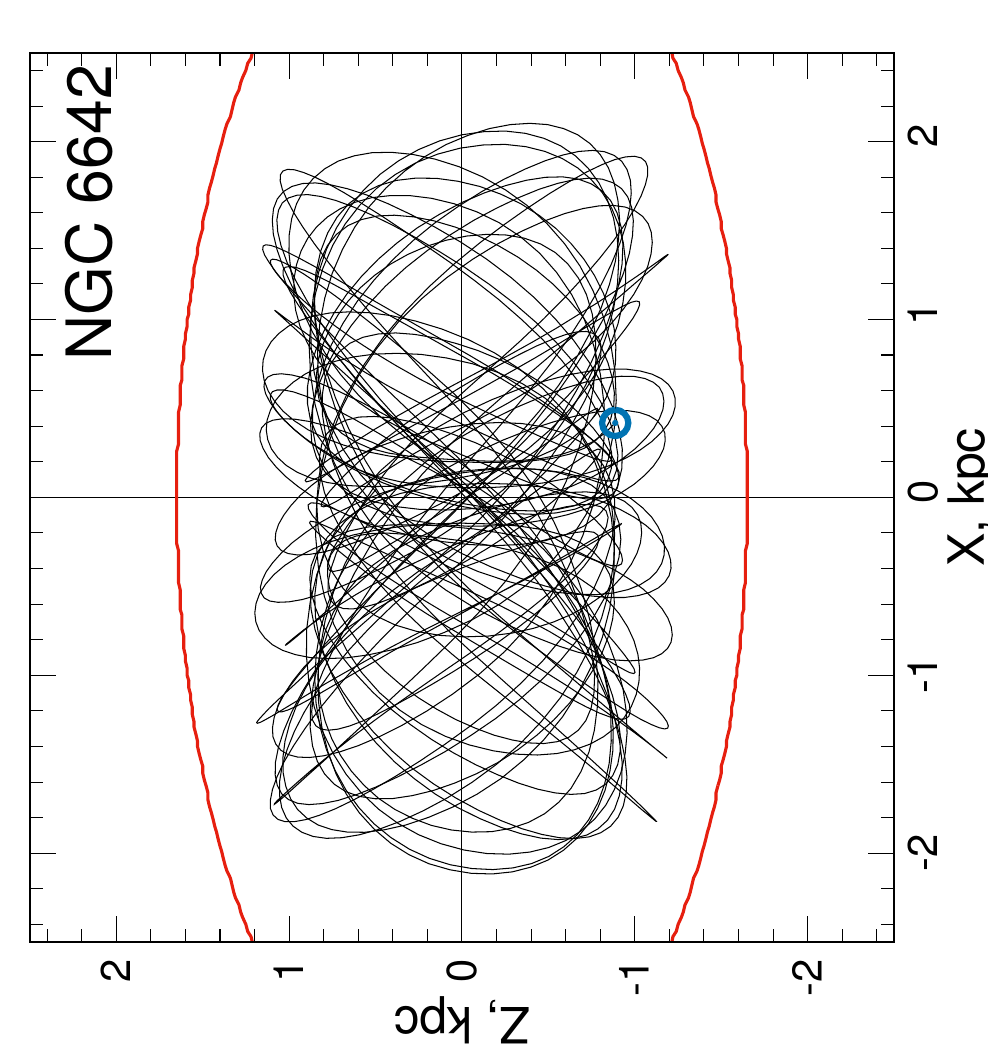}\
        \includegraphics[width=0.225\textwidth,angle=-90]{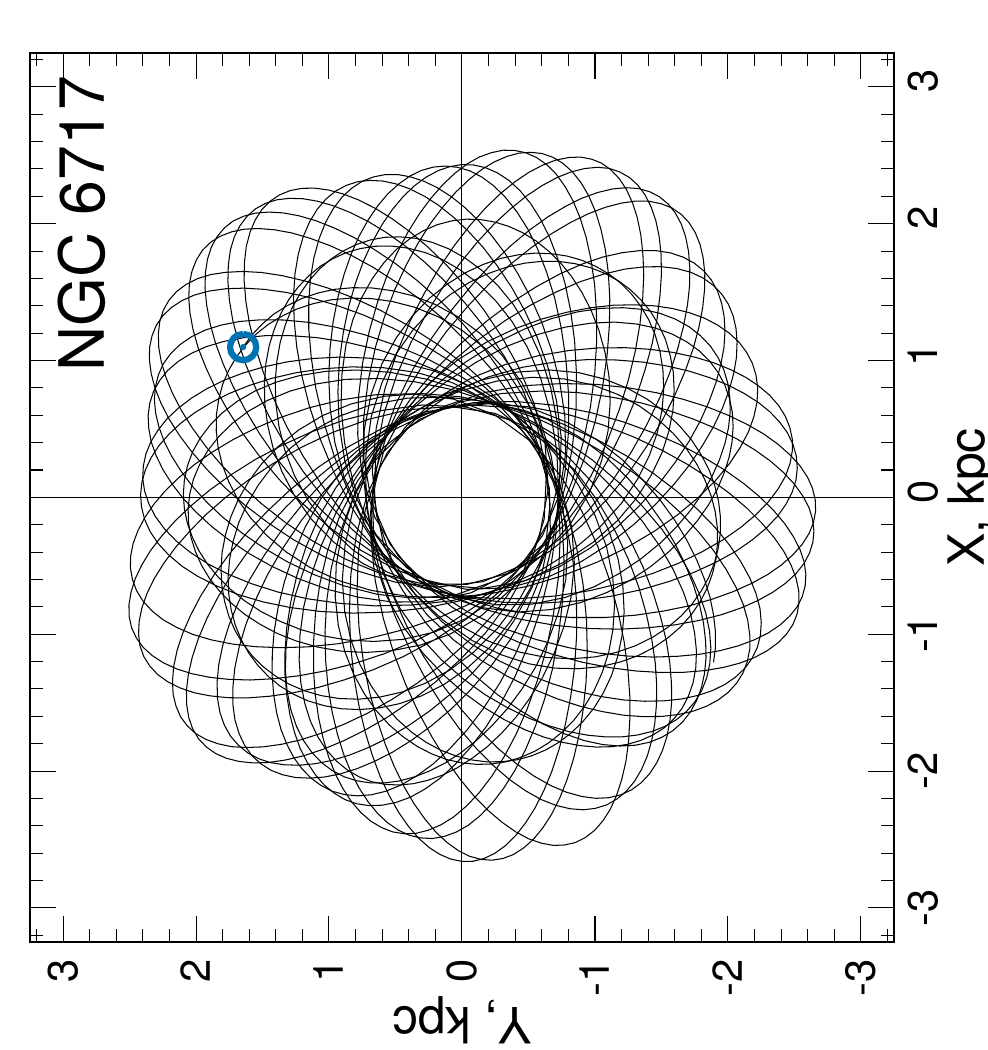}
     \includegraphics[width=0.225\textwidth,angle=-90]{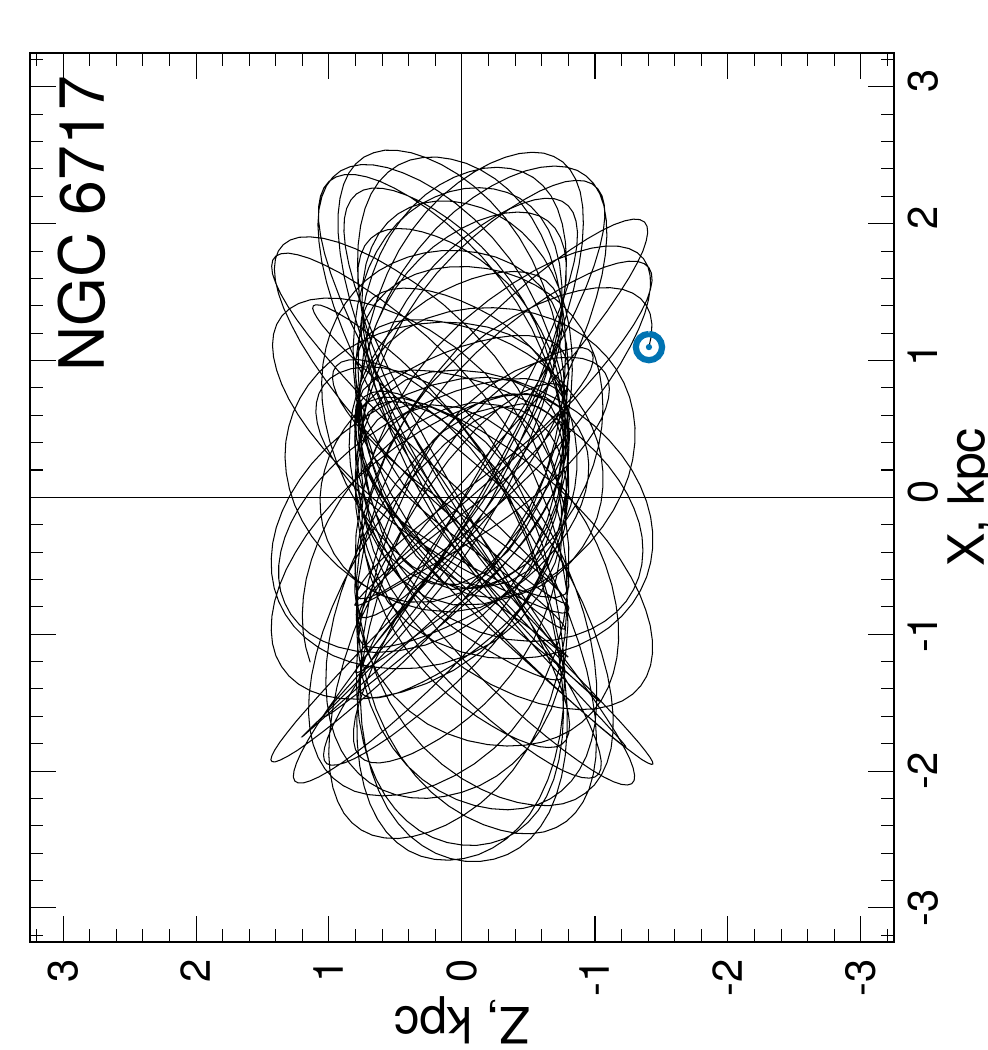}
        \includegraphics[width=0.225\textwidth,angle=-90]{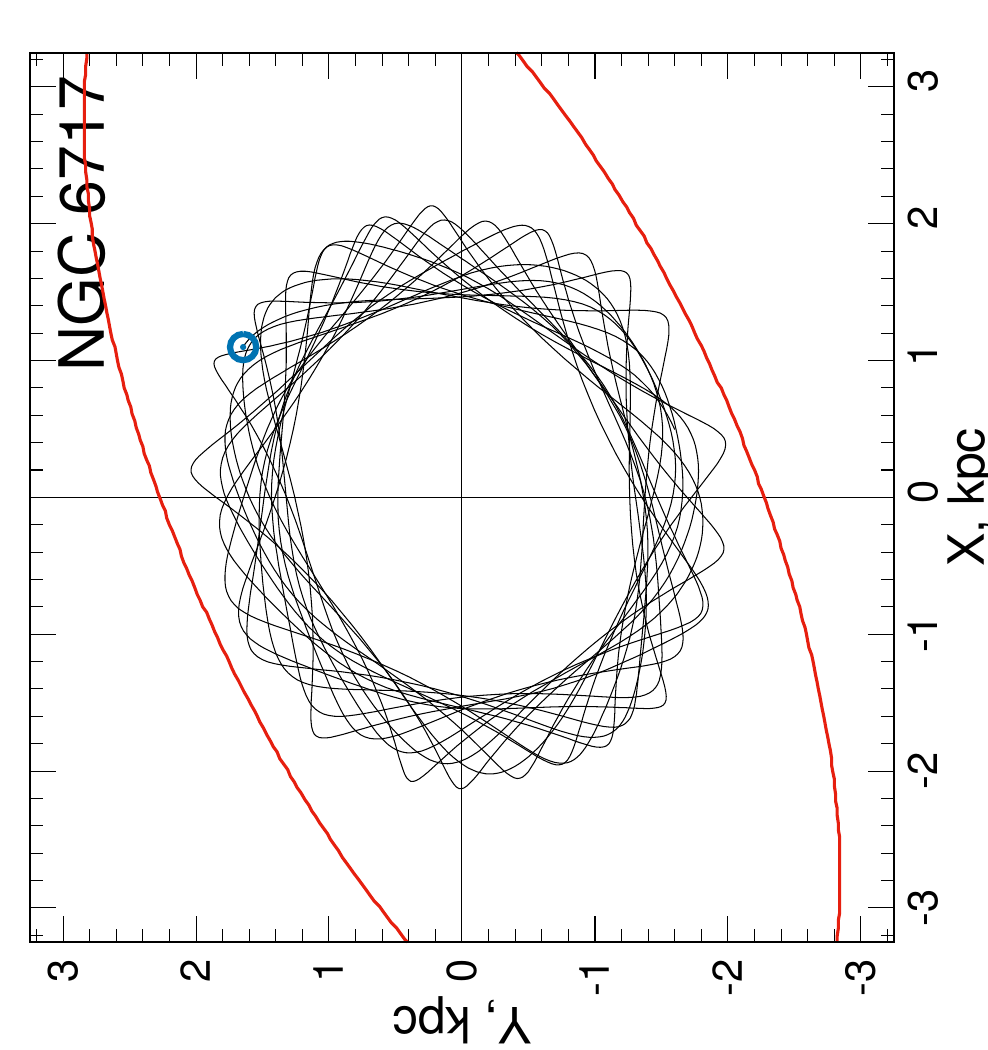}
     \includegraphics[width=0.225\textwidth,angle=-90]{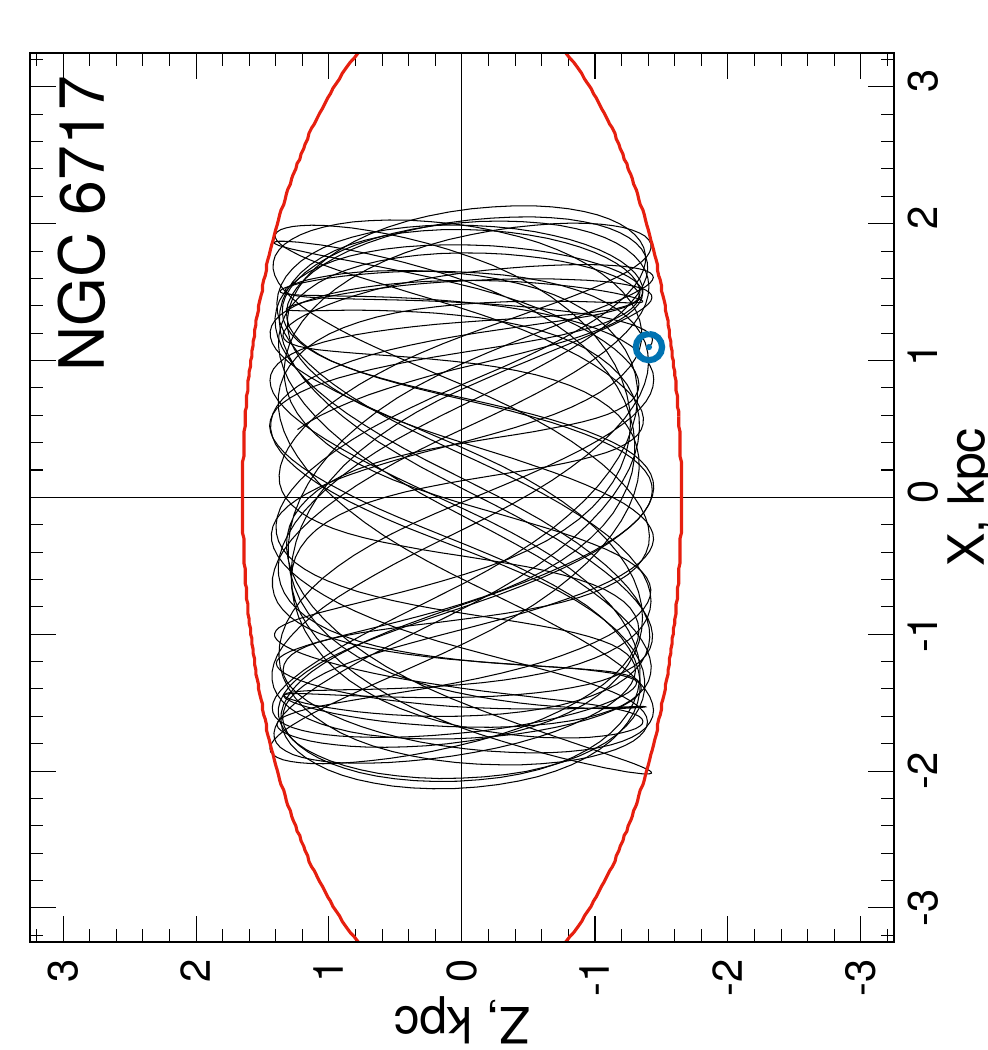}\

\medskip

 \centerline{APPENDIX. Continued}
\label{fD}
\end{center}}
\end{figure*}

\begin{figure*}
{\begin{center}
 \includegraphics[width=0.225\textwidth,angle=-90]{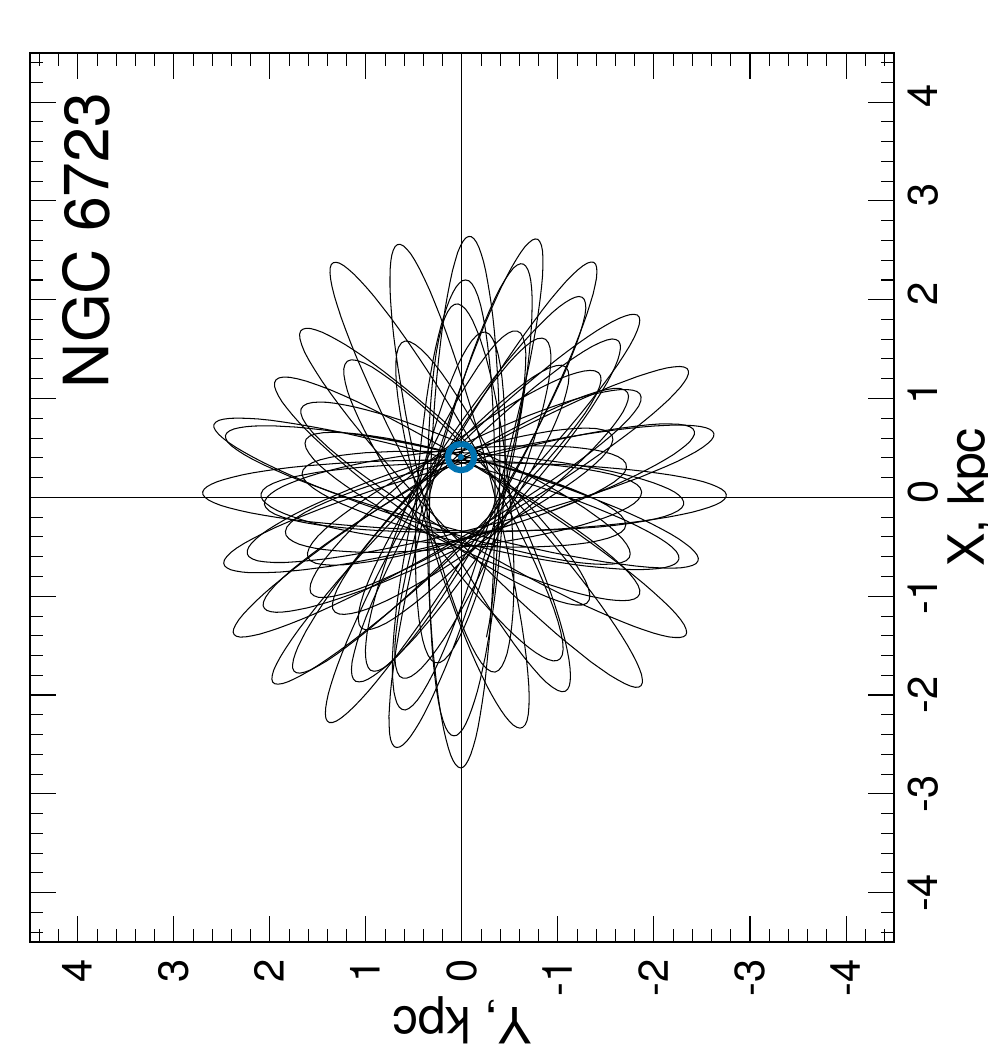}
     \includegraphics[width=0.225\textwidth,angle=-90]{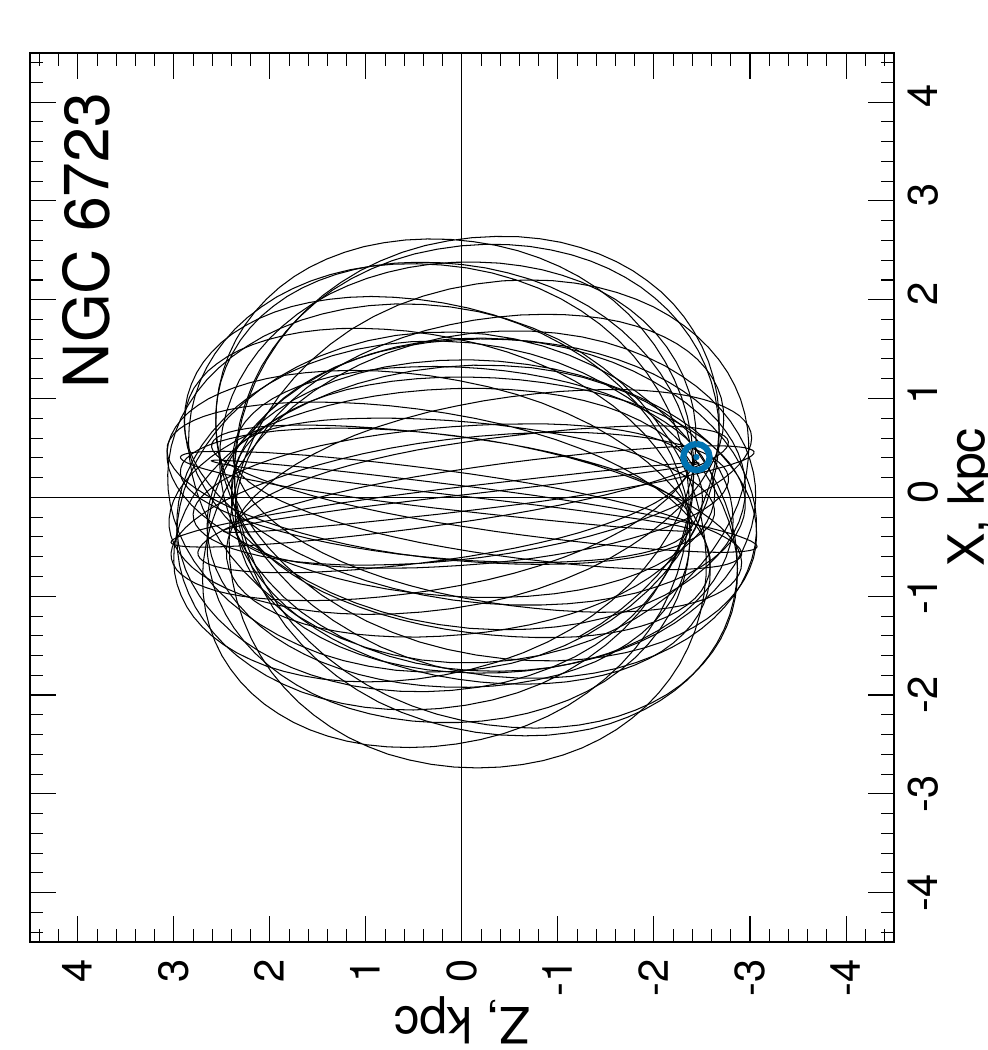}
     \includegraphics[width=0.225\textwidth,angle=-90]{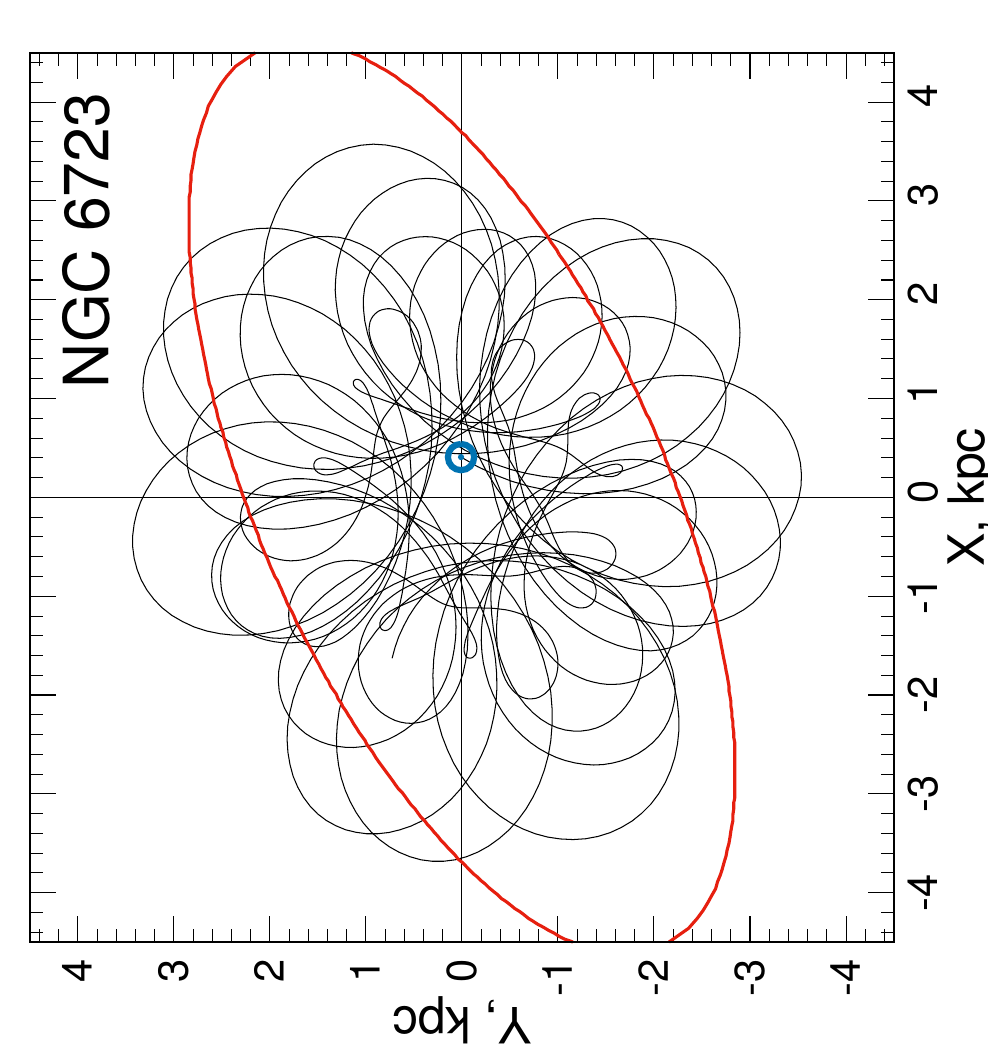}
     \includegraphics[width=0.225\textwidth,angle=-90]{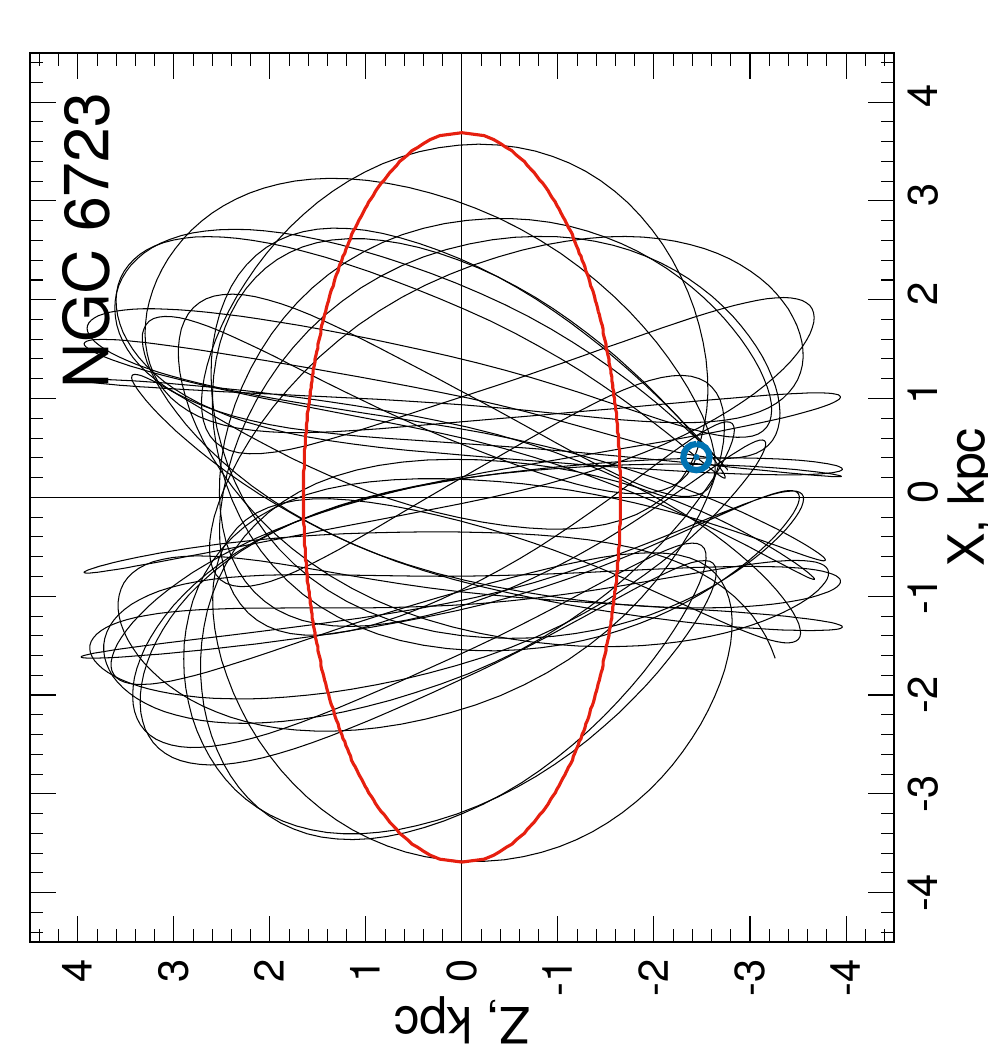}\
      \includegraphics[width=0.225\textwidth,angle=-90]{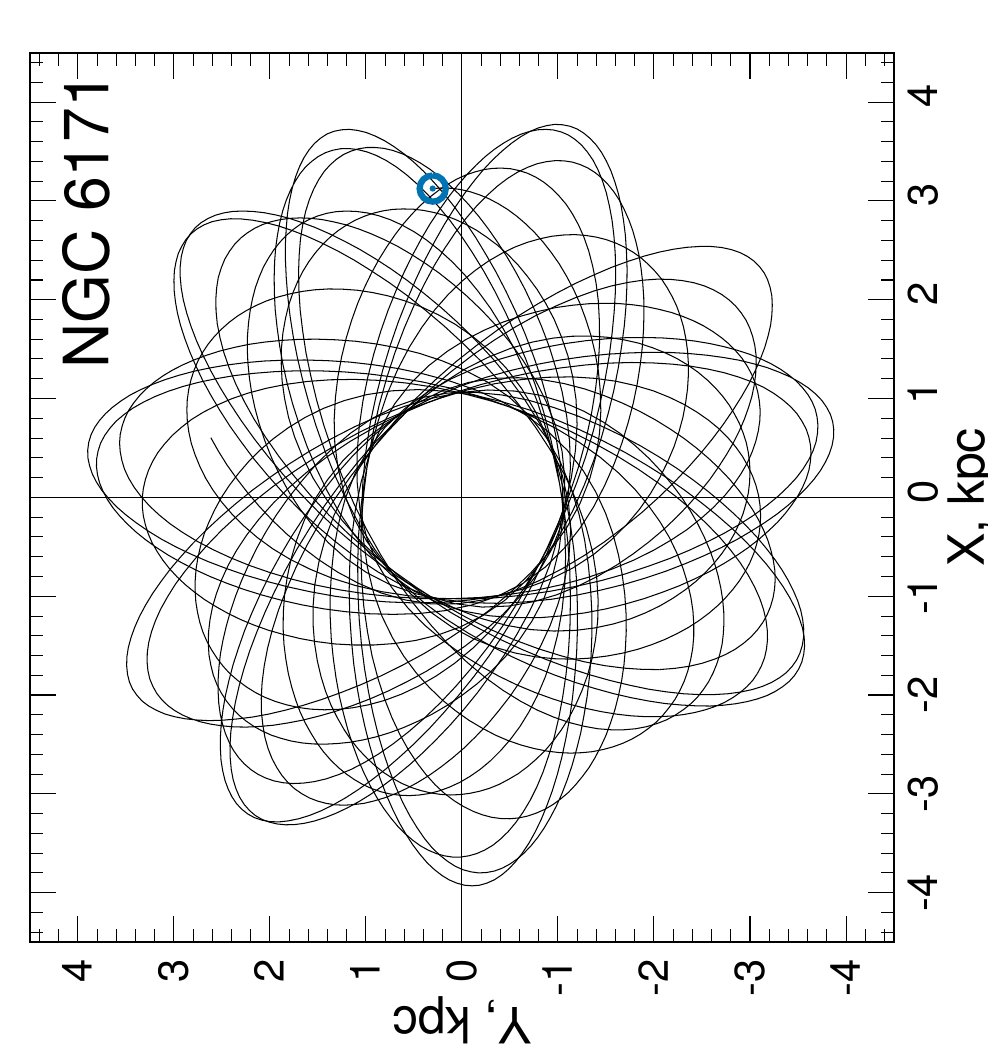}
     \includegraphics[width=0.225\textwidth,angle=-90]{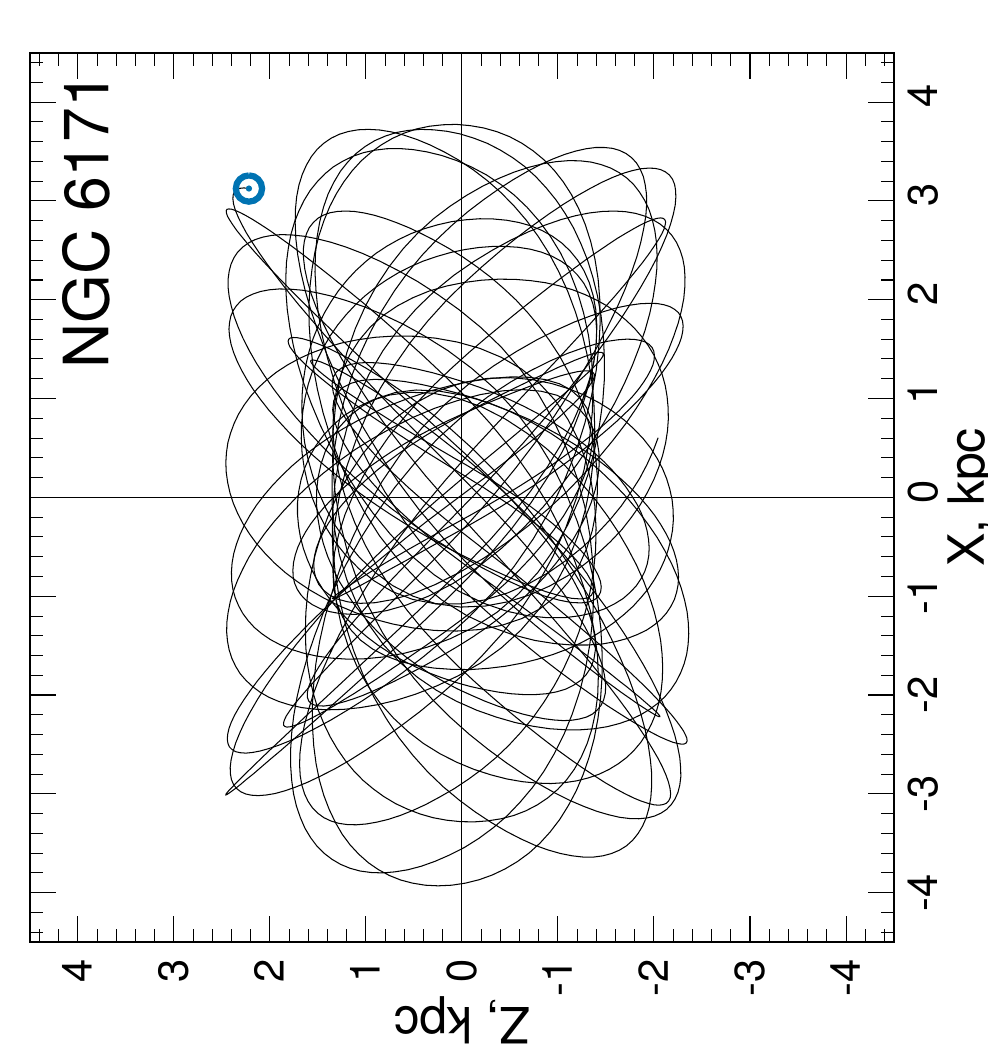}
      \includegraphics[width=0.225\textwidth,angle=-90]{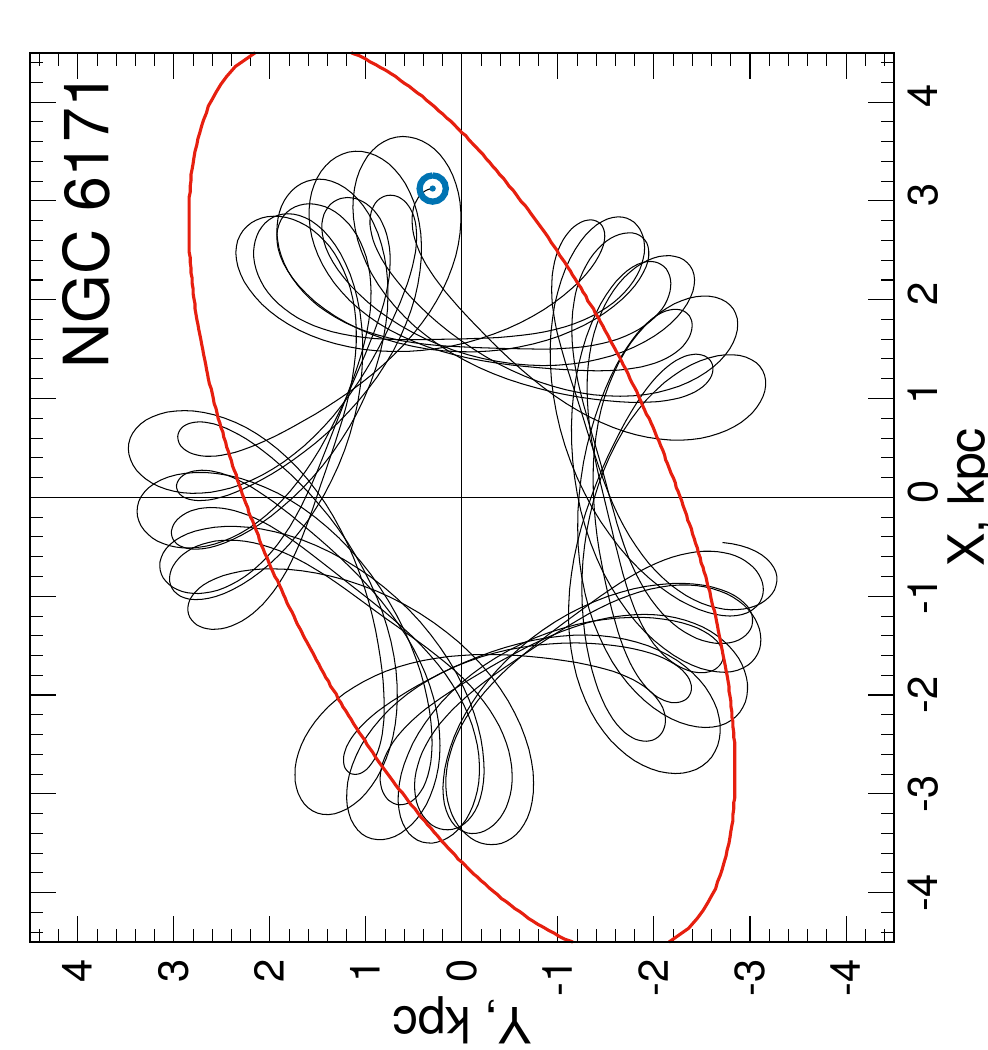}
     \includegraphics[width=0.225\textwidth,angle=-90]{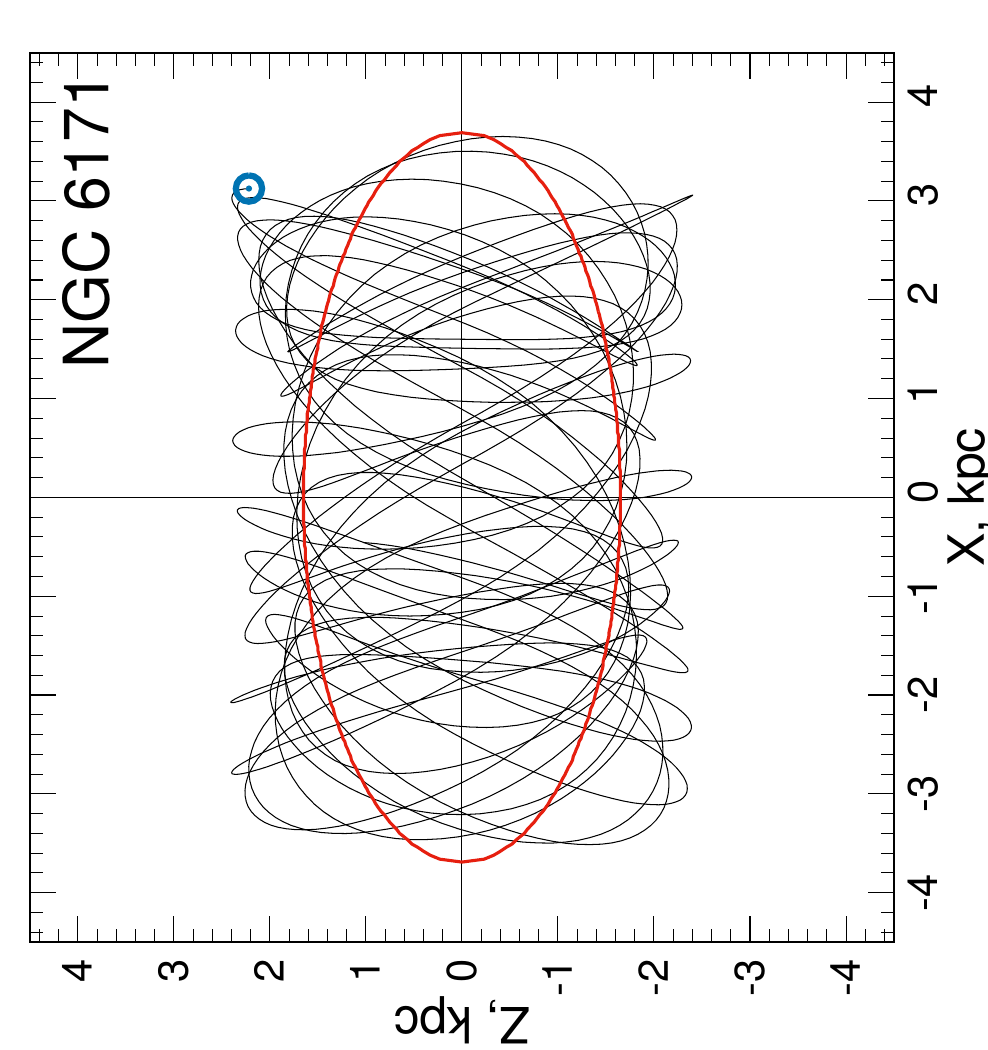}\
   \includegraphics[width=0.225\textwidth,angle=-90]{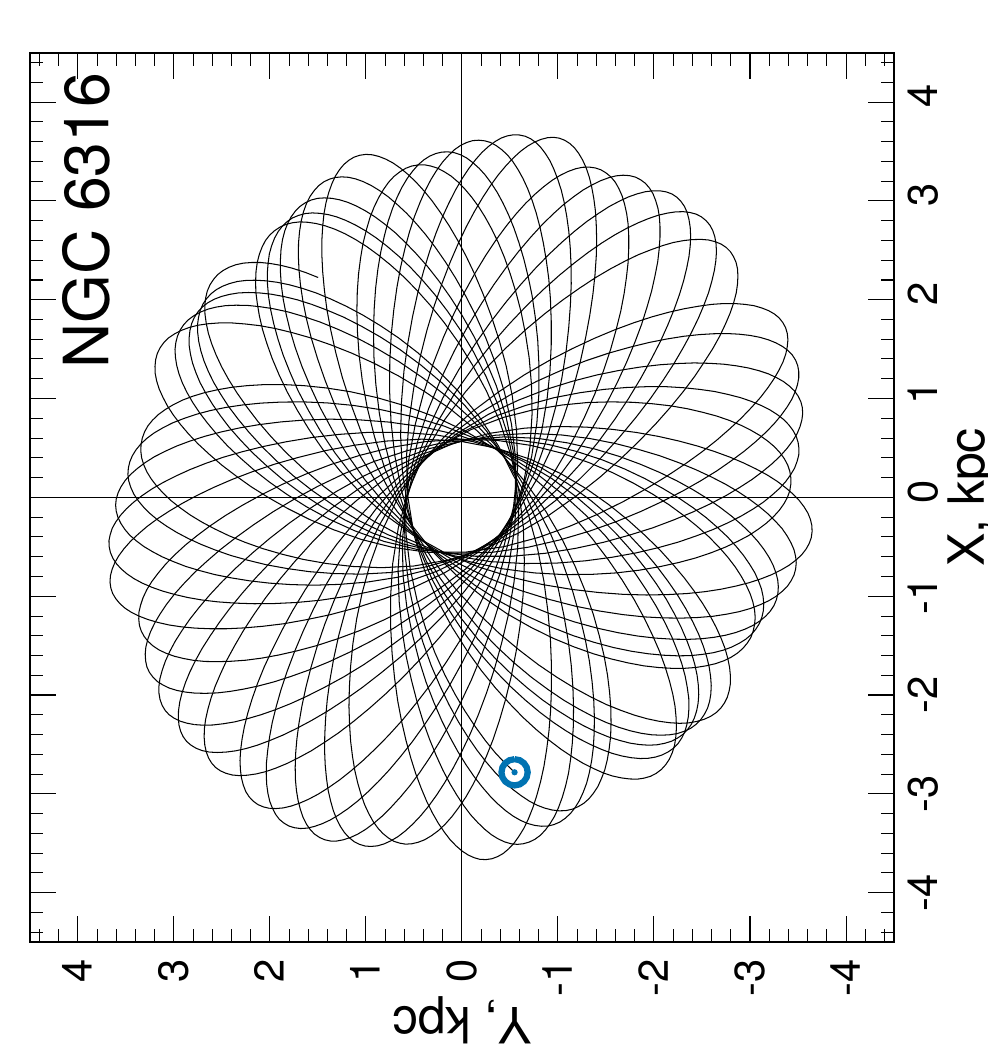}
     \includegraphics[width=0.225\textwidth,angle=-90]{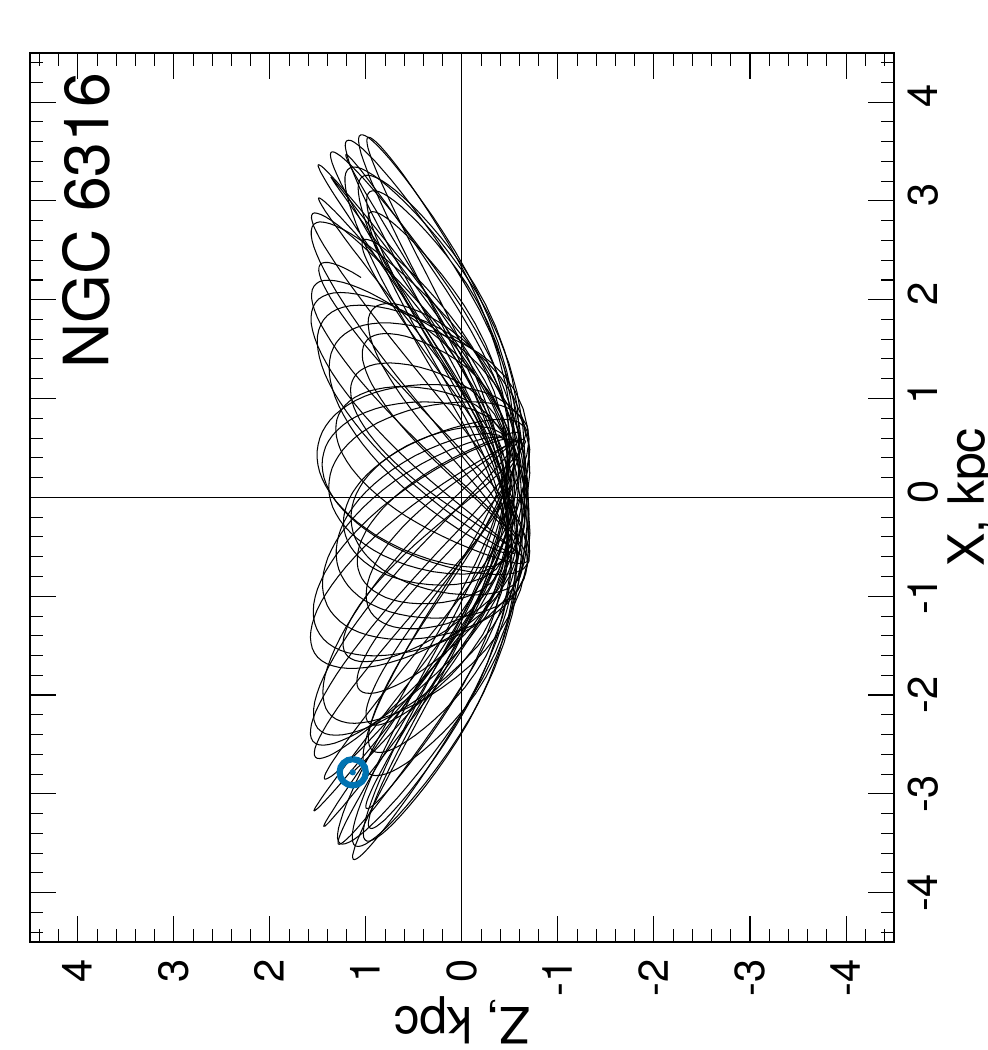}
        \includegraphics[width=0.225\textwidth,angle=-90]{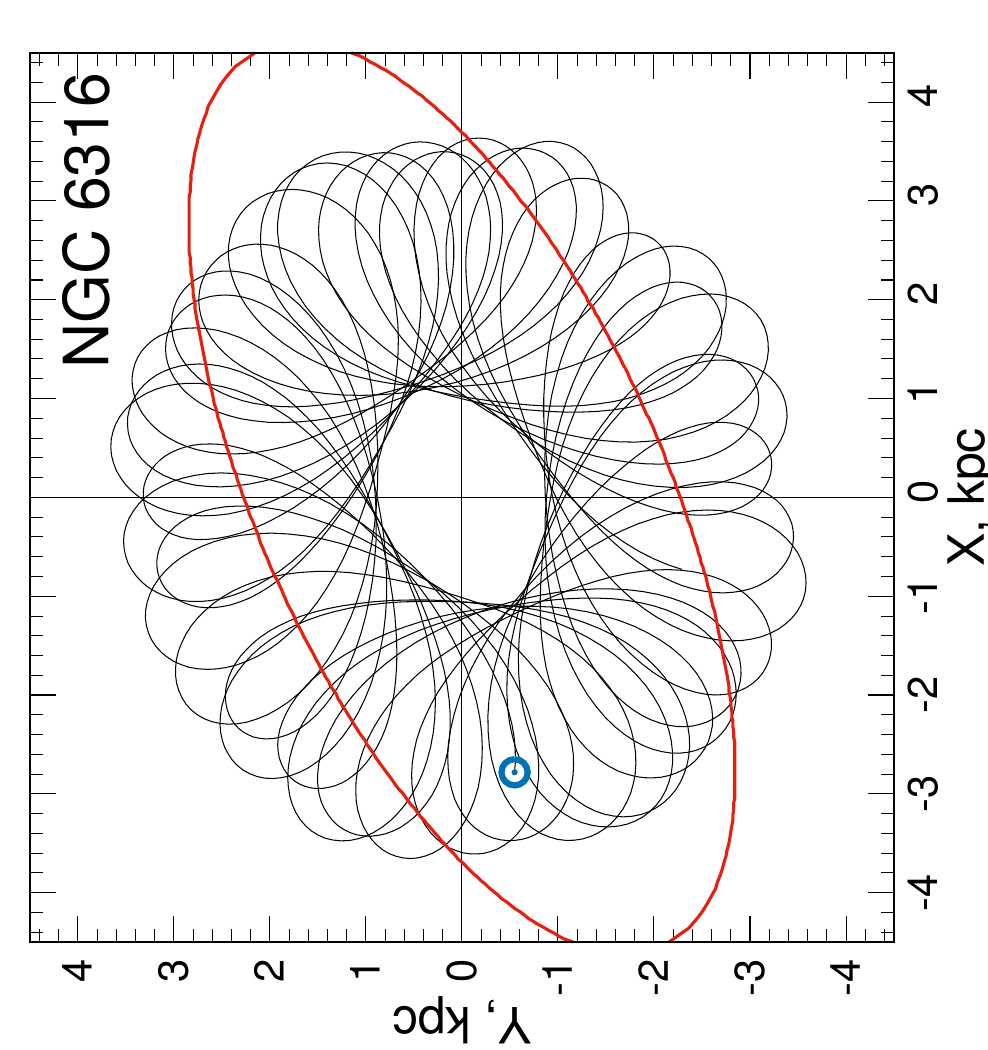}
     \includegraphics[width=0.225\textwidth,angle=-90]{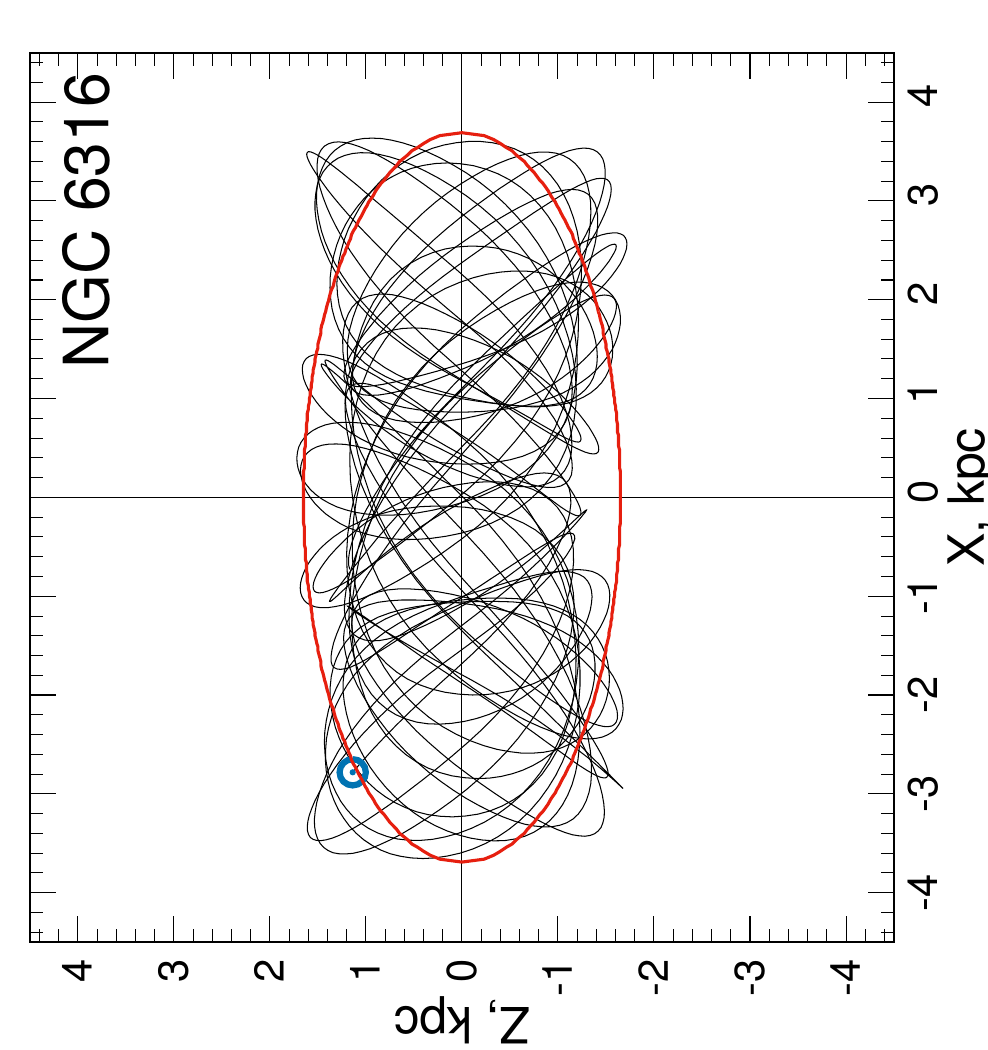}\
    \includegraphics[width=0.225\textwidth,angle=-90]{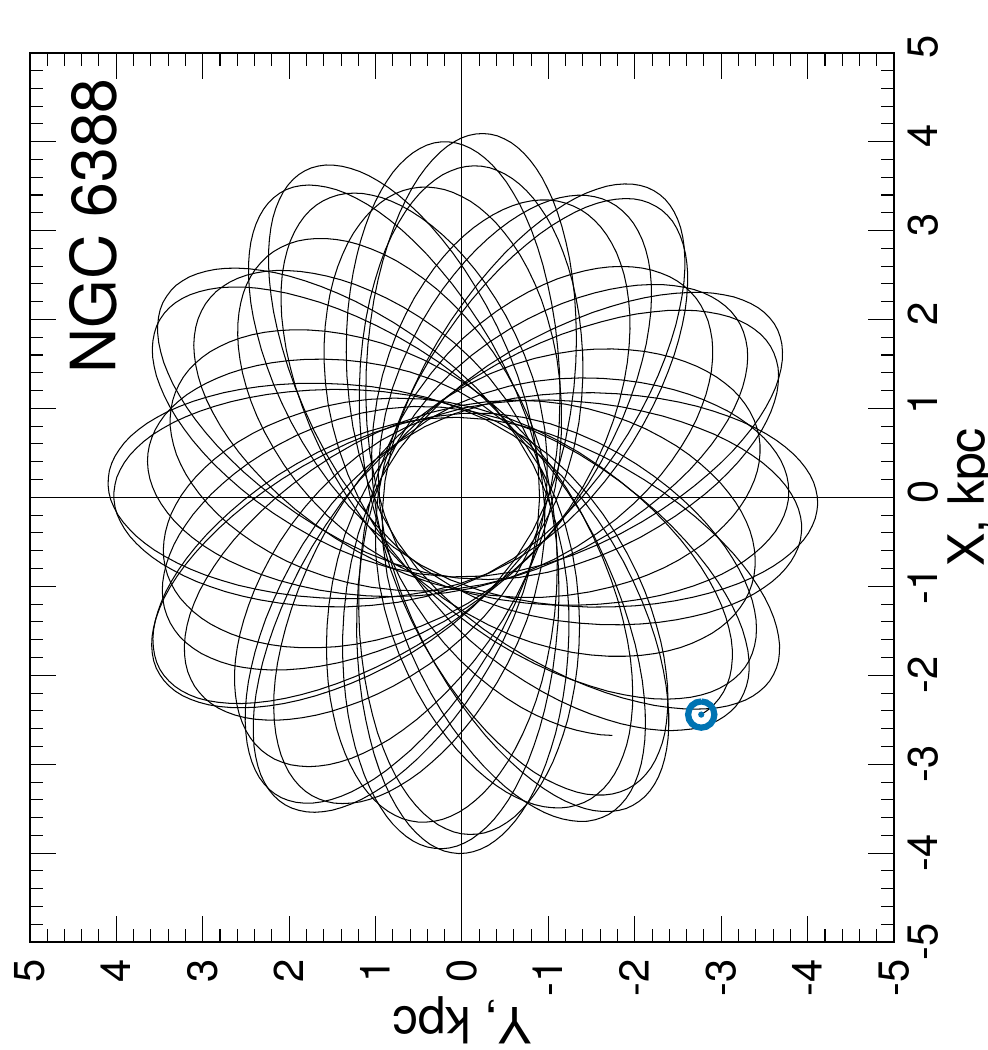}
     \includegraphics[width=0.225\textwidth,angle=-90]{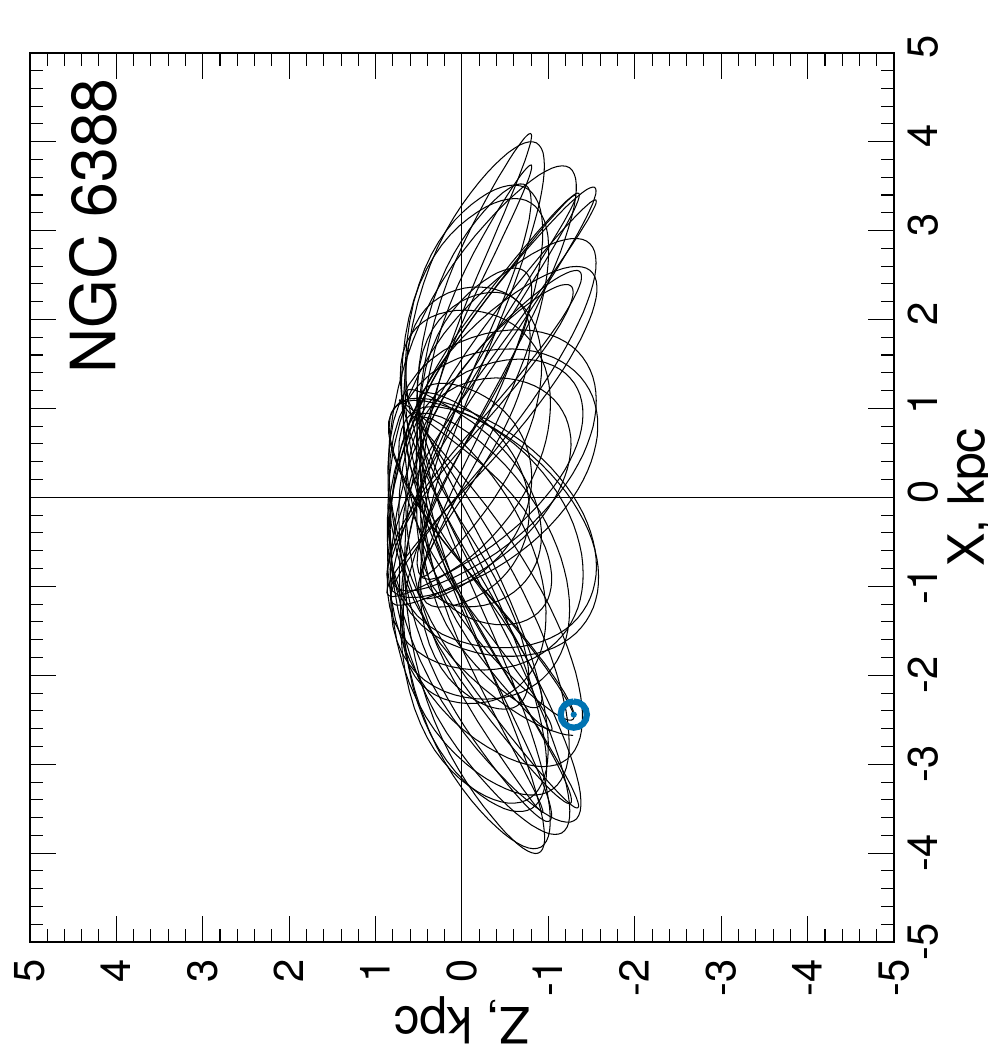}
    \includegraphics[width=0.225\textwidth,angle=-90]{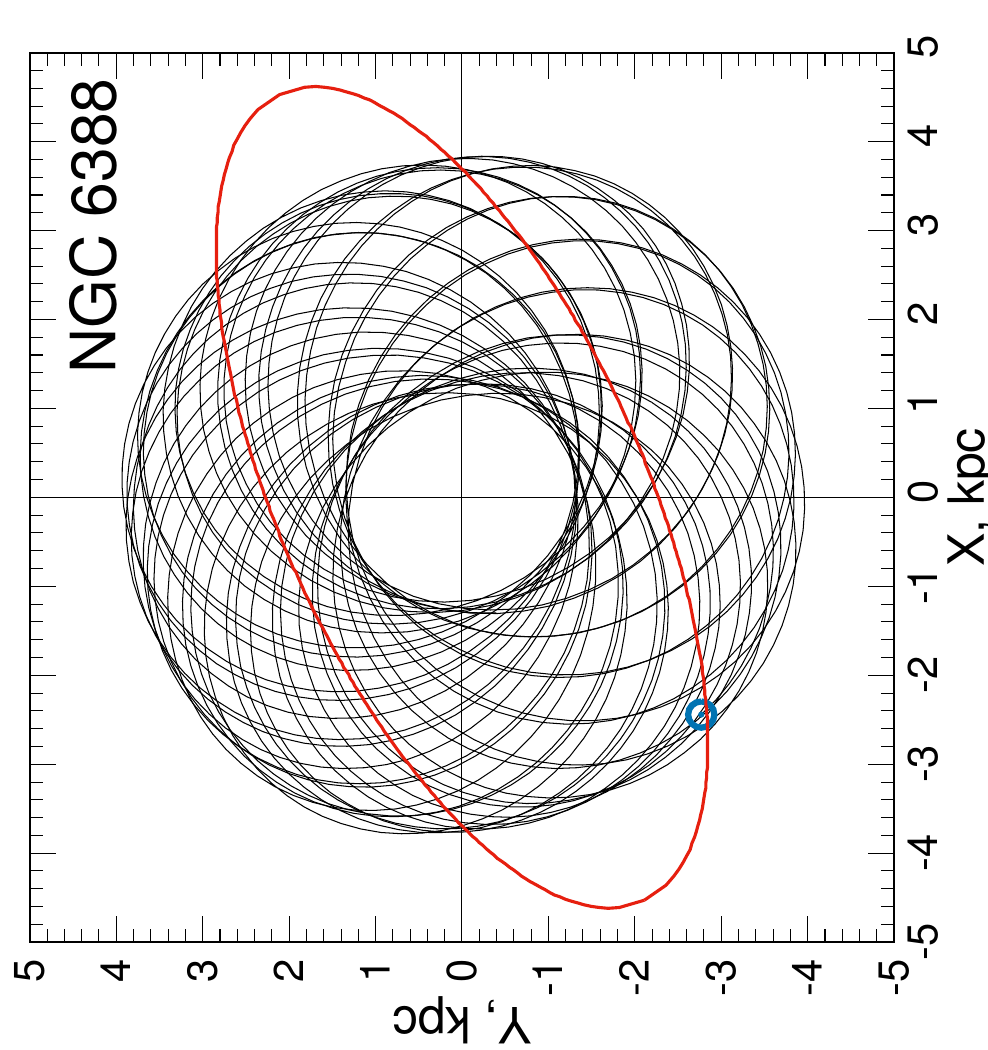}
     \includegraphics[width=0.225\textwidth,angle=-90]{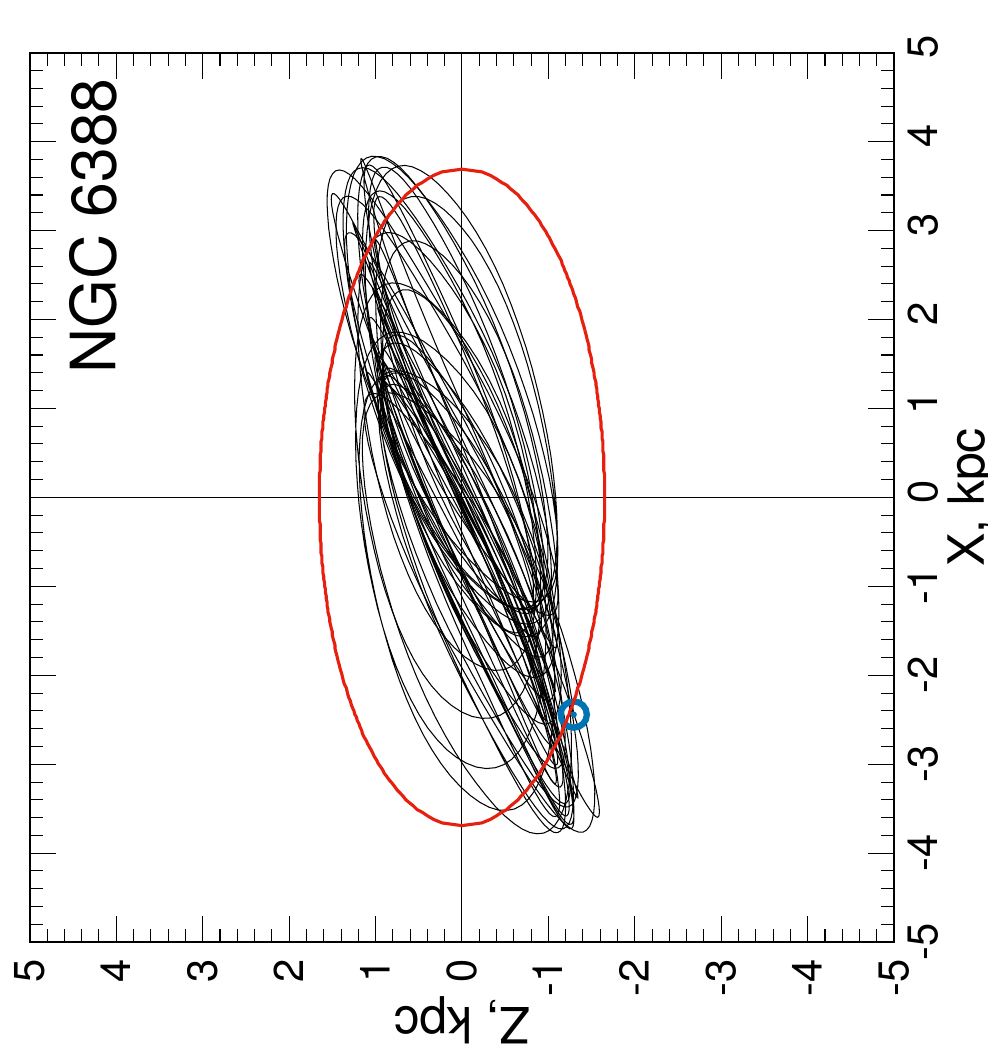}\
    \includegraphics[width=0.225\textwidth,angle=-90]{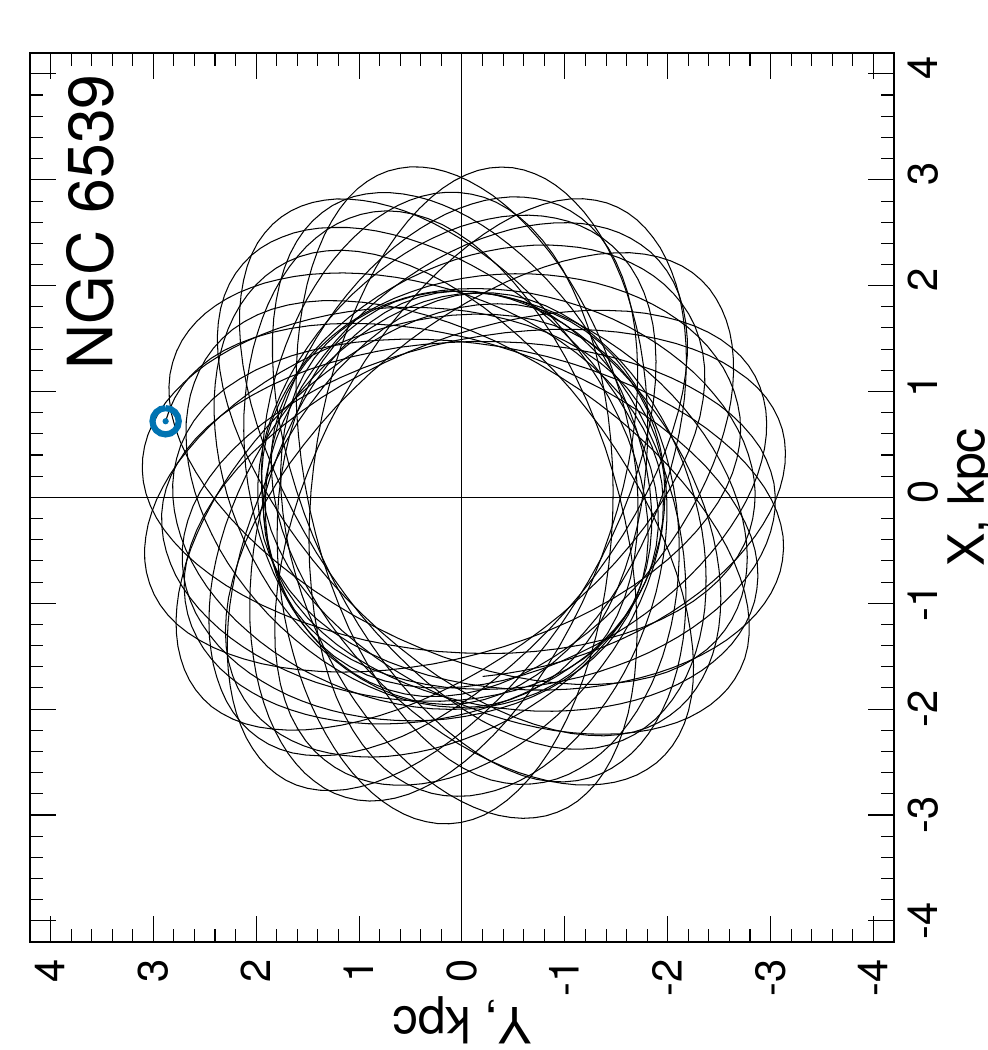}
     \includegraphics[width=0.225\textwidth,angle=-90]{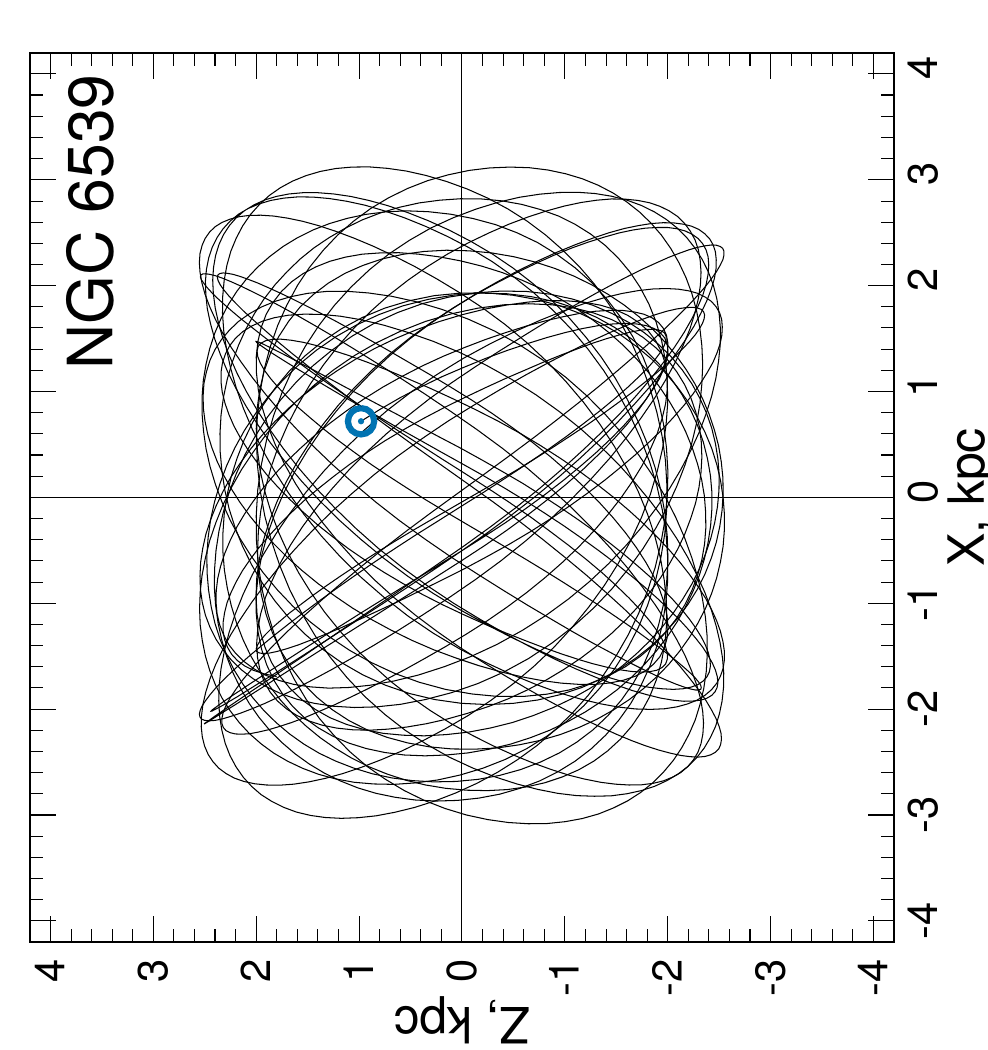}
     \includegraphics[width=0.225\textwidth,angle=-90]{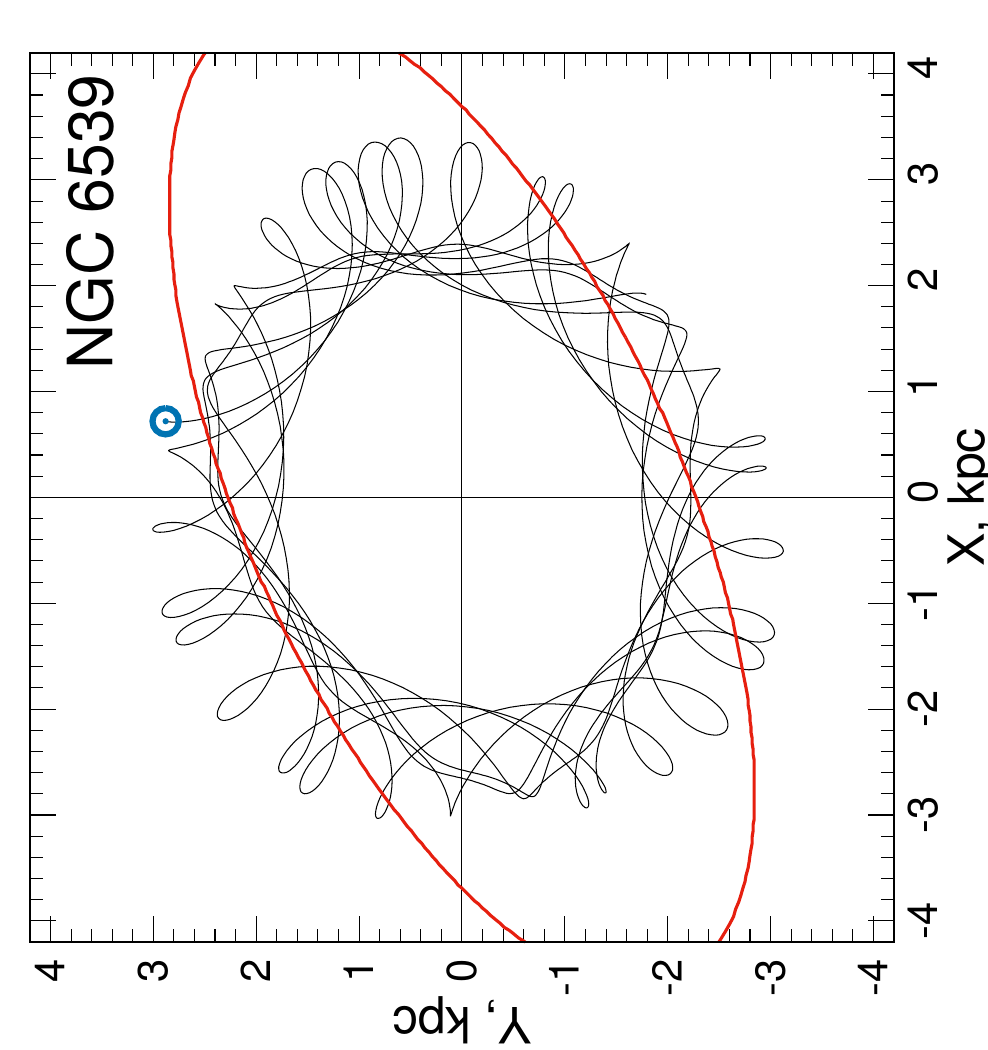}
     \includegraphics[width=0.225\textwidth,angle=-90]{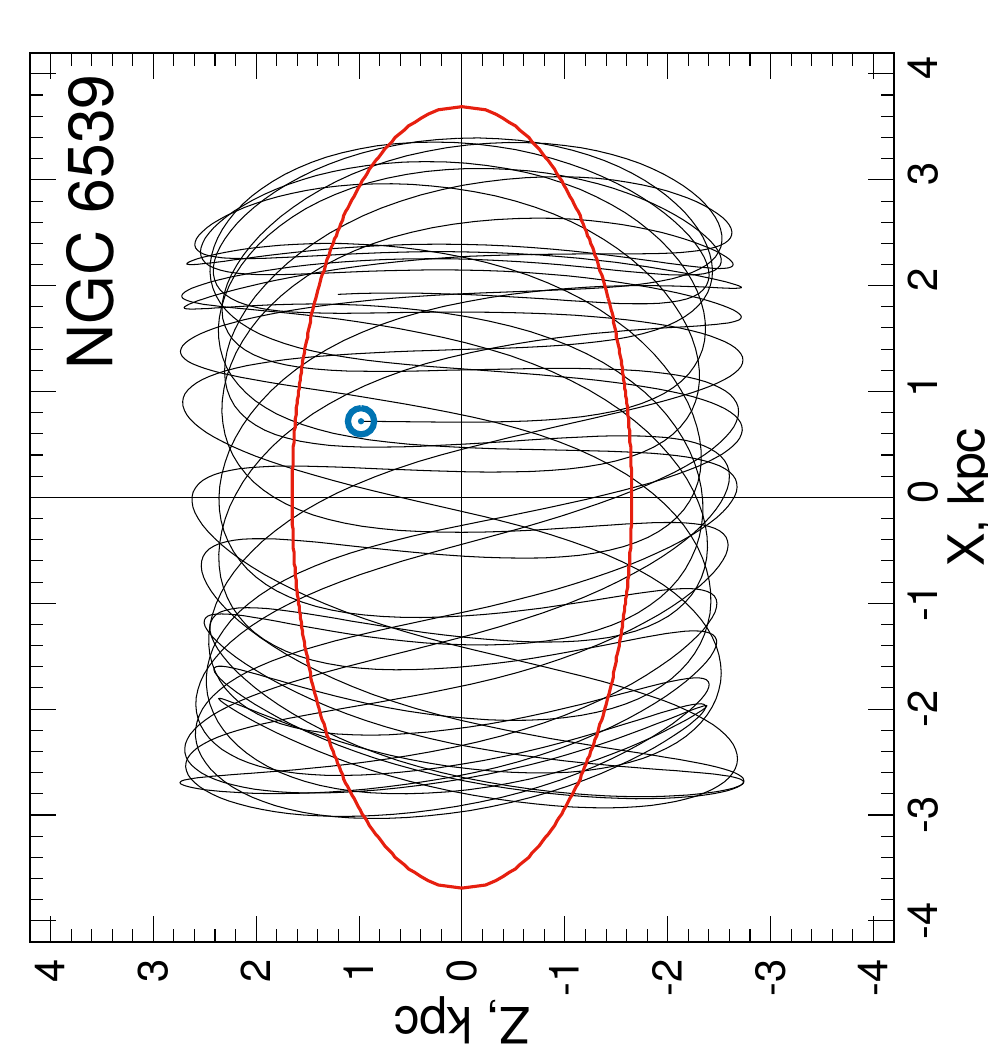}\
        \includegraphics[width=0.225\textwidth,angle=-90]{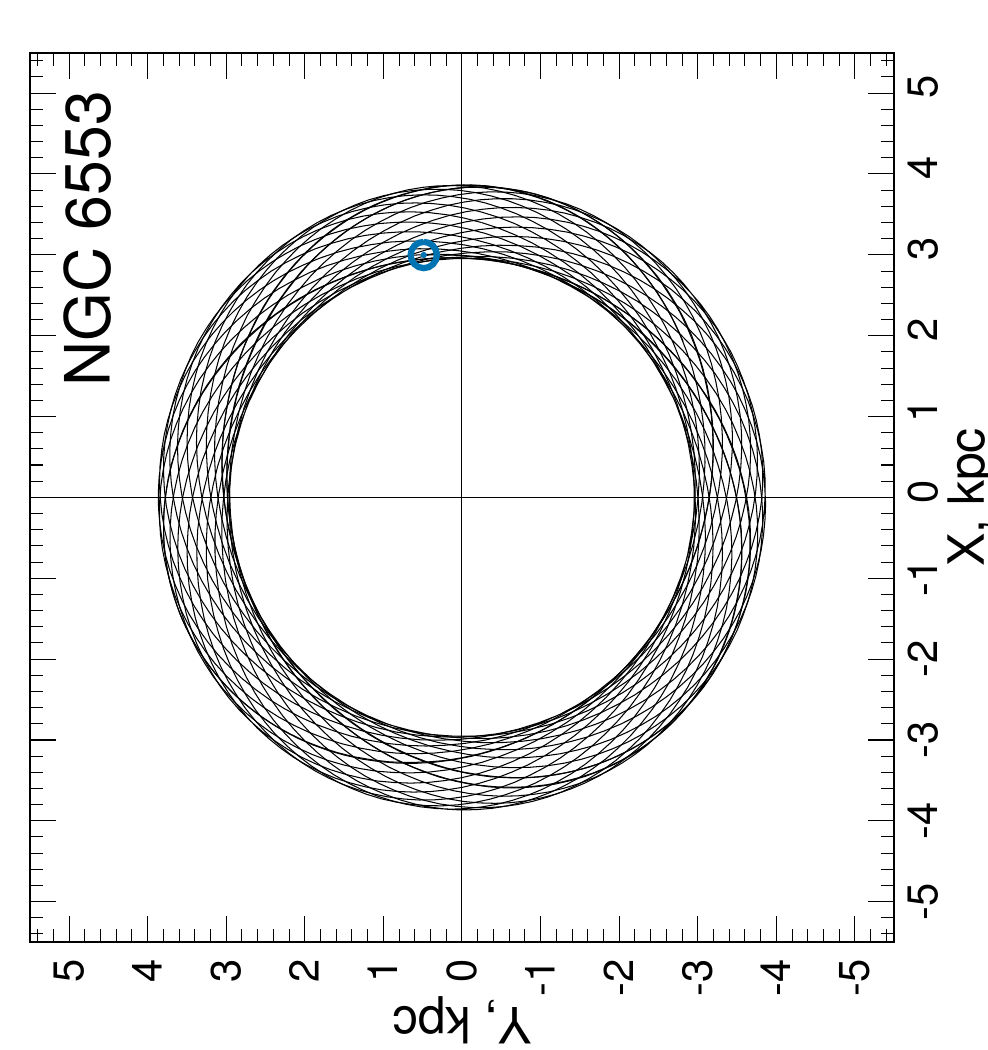}
     \includegraphics[width=0.225\textwidth,angle=-90]{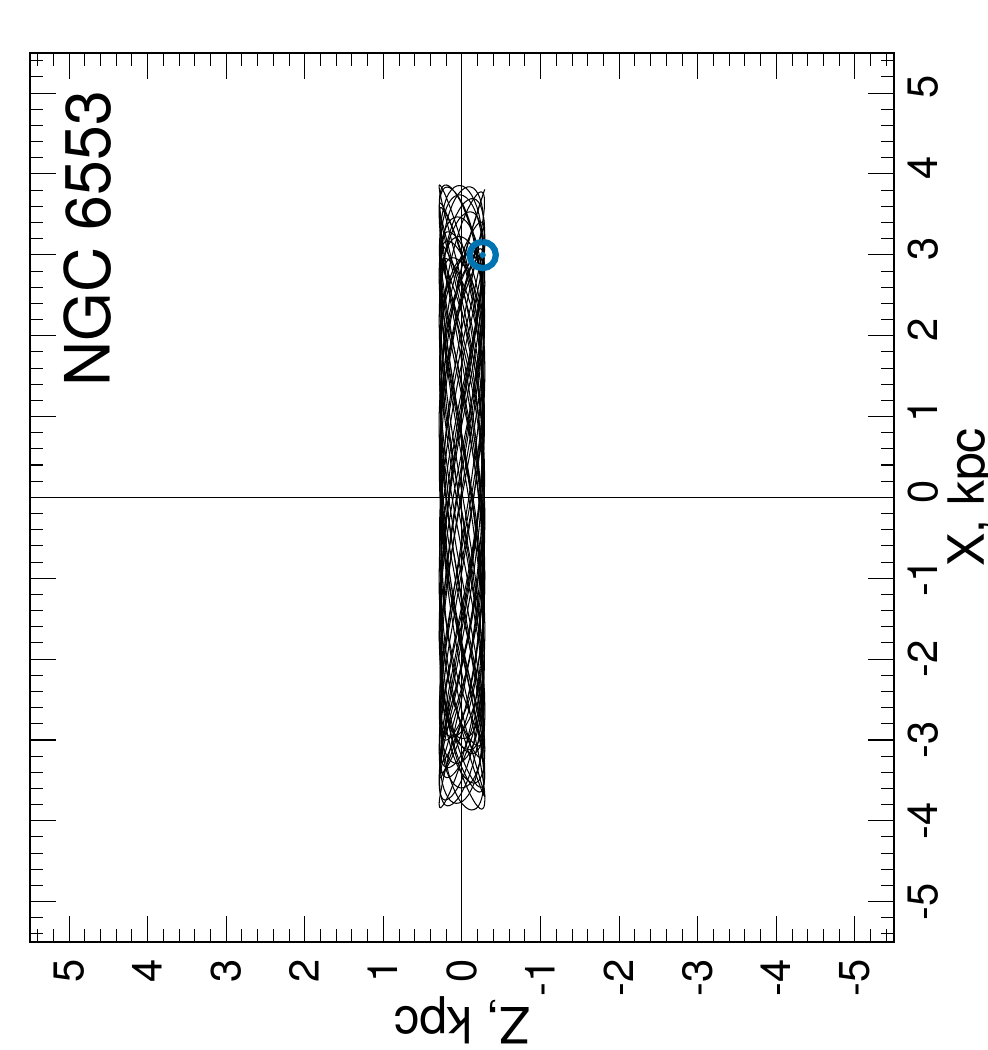}
         \includegraphics[width=0.225\textwidth,angle=-90]{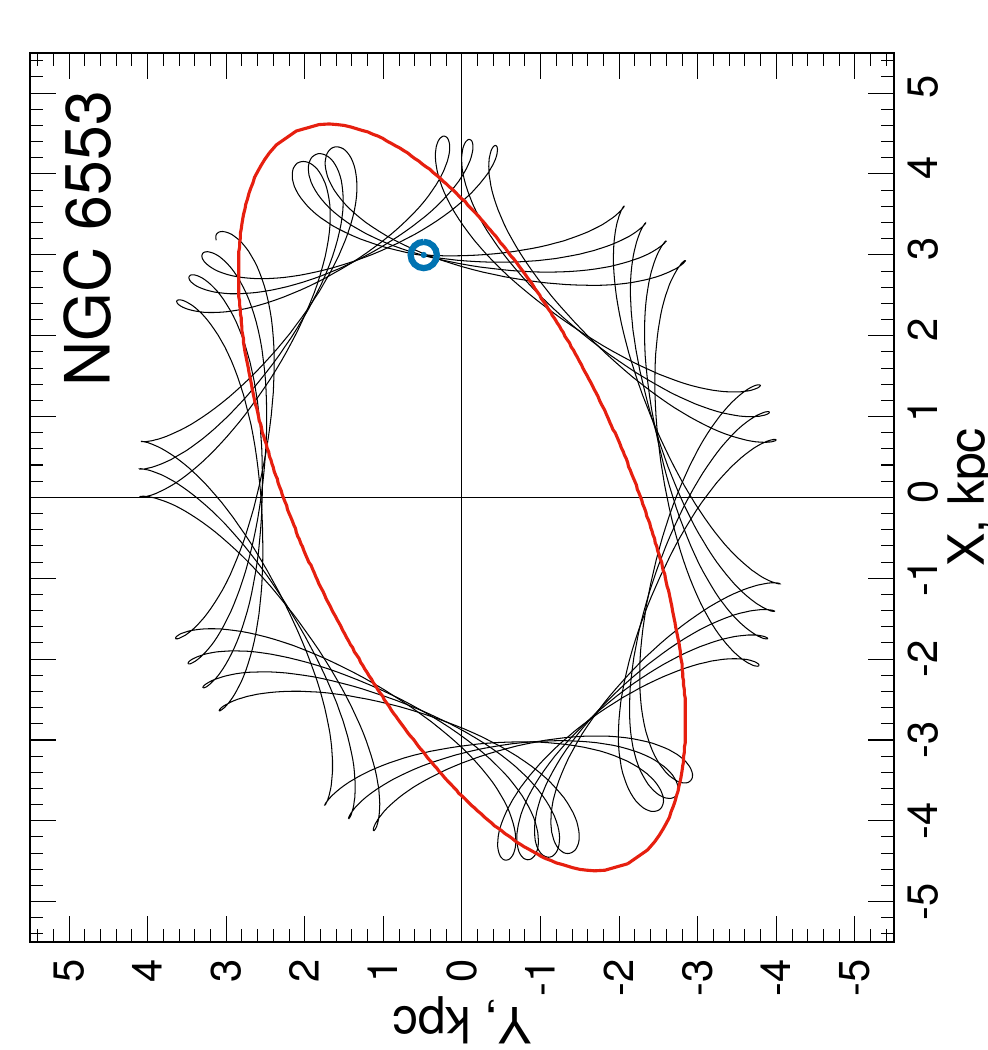}
     \includegraphics[width=0.225\textwidth,angle=-90]{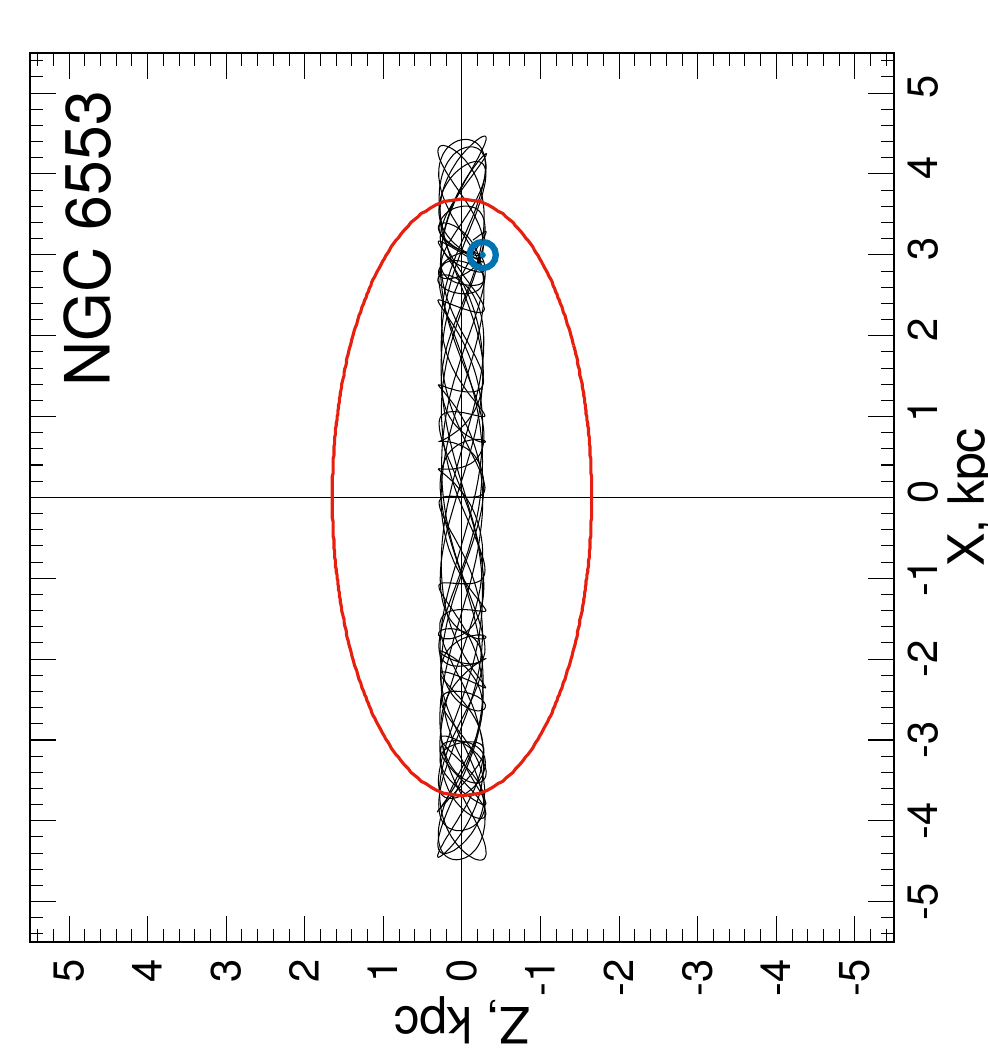}\

\medskip

 \centerline{APPENDIX. Continued}
\label{fD}
\end{center}}
\end{figure*}

\begin{figure*}
{\begin{center}
 \includegraphics[width=0.225\textwidth,angle=-90]{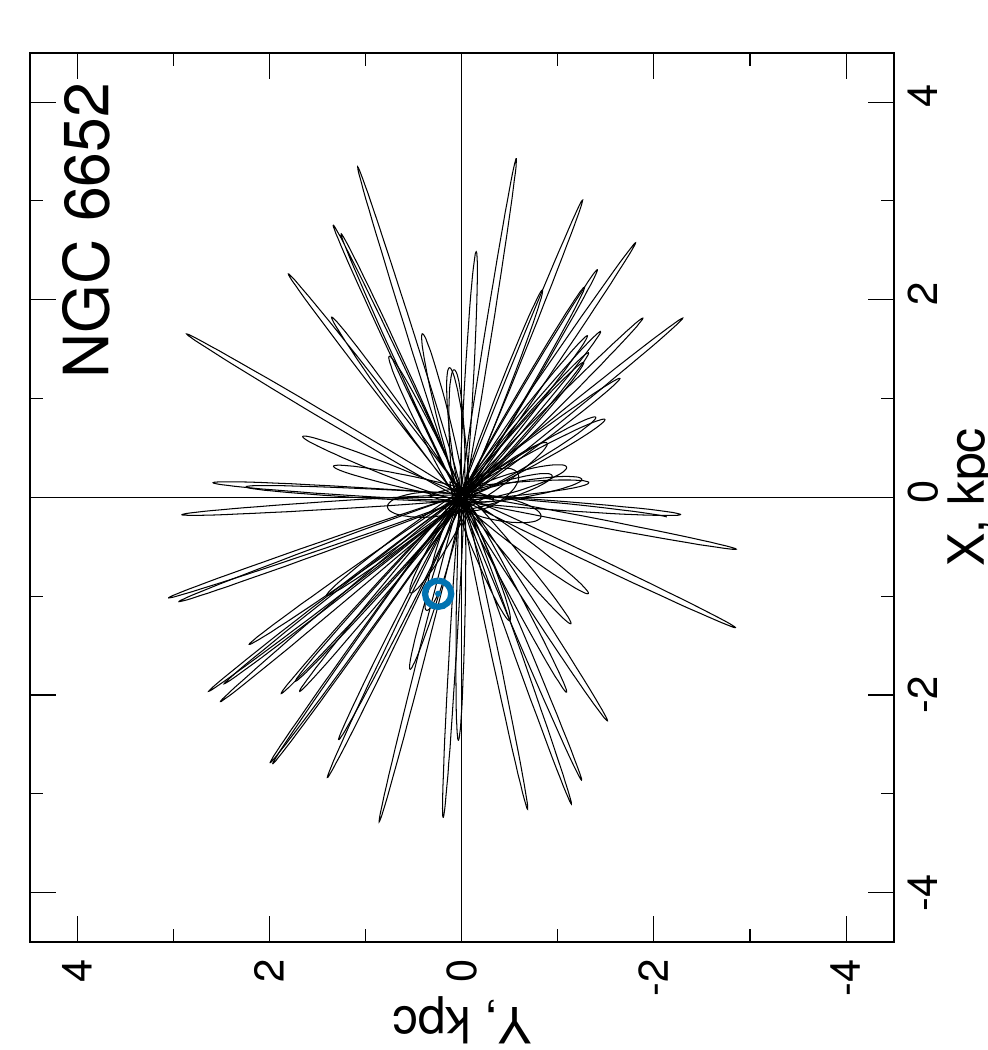}
     \includegraphics[width=0.225\textwidth,angle=-90]{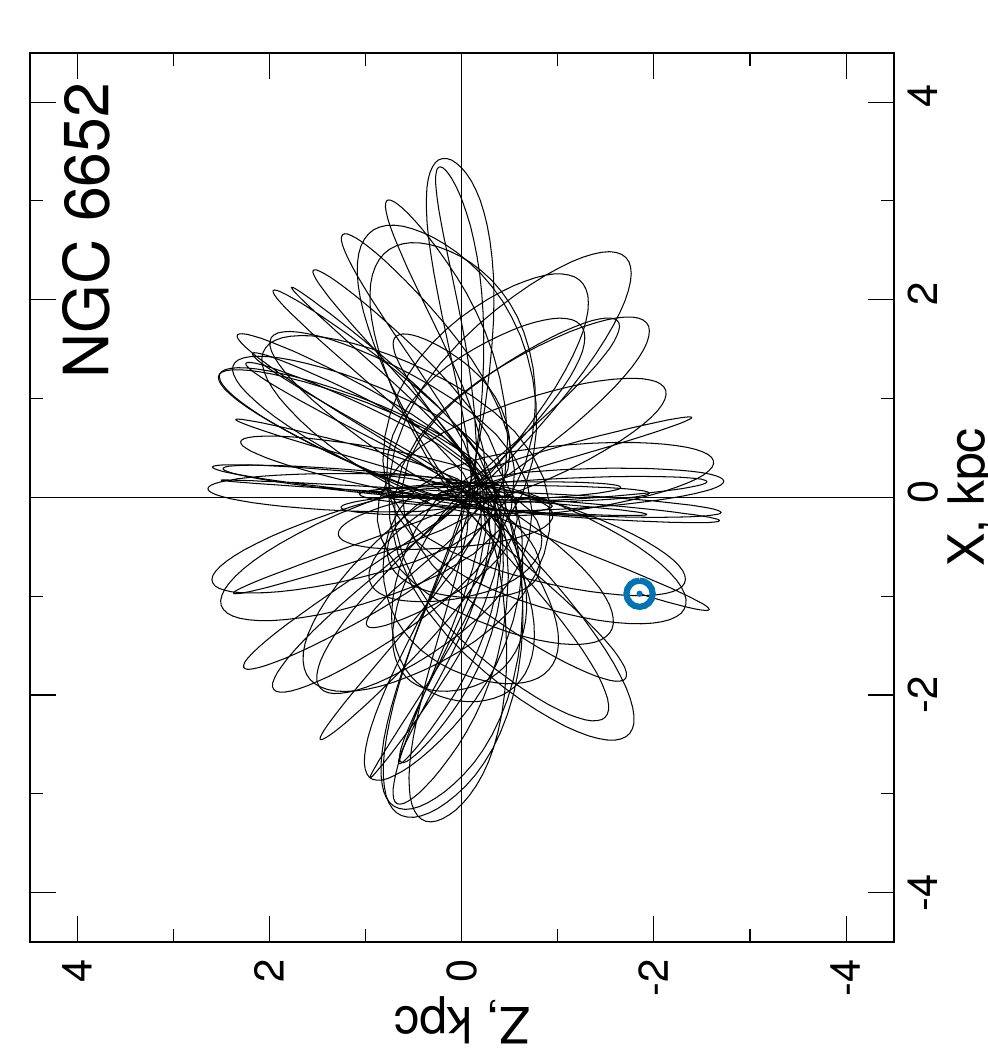}
     \includegraphics[width=0.225\textwidth,angle=-90]{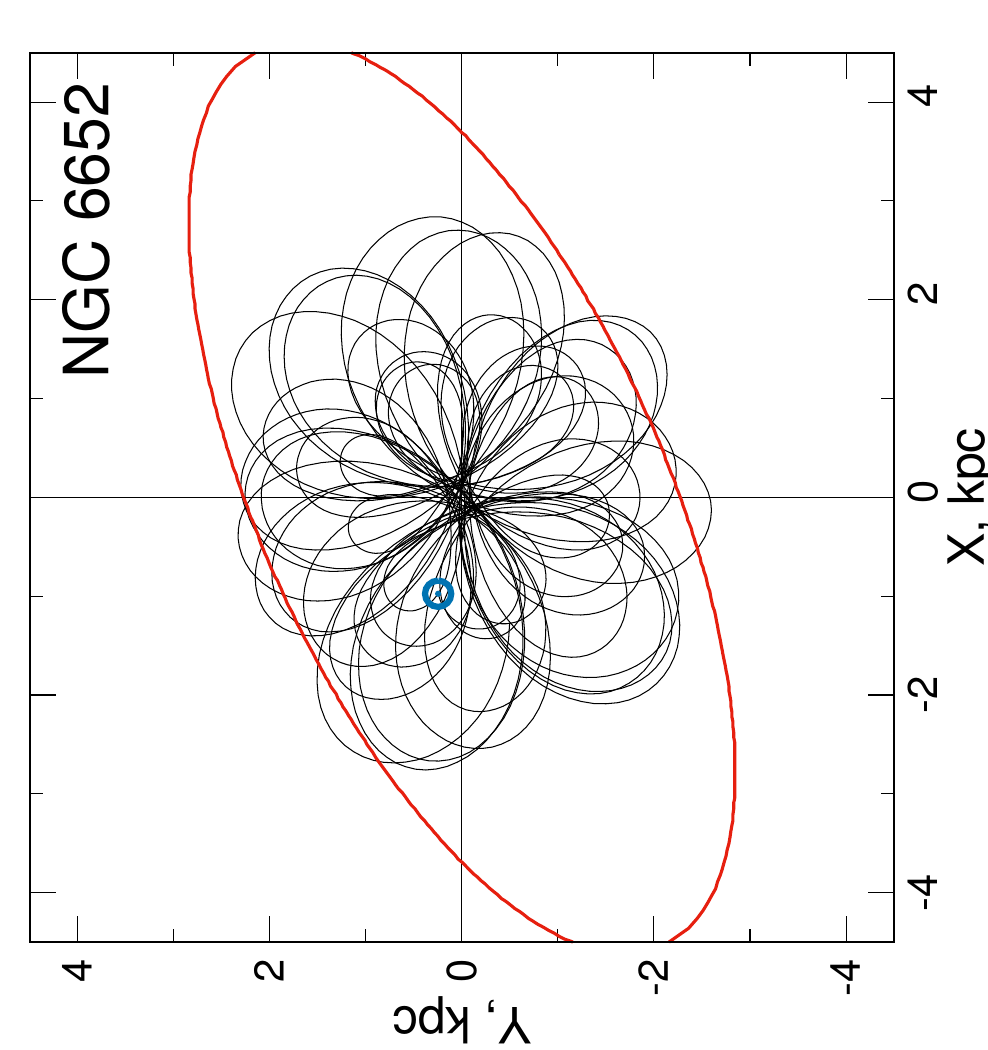}
     \includegraphics[width=0.225\textwidth,angle=-90]{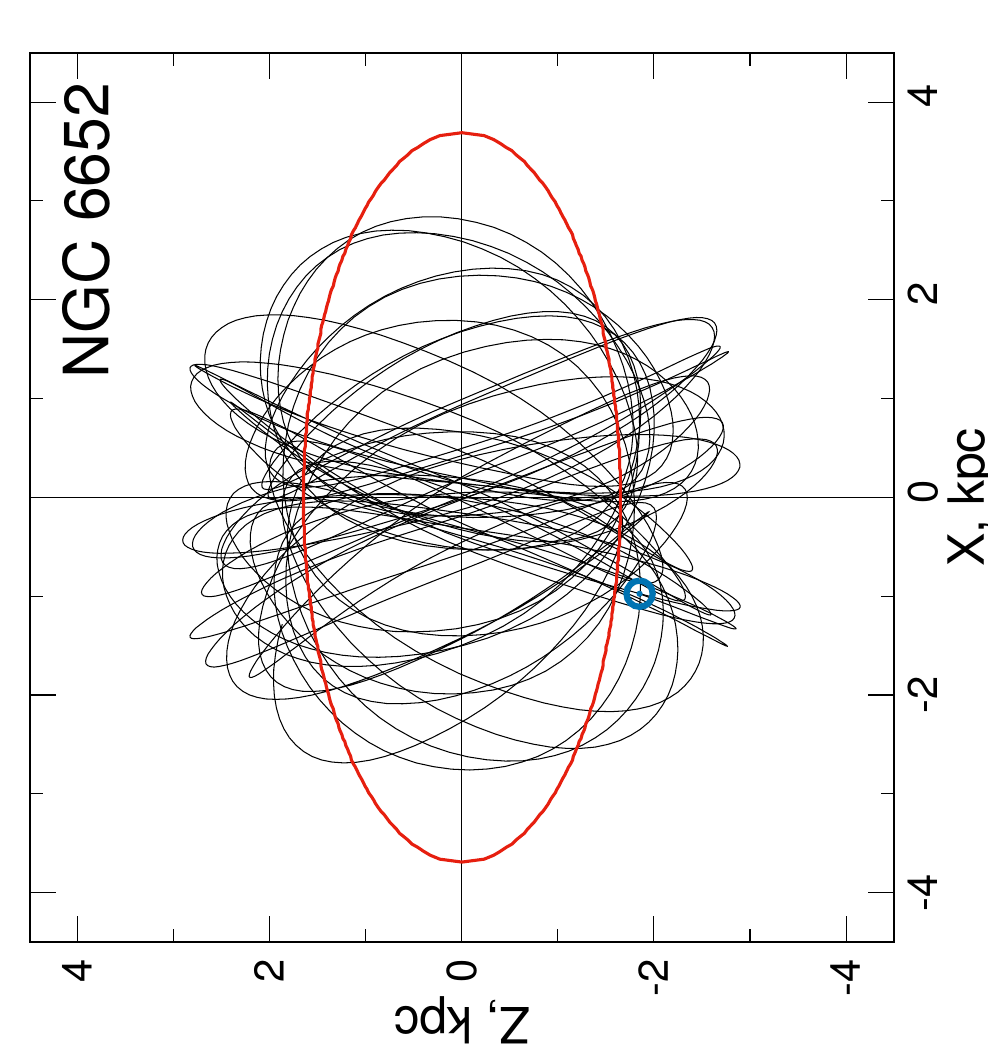}\
   \includegraphics[width=0.225\textwidth,angle=-90]{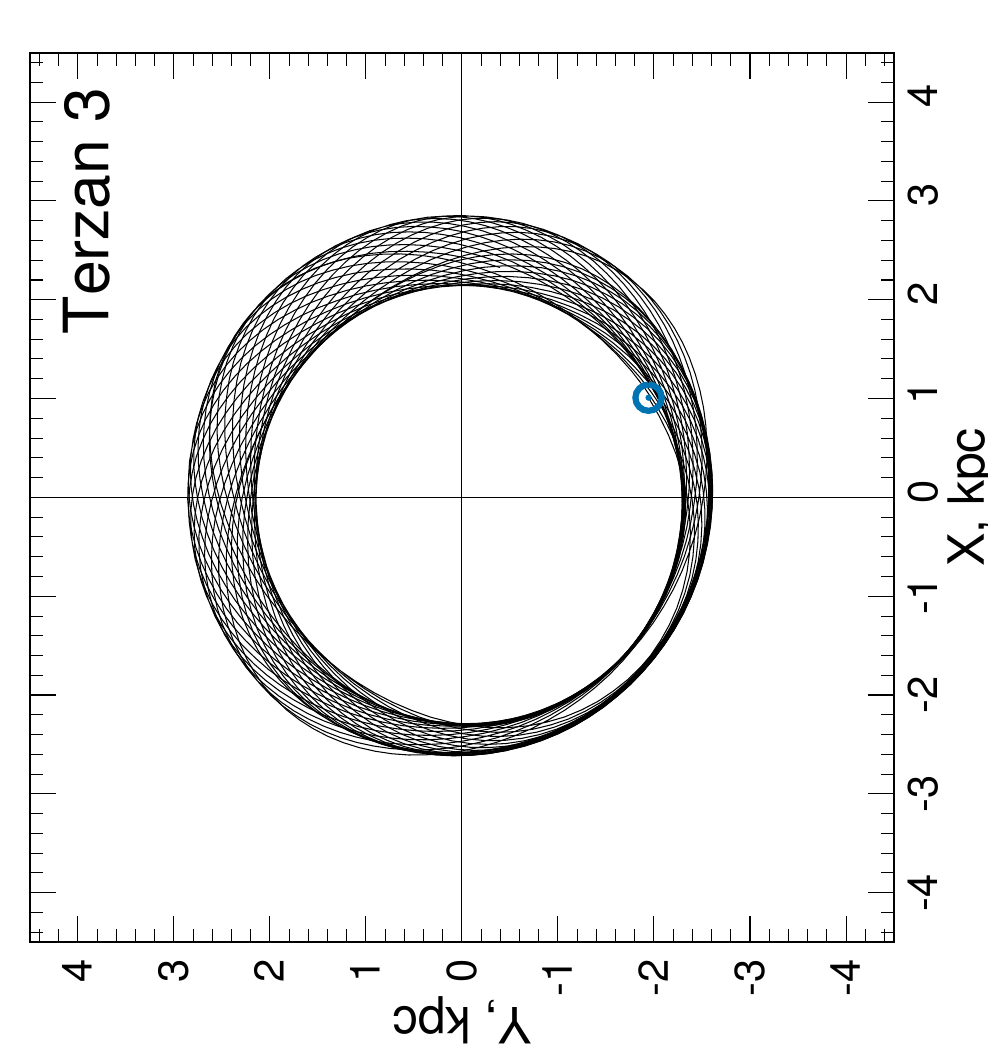}
     \includegraphics[width=0.225\textwidth,angle=-90]{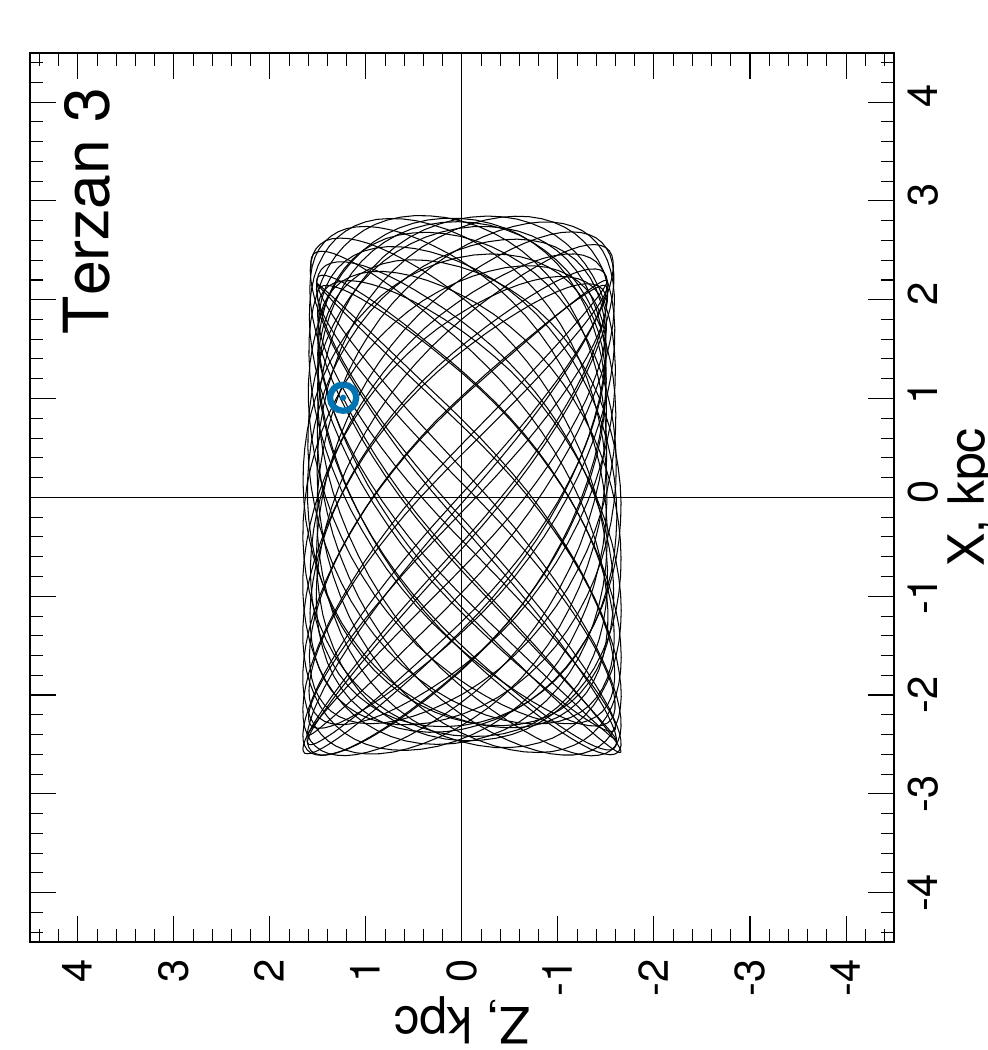}
   \includegraphics[width=0.225\textwidth,angle=-90]{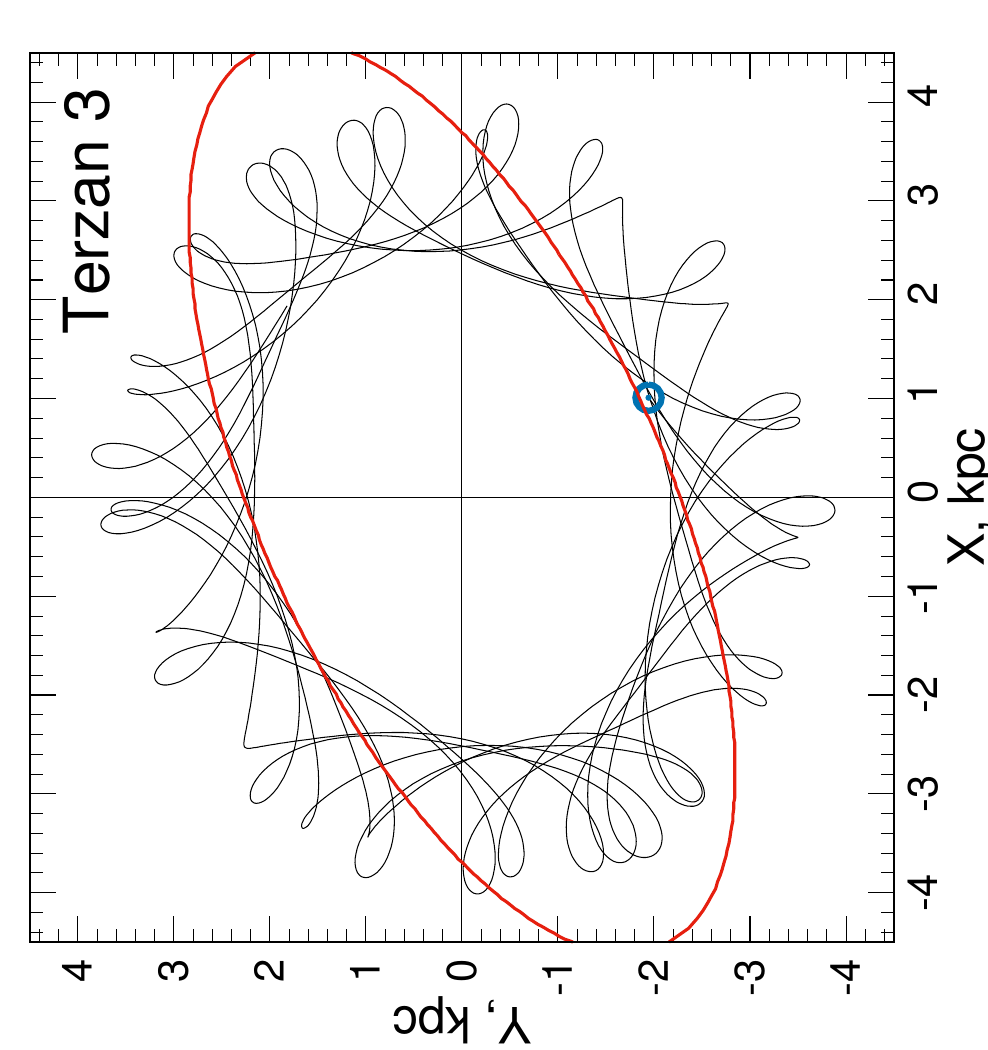}
     \includegraphics[width=0.225\textwidth,angle=-90]{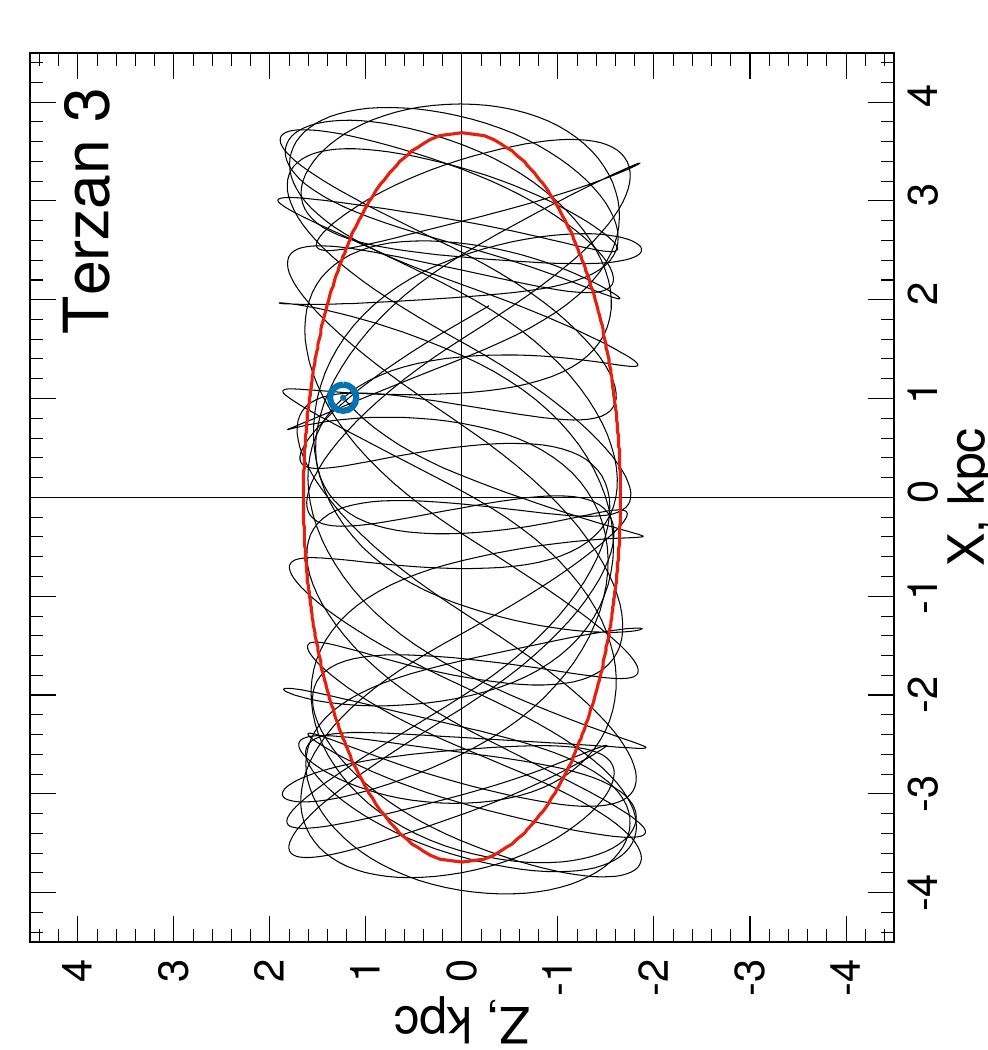}\
   \includegraphics[width=0.225\textwidth,angle=-90]{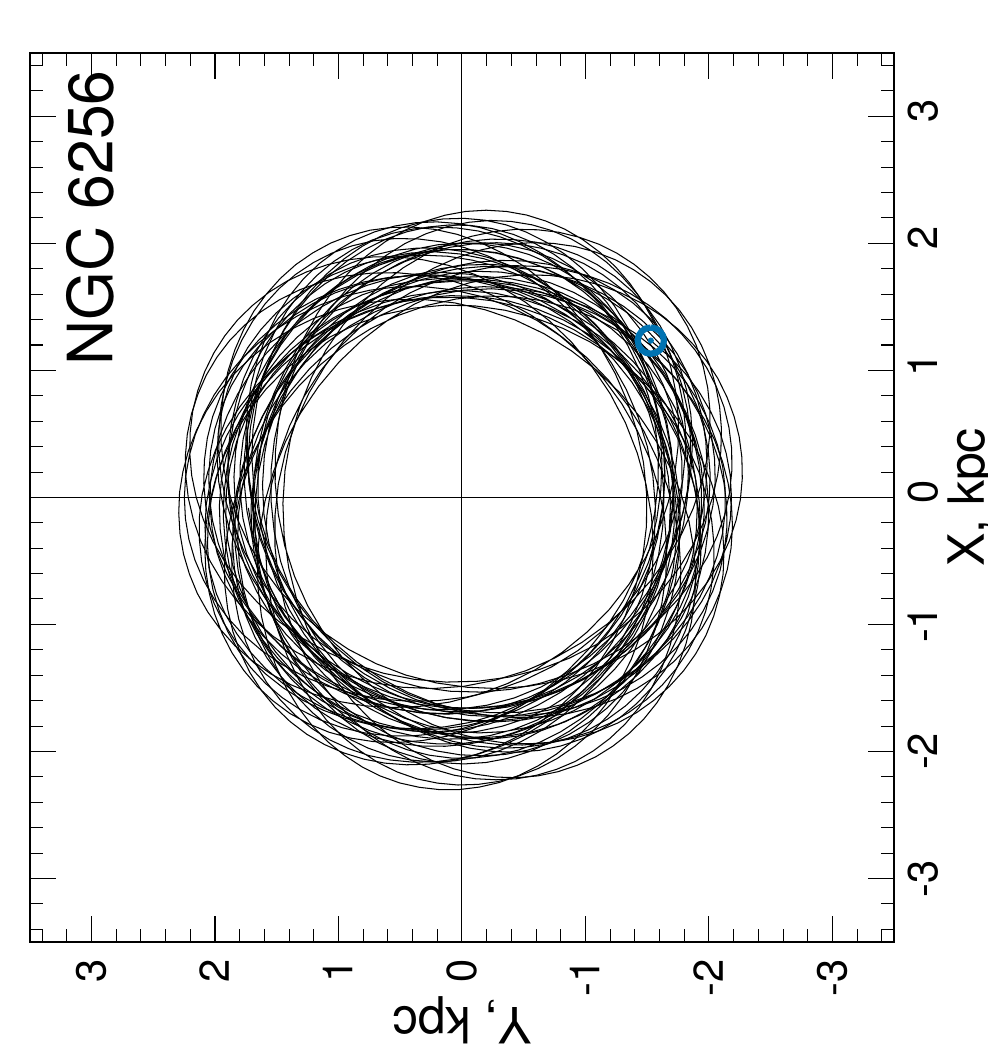}
     \includegraphics[width=0.225\textwidth,angle=-90]{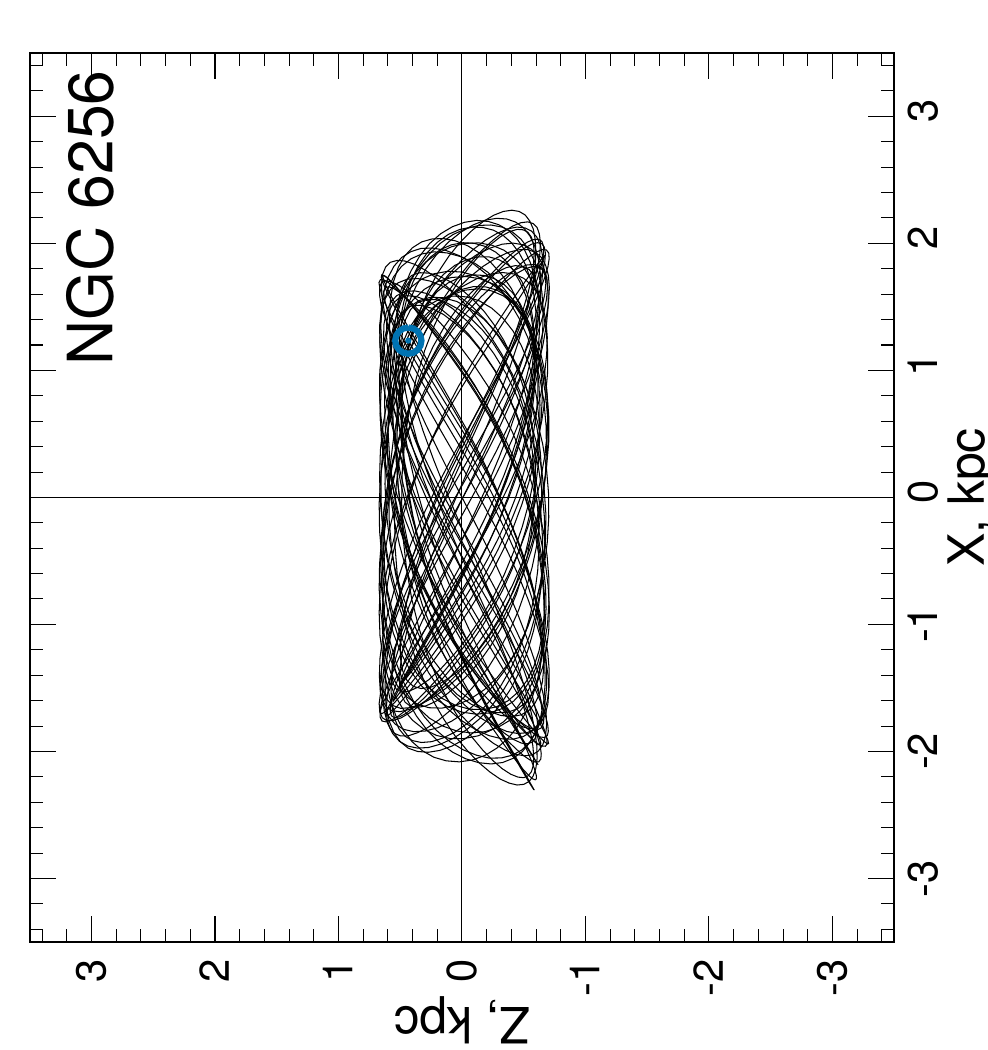}
        \includegraphics[width=0.225\textwidth,angle=-90]{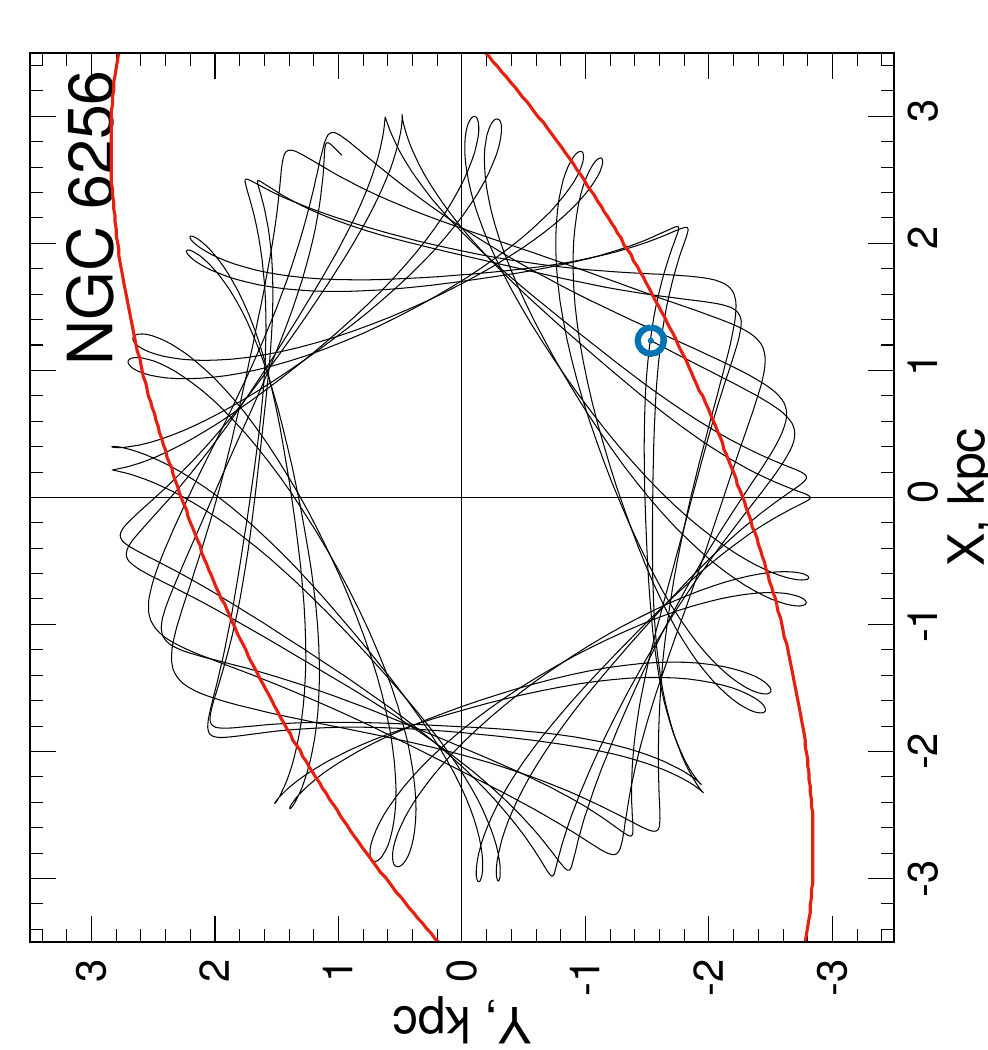}
     \includegraphics[width=0.225\textwidth,angle=-90]{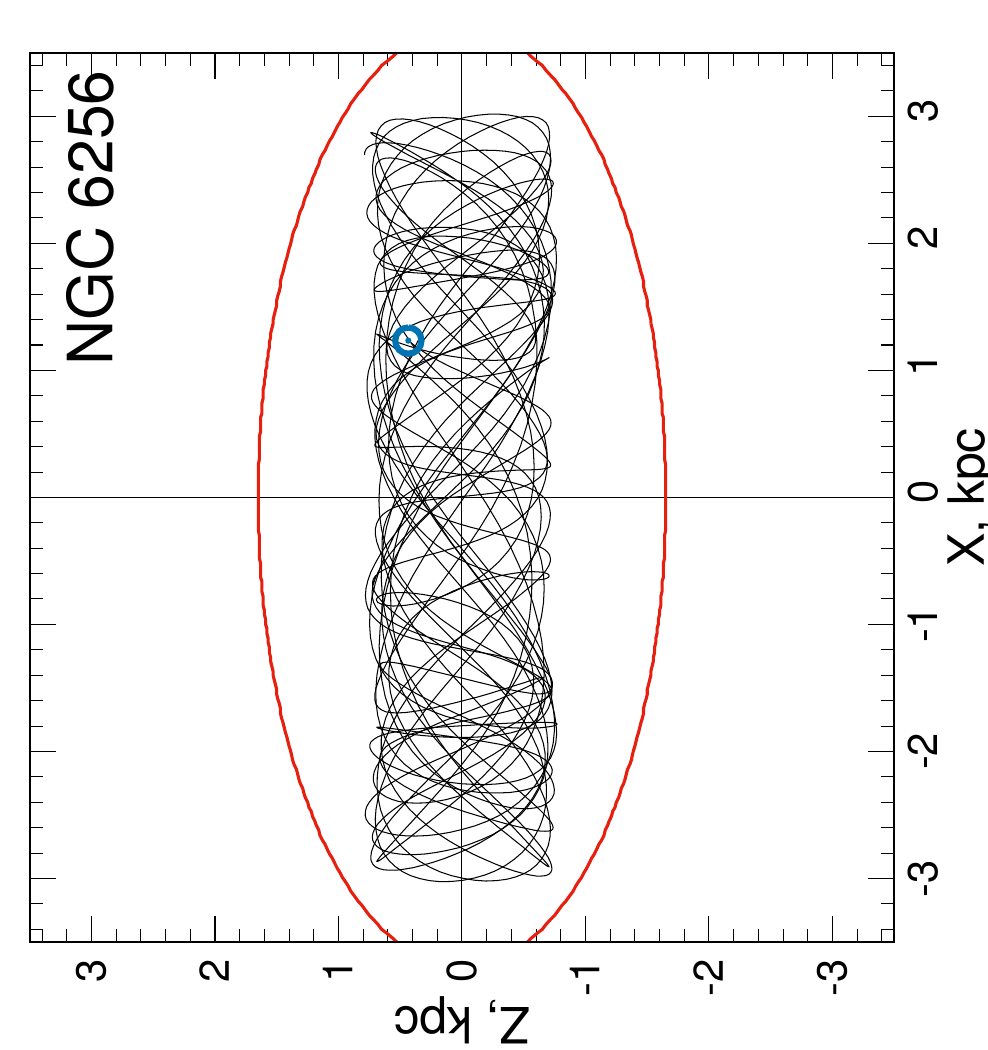}\
   \includegraphics[width=0.225\textwidth,angle=-90]{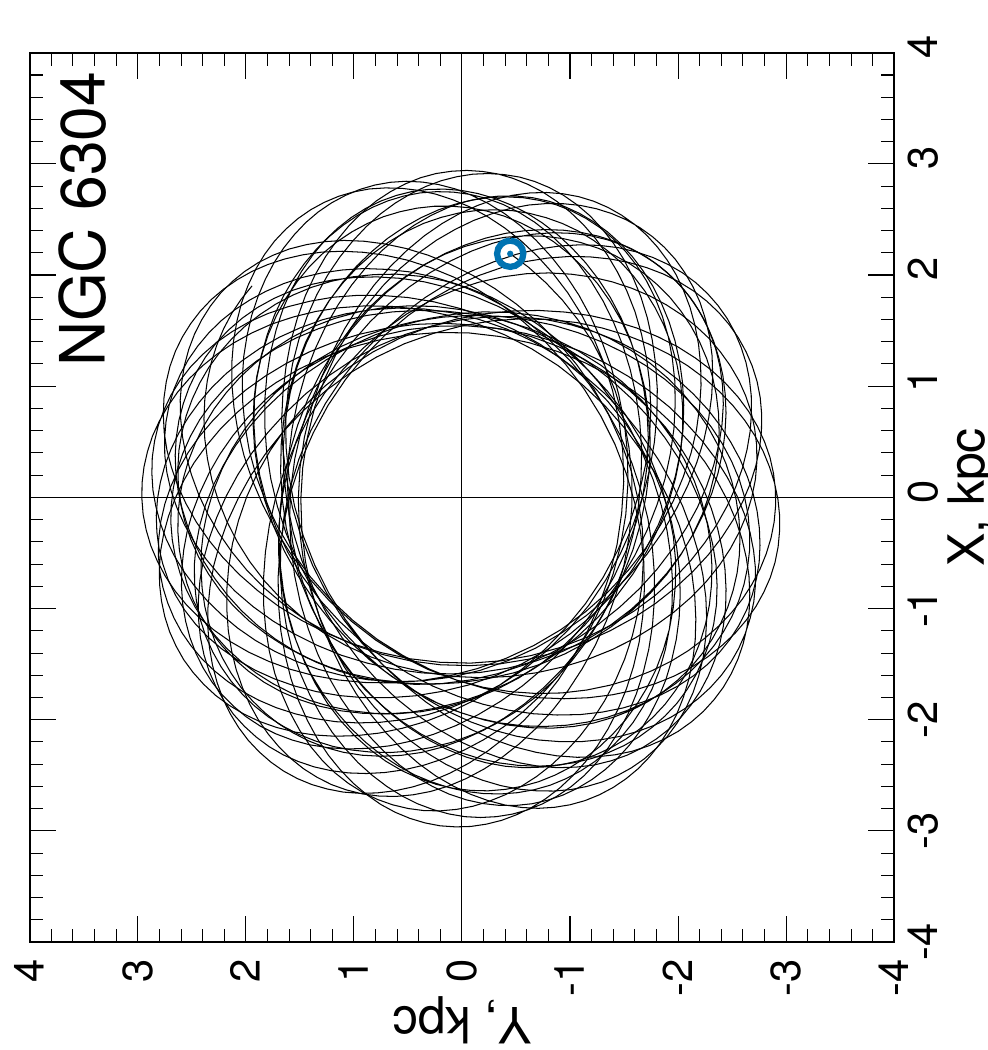}
     \includegraphics[width=0.225\textwidth,angle=-90]{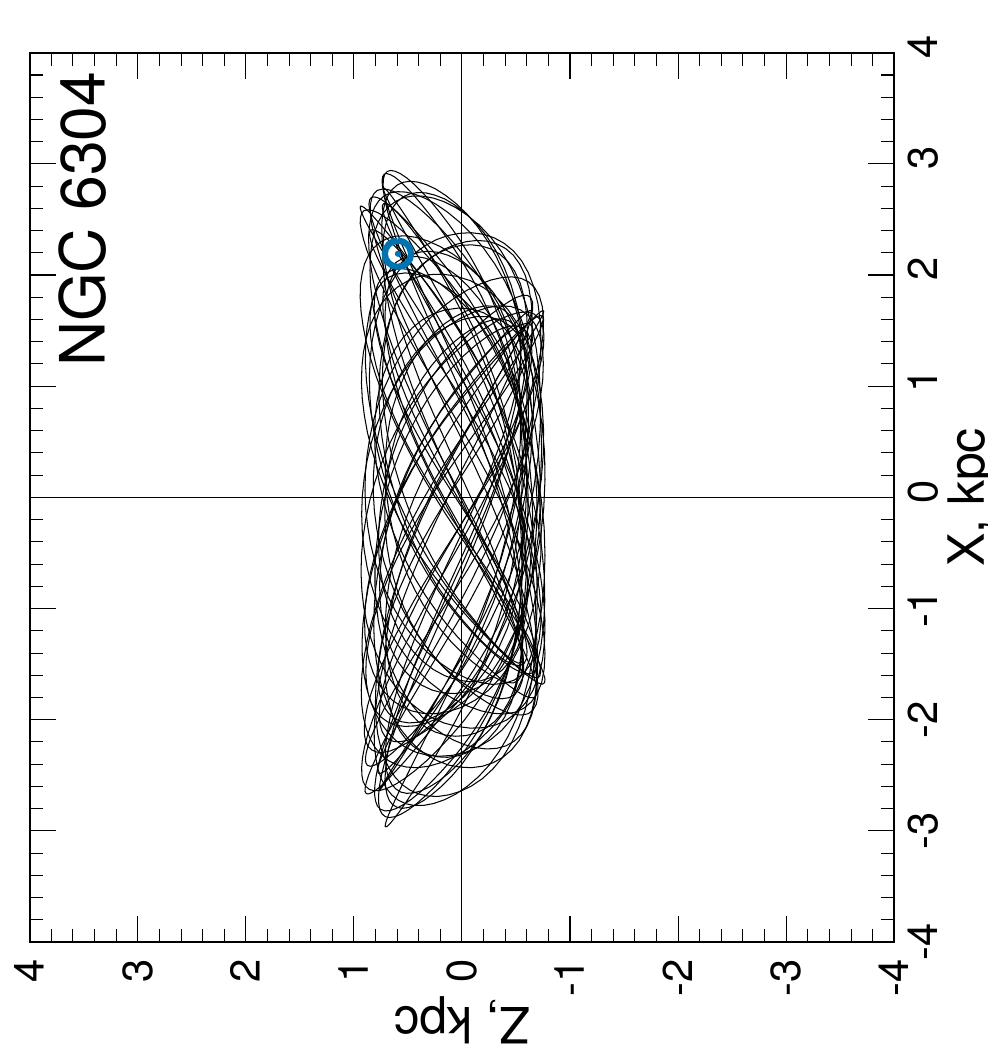}
   \includegraphics[width=0.225\textwidth,angle=-90]{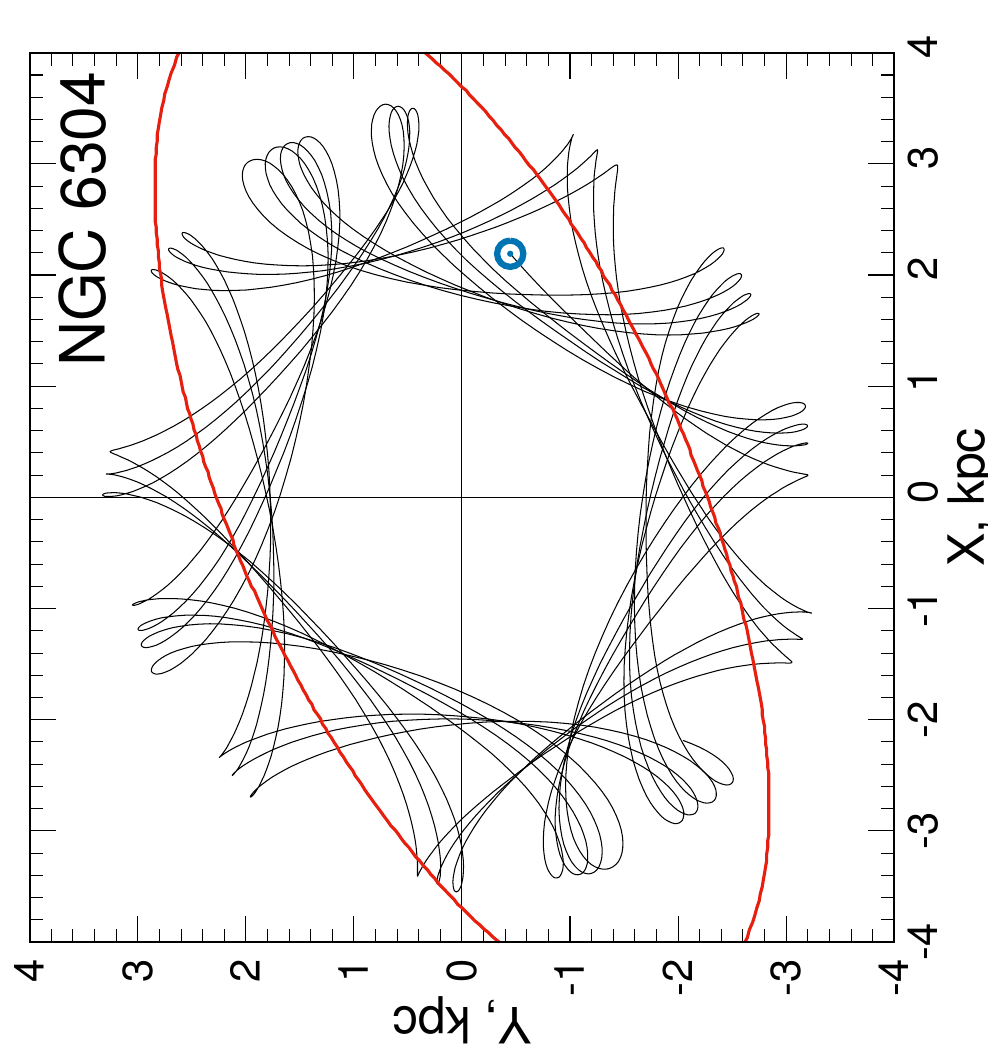}
     \includegraphics[width=0.225\textwidth,angle=-90]{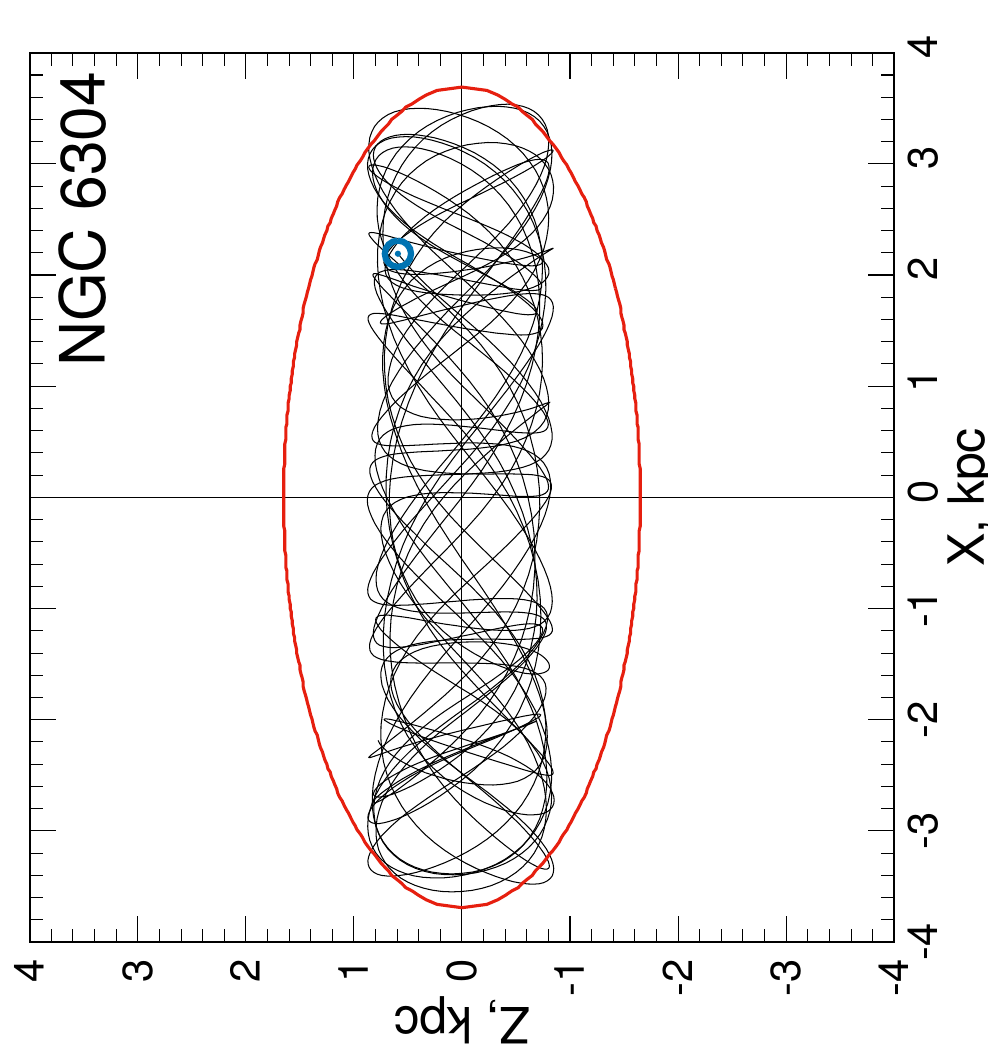}\
 \includegraphics[width=0.225\textwidth,angle=-90]{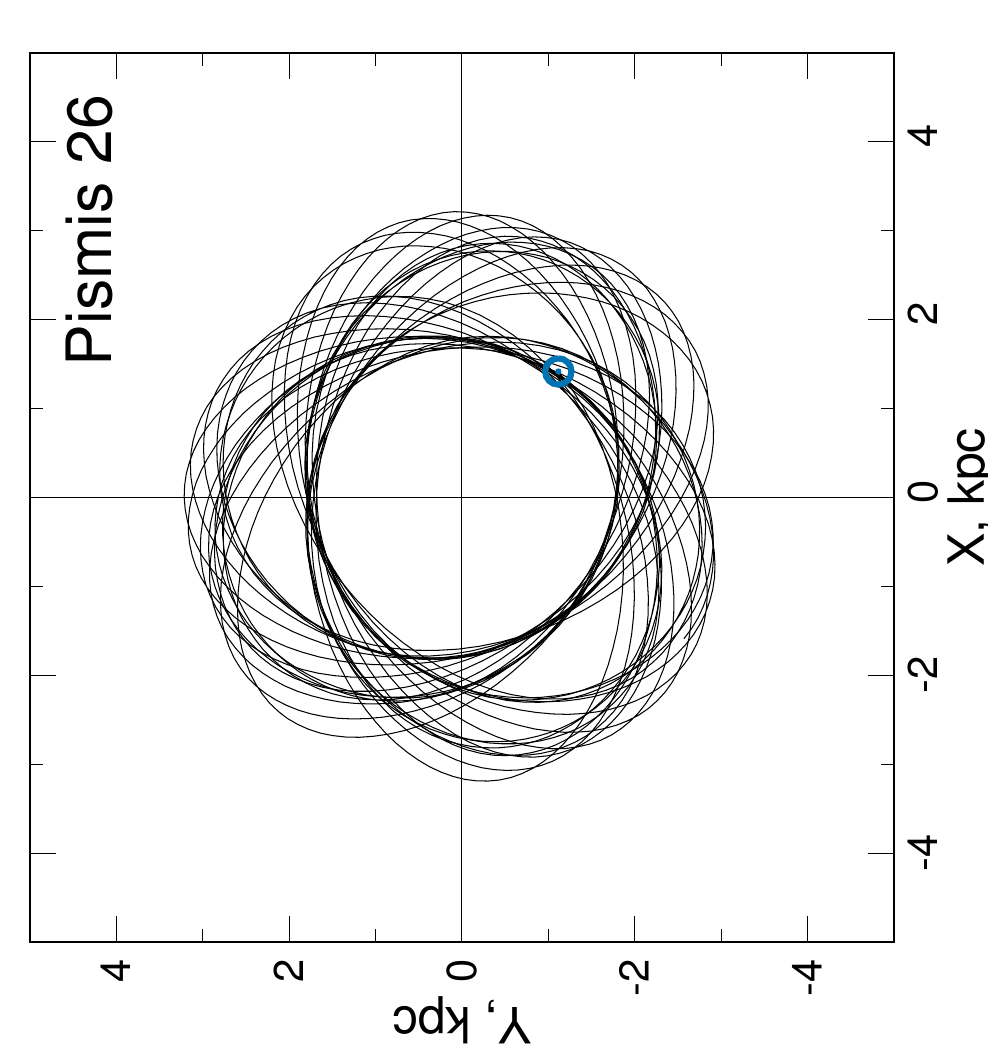}
     \includegraphics[width=0.225\textwidth,angle=-90]{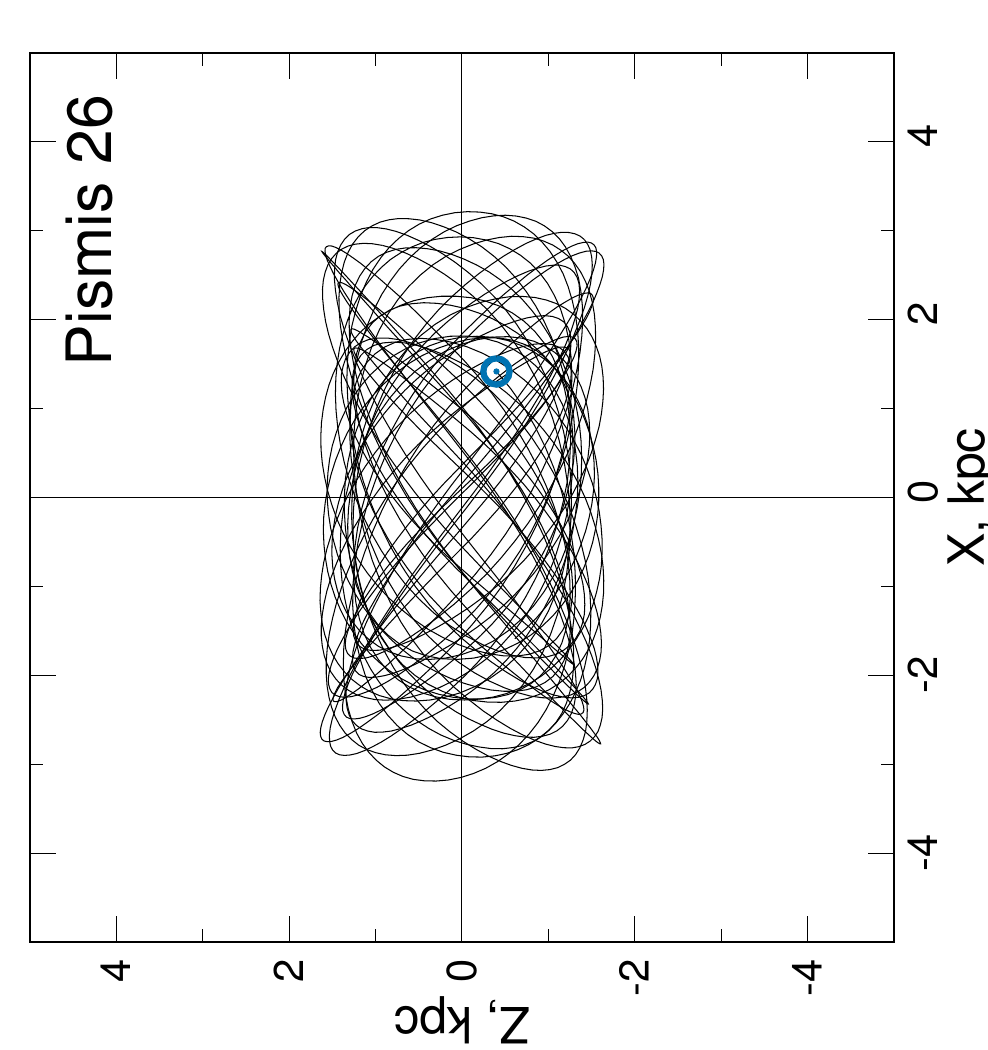}
      \includegraphics[width=0.225\textwidth,angle=-90]{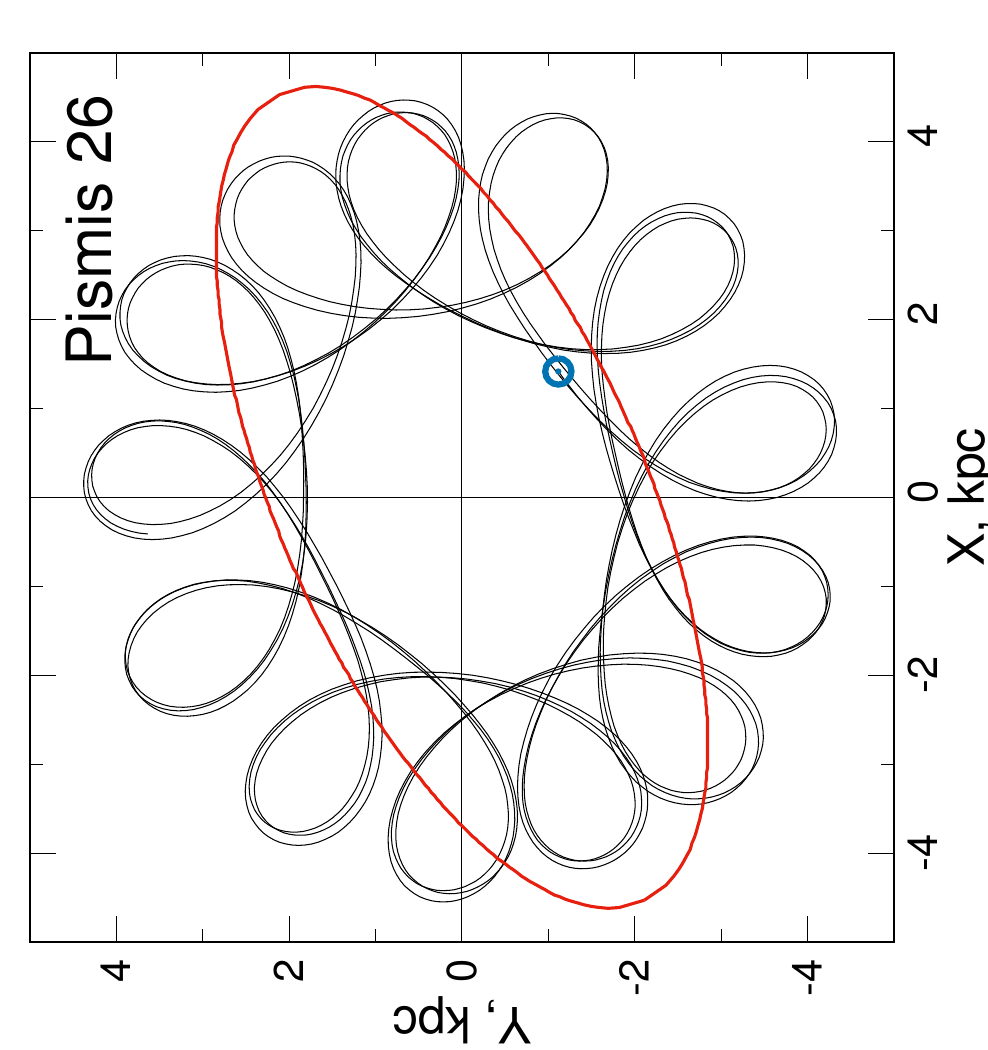}
     \includegraphics[width=0.225\textwidth,angle=-90]{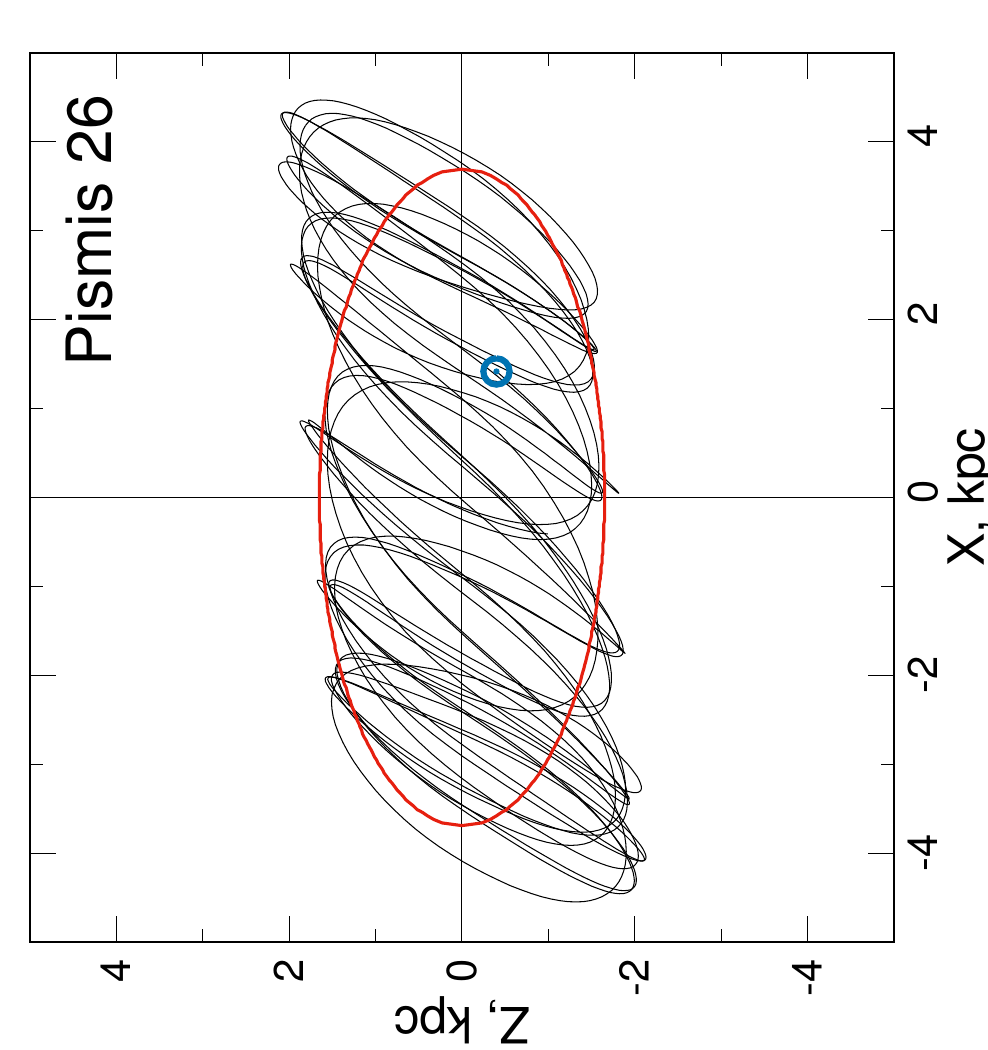}\
   \includegraphics[width=0.225\textwidth,angle=-90]{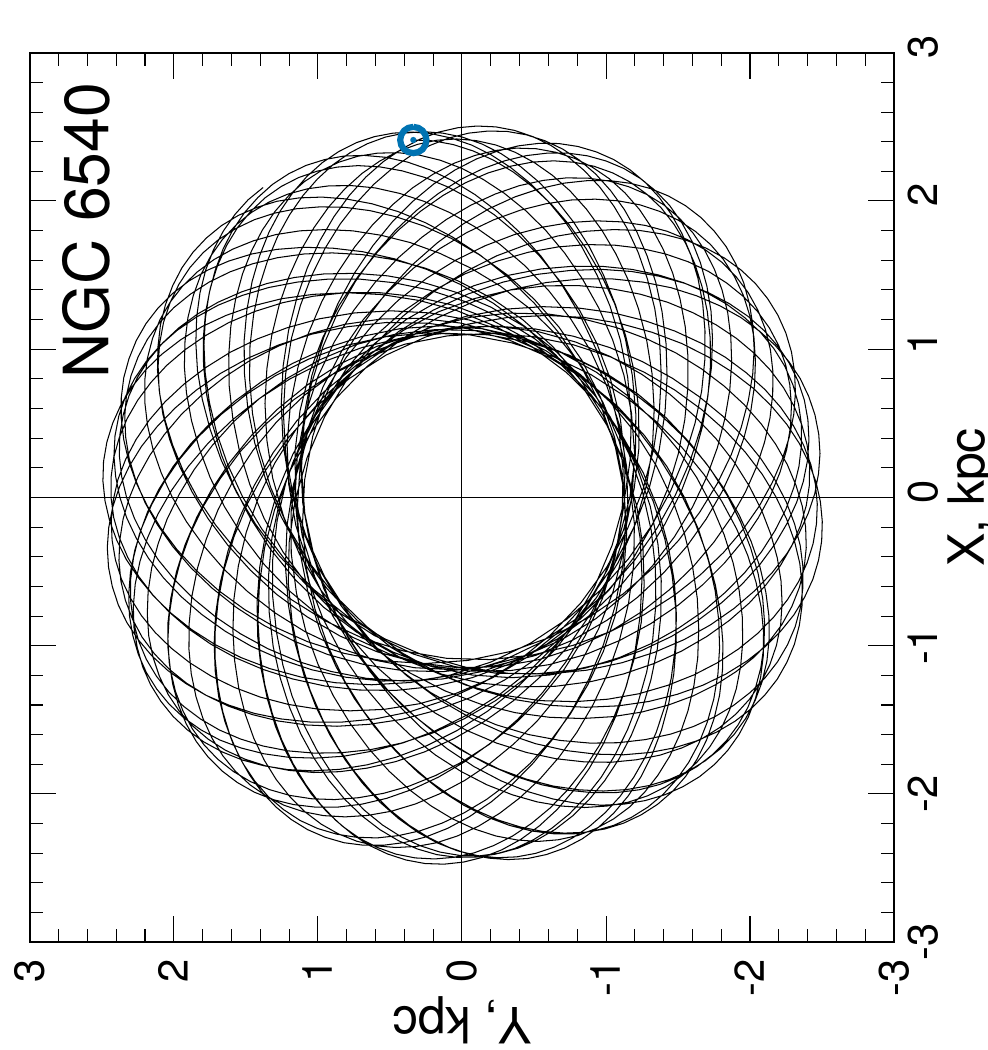}
     \includegraphics[width=0.225\textwidth,angle=-90]{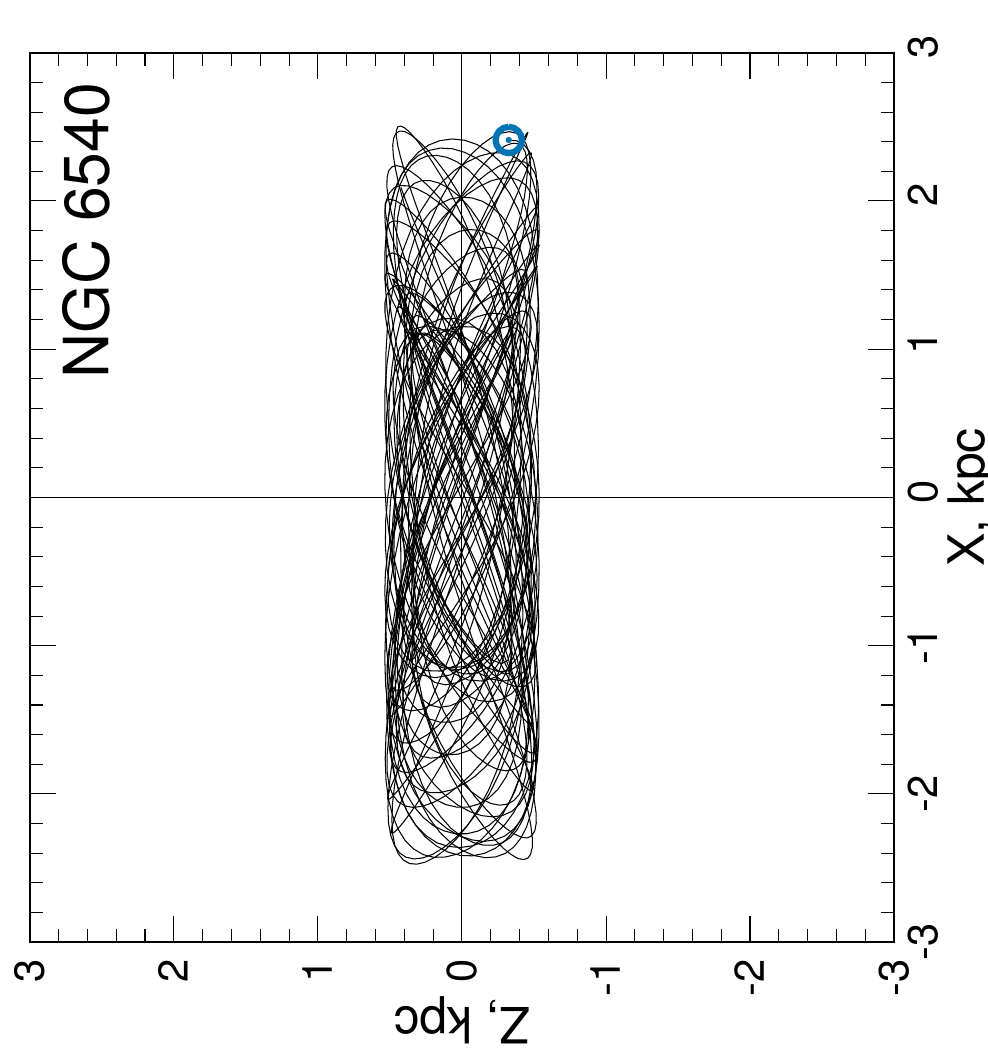}
    \includegraphics[width=0.225\textwidth,angle=-90]{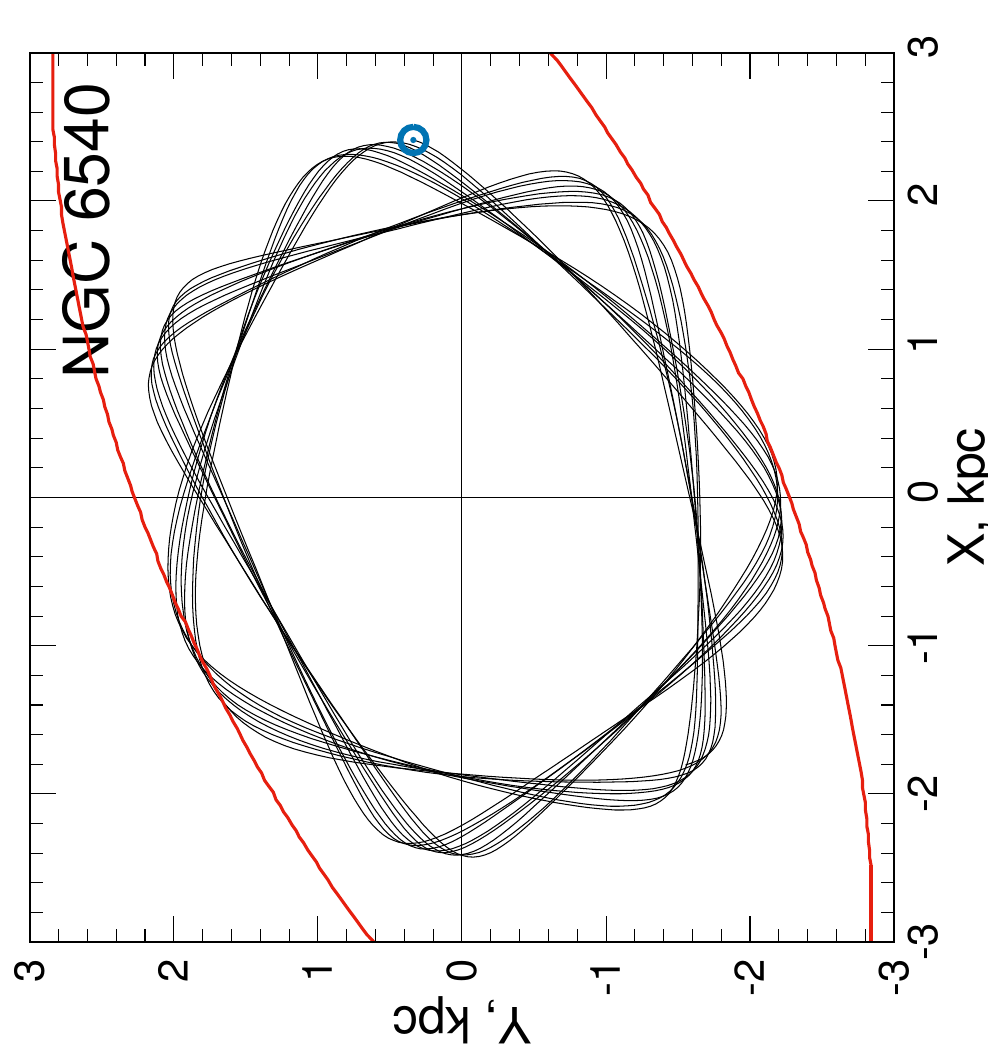}
     \includegraphics[width=0.225\textwidth,angle=-90]{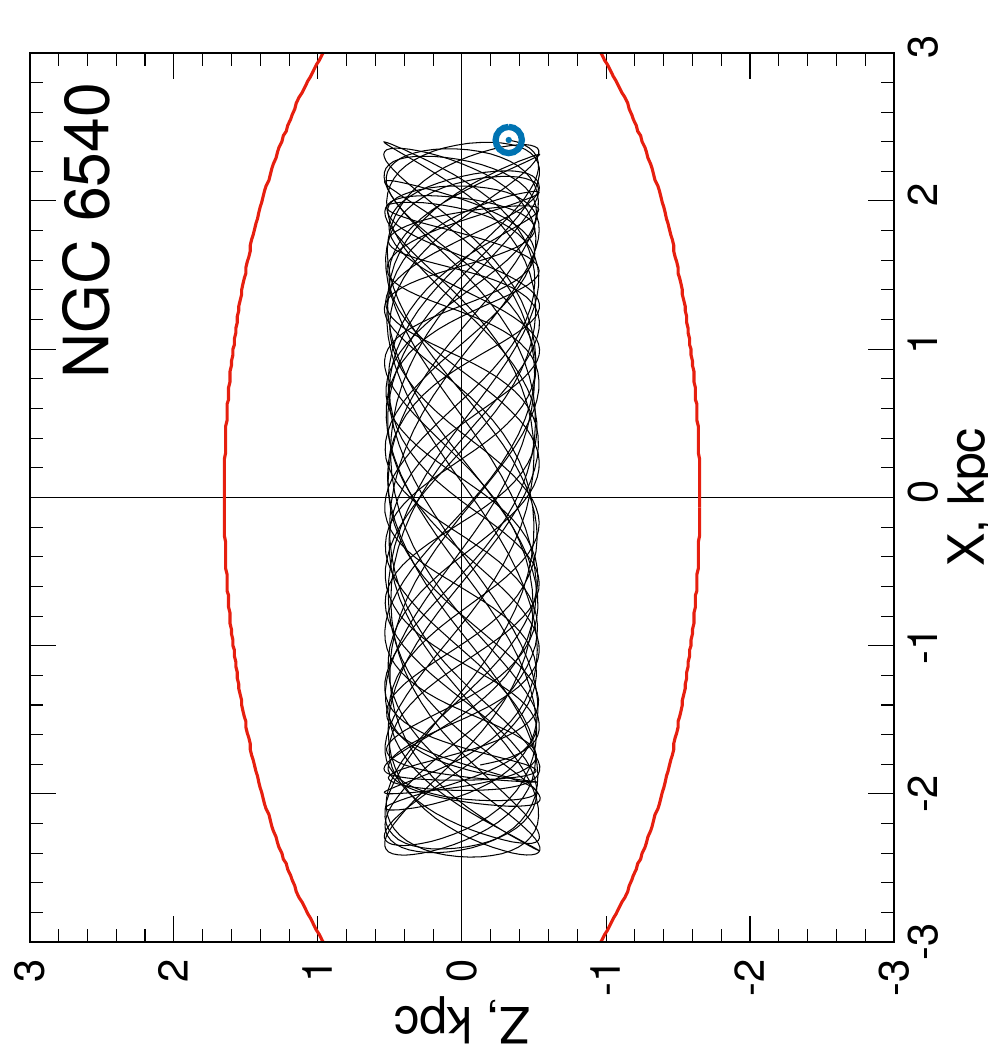}\

\medskip

 \centerline{APPENDIX. Continued}
\label{fB}
\end{center}}
\end{figure*}

\begin{figure*}
{\begin{center}
   \includegraphics[width=0.225\textwidth,angle=-90]{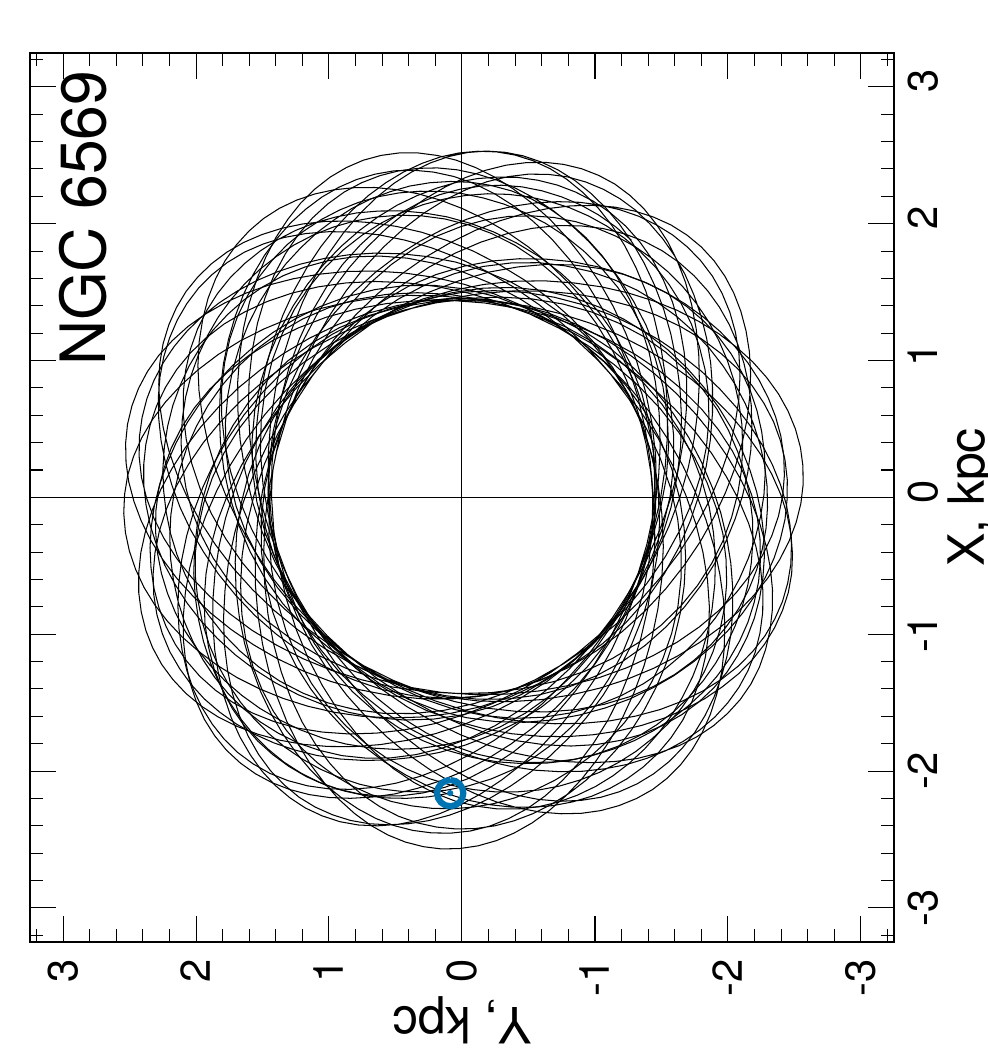}
     \includegraphics[width=0.225\textwidth,angle=-90]{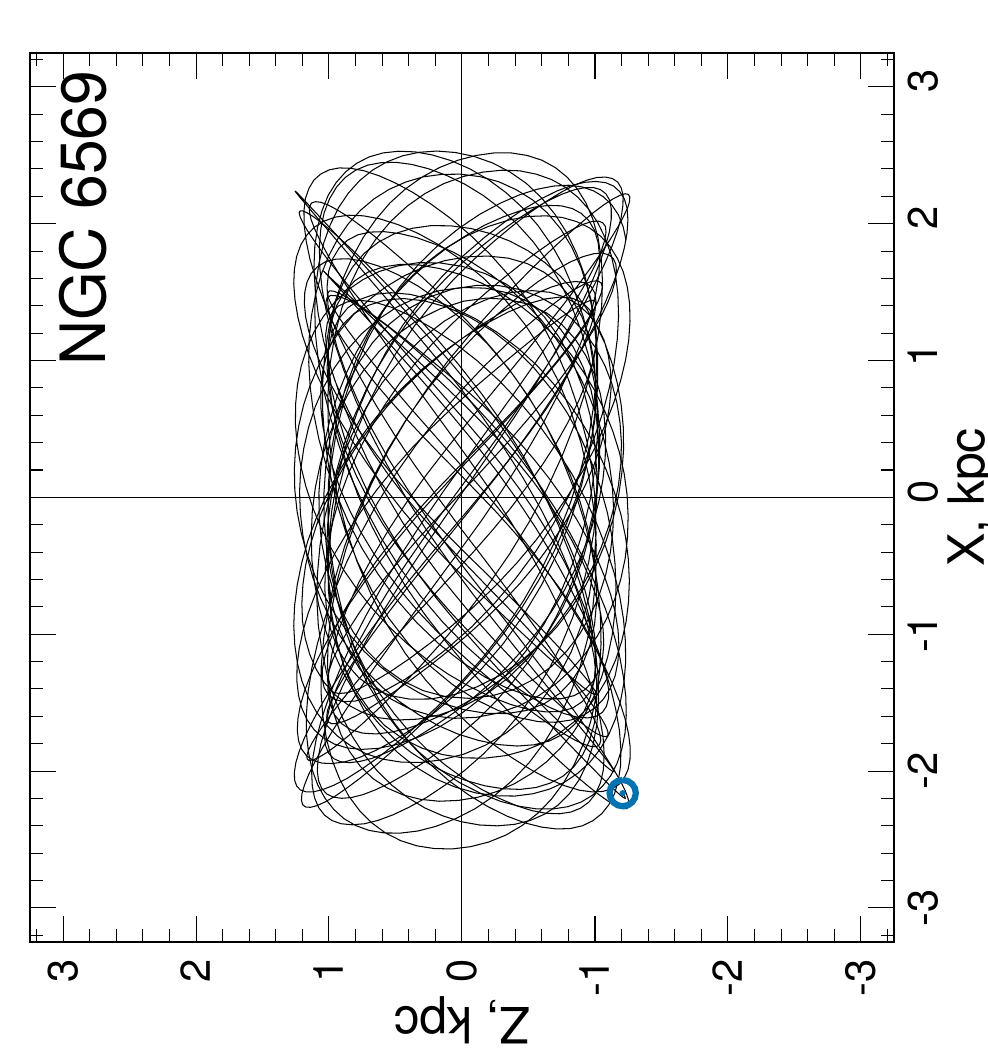}
        \includegraphics[width=0.225\textwidth,angle=-90]{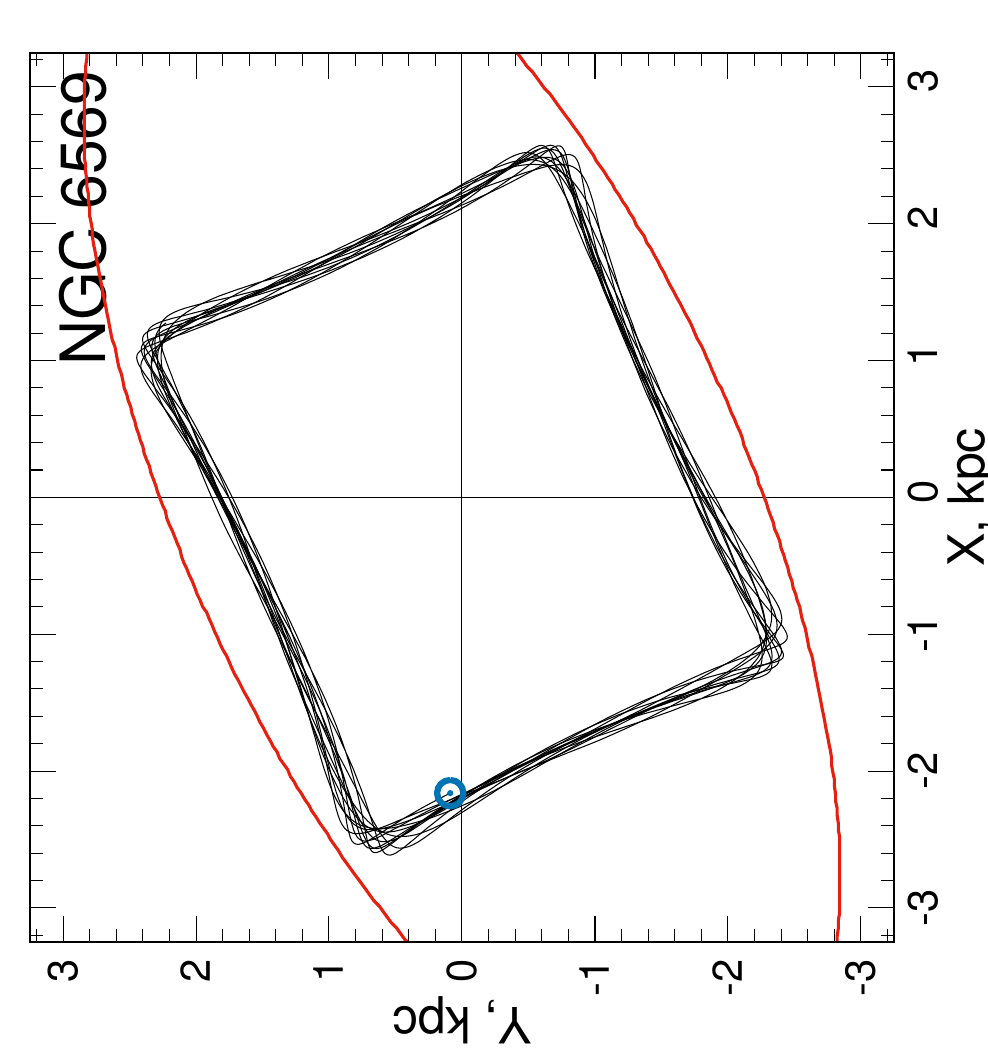}
     \includegraphics[width=0.225\textwidth,angle=-90]{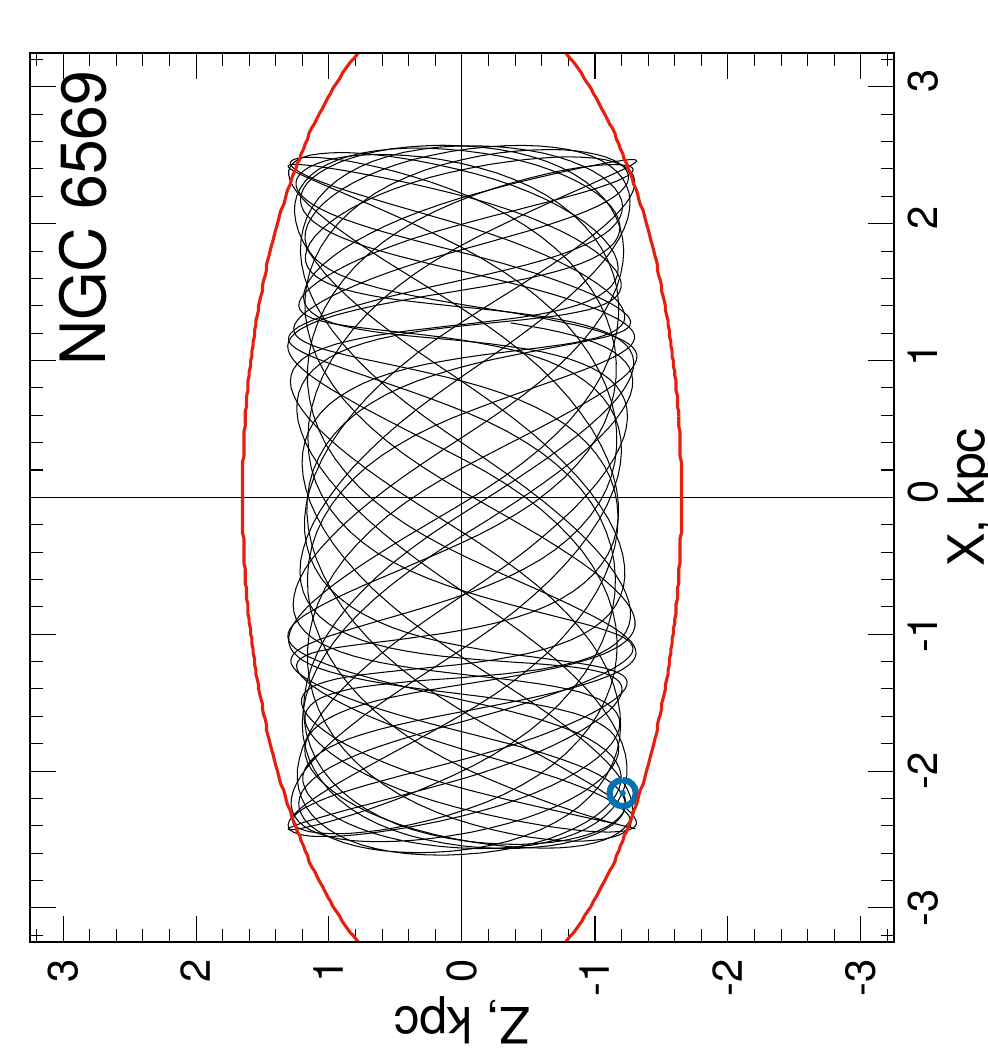}\
 \includegraphics[width=0.225\textwidth,angle=-90]{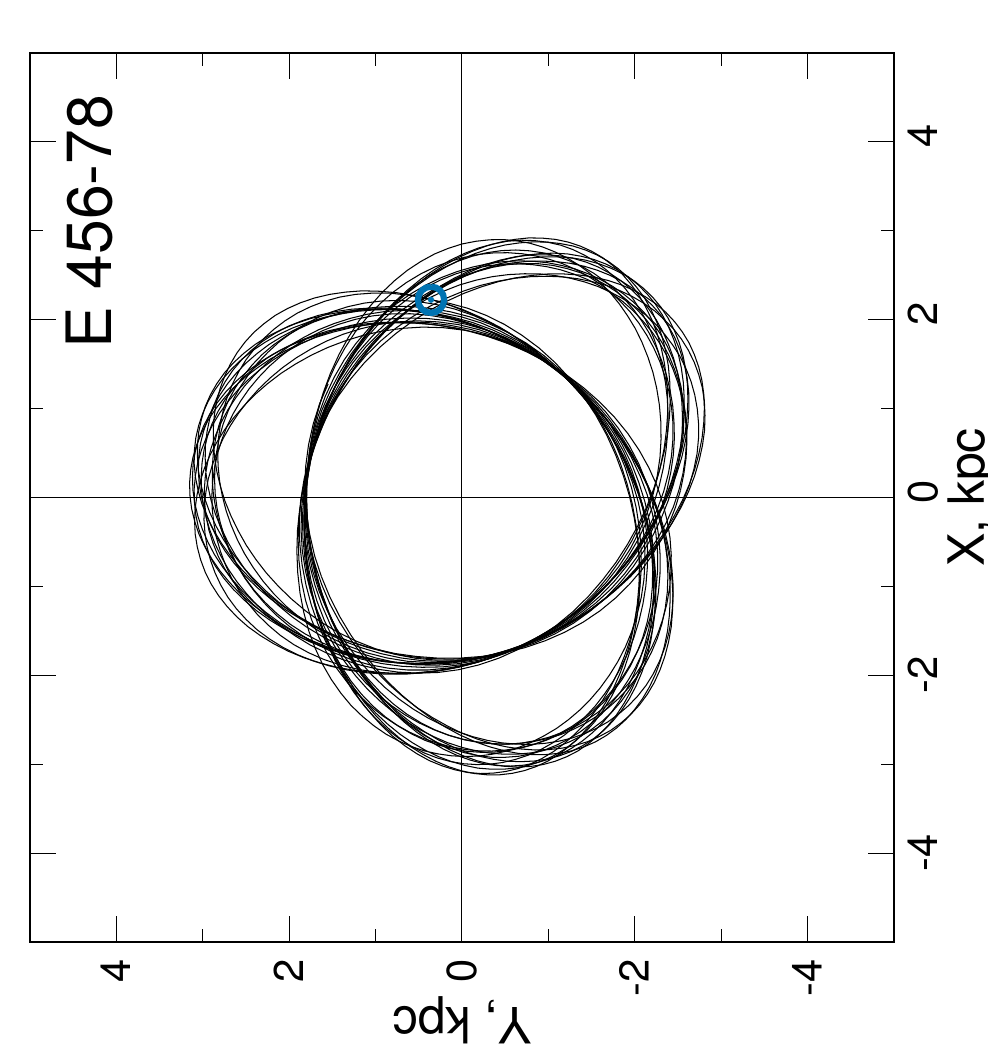}
     \includegraphics[width=0.225\textwidth,angle=-90]{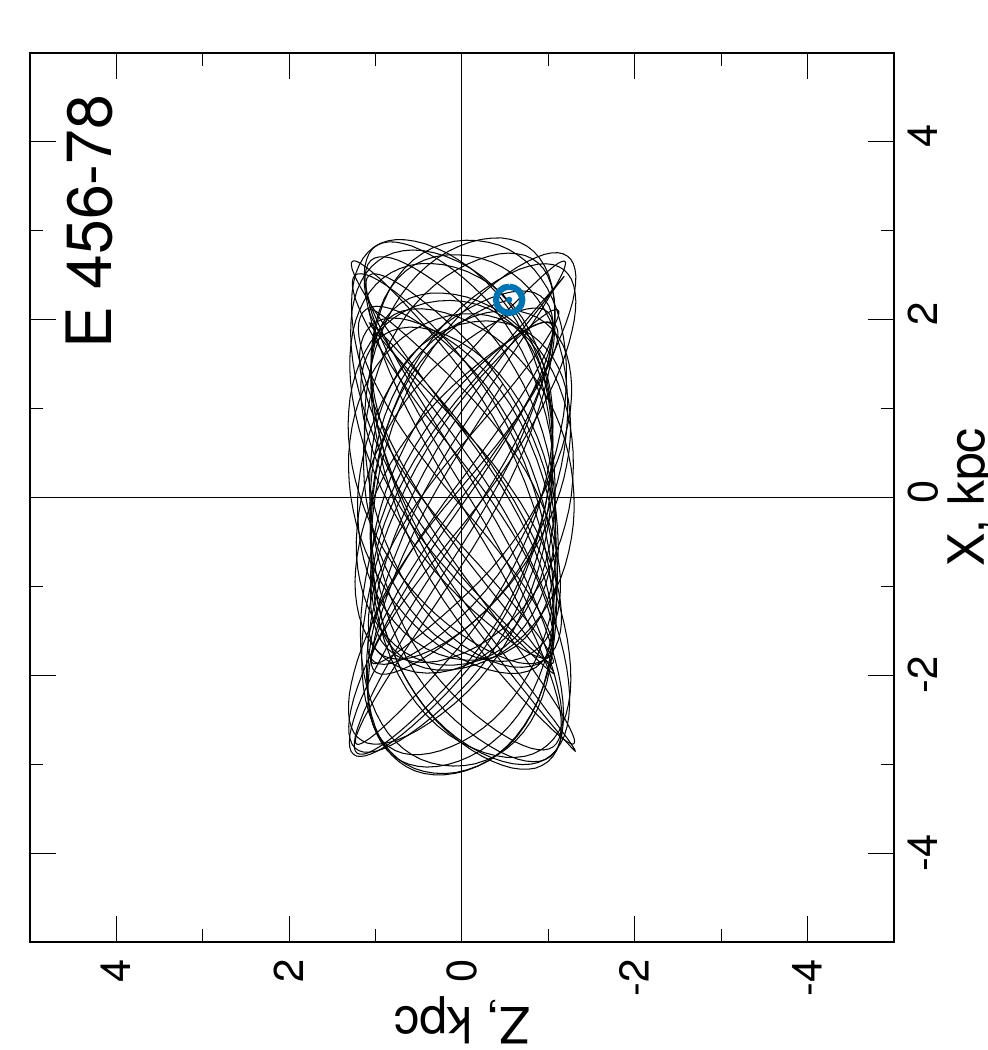}
 \includegraphics[width=0.225\textwidth,angle=-90]{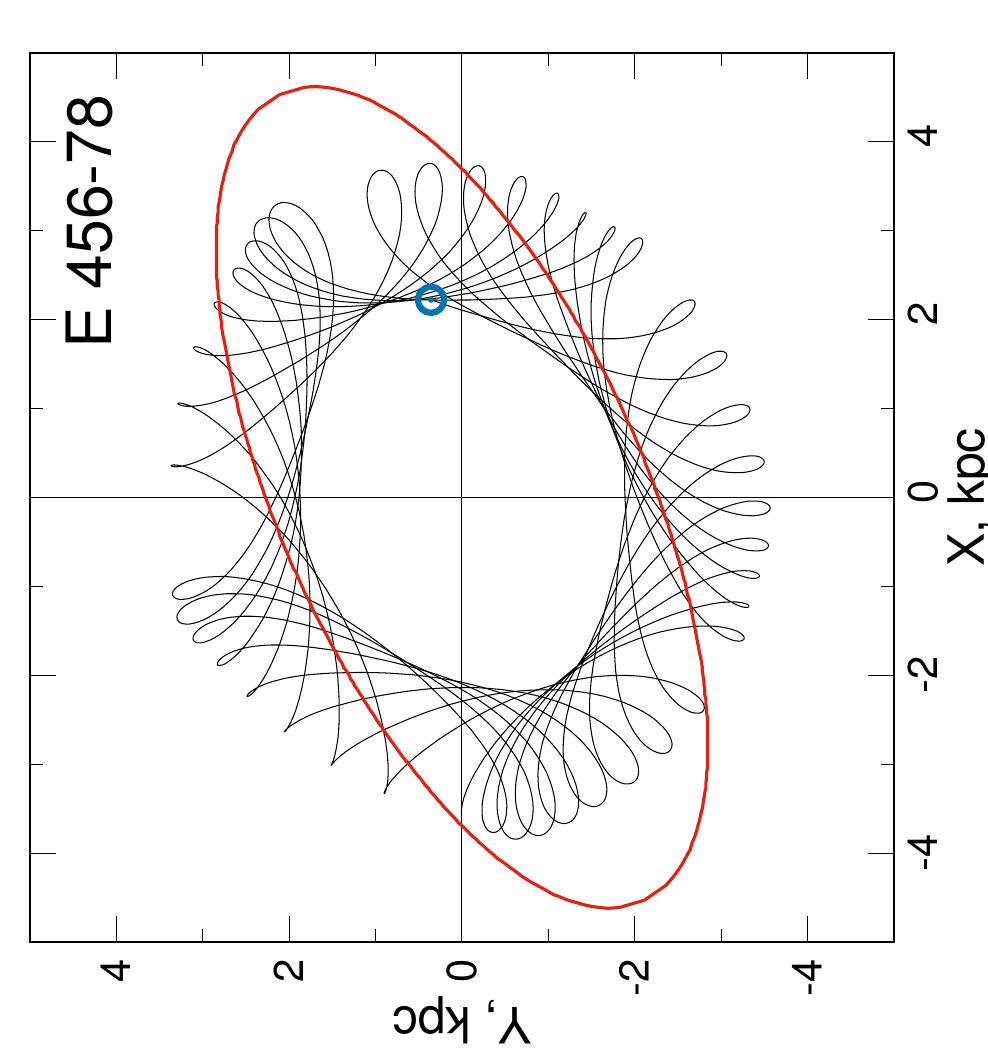}
     \includegraphics[width=0.225\textwidth,angle=-90]{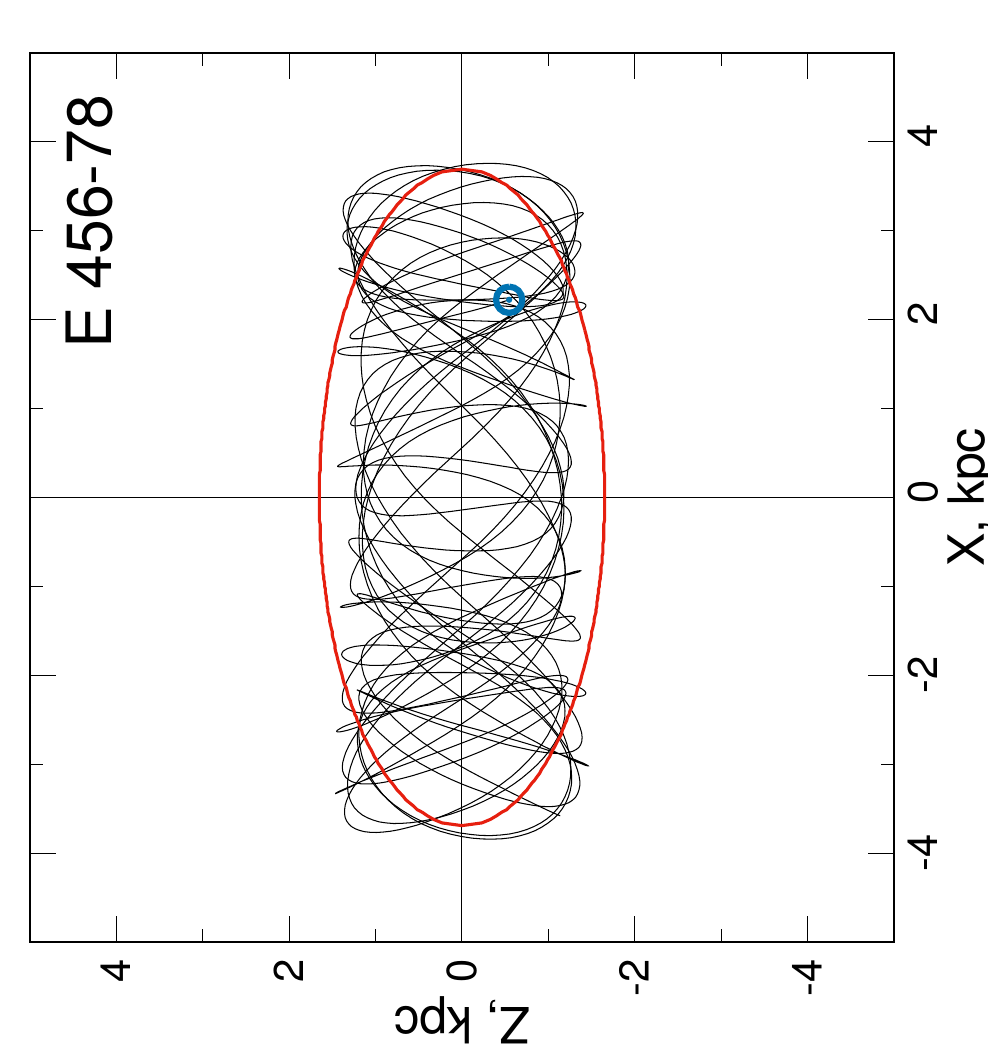}\
      \includegraphics[width=0.225\textwidth,angle=-90]{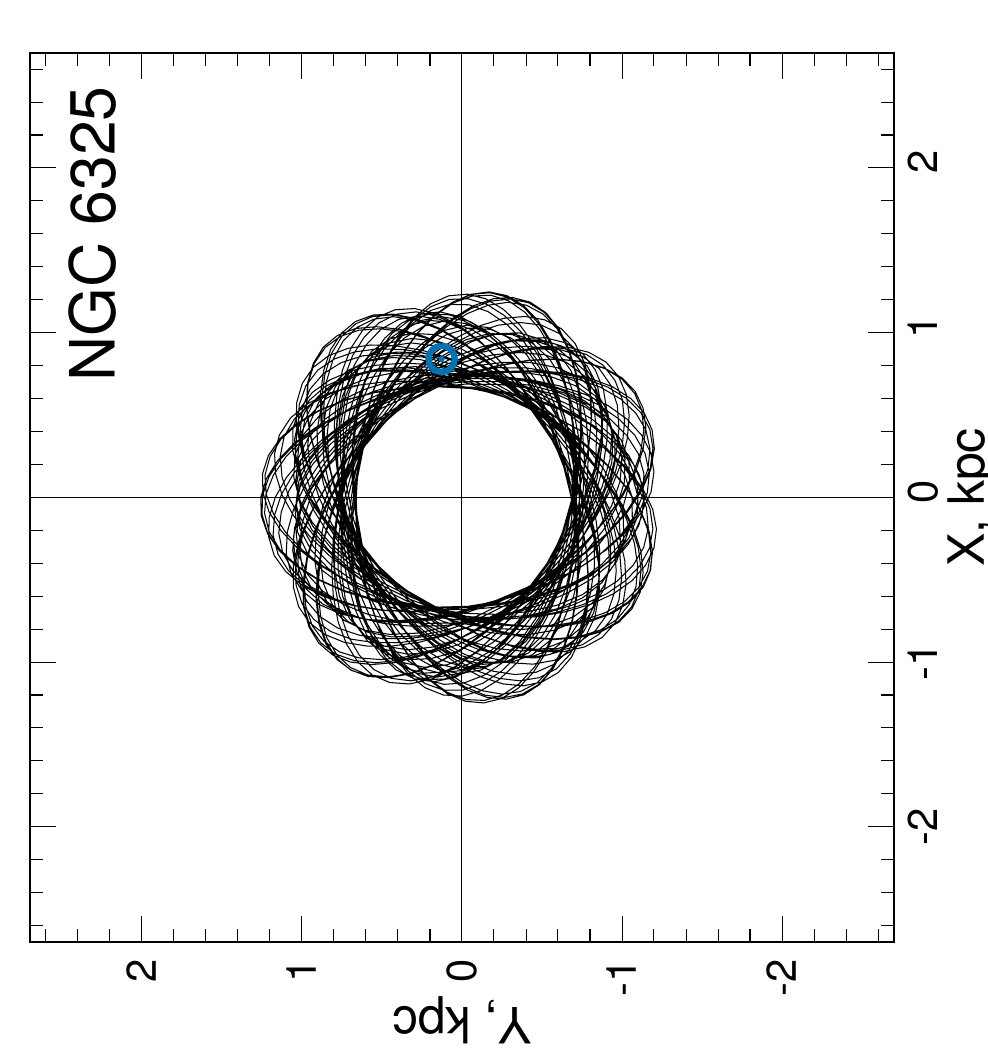}
     \includegraphics[width=0.225\textwidth,angle=-90]{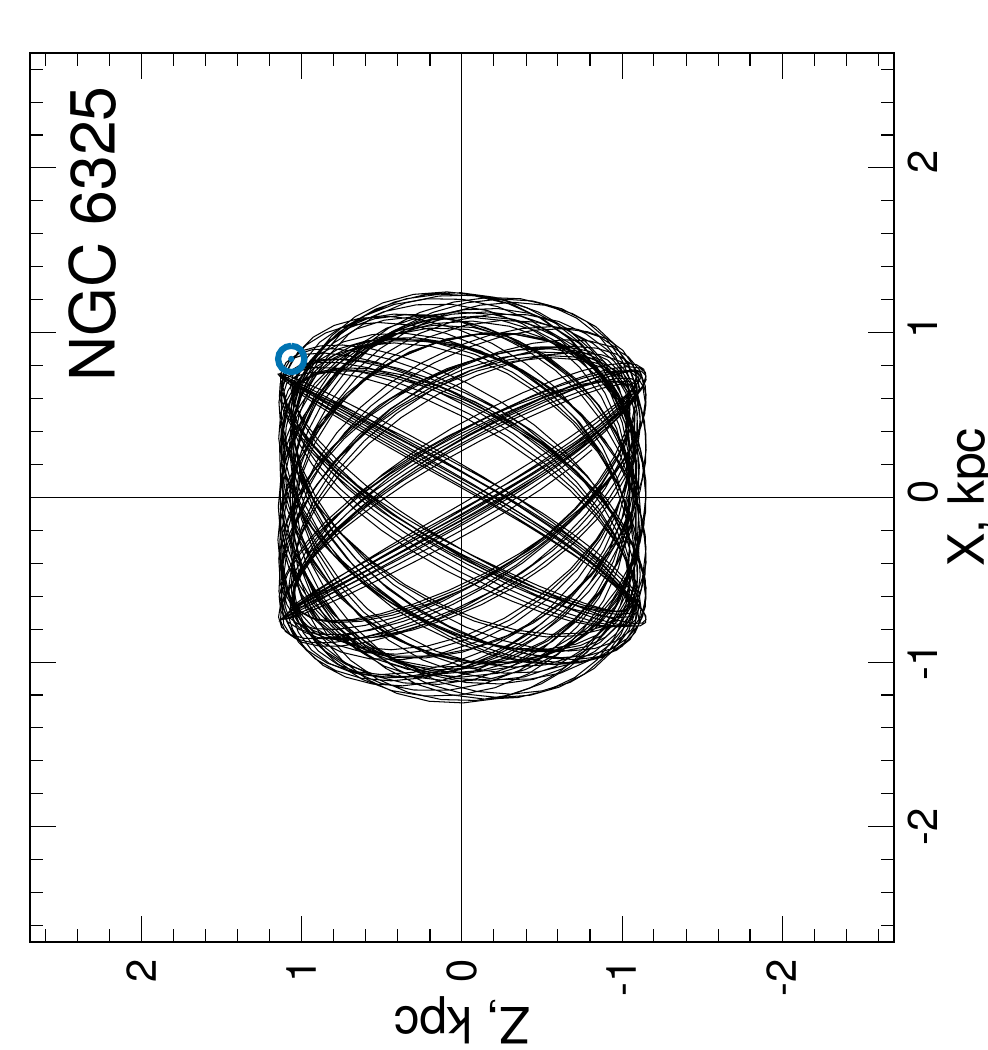}
     \includegraphics[width=0.225\textwidth,angle=-90]{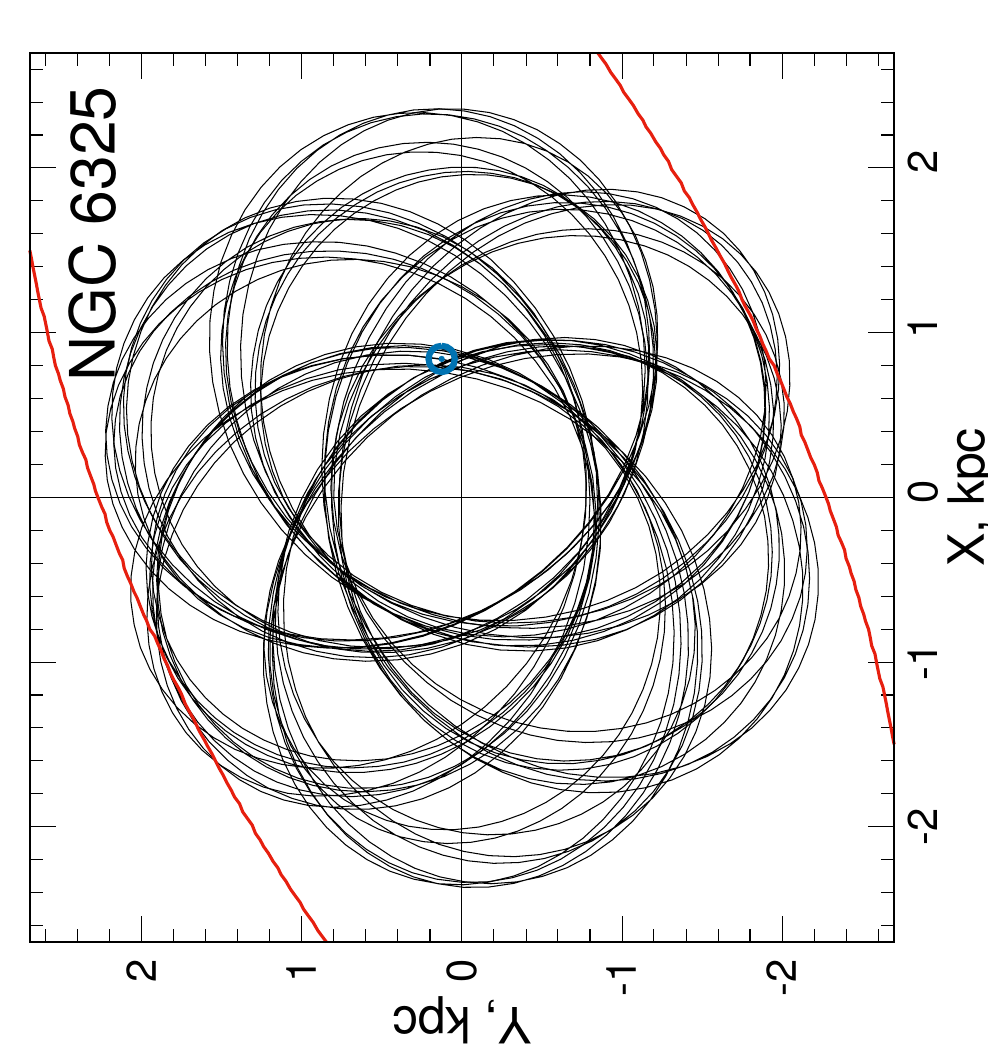}
     \includegraphics[width=0.225\textwidth,angle=-90]{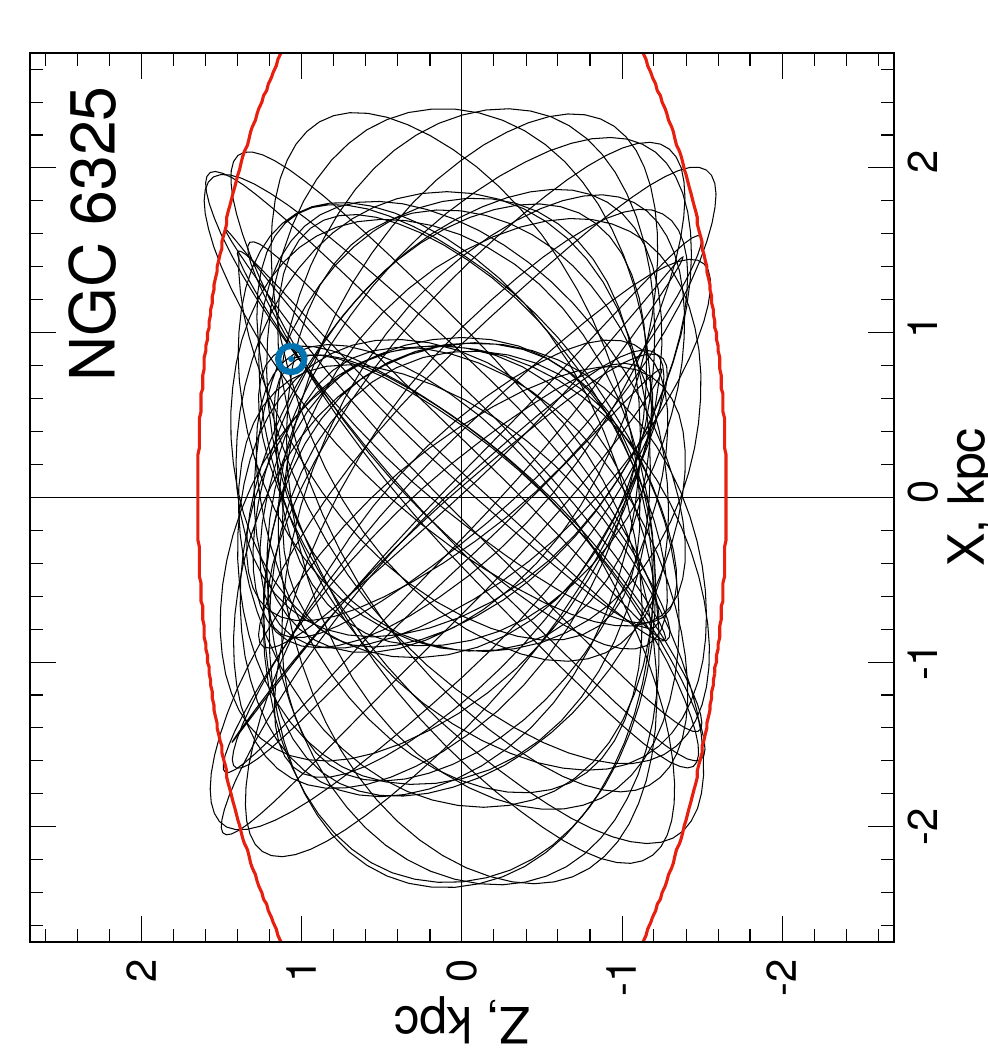}\
       \includegraphics[width=0.225\textwidth,angle=-90]{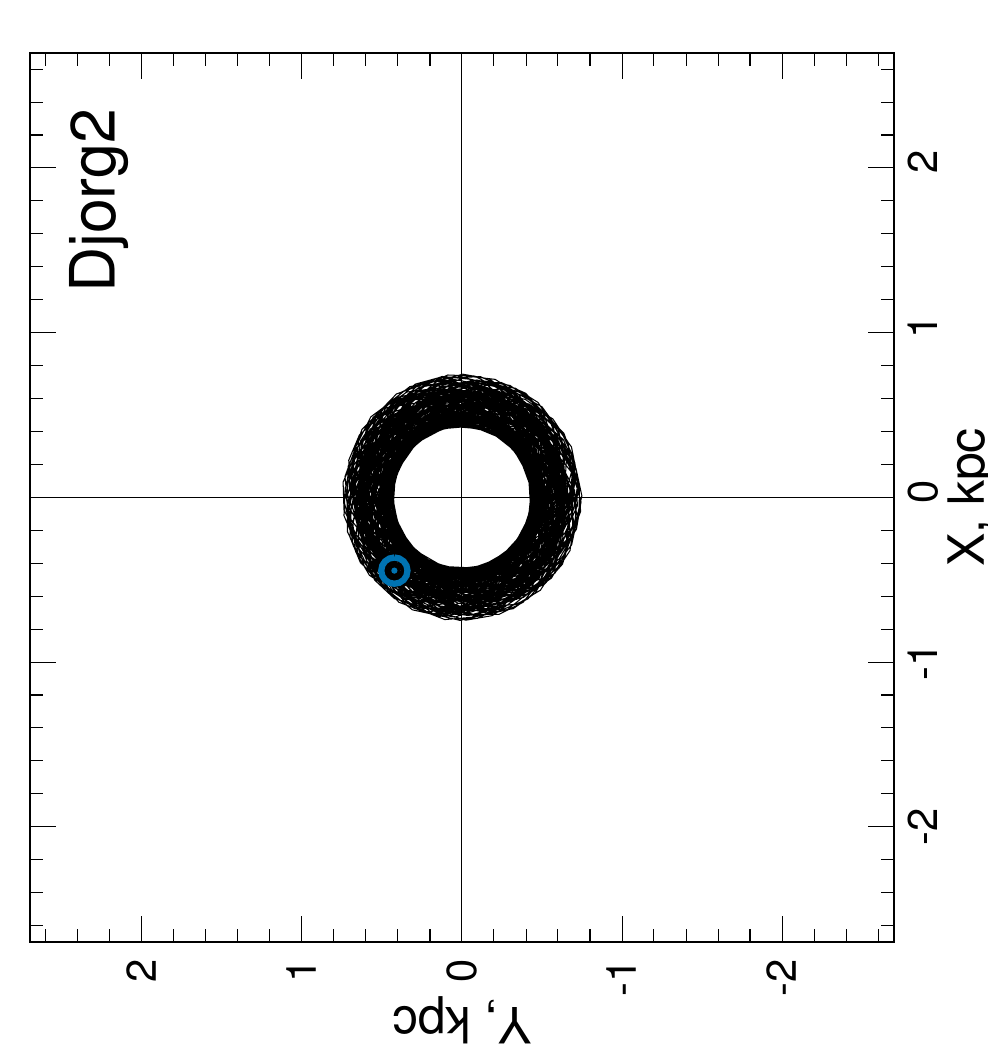}
     \includegraphics[width=0.225\textwidth,angle=-90]{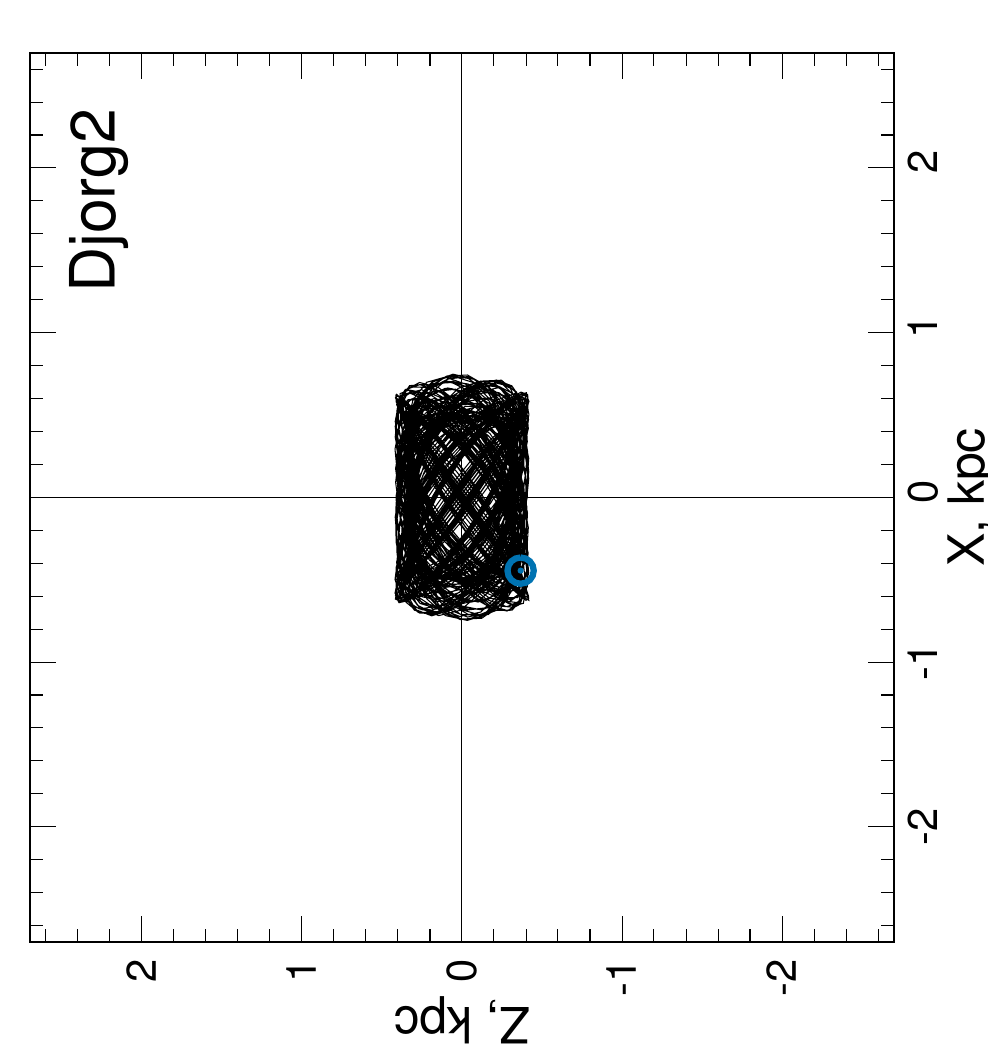}
     \includegraphics[width=0.225\textwidth,angle=-90]{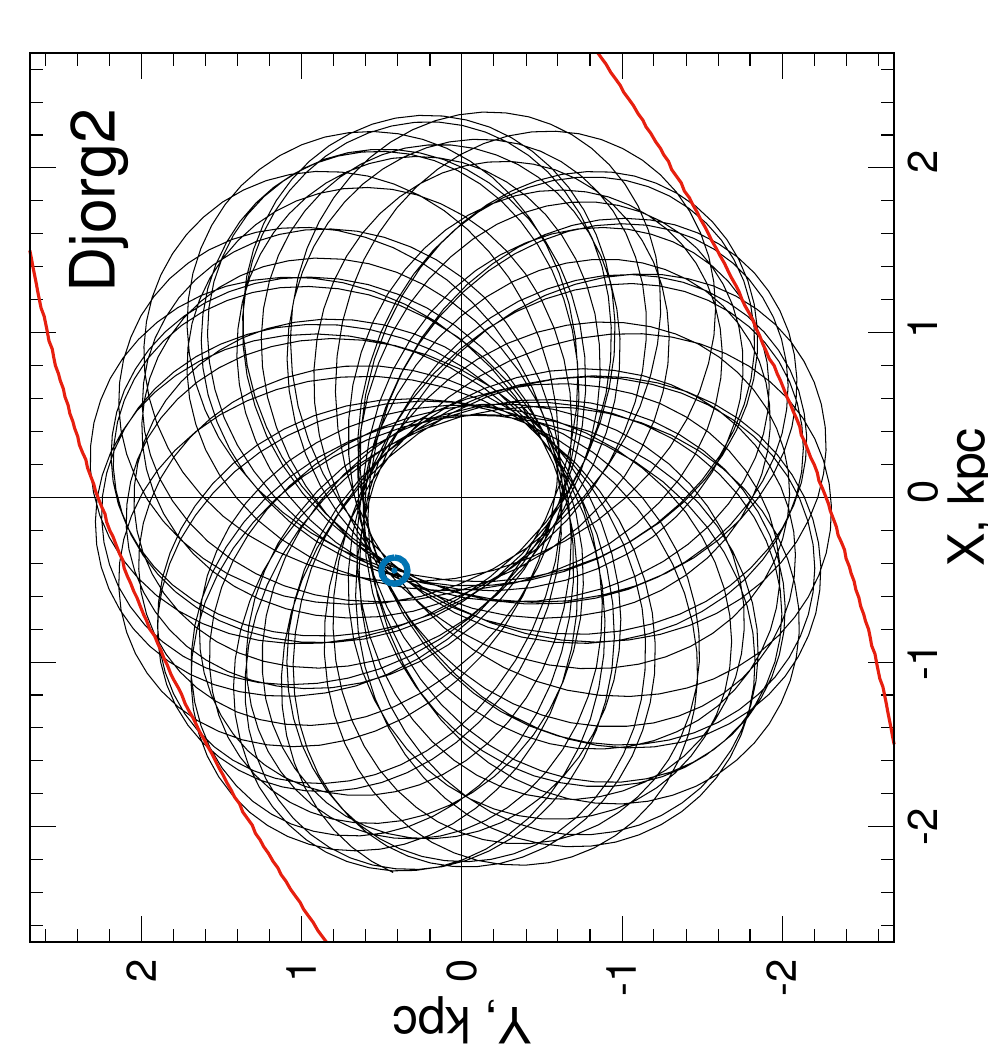}
     \includegraphics[width=0.225\textwidth,angle=-90]{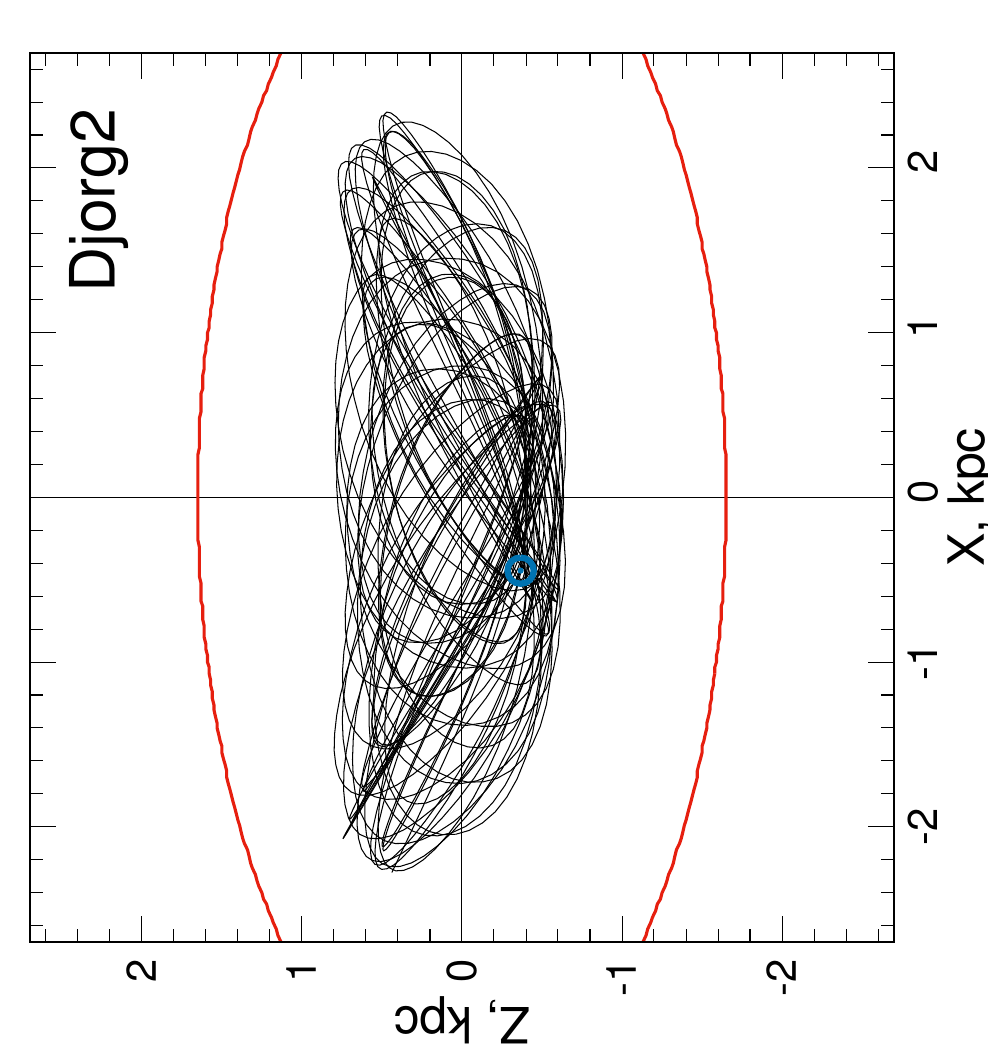}\

\medskip

 \centerline{APPENDIX. Continued}
\label{fB}
\end{center}}
\end{figure*}

\end{document}